\documentclass[12pt]{report}
\usepackage{etoolbox}
\usepackage[colorlinks=true,urlcolor=blue,citecolor=red,linkcolor=blue]{hyperref}
\usepackage[left=2.5cm,right=2.5cm,top=3cm,bottom=3cm]{geometry}
\usepackage{dcolumn,cite,nicefrac}
\usepackage{amsmath,amssymb,amsfonts,mathtools}
\usepackage{mathrsfs,bbold,bm}
\usepackage{graphicx}
\usepackage{epstopdf}
\usepackage{accents}
\usepackage{nicefrac}
\usepackage{pdfpages,setspace}
\newlength{\dhatheight}
\usepackage{wrapfig}
\usepackage{flushend,BOONDOX-cal,BOONDOX-frak}

\makeatletter
\newsavebox{\@brx}
\newcommand{\llangle}[1][]{\savebox{\@brx}{\(\m@th{#1\langle}\)}%
  \mathopen{\copy\@brx\kern-0.5\wd\@brx\usebox{\@brx}}}
\newcommand{\rrangle}[1][]{\savebox{\@brx}{\(\m@th{#1\rangle}\)}%
  \mathclose{\copy\@brx\kern-0.5\wd\@brx\usebox{\@brx}}}
\makeatother
\newcommand{\doublehat}[1]{%
    \settoheight{\dhatheight}{\ensuremath{\hat{#1}}}%
    \addtolength{\dhatheight}{-0.35ex}%
    \hat{\vphantom{\rule{1pt}{\dhatheight}}%
    \smash{\hat{#1}}}}

\begin{document}
\begin{titlepage}
\begin{center}
\LARGE{\textbf{Signatures Of Quantum Gravity In Relativistic Quantum Systems}}\\
\vspace{5cm}
\large{\textbf{Thesis submitted for the degree of}}\\
\vspace{0.5cm}
\large{\textbf{Doctor of Philosophy (Science)}}\\
\vspace{0.5cm}
\large{\textbf{in Physics (Theoretical)}}\\
\vspace{7cm}
\textbf{\large{by}}\\
\vspace{0.5cm}
\large{\textbf{Soham Sen}}\\
%\vspace{0.5cm}
%\large{\textbf{Supervisor: Prof Sunandan Gangopadhyay}}\\
\vspace{0.5cm}
%\large{Department of Astrophysics and High Energy Physics}\\
\textbf{\large{Department of Physics}}\\
%\vspace{0.5cm}
%\large{S. N. Bose National Centre for Basic Sciences}\\
\vspace{0.5cm}
\large{\textbf{University of Calcutta}}\\
\vspace{0.5cm}
\large{\textbf{2025}}
\end{center}
\bigskip
\end{titlepage}
\pagenumbering{Roman}
\begin{center}
\includegraphics[scale=0.65]{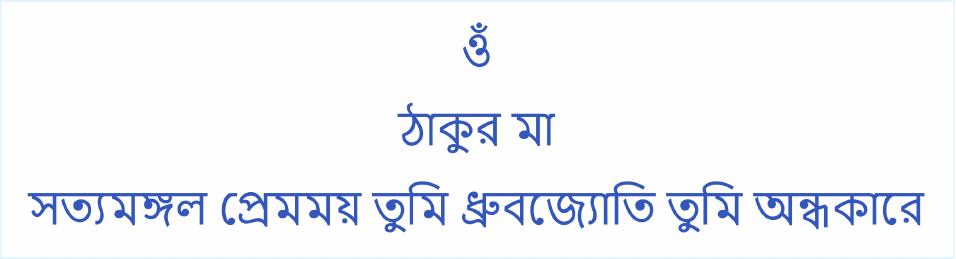}\\\vspace{1cm}
\includegraphics[scale=0.75]{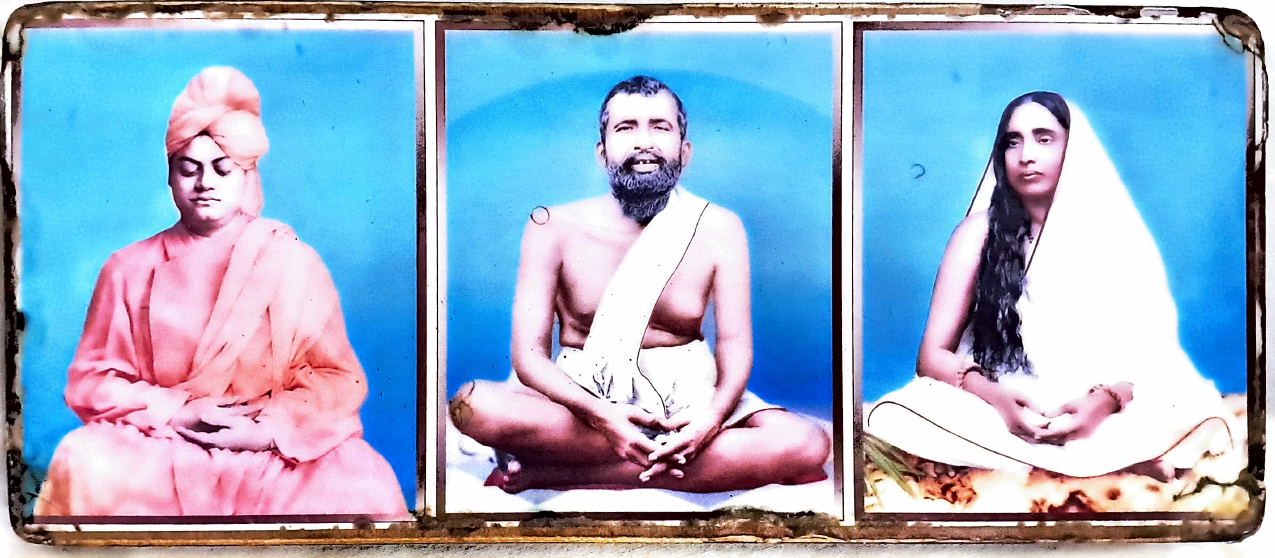}\\
\vspace{3cm}
\includegraphics[scale=0.75]{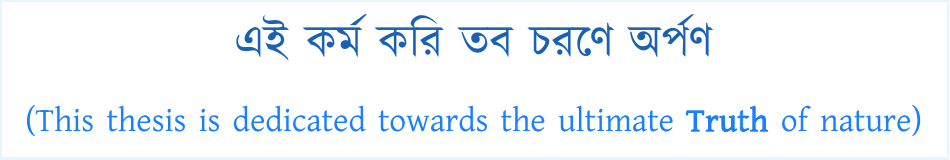}
%\large{\textit{This thesis is dedicated towards a new way for the search of the ultimate ``\textbf{Truth}" of nature}}
\end{center}\pagebreak
\begin{center}
\LARGE{Acknowledgement}\\
\medskip
\medskip
\includegraphics[scale=0.9]{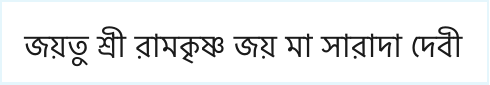}\\
\large{\textit{A path is incomplete without the blessing of thy two}}
\end{center}
\begin{onehalfspace}
I heartily thank \textbf{Prof. Sunandan Gangopadhyay} for his continuous support and efforts throughout the course of my entire tenure of Ph.D. He is a very supportive person in all of my research endeavours and has provided me with complete freedom of research. He has continuously encouraged my enthusiasm and new ideas. As a result of which this thesis is successfully submitted before the completion of my Ph.D. tenure. He has also been a person very close to me throughout my entire time here and I wish to pursue this close bonding with him. I express my sincere gratitude towards my supervisor Prof. Sunandan Gangopadhyay.

\noindent I also want to sincerely thank the S. N. Bose National Centre for Basic Sciences for giving me the opportunity to join as an integrated Ph.D. student and continue my Ph.D. here. I thank all of my teachers, academic, non-academic, and non-teaching staff who have supported me in several ways and assisted me during my journey here. I also thank all of my group members as well as all of my collaborators who have with whom I have shared a lot of memories and spent a lot of time. Among the group members, I would like to recall the names of  Dr. Sukanta Bhattacharyya, Ms. Manjari Dutta, Mr. Gopinath Guin, Mr. Arnab Mukherjee, Ms. Arpita Jana, Dr. Rituparna Mandal, and Dr. Abhijit Dutta with whom  I have shared a close bonding and a very friendly relationship. 
I am also lucky enough to have a rich cultural circle with all my juniors, colleagues, and seniors involved. I thank all of the persons with whom I have performed several programs on stage and shared fond memories with.  

\noindent Finally, I shed my heartfelt, love, respect, and gratitude towards my family members who were always beside me. I cannot imagine my journey especially without my ma (mother) Smt. Isita Sen, baba (father) Sri Haridas Sen, dida (grandmother) Smt. Shyamali Bosu, and my dadu (grandfather) Sri Amal Kr. Bosu. They have been a constant support throughout my entire journey who have never left my hand irrespective of the situation. I would specifically mention the name of ma who has been a constant friend, philosopher, and guide throughout my entire journey and my father always try to fulfil my needs without a single hesitation. 

\noindent I feel blessed to be surrounded by such exceptional personalities and figures.

\end{onehalfspace}
\noindent 
\pagebreak
\begin{center}
\textbf{\large{List of publications}}\\
\end{center}
\begin{enumerate}
\item  \textbf{Soham Sen} and Sunandan Gangopadhyay, ``\textit{Minimal length scale correction in the noise of gravitons}", \href{https://doi.org/10.1140/epjc/s10052-023-12230-2}{Eur. Phys. J. C 83 (2023) 1044}.
\item \textbf{Soham Sen} and Sunandan Gangopadhyay, ``\textit{Uncertainty principle from the noise of gravitons}", \href{https://doi.org/10.1140/epjc/s10052-024-12481-7}{Eur. Phys. J. C 84 (2024) 116}.
\item \textbf{Soham Sen}, Sunandan Gangopadhyay, and Sukanta Bhattacharyya, ``\textit{Quantum gravity signatures in gravitational wave detectors placed inside a harmonic trap potential}", \href{https://link.aps.org/doi/10.1103/PhysRevD.110.026008}{Phys. Rev. D 110 (2024) 026008}.
\item \textbf{Soham Sen} and Sunandan Gangopadhyay, ``\textit{Probing the quantum nature of gravity using a Bose-Einstein condensate}", \href{https://link.aps.org/doi/10.1103/PhysRevD.110.026014}{Phys. Rev. D 110 (2024) 026014}.
\item \textbf{Soham Sen} and Sunandan Gangopadhyay, ``\textit{Quantum nature of gravity in a Bose-Einstein condensate}", \href{https://doi.org/10.1103/PhysRevD.111.066002}{Phys. Rev. D 111 (2025) 066002}.
\end{enumerate}
\vspace{0.5cm}
\begin{center}
\textbf{\large{List of publications (not included in the thesis)}}\\
\end{center}
\begin{enumerate}
\item \textbf{Soham Sen}, Sukanta Bhattacharyya, and Sunandan Gangopadhyay, ``\textit{Probing the generalized uncertainty principle through quantum noises in optomechanical systems}", \href{https://doi.org/10.1088/1361-6382/ac55ab}{Class. Quant. Grav. 39 (2022) 075020}.
\item \textbf{Soham Sen}, Rituparna Mandal, and Sunandan Gangopadhyay, ``\textit{Equivalence principle and HBAR entropy of an atom falling into a quantum corrected black hole}", \href{https://link.aps.org/doi/10.1103/PhysRevD.105.085007}{Phys. Rev. D 105 (2022) 085007}.
\item \textbf{Soham Sen}, Rituparna Mandal, and Sunandan Gangopadhyay, ``\textit{Near horizon aspects of acceleration radiation of an atom falling into a class of static spherically symmetric black hole geometries}", \href{https://link.aps.org/doi/10.1103/PhysRevD.106.025004}{Phys. Rev. D 106 (2022) 025004}.
\item Neeraj Kumar, \textbf{Soham Sen}, and Sunandan Gangopadhyay, ``\textit{Phase transition structure and breaking of universal nature of central charge criticality in a Born-Infeld AdS black hole}", \href{https://link.aps.org/doi/10.1103/PhysRevD.106.026005}{Phys. Rev. D 106 (2022)  026005}.
\item \textbf{Soham Sen}, Sukanta Bhattacharyya, and Sunandan Gangopadhyay, ``\textit{Path Integral Action for a Resonant Detector of Gravitational Waves in the Generalized Uncertainty Principle Framework}", \href{https://doi.org/10.3390/universe8090450}{Universe 8 (2022) 450} (Special issue: Quantum Gravity Phenomenology II).
\item Ashmita Das, \textbf{Soham Sen}, and Sunandan Gangopadhyay, ``\textit{Virtual transitions in an atom-mirror system in the presence of two scalar photons}", \href{https://link.aps.org/doi/10.1103/PhysRevD.107.025009}{Phys. Rev. D 107 (2023) 025009}.
\item Neeraj Kumar, \textbf{Soham Sen}, and Sunandan Gangopadhyay, ``\textit{Breaking of the universal nature of the central charge criticality in AdS black holes in Gauss-Bonnet gravity}", \href{https://link.aps.org/doi/10.1103/PhysRevD.107.046005}{Phys. Rev. D 107 (2023) 046005}.
\item \textbf{Soham Sen}, Rituparna Mandal, and Sunandan Gangopadhyay, ``\textit{Near horizon approximation and beyond for a two-level atom falling into a Kerr–Newman black hole}", \href{https://doi.org/10.1140/epjp/s13360-023-04482-4}{Eur. Phys. J. Plus 138 (2023) 855}.
\item Sunandan Gangopadhyay, \textbf{Soham Sen}, and Rituparna Mandal, ``\textit{Interference and reflection from the event horizon of a quantum corrected black hole}", \href{https://iopscience.iop.org/article/10.1209/0295-5075/acb80f}{Eur. Phys. Lett. 141 (2023) 49001}.
\item \textbf{Soham Sen}, Manjari Dutta, and Sunandan Gangopadhyay, ``\textit{Lewis and berry phases for a gravitational wave interacting with a quantum harmonic oscillator}", \href{https://doi.org/10.1088/1402-4896/ad1234}{Phys. Scr. 99 (2024) 015007}.
\item \textbf{Soham Sen}, Arnab Mukherjee, and Sunandan Gangopadhyay, ``\textit{Entanglement degradation as a tool to detect signatures of modified gravity}", \href{https://link.aps.org/doi/10.1103/PhysRevD.109.046012}{Phys. Rev. D 109 (2024) 046012}.
\item Sukanta Bhattacharyya, \textbf{Soham Sen}, and Sunandan Gangopadhyay, ``\textit{Resonant detectors of gravitational wave in the linear and quadratic generalized uncertainty principle framework}", \href{https://doi.org/10.1140/epjc/s10052-024-12786-7}{Eur. Phys. J. C 84 (2024) 425}.
\item Ashmita Das, \textbf{Soham Sen}, and Sunandan Gangopadhyay, ``\textit{Horizon brightened accelerated radiation in the background of braneworld black holes}", \href{https://link.aps.org/doi/10.1103/PhysRevD.109.064087}{Phys. Rev. D 109 (2024) 064087}.
\item \textbf{Soham Sen}, Ashis Saha, and Sunandan Gangopadhyay, ``\textit{Signatures of quantum geometry from exponential corrections to the black hole entropy}", \href{https://doi.org/10.1007/s10714-024-03241-9}{Gen. Rel. Grav. 56 (2024) 57}.
\item Arpita Jana, \textbf{Soham Sen}, and Sunandan Gangopadhyay, ``\textit{Atom falling into a quantum corrected charged black hole and HBAR entropy}", \href{https://link.aps.org/doi/10.1103/PhysRevD.110.026029}{Phys. Rev. D 110 (2024) 026029}.
\item Arnab Mukherjee, \textbf{Soham Sen}, and Sunandan Gangopadhyay, ``\textit{Quantum coherence measures for generalized Gaussian wave packets under a Lorentz boost}", \href{https://doi.org/10.48550/arXiv.2407.06599}{Phys. Rev. A 110 (2024) 052413}.
\item \textbf{Soham Sen}, Abhijit Dutta, and Sunandan Gangopadhyay, ``\textit{Thermodynamics of a Schwarzschild black hole surrounded by quintessence in the generalized uncertainty principle framework}", \href{https://doi.org/10.1140/epjc/s10052-025-13811-z}{Eur. Phys. J. C 85 (2025) 117}.
\item Arpita Jana, \textbf{Soham Sen}, and Sunandan Gangopadhyay, ``\textit{Inverse logarithmic correction in the HBAR entropy of an atom falling into a renormalization group improved charged black hole}", \href{https://doi.org/10.1103/PhysRevD.111.085017}{Phys. Rev. D 111 (2025) 085017}.
\item Ashmita Das, Anjana Krishnan, \textbf{Soham Sen}, and Sunandan Gangopadhyay, ``\textit{Derivative coupling in horizon brightened acceleration radiation: A quantum optics approach}", \href{https://doi.org/10.48550/arXiv.2505.16897}{arXiv:2505.16897 [gr-qc]} (Accepted for publication in Physical Review D).
%\item Gopinath Guin, \textbf{Soham Sen}, and Sunandan Gangopadhyay, ``\textit{Renormalization group improved cosmology in the presence of a stiff matter era}", \href{https://doi.org/10.48550/arXiv.2411.03693}{arXiv:2411.03693 [gr-qc]} (Communicated to a peer-reviewed journal).
%\item A. Das, A. Krishnan, \textbf{Soham Sen}, and S. Gangopadhyay, ``\textit{Derivative coupling in horizon brightened acceleration radiation: a quantum optics approach}", \href{https://arxiv.org/abs/2505.16897}{arXiv:2505.16897 [gr-qc]} (Communicated to a peer-reviewed journal).
%\item \textbf{Soham Sen}, M. Dutta, and S. Gangopadhyay, ``\textit{Density matrix analysis of systems influenced by periodic Hamiltonian}", \href{https://doi.org/10.48550/arXiv.2507.06355}{arXiv:2507.06355 [quant-ph]}.
%\item \textbf{Soham Sen}, S. Bhattacharyya, and S. Gangopadhyay, ``\textit{Can resonating systems amplify the noise induced by gravitons?}".
\end{enumerate}
\begin{abstract}
In this thesis, we have used a linearized quantum gravity setting to investigate the effects of gravitons on matter systems. Based on the graviton-matter interaction, we have then proposed detector models that may be able to pick up graviton-induced signatures in a matter of a few years. We start with the simple model of a two-particle model detector system interacting with quantized gravitational fluctuations while the detector degrees of freedom obey the generalized uncertainty principle (GUP). Using a path integral quantization technique for quantizing the gravitational fluctuation, we obtain the transition probability for the system to go from an initial state to some final state and after quite a bit of analytical calculations, we arrive at the final form of the transition probability where the graviton effect is captured by the Feynman-Vernon influence functional. Extracting the action part from the transition probability and then extremizing the action, we arrive at the GUP modified geodesic deviation equation infused by graviton noise fluctuations. Obtaining the analytical solution and calculating the standard deviation, we obtain a time-dependent standard deviation of the geodesic separation, which is indicative of the GUP effect in a quantum gravity scenario. We have then considered the simple model of a freely-falling point-particle under the effect of Earth's gravitational field while it interacts with quantum gravity fluctuations. It is observed that the standard deviation in the position as well as the momentum of the freely-falling particle depends on the two-point graviton noise-noise correlator. From the uncertainty product of the position and momentum of the particle, one can write down an inequality, which eventually leads to a quantum gravity modified uncertainty relation. We also obtain an upper bound to the uncertainty relation and find out that the analytical structure of the uncertainty relation is the same for different graviton states. We also observe from the lower bound of the uncertainty relation that in the Planck mass limit, the quantum gravity modified uncertainty relation reduces to the generalized uncertainty principle with a pre-determined GUP coefficient. For the next part of the thesis, we have mainly focused on the phenomenological aspects of a linearized quantum gravity theory. For the initial phenomenological model, we have considered the same two-particle model detector system interacting with gravitational fluctuations where the entire set-up is placed inside a harmonic trap potential. We now quantize the detector 
as well as the gravitational part of the system and investigate the transition probability for the model system to go from some initial state to a different final state. We find out that the transition probability for the detector to go from an initial lower energy state to an excited state via the absorption of a graviton is identical to the case of a detector making the same transition by the energy absorption from a classical gravitational wave when the gravitational wave is considered as a combination of a finite number of gravitons and the energy-flux relation corresponding to a gravitational wave is implemented. However, for the de-excitation scenario, we observe that even if the initial state has no gravitons in it, it can still come from a higher excited state to a lower excited state via the spontaneous emission of a single graviton. This spontaneous emission of a graviton is a pure signature of quantum gravity, and if observed in an experimental scenario, will be indicative of the existence of gravitons in nature. In order to inspect a more involved phenomenological aspect, instead of the standard matter-detector system, we make use of a relativistic scalar Bose-Einstein condensate (BEC). At first, we have investigated the response of relativistic BEC towards incoming graviton fluctuations. We make use of quantum metrological techniques to obtain the quantum gravity modified Fisher information, which is constructed out of the matrix elements of the single-mode BEC covariance matrix where the matrix elements are influenced by graviton-induced noise fluctuations. Using the quantum gravitational Fisher information, we have then obtained the quantum gravity modified Cram\'{e}r-Rao bound from which it is possible to obtain the minimum value in the measurement of the standard deviation of the gravitational wave amplitude. We observe that for gravitons with high squeezing, the BEC, even with very small phonon squeezing, is able to pick up the signatures of quantum gravity. However, for a classical gravitational wave, the BEC is not sensitive to the incoming gravitational fluctuations. It therefore indicates that if a graviton comes and interacts with a BEC where the graviton states are highly squeezed, then a detection of the gravitational fluctuation by the BEC will be a direct evidence of graviton detection by a BEC-based graviton detector. Based on this model detector system, we further extend our analysis to the case where the wave vectors corresponding to the single mode of the BEC are also quantized. From the analytical form of the reduced density matrix of the Bose-Einstein condensate where the graviton degrees of freedom are traced out, we observe that the off-diagonal elements of the density matrix corresponding to the BEC part undergo a time-dependent decoherence which is instigated by bremsstrahlung induced by the graviton noise-fluctuations. We then consider a maximally entangled momentum state of the Bose-Einstein condensate. We find out that the initial tensor product state of the BEC-graviton system gets entangled over time due to graviton-induced decoherence, which we term as a Bose-Einstein ``supercondensate". Based on this principle, we propose a BEC-based graviton detector, which is a two-phase Mach-Zehnder type atom interferometer where the atom-lasers generated from a continuous wave Bose-Einstein condensate play the role of matter wave packets. If this experimental model is implemented successfully, it is evident that gravitons will be detected in a matter of decades. This thesis, therefore, paves the way for the idealization of new graviton detector models and also directs towards the investigation of new and significant phenomenological aspects of a quantum theory of gravity, which may be possible to implement using standard and future experimental advancements. 
\end{abstract}
\pagebreak


## মাতৃভাষায় সারাংশ

এই প্রবন্ধে, আমরা একটি পদার্থের উপর গ্র্যাভিটন কণার প্রভাব অনুসন্ধান করি যেখানে সমগ্র ব্যবস্থাটি একটি রৈখিক বা লিনিয়ার কোয়ান্টাম মহাকর্ষ তত্ত্বকে মেনে চলে। গ্র্যাভিটন ও পদার্থের মধ্যের এই মিথস্ক্রিয়ার উপর নির্ভর করে, আমরা একটি অনুসন্ধানকারী যন্ত্রের বা সংবেদকের প্রস্তাব রাখি, যেটি সম্ভবত কিছু বছরের মধ্যেই গ্র্যাভিটন কণার অস্তিত্ব খুঁজে পেতে সক্ষম হবে। আমরা এখানে একটি সাধারণ মডেল নিয়ে আমাদের কাজ শুরু করি যেখানে দুইটি কণার সঙ্গমে তৈরী একটি সংবেদক বা ডিটেক্টর কোয়ান্টাম মহাকর্ষীয় ফ্লাকচুয়েশনের সাথে মিথস্ক্রিয়া করে যখন ডিটেক্টরের অবস্থান ও ভরবেগ সাধারণ (জেনেরালাইজড) অনিশ্চয়তা নীতিকে অনুসরণ করে। পথনির্ভর সমাকলনীয় কোয়ান্টাইজেশন প্রক্রিয়া বা পাথ ইন্টিগ্রাল কোয়ান্টাইজেশন প্রক্রিয়ার প্রয়োগ করে আমরা মহাকর্ষীয় ফ্লাকচুয়েশনকে কোয়ান্টাইজ করি এবং যথোপযুক্ত গাণিতিক সরলীকরণের পর সমগ্র ব্যবস্থাটির একটি প্রাথমিক অবস্থা বা স্টেট থেকে একটি অন্তিম অবস্থায় যাবার সম্ভাবনাকে নিরূপণ করি যেখানে গ্র্যাভিটনের সমগ্র প্রভাবকেই আবদ্ধ করে রাখে ফাইনম্যান-ভারনন ইনফ্লুয়েন্স ফাংশনাল। স্থানান্তরীকরণের সম্ভাবনা থেকে অ্যাকশন বা ক্রিয়াকে সনাক্ত করি এবং তারপর সেই অ্যাকশনের এক্সট্রিমাইজেশন বা চরমীকরণের মাধ্যমে আমরা সাধারণ অনিশ্চয়তা তত্ত্ব দ্বারা প্রভাবিত এবং গ্র্যাভিটন কণার বিশৃঙ্খলা দ্বারা অনুপ্রবিষ্ট অন্তিম জিওডেসিক বিচ্যুতি সমীকরণে অবতীর্ণ হই। উপরুক্ত সমীকরণের গাণিতীক সমাধান এবং তার থেকে সাধারণ বিচ্যুতি বের করার পর আমরা জিওডেসিক ব্যবধানের একটি সময়নির্ভর সাধারণ বিচ্যুতি খুঁজে পাই যেটি এই কোয়ান্টাম মহাকর্ষীয় বিন্যাসের মধ্যে সাধারণ অনিশ্চয়তা তত্ত্বের অস্তিত্বের ইঙ্গিত দেয়। এরপর আমরা একটি সাধারণ মডেল ধরে নি যেখানে একটি বিন্দুর মতন কণা পৃথিবীর অভিকর্ষের প্রভাবে অবাধে অবতরণ করে এবং একইসাথে কোয়ান্টাম মহাকর্ষীয় ফ্লাকচুয়েশনের সাথে মিথস্ক্রিয়া করে। আমরা খুঁজে পাই যে কণাটির অবস্থান এবং ভরবেগের সাধারণ বিচ্যুতি গ্র্যাভিটনের বিশৃঙ্খলার পারস্পরিক সম্পর্ক বা নয়েজ্-নয়েজ্ কোরিলেটরের উপর নির্ভরশীল। কণাটির অবস্থান এবং ভরবেগের অনিশ্চয়তার গুণফল থেকে একটি গানিতীক অসমতা লেখা সম্ভব যেটা পরবর্তীকালে একটি মহাকর্ষীয় কোয়ান্টাম তত্ত্ব দ্বারা পরিবর্তিত অনিশ্চয়তা নীতির দিকে আমাদের এগিয়ে নিয়ে যায়। এই অনিশ্চয়তা সম্পর্কের আমরা একটি ঊর্ধ্বসীমা খুঁজে পাই এবং গ্র্যাভিটনের বিভিন্ন প্রাথমিক অবস্থার জন্য আমরা এই অনিশ্চয়তা সম্পর্কের অভিন্ন বিশ্লেষণাত্মক গঠন খুঁজে পাই। আমরা যদি এই অনিশ্চয়তা সম্পর্কের নিম্নসীমায় থাকা কণার ভরকে প্ল্যাঙ্ক-এর ভর দিয়ে প্রতিস্থাপন করি তাহলে আমরা পূর্বোক্ত সাধারণ অনিশ্চয়তা নীতিকে ফিরে পাই যেখানে সাধারণ অনিশ্চয়তা গুণাঙ্কের মান পুরবনির্ধারিত। আমাদের কাজের পরবর্তী ভাগের জন্য আমরা প্রধানত রৈখিক কোয়ান্টাম মহাকর্ষ তত্ত্বের ফেনমেনলজিক্যাল বা প্রপঞ্চবৈজ্ঞানিক (প্রতিফলনমূলক) দিকটিতে মনোনিবেশ করি। প্রাথমিক প্রতিফলনমূলক মডেলের জন্য আমরা পূর্বোক্ত দ্বৈত কণার সঙ্গমে তৈরী অনুসন্ধানী যন্ত্রকে বিবেচনা করি যেখানে সমগ্র বিন্যাসটি একটি ঐকতানসম্বন্ধীয় জাল বা হারমোনিক ট্র্যাপের মধ্যে রাখা। এক্ষণে আমরা সমগ্র বিন্যাসের সংবেদক অংশ এবং মহাকর্ষীয় অংশকে কোয়ান্টাইজ করে, সমগ্র ব্যবস্থাটির একটি প্রাথমিক অবস্থা থেকে ভিন্ন অন্তিম অবস্থায় যাওয়ার সম্ভাবনাকে অনুসন্ধান এবং অবশেষে নির্ধারণ করি। আমরা আবিষ্কার করি যে সংবেদকটির কম ক্ষমতাসম্পন্ন স্তর থেকে একটি গ্র্যাভিটন কণাকে শোষণ করে একটি উত্তেজিত স্তরে যাওয়ার প্রবণতা, যন্ত্রটির একটি ক্ল্যাসিক্যাল মহাকর্ষীয় তরঙ্গের থেকে শক্তিলাভ করে একইরকম অবস্থান্তরের প্রবণতার সাথে অভিন্ন যেখানে এই সম্পর্ক স্থাপনের জন্য আমাদের ক্ল্যাসিক্যাল মহাকর্ষীয় তরঙ্গকে সসীম সংখ্যক গ্র্যাভিটন কণার সমন্বয় হিসেবে ধরি এবং একইসঙ্গে মহাকর্ষীয় তরঙ্গের শক্তিপ্রবাহ সম্পর্ক ব্যবহার করি। এর বিপরীত প্রক্রিয়ায় যখন অনুসন্ধানী যন্ত্রটি উত্তেজিত স্তর থেকে স্বল্প ক্ষমতাসম্পন্ন স্তরে অবতীর্ণ হয়, সেক্ষত্রে যদি প্রাথমিক অবস্থায়

কোন গ্র্যাভিটন কণা না থাকে তৎসত্ত্বেও অনুসন্ধানী যন্ত্রটি স্বতঃস্ফূর্তভাবে একটি গ্র্যাভিটন কণা নির্গমন করে স্বল্প ক্ষমতাসম্পন্ন স্তরে অবতীর্ণ হয়। গ্র্যাভিটন কণার এই স্বতঃস্ফূর্ত নির্গমন মহাকর্ষীয় কোয়ান্টাম তত্ত্বের অস্তিত্বের স্বাক্ষর এবং একটি পরীক্ষামূলক দৃশ্যকল্পে এই ঘটনার পর্যবেক্ষণ, প্রকৃতিতে গ্র্যাভিটন কণার অস্তিত্বের প্রমাণের নির্দেশক। আরও গভীর ফেনমেনলজিক্যাল দিক অনুসন্ধান করার স্বার্থে আমরা গতানুগতিক পদার্থনির্মিত অনুসন্ধানী যন্ত্রের পরিবর্তে একটি বোস-আইনস্টাইন কন্ডেনসেট (সজ্ঘনন) বিবেচনা করি যেখানে কন্ডেনসেটটি বিশেষ অপেক্ষবাদ তত্ত্ব মেনে চলে (রিলেটিভিস্টিক বোস-আইনস্টাইন কন্ডেনসেট)। সর্বপ্রথমে আমরা রিলেটিভিস্টিক বোস-আইনস্টাইন কনডেনসেটের গ্র্যাভিটন কণার ফ্লাকচুয়েশনের প্রতি সাড়া বা প্রতিক্রিয়াকে অনুস্বন্ধান করি। কোয়ান্টাম পরিমাপণবিদ্যার কৌশল ব্যাবহার করে আমরা কোয়ান্টাম মহাকর্ষবিদ্যা দ্বারা পরিবর্তিত কোয়ান্টাম ফিশার তথ্য (ইনফরমেশন) নিরূপণ করি যেটি একক-মোড বোস-আইনস্টাইন কনডেনসেটের সহপ্রভৃতি বা কোভেরিয়েন্স ম্যাট্রিক্সের এলিমেন্ট বা উপাদান দিয়ে তৈরী যেখানে কোভেরিয়েন্স ম্যাট্রিক্সটি গ্র্যাভিটন কণার গোলমাল বা নয়েজ্ দ্বারা প্রভাবিত। কোয়ান্টাম মহাকর্ষীয় ফিশার তথ্য ব্যবহার করে আমরা কোয়ান্টাম মহাকর্ষবিদ্যার দ্বারা প্রভাবিত ক্রামার-রাও সীমা অর্জন করি যেটি ব্যবহার করে আমরা মহাকর্ষীয় তরঙ্গের প্রশস্ততার সাধারণ বিচ্যুতির সর্বনিম্ন মানের পরিমাপ করতে পারি। আমরা নিরীক্ষণ করি যে ফোনন কণাগুলির যৎসামান্য নিষ্পেষণ সত্ত্বেও, অতি নিষ্পেষিত গ্র্যাভিটন কণার জন্য বোস-আইনস্টাইন কনডেনসেটটি কোয়ান্টাম মহাকর্ষীয় তত্ত্বের স্বাক্ষর খুঁজে পেতে সক্ষম। যদিও ক্ল্যাসিক্যাল মহাকর্ষীয় তরঙ্গের জন্য বোস-আইনস্টাইন কনডেনসেটটি আগত মহাকর্ষীয় ফ্লাকচুয়েশনের প্রতি সংবেদনশীল নয়। অতএব আমাদের অনুসন্ধান এটি নির্দেশ করে যে যদি একটি গ্র্যাভিটন কণা আসে এবং বোস-আইনস্টাইন কনডেনসেটের সাথে মিথষ্ক্রিয়া করে যেখানে গ্র্যাভিটন কণাটি অতীব নিষ্পেষিত সেক্ষেত্রে এই গ্র্যাভিটন প্রভঙ্গ বা ফ্লাকচুয়েশনের কনডেনসেটটির দ্বারা অনুসন্ধান, গ্র্যাভিটনের অস্তিত্বের প্রত্যক্ষ প্রমাণের নির্দেশক। এই মডেলটির উপর নির্ভর করে আমরা আমাদের এই অনুসন্ধানকারী যন্ত্রভিত্তিক মডেলটিকে আরও অগ্রসর করি যেখানে একক-মোড বোস-আইনস্টাইন কনডেনসেটটির তরঙ্গ ভেক্টরগুলিকেও কোয়ান্টাইজ করা হয়। বোস-আইনস্টাইন কন্ডেনসেটটির সংকুচিত ঘনত্ব ম্যাট্রিক্সের বিশ্লেষণাত্মক আকার থেকে যেখানে গ্র্যাভিটনের অংশকে ট্রেস করে বের করে দেওয়া হয়েছে, আমরা নিরীক্ষণ করি যে ম্যাট্রিক্সের তির্যক উপাদানগুলি একটি সময়নির্ভর অসঙ্গতির শিকার হয় যার প্রধান কারণ হল গ্র্যাভিটন কণার বিশৃঙ্খলাজনিত ব্রেম্স্ট্রাহলুং। আমরা বোস-আইনস্টাইন কন্ডেনসেটটির একটি সর্বাধিকভাবে জড়িয়ে বা এনট্যাঙ্গেল হয়ে থাকা ভরবেগ-অবস্থাকে (মোমেন্টাম-স্টেট) বিবেচনা করি। আমরা খুঁজে পাই যে কনডেনসেট-গ্র্যাভিটনের প্রাথমিক টেনসর-প্রোডাক্ট স্টেটটি গ্র্যাভিটনসৃষ্ট অসঙ্গতির (ডিকোহেরেন্স) জন্য এনট্যাঙ্গেল হয়ে যায় যাকে এই প্রবন্ধে আমরা বোস-আইনস্টাইন "সুপারকন্ডেনসেট" নামকরণ করেছি। এই নীতির উপর ভিত্তি করে আমরা বোস-আইনস্টাইন কন্ডেনসেট দ্বারা নির্মিত একটি গ্র্যাভিটন অনুসন্ধানকারী যন্ত্রের প্রস্তাব রাখি যেটি একটি দ্বিস্তরীয় মাক-জেন্ডার ইন্টারফেরোমিটার যেখানে একটি নিরন্তর-তরঙ্গ বোস-আইনস্টাইন কন্ডেনসেট থেকে সৃষ্ট পরমাণু-লেজার পদার্থের তরঙ্গ-প্যাকেটের ভূমিকা পালন করে। যদি এই পরীক্ষামূলক মডেলের বাস্তবীকরন সম্ভব হয়, গ্র্যাভিটন কণার অস্তিত্ব এক দশকের মধ্যে খুঁজে পাওয়া সম্ভব হবে। এই প্রবন্ধটি তাই নূতন গ্র্যাভিটন-অনুসন্ধানী যন্ত্রের আদর্শায়নের সাথে সাথে কোয়ান্টাম মহাকর্ষ তত্ত্বের নূতন এবং গুরুত্বপূর্ণ ঘটনামূলক দিকের (ফেনমেনলজিক্যাল) নির্দেশ দেয় যেগুলি বর্তমান এবং ভবিষ্যতের পরীক্ষামূলক অগ্রগতির সাহায্যে বাস্তবায়ন করা সম্ভব হবে।


\pagenumbering{arabic}
\tableofcontents
\listoffigures
%Quantum gravity and its fundamental aspects
\chapter{Introduction}\label{C.1.OTM}
At the end of the nineteenth century the understanding of reality completely changed. Maxwell has just combined four of the distinct laws of electrostatics and electrodynamics creating the combined theory of electromagnetism. These Maxwell's equations are also considered as one of the first and most beautiful unifications achieved till date. In the beginning of the twentieth century, Albert Einstein proposed the idea of  special relativity \cite{AnnusMirabilis} which shattered the idea of an universal time proposed by Isaac Newton in the seventeenth century and this is also considered as one of the biggest landmarks in theoretical physics. In the year 1900-1901, Max Planck wrote several papers introducing the fact that energy is not transferred continuously rather it moves in packets or ``\textit{quanta}" with fixed energy value which is equal to the value of Planck's constant multiplied by the frequency of the emitted radiation \cite{MPlanck1,MPlanck2,MPlanck3,MPlanck4}. In the mean time Albert Einstein also had given the first theoretical analysis of the photoelectric effect by considering light as a collection of discrete quanta of energy \cite{EinsteinPhotoElectric}, the energy of which is equal to $h\nu$ ($\nu$ being the frequency of light). All these new experimental and theoretical findings directed towards a new fundamental theory of nature, also known later as quantum mechanics. This new theoretical aspect mainly told a completely different story of the microscopic world where suddenly from the classical determinism physicists jump into the world of indeterminism and probabilistic outcomes. In the years 1911-1915, Albert Einstein wrote another series of papers which led to the generalization of special relativity also known as the theory of general relativity \cite{EinsteinGRelativity,EinsteinGRelativity2,
EinsteinGRelativity3,EinsteinGRelativity4,
EinsteinGRelativity5}. The general theory of relativity, although being a classical field theory of gravitation is extremely accurate and is considered to be one of the most accurate theoretical descriptions of the universe at larger length scales. For a many particle description of Quantum mechanics, when considered in a special relativistic setting, one needs to generalize the idea of quantum mechanics and it led to the birth of quantum field theory. With the increasing theoretical and experimental advancements, it was observed that it is possible to write down a self-consistent and fully renormalizable quantum field theory for three of the four fundamental forces of nature, namely electromagnetic force, weak nuclear force, and the strong nuclear force. However, when trying to write down a quantum field theory of gravity, it led to several fundamental problems. One of the main problems being the non-renormalizability of the quantum field theory of gravity. It was observed that when one approaches a very high energy, the higher order loop corrections start to diverge resulting in the non-renormalizability of the theory. It is postulated that quantum gravity effects, if exists, will be dominant in very small length scales. In \cite{GravityClassical1} it was argued that at such small length scales, gravity becomes so weak that all of the other three fundamental forces hugely dominate and the effect of gravity becomes negligible. One of the other consideration is that the the quantum theory completely breaks down at larger length scales and the gravity behaves classically \cite{GravityClassical2,
GravityClassical3,GravityClassical4,GravityClassical5,GravityClassical6}.
There are although some real problems in a classical theory of gravity, which are the existence of singularities in some solutions of the Einstein's field equations. Places like the Big bang singularity and the singularity of a black hole, the classical field theory of gravity breaks down indicating the possibility of the existence of a more involved theoretical structure lying at a length scale which is of the order of the Planck length, given by the analytical expression $l_{\text{Pl}}=\sqrt{\frac{\hbar G}{c^3}}\sim 10^{-35}$ m. There have been several attempts to write down a full quantum field theory of gravity. The currently considered contenders for a theory of quantum gravity are string theory \cite{String1,String2, String3, String4, String5,String6,String7}, loop quantum gravity \cite{lqg1,lqg2,lqg3,lqg4}, and noncommutative geometry \cite{ncg1,ncg2,ncg3}. String theory considers the fundamental object of the universe being a particle instead of a string and as its low energy limit the general theory of relativity comes out whereas loop quantum gravity considers the spacetime metric as an operator and the quantization of the background spacetime leads to a quantum gravity theory. Noncommutative geometry on the other hand relies on the fact that the operator-valued spacetime coordinates have a non-vanishing canonical commutation relation between them. All of these results indicate that spacetime at very small length scales is not continuous and have a discrete mesh like structure. Although these theories are successful in explaining some of the fundamental aspects of gravity and spacetime itself at very high energies, they cannot give a very robust and a solid picture and suffer from some serious issues. Hence, instead of going for a full quantum gravity theory, the easiest way is to look for quantum gravity signatures in a linearized quantum gravity model. In principle, Einstein's field equations are highly non-linear in nature which means that for two different analytical forms of the energy momentum tensor, if  certain analytical forms of the background spacetime is obtained, then the sum of the two solutions does not produce the Einstein's field equations for the combined geometrical effect. The linearized quantum gravity model considers small gravitational fluctuations upon a background spacetime metric. In his 1916 article \cite{EinsteinGWave}, Albert Einstein considered a flat Minkowski background upon which he took small gravitational fluctuations. Substituting this metric in the field equations, he obtained a wave equation that propagates with the speed of light. These waves therefore are also termed as the gravitational waves. In 2015, the detection of gravitational waves \cite{GWave1,GWave2,GWave3} after almost a hundred years of the prediction by Einstein led to an upsurge in the research related to gravitational wave physics and its detection scenario. The currently operating gravitational wave observatories include LIGO, VIRGO, KAGRA\footnote{LIGO: Laser Interferometer Gravitational-wave Observatory; VIRGO: Virgo Interferometer for the Detection of Gravitational Waves; KAGRA: Kamioka Gravitational Wave Detector.}, and GEO600. The standard linearized quantum gravity theory stands upon this quantization of the small gravitational fluctuations over the background metric. The first proposals regarding the detection of the quantum nature of gravity were given initially by M. P. Bronstein \cite{BronsteinQuantumGravity} and later by R. P. Feynman \cite{FeynmanQuantumGravity} in the Chappel Hill conference \cite{ChapelHillQuantumGravity}. The formalism for quantizing a linearized gravity theory was initially postulated by M. P. Bronstein \cite{BronsteinQuantumGravity3,BronsteinQuantumGravity4} and later it was further developed by S. N. Gupta \cite{GuptaQuantumGravity1,GuptaQuantumGravity2}. In \cite{GuptaQuantumGravity1,GuptaQuantumGravity2}, it was first shown that a ``graviton", the quanta of linearized theory of gravity, is either a spin-zero or a spin-two particle. It is way easier to quantize a linearized gravity theory in a consistent way and in this procedure some of the standard problems in renormalization go away. Hence, considering graviton interaction with standard physical systems and trying to unearth signatures of quantum gravity may lead to the correct way of approaching a full quantum theory of gravity\footnote{For a detailed review on the evolution of the quantum gravity theory in the first half of the twentieth century please refer to \cite{EvolutionQuantumGravity}.}. The main problem with any such approach lies in the fact that the theory of linearized quantum gravity primarily resides in the low energy limit and as a result any kind of matter-graviton interaction signatures become very small rendering it impossible for an experimental detection.  For finding signatures of quantum gravity there are several directions; a direct and an indirect approach. The indirect approach relies on looking at physical systems where the quantum gravity signatures arise as a subtle change in a physical observable and it does not require a full quantization of the background spacetime. One of such examples being the modification of the standard Heisenberg's uncertainty principle which is also known as the generalized uncertainty principle \cite{KempfManganoMann,AKempf1,AKempf2}. Instead of directly quantizing the background gravitational fluctuations, one can just modify the standard Heisenberg's uncertainty principle and calculate different observables to search for any changes due to the inclusion of the generalized uncertainty principle compared to the standard results obtained using the Heisenberg's uncertainty principle. Such detection will lead to an indirect signature of quantum gravity. Another way is the direct approach where the background spacetime fluctuations are quantized to invoke quantum gravity effects in the physical system. There have been a series of investigations that aim to unearth signatures of quantum gravity via investigating the generation of entanglement due to graviton interaction between two massive bodies \cite{MarlettoVedral,EntanglementQuantumGravity,
MarlettoVedral2,EntanglementQuantumGravity2,
EntanglementQuantumGravity3}. As two gravitational fluctuations can superpose with each other in a  quantum gravity set-up, it is possible to mimic the experimental set-up proposed in \cite{MarlettoVedral,EntanglementQuantumGravity,
MarlettoVedral2,EntanglementQuantumGravity2} by creating 
a pair of coherent quantum matter source which is very difficult to achieve experimentally. Such experiments rely primary upon the gravitational fluctuations created by quantum matter sources. A different kind of experimental scenario is possible where the gravitational fluctuations come directly from gravitational waves. In a series of works \cite{QGravNoise,QGravLett,QGravD,KannoSodaTokuda,
KannoSodaTokuda2}, the interaction of a gravitational wave with two-particle model detector systems have been considered. The model considers a particle with a very small mass and the other particle with a heavier much. The mass of the heavier particle is significant enough for its dynamics to be neglected. This model imitates one arm of the `L' shaped interferometric detectors currently in existence.  Now from the model proposed in \cite{QGravNoise,QGravLett,QGravD,
KannoSodaTokuda,KannoSodaTokuda2}, by extremizing the quantum gravity modified total action of the model system, one obtains the graviton induced modified geodesic deviation equation. The geodesic deviation equation now gets infused by a quantum gravitational stochastic term that gives it a Langevin-like structure. It is a very important observation as the analytical form of the geodesic separation becomes dependent on the stochastic graviton-induced noise term and as a result there is a non-vanishing uncertainty in the measurement of the geodesic separation between the two particles. This is equivalent to measuring an overall uncertainty in the measurement of the detector arm length which is along one of the polarization of the gravitational wave. The standard deviation in the measurement of the geodesic separation is observed to be of the scale of that of the Planck length. As has been discussed in \cite{QGravLett,QGravD,KannoSodaTokuda
,KannoSodaTokuda2}, one of the ways to get an enhanced signal is if the gravitons are thermal in nature or carry an inherent squeezing. Such high value of the squeezing or the temperature can only be found in primordial gravitational waves or very high temperature gravitational explosions. Now, primordial gravitational waves are considered to generate during the inflation period and lies in the $0.0001-10$ Hz frequency range. At this moment there are no available gravitational wave observatories that work in such low frequency ranges however space based gravitational wave observatories like LISA ($0.0001-1$ Hz) and DECIGO ($0.1-10$ Hz) are already in consideration as the next step towards gravitational wave astronomy and astrophysics \footnote{LISA: Laser Interferometer Space Antenna; DECIGO: DECI-hertz Interferometer Gravitational wave Observatory.}. As a result, the possibility for finding a signature of graviton interaction with this future generation of gravity wave detector is not very faint. As has been discussed in \cite{QGravD}, for high initial squeezing of the gravitons, the uncertainty in the measurement of the arm length does lie within the sensitivity range of the future generation space based gravitational wave detectors. The aim of this thesis is quite simple but involved, to find fundamental aspects of a quantum gravity setting and to finally propose an experimental model which shall be sufficient for picking up graviton signatures in future. The thesis is separated into two parts. The first part, ``\textit{The fundamental minimal length in quantum gravity}", revolves around few of the fundamental aspects of quantum gravity when considered in simple physical set-ups. There are three chapters in the first part and the contents covered in the chapters go as follows.
\begin{itemize}
\item[$\circ$] In chapter(\ref{C.2.OTM}), we start with the basic discussion of the existence of a fundamental length scale in nature if a quantum theory of gravity exists. This helps one to modify the Heisenberg's uncertainty relation in a way that the uncertainty in the measurement of the position observable can not go below the value of the fundamental length. With this basic discussion of the generalized uncertainty principle and its $n$-dimensional generalization, we finally note down some of the drawbacks present in the used standard structure of the generalized uncertainty principle.
\item[$\circ$] In chapter(\ref{C.3.OTM}), we consider the two-particle interferometer model proposed in \cite{QGravNoise,QGravLett,QGravD}, and make use of a path integral approach to quantize the background gravitational fluctuations where the detector phase space variables obey the generalized uncertainty principle. We start with the basic derivation of the Fermi normal coordinates and describe the methodology presented in \cite{QGravD}. Making use of the path integral approach, we arrive at the action for the joint graviton-detector model system where the geodesic deviation and its conjugate momentum obey the generalized uncertainty principle (GUP). Finally extremizing the action with respect to the geodesic separation, we arrive at the GUP modified stochastic geodesic deviation equation. From their, we calculate the bound on the undetermined dimensionless GUP parameter by claiming that the GUP corrected term cannot be larger than the standard term in the analytical  form of the geodesic separation. We also plot the time dependent uncertainty in the geodesic deviation and investigate whether the GUP infusion lead to any significant insight. This chapter is based on the publication S. Sen and S. Gangopadhyay, ``\textit{Minimal length scale correction in the noise of gravitons}", \href{https://doi.org/10.1140/epjc/s10052-023-12230-2}{Eur. Phys. J. C 83 (2023) 1044}.
\item[$\circ$] In chapter(\ref{C.4.OTM}), we take the model of a point-particle freely falling under the effect of Earth's gravitational field. This model was first proposed in \cite{ChawlaParikh}. Here, we primarily aim to use this simplified model where as the background, the Earth's gravitational field is considered (upto the Newtonian correction) and the fluctuations upon this gravitational background is quantized using a path integral approach. We then calculate the uncertainty in the position corresponding to the particle part and multiply it with its corresponding uncertainty in momentum. This helps us to write down a true quantum gravity modified uncertainty relation using the Planck constants. We analyze the uncertainty relation for the gravitons initially being in a vacuum, squeezed, and a thermal state and from there investigate whether such a quantum gravity modified uncertainty relation is universal in nature. We also compare our result with the standard generalized uncertainty principle and comment on the similarities and whether it is possible to obtain the generalized uncertainty principle using the quantum gravitational uncertainty relation. This chapter is based on the publication S. Sen and S. Gangopadhyay, ``\textit{Uncertainty principle from the noise of gravitons}", \href{https://doi.org/10.1140/epjc/s10052-024-12481-7}{Eur. Phys. J. C 84 (2024) 116}.
\end{itemize}
These three chapters conclude the first part of our thesis, where we aim to investigate the role of fundamental minimal length in detector model systems and also aim to derive the uncertainty relation based on which it is possible to investigate the important aspects of linearized quantum gravity in simple physical systems. 

\noindent For the next part of our thesis, ``\textit{Quantum gravity phenomenology}", we primarily aim on the phenomenological aspects of a linearized quantum gravity model and we also aim to propose an experiment based on the fundamental properties of exotic matter systems which may lead towards a detection of gravitons in the very near future. In this part, we use the canonical quantization approach of linearized gravity. The second part of the thesis also consists of three chapters and the contents covered in the chapters go as follows.
\begin{itemize}
\item [$\circ$] In chapter(\ref{C.5.OTM}), we consider the model of the two particle system and place it inside a harmonic trap potential. This model mimics the set-up of an interferometer detector placed inside a harmonic trap potential which can also be considered as a resonant bar gravitational wave detector. We quantize the detector part as well as the gravitational fluctuation part of the system. We then obtain the transition probability of the system for going from an initial state to some final state. We then make use of the energy-flux relation for a gravitational wave and we consider that this energy is transferred by a finite number of gravitons. Finally, for the resonant absorption case, we compare the quantum gravitational absorption probability with the semi-classical case where the gravitational wave is treated classically. We also compare the quantum gravitational result for the emission scenario with the semi-classical result. Based on the observed differences between the two cases, we comment on the possible detectability of a graviton-signature in such class of gravitational wave observatories in future. This chapter is based on the publication, S. Sen, S. Gangopadhyay, and S. Bhattacharyya, ``\textit{Quantum gravity signatures in gravitational wave detectors placed inside a harmonic trap potential}", \href{https://doi.org/10.1103/PhysRevD.110.026008}{Phys. Rev. D 110 (2024) 026008}.
\item[$\circ$] In chapter(\ref{C.6.OTM}), we use the model of a Bose-Einstein condensate interacting with gravitational fluctuations where the gravitational fluctuations are quantized. We start with a brief discussion of quantum metrological techniques which are required for this analysis and obtain the covariance matrix for the single mode of a Bose-Einstein condensate at zero temperature. We next discuss the nonrelativistic Bose-Einstein condensate for an ideal Bose gas and obtain the condensate temperature for a relativistic model of the Bose-Einstein condensate. We obtain the Lagrangian for the relativistic Bose-Einstein condensate in a flat background with gravitational fluctuations. Using a canonical quantization technique the gravitational fluctuations are quantized which then help us to write the stochastic equation of motion corresponding to the time-dependent part of the Goldstone bosons. One then obtain the equation of motion solving the stochastic Langevin-like equation of motion from which it is possible to read the Bogoliubov coefficients. Using quantum metrological techniques and the analytical forms of the Bogoliubov coefficients, it is possible to obtain the quantum gravity modified Fisher information which is related to the uncertainty in the measurement of the gravitational wave amplitude via the Cram\'{e}r-Rao bound in this quantum gravitational set-up. We then compare the sensitivity of the Bose-Einstein condensate with the proposed sensitivity curve of the upcoming space based gravitational wave observatories. The frequency range in consideration are in the $0.0001-10$ Hz range which is the exact range for primordial gravitational waves generated during the beginning of the inflation. We also look into the case when the phonon modes are interacting with each other resulting in an overall decoherence effect and compare it with the case when the phonons are not-interacting with each other. This chapter is based on the publication, S. Sen and S. Gangopadhyay, ``\textit{Probing the quantum nature of gravity using a Bose-Einstein condensate}", \href{https://doi.org/10.1103/PhysRevD.110.026014}{Phys. Rev. D 110 (2024) 026014}.

\item[$\circ$] In chapter(\ref{C.7.OTM}), we extend the analysis presented in chapter(\ref{C.6.OTM}) and aim to propose an experimental set-up which shall be able to detect quantum gravity signatures in future. We consider a maximally entangled momentum state for the Bose-Einstein condensate and investigate the decoherence effect generated due to the graviton induced noise in the Bose-Einstein condensate. To observe the decoherence effect, we start with the von-Neumann equation of motion corresponding to the density matrix of the single mode Bose-Einstein condensate for a maximally entangled momentum state. Taking trace over the gravitational field degrees of freedom, we observe from the analytical form of the reduced density matrix, the consequences generated due to gravitational bremsstrahlung. We also analyze entanglement degradation due to graviton interaction. Based on the fundamental results of this analysis, we propose an atom-interferometry based experimental model where for interference coherent atom laser beams generated from a continuously generated Bose-Einstein condensate inside of a harmonic trap potential is used. The fundamental principle behind this experimental model is that decoherence will lead to blurring of the atom interference pattern leading to a more mixed distribution of the bosons. The important thing to note that the experimental model proposed here is a combination of different ground-breaking experimental models which are implementable currently. As a result the experimental model proposed here is implementable in a matter of a decade with the use of some very complicated engineering marvels. This chapter is based on the publication, S. Sen and S. Gangopadhyay, ``\textit{Quantum nature of gravity in a Bose-Einstein condensate}", \href{https://doi.org/10.1103/PhysRevD.111.066002}{Phys. Rev. D 111 (2025) 066002}.
\item[$\circ$] Finally, in chapter(\ref{C.8.OTM}), we conclude our results and discuss the important aspects and novel findings.
\end{itemize} 

\part{The fundamental minimal length in quantum gravity}

\chapter{The fundamental minimal length}\label{C.2.OTM}
In 1927, Werner Heisenberg for the first time postulated that if the position is measured for an object precisely then it is impossible to precisely measure the momentum of the particle simultaneously \cite{HeisenbergUncertainty}. The multiplication of the standard deviations in the position and momentum observables obey an inequality, which reads
\begin{equation}\label{HeisenbergUncertainty}
\Delta x\Delta p\geq\frac{\hbar}{2}
\end{equation} 
where $\hbar=\frac{h}{2\pi}$ is the reduced Planck's constant with $h$ denoting the Planck's constant. The above inequality is also known as the Heisenberg's uncertainty relation. From the equality it is easy to check some important aspects between the uncertainties in the position and momentum. It is easy to check that for a precise measurement of $x$ ($\Delta x=0$), the uncertainty in the momentum becomes infinite whereas for a precise momentum measurement ($\Delta p=0$), the uncertainty in position becomes infinite. From eq.(\ref{HeisenbergUncertainty}), it is possible to write down a commutator bracket between the position and the momentum operator as $[\hat{x},\hat{p}]=i\hbar$. Now, the uncertainty principle in eq.(\ref{HeisenbergUncertainty}) does not have any influence from a gravitational perspective and for considering true quantum gravitational effects the form of the uncertainty principle in eq.(\ref{HeisenbergUncertainty}) should change.
In this chapter, we shall start with the basic discussion of the modified Heisenberg's uncertainty relation which is considered to be a direct effect of the quantization of the gravitational field.
Now, the primary drawback of  unifying quantum field theory and general relativity lies in the fact that any quantum field theory of gravitation is non-renormalizable as has also been discussed in chapter(\ref{C.1.OTM}).
Any standard quantum gravity theory, in general, gives an ultraviolet cut-off which leads to the existence of a fundamental minimal length in nature. Thorough investigations in string theory \cite{String1,String2, String3, String4, String5,String6,String7}, loop quantum gravity \cite{lqg1,lqg2,lqg3,lqg4}, and noncommutative geometry \cite{ncg1,ncg2,ncg3} propose that this fundamental minimal length is of the order of the Planck length given by $l_{\text{Pl}}=\sqrt{\frac{\hbar G}{c^3}}\sim 10^{-35}$ m. As a theory of quantum gravity introduces a fundamental minimal length, in principle, it should solve the singularity problem in classical general relativity. This fundamental minimal length should be represented by the uncertainty in the measurement of the position parameter ($\Delta x$) from a quantum mechanical perspective. The most straight forward way to incorporate this fundamental minimal length in the theory is by modifying the standard Heisenberg's uncertainty relation. The modified uncertainty relation in one dimension incorporating a non-vanishing minimum uncertainty $\Delta x$ reads \cite{KempfManganoMann}
\begin{equation}\label{Uncertainty.1}
\Delta x\Delta p\geq \frac{\hbar}{2}\left[1+\beta_p(\Delta p)^2+\zeta^p_0\right]
\end{equation} 
where $\beta_p$ and $\zeta^p_0$ neither depends on the uncertainty in the position parameter nor in the uncertainty in the momentum parameter. The subscript $p$ in $\beta_p$ denotes that it is a coefficient corresponding to the $(\Delta p)^2$ correction in the uncertainty relation. A more generalized version of the above uncertainty relation given in eq.(\ref{Uncertainty.1}) can be written where the right hand side also consists of an uncertainty in the position variable \cite{AKempf1,AKempf2} which is given by\footnote{For a straightforward but simple derivation of the generalized uncertainty principle see \cite{KuzmichevKuzmichev}.}
\begin{equation}\label{Uncertainty.2}
\Delta x\Delta p\geq \frac{\hbar}{2}\left[1+\beta_x(\Delta x)^2+\beta_p(\Delta p)^2+\zeta_0\right]
\end{equation}
with $\beta_x$ and $\beta_p$ denoting the coefficients corresponding to the square of the uncertainties in position and momentum variables in the uncertainty principle and $\zeta_0\equiv \zeta_0^x+\zeta_0^p$. 
\section{Generalized commutation relation}
We already have discussed the proposed form of the modified Heisenberg's uncertainty relation when there is an existence of the fundamental minimal length in nature. The next aim is to write down the modified commutation relation between the position operator $\hat{x}$ and the conjugate momentum operator $\hat{p}$ from the form of the generalized uncertainty relation in eq.(\ref{Uncertainty.2}). 
We start by considering two Hermitian operators $\hat{\mathcal{A}}$ and $\hat{\mathcal{B}}$ where the state of the system in consideration is given by the ket $|\psi\rangle$. The variances in $\mathcal{A}$ and $\mathcal{B}$ with respect to the state $|\psi\rangle$ reads
\begin{equation}\label{Uncertainty.3}
\langle(\Delta \hat{\mathcal{A}})^2\rangle= \langle \psi|(\hat{\mathcal{A}}-\langle\hat{\mathcal{A}}\rangle)^2|\psi\rangle~, ~~\langle(\Delta \hat{\mathcal{B}})^2\rangle= \langle \psi|(\hat{\mathcal{B}}-\langle\hat{\mathcal{B}}\rangle)^2|\psi\rangle~.
\end{equation}
To obtain the uncertainty relation between $\hat{\mathcal{A}}$ and $\hat{\mathcal{B}}$, we first define two new states $|\psi_\mathcal{A}\rangle$ and $|\psi_\mathcal{B}\rangle$ which are defined as $|\psi_\mathcal{A}\rangle\equiv (\hat{\mathcal{A}}-\langle\hat{\mathcal{A}}\rangle)|\psi\rangle$ and $|\psi_\mathcal{B}\rangle\equiv (\hat{\mathcal{B}}-\langle\hat{\mathcal{B}}\rangle)|\psi\rangle$. One can then rewrite the variances of the Hermitian operators $\hat{\mathcal{A}}$ and $\hat{\mathcal{B}}$ in terms of the newly defined states $|\psi_\mathcal{A}\rangle$ and $|\psi_\mathcal{B}\rangle$ as
\begin{equation}\label{Uncertainty.4}
\langle (\Delta \hat{\mathcal{A}})^2\rangle=\langle \psi_\mathcal{A}|\psi_\mathcal{A}\rangle~,~~\langle (\Delta \hat{\mathcal{B}})^2\rangle=\langle \psi_\mathcal{B}|\psi_\mathcal{B}\rangle~.
\end{equation}
Using the above equation, one can write down the Cauchy-Schwarz inequality as
\begin{equation}\label{Uncertainty.5}
\langle (\Delta \hat{\mathcal{A}})^2\rangle\langle (\Delta \hat{\mathcal{B}})^2\rangle=\langle \psi_\mathcal{A}|\psi_\mathcal{A}\rangle\langle \psi_\mathcal{B}|\psi_\mathcal{B}\rangle\geq |\langle \psi_\mathcal{A}|\psi_\mathcal{B}\rangle|^2~.
\end{equation}
It is now possible to write down the analytical expression for $|\langle \psi_\mathcal{A}|\psi_\mathcal{B}\rangle|$ as
\begin{equation}\label{Uncertainty.6}
\begin{split}
|\langle \psi_\mathcal{A}|\psi_\mathcal{B}\rangle|&=|\langle \psi|(\hat{\mathcal{A}}-\langle\hat{\mathcal{A}}\rangle)(\hat{\mathcal{B}}-\langle\hat{\mathcal{B}}\rangle)|\psi\rangle|\\
&=\frac{1}{2}\left|\langle [\hat{\mathcal{A}},\hat{\mathcal{B}}]\rangle+\langle\{(\hat{\mathcal{A}}-\langle\hat{\mathcal{A}}\rangle),(\hat{\mathcal{B}}-\langle\hat{\mathcal{B}}\rangle)\}\rangle
\right|~.\end{split}
\end{equation}
%=|\langle \hat{\mathcal{A}}\hat{\mathcal{B}}\rangle-\langle\hat{\mathcal{A}}\rangle\langle\hat{\mathcal{B}}\rangle|\\
Now, we can further simplify the above equation as
\begin{equation}\label{Uncertainty.7}
\begin{split}
|\langle \psi_\mathcal{A}|\psi_\mathcal{B}\rangle|^2&=\frac{1}{4}\left(|\langle [\hat{\mathcal{A}},\hat{\mathcal{B}}]\rangle|^2+|\langle\{(\hat{\mathcal{A}}-\langle\hat{\mathcal{A}}\rangle),(\hat{\mathcal{B}}-\langle\hat{\mathcal{B}}\rangle)\}\rangle|^2\right)~.
\end{split}
\end{equation}
%\frac{1}{4}\Bigr(\langle [\hat{\mathcal{A}},\hat{\mathcal{B}}]\rangle\langle [\hat{\mathcal{B}},\hat{\mathcal{A}}]\rangle+\langle [\hat{\mathcal{A}},\hat{\mathcal{B}}]\rangle \langle\{(\hat{\mathcal{B}}-\langle\hat{\mathcal{B}}\rangle),(\hat{\mathcal{A}}-\langle\hat{\mathcal{A}}\rangle)\}\rangle\\&+\langle [\hat{\mathcal{B}},\hat{\mathcal{A}}]\rangle\langle\{(\hat{\mathcal{A}}-\langle\hat{\mathcal{A}}\rangle),(\hat{\mathcal{B}}-\langle\hat{\mathcal{B}}\rangle)\}\rangle+|\langle\{(\hat{\mathcal{A}}-\langle\hat{\mathcal{A}}\rangle),(\hat{\mathcal{B}}-\langle\hat{\mathcal{B}}\rangle)\}\rangle|^2\Bigr)\\
%\implies |\langle \psi_\mathcal{A}|\psi_\mathcal{B}\rangle|^2&=
Here, $|\langle\{(\hat{\mathcal{A}}-\langle\hat{\mathcal{A}}\rangle),(\hat{\mathcal{B}}-\langle\hat{\mathcal{B}}\rangle)\}\rangle|^2$ is a real and positive number, hence, from the above equation, it is possible to write down the inequality as\footnote{This is a straight forward derivation and  can be found in any standard quantum mechanics books \cite{QuantumMechanics1,QuantumMechanics2,QuantumMechanics3}.}
\begin{equation}\label{Uncertainty.8}
|\langle \psi_\mathcal{A}|\psi_\mathcal{B}\rangle|^2\geq \frac{1}{4} |\langle [\hat{\mathcal{A}},\hat{\mathcal{B}}]\rangle|^2~.
\end{equation}
Denoting the standard deviations in $\mathcal{A}$ and $\mathcal{B}$ as $\Delta\mathcal{A}\equiv\sqrt{\langle (\Delta \hat{\mathcal{A}})^2\rangle}$ and $\Delta\mathcal{B}\equiv\langle (\Delta \hat{\mathcal{B}})^2\rangle$, we can write down the final form of the inequality as
\begin{equation}\label{Uncertainty.9}
\Delta\mathcal{A}\Delta\mathcal{B}\geq \frac{1}{2}|\langle [\hat{\mathcal{A}},\hat{\mathcal{B}}]\rangle|~.
\end{equation}
We already know that the variance in $x$ is given as $(\Delta x)^2=\langle \hat{x}^2\rangle-\langle\hat{x}\rangle^2$. Now, if $\zeta_0=\beta_x\langle \hat{x}\rangle^2+\beta_p\langle \hat{p}\rangle^2$ then one can rewrite eq.(\ref{Uncertainty.2}) as $\Delta x\Delta p\geq \frac{\hbar}{2}[1+\beta_x\langle \hat{x}^2\rangle+\beta_p\langle \hat{p}^2\rangle]$ which can be further simplified as
\begin{equation}\label{Uncertainty.10}
\begin{split}
\Delta x\Delta p&\geq\frac{1}{2}|\langle i\hbar (1+\beta_x\hat{x}^2+\beta_p\hat{p}^2)\rangle|~.
\end{split}
\end{equation}
Again from eq.(\ref{Uncertainty.9}), we already know that $\Delta x\Delta p\geq \frac{1}{2}|\langle [\hat{x},\hat{p}]\rangle|$. Comparing the right hand side of this uncertainty relation from the right hand side of eq.(\ref{Uncertainty.10}), we obtain the commutator between the modified phase space variables as\cite{KempfManganoMann}
\begin{equation}\label{Uncertainty.11}
[\hat{x},\hat{p}]=i\hbar\left(1+\beta_x\hat{x}^2+\beta_p\hat{p}^2\right) ~.
\end{equation}
In general while considering such modified uncertainty relations, one can always set the constant $\beta_x$ to have the value zero. As $\beta_p\langle\hat{p}^2\rangle$ is a dimensionless quantity, $\beta_p$ must have the dimension equal to the inverse-square of the dimension of momentum which can be fixed by making use of the Planckian constants. One can write $\beta_p=\frac{\beta}{m_{\text{Pl}}^2c^2}$ where $m_{\text{Pl}}$ denotes the Planck's mass given by $m_{\text{Pl}}=\sqrt{\frac{\hbar c}{G}}$ with $\beta$ denoting an undefined dimensionless constant. The generalized commutation relation in eq.(\ref{Uncertainty.11}) for $\beta_x=0$ and for the obtained analytical form of $\beta_p$ then reads
\begin{equation}\label{Uncertainty.12}
[\hat{x},\hat{p}]=i\hbar\left(1+\frac{\beta}{m_{\text{Pl}}^2c^2}\hat{p}^2\right)~. 
\end{equation}
\subsection{Some important aspects of the generalized uncertainty principle}
We shall, for the moment, consider the form of the generalized uncertainty relation to be empirical in nature with the analytical form of the uncertainty principle being (as can be seen from eq.(\ref{Uncertainty.2}) as well) given as
\begin{equation}\label{Uncertainty.13}
\Delta x\Delta p\geq \frac{\hbar}{2}\left(1+\beta_p(\Delta p)^2+\beta_p\langle\hat{p}\rangle^2\right)
\end{equation}
where we have set $\beta_x=0$ in eq.(\ref{Uncertainty.2}) and substituted the value of $\zeta_0$ to be $\beta_p\langle\hat{p}\rangle^2$ as has been used before to obtain eq.(\ref{Uncertainty.10}) (for $\beta_x=0$). For the case of minimum uncertainty, we can simplify and write down eq.(\ref{Uncertainty.13}) as
\begin{equation}\label{Uncertainty.14}
\begin{split}
 (\Delta p)^2=\frac{2}{\hbar\beta_p} \Delta x\Delta p-\frac{1 }{\beta_p}- \langle\hat{p}\rangle^2
\end{split}
\end{equation}
which is a quadratic equation in $\Delta p$. Following standard procedure, one can solve the above equation as
\begin{equation}\label{Uncertainty.15}
\Delta p=\frac{\Delta x}{\hbar\beta_p}\pm\sqrt{\frac{(\Delta x)^2}{\hbar^2\beta_p^2}-\frac{1}{\beta_p}-\langle\hat{p}\rangle^2}~.
\end{equation}
From the above equation it is evident that if the term insider of the square-root becomes negative then the uncertainty in the momentum will be a complex number. Hence, the minimum value in the uncertainty of the position variable can be obtained as
\begin{equation}\label{Uncertainty.16}
\begin{split}
\frac{(\Delta x_{\text{min}})^2}{\hbar^2\beta_p^2}&=\frac{1}{\beta_p}+\langle\hat{p}\rangle^2\\
\implies \Delta x_{\text{min}}&=\hbar\sqrt{\beta_p}\left(1+\beta_p\langle\hat{p}\rangle^2\right)^\frac{1}{2}~.
\end{split}
\end{equation}
We already have argued that $\beta_p=\frac{\beta}{m_{\text{Pl}}^2}$ and if the expectation value in the momentum is zero then we obtain the absolute minimum value in the uncertainty in the position to be
\begin{equation}\label{Uncertainty.17}
\Delta x_{\text{min}}=l_{\text{Pl}}\sqrt{\beta}
\end{equation}
where $l_{\text{Pl}}=\sqrt{\frac{\hbar G}{c^3}}\sim 10^{-35} \text{m}$ denotes the Planck length with $\beta$ being a dimensionless number. It is therefore straightforward to claim that if the Heisenberg's uncertainty relation takes the form given in eq.(\ref{Uncertainty.13}) (when the expectation value of the momentum operator vanishes) then the minimum length is of the order of the Planck length beyond which one can not probe experimentally.
\subsection{Generalization to \textit{n}-dimensions}
One can now further generalize the ``generalized uncertainty relation" in one dimension given in eq.(\ref{Uncertainty.13}) to $n$ dimensions. For $n$ spatial dimensions, the generalized uncertainty principle reads \cite{KempfManganoMann}
\begin{equation}\label{Uncertainty.18}
\Delta x_j\Delta p_k\geq\frac{\hbar}{2}\delta_{jk}\left(1+\frac{\beta}{m_{\text{Pl}}^2c^2}\sum_{l=1}^n\left((\Delta p_l)^2+\langle\hat{p}_l\rangle^2\right)\right)~.
\end{equation}
It is possible to write down the modified commutator for $n$ spatial dimensions from the above uncertainty relation as
\begin{equation}\label{Uncertainty.19}
[\hat{x}_j,\hat{p}_k]=i\hbar \delta_{jk}\left(1+\frac{\beta}{m_{\text{Pl}}^2c^2}\hat{p}^2\right)
\end{equation}
where $\hat{p}^2=\sum_{k=1}^n\hat{p}_k\hat{p}_k$. If one requires $[\hat{p}_j,\hat{p}_k]$ to vanish then for a momentum representation $\psi(p)=\langle p|\psi\rangle $ (where $p=\sqrt{p_kp^k}$), one can write down the representation for the conjugate operators as
\begin{equation}\label{Uncertainty.20}
\langle p|\hat{p}_j|\psi\rangle=p_j\psi(p)~,~~\langle p|\hat{x}_k|\psi\rangle=i\hbar(1+\beta_pp^2)\partial_{p_k}\psi(p)~.
\end{equation}
Making use of the above representations, one can write down a commutation relation between two position operators as
\begin{equation}\label{Uncertainty.21}
[\hat{x}_j,\hat{x}_k]=-2i\hbar \beta_p(1+\beta_p\hat{p}^2)\hat{L}_{jk}
\end{equation}
where $\hat{L}_{jk}$ is the generator of rotations in $n$ dimensions with the analytical form 
\begin{equation}\label{Uncertainty.22}
\hat{L}_{jk}=\frac{1}{1+\beta_p\hat{p}^2}(\hat{x}_j\hat{p}_k-\hat{x}_k\hat{p}_j)~.
\end{equation}
\subsection{Some underlying problems with the generalized uncertainty principle}
The correction to the standard Heisenberg's uncertainty principle depicted in eq.(\ref{Uncertainty.1}) has some drawbacks.
\begin{enumerate}
\item At first the dimensionless coefficient $\beta$ is an undefined constant and therefore it is always required to obtain a bound on the dimensionless coefficient before working with the generalized uncertainty principle.
\item The second issue lies in the fact that the coefficient $\beta_p$ has the analytical form $\beta_p=\frac{\beta}{m_{Pl^2}c^2}$. Substituting the analytical form of the Planck mass, we can write down $\beta_p$ as $\beta_p=\frac{\beta G}{\hbar c^3}$. Now we shall again write down the generalized uncertainty principle in eq.(\ref{Uncertainty.1}) in an expanded form as
\begin{equation}\label{Uncertainty.23}
\Delta x\Delta p\geq \frac{\hbar}{2}+\frac{\beta G}{2c^3}(\Delta p)^2
\end{equation}
where we have set $\zeta_0^p$ to zero. The interesting thing to observe is that the quantum gravity correction term is dependent on two of the fundamental constants, namely Newton's gravitational constant and the speed of light. The primary problem is that the correction term is a quantum gravitational contribution and as a result should involve both the Planck's constant $\hbar$ and the Newton's gravitational constant $G$.  In eq.(\ref{Uncertainty.23}), we explicitly observe the coefficient of $(\Delta p)^2$ in the right hand side of the inequality does not involve any $\hbar$ contribution rendering it as not sufficient to be called a true quantum gravitational correction. 
\end{enumerate}
In the next chapter, we shall work in a generalized uncertainty principle modified background where the response of an interferometric detector will be analytically observed when the detector is interacting with gravitons. We have then considered a freely falling particle under the effect of the gravitational field of the Earth in chapter(\ref{C.4.OTM}). With this simple model, we then aim to obtain a true quantum gravity modified uncertainty relation that gets rid of the problems discussed in this subsection.
\chapter{The minimal length scale correction in linearized quantum gravity}\label{C.3.OTM}
For a single point-particle which falls under the effect of the gravitational field of any massive object, the standard trajectory is not a straight-line and the path followed by the particle is known as the geodesic equation. Now for a pair of particle falling freely under a gravitational field follows the geodesic deviation equation. The curvature effect comes from the consideration of standard classical general theory of relativity. In a series of works \cite{QGravNoise,QGravLett,QGravD,
KannoSodaTokuda,KannoSodaTokuda2}, the authors have considered the case of a pair of point particles where one of the particle is way less massive than the other counter part. The background is considered as a flat Minkowski background with gravitational fluctuations upon it. In order to incorporate quantum gravitational effects in the model system, the background fluctuation is quantized. In \cite{QGravLett,QGravD}, a path integral approach is taken to incorporate quantization effects in the detector arm which is modelled by the two point-particles, separated by a finite spatial distance. It is observed that the geodesic separation of  the two freely falling point-particles gets infused by graviton induced noise fluctuations. It was observed in \cite{QGravD} that for the graviton initially being in a vacuum state, the uncertainty in the measurement of the geodesic separation or the arm length of the detector comes out to be of the order of the Planck length. The important thing to remember from the discussion of chapter(\ref{C.2.OTM}) as well as the introduction is that any true quantum gravity theories like \cite{String1,String2,String3,String4,String5,String6}, loop quantum gravity \cite{lqg1,lqg2,lqg3}, and noncommutative geometry \cite{ncg1,ncg2,ncg3} (which incorporates a minimal length in its formalism), indicate towards the existence of a fundamental length which is of the order of the Planck length\footnote{The existence of such minimal length can although be avoided if the strings have dynamical string tensions in a string theory. In \cite{TimeAvoidance}, the existence of a fundamental time was avoided which can in principle be done for the fundamental length as well.} and as a result the standard Heisenberg's uncertainty principle gets modified and the new uncertainty relation is termed as the generalized uncertainty relation. This connection between the minimal length and gravity was first theoretically demonstrated in \cite{BronsteinQuantumGravity3,BronsteinQuantumGravity4}.
Later in \cite{MeadGravity}, it was argued that introducing a fundamental minimal length by taking resort to classical gravitational effects is not possible where a gedanken experiment was considered which strongly indicated towards the existence of such a minimal length in nature in a quantum gravity set-up. The theoretical and phenomenological aspects of this generalized uncertainty principle has been extensively reviewed in the literature in areas related to black holes \cite{MaggioreAlgebraic,ScardigliGedanken,AdlerSantiago,
AdlerSantiago2,SGADAS,ScardigliCasadio,RMSBSG,OngJCAP,
CorpuscularGUP,MajumderGUP}, optomechanical system \cite{GUPOpto1,GUPOpto2,GUPOpto3,GUPOpto4} (including laboratory based experimental investigations \cite{GUPOpto5,GUPOpto6}), and harmonic oscillators (including gravitational wave bar detectors) \cite{GUPOscillator1,GUPOscillator2,GUPResonantDetector1,
GUPResonantDetector2,GUPResonantDetector3}. Following the discussion of the earlier chapter, we already know that if $\hat{\xi}^{\text{G}}$ and $\hat{\pi}^{\text{G}}$\footnote{Here the superscript $\text{G}$ in the phase space variables denote that the phase operators obeys the generalized uncertainty principle.} denotes position and momentum operators corresponding to some system then the uncertainties in the $j$-th spatial direction follows the following inequality
\begin{equation}\label{Minimal.1}
\Delta \xi^{\text{G}}_j\Delta \pi^{\text{G}}_j\geq \frac{\hbar}{2}\left[1+\beta\left(\Delta {\left(\pi^{\text{G}}\right)}^2+\langle \hat{\pi}^{\text{G}}\rangle^2\right)+2\beta \left(\Delta {\left(\pi^{\text{G}}_j\right)}^2+\langle \hat{\pi}^{\text{G}}_j\rangle^2\right)\right]
\end{equation}
where, the index $j$ can take the values $j\in\{1,2,3\}$ and $\beta$\footnote{As has already been discussed in chapter(\ref{C.2.OTM}), $\beta$ can be represented in terms of the dimensionless GUP parameter $\beta_0$ as $\beta=\frac{\beta_0}{m_{\text{Pl}}^2c^2}$.} denotes the undetermined generalized uncertainty principle parameter. The phase space operators $\hat{\xi}^\text{G}$ and $\hat{\pi}^{\text{G}}$ then satisfies the commutation relation \cite{AliDasVagenas,AliDasVagenas2,BhattacharyyaGangopadhyay}
\begin{equation}\label{Minimal.2}
[\hat{\xi}^{\text{G}}_j,\hat{\pi}^\text{G}_k]=i\hbar\left[\delta_{jk}+\beta\left(\delta_{jk}{\left(\hat{\pi}^{\text{G}}\right)}^2+2\hat{\pi}^\text{G}_j\hat{\pi}^\text{G}_k\right)\right]~.
\end{equation}
If $j=k=1$, in that case the above commutation relation can be expressed simply as
\begin{equation}\label{Minimal.3}
\left[\hat{\xi}^\text{G},\hat{\pi}^{\text{G}}\right]=i\hbar\left[1+3\beta{\left(\hat{\pi}^{\text{G}}\right)}^2\right]~.
\end{equation}
If the standard phase space operators obeying the Heisenberg's uncertainty principle (HUP) are given by $\hat{\xi}^{\text{H}}$ and $\hat{\pi}^{\text{H}}$ such that $\left[\hat{\xi}^{\text{H}},\hat{\pi}^{\text{H}}\right]=i\hbar$, then it is possible to establish a relation between the pair of canonically conjugate operators in Heisenberg as well as generalized uncertainty principle framework as
\begin{equation}\label{Minimal.4}
\hat{\xi}^{\text{G}}=\hat{\xi}^{\text{H}}~,~~\hat{\pi}^{\text{G}}=\hat{\pi}^{\text{H}}(1+\beta\left(\hat{\pi}^{\text{H}}\right)^2)~.
\end{equation}
As we shall be executing a purely path integral analysis, we shall therefore not be using the `$~\hat{}~$' notation for the detector phase space `variables' explicitly. 

\noindent In this chapter\footnote{This chapter is based on the publication S. Sen and S. Gangopadhyay, ``\textit{Minimal length scale correction in the noise of gravitons}", \href{https://doi.org/10.1140/epjc/s10052-023-12230-2}{Eur. Phys. J. C 83 (2023) 1044}.}, we shall be concerned about the implementation of the graviton-detector model proposed in \cite{QGravD,QGravLett,QGravNoise}, in the generalized uncertainty principle framework, where the primary reason of choosing this framework lies in the fact that there exists a fundamental minimal length scale in a true quantum gravitational analysis. We implement the generalized uncertainty principle in the detector phase space variable incorporate indirect effects of quantization of gravity in the detector part of the analysis. We have then used the path-integral formalism proposed in \cite{QGravNoise,QGravD,QGravLett}. Using the generalized uncertainty principle modified Hamiltonian for the graviton-detector system, we calculate the Feynman-Vernon influence functional \cite{FeynmanVernon}. This Feynman-Vernon influence functional captures the effect of the environment on the detector system. Here the environment can be considered as the linearized gravitational field of which the quanta are the gravitons. As a result, proper calculation of the influence functional will give us the effect of the graviton induced fluctuations on the detector. This chapter is organized as follows.

\noindent At first, we provide a basic discussion of the Fermi-normal coordinate system and then with the basic discussion of the background model, we move towards the path integral quantization of the linearized gravity theory and its implementation. We, have then investigated the effect of the graviton fluctuation on the geodesic separation of the detector (indicating the arm length of the detector).  Then finally, we have investigated analytically the uncertainty in the measurement of the detector arm length and investigate the effect of a generalized uncertainty relation in the detector part of the system and then look into some its phenomenological implications and the possibility of detection of such effects in future generation of gravitational wave detectors.
\section{Fermi-normal coordinate system}
\begin{figure}[t!]
\begin{center}
\includegraphics[scale=1.2]{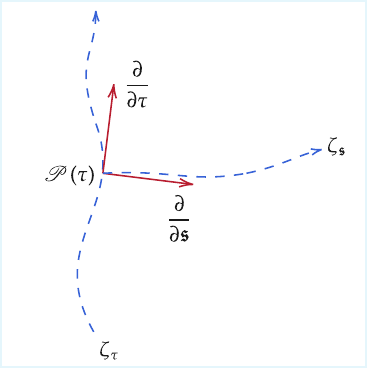}
\caption{A time-like geodesic $\zeta_\tau$ is perpendicular to a space-like geodesic $\zeta_\mathfrak{s}$ at the point $\mathcal{P}(\tau)$ where the space-like geodesic is parametrized by the proper distance $\mathfrak{s}$ and the time-like geodesic is parametrized by the proper time $\tau$. Here, $\frac{\partial}{\partial \tau}$ is the tangent vector to $\zeta_\tau$ at $\mathcal{P}(\tau)$ and $\frac{\partial}{\partial\mathfrak{s}}$ gives the tangent normal vector at the crossing point.\label{Fermi_Normal_OTM}}
\end{center}
\end{figure}
In this section, we shall be discussing the Fermi-normal coordinates in relation to our current model system\footnote{For a discussion on this matter one can also look at \cite{ItoSoda}.}. If a particle is following a geodesic then it is possible to create a coordinate system which is local and moves with the particle along the geodesic. Such a coordinate system is known as a Fermi-normal coordinate system \cite{FermiNormalCoordinate}. In this section, we proceed with the analysis of how to properly construct the Fermi-normal coordinates when the particle is moving along a time-like geodesic. The reason for considering the Fermi-normal coordinate system lies in the fact that, the model of our system consists of two freely falling particle representative of the arm of an interferometric detector where the heavier mass particle follows a time-like geodesic. In Fig.(\ref{Fermi_Normal_OTM}), we observe a time-like geodesics $\zeta_{\tau}$ which is parametrized by the proper time $\tau$. We consider another geodesic $\zeta_{\mathfrak{s}}$ (parametrized by the proper distance $\mathfrak{s}$) which is perpendicular to the time like geodesic at the point $\mathcal{P}(\tau)$. The choice of the crossing point is taken in a way such that at $\mathcal{P}(\tau)$, $\mathfrak{s}$ is zero. It is then possible to write down the local inertial frame as
\begin{equation}\label{Minimal.5}
x^0=\tau~,~~x^j=\kappa^j \mathfrak{s}
\end{equation}
where $\kappa^j$ denotes the coefficients of $\frac{\partial}{\partial \mathfrak{s}}$ which is the tangent vector ($\frac{\partial}{\partial \mathfrak{s}}=\kappa^j\frac{\partial}{\partial x^j}$) and can be seen from Fig.(\ref{Fermi_Normal_OTM}).
These coordinates in eq.(\ref{Minimal.5}) are the Fermi-normal coordinates. The base vectors $\frac{\partial}{\partial x^{\beta}}$ are parallel-transported along the time like geodesic. If it is possible to find a point on $\zeta_{\tau}$ such that the orthonormality condition gets satisfied (taking into consideration the degree of rescaling for the coefficients $\kappa^j$) then it will hold true along the geodesic as the bases are parallelly transported. It can therefore be safely claimed that in the Fermi-normal coordinates the metric on the time-like geodesic is the standard Minkowski metric $\eta_{\alpha\beta}$. The next step is to show that the Christoffel symbols vanish along the time-like geodesic $\zeta_\tau$. As the bases are parallelly transported along the time-like geodesic, it is easy to show that $
\Gamma^{\alpha}_{~\beta 0}(x^0=\tau,x^j=0)=\Gamma^{\alpha}_{~\beta 0}\rvert_{\zeta_{\tau}}=0$. The geodesic equation corresponding to the space-like geodesic reads 
\begin{equation}\label{Minimal.6}
\begin{split}
&\frac{d^2x^\alpha}{d\mathfrak{s}^2}+\Gamma^\alpha_{\mu\nu}\frac{dx^\mu}{d\mathfrak{s}}\frac{dx^\nu}{d\mathfrak{s}}=0\\
\implies &\frac{d^2x^\alpha}{d\mathfrak{s}^2}+\Gamma^\alpha_{~00}\frac{d\tau}{d\mathfrak{s}}\frac{d\tau}{d\mathfrak{s}}+2\Gamma^\alpha_{~0j}\frac{d\tau}{d\mathfrak{s}}\kappa^j+\Gamma^{\alpha}_{~ij}\kappa^i\kappa^j=0~.
\end{split}
\end{equation}
Now, $x^\alpha=\{\tau, \kappa^j \mathfrak{s}\}$ which gives $\frac{d^2x^\alpha}{d\mathfrak{s}^2}=0$. Substituting this result in the above equation, we obtain
\begin{equation}\label{Minimal.7}
\Gamma^\alpha_{~ij}\kappa^i\kappa^j=0~.
\end{equation}
Now, both $\kappa^i$ and $\kappa^j$ can pick up arbitrary values, which leaves us with the result 
\begin{equation}\label{Minimal.8}
\Gamma^\alpha_{~ij}(x^0=\tau,x^j=\kappa^j \mathfrak{s})=0~.
\end{equation}
If now, $\mathfrak{s}=0$ then the above Christoffel symbol is evaluated on the time-like geodesic at the point $\mathcal{P}(\tau)=0$ which gives us
\begin{equation}\label{Minimal.9}
\Gamma^\alpha_{~ij}(x^0=\tau,x^j=0)=\Gamma^{\alpha}_{~ij}\rvert_{\zeta_\tau}=0~.
\end{equation}
Combining the above result with $\Gamma^{\alpha}_{~0\beta}\rvert_{\zeta_\tau}$, we can conclude that the Christoffel symbols do vanish along $\zeta_\tau$. We shall now need to calculate the components of the metric about the time-like geodesic such that the system itself does not feel the effect of the background curvature. This will help one to expand the metric in terms of the coordinates. We start with the geodesic deviation equation 
\begin{equation}\label{Minimal.10}
\begin{split}
\frac{\mathcal{D}^2\xi^\alpha}{d\sigma^2}&=R^{\alpha}_{~\beta\mu\nu} u^{\beta}u^{\mu}\xi^\nu
\end{split}
\end{equation}
where $\frac{\mathcal{D}}{d\sigma}$ denotes a covariant derivative where $\sigma\in\{\tau,\mathfrak{s}\}$. The left hand side of the above equation can be expanded as
\begin{equation}\label{Minimal.11}
\begin{split}
\frac{\mathcal{D}^2\xi^\alpha}{d\sigma^2}&=\frac{\mathcal{D}}{d\sigma}\left(\frac{d\xi^\alpha}{d\sigma}+\Gamma^\alpha_{~\beta\mu}u^\beta \xi^\mu\right)\\
&=\frac{\mathcal{D}}{d\sigma}\left(\frac{d\xi^\alpha}{d\sigma}\right)+\frac{\mathcal{D}\Gamma^\alpha_{~\beta\nu}}{d\sigma}u^\beta \xi^\nu+\Gamma^\alpha_{~\beta\nu}\frac{\mathcal{D}u^\beta}{d\sigma} \xi^\nu+\Gamma^\alpha_{~\beta\nu}u^\beta\frac{\mathcal{D}\xi^\nu}{d\sigma} \\
&=\frac{d^2\xi^\alpha}{d\sigma^2}+\Gamma^\alpha_{~\mu\nu}\frac{d\xi^\mu}{d\sigma}u^\nu+\frac{d\Gamma^{\alpha}_{~\beta \nu}}{d\sigma}u^\beta\xi^\nu+\Gamma^\alpha_{~\mu\rho}u^\mu\Gamma^\rho_{~\beta\nu}u^\beta\xi^\nu-\Gamma^\rho_{~\beta\mu}\Gamma^\alpha_{~\rho\nu}u^\mu u^\beta \xi^\nu\\&-\Gamma^\rho_{~\mu\nu}\Gamma^\alpha_{~\beta\rho}u^\mu u^\beta \xi^\nu+\Gamma^{\alpha}_{~\beta\nu}u^\beta \frac{d\xi^\nu}{d\sigma}+\Gamma^{\alpha}_{~\beta\nu}u^\beta\Gamma^\nu_{~\rho\mu}\xi^\rho u^\mu\\
&=\frac{d^2\xi^\alpha}{d\sigma^2}+2\Gamma^\alpha_{~\mu\nu}\frac{d\xi^\mu}{d\sigma}u^\nu+\frac{d\Gamma^\alpha_{~\beta\nu}}{dx^\mu}\frac{dx^\mu}{d\sigma}u^\beta\xi^\nu+\Gamma^{\alpha}_{~\mu\rho}\Gamma^{\rho}_{~\beta\nu}u^\mu u^\beta\xi^\nu-\Gamma^\alpha_{~\rho\nu}\Gamma^\rho_{~\beta\mu}u^\mu u^\beta\xi^\nu
\end{split}
\end{equation}
where, we have made use of the fact that the covariant derivative of the four-velocity $u^\beta=\frac{d x^\beta}{d\sigma}$ is zero. Using the final line of the above expression and substituting it back in eq.(\ref{Minimal.10}), we arrive at the form of the geodesic deviation equation as
\begin{equation}\label{Minimal.12}
\frac{d^2\xi^\alpha}{d\sigma^2}+2\Gamma^\alpha_{~\mu\nu}\frac{d\xi^\mu}{d\sigma}u^\nu+\left(R^{\alpha}_{~\beta\nu\mu}+\frac{d\Gamma^\alpha_{~\beta\nu}}{dx^\mu}+\Gamma^{\alpha}_{~\mu\rho}\Gamma^{\rho}_{~\beta\nu}-\Gamma^\alpha_{~\rho\nu}\Gamma^\rho_{~\beta\mu}\right)u^\mu u^\beta\xi^\nu=0
\end{equation}
where we have used the anti-symmetric property of the Riemann curvature tensor given by $R^\alpha_{~\beta\mu\nu}=-R^\alpha_{~\beta\nu\mu}$. Any point on the space-like geodesic $\zeta_\mathfrak{s}$ can be determined by three parameters, namely $\tau$, $\kappa^j$, and $\mathfrak{s}$. It is then possible to consider two vectors $\left(\frac{\partial}{\partial\tau}\right)_{\mathfrak{s},\kappa^j}$ and $\left(\frac{\partial}{\partial\kappa^j}\right)_{\tau,\mathfrak{s}}$ which create deviations with respect to the space-like geodesic $\zeta_{\mathfrak{s}}$. If  two space-like geodesics start from two different points on the time-like geodesic $\zeta_\tau$ then the deviation vector $\left(\frac{\partial}{\partial\tau}\right)_{\mathfrak{s},\kappa^j}$ measures the deviation between two such geodesics whereas the deviation vector $\left(\frac{\partial}{\partial\kappa^j}\right)_{\tau,\mathfrak{s}}$, measures the deviation when the space-like geodesics has generated or passed through the same point $\mathcal{P}(\tau)$. We can now write down the geodesic separation $\xi^\mu$ in terms of the deviation vector $\left(\frac{\partial}{\partial\tau}\right)_{\mathfrak{s},\kappa^j}$ as
$\xi^\alpha=\left(\frac{\partial}{\partial\tau}\right)^\alpha_{\mathfrak{s},\kappa^j}=\left(\frac{\partial x^\alpha}{\partial\tau}\right)=\delta^\alpha_{~0}$. Substituting the above expression for $\xi^\alpha$ in eq.(\ref{Minimal.12}), we obtain
\begin{equation}\label{Minimal.13}
\begin{split}
&\frac{d^2\delta^\alpha_{~0}}{d\sigma^2}+2\left.\Gamma^\alpha_{~\mu\nu}\right\rvert_{\zeta_\tau}\frac{d\delta^\mu_{~0}}{d\sigma}u^\nu+\left.\left(R^{\alpha}_{~\beta\nu\mu}+\frac{d\Gamma^\alpha_{~\beta\nu}}{dx^\mu}+\Gamma^{\alpha}_{~\mu\rho}\Gamma^{\rho}_{~\beta\nu}-\Gamma^\alpha_{~\rho\nu}\Gamma^\rho_{~\beta\mu}\right)\right\rvert_{\zeta_\tau}u^\mu u^\beta\delta^\nu_{~0}=0\\
\implies&\left.\left(R^\alpha_{~\beta 0\mu}+\frac{d\Gamma^\alpha_{~\beta 0}}{dx^\mu}+\Gamma^{\alpha}_{~\mu\rho}\Gamma^{\rho}_{~\beta 0}-\Gamma^\alpha_{~\rho 0}\Gamma^\rho_{~\beta\mu}\right)\right\rvert_{\zeta_\tau}u^\mu u^\beta=0~.
\end{split}
\end{equation}
We already know that $\Gamma^{\alpha}_{~0\beta}\rvert_{\zeta_\tau}=0$ and using this expression and the anti-symmetric property of the Riemann curvature tensor, we arrive at the equation
\begin{equation}\label{Minimal.14}
\left(\left.\frac{d\Gamma^\alpha_{~\beta 0}}{dx^\mu}\right\rvert_{\zeta_\tau}-\left.R^{\alpha}_{~\beta\mu 0}\right\rvert_{\zeta_\tau}\right)u^\mu u^\beta=0~.
\end{equation}
We can consider only the spatial degrees of freedom corresponding to the indices $\beta$ and $\mu$, which will give us the expression
\begin{equation}\label{Minimal.15}
\left(\left.\frac{d\Gamma^\alpha_{~i 0}}{dx^j}\right\rvert_{\zeta_\tau}-\left.R^{\alpha}_{~ij 0}\right\rvert_{\zeta_\tau}\right)\kappa^i \kappa^j=0
\end{equation}
where we have made use of the fact that $u^j\rvert_{\zeta_\tau}=\kappa^j$ and the zeroth component vanishes. 
The next step is to express the geodesic deviation $\xi^\alpha$ in terms of the deviation vector $\left(\frac{\partial}{\partial\kappa^j}\right)_{\tau,\mathfrak{s}}$ which can be expressed as $\xi^\alpha=\left(\frac{\partial}{\partial\kappa^j}\right)^\alpha_{\tau,\mathfrak{s}}=\frac{\partial x^\alpha}{\partial \kappa^j}=\mathfrak{s}\frac{\partial\kappa^\alpha}{\partial\kappa^j}=\mathfrak{s}\delta^\alpha_{~j}$. Substituting the analytical form of $\xi^\alpha=\mathfrak{s}\delta^\alpha_{~j}$
in eq.(\ref{Minimal.12}), we arrive at the geodesic deviation equation for two spatial geodesics stemming from $\mathcal{P}(\tau)$ as
\begin{equation}\label{Minimal.16}
\frac{d^2(\mathfrak{s}\delta^{\alpha}_{~j})}{d\sigma^2}+2\Gamma^\alpha_{~\mu\nu}\frac{d(\mathfrak{s}\delta^{\mu}_{~j})}{d\sigma}u^\nu+\left(R^{\alpha}_{~\beta\nu\mu}+\frac{d\Gamma^\alpha_{~\beta\nu}}{dx^\mu}+\Gamma^{\alpha}_{~\mu\rho}\Gamma^{\rho}_{~\beta\nu}-\Gamma^\alpha_{~\rho\nu}\Gamma^\rho_{~\beta\mu}\right)u^\mu u^\beta\mathfrak{s}\delta^{\nu}_{~j}=0~.
\end{equation}
It is easy to check from the above equation that the first term vanishes ($\sigma=\mathfrak{s}$)which helps to simplify the above equation further as
\begin{equation}\label{Minimal.17}
2\Gamma^\alpha_{~j\nu}u^\nu+\left(R^{\alpha}_{~\beta j\mu}+\frac{d\Gamma^\alpha_{~\beta j}}{dx^\mu}+\Gamma^{\alpha}_{~\mu\rho}\Gamma^{\rho}_{~\beta j}-\Gamma^\alpha_{~\rho j}\Gamma^\rho_{~\beta\mu}\right)\mathfrak{s} u^\mu u^\beta=0~.
\end{equation}
 It is now possible to expand the Christoffel symbol in powers of $\mathfrak{s}$ as
\begin{equation}\label{Minimal.18}
\begin{split}
\Gamma^{\alpha}_{~jk}&=\Gamma^\alpha_{~jk}\rvert_{\zeta_\tau}+(\mathfrak{s}-\mathfrak{s}\rvert_{\mathcal{P}(\tau})\left.\frac{\partial \Gamma^{\alpha}_{~jk}}{\partial\mathfrak{s}}\right\rvert_{\zeta_\tau}+\mathcal{O}(\mathfrak{s}^2)\simeq\mathfrak{s}\left.\frac{\partial \Gamma^{\alpha}_{~jk}}{\partial\mathfrak{s}}\right\rvert_{\zeta_\tau}
\end{split}
\end{equation} 
where we have truncated $\mathcal{O}(\mathfrak{s}^2)$ and higher order terms and used the value of $\mathfrak{s}$ at the point $\mathcal{P}(\tau)$ which is zero. We have also made use of the fact that along the time-like geodesic the Christoffel symbol vanishes. Using the Taylor-series expansion of the Christoffel symbol from eq.(\ref{Minimal.18}) and substituting back in eq.(\ref{Minimal.17}), we obtain the modified geodesic deviation equation as
\begin{equation}\label{Minimal.19}
\begin{split}
&2\mathfrak{s}\left.\frac{\partial \Gamma^\alpha_{~ji}}{\partial \mathfrak{s}}\right\rvert_{\zeta_\tau}\kappa ^i+\left(\left. R^{\alpha}_{~ijk}\right\rvert_{\zeta_\tau}+\left.\frac{d\Gamma^\alpha_{~ij}}{\partial x^k}\right\rvert_{\zeta_\tau}\right)\mathfrak{s}\kappa^k\kappa^i+\mathcal{O}(\mathfrak{s}^2)=0\\
\implies &2\mathfrak{s}\left.\frac{\partial \Gamma^\alpha_{~ji}}{\partial x^k}\right\rvert_{\zeta_\tau}\kappa ^i\kappa^k+\left(\left. R^{\alpha}_{~ijk}\right\rvert_{\zeta_\tau}+\left.\frac{d\Gamma^\alpha_{~ij}}{\partial x^k}\right\rvert_{\zeta_\tau}\right)\mathfrak{s}\kappa^k\kappa^i+\mathcal{O}(\mathfrak{s}^2)=0\\
\implies &\left(\left.\frac{d\Gamma^\alpha_{~ij}}{\partial x^k}\right\rvert_{\zeta_\tau}+\frac{1}{3}\left. R^{\alpha}_{~ijk}\right\rvert_{\zeta_\tau}\right)\kappa^i\kappa^k=0
\end{split}
\end{equation}
Using the above relation and executing the operation $i\leftrightarrow k$, we arrive at the second relation
\begin{equation}\label{Minimal.20}
\left(\left.\frac{d\Gamma^\alpha_{~kj}}{\partial x^i}\right\rvert_{\zeta_\tau}+\frac{1}{3}\left. R^{\alpha}_{~kji}\right\rvert_{\zeta_\tau}\right)\kappa^i\kappa^k=0~.
\end{equation}
Combining the above two relations, we arrive at the expression (keeping in mind that $\kappa^i$ and $\kappa^k$ can take any values)
\begin{equation}\label{Minimal.21}
\left.\frac{d\Gamma^\alpha_{~ij}}{\partial x^k}\right\rvert_{\zeta_\tau}+\left.\frac{d\Gamma^\alpha_{~kj}}{\partial x^i}\right\rvert_{\zeta_\tau}=-\frac{1}{3}\left(\left. R^{\alpha}_{~ijk}\right\rvert_{\zeta_\tau}+\left. R^{\alpha}_{~kji}\right\rvert_{\zeta_\tau}\right)~.
\end{equation}
Now, in terms of the metric tensor, the Christoffel symbol is defined as
\begin{equation}\label{Minimal.22}
\Gamma^\mu_{~\beta\rho}=\frac{1}{2}g^{\mu\lambda}\left(\frac{\partial g_{\lambda \beta}}{\partial x^\rho}+\frac{\partial g_{\lambda \rho}}{\partial x^\beta}-\frac{\partial g_{\beta\rho}}{\partial x^\lambda}\right)~.
\end{equation}
With the action of the metric tensor $g_{\alpha\mu}$, we can write down from the above equation
\begin{equation}\label{Minimal.23}
g_{\alpha\mu}\Gamma^\mu_{~\beta\rho}=\frac{1}{2}\left(\frac{\partial g_{\alpha \beta}}{\partial x^\rho}+\frac{\partial g_{\alpha \rho}}{\partial x^\beta}-\frac{\partial g_{\beta\rho}}{\partial x^\alpha}\right)~.
\end{equation}
Combining the above expression with $g_{\beta\mu}\Gamma^{\mu}_{~\alpha\rho}$, we obtain the following expression 
\begin{equation}\label{Minimal.24}
g_{\alpha\mu}\Gamma^\mu_{~\beta\rho}+g_{\beta\mu}\Gamma^\mu_{~\alpha\rho}=\frac{\partial g_{\alpha \beta}}{\partial x^\rho}~.
\end{equation}
Taking a second differential of the right hand side of the above expression, we arrive at the following expression
\begin{equation}\label{Minimal.25}
\begin{split}
\frac{\partial^2 g_{\alpha\beta}}{\partial x^\lambda \partial x^\rho}&=\frac{\partial}{\partial x^\lambda}\left(g_{\alpha\mu}\Gamma^\mu_{~\beta\rho}+g_{\beta\mu}\Gamma^\mu_{~\alpha\rho}\right)\\
&=\frac{\partial g_{\alpha\mu}}{\partial x^\lambda}\Gamma^{\mu}_{~\beta \rho}+g_{\alpha\mu}\frac{\partial \Gamma^\mu_{~\beta\rho}}{\partial x^\lambda}+\frac{\partial g_{\beta\mu}}{\partial x^\lambda}\Gamma^{\mu}_{~\alpha \rho}+g_{\beta\mu}\frac{\partial \Gamma^\mu_{~\alpha\rho}}{\partial x^\lambda}\\
&\simeq \eta_{\alpha\mu}\left.\frac{\partial \Gamma^\mu_{~\beta\rho}}{\partial x^\lambda}\right\rvert_{\zeta_\tau}+\eta_{\beta\mu}\left.\frac{\partial \Gamma^\mu_{~\alpha\rho}}{\partial x^\lambda}\right\rvert_{\zeta_\tau}+\mathcal{O}(x,\mathfrak{s}) 
\end{split}
\end{equation}
where we have expanded the metric about the metric along the time-like geodesic $\zeta_\tau$ which is the Minkowski metric and we have also made use of the fact that along $\zeta_\tau$ all the Christoffel symbols vanish. Dropping terms which will contribute to $\mathcal{O}(x^3)$ terms in the metric tensor, we can write down the expression for the second derivative of the metric when $\alpha=\beta=0$, and $\rho=i$ and $\lambda=j$ as
\begin{equation}\label{Minimal.26}
\frac{\partial^2g_{00}}{\partial x^j\partial x^i}\simeq 2\eta_{0\mu}\left.\frac{\partial \Gamma^\mu_{~0 i}}{\partial x^j}\right\rvert_{\zeta_\tau}~.
\end{equation}
From eq.(\ref{Minimal.15}), we have $\left.\frac{\partial \Gamma^\mu_{~0 i}}{\partial x^j}\right\rvert_{\zeta_\tau}=\left.R^\mu_{~ij0}\right\rvert_{\zeta_\tau}$  and from this, we obtain $\left.\frac{\partial \Gamma^\mu_{~i0}}{\partial x^j}\right\rvert_{\zeta_\tau}=-\left.R^\mu_{~i0j}\right\rvert_{\zeta_\tau}$. This helps us to write down the expression in eq.(\ref{Minimal.26}) as
\begin{equation}\label{Minimal.27}
\begin{split}
\frac{\partial^2g_{00}}{\partial x^j\partial x^i}&\simeq -2\eta_{0\mu}\left.R^\mu_{~i0j}\right\rvert_{\zeta_\tau}\\
&=-2\left.R_{0i0j}\right\rvert_{\zeta_\tau}~.
\end{split}
\end{equation}
From the above expression, we can now proceed to obtain the $\{0,0\}$ component of the metric tensor upto $\mathcal{O}(x^2)$. Integrating the above expression single time, we obtain
\begin{equation}\label{Minimal.28}
\begin{split}
\frac{\partial g_{00}}{\partial x^i}&=-2\int\left.R_{0i0j}\right\rvert_{\zeta_\tau} dx^j+\mathfrak{C}_i
\end{split}
\end{equation}
where $\mathfrak{C}_i$ is a constant vector. If we check the entire equation along the time-like geodesic then using eq.(\ref{Minimal.24}), one can set $\mathfrak{C}_i=0$. We can further integrate eq.(\ref{Minimal.28}) to obtain
\begin{equation}\label{Minimal.29}
\begin{split}
g_{00}&=\mathfrak{C}_0-2\int dx^i\int dx^j \left.R_{0i0j}\right\rvert_{\zeta_\tau}\\
&=\mathfrak{C}_{0}-2\int dx^i x^j\left.R_{0i0j}\right\rvert_{\zeta_\tau}\\
&= \mathfrak{C}_0-\left.R_{0i0j}\right\rvert_{\zeta_\tau} x^i x^j
\end{split}
\end{equation}
where we have made use of the fact that along the time-like geodesic the Riemann-curvature tensor does not depend on the spatial degrees of freedom. The integration constant $\mathfrak{C}_0$ can be fixed by setting $x^i=0$, which requires $g_{00}$ to lie on the time-like geodesic making it equal to $\eta_{00}$. Hence, $\mathfrak{C}_0=\eta_{00}=-1$. As a result, we can write down the expression for $g_{00}$ in the Fermi-normal coordinates as
\begin{equation}\label{Minimal.30}
g_{00}=-1-\left.R_{0i0j}\right\rvert_{\zeta_\tau}x^i x^j~.
\end{equation}
Similarly, from eq.(\ref{Minimal.25}), one can write down two new expressions given by
\begin{align}
\frac{\partial^2g_{0i}}{\partial x^k\partial x^j}
&\simeq-\frac{4}{3}\left.R_{0jik}\right\rvert_{\zeta_\tau}~,~~\frac{\partial^2g_{ij}}{\partial x^l\partial x^k}
\simeq-\frac{2}{3}\left.R_{ikjl}\right\rvert_{\zeta_\tau}~.\label{Minimal.31}
\end{align}
Integrating both the equations in eq.(\ref{Minimal.31}), we arrive at the following two expressions given by
\begin{align}
g_{0i}&=-\frac{2}{3}\left.R_{0jik}\right\rvert_{\zeta_\tau}x^jx^k~,~~
g_{ij}=\delta_{ij}-\frac{1}{3}\left.R_{ikjl}\right\rvert_{\zeta_\tau}x^kx^l\label{Minimal.32}~.
\end{align}
Eq.(\ref{Minimal.30}) combined with eq.(\ref{Minimal.32}) gives the complete expression for the non-vanishing components of the metric tensor up to $\mathcal{O}(x^2)$ in the Fermi-normal coordinates. If one considers small spacetime fluctuations about the Minkowski background then under gauge transformations, the Riemann curvature tensor remains invariant. As a result the Riemann curvature tensor remains same in the transverse-traceless gauge when transformed from the Fermi-normal coordinates leaving the structure of the metric tensor in eq.(s)(\ref{Minimal.30},\ref{Minimal.32}) same. In the next section, we start with the background model and the path integral quantization technique for a freely-falling two particle system interacting with quantized gravitational wave in the generalized uncertainty principle framework.
\section{Background model and the path integral quantization of linearized gravity}
We start with the consideration of the basic model-system where the arm of an interferometric detector is modelled by a freely-falling two-atom system with one of the  particles being considerably heavier compared to the other one. The heavier mass particle follows a time-like geodesic with respect to which a coordinate system is needed to be implemented to measure the geodesic separation between the two particles over time. The interesting thing to note that this geodesics separation is representing the arm of an interferometric detector where the particle with the lighter mass models the suspending light mirror at the end of the interferometric detector. The first step is to obtain the model Lagrangian when the detector is interacting with gravitational waves when the gravitational fluctuations are quantized where the phase space variables corresponding to the detector part obeys the generalized uncertainty principle.
\subsection{Lagrangian for the gravitational wave-detector system}
\noindent To write down the Lagrangian for the model system, we need to consider the action for the gravitational fluctuation part and the detector part separately and then we shall combine them both to obtain the total action for the model system. We can in principle work in a curved background over which gravitational fluctuations will be considered as $g_{\mu\nu}=g^\oplus_{\mu\nu}+h_{\mu\nu}$ with $g^\oplus_{\mu\nu}$ denoting the metric corresponding to the Earth and $h_{\mu\nu}$ denoting the background fluctuations. However, as we shall see in the later part of this chapter, we will primarily consider detection using the future space-base gravitational wave observatories. (as has also been done in \cite{QGravNoise,QGravLett,QGravD,KannoSodaTokuda,
KannoSodaTokuda2}). It is therefore quite intuitive to consider the background metric as a small perturbation on the flat Minkowski background. The background metric for our model is then given by
\begin{equation}\label{QGMinimal.1.33}
g_{\mu\nu}=\eta_{\mu\nu}+h_{\mu\nu}
\end{equation} 
where the flat Minkowski metric reads $\eta_{\mu\nu}=\text{Diag}\{-1,1,1,1\}$. Now, the gravitational part of the action will be dictated by the Einstein-Hilbert action which reads\footnote{We set the speed of light to be unity which has been restored later.}
\begin{equation}\label{QGMinimal.2.34}
S_{\text{EH}}=\frac{1}{16\pi G}\int d^4 x \sqrt{-g} R
\end{equation}
with $R=g_{\mu\nu}R^{\mu\nu}$ being the Ricci scalar, $R^{\mu\nu}$ denoting the Ricci tensor, and $g$ denoting the determinant of the metric tensor $g=\text{det}(g_{\mu\nu})$. Substituting the analytical form of the metric tensor from eq.(\ref{QGMinimal.1.33}) in the Einstein Hilbert action, we obtain the form of the action by keeping terms up to quadratic order in the perturbation term $h_{\mu\nu}$ in eq.(\ref{QGMinimal.1.33}) as
\begin{equation}\label{QGMinimal.3.35}
\begin{split}
S_{\text{EH}}\simeq&\frac{1}{64\pi G}\int d^4x ~(h_{\mu\nu}\Box h^{\mu\nu}-h\Box h+2 h^{\mu\nu}\partial_\mu\partial_\nu h-2h_{\mu\alpha}\partial_\kappa\partial^{\alpha}h^{\mu\kappa})
\end{split}
\end{equation}
where $h\equiv \eta_{\mu\nu}h^{\mu\nu}$. We can now make use of the underlying gauge symmetry and express the perturbation term as
\begin{equation}\label{QGMinimal.4.36}
h_{\mu\nu}={h}^{\text{TT}}_{\mu\nu}+\partial_\mu\zeta_\nu+\partial_\nu\zeta_\mu~. 
\end{equation}
For notational simplicity, we now denote $h^{\text{TT}}$ as $\mathfrak{h}$.
To get rid of the redundant degrees of freedom, one can finally choose the transverse-traceless gauge as
\begin{align}\label{QGMinimal.5.37}
\partial_{\alpha}\mathfrak{h}^{\alpha\beta}=0~,\mathfrak{h}^{\alpha}_{~\alpha}=0~,~k_\beta\mathfrak{h}^{\beta\rho}=0
\end{align}
where a constant time-like vector $k_\beta$ is denoted by $k_\beta=\delta_{~\beta}^{0}$. In this transverse-traceless gauge the Einstein-Hilbert action in eq.(\ref{QGMinimal.2.34}) can be expressed as
\begin{equation}\label{QGMinimal.6.38}
S_{\text{EH}}=-\frac{1}{64\pi G}\int d^4 x~ \partial_{\rho}\mathfrak{h}_{jk}\partial^{\rho}\mathfrak{h}^{jk}.
\end{equation}
With the analytical form of the gravitational part of the action, we will now proceed to obtain the action for the detector part of the system which consists of a two-particle system where one of the particles is significantly more massive than the other one. One can consider the particle with the higher mass to be on-shell. In the Fermi-normal coordinates the origin of the coordinate system will move with the  heavier-mass particle following a time-like geodesic during the free-fall and the distance of the smaller-mass particle from the origin will represent the geodesic separation. In the Fermi-normal coordinates $x^i$ can thus be replaced by $\xi^i$ where $\xi$ denotes the coordinate separation between the two particles. From eq.(s)(\ref{Minimal.30},\ref{Minimal.32}), it is possible express the metric tensor as
\begin{align}
g_{00}(t,\xi)&=-1-R_{i0j0}(t,0)\xi^{i}\xi^{j}+\mathcal{O}(\xi^3)\label{QGMinimal.7.39}\\
g_{0k}(t,\xi)&=-\frac{2}{3}R_{0jkl}(t,0)\xi^{j}\xi^{l}+\mathcal{O}(\xi^3)\label{QGMinimal.8.40}\\
g_{jk}(t,\xi)&=\delta_{jk}-\frac{1}{3}R_{jlkp}(t,0)\xi^{l}\xi^{p}+\mathcal{O}(\xi^3)\label{QGMinimal.9.41}~.
\end{align}
Now, one can write down the relativistic action for the particle with the smaller mass $m_0$ as 
\begin{equation}\label{QGMinimal.10.42}
S_{\text{P}}=-m_0\int d\tau\sqrt{-g_{\mu\nu}\dot{\mathfrak{Y}}^{\mu}\dot{\mathfrak{Y}}^{\nu}}
\end{equation}
with $\mathfrak{Y}^\mu=\{t,\xi^j\}$ denoting the coordinates of the particle with mass $m_0$. In the above equation for the action, the dot represents derivative with respect to the proper time $\tau$. We can now recast the above action in the following way
\begin{equation}\label{QGMinimal.11.43}
\begin{split}
S_\text{P}&=-m_0\int d\tau\sqrt{-g_{\mu\nu}\frac{d\mathfrak{Y}^{\mu}}{d\tau}\frac{d\mathfrak{Y}^{\nu}}{d\tau}}\\
&=-m_0\int d\tau \sqrt{-g_{\mu\nu}\frac{d\mathfrak{Y}^{\mu}}{dt}\frac{d\mathfrak{Y}^{\nu}}{dt}\frac{dt}{d\tau}\frac{dt}{d\tau}}\\
&=-m_0\int dt \sqrt{-g_{\mu\nu}\frac{d\mathfrak{Y}^\mu}{dt}\frac{d\mathfrak{Y}^\nu}{dt}}
\end{split}
\end{equation}
which confirms that the action $S_{\text{P}}$ is reparametrization invariant. Substituting the analytical form of the metric tensor in the Fermi-normal coordinates from eq.(\ref{QGMinimal.7.39},\ref{QGMinimal.8.40},\ref{QGMinimal.9.41}), and keeping terms up to second order in $\xi$, we arrive at the modified form of the action as
\begin{equation}\label{QGMinimal.12.44}
\begin{split}
S_{\text{P}}&\simeq-m_0\int dt \left(1+\frac{1}{2}R_{i0j0}(t,0)\xi^i\xi^j-\frac{1}{2}\delta_{ij}\dot{\xi}^i\dot{\xi}^j\right)~.
\end{split}
\end{equation}
We have already discussed that the Riemann curvature tensor remains invariant in the transverse-traceless gauge which helps us to recast $R_{i0j0}(t,0)$ as $R_{i0j0}(t,0)\simeq -\frac{1}{2}\ddot{\mathfrak{h}}_{ij}(t,0)$. Using the analytical form of the curvature tensor in the transverse-traceless gauge and substituting it back in eq.(\ref{QGMinimal.12.44}), we can write down the analytical form of the detector part of the action as
\begin{equation}\label{QGMinimal.13.45}
S_\text{P}\simeq-m_0\int dt \left(1-\frac{1}{4}\ddot{\mathfrak{h}}_{jk}(t,0)\xi^j\xi^k-\frac{1}{2}\delta_{jk}\dot{\xi}^j\dot{\xi}^k\right)
\end{equation}
where it is important to note that the first term inside of the parenthesis does not contribute towards the overall dynamics of the particle. It is important to note that the action remains unchanged while going from Fermi normal coordinates to the transverse-traceless gauge. At first the gravitational perturbation in the transverse-traceless gauge needs to be decomposed into its individual Fourier modes and the mode decomposition reads \cite{QGravNoise,QGravLett,QGravD}
\begin{equation}\label{QGMinimal.14.46}
\mathfrak{h}_{ij}(t,\vec{x})=\frac{1}{l_\text{Pl}}\sum\limits_{\vec{k},s}q_s(\vec{k})e^{i\vec{k}\cdot\vec{x}}\epsilon^{s}_{ij}(\vec{k})
\end{equation}
with $q_s(\vec{k})$ denoting the Fourier mode amplitude corresponding to the mode with frequency $\omega=|\vec{k}|$ and the wave vector $\vec{k}$ being denoted as $\vec{k}=\frac{2\pi\vec{n}}{L}$ ($\vec{n}\in\mathbb{Z}^3$) when the system is quantized inside of a box of length $L$. Here, $\epsilon^{s}_{ij}(\vec{k})$ denotes the polarization tensor with $s=+,\times$ giving the two distinct polarizations of the gravitational wave. The polarization tensor satisfies the following conditions
\begin{align}\label{QGMinimal.15.47}
\epsilon^s_{jp}(\vec{k})\epsilon^{jp}_{\bar{s}}(\vec{k})&=\delta^s_{~\bar{s}},~
\delta^{jp}\epsilon^s_{jp}(\vec{k})=0,~\text{and}~
k^j\epsilon^s_{jp}(\vec{k})=0
\end{align}
where the first relation gives the normalization condition, the second relation gives the tracelessness condition, and the final equation gives the transversality of the polarization tensor. Substituting the discrete mode decomposition of the gravitational fluctuation from eq.(\ref{QGMinimal.14.46}) in the gravitational part as well as the detector part of the action in eq.(s)(\ref{QGMinimal.6.38},\ref{QGMinimal.13.45}), and making use of the tracelessness, transversality,  and normalization conditions for the polarization tensor from eq.(\ref{QGMinimal.15.47}), one can write the full gauge fixed action combining both the detector and gravitational wave part as \cite{QGravD} 
\begin{equation}\label{QGMinimal.16.48}
\begin{split}
S=&S_{\text{EH}}+S_\text{P}\\=&\frac{m}{2}\sum\limits_{\vec{k},s}\int dt\left(\dot{q}_{s}^2(\vec{k})-\vec{k}^2q^2_{s}(\vec{k})\right)+\frac{m_0}{2}\int dt\left(\delta_{ij}\dot{\xi}^i\dot{\xi}^j-\frac{1}{\sqrt{\hbar G}}\sum\limits_{\vec{k},s}\dot{q}_s(\vec{k})\epsilon^{s}_{ij}(\vec{k})\dot{\xi}^{i}\xi^{j}\right)
\end{split}
\end{equation}
where the effective mass $m$ corresponding to the gravitational part is defined as $m\equiv\frac{L^3}{16\pi\hbar G^2}$.  We consider the gravitational wave to propagate along the $z$ axis with magnitude $\omega=|\vec{k}|$ and as a result the $xy$ plane denotes the polarization plane. One can now consider a single mode of the gravitational wave and the polarization can be restricted to plus polarization only where one can also consider the fluctuation of the physical separation between the two masses to be zero in the $y$ and $z$ direction. The modified action for the detector-gravitational wave system can then be expressed from eq.(\ref{QGMinimal.16.48}) as
\begin{equation}\label{QGMinimal.17.49}
S_{\mathcal{A}}=\int dt\left(\frac{m}{2}\left(\dot{q}^2-\omega^2q^2\right)+\frac{m_0}{2}\left({\dot{\xi}^\text{H}}\right)^2-\mathfrak{g}\dot{q}\dot{\xi}^{\text{H}}\xi^{\text{H}}\right)
\end{equation}  
where $\xi_{x}=\xi^\text{H}$ and $q_+(k_z)=q$. In the above equation $\mathfrak{g}=\frac{m_0}{2l_{\text{Pl}}}$ denotes the detector-field coupling constant. In our model, we consider that the detector phase space operators obey the generalized commutator in eq.(\ref{Minimal.1}) and then the action in eq.(\ref{QGMinimal.17.49}) takes the form
\begin{equation}\label{QGMinimal.18.50}
S^{\text{G}}_{\mathcal{A}}=\int dt\left(\frac{m}{2}\left(\dot{q}^2-\omega^2q^2\right)+\frac{m_0}{2}\left({\dot{\xi}^\text{G}}\right)^2-\mathfrak{g}\dot{q}\dot{\xi}^{\text{G}}\xi^{\text{G}}\right)~.
\end{equation}
From the above equation, we can write down the Lagrangian for the model system as
\begin{equation}\label{QGMinimal.19.51}
L=\frac{m}{2}\left(\dot{q}^2-\omega^2q^2\right)+\frac{m_0}{2}\left({\dot{\xi}^\text{G}}\right)^2-\mathfrak{g}\dot{q}\dot{\xi}^{\text{G}}\xi^{\text{G}}~.
\end{equation}
Using the above Lagrangian, one can obtain the conjugate momentum to the position corresponding to the gravitational part $q$ as well as the conjugate momentum to the position corresponding to the detector part $\xi^{\text{G}}$ as
\begin{equation}\label{QGMinimal.20.52}
p=\frac{\partial L}{\partial \dot{q}}=m\dot{q}-\mathfrak{g}\dot{\xi}^{\text{G}}\xi^{\text{G}}~\text{and}~~\pi^{\text{G}}=\frac{\partial L}{\partial \dot{\xi}^{\text{G}}}=m_0\dot{\xi}^{\text{G}}-\mathfrak{g}\dot{q}\xi^{\text{G}}~.
\end{equation}
From the Lagrangian for the model system in eq.(\ref{QGMinimal.19.51}) and the analytical forms of the conjugate momentum in the above equation, we obtain the Hamiltonian for the model system as
\begin{equation}\label{QGMinimal.21.53}
\begin{split}
H&=\left(\frac{p^2}{2m}+\frac{{\pi^\text{G}}^2}{2m_0}+\frac{\mathfrak{g}p\pi^\text{G}\xi^\text{G}}{mm_0}\right)\left(1-\frac{\mathfrak{g}^2{\xi^\text{G}}^2}{mm_0}\right)^{-1}+\frac{1}{2}m\omega^2q^2~.
\end{split}
\end{equation}
To quantize the model system one needs to raise the position and corresponding conjugate momentum variables to operator status. As here, we shall be using a path-integral quantization technique, we shall not explicitly use the operatorial notation. The position $\xi^\text{G}$ and momenta $\pi^\text{G}$ in the operator representation are expressed as $\hat{\xi}^\text{G}$ and $\hat{\pi}^{\text{G}}$ which denotes the GUP modified position and momentum operators. The GUP modified operators in terms of the unmodified phase space operators are expressed as $\hat{\xi}^\text{G}=\hat{\xi}^\text{H}$ and $\hat{\pi}^\text{G}=\hat{\pi}^\text{H}(1+\beta {\hat{\pi}^\text{H}}^2)$ as has also been discussed in eq.(\ref{Minimal.4}). Using the standard notation (instead of the operatorial notation), we can express the Hamiltonian in eq.(\ref{QGMinimal.21.53}) using the expression for the GUP modified position and momentum operators corresponding to the detector in terms of the standard canonically conjugate operators obeying the Heisenberg's uncertainty relation up to first order in the GUP parameter as
\begin{equation}\label{QGMinimal.22.54}
H=\frac{\frac{p^2}{2m}+\frac{{\pi^\text{H}}^2}{2m_0}+\frac{\mathfrak{g}p\pi^\text{H}\xi^\text{H}}{mm_0}+\beta\left(\frac{{\pi^\text{H}}^4}{m_0}+\frac{\mathfrak{g}p{\pi^\text{H}}^3\xi^\text{H}}{mm_0}\right)}{1-\frac{\mathfrak{g}^2{\xi^\text{H}}^2}{mm_0}}+\frac{1}{2}m\omega^2q^2~.
\end{equation}
In order to denote the model system, we shall be using the $\text{s}$ in the superscript instead of the $\text{H}$ in the superscript. Hence, we shall be using $\{\xi^{\text{s}},\pi^{\text{s}}\}$ instead of $\{\xi^{\text{H}},\pi^{\text{H}}\}$. With the analytical form of the model Hamiltonian in hand, we are now in a position to obtain the transition probability of the model system to go from an initial state to some final state.
\subsection{The analytical form of the transition probability and the Feynman-Vernon influence functional}
\noindent At infinite past ($t=-\infty$), it can be safely assumed that the detector state along with the gravitational wave state forms a tensor product state. It can also be intuitively argued as even before the interaction was switched on, the gravitational fluctuation (or field) was generated. To truly give control over the model system, one can redefine the coupling constant $\mathfrak{g}$ as $\mathfrak{g}\rightarrow\mathfrak{g}(t)=\mathfrak{s}(t)\mathfrak{g}$, where the time-dependent switching function $\mathfrak{s}(t)$ is expressed as
\begin{equation}\label{QGMinimal.23.55}
\mathfrak{s}(t)=\left.\begin{matrix}
0&&t\leq-\Delta ~\text{and}~t\geq T+\Delta\\
1&& 0\leq t\leq T
\end{matrix}\right\}
\end{equation}
where $\Delta\ll T$. The switching function makes sure that the time-dependent coupling $\mathfrak{g}(t)$ is being switched on and off adiabatically. Consider that at some time $t=-\Delta$, the particle (or the detector) is in a state $|\phi_i^\text{s}\rangle$ whereas the gravitational wave mode is in a state $|\psi^\text{g}_\omega\rangle$. It is evident from eq.(\ref{QGMinimal.23.55}) that the interaction happens in the time interval $t\in[0,T]$. As a result, at a later time $t=T+\Delta$, the system can also be considered as a tensor product between the final state of the detector and the state corresponding to the gravitational wave mode. One can consider the final state of the detector at time $t=t+\Delta$ to be $|\phi^\text{s}_f\rangle$ and the gravitational wave mode to be in the Harmonic oscillator state $|\mathcal{F}^\text{g}\rangle$. Hence, for an interaction between the gravitational fluctuations and the detector in the time interval $0\leq t\leq T$, one can write down the transition probability for the system going from an initial state $|\psi^\text{g}_{\omega},\phi^\text{s}_{i}\rangle\equiv |\psi^\text{g}_{\omega}\rangle\otimes|\phi^\text{s}_{i}\rangle$ to the final state $|\mathcal{F}^\text{g},\phi^\text{s}_{f}\rangle\equiv |\mathcal{F}^\text{g}\rangle\otimes|\phi^\text{s}_{f}\rangle$ as
\begin{equation}\label{QGMinimal.24.56}
\begin{split}
P_{\psi^\text{g}_{\omega}}^{[\phi^\text{s}_i\rightarrow\phi^\text{s}_f]}=&\sum\limits_{\lvert\mathcal{F}^{\text{g}}\rangle}\left|\langle \mathcal{F}^\text{g},\phi^\text{s}_f|\hat{U}(T+\Delta,-\Delta)|\psi^\text{g}_\omega,\phi^\text{s}_i\rangle\right|^2\\
=&\sum\limits_{\lvert\mathcal{F}^\text{g}\rangle}\langle\psi^\text{g}_\omega,\phi^\text{s}_i |\hat{U}^\dagger(T+\Delta,-\Delta)|\mathcal{F}^\text{g},\phi^\text{s}_f\rangle\langle \mathcal{F}^\text{g},\phi^\text{s}_f|\hat{U}(T+\Delta,-\Delta)|\psi^\text{g}_\omega,\phi^\text{s}_i\rangle~.
\end{split}
\end{equation}
Inserting the completeness relation corresponding to the base kets of the detector and the gravitational part at the initial and final times\footnote{ The eight completeness relations inserted are $\int dq_i|q_i\rangle\langle q_i|=\hat{\mathbb{1}}$, $\int dq_f|q_f\rangle\langle q_f|=\hat{\mathbb{1}}$, $\int d\xi^\text{s}_i|\xi^\text{s}_i\rangle\langle \xi^\text{s}_i|=\hat{\mathbb{1}}$, and $\int d\xi^\text{s}_f|\xi^\text{s}_f\rangle\langle \xi^\text{s}_f|=\hat{\mathbb{1}}$, and their primed counterparts.}, we can write down the above transition probability as
\begin{equation}\label{QGMinimal.25.57}
\begin{split}
P_{\psi^\text{g}_{\omega}}^{[\phi^\text{s}_i\rightarrow\phi^\text{s}_f]}=&\sum\limits_{\lvert\mathcal{F}^\text{g}\rangle}\idotsint d\xi^\text{s}_id\acute{\xi}^\text{s}_id\xi^\text{s}_fd\acute{\xi}^\text{s}_fdq_id\acute{q}_idq_fd\acute{q}_f\langle\psi^\text{g}_\omega,\phi^\text{s}_i |\acute{q}_i,\acute{\xi}^\text{s}_i\rangle\langle \acute{q}_i,\acute{\xi}^\text{s}_i|\hat{U}^\dagger(T+\Delta,-\Delta)|\acute{q}_f,\acute{\xi}^\text{s}_f\rangle\\&\times\langle \acute{q}_f,\acute{\xi}^\text{s}_f|\mathcal{F}^\text{g},\phi^\text{s}_f\rangle\langle \mathcal{F}^\text{g},\phi^\text{s}_f|q_f,\xi^\text{s}_f\rangle\langle q_f,\xi^\text{s}_f|\hat{U}(T+\Delta,-\Delta)|q_i,\xi^\text{s}_i\rangle\langle q_i,\xi^\text{s}_i|\psi^\text{g}_\omega,\phi^\text{s}_i\rangle\\
=&\idotsint d\xi^\text{s}_id\acute{\xi}^\text{s}_id\xi^\text{s}_fd\acute{\xi}^\text{s}_fdq_id\acute{q}_idq_fd\acute{q}_f\langle\psi^\text{g}_\omega|\acute{q}_i\rangle\langle\phi^\text{s}_i |\acute{\xi}^\text{s}_i\rangle\langle \acute{q}_i,\acute{\xi}^\text{s}_i|\hat{U}^\dagger(T+\Delta,-\Delta)|\acute{q}_f,\acute{\xi}^\text{s}_f\rangle\\&\times\langle \acute{\xi}^\text{s}_f|\phi^\text{s}_f\rangle\langle \phi^\text{s}_f|\xi^\text{s}_f\rangle\delta(q_f-\acute{q}_f)\langle q_f,\xi^\text{s}_f|\hat{U}(T+\Delta,-\Delta)|q_i,\xi^\text{s}_i\rangle\langle q_i|\psi^\text{g}_\omega\rangle\langle \xi^\text{s}_i|\phi^\text{s}_i\rangle
\end{split}
\end{equation}
where in the last-line of the above equation, we have made use of the fact that the state $|q_f\rangle$ is orthogonal to $|\acute{q}_f\rangle$ () and the completeness of the final states of the gravitational wave modes (or gravitons), considering that they create a complete set of eigenstates such that $\sum_{|\mathcal{F}^\text{g}\rangle}|\mathcal{F}^\text{g}\rangle\langle\mathcal{F}^\text{g}|=\hat{\mathbb{1}}$. Now using the representation of the overlap between base kets (or bras) and the state bras (or kets) corresponding to the detector part as well as the graviton part, we can recast the above expression for the transition probability in eq.(\ref{QGMinimal.25.57}) after executing the $\acute{q}_f$ integral as\footnote{Here the multiple integrals $\idotsint$ is replaced by a single integral $\int$ sign to avoid cluttering in the expression for the transition probability.}
\begin{equation}\label{QGMinimal.26.58} 
\begin{split}
P_{\psi^\text{g}_{\omega}}^{[\phi^\text{s}_i\rightarrow\phi^\text{s}_f]}=&\int dq_id\acute{q}_idq_fd\xi^\text{s}_id\acute{\xi}^\text{s}_id\xi^\text{s}_fd\acute{\xi}^\text{s}_f\psi^\text{g}_\omega(q_i)\psi^{\text{g}^*}_\omega(\acute{q}_i)\phi^\text{s}_i(\xi^\text{s}_i)\phi^{\text{s}^*}_i(\acute{\xi}^\text{s}_i)\phi^{\text{s}^*}_f(\xi^\text{s}_f)\phi^\text{s}_f(\acute{\xi}^\text{s}_f)\\&\times\langle \acute{q}_i,\acute{\xi}^\text{s}_i|\hat{U}^\dagger(T+\Delta,-\Delta)|q_f,\acute{\xi}^\text{s}_f\rangle\langle q_f,\xi^\text{s}_f|\hat{U}(T+\Delta,-\Delta)|q_i,\xi^\text{s}_i\rangle~.
\end{split}
\end{equation}
The important thing to note that, all of the final graviton states have been summed over as after the graviton is absorbed and released by the detector, it is no longer important to look at the final graviton state as we can only observe the change in the detector when the gravitons get released. Obtaining the form of the transition probability now boils down to the calculation of the two propagators $\langle \acute{q}_i,\acute{\xi}^\text{s}_i|\hat{U}^\dagger(T+\Delta,-\Delta)|q_f,\acute{\xi}^\text{s}_f\rangle$ and $\langle q_f,\xi^\text{s}_f|\hat{U}(T+\Delta,-\Delta)|q_i,\xi^\text{s}_i\rangle$. In a path integral representation, one can now  express the propagator $\langle q_f,\xi_f|\hat{U}(T+\Delta,-\Delta)|q_i,\xi_i\rangle$ as
\begin{equation}\label{QGMinimal.27.59}
\langle q_f,\xi^\text{s}_f|\hat{U}(T+\Delta,-\Delta)|q_i,\xi^\text{s}_i\rangle=\int\left[\mathcal{D}\xi^\text{s}\right]_{\xi^\text{s}_i,-\Delta}^{\xi^\text{s}_f,T+\Delta}\left[\mathcal{D}q\right]_{q_i,-\Delta}^{q_f,T+\Delta}\mathcal{D}\pi^\text{s}\mathcal{D}p~e^{\frac{i}{\hbar}\int_{-\Delta}^{T+\Delta}dt\left[\pi^\text{s}\dot{\xi}^\text{s}+p\dot{q}-H_{\beta}\left(q,p,\xi^\text{s},\pi^\text{s}\right)\right]}
\end{equation}
where the Hamiltonian $H_\beta(q,p,\xi^\text{s},\pi^\text{s})$
is given in eq.(\ref{QGMinimal.22.54}) and has the analytical form
\begin{equation}\label{QGMinimal.28.60}
\begin{split}
H_\beta(q,p,\xi^\text{s},\pi^\text{s})=\frac{\frac{p^2}{2m}+\frac{{\pi^\text{s}}^2}{2m_0}+\frac{\mathfrak{g}p\pi^\text{s}\xi^\text{s}}{mm_0}+\beta\left(\frac{{\pi^\text{s}}^4}{m_0}+\frac{\mathfrak{g}p{\pi^\text{s}}^3\xi^\text{s}}{mm_0}\right)}{1-\frac{\mathfrak{g}^2{\xi^\text{s}}^2}{mm_0}}+\frac{1}{2}m\omega^2q^2~.
\end{split}
\end{equation}
Executing the path integral over $\pi^\text{s}$ in eq.(\ref{QGMinimal.27.59}), we can recast the propagator $\langle q_f,\xi^\text{s}_f|\hat{U}(T+\Delta,-\Delta)|q_i,\xi^\text{s}_i\rangle$ as
\begin{equation}\label{QGMinimal.29.61}
\begin{split}
&\langle q_f,\xi^\text{s}_f|\hat{U}(T+\Delta,-\Delta)|q_i,\xi^\text{s}_i\rangle=\int \tilde{\mathcal{D}}\xi^\text{s}~e^{\frac{im_0}{2\hbar}\int_{-\Delta}^{T+\Delta}dt\left(\dot{\xi}^{\text{s}^2}-2\beta m_0^2\dot{\xi}^{\text{s}^4}\right)}\langle q_f|\hat{U}^{\beta}_{\xi^\text{s}}(T+\Delta,-\Delta)|q_i\rangle
\end{split}
\end{equation}
where $\langle q_f|\hat{U}^{\beta}_{\xi^\text{s}}(T+\Delta,-\Delta)|q_i\rangle$ is defined as
\begin{equation}\label{QGMiniam.30.62}
\langle q_f|\hat{U}^{\beta}_{\xi^\text{s}}(T+\Delta,-\Delta)|q_i\rangle\equiv \int\mathcal{D}q\mathcal{D}p ~e^{\frac{i}{\hbar}\int_{-\Delta}^{T+\Delta}dt\left(p\dot{q}-H_{\beta\xi^\text{s}}(q,p)\right)}
\end{equation}
with the analytical form of the modified Hamiltonian $H_{\beta\xi^\text{s}}(q,p)$ being given as
\begin{equation}\label{QGMinimal.31.63}
H_{\beta\xi^\text{s}}(q,p)=\frac{\left(p+\mathfrak{g}\xi^\text{s}\dot{\xi}^\text{s}(1-3\beta m_0^2\dot{\xi}^{\text{s}^2})\right)^2}{2m}+\frac{1}{2}m\omega^2q^2~.
\end{equation}
The important thing to observe is that the $\xi^\text{s}$ dependent terms after the execution of the path integral over $\pi^\text{s}$ in eq.(\ref{QGMinimal.27.59}) has been absorbed in the path integral measure $\mathcal{D}\xi^\text{s}$ and one can write it as $\mathcal{D}\xi^\text{s}\rightarrow \tilde{\mathcal{D}}\xi^\text{s}$ as has been done in eq.(\ref{QGMinimal.29.61}) \cite{QGravD}. Similarly, we can write down the other propagator $\langle \acute{q}_i,\acute{\xi}^\text{s}_i|\hat{U}^\dagger(T+\Delta,-\Delta)|q_f,\acute{\xi}^\text{s}_f\rangle$ as
\begin{equation}\label{QGMinimal.32.64}
\begin{split}
&\langle \acute{q}_i,\acute{\xi}^\text{s}_i|\hat{U}^\dagger(T+\Delta,-\Delta)|q_f,\acute{\xi}^\text{s}_f\rangle=\int \tilde{\mathcal{D}}\acute{\xi}^\text{s}~e^{-\frac{im_0}{2\hbar}\int_{-\Delta}^{T+\Delta}\left(\dot{\acute{\xi}}^{\text{s}^2}-2\beta m_0^2\dot{\acute{\xi}}^{\text{s}^4}\right)}\langle \acute{q}_i|\hat{U}_{\acute{\xi}^\text{s}}^{\beta^\dagger}(T+\Delta,-\Delta)|q_f\rangle~.
\end{split}
\end{equation}
We can now really simplify the form of the transition probability in eq.(\ref{QGMinimal.26.58}) by executing the `$q$' integrals. We start with simplifying the following integral
\begin{equation}\label{QGMinimal.33.65}
\begin{split}
&\int dq_id\acute{q}_idq_f~\psi^\text{g}_\omega(q_i)\psi^{\text{g}^*}_\omega(\acute{q}_i)\langle \acute{q}_i,\acute{\xi}^\text{s}_i|\hat{U}^\dagger(T+\Delta,-\Delta)|q_f,\acute{\xi}^\text{s}_f\rangle\langle q_f,\xi^\text{s}_f|\hat{U}(T+\Delta,-\Delta)|q_i,\xi^\text{s}_i\rangle\\
=&\int\tilde{\mathcal{D}}\xi^\text{s}\tilde{\mathcal{D}}\acute{\xi}^\text{s}e^{\frac{im_0}{2\hbar}\int_{-\Delta}^{T+\Delta}dt\left[\left[\dot{\xi}^{\text{s}^2}-\dot{\acute{\xi}}^{\text{s}^2}\right]-2\beta m_0^2\left[\dot{\xi}^{\text{s}^4}-\dot{\acute{\xi}}^{\text{s}^4}\right]\right]}\langle \psi_\omega^\text{g}|\int d\acute{q}_i |\acute{q}_i\rangle\langle \acute{q}_i|\hat{U}^{\beta^\dagger}_{\acute{\xi}^\text{s}}(T+\Delta,-\Delta)\int dq_f |q_f\rangle\\
\times&\langle q_f|\hat{U}^{\beta}_{\xi^\text{s}}(T+\Delta,-\Delta)\int dq_i|q_i\rangle\langle q_i|\psi_\omega^\text{g}\rangle\\
=&\int\tilde{\mathcal{D}}\xi^\text{s}\tilde{\mathcal{D}}\acute{\xi}^\text{s}e^{\frac{im_0}{2\hbar}\int_{-\Delta}^{T+\Delta}dt\left[\left[\dot{\xi}^{\text{s}^2}-\dot{\acute{\xi}}^{\text{s}^2}\right]-2\beta m_0^2\left[\dot{\xi}^{\text{s}^4}-\dot{\acute{\xi}}^{\text{s}^4}\right]\right]}\langle \psi_\omega^\text{g}|\hat{U}^{\beta^\dagger}_{\acute{\xi}^\text{s}}(T+\Delta,-\Delta)\hat{U}^{\beta}_{\xi^\text{s}}(T+\Delta,-\Delta)|\psi_\omega^\text{g}\rangle
\end{split}
\end{equation}
where in the last line of the above equation, we have made use of the completeness relation for the $|q\rangle$ states. Substituting the above analytical expression for the integral in eq.(\ref{QGMinimal.33.65}) and substituting it back in eq.(\ref{QGMinimal.26.58}), we can write down the simplified analytical form of the transition probability as 
\begin{equation}\label{QGMinimal.34.66}
\begin{split}
P_{\psi^\text{g}_{\omega}}^{[\phi^\text{s}_i\rightarrow\phi^\text{s}_f]}&=\int d\xi^\text{s}_id\acute{\xi}^\text{s}_id\xi^\text{s}_fd\acute{\xi}^\text{s}_f\phi^\text{s}_i(\xi^\text{s}_i)\phi_i^{\text{s}^*}(\acute{\xi}^\text{s}_i)\phi^{\text{s}^*}_f(\xi^\text{s}_f)\phi^\text{s}_f(\acute{\xi}^\text{s}_f)\\&\times\int\left[\tilde{\mathcal{D}}\xi^\text{s}\right]_{\xi^\text{s}_i,-\Delta}^{\xi^\text{s}_f,T+\Delta}\left[\tilde{\mathcal{D}}\acute{\xi}^\text{s}\right]_{\acute{\xi}^\text{s}_i,-\Delta}^{\acute{\xi}^\text{s}_f,T+\Delta}~e^{\frac{im_0}{2\hbar}\int_{-\Delta}^{T+\Delta}dt\left[\left[\dot{\xi}^{\text{s}^2}-\dot{\acute{\xi}}^{\text{s}^2}\right]-2\beta m_0^2\left[\dot{\xi}^{\text{s}^4}-\dot{\acute{\xi}}^{\text{s}^4}\right]\right]} F^{\beta}_{\psi^\text{g}_\omega}\left[\xi^\text{s},\acute{\xi}^\text{s}\right]
\end{split}
\end{equation}
where $F^{\beta}_{\psi^\text{g}_\omega}\left[\xi^\text{s},\acute{\xi}^\text{s}\right]$ denotes the Feynman-Vernon influence functional \cite{FeynmanVernon} which is given by 
\begin{equation}\label{QGMinimal.35.67}
F^{\beta}_{\psi^\text{g}_\omega}\left[\xi^\text{s},\acute{\xi}^\text{s}\right]=\langle \psi_\omega^\text{g}|\hat{U}^{\beta^\dagger}_{\acute{\xi}^\text{s}}(T+\Delta,-\Delta)\hat{U}^{\beta}_{\xi^\text{s}}(T+\Delta,-\Delta)|\psi_\omega^\text{g}\rangle~.
\end{equation}
Inspecting the analytical form of the transition probability carefully, we find out that the $F^{\beta}_{\psi^\text{g}_\omega}\left[\xi^\text{s},\acute{\xi}^\text{s}\right]$ term only captures the influence of the gravitons on the detector and as a result it is termed as the influence functional. We shall now be explicitly analyzing the influence functional in the next section.
\section{The influence functional analysis}
In this section, we shall be explicitly analyze the influence functional in eq.(\ref{QGMinimal.35.67}) and obtain its explicit form for the gravitons being in different initial states. At first, it is evident that one can separate the unitary time evolution operator as
\begin{equation}\label{Influence.1.68}
\hat{U}^{\beta}_{\xi^\text{s}}(T+\Delta,-\Delta)=\hat{U}^{\beta}_{\xi^\text{s}}(T+\Delta,T)\hat{U}^{\beta}_{\xi^\text{s}}(T,0)\hat{U}^{\beta}_{\xi^\text{s}}(0,-\Delta)~.
\end{equation}
The important thing to remember is that the switching here is an adiabatic one and one needs to therefore work with instantaneous eigenstates of the Hamiltonian as soon as the interaction starts. Now eigenstates corresponding to the Hamiltonian in eq.(\ref{QGMinimal.31.63}) can be obtained by shifting the momentum $p$ by $p+\mathfrak{g}\xi^\text{s}\dot{\xi}^\text{s}(1-3\beta m_0^2\dot{\xi}^{\text{s}^2})$ for the eigenstate of a simple harmonic oscillator with Hamiltonian $H_0=\frac{p^2}{2m}+\frac{1}{2}m\omega^2q^2$. As has also been discussed earlier in \cite{QGravD}, position operators generate shifts in the momentum space and as a result one can write down the analytical forms of $\hat{U}^{\beta}_{\xi^\text{s}}(0,-\Delta)$ and $\hat{U}^{\beta}_{\xi^\text{s}}(T+\Delta,T)$ as
\begin{align}
\hat{U}^{\beta}_{\xi^\text{s}}(0,-\Delta)&=e^{-\frac{i}{\hbar}\hat{q}\mathfrak{g}\xi^\text{s}(0)\dot{\xi}^\text{s}(0)\left(1-3\beta m_0^2\dot{\xi}^{\text{s}^2}(0)\right)}e^{-\frac{i}{\hbar}\hat{H}_0\Delta}\label{Influence.2.69}\\
\hat{U}^{\beta}_{\xi^\text{s}}(T+\Delta,T)&=e^{-\frac{i}{\hbar}\hat{H}_0\Delta}e^{\frac{i}{\hbar}\hat{q}\mathfrak{g}\xi^\text{s}(T)\dot{\xi}^\text{s}(T)\left(1-3\beta m_0^2\dot{\xi}^{\text{s}^2}(T)\right)}\label{Influence.3.70}~.
\end{align}
Further defining the graviton state of the system as
$|\tilde{\psi}_\omega^\text{g}\rangle=e^{-\frac{i}{\hbar}\hat{H}\Delta}|\psi^\text{g}_\omega\rangle$ where $|\tilde{\psi}_\omega^\text{g}\rangle$ denotes the initial state of the graviton at $t=0$. The analytical form of the influence functional in eq.(\ref{QGMinimal.35.67}) can be recast as 
\begin{equation}\label{Influence.4.71}
\begin{split}
F^{\beta}_{\psi^\text{g}_\omega}\left[\xi^\text{s},\acute{\xi}^\text{s}\right]=&\langle \psi_\omega^\text{g}|\hat{U}^{\beta^\dagger}_{\acute{\xi}^\text{s}}(T+\Delta,-\Delta)\hat{U}^{\beta}_{\xi^\text{s}}(T+\Delta,-\Delta)|\psi_\omega^\text{g}\rangle\\
=&\langle \psi_\omega^\text{g}|\hat{U}^{\beta^\dagger}_{\acute{\xi}^\text{s}}(0,-\Delta)\hat{U}^{\beta^\dagger}_{\acute{\xi}^\text{s}}(T,0)\hat{U}^{\beta^\dagger}_{\acute{\xi}^\text{s}}(T+\Delta,T)\hat{U}^{\beta}_{\xi^\text{s}}(T+\Delta,T)\hat{U}^{\beta}_{\xi^\text{s}}(T,0)\hat{U}^{\beta}_{\xi^\text{s}}(0,-\Delta)|\psi^\text{g}_\omega\rangle\\
=&\langle\tilde{\psi}^\text{g}_\omega|e^{\frac{i}{\hbar}\hat{q}\mathfrak{g}\acute{\xi}^\text{s}(0)\dot{\acute{\xi}}^\text{s}(0)\left(1-3\beta m_0^2\dot{\acute{\xi}}^{\text{s}^2}(0)\right)}\hat{U}^{\beta^\dagger}_{\acute{\xi}^\text{s}}(T,0)e^{-\frac{i}{\hbar}\hat{q}\mathfrak{g}\acute{\xi}^\text{s}(T)\dot{\acute{\xi}}^\text{s}(T)\left(1-3\beta m_0^2\dot{\acute{\xi}}^{\text{s}^2}(T)\right)}\\
&\times e^{\frac{i}{\hbar}\hat{q}\mathfrak{g}\xi^\text{s}(T)\dot{\xi}^\text{s}(T)\left(1-3\beta m_0^2\dot{\xi}^{\text{s}^2}(T)\right)}\hat{U}^{\beta}_{\xi^\text{s}}(T,0)e^{-\frac{i}{\hbar}\hat{q}\mathfrak{g}\xi^\text{s}(0)\dot{\xi}^\text{s}(0)\left(1-3\beta m_0^2\dot{\xi}^{\text{s}^2}(0)\right)}|\tilde{\psi}^\text{g}_\omega\rangle~.
\end{split}
\end{equation}
The next step is to reduce the path integral of the geodesics separations from $t\in[-\Delta,T+\Delta]$ to $t\in[0,T]$. We start with considering the path integral
\begin{equation}\label{Influence.5.72}
\mathcal{I}_\mathcal{P_1}=\int\left[\tilde{\mathcal{D}}\xi^\text{s}\right]_{\xi^\text{s}_i,-\Delta}^{\xi^\text{s}_f,T+\Delta}e^{\frac{im_0}{2\hbar}\int_{-\Delta}^{T+\Delta}dt\left(\dot{\xi}^{\text{s}^2}-2\beta m_0^2\dot{\xi}^{\text{s}^4}\right)}
\end{equation}
where $\xi_i^\text{s}=\xi^\text{s}(-\Delta)$ and $\xi_f^\text{s}=\xi^\text{s}(T+\Delta)$. Total time interval is $T+\Delta-(-\Delta)=T+2\Delta$. Here, we break $T$ in $N$ intervals such that $T=N\Delta t$ with $\Delta t=\varepsilon$ denoting the small time segments and we break $\Delta$ into $n$ time intervals. The path integral in eq.(\ref{Influence.5.72}) can be expressed in the discrete form as
\begin{equation}\label{Influence.6.73}
\begin{split}
\mathcal{I}^{\text{D}}_{\mathcal{P}_1}&=\int \mathfrak{N}(\xi^\text{s})d\xi^{\text{s}}_{N+n-1}\cdots d\xi^{\text{s}}_{N+1}d\xi^{\text{s}}_{N}d\xi^{\text{s}}_{N-1}\cdots d\xi^{\text{s}}_{1}d\xi^{\text{s}}_{0}d\xi^{\text{s}}_{-1}\cdots d\xi^{\text{s}}_{-n+1}\\
&\times\exp\left[\sum\limits_{k=-n}^{N+n-1}\frac{i\varepsilon m_0}{2\hbar}\left[\left(\frac{\xi^{\text{s}}_{k+1}-\xi^{\text{s}}_{k}}{\varepsilon}\right)^2-2\beta m_0^2\left(\frac{\xi^{\text{s}}_{k+1}-\xi^{\text{s}}_{k}}{\varepsilon}\right)^4\right]\right]
\end{split}
\end{equation}
where $\mathfrak{N}(\xi^\text{s})$ is a function of $\xi^\text{s}$. One can now recast the above integral as
\begin{equation}\label{Influence.7.74}
\begin{split}
\mathcal{I}^\text{D}_{\mathcal{P}_1}&= \int d\xi^{\text{s}}_{N}\int d\xi^{\text{s}}_{0} \int \mathfrak{N}_0(\xi^\text{s})d\xi^{\text{s}}_{N+n-1}\cdots d\xi^{\text{s}}_{N+1}e^{\sum\limits_{k=N}^{N+n-1}\frac{i\varepsilon m_0}{2\hbar}\left[\left(\frac{\xi^{\text{s}}_{k+1}-\xi^{\text{s}}_{k}}{\varepsilon}\right)^2-2\beta m_0^2\left(\frac{\xi^{\text{s}}_{k+1}-\xi^{\text{s}}_{k}}{\varepsilon}\right)^4\right]}\\
&\times\int \mathfrak{N}_1(\xi^\text{s}) d\xi^{\text{s}}_{N-1}\cdots d\xi^\text{s}_1 e^{\sum\limits_{k=0}^{N-1}\frac{i\varepsilon m_0}{2\hbar}\left[\left(\frac{\xi^{\text{s}}_{k+1}-\xi^{\text{s}}_{k}}{\varepsilon}\right)^2-2\beta m_0^2\left(\frac{\xi^{\text{s}}_{k+1}-\xi^{\text{s}}_{k}}{\varepsilon}\right)^4\right]}\\
&\times\int \mathfrak{N}_2(\xi^\text{s}) d\xi^{\text{s}}_{-1}\cdots d\xi^\text{s}_{-n+1} e^{\sum\limits_{k=-n}^{1}\frac{i\varepsilon m_0}{2\hbar}\left[\left(\frac{\xi^{\text{s}}_{k+1}-\xi^{\text{s}}_{k}}{\varepsilon}\right)^2-2\beta m_0^2\left(\frac{\xi^{\text{s}}_{k+1}-\xi^{\text{s}}_{k}}{\varepsilon}\right)^4\right]}~.
\end{split}
\end{equation}
where $\mathfrak{N}(\xi^\text{s})=\mathfrak{N}_0(\xi^\text{s})\mathfrak{N}_1(\xi^\text{s})\mathfrak{N}_2(\xi^\text{s})$.
Taking the $N,n\rightarrow\infty$ and the $\varepsilon\rightarrow 0$ limit, we can rewrite the above equation as
\begin{equation}\label{Influence.8.75}
\begin{split}
\mathcal{I}_{\mathcal{P}_1}&=\lim\limits_{\substack{{\varepsilon\rightarrow 0}\\{N,n\rightarrow\infty}}}\mathcal{I}^\text{D}_{\mathcal{P}_1}\\
&= \int d\tilde{\xi}^{\text{s}}_f\int d\tilde{\xi}^{\text{s}}_i \int \left[\tilde{\mathcal{D}}\xi^\text{s}\right]_{\tilde{\xi}^\text{s}_f,T}^{\xi^\text{s}_f,T+\Delta}e^{\frac{im_0}{2\hbar}\int_{T}^{T+\Delta}dt\left(\dot{\xi}^{\text{s}^2}-2\beta m_0^2\dot{\xi}^{\text{s}^4}\right)}\\
&\times\int \left[\tilde{\mathcal{D}}\xi^\text{s}\right]_{\tilde{\xi}^\text{s}_i,0}^{\tilde{\xi}^\text{s}_f,T}e^{\frac{im_0}{2\hbar}\int_{0}^{T}dt\left(\dot{\xi}^{\text{s}^2}-2\beta m_0^2\dot{\xi}^{\text{s}^4}\right)} \int \left[\tilde{\mathcal{D}}\xi^\text{s}\right]_{\xi^\text{s}_i,-\Delta}^{\tilde{\xi}^\text{s}_i,0}e^{\frac{im_0}{2\hbar}\int_{-\Delta}^{0}dt\left(\dot{\xi}^{\text{s}^2}-2\beta m_0^2\dot{\xi}^{\text{s}^4}\right)}
\end{split}
\end{equation}
where $\xi^\text{s}(0)=\tilde{\xi}^\text{s}_i=\xi^\text{s}_0$ and $\xi^\text{s}(T)=\tilde{\xi}^\text{s}_f=\xi^\text{s}_N$. Similarly, it is possible to write down the second path integral as
\begin{equation}\label{Influence.9.76}
\begin{split}
\mathcal{I}_\mathcal{P_2}&=\int\left[\tilde{\mathcal{D}}\acute{\xi}^\text{s}\right]_{\acute{\xi}^\text{s}_i,-\Delta}^{\acute{\xi}^\text{s}_f,T+\Delta}e^{-\frac{im_0}{2\hbar}\int_{-\Delta}^{T+\Delta}dt\left(\dot{\acute{\xi}}^{\text{s}^2}-2\beta m_0^2\dot{\acute{\xi}}^{\text{s}^4}\right)}\\
&=\int d\tilde{\acute{\xi}}^{\text{s}}_f\int d\tilde{\acute{\xi}}^{\text{s}}_i \int \left[\tilde{\mathcal{D}}\acute{\xi}^\text{s}\right]_{\acute{\xi}^\text{s}_i,-\Delta}^{\tilde{\acute{\xi}}^\text{s}_i,0}e^{-\frac{im_0}{2\hbar}\int_{-\Delta}^{0}dt\left(\dot{\acute{\xi}}^{\text{s}^2}-2\beta m_0^2\dot{\acute{\xi}}^{\text{s}^4}\right)}\\
&\times\int \left[\tilde{\mathcal{D}}\acute{\xi}^\text{s}\right]_{\tilde{\acute{\xi}}^\text{s}_i,0}^{\tilde{\acute{\xi}}^\text{s}_f,T}e^{-\frac{im_0}{2\hbar}\int_{0}^{T}dt\left(\dot{\acute{\xi}}^{\text{s}^2}-2\beta m_0^2\dot{\acute{\xi}}^{\text{s}^4}\right)} \int \left[\tilde{\mathcal{D}}\acute{\xi}^\text{s}\right]_{\tilde{\acute{\xi}}^\text{s}_f,T}^{\acute{\xi}^\text{s}_f,T+\Delta}e^{-\frac{im_0}{2\hbar}\int_{T}^{T+\Delta}dt\left(\dot{\acute{\xi}}^{\text{s}^2}-2\beta m_0^2\dot{\acute{\xi}}^{\text{s}^4}\right)}
\end{split}
\end{equation}
where $\acute{\xi}^\text{s}(0)=\tilde{\acute{\xi}}^\text{s}_i=\acute{\xi}^\text{s}_0$ and $\acute{\xi}^\text{s}(T)=\tilde{\acute{\xi}}^\text{s}_f=\acute{\xi}^\text{s}_N$. We now define two new wave functions as
\begin{align}
\tilde{\phi}_i^\text{s}(\tilde{\xi}^\text{s}_i)&=\int d\xi^\text{s}_i \phi_i^\text{s}(\xi^\text{s}_i)\int \left[\tilde{\mathcal{D}}\xi^\text{s}\right]_{\xi^\text{s}_i,-\Delta}^{\tilde{\xi}^\text{s}_i,0}e^{\frac{im_0}{2\hbar}\int_{-\Delta}^{0}dt\left(\dot{\xi}^{\text{s}^2}-2\beta m_0^2\dot{\xi}^{\text{s}^4}\right)}\label{Influence.10.77}\\
\tilde{\phi}_f^\text{s}(\tilde{\xi}^\text{s}_f)&=\int d\xi^\text{s}_f \phi_f^\text{s}(\xi^\text{s}_f)\int \left[\tilde{\mathcal{D}}\xi^\text{s}\right]_{\xi^\text{s}_f,T}^{\tilde{\xi}^\text{s}_f,T+\Delta}e^{-\frac{im_0}{2\hbar}\int_{T}^{T+\Delta}dt\left(\dot{\xi}^{\text{s}^2}-2\beta m_0^2\dot{\xi}^{\text{s}^4}\right)}\label{Influence.11.78}~.
\end{align}
Making use of eq.(s)(\ref{Influence.8.75},\ref{Influence.9.76},\ref{Influence.10.77},\ref{Influence.11.78}), it is possible to express the transition probability in eq.(\ref{QGMinimal.34.66}) as
\begin{equation}\label{Influence.12.79}
\begin{split}
P_{\tilde{\psi}^\text{g}_{\omega}}^{[\tilde{\phi}^\text{s}_i\rightarrow\tilde{\phi}^\text{s}_f]}&=\int d\tilde{\xi}^\text{s}_id\acute{\tilde{\xi}}^\text{s}_id\tilde{\xi}^\text{s}_fd\acute{\tilde{\xi}}^\text{s}_f\tilde{\phi}^\text{s}_i(\tilde{\xi}^\text{s}_i)\tilde{\phi}_i^{\text{s}^*}(\tilde{\acute{\xi}}^\text{s}_i)\tilde{\phi}^{\text{s}^*}_f(\tilde{\xi}^\text{s}_f)\tilde{\phi}^\text{s}_f(\tilde{\acute{\xi}}^\text{s}_f)\\&\times\int\left[\tilde{\mathcal{D}}\xi^\text{s}\right]_{\tilde{\xi}^\text{s}_i,0}^{\tilde{\xi}^\text{s}_f,T}\left[\tilde{\mathcal{D}}\acute{\xi}^\text{s}\right]_{\tilde{\acute{\xi}}^\text{s}_i,0}^{\tilde{\acute{\xi}}^\text{s}_f,T}~e^{\frac{im_0}{2\hbar}\int_{-\Delta}^{T+\Delta}dt\left[\left[\dot{\xi}^{\text{s}^2}-\dot{\acute{\xi}}^{\text{s}^2}\right]-2\beta m_0^2\left[\dot{\xi}^{\text{s}^4}-\dot{\acute{\xi}}^{\text{s}^4}\right]\right]} F^{\beta}_{\tilde{\psi}^\text{g}_\omega}\left[\xi^\text{s},\acute{\xi}^\text{s}\right]~.
\end{split}
\end{equation}
The important thing to note is that $F^{\beta}_{\tilde{\psi}^\text{g}_\omega}\left[\xi^\text{s},\acute{\xi}^\text{s}\right]=F^{\beta}_{\psi^\text{g}_\omega}\left[\xi^\text{s},\acute{\xi}^\text{s}\right]$ from eq.(\ref{Influence.4.71}) where we have just noted the fact that $\xi^\text{s}$ and $\acute{\xi}^{\text{s}}$ vary in the range $t\in[0,T]$.
Instead of carrying the tilde symbol, one can genuinely simplify the expression by dropping the tilde symbol and write down the expression of the transition probability in a way simpler form as
\begin{equation}\label{Influence.13.80}
\begin{split}
P_{\psi^\text{g}_{\omega}}^{[\phi^\text{s}_i\rightarrow\phi^\text{s}_f]}&=\int d\xi^\text{s}_id\acute{\xi}^\text{s}_id\xi^\text{s}_fd\acute{\xi}^\text{s}_f\phi^\text{s}_i(\xi^\text{s}_i)\phi_i^{\text{s}^*}(\acute{\xi}^\text{s}_i)\phi^{\text{s}^*}_f(\xi^\text{s}_f)\phi^\text{s}_f(\acute{\xi}^\text{s}_f)\\&\times\int\left[\tilde{\mathcal{D}}\xi^\text{s}\right]_{\xi^\text{s}_i,0}^{\xi^\text{s}_f,T}\left[\tilde{\mathcal{D}}\acute{\xi}^\text{s}\right]_{\acute{\xi}^\text{s}_i,0}^{\acute{\xi}^\text{s}_f,T}~e^{\frac{im_0}{2\hbar}\int_{0}^{T}dt\left[\left[\dot{\xi}^{\text{s}^2}-\dot{\acute{\xi}}^{\text{s}^2}\right]-2\beta m_0^2\left[\dot{\xi}^{\text{s}^4}-\dot{\acute{\xi}}^{\text{s}^4}\right]\right]} F^{\beta}_{\psi^\text{g}_\omega}\left[\xi^\text{s},\acute{\xi}^\text{s}\right]
\end{split}
\end{equation}
where we have redefined the initial state of the graviton $|\tilde{\psi}^\text{g}_\omega\rangle$ at time $t=0$ by $|\psi^\text{g}_\omega\rangle$. Following the arguments presented in \cite{QGravD}, we can also claim that at time $t=0$ and $t=T$, $\dot{\xi}^{\text{s}}(t)$, $\dot{\acute{\xi}}^{\text{s}}(t)$, $\ddot{\xi}^{\text{s}}(t)$, and $\ddot{\acute{\xi}}^{\text{s}}(t)$ all vanishes. We shall now again shift our focus back to the analytical form of the influence functional in eq.(\ref{Influence.4.71}) as
\begin{equation}\label{Influence.14.81}
\begin{split}
F^{\beta}_{\psi^\text{g}_\omega}\left[\xi^\text{s},\acute{\xi}^\text{s}\right]
=&\langle\psi^\text{g}_\omega|e^{\frac{i}{\hbar}\hat{q}\mathfrak{g}\acute{\xi}^\text{s}_i\dot{\acute{\xi}}^\text{s}_i\left(1-3\beta m_0^2\dot{\acute{\xi}}^{\text{s}^2}_i\right)}\hat{U}^{\beta^\dagger}_{\acute{\xi}^\text{s}}(T,0)e^{-\frac{i}{\hbar}\hat{H}_0 T}e^{\frac{i}{\hbar}\hat{H}_0 T}e^{-\frac{i}{\hbar}\hat{q}\mathfrak{g}\acute{\xi}^\text{s}_f\dot{\acute{\xi}}^\text{s}_f\left(1-3\beta m_0^2\dot{\acute{\xi}}^{\text{s}^2}_f\right)}e^{-\frac{i}{\hbar}\hat{H}_0 T}\\
&\times e^{\frac{i}{\hbar}\hat{H}_0 T} e^{\frac{i}{\hbar}\hat{q}\mathfrak{g}\xi^\text{s}_f\dot{\xi}^\text{s}_f\left(1-3\beta m_0^2\dot{\xi}^{\text{s}^2}_f\right)}e^{-\frac{i}{\hbar}\hat{H}_0 T}e^{\frac{i}{\hbar}\hat{H}_0 T}\hat{U}^{\beta}_{\xi^\text{s}}(T,0)e^{-\frac{i}{\hbar}\hat{q}\mathfrak{g}\xi^\text{s}_i\dot{\xi}^\text{s}_i\left(1-3\beta m_0^2\dot{\xi}^{\text{s}^2}_i\right)}|]\psi^\text{g}_\omega\rangle\\
=&\langle\psi^\text{g}_\omega|e^{\frac{i}{\hbar}\hat{q}\mathfrak{g}\acute{\xi}^\text{s}_i\dot{\acute{\xi}}^\text{s}_i\left(1-3\beta m_0^2\dot{\acute{\xi}}^{\text{s}^2}_i\right)}\hat{U}^{\text{IP}^\dagger}_{\beta\acute{\xi}^\text{s}}(T,0)e^{-\frac{i}{\hbar}\hat{q}_{\text{IP}}(T)\mathfrak{g}\acute{\xi}^\text{s}_f\dot{\acute{\xi}}^\text{s}_f\left(1-3\beta m_0^2\dot{\acute{\xi}}^{\text{s}^2}_f\right)}\\
&\times  e^{\frac{i}{\hbar}\hat{q}_{\text{IP}}(T)\mathfrak{g}\xi^\text{s}_f\dot{\xi}^\text{s}_f\left(1-3\beta m_0^2\dot{\xi}^{\text{s}^2}_f\right)}\hat{U}^{\text{IP}}_{\beta\xi^\text{s}}(T,0)e^{-\frac{i}{\hbar}\hat{q}\mathfrak{g}\xi^\text{s}_i\dot{\xi}^\text{s}_i\left(1-3\beta m_0^2\dot{\xi}^{\text{s}^2}_i\right)}|\psi^\text{g}_\omega\rangle
\end{split}
\end{equation}
where $\hat{q}_{\text{IP}}(T)=e^{\frac{i}{\hbar}\hat{H}_0T}\hat{q}e^{-\frac{i}{\hbar}\hat{H}_0T}$ denotes the graviton position operator in the interaction picture and $\hat{U}^{\text{IP}}_{\beta\xi^\text{s}}(T,0)$ denotes the unitary time evolution operator in the interaction picture (IP). The unitary time evolution operator in the interaction picture $\hat{U}^{\text{IP}}_{\beta\xi^\text{s}}(T,0)$ is expressed in terms of the unitary time evolution operator $\hat{U}^{\beta}_{\xi^\text{s}}(T,0)$ as
\begin{equation}\label{Influence.15.82}
\hat{U}^{\text{IP}}_{\beta\xi^\text{s}}(T,0)=e^{\frac{i}{\hbar}\hat{H}_0T}\hat{U}^{\beta}_{\xi^\text{s}}(T,0)
\end{equation}
where $\hat{U}^{\beta}_{\xi^\text{s}}(T,0)$ has the analytical expression $\hat{U}^{\beta}_{\xi^\text{s}}(T,0)=e^{-\frac{i}{\hbar}\hat{H}^{\text{int}}_{\beta\acute{\xi}^\text{s}}T}$ with the analytical form of the interaction part of the Hamiltonian being given as\footnote{It is possible to divide the Hamiltonian operator $\hat{H}_{\beta\xi^\text{s}}$ from eq.(\ref{QGMinimal.31.63}) in a base harmonic oscillator part $\hat{H}_0$ and an interaction part $\hat{H}^{\text{int}}_{\beta\xi^\text{s}}$.} (upto $\mathcal{O}(\beta)$)
\begin{equation}\label{Influence.16.83}
\hat{H}^{\text{int}}_{\beta\xi^\text{s}}\simeq\frac{\mathfrak{g}\hat{p}\xi^\text{s}\dot{\xi}^{\text{s}}}{m}+\frac{\mathfrak{g}^2\xi^{\text{s}^2}\dot{\xi}^{\text{s}^2}}{2m}-\frac{3\beta m_0^2\mathfrak{g}\hat{p}\xi^\text{s}\dot{\xi}^{\text{s}^3}}{m}-\frac{3\beta m_0^2\mathfrak{g}^2\xi^{\text{s}^2}\dot{\xi}^{\text{s}^4}}{m}~.
\end{equation}
In the next subsection, we shall investigate the unitary time evolution operator in details.
\subsection{Unitary time evolution operator}
We start with the simple analytical form of the unitary time evolution operator in the interaction picture as
\begin{equation}\label{Unitary.1.84}
\hat{U}^{\text{IP}}_{\beta\xi^\text{s}}(T,0)=e^{\frac{i}{\hbar}\hat{H}_0T}e^{-\frac{i}{\hbar}\hat{H}^{\text{int}}_{\beta\acute{\xi}^\text{s}}T}~.
\end{equation}

\noindent Now, consider a state $|\psi\rangle_H$ in the Heisenberg picture. In the Schr\"{o}dinger picture the state at time $t$ is expressed as $|\psi_{\beta\xi^\text{s}}(t)\rangle=e^{-\frac{i}{\hbar}\hat{H}_{\beta\xi^\text{s}}t}|\psi\rangle_H$. In the interaction picture, we can express the state as
\begin{equation}\label{Unitary.2.85}
|\psi^{\text{IP}}_{\beta\xi^\text{s}}(t)\rangle=e^{\frac{i}{\hbar}\hat{H}_0t}|\psi_{\beta\xi^\text{s}}(t)\rangle~.
\end{equation}
Operating with the $i\hbar\frac{\partial}{\partial t}$ in the above equation, we obtain
\begin{equation}\label{Unitary.3.86}
\begin{split}
i\hbar\frac{\partial|\psi^{\text{IP}}_{\beta\xi^\text{s}}(t)\rangle}{\partial t}&=i\hbar\frac{\partial}{\partial t}\left(e^{\frac{i}{\hbar}\hat{H}_0t}|\psi_{\beta\xi^\text{s}}(t)\rangle\right)\\
&=-\hat{H}_0e^{\frac{i}{\hbar}\hat{H}_0t}|\psi_{\beta\xi^\text{s}}(t)\rangle+i\hbar e^{\frac{i}{\hbar}\hat{H}_0 t}\frac{\partial}{\partial t}\left(e^{-\frac{i}{\hbar}\hat{H}_{\beta\xi^\text{s}}t}|\psi\rangle_H\right)\\
&=-\hat{H}_0|\psi^{\text{IP}}_{\beta\xi^\text{s}}(t)\rangle+e^{\frac{i}{\hbar}\hat{H}_0 t}\hat{H}_{\beta\xi^\text{s}}e^{-\frac{i}{\hbar}\hat{H}_0 t}e^{\frac{i}{\hbar}\hat{H}_0 t}e^{-\frac{i}{\hbar}\hat{H}_{\beta\xi^\text{s}}t}|\psi\rangle_H\\
&=e^{\frac{i}{\hbar}\hat{H}_0t}\left(\hat{H}_{\beta\xi^\text{s}}-\hat{H}_0\right)e^{-\frac{i}{\hbar}
\hat{H}_0 t}|\psi^{\text{IP}}_{\beta\xi^\text{s}}(t)\rangle~.
\end{split}
\end{equation}
We already know from the Hamiltonian in eq.(\ref{QGMinimal.31.63}) that $\hat{H}_{\beta\xi^\text{s}}=\hat{H}_0+\hat{H}^{\text{int}}_{\beta\xi^\text{s}}$ and we can further simplify the above equation as
\begin{equation}\label{Unitary.4.87}
i\hbar\frac{\partial|\psi^{\text{IP}}_{\beta\xi^\text{s}}(t)\rangle}{\partial t}=\left(\hat{H}^{\text{int}}_{\beta\xi^\text{s}}\right)^{\text{IP}}|\psi^{\text{IP}}_{\beta\xi^\text{s}}(t)\rangle
\end{equation}
where $\left(\hat{H}^{\text{int}}_{\beta\xi^\text{s}}\right)^{\text{IP}}$ denotes the interaction Hamiltonian in the interaction picture. It is also possible to write down the state $|\psi^{\text{IP}}_{\beta\xi^\text{s}}(t)\rangle$ as $|\psi^{\text{IP}}_{\beta\xi^\text{s}}(t)\rangle=\hat{U}^{\text{IP}}_{\beta\xi^\text{s}}(t,0)|\psi\rangle_H$ with $|\psi\rangle_H$ being time-independent. It is then possible to simplify eq.(\ref{Unitary.4.87}) as
\begin{equation}\label{Unitary.5.88}
i\hbar\frac{\partial \hat{U}^{\text{IP}}_{\beta\xi^\text{s}}(t,0)}{\partial t}=\left(\hat{H}^{\text{int}}_{\beta\xi^\text{s}}\right)^{\text{IP}}\hat{U}^{\text{IP}}_{\beta\xi^\text{s}}(t,0)~.
\end{equation}
To obtain an analytical solution of the unitary time evolution operator, we need to integrate the above equation, after which we obtain
\begin{equation}\label{Unitary.6.89}
\begin{split}
\hat{U}^{\text{IP}}_{\beta\xi^\text{s}}(t,0)=\hat{\mathbb{1}}-\frac{i}{\hbar}\int_0^t dt'\hat{H}^{\text{int}^{\text{IP}}}_{\beta\xi^\text{s}}(t')\hat{U}^{\text{IP}}_{\beta\xi^\text{s}}(t',0)
\end{split}
\end{equation}
where we have made use of the fact that $\hat{U}^{\text{IP}}_{\beta\xi^\text{s}}(0,0)=\hat{\mathbb{1}}$.
Using an iterative procedure and substituting the expression for the unitary time evolution operator from the above equation inside of the integrand in the right hand side of the above equation and continuing this procedure, we arrive at the following expression for the unitary time evolution operator
\begin{equation}\label{Unitary.7.90}
\begin{split}
\hat{U}^{\text{IP}}_{\beta\xi^\text{s}}(t,0)&=\hat{\mathbb{1}}-\frac{i}{\hbar}\int_0^t dt'\hat{H}^{\text{int}^{\text{IP}}}_{\beta\xi^\text{s}}(t')+\left(-\frac{i}{\hbar}\right)^2\int_0^t dt'\hat{H}^{\text{int}^{\text{IP}}}_{\beta\xi^\text{s}}(t')\int_0^{t'} dt''\hat{H}^{\text{int}^{\text{IP}}}_{\beta\xi^\text{s}}(t'')+\cdots\\
&=\hat{\mathbb{1}}-\frac{i}{\hbar}\int_0^t dt'\hat{H}^{\text{int}^{\text{IP}}}_{\beta\xi^\text{s}}(t')+\frac{1}{2}\left(-\frac{i}{\hbar}\right)^2\int_0^t dt'\int_0^{t} dt''\mathcal{T}\left[\hat{H}^{\text{int}^{\text{IP}}}_{\beta\xi^\text{s}}(t')\hat{H}^{\text{int}^{\text{IP}}}_{\beta\xi^\text{s}}(t'')\right]+\cdots~.
\end{split}
\end{equation}
Combining all the higher order terms and setting $t=T$, we can obtain the analytical form of the unitary time evolution operator as
\begin{equation}\label{Unitary.8.91}
\begin{split}
\hat{U}^{\text{IP}}_{\beta\xi^\text{s}}(T,0)=\mathcal{T}\left[e^{-\frac{i}{\hbar}\int_0^Tdt \hat{H}^{\text{int}^{\text{IP}}}_{\beta\xi^\text{s}}(t)}\right]~.
\end{split}
\end{equation}
The time ordered expression in the right hand side of the above equation can be further simplified. We now again move to discrete limits where we divide the time interval $T$ into $N$ equal time segments of value $\varepsilon$ such that $t_N=N\varepsilon=T$ and $t_0=0$. In this discrete limit, it is possible to recast the right hand side of the above equation as
\begin{equation}\label{Unitary.9.92}
\begin{split}
&\mathcal{T}\left[e^{-\frac{i}{\hbar}\int_0^Tdt \hat{H}^{\text{int}^{\text{IP}}}_{\beta\xi^\text{s}}(t)}\right]\\&=\mathcal{T}\left[\lim\limits_{\substack{{\varepsilon\rightarrow 0}\\{N\rightarrow\infty}}}\left[e^{-\frac{i\varepsilon}{\hbar}\hat{H}^{\text{int}^{\text{IP}}}_{\beta\xi^\text{s}}(t_0)}e^{-\frac{i\varepsilon}{\hbar}\hat{H}^{\text{int}^{\text{IP}}}_{\beta\xi^\text{s}}(t_1)}\cdots e^{-\frac{i\varepsilon}{\hbar}\hat{H}^{\text{int}^{\text{IP}}}_{\beta\xi^\text{s}}(t_N)}\right]\right]\\
&=\lim\limits_{\substack{{\Delta t\rightarrow 0}\\{N\rightarrow\infty}}}\left[e^{-\frac{i\varepsilon}{\hbar}\hat{H}^{\text{int}^{\text{IP}}}_{\beta\xi^\text{s}}(t_N)}e^{-\frac{i\varepsilon}{\hbar}\hat{H}^{\text{int}^{\text{IP}}}_{\beta\xi^\text{s}}(t_{N-1})}\cdots e^{-\frac{i\varepsilon}{\hbar}\hat{H}^{\text{int}^{\text{IP}}}_{\beta\xi^\text{s}}(t_0)}\right]\\
&=\lim\limits_{\substack{{\Delta t\rightarrow 0}\\{N\rightarrow\infty}}}\left[e^{-\frac{i\varepsilon}{\hbar} t\hat{H}^{\text{int}^{\text{IP}}}_{\beta\xi^\text{s}}(t_N)}\cdots e^{-\frac{i\varepsilon}{\hbar}\hat{H}^{\text{int}^{\text{IP}}}_{\beta\xi^\text{s}}(t_2)}e^{-\frac{i\varepsilon}{\hbar}\left[\hat{H}^{\text{int}^{\text{IP}}}_{\beta\xi^\text{s}}(t_1)+\hat{H}^{\text{int}^{\text{IP}}}_{\beta\xi^\text{s}}(t_0)\right]-\frac{\varepsilon^2}{2\hbar^2}\left[\hat{H}^{\text{int}^{\text{IP}}}_{\beta\xi^\text{s}}(t_1),\hat{H}^{\text{int}^{\text{IP}}}_{\beta\xi^\text{s}}(t_0)\right]}\right]
\end{split}
\end{equation}
where in the last line of the above equation we have made use of the Baker-Campbell-Hausdorff formula and the commutator brackets just result in numbers. Via repeated application of the Baker-Campbell-Hausdorff formula, we arrive at the result as
\begin{equation}\label{Unitary.10.93}
\begin{split}
&\mathcal{T}\left[e^{-\frac{i}{\hbar}\int_0^Tdt \hat{H}^{\text{int}^{\text{IP}}}_{\beta\xi^\text{s}}(t)}\right]
\\&=\lim\limits_{\substack{{\Delta t\rightarrow 0}\\{N\rightarrow\infty}}}\left[e^{-\frac{i\varepsilon}{\hbar}\left[\hat{H}^{\text{int}^{\text{IP}}}_{\beta\xi^\text{s}}(t_N)+~\cdots~+\hat{H}^{\text{int}^{\text{IP}}}_{\beta\xi^\text{s}}(t_0)\right]}e^{-\frac{\varepsilon^2}{2\hbar^2}\left(\left[\hat{H}^{\text{int}^{\text{IP}}}_{\beta\xi^\text{s}}(t_N),\hat{H}^{\text{int}^{\text{IP}}}_{\beta\xi^\text{s}}(t_{N-1})\right]+~\cdots~+\left[\hat{H}^{\text{int}^{\text{IP}}}_{\beta\xi^\text{s}}(t_N),\hat{H}^{\text{int}^{\text{IP}}}_{\beta\xi^\text{s}}(t_{0})\right]\right)}\right.\\&\times\left.e^{-\frac{\varepsilon^2}{2\hbar^2}\left(\left[\hat{H}^{\text{int}^{\text{IP}}}_{\beta\xi^\text{s}}(t_{N-1}),\hat{H}^{\text{int}^{\text{IP}}}_{\beta\xi^\text{s}}(t_{N-2})\right]+~\cdots~+\left[\hat{H}^{\text{int}^{\text{IP}}}_{\beta\xi^\text{s}}(t_{N-1}),\hat{H}^{\text{int}^{\text{IP}}}_{\beta\xi^\text{s}}(t_{0})\right]\right)+~\cdots~-\frac{\varepsilon^2}{2\hbar^2}\left[\hat{H}^{\text{int}^{\text{IP}}}_{\beta\xi^\text{s}}(t_{1}),\hat{H}^{\text{int}^{\text{IP}}}_{\beta\xi^\text{s}}(t_{0})\right]} \right]\\
&=\lim\limits_{\substack{{\Delta t\rightarrow 0}\\{N\rightarrow\infty}}}\exp\left[-\frac{i\varepsilon}{\hbar}\sum_{j=0}^N\hat{H}^{\text{int}^\text{IP}}_{\beta\xi^\text{s}}(t_j)-\frac{\varepsilon^2}{2\hbar^2}\sum_{k=1}^N\sum_{j<k}\left[\hat{H}^{\text{int}^\text{IP}}_{\beta\xi^\text{s}}(t_k),\hat{H}^{\text{int}^\text{IP}}_{\beta\xi^\text{s}}(t_j)\right]\right]~.
\end{split}
\end{equation}
In the continuum limits, we can then express the unitary time evolution operator in eq.(\ref{Unitary.8.91}) as
\begin{equation}\label{Unitary.11.94}
\begin{split}
\hat{U}^{\text{IP}}_{\beta\xi^\text{s}}(T,0)=\exp\left[-\frac{i}{\hbar}\int_0^T dt' \hat{H}^{\text{int}^\text{IP}}_{\beta\xi^\text{s}}(t')-\frac{1}{2\hbar^2}\int_0^Tdt'\int_0^{t'}dt''\left[\hat{H}^{\text{int}^\text{IP}}_{\beta\xi^\text{s}}(t'),\hat{H}^{\text{int}^\text{IP}}_{\beta\xi^\text{s}}(t'')\right]\right]~.
\end{split}
\end{equation}
Using the analytical form of the interaction Hamiltonian in eq.(\ref{Influence.16.83}) and a tedious amount of analytical calculations and simplifications, we arrive at the final form of the unitary time evolution operator from eq.(\ref{Unitary.11.94}) as
\begin{equation}\label{Unitary.12.95}
\begin{split}
&\hat{U}^{\text{IP}}_{\beta\xi^\text{s}}(T,0)= \exp\left[-\frac{i\mathfrak{g}}{\hbar}\hat{q}_{\text{IP}}(T)\xi^\text{s}_f\dot{\xi}^\text{s}_f\left[1-3\beta m_0^2\dot{\xi}_f^{\text{s}^2}\right]\right] \exp\left[\frac{i\mathfrak{g}}{2\hbar}\int_0^Tdt~\hat{q}_{\text{IP}}(t)\mathcal{Z}(t)\right]\\\times&\exp\left[\frac{i\mathfrak{g}}{\hbar}\hat{q}_{\text{IP}}(0)\xi^\text{s}_i\dot{\xi}^\text{s}_i\left[1-3\beta m_0^2\dot{\xi}^{\text{s}^2}_i\right]\right]\exp\left[-\frac{\mathfrak{g}^2}{8\hbar^2}\int_0^Tdt\int_0^tdt'\mathcal{Z}(t)\mathcal{Z}(t')\left[\hat{q}_{\text{IP}}(t),\hat{q}_{\text{IP}}(t')\right]\right]
\end{split}
\end{equation}
where $\hat{q}_{\text{IP}}(0)=\hat{q}$ and $\mathcal{Z}(t)$ is defined by the expression
\begin{equation}\label{Unitary.13.96}
\begin{split}
\mathcal{Z}(t)\equiv\mathcal{X}(t)-3\beta m_0^2\mathcal{Y}(t)
\end{split}
\end{equation}
with $\mathcal{X}(t)$ and $\mathcal{Y}(t)$ being defined as
\begin{align}\label{Unitary{14.97}}
\mathcal{X}(t)\equiv\frac{d^2}{dt^2}\left(\xi^{\text{s}^2}(t)\right)~\text{and}~\mathcal{Y}(t)\equiv \frac{d}{dt}\left(\dot{\xi}^{\text{s}^2}(t)\frac{d}{dt}\left(\xi^{\text{s}^2}(t)\right)\right)~.
\end{align}
In eq.(\ref{Unitary.12.95}), $\hat{q}_{\text{IP}}(T)$ denotes the graviton position operator in the interaction picture. In terms of the graviton phase space operators $\{\hat{q},\hat{p}\}$, we can write down the creation and annihilation operators corresponding to the graviton states as
\begin{equation}\label{Unitary.15.98}
\hat{a}=\sqrt{\frac{m\omega}{2\hbar}}\left(\hat{q}+\frac{i}{m\omega}\hat{p}\right)~\text{and}~\hat{a}^\dagger=\sqrt{\frac{m\omega}{2\hbar}}\left(\hat{q}-\frac{i}{m\omega}\hat{p}\right)~.
\end{equation}
One can then write down the position operator for the graviton in the interaction picture as
\begin{equation}\label{Unitary.16.99}
\hat{q}_{\text{IP}}(t)=e^{\frac{i}{\hbar}\hat{H}_0T}\hat{q}e^{-\frac{i}{\hbar}\hat{H}_0T}=\sqrt{\frac{\hbar}{2m\omega}}\left(\hat{a}e^{-i\omega t}+\hat{a}^\dagger e^{i\omega t}\right)~.
\end{equation}
We can now obtain the analytical form of the commutator bracket $\left[\hat{q}_{\text{IP}}(t),\hat{q}_{\text{IP}}(t')\right]$ as
\begin{equation}\label{Unitary.17.100}
\left[\hat{q}_{\text{IP}}(t),\hat{q}_{\text{IP}}(t')\right]=\frac{\hbar}{2m\omega}\left(\exp\left[-\omega(t-t')\right]-\exp\left[i\omega(t-t')\right]\right)
\end{equation}
which indeed is a number.
\subsection{Graviton coherent state and the influence functional functional analysis}
We are now in a position to write down the final analytical form of the Feynman-Vernon influence functional using the analytical form of the unitary time-evolution operator in eq.(\ref{Unitary.12.95}). Taking the analytical form of the unitary time evolution operator from eq.(\ref{Unitary.12.95}) and substituting it back in the expression for the Feynman-Vernon influence functional from eq.(\ref{Influence.14.81}), we can arrive at a very beautiful expression for the influence functional as
\begin{equation}\label{QGInfluence.1.101}
F^\beta_{\psi^\text{g}_\omega}[\xi^\text{s},\acute{\xi}^\text{s}]=\langle \psi^\text{g}_\omega |e^{-\mathcal{V}^*\hat{a}^\dagger}e^{\mathcal{V}\hat{a}}|\psi^\text{g}_\omega\rangle e^{i\Phi^\beta_{0^\text{g}_\omega}[\xi^\text{s},\acute{\xi}^\text{s}]}
\end{equation}
where $\mathcal{V}$ and $i\Phi^\beta_{0^\text{g}_\omega}[\xi^\text{s},\acute{\xi}^\text{s}]$ are defined as
\begin{align}
\mathcal{V}\equiv&\frac{i\mathfrak{g}}{\sqrt{8\hbar m\omega}}\int_0^Tdt~ e^{-i\omega t}(\mathcal{Z}(t)-\acute{\mathcal{Z}}(t))~,\label{QGInfluence.2.102}\\
i\Phi^\beta_{0^\text{g}_\omega}[\xi^\text{s},\acute{\xi}^\text{s}]\equiv&-\frac{\mathfrak{g}^2}{8\hbar m\omega}\int_0^Tdt\int_0^t dt'\left(\mathcal{Z}(t)-\mathcal{\acute{Z}}(t)\right)\left(\mathcal{Z}(t')e^{-i\omega(t-t')}-\acute{\mathcal{Z}}(t')e^{i\omega(t-t')}\right)\label{QGInfluence.3.103}
\end{align}
with $\mathcal{Z}(t)$ and $\acute{\mathcal{Z}}(t)$being defined as $\mathcal{Z}(t)=\mathcal{X}(t)-3\beta m_0^2\mathcal{Y}(t)$ and $\acute{\mathcal{Z}}(t)=\acute{\mathcal{X}}(t)-3\beta m_0^2\acute{\mathcal{Y}}(t)$. Here, $\mathcal{X}(t)$, $\acute{\mathcal{X}}(t)$, $\mathcal{Y}(t)$, and $\acute{\mathcal{Y}}(t)$ are defined as 
\begin{align}
\mathcal{X}(t)&\equiv\frac{d^2}{dt^2}\left(\xi^{\text{s}^2}(t)\right)~\text{and}~\mathcal{Y}(t)\equiv \frac{d}{dt}\left(\dot{\xi}^{\text{s}^2}(t)\frac{d}{dt}\left(\xi^{\text{s}^2}(t)\right)\right)\label{QGInfluence.4.104}~,\\
\acute{\mathcal{X}}(t)&\equiv\frac{d^2}{dt^2}\left(\acute{\xi}^{\text{s}^2}(t)\right)~\text{and}~\acute{\mathcal{Y}}(t)\equiv \frac{d}{dt}\left(\dot{\acute{\xi}}^{\text{s}^2}(t)\frac{d}{dt}\left(\acute{\xi}^{\text{s}^2}(t)\right)\right)\label{1QGInfluence.5.105}~.
\end{align}
In order to obtain the analytical form of the influence functional, we start by considering the single-mode gravitational wave state or the graviton state to be a coherent state. Now a coherent state denotes the minimal uncertainty state and as a result most closely resembles a classical scenario. For the single mode  of the gravitational wave (graviton) carrying an energy $\hbar \omega$, we can write down the harmonic oscillator state as $|\psi^\text{g}_\omega\rangle=|\alpha^\text{g}_\omega\rangle.$ The coherent state $|\alpha^\text{g}_\omega\rangle$ is defined as
\begin{equation}\label{QGInfluence.6.106}
\hat{a}|\alpha^\text{g}_\omega\rangle=\alpha^\text{g}_\omega|\alpha^\text{g}_\omega\rangle~,~\langle\alpha^\text{g}_\omega|\hat{a}^\dagger=\langle\alpha^\text{g}_\omega|\alpha^{\text{g}^*}_\omega~.
\end{equation}
It is now possible to consider a classical gravitational wave mode given as 
\begin{equation}\label{QGInfluence.7.107}
q_{\text{cl}}(t)=\mathcal{f}^{\text{g}}_\omega\cos(\omega t+\phi^\text{g}_\omega)
\end{equation}
where $\mathcal{f}^\text{g}_\omega$ gives the gravitational wave amplitude and $\phi^\text{g}_\omega$ introduces an additional time independent phase factor. One can now obtain the analytical form of $\alpha^{\text{g}}_\omega$ as
\begin{equation}\label{QGInfluence.8.108}
\begin{split}
\alpha^\text{g}_\omega&=\sqrt{\frac{m\omega}{2\hbar}}\left<\alpha^\text{g}_\omega\right|\hat{q}+\frac{i}{m\omega}\hat{p}\left|\alpha^\text{g}_\omega\right>=\sqrt{\frac{m\omega}{2\hbar}}\mathcal{f}^\text{g}_\omega \exp\left[-i\phi^\text{g}_\omega\right]~.
\end{split}
\end{equation} 
For the gravitons being in an initial coherent state, it is possible to write down the Feynman-Vernon influence functional in eq.(\ref{QGInfluence.1.101}) as
\begin{equation}\label{QGInfluence.9.109}
\begin{split}
\left.F^\beta_{\psi^\text{g}_\omega}[\xi^\text{s},\acute{\xi}^\text{s}]\right\rvert_{\psi^\text{g}_\omega=\alpha^{\text{g}}_\omega}&=F^\beta_{\alpha^\text{g}_\omega}[\xi^\text{s},\acute{\xi}^\text{s}]\\&=\langle \alpha^\text{g}_\omega |e^{-\mathcal{V}^*\hat{a}^\dagger}e^{\mathcal{V}\hat{a}}|\alpha^\text{g}_\omega\rangle e^{i\Phi^\beta_{0^\text{g}_\omega}[\xi^\text{s},\acute{\xi}^\text{s}]}
\\\implies F^\beta_{\alpha^\text{g}_\omega}[\xi^\text{s},\acute{\xi}^\text{s}] &=F^\beta_{0^\text{g}_\omega}[\xi^\text{s},\acute{\xi}^\text{s}]e^{-\mathcal{V}^*\alpha^{\text{g}^*}_\omega+\mathcal{V}\alpha^\text{g}_\omega}
\end{split}
\end{equation}
where the single-mode vacuum influence functional is defined as
\begin{equation}\label{QGInfluence.10.110}
F^\beta_{0^\text{g}_\omega}[\xi^\text{s},\acute{\xi}^\text{s}]\equiv \exp\left[i\Phi^\beta_{0^\text{g}_\omega}[\xi^\text{s},\acute{\xi}^\text{s}]\right]~.
\end{equation}
Using the analytical expressions of $\mathcal{V}$ from eq.(\ref{QGInfluence.2.102}) and $\alpha^{\text{g}}_\omega$ from eq.(\ref{QGInfluence.8.108}), we can recast the form of the influence functional in eq.(\ref{QGInfluence.10.110}) as
\begin{equation}\label{QGInfluence.11.111}
\begin{split}
F_{\alpha^\text{g}_\omega}^{\beta}[\xi^\text{s},\acute{\xi}^\text{s}]=F^\beta_{0^\text{g}_\omega}[\xi^\text{s},\acute{\xi}^\text{s}]\exp\left[\frac{i\mathfrak{g}\mathcal{f}^\text{g}_\omega}{2\hbar}\int_0^Tdt \cos(\omega t+\phi^\text{g}_\omega)(\mathcal{Z}(t)-\acute{\mathcal{Z}}(t))\right]~.
\end{split}
\end{equation}
Substituting the above form of the influence functional in the transition probability obtained in eq.(\ref{Influence.13.80}), we arrive at the transition probability for the detector going from its initial state $|\phi_i^\text{s}\rangle$ to its final state $|\phi_f^\text{s}\rangle$, while the graviton initially is in a coherent state, as
\begin{equation}\label{QGInfluence.12.112}
\begin{split}
&P_{\psi^\text{g}_{\omega}}^{[\phi^\text{s}_i\rightarrow\phi^\text{s}_f]}=\int d\xi^\text{s}_id\acute{\xi}^\text{s}_id\xi^\text{s}_fd\acute{\xi}^\text{s}_f\phi^\text{s}_i(\xi^\text{s}_i)\phi_i^{\text{s}^*}(\acute{\xi}^\text{s}_i)\phi^{\text{s}^*}_f(\xi^\text{s}_f)\phi^\text{s}_f(\acute{\xi}^\text{s}_f)\int\left[\tilde{\mathcal{D}}\xi^\text{s}\right]_{\xi^\text{s}_i,0}^{\xi^\text{s}_f,T}\left[\tilde{\mathcal{D}}\acute{\xi}^\text{s}\right]_{\acute{\xi}^\text{s}_i,0}^{\acute{\xi}^\text{s}_f,T}F^{\beta}_{0^\text{g}_\omega}\left[\xi^\text{s},\acute{\xi}^\text{s}\right]\\&\times \exp\left[\frac{i}{\hbar}\int_{0}^{T}dt\left[\frac{m_0}{2}\left(\left(\dot{\xi}^{\text{s}^2}-\dot{\acute{\xi}}^{\text{s}^2}\right)-2\beta m_0^2\left(\dot{\xi}^{\text{s}^4}-\dot{\acute{\xi}}^{\text{s}^4}\right)\right)+\frac{\mathfrak{g}\mathcal{f}^\text{g}_\omega}{2}\cos(\omega t+\phi^\text{g}_\omega)(\mathcal{Z}(t)-\acute{\mathcal{Z}}(t))\right]\right] 
\end{split}
\end{equation}
where in the above form of the transition probability only a single mode of the gravitational wave has been considered. It is now possible to extend this analysis to all possible modes which will imitate the existence and interaction of a  gravitational field with the detector. Considering all possible modes, one can now write down the expression for the state of the gravitational field as
\begin{equation}\label{QGInfluence.13.113}
|\Psi_{\text{G}}\rangle=\mathop{\otimes}\limits_{\mathbf{k}}\left|\psi^{\text{g}}_{\omega(\mathbf{k})}\right\rangle
\end{equation}
which is tensor products of individual graviton states corresponding to individual mode frequencies. Now the complete Feynman-Vernon influence functional for the gravitational field can be represented as a product of individual Feynman-Vernon influence functional for each gravitational field modes and this is possible because the total state of the gravitational field is a tensor product of single-mode graviton states as can also be seen from eq.(\ref{QGInfluence.13.113}). It is now possible to write down the influence functional corresponding to the gravitational field state $|\Psi_\text{G}\rangle$ as
\begin{equation}\label{QGInfluence.14.114}
\begin{split}
F^{\beta}_{{\Psi}_{\text{G}}}[\xi^\text{s},\acute{\xi}^\text{s}]&=\prod\limits_{\mathbf{k}}F^{\beta}_{\psi^\text{g}_{\omega(\mathbf{k})}}[\xi^\text{s},\acute{\xi}^\text{s}]\\
&=\prod\limits_{\mathbf{k}}\left[\langle \psi^\text{g}_{\omega(\mathbf{k})} |e^{-\mathcal{V}^*\hat{a}^\dagger}e^{\mathcal{V}\hat{a}}|\psi^\text{g}_{\omega(\mathbf{k})}\rangle \exp\left[i\Phi^\beta_{0^\text{g}_{\omega(\mathbf{k})}}[\xi^\text{s},\acute{\xi}^\text{s}]\right]\right]\\
&=\left[\prod\limits_{\mathbf{k}}\langle \psi^\text{g}_{\omega(\mathbf{k})} |e^{-\mathcal{V}^*\hat{a}^\dagger}e^{\mathcal{V}\hat{a}}|\psi^\text{g}_{\omega(\mathbf{k})}\rangle \right]\exp\left[\sum\limits_{\mathbf{k}}i\Phi^\beta_{0^\text{g}_{\omega(\mathbf{k})}}[\xi^\text{s},\acute{\xi}^\text{s}]\right]\\
\implies F^{\beta}_{{\Psi}_{\text{G}}}[\xi^\text{s},\acute{\xi}^\text{s}] &=\left[\prod\limits_{\mathbf{k}}\langle \psi^\text{g}_{\omega(\mathbf{k})} |e^{-\mathcal{V}^*\hat{a}^\dagger}e^{\mathcal{V}\hat{a}}|\psi^\text{g}_{\omega(\mathbf{k})}\rangle \right] F^{\beta}_{{0}_{\text{G}}}[\xi^\text{s},\acute{\xi}^\text{s}]
\end{split}
\end{equation}
where, corresponding to the gravitational field part, the form of the vacuum influence functional reads
\begin{equation}\label{QGInfluence.15.115}
F^{\beta}_{{0}_{\text{G}}}[\xi^\text{s},\acute{\xi}^\text{s}]=\exp\left[i\Phi_{0_\text{G}}^\beta[\xi^\text{s},\acute{\xi}^\text{s}]\right]=\exp\left[\sum\limits_{\mathbf{k}}i\Phi^\beta_{0^\text{g}_{\omega(\mathbf{k})}}[\xi^\text{s},\acute{\xi}^\text{s}]\right]
\end{equation}
where the $i\Phi_{0_\text{G}}^\beta[\xi^\text{s},\acute{\xi}^\text{s}]$ is defined as
\begin{equation}\label{QGInfluence.16.1116}
i\Phi^\beta_{0_{\text{G}}}[\xi^\text{s},\acute{\xi}^\text{s}]\equiv\sum\limits_{\mathbf{k}}i\Phi^\beta_{0^\text{g}_{\omega(\mathbf{k})}}[\xi^\text{s},\acute{\xi}^\text{s}]~.
\end{equation}
Using the analytical form of the influence phase term $i\Phi^\beta_{0^\text{g}_{\omega(\mathbf{k})}}[\xi^\text{s},\acute{\xi}^\text{s}]$ from eq.(\ref{QGInfluence.3.103}), we can obtain the analytical expression for the total phase term in the left hand side of the above equation as
\begin{equation}\label{QGInfluence.17.117}
\begin{split}
&i\Phi^\beta_{0_{\text{G}}}[\xi^\text{s},\acute{\xi}^\text{s}]\\=&-\sum\limits_{\mathbf{k}}\frac{\mathfrak{g}^2}{8\hbar m\omega(\mathbf{k})}\int_0^Tdt\int_0^t dt'\left[\mathcal{Z}(t)-\mathcal{\acute{Z}}(t)\right]\left[\mathcal{Z}(t')e^{-i\omega(\mathbf{k})[t-t']}-\acute{\mathcal{Z}}(t')e^{i\omega(\mathbf{k})[t-t']}\right]\\
=&-\frac{m_0^2G}{16\pi^2\hbar}\sum\limits_{\mathbf{k}}\left(\frac{2\pi}{L}\right)^3\frac{1}{\omega(\mathbf{k})}\int_0^Tdt\int_0^t dt'\left[\mathcal{Z}(t)-\mathcal{\acute{Z}}(t)\right]\left[\mathcal{Z}(t')e^{-i\omega(\mathbf{k})[t-t']}-\acute{\mathcal{Z}}(t')e^{i\omega(\mathbf{k})[t-t']}\right]\end{split}
\end{equation}
where in the last line of the above equation, we have substituted the analytical forms of the coupling constant $\mathfrak{g}$ and the effective mass $m$.
Up to the above equation, we have considered discrete mode sums over the wave vector $\vec{k}=\mathbf{k}$ and we can now come to the continuous picture by implementing the substitution $\sum\limits_{\mathbf{k}}\left(\frac{2\pi}{L}\right)^3\rightarrow \int d^3\omega$. In the continuum limit, we can recast eq.(\ref{QGInfluence.17.117}) as
\begin{equation}\label{QGInfluence.18.118}
\begin{split}
i\Phi^\beta_{0_{\text{G}}}[\xi^\text{s},\acute{\xi}^\text{s}]=&-\frac{m_0^2G}{4\pi\hbar}\int_0^\infty \omega d\omega\int_0^T dt\int_0^{t}dt'\cos[\omega(t-t')]\left[\mathcal{Z}(t)-\acute{\mathcal{Z}}(t)\right]\left[\mathcal{Z}(t')-\acute{\mathcal{Z}}(t')\right]\\&+\frac{im_0^2G}{4\pi\hbar}\int_0^\infty \omega d\omega\int_0^T dt\int_0^{t}dt'\sin[\omega(t-t')]\left[\mathcal{Z}(t)-\acute{\mathcal{Z}}(t)\right]\left[\mathcal{Z}(t')+\acute{\mathcal{Z}}(t')\right]~.
\end{split}
\end{equation}
In the above equation, we obtain the phase factor corresponding to the vacuum influence functional for a gravitational field. The important thing to observe is that the frequency integrals are indeed divergent. This can be solved by giving a cut-off frequency to the frequency integrals such that the integrals get regularized. From the last line of eq.(\ref{QGInfluence.18.118}), it is possible to write down the frequency integral via using the integral representation of the Dirac-Delta function as
\begin{equation}\label{QGInfluence.19.119}
\int_{0}^\infty d\omega~\omega \sin[\omega(t-t')]=-\pi\dot{\delta}(t-t')
\end{equation} 
where the $\dot{\delta}(t-t')=\frac{d}{d(t-t')}\delta(t-t')$.
Using eq.(\ref{QGInfluence.19.119}), we can simplify the second integral in the expression for the phase term in eq.(\ref{QGInfluence.17.117}) as
\begin{equation}\label{QGInfluence.20.120}
\begin{split}
&\int_0^\infty \omega d\omega\int_0^T dt\int_0^{t}dt'\sin[\omega(t-t')]\left[\mathcal{Z}(t)-\acute{\mathcal{Z}}(t)\right]\left[\mathcal{Z}(t')+\acute{\mathcal{Z}}(t')\right]\\=&-\pi\int_0^Tdt\int_0^t dt'\dot{\delta}(t-t')\left[\mathcal{Z}(t)-\acute{\mathcal{Z}}(t)\right]
\left[\mathcal{Z}(t')+\acute{\mathcal{Z}}(t')\right]\\
=&-\frac{\pi}{2}~\int_0^T dt \left[\mathcal{Z}(t)-\acute{\mathcal{Z}}(t)\right]\left[\dot{\mathcal{Z}}(t)+\dot{\acute{\mathcal{Z}}}(t)\right]+\pi \int_0^T dt~\delta(0) \left[\mathcal{Z}^2(t)-\acute{\mathcal{Z}}^{2}(t)\right]\\&-\frac{\pi}{2}(\mathcal{Z}^2(0)-\acute{\mathcal{Z}}^{2}(0))
\end{split}
\end{equation}
where we have made use of the identity, $\int_0^tdt'\delta(t-t')f(t')=f(t)\Theta(0)$ to arrive at the last line of the above equation. After eq.(\ref{Influence.13.80}), we have made the assumption that $\dot{\xi}^{\text{s}}(t)$, $\dot{\acute{\xi}}^{\text{s}}(t)$, $\ddot{\xi}^{\text{s}}(t)$, and $\ddot{\acute{\xi}}^{\text{s}}(t)$ all vanishes at $t=0$ and $t=T$. This implies that $\mathcal{Z}(0)=\mathcal{\acute{Z}}(0)=0$ which results in the vanishing of the final term in the last line of eq.(\ref{QGInfluence.20.120}). The second term $\pi \int_0^T dt~\delta(0) \left[\mathcal{Z}^2(t)-\acute{\mathcal{Z}}^{2}(t)\right]$ in the above equation resembles the difference of two actions and as has been argued in \cite{QGravD}, it is also possible to add a suitable counter-term in the point-particle action in our analysis to get rid of the second term as well. It is important to notice that throughout the calculation, some higher order terms in the GUP parameter $\beta$ has been kept to write all the results in a more elegant form, however, one can also drop the higher order terms. In any case, while obtaining the standard deviation in the geodesic separation, we shall drop any higher order contributions resulting in the same result. In the expression for the total influence phase term from eq.(\ref{QGInfluence.18.118}), we shall now look at the first term and recast it in the following way
\begin{equation}\label{QGInfluence.21.121}
\begin{split}
&-\frac{m_0^2 G}{4\pi\hbar}\int_0^\infty d\omega~\omega\int_0^T dt\int_0^t dt' \cos[\omega(t-t')]\left[\mathcal{Z}(t)-\acute{\mathcal{Z}}(t)\right]\left[\mathcal{Z}(t')-\acute{\mathcal{Z}}(t')\right]\\
=&-\frac{m_0^2 G}{8\pi\hbar}\int_0^\infty d\omega~\omega\int_0^T dt\int_0^T dt' \cos[\omega(t-t')]\left[\mathcal{Z}(t)-\acute{\mathcal{Z}}(t)\right]\left[\mathcal{Z}(t')-\acute{\mathcal{Z}}(t')\right]\\
=&-\frac{m_0^2}{32\hbar^2}\int_0^T dt\int_0^T dt'\left[\frac{4\hbar G}{\pi}\int_0^\infty d\omega~\omega \cos[\omega(t-t')]\right]\left[\mathcal{Z}(t)-\acute{\mathcal{Z}}(t)\right]\left[\mathcal{Z}(t')-\acute{\mathcal{Z}}(t')\right]
\end{split}
\end{equation}
where in the second line of the above equation, we have made use of the symmetry of the cosine function. We now defined the term inside of the parenthesis (involving the frequency integral) as a new function and write it as
\begin{equation}\label{QGInfluence.22.122}
\mathcal{A}_0(t-t')\equiv \frac{4\hbar G}{\pi}\int_0^\infty d\omega~\omega \cos[\omega(t-t')]~.
\end{equation}
Surely the function $\mathcal{A}_0(t-t')$ is divergent and one can impose a cut-off frequency as a gravitational wave detector is not sensitive to very high frequencies. Again the entire formalism also does not hold above frequencies beyond the Planck regime which enables one to put a hard cut-off frequency. Implementing the cut-off frequency, one can rewrite $\mathcal{A}_0(t-t')$ as 
\begin{equation}\label{QGInfluence.23.123}
\mathcal{A}^{\text{R}}_0(t-t')=\frac{4\hbar G}{\pi}\int_0^{\omega_\text{max}} d\omega~\omega \cos[\omega(t-t')]
\end{equation}
with $\omega_\text{max}$ introducing the hard cut-off frequency. The autocorrelation function $\mathcal{A}^{\text{R}}_0(t-t')$ obeys the identity, 
$\int_0^T d\tau \mathcal{A}^\text{R}_0(t-\tau)\mathcal{A}^{\text{R}^{-1}}_0(\tau-t')=\delta(t-t').$
With the implementation of the cut-off frequency, the integral in eq.(\ref{QGInfluence.21.121}) is now non-divergent and with the implementation of the Feynman-Vernon trick (which is a disguised reverse version of the Gaussian integration), we can express the exponential term with the exponent being the expression in eq.(\ref{QGInfluence.21.121}) as
\begin{equation}\label{QGInfluence.24.124}
\begin{split}
&\exp\left[-\frac{m_0^2}{32\hbar^2}\int_0^T dt\int_0^T dt'\mathcal{A}^\text{R}_0(t-t')\left[\mathcal{Z}(t)-\acute{\mathcal{Z}}(t)\right]\left[\mathcal{Z}(t')-\acute{\mathcal{Z}}(t')\right]\right]\\
=& \int\tilde{\mathcal{D}}\mathcal{N}_0~\exp\left[-\frac{1}{2}\int_0^Tdt\int_0^Tdt'\mathcal{A}^{\text{R}^{-1}}_0(t,t')\mathcal{N}_0(t)\mathcal{N}_0(t')+\frac{im_0}{4\hbar}\int_0^T dt~\mathcal{N}_0(t)\left[\mathcal{Z}(t)-\acute{\mathcal{Z}}(t)\right]\right]
\end{split}
\end{equation}
where the additional coefficients have been absorbed in the path integral measure $\mathcal{D}\mathcal{N}_0$ and expressed as $\tilde{\mathcal{D}}\mathcal{N}_0$. 
We now need to properly interpret the newly invoked function $\mathcal{N}_0(t)$ in eq.(\ref{QGInfluence.24.124}).  One can now define the average of $\mathcal{N}_0(t)$ as \cite{QGravD}
\begin{equation}\label{QGInfluence.25.125}
\langle\mathcal{N}_0(t)\rangle\equiv\int\mathcal{D}\mathcal{N}_0\exp\Bigr[-\frac{1}{2}\int_0^T\int_0^Tdt' dt''\mathcal{A}^{\text{R}^{-1}}_0(t',t'')\mathcal{N}_0(t')\mathcal{N}_0(t'')\Bigr]\mathcal{N}_0(t)=0~.
\end{equation}
Using the above definition, one can obtain the two point correlator as
\begin{equation}\label{QGInfluence.26.126}
\begin{split}
\langle \mathcal{N}_0(t)\mathcal{N}_0(t')\rangle&\equiv \int\mathcal{D}\mathcal{N}_0\exp\Bigr[-\frac{1}{2}\int_0^T\int_0^Tdt'' dt'''\mathcal{A}^{\text{R}^{-1}}_0(t'',t''')\mathcal{N}_0(t'')\mathcal{N}_0(t''')\Bigr]\mathcal{N}_0(t)\mathcal{N}_0(t')\\&=\mathcal{A}^{\text{R}}_0(t,t')~.
\end{split}
\end{equation}
The important thing to note is that the two-point correlator of $\mathcal{N}_0(t)$ has a non-zero value whereas the one-point correlator vanishes giving $\mathcal{N}_0(t)$ the interpretation of a stochastic function. This stochastic function generates solely due to the consideration of gravitons and their interaction with the particle in the detector-graviton model system and hence the auxiliary function $\mathcal{N}_0(t)$ (having a Gaussian probability density) can be considered as a noise term which are directly related to the graviton induced quantum gravitational fluctuations. Making use of the surviving terms in eq.(\ref{QGInfluence.20.120}) and using eq.(\ref{QGInfluence.24.124}), one can write down the full vacuum influence functional corresponding to the gravitational field as
\begin{equation}\label{QGInfluence.27.127}
\begin{split}
&F_{0_\text{G}}^\beta[\xi^\text{s},\acute{\xi}^\text{s}]\\=& \int\tilde{\mathcal{D}}\mathcal{N}_0~\exp\left[-\frac{1}{2}\int_0^Tdt\int_0^Tdt'\mathcal{A}^{\text{R}^{-1}}_0(t,t')\mathcal{N}_0(t)\mathcal{N}_0(t')+\frac{im_0}{4\hbar}\int_0^T dt~\mathcal{N}_0(t)\left[\mathcal{Z}(t)-\acute{\mathcal{Z}}(t)\right]\right]\\&\times\exp\left[-\frac{im_0^2G}{8\hbar}\int_0^Tdt \left[\mathcal{Z}(t)-\acute{\mathcal{Z}}(t)\right]\left[\dot{\mathcal{Z}}(t)+\dot{\acute{\mathcal{Z}}}(t)\right]\right]~.
\end{split}
\end{equation}
With the form of the vacuum influence functional in hand, we need the analytical form of the term inside of the parenthesis in the right hand side of eq.(\ref{QGInfluence.14.114}). Considering each of the gravitational wave modes to be in a coherent state, we can simplify $\prod\limits_{\mathbf{k}}\langle \psi^\text{g}_{\omega(\mathbf{k})} |e^{-\mathcal{V}^*\hat{a}^\dagger}e^{\mathcal{V}\hat{a}}|\psi^\text{g}_{\omega(\mathbf{k})}\rangle$ as
\begin{equation}\label{QGInfluence.28.128}
\begin{split}
\prod\limits_{\mathbf{k}}\langle \alpha^\text{g}_{\omega(\mathbf{k})} |e^{-\mathcal{V}^*\hat{a}^\dagger}e^{\mathcal{V}\hat{a}}|\alpha^\text{g}_{\omega(\mathbf{k})}\rangle =&\exp\left[\sum\limits_\omega(-\mathcal{V}^*\alpha^{\text{g}^*}_{\omega}+\mathcal{V}\alpha^\text{g}_\omega)\right]\\
=&\exp\left[\frac{im_0}{4\hbar}\int_0^Tdt~\bar{\mathfrak{h}}(t)\left[\mathcal{X}(t)-\acute{\mathcal{X}}(t)-3\beta m_0^2(\mathcal{Y}(t)-\acute{\mathcal{Y}}(t))\right]\right]
\end{split}
\end{equation}
where $\bar{\mathfrak{h}}(t)$ is defined as
\begin{equation}\label{QGInfluence.29.129}
\bar{\mathfrak{h}}(t)\equiv\sum\limits_\omega\frac{\mathcal{f}^\text{g}_\omega}{\sqrt{\hbar G}}\cos(\omega t+\phi^\text{g}_\omega)~.
\end{equation}
Multiplying eq.(\ref{QGInfluence.27.127}) with eq.(\ref{QGInfluence.28.128}), we arrive at the total influence functional $F^\beta_{\Psi_\text{G}}[\xi^\text{s},\acute{\xi}^\text{s}]$ and using that, we can write down the transition probability for the detector to go from its initial state to the final state where the gravitational field is initially in the state $|\Psi_{\text{G}}\rangle$ (where all of its individual field-modes are in coherent states) as 
\begin{equation}\label{QGInfluence.30.130}
\begin{split}
&P_{\Psi_\text{G}}^{[\phi^\text{s}_i\rightarrow\phi^\text{s}_f]}=\int d\xi^\text{s}_id\acute{\xi}^\text{s}_id\xi^\text{s}_fd\acute{\xi}^\text{s}_f\phi^\text{s}_i(\xi^\text{s}_i)\phi_i^{\text{s}^*}(\acute{\xi}^\text{s}_i)\phi^{\text{s}^*}_f(\xi^\text{s}_f)\phi^\text{s}_f(\acute{\xi}^\text{s}_f)\int\tilde{\mathcal{D}}\mathcal{N}_0~e^{-\frac{1}{2}\int_0^Tdt\int_0^Tdt'\mathcal{A}^{\text{R}^{-1}}_0(t,t')\mathcal{N}_0(t)\mathcal{N}_0(t')}\\&\times\int\left[\tilde{\mathcal{D}}\xi^\text{s}\right]_{\xi^\text{s}_i,0}^{\xi^\text{s}_f,T}\left[\tilde{\mathcal{D}}\acute{\xi}^\text{s}\right]_{\acute{\xi}^\text{s}_i,0}^{\acute{\xi}^\text{s}_f,T}~e^{\frac{im_0}{2\hbar}\int_{0}^{T}dt\left[\left[\dot{\xi}^{\text{s}^2}-\dot{\acute{\xi}}^{\text{s}^2}\right]-2\beta m_0^2\left[\dot{\xi}^{\text{s}^4}-\dot{\acute{\xi}}^{\text{s}^4}\right]+\left[\mathcal{Z}(t)-\acute{\mathcal{Z}}(t)\right]\left[\frac{\bar{\mathfrak{h}}(t)+\mathcal{N}_0(t)}{2}-\frac{m_0G}{4}\left[\dot{\mathcal{Z}}(t)+\dot{\acute{\mathcal{Z}}}(t)\right]\right]\right]}~.
\end{split} 
\end{equation}
With the analytical form of the transition probability in hand, we are now in a position to make use of the saddle-point approximation which shall give us the quantum gravity modified geodesic deviation equation in the generalized uncertainty principle framework. The reason behind the implementation of the saddle-point approximation lies in the fact that the path integrals over $\xi^{\text{s}}$ and $\acute{\xi}^\text{s}$ will have dominant contributions from the saddle-points corresponding to eq.(\ref{QGInfluence.30.130}). After a tedious amount of simplifications, we can finally obtain two coupled differential equations given by
\begin{equation}\label{QGInfluence.31.131}
\begin{split}
&\ddot{\xi}^\text{s}(1-12\beta m_0^2\dot{\xi}^{\text{s}^2})-\frac{\xi^{\text{s}}}{2}\left[(\ddot{\bar{\mathfrak{h}}}(t)+\ddot{\mathcal{N}}_0(t))(1-9\beta m_0^2\dot{\xi}^{\text{s}^2})-m_0G\left(\dddot{\acute{\mathcal{X}}}-3\beta m_0^2 \dddot{\acute{\mathcal{Y}}}-9\beta m_0^2\dot{\xi}^{\text{s}^2} \dddot{\acute{\mathcal{X}}}\right.\right.\\&\left.\left.-9\beta m_0^2\dot{\xi}^{\text{s}^2}(\dot{\mathcal{X}}-\dot{\acute{\mathcal{X}}})\right)\right]+3\beta m_0^2 (\dot{\xi}^{\text{s}^2}+3\xi^\text{s}\ddot{\xi}^\text{s})\left[\dot{\bar{\mathfrak{h}}}(t)+\dot{\mathcal{N}}_0(t)-m_0G\ddot{\acute{\mathcal{X}}}\right]\dot{\xi}^\text{s}\\&+\frac{9\beta m_0^3G}{2}(\mathcal{X}-\acute{\mathcal{X}})(\dot{\xi}^\text{s}\ddot{\xi}^\text{s}+2\xi^\text{s}\dddot{\xi}^\text{s})\ddot{\xi}^\text{s}=0
\end{split}
\end{equation}
and the counterpart equation with $\xi^\text{s}(t)$ replaced by $\acute{\xi}^\text{s}(t)$ and vice versa. Using the arguments in \cite{QGravD}, one can further simplify the above stochastic Langevin-like geodesic deviation equation by setting $\xi^\text{s}=\acute{\xi}^\text{s}$ which gives $\mathcal{X}=\acute{\mathcal{X}}$ and $\mathcal{Y}=\acute{\mathcal{Y}}$\footnote{As suggested in \cite{QGravD}, we can also define two new variables $\bar{\xi}=\frac{\xi+\xi'}{2}$ and $\Delta \xi=\xi-\xi'$, and keep terms up to $\mathcal{O}(\Delta \xi).$}. Making use of the above simplification along with an appropriate dimensional reconstruction, we arrive at the simplified Langevin-like stochastic geodesic deviation equation as
\begin{equation}\label{QGInfluence.32.132}
\begin{split}
&\ddot{\xi}^\text{s}-\frac{1}{2}\left[\left(\ddot{\bar{\mathfrak{h}}}(t)+\ddot{\mathcal{N}}_0(t)-\frac{m_0G}{c^5}\frac{d^5}{dt^5}(\xi^{\text{s}^2})\right)\left(1+3\beta m_0^2\dot{\xi}^{\text{s}^2}\right)+\frac{3\beta m_0^3G }{c^5}\frac{d^4}{dt^4}\left(\frac{d}{dt}\left(\xi^{\text{s}^2}\right)\dot{\xi}^{\text{s}^2}\right)\right]\xi^\text{s}\\&+3\beta m_0^2\left(\dot{\xi}^{\text{s}^3}+3\xi^\text{s}\dot{\xi}^\text{s}\ddot{\xi}^\text{s}\right)\left[\dot{\bar{\mathfrak{h}}}(t)+\dot{\mathcal{N}}_0(t)-\frac{m_0G}{c^5}\frac{d^4}{dt^4}\left(\xi^{\text{s}^2}\right)\right]=0~.
\end{split}
\end{equation}
The above equation is one of the main results in our analysis. Our next aim is to obtain the Langevin-like stochastic differential equation when the individual graviton states are in squeezed states.
\subsection{Graviton squeezed state and the stochastic geodesic deviation equation}\label{C3.3.3.Squeezing}
\noindent One perspective is evident that even if one calculates the standard deviation in the geodesic separation, the quantum gravity induced effect would be very small as we are working in the low-energy limit. One of the ways to enhance such signals is to incorporate squeezing in the picture via considering squeezed graviton states \cite{QGravD,QGravLett,QGravNoise,KannoSodaTokuda,
KannoSodaTokuda2}. We start by considering that the initial single-mode state of the gravitational wave is in a squeezed state $|\psi^{\text{Sq.}}_\omega\rangle=\hat{S}(\mathcal{z}_{\omega})|0^\text{g}_\omega\rangle$ where the analytical form of the squeezing operator in terms of graviton raising and lowering operators is given as $\hat{S}(\mathcal{z}_{\omega})=e^{\frac{1}{2}(\mathcal{z}_\omega^*\hat{a}^2-\mathcal{z}_\omega{\hat{a}^{\dagger2})}}$ with $\mathcal{z}_\omega$ being given as $\mathcal{z}_\omega=\mathfrak{r}_\omega e^{i\varphi_\omega}$. Here, $\mathfrak{r}_\omega$ denotes the squeezing parameter and $\varphi_\omega$ denotes the squeezing angle. One can now make use of the following identities involving the squeezing operator
\begin{align}
\hat{S}^\dagger(\mathcal{z}_\omega)\hat{a}\hat{S}(\mathcal{z}_\omega)&=\cosh \mathfrak{r}_\omega\hat{a}-e^{i\varphi_\omega}\sinh \mathfrak{r}_\omega\hat{a}^\dagger\label{QGInfluence.33.133}~,\\
\hat{S}^\dagger(\mathcal{z}_\omega)\hat{a}^\dagger \hat{S}(\mathcal{z}_\omega)&=\cosh \mathfrak{r}_\omega\hat{a}^\dagger-e^{-i\varphi_\omega}\sinh \mathfrak{r}_\omega\hat{a}\label{QGInfluence.34.134}
\end{align}
to obtain the analytical form of the influence functional for a single gravitational wave mode when the graviton is initially in a squeezed vacuum state. The single mode influence functional then reads
\begin{equation}\label{QGInfluence.35.135}
\begin{split}
F^\beta_{\mathcal{z}_\omega}\left[\xi^\text{s},\acute{\xi}^\text{s}\right]=&F^\beta_{0^\text{g}_\omega}\left[\xi^\text{s},\acute{\xi}^\text{s}\right]\langle 0^\text{g}_\omega |\hat{S}(\mathcal{z}_\omega)^\dagger e^{-\mathcal{V}^*\hat{a}^\dagger}e^{\mathcal{V}\hat{a}}\hat{S}(\mathcal{z}_\omega)|0^\text{g}_\omega\rangle\\
=&F^\beta_{0^\text{g}_\omega}\left[\xi^\text{s},\acute{\xi}^\text{s}\right]e^{i\Phi_{\mathcal{z}_\omega}\left[\xi^\text{s},\acute{\xi}^\text{s}\right]}\\
=&F^\beta_{0^\text{g}_\omega}\left[\xi^\text{s},\acute{\xi}^\text{s}\right]e^{-\frac{1}{2}|\mathcal{V}|^2(\cosh 2\mathfrak{r}_\omega-1)-\frac{1}{4}(\mathcal{V}^{2}e^{i\varphi_\omega}+\mathcal{V}^{*2}e^{-i\varphi_\omega})\sinh 2\mathfrak{r}_\omega}
\end{split}
\end{equation}
where the influence phase $i\Phi^\beta_{\mathcal{z}_\omega}[\xi^\text{s},\acute{\xi}^\text{s}]$ can be expressed as
\begin{equation}\label{QGInfluence.36.136}
\begin{split}
i\Phi^\beta_{\mathcal{z}_\omega}[\xi^\text{s},\acute{\xi}^\text{s}]&=\frac{\mathfrak{g}^2}{16\hbar m\omega}\int_0^T dt\int_0^T dt'\cos[\omega(t+t')-\varphi_\omega][\mathcal{Z}(t)-\acute{\mathcal{Z}}(t)][\mathcal{Z}(t')-\acute{\mathcal{Z}}(t')]\sinh 2\mathfrak{r}_\omega \\
&-\frac{\mathfrak{g}^2}{16\hbar m\omega}\int_0^Tdt\int_0^Tdt'\cos[\omega(t-t')]
[\mathcal{Z}(t)-\acute{\mathcal{Z}}(t)][\mathcal{Z}(t')-\acute{\mathcal{Z}}(t')](\cosh 2\mathfrak{r}_\omega -1).
\end{split}
\end{equation}
The graviton squeezing parameter however can be considered to be independent of the gravitational mode frequency. In general squeezed graviton state can come via primordial gravitational waves which were generated during the inflationary period \cite{KannoSodaTokuda}. As a result, it is not quite required that the squeezing will depend on the mode frequency rather we can consider $\mathfrak{r}_\omega\rightarrow\mathfrak{r}$ \cite{QGravD}. Again summing over all mode frequencies (similar to the coherent state case), we obtain the analytical form of the total influence phase factor $i\Phi^{\beta}_{\mathcal{z}}[\xi,\xi']$ as
\begin{equation}\label{QGInfluence.37.137}
\begin{split}
&i\Phi^{\beta}_\mathcal{z}[\xi,\xi']\\=&-\frac{m_0^2G}{8\pi\hbar}(\cosh 2\mathfrak{r}-1)\int _0^\infty \omega d\omega\int_0^Tdt\int_0^Tdt'\cos(\omega(t-t'))(\mathcal{Z}(t)-\acute{\mathcal{Z}}(t))(\mathcal{Z}(t')-\acute{\mathcal{Z}}(t'))\\
&+\frac{m_0^2G}{8\pi\hbar}\sinh 2\mathfrak{r}\int_0^\infty\omega d\omega\int_0^Tdt\int_0^Tdt'\cos(\omega(t+t')-\varphi_\omega)(\mathcal{Z}(t)-\acute{\mathcal{Z}}(t))(\mathcal{Z}(t')-\acute{\mathcal{Z}}(t'))~.
\end{split}
\end{equation}
With the form of the influence phase obtained in the above equation, we shall now write down the analytical form of the total influence functional as
\begin{equation}\label{QGInfluence.38.138}
\begin{split}
&F^{\beta}_\mathcal{z}[\xi^{\text{s}},\acute{\xi}^\text{s}]\\=&F^\beta_{0_{\text{G}}}[\xi^\text{s},\acute{\xi}^\text{s}]\exp\left[i\Phi^{\beta}_\mathcal{z}[\xi^\text{s},\acute{\xi}^\text{s}]\right]\\
=&\exp\left[-\frac{m_0^2G}{8\pi\hbar}\cosh 2\mathfrak{r}\int _0^\infty \omega d\omega\int_0^Tdt\int_0^Tdt'\cos(\omega(t-t'))(\mathcal{Z}(t)-\acute{\mathcal{Z}}(t))(\mathcal{Z}(t')-\acute{\mathcal{Z}}(t'))\right.\\
&\left.+\frac{m_0^2G}{8\pi\hbar}\sinh 2\mathfrak{r}\int_0^\infty\omega d\omega\int_0^Tdt\int_0^Tdt'\cos(\omega(t+t')-\varphi_\omega)(\mathcal{Z}(t)-\acute{\mathcal{Z}}(t))(\mathcal{Z}(t')-\acute{\mathcal{Z}}(t'))\right]\\
\times& \exp\left[-\frac{im_0^2G}{8\pi\hbar}\int_0^Tdt (\mathcal{Z}(t)-\acute{\mathcal{Z}}(t))(\dot{\mathcal{Z}}(t)+\dot{\acute{\mathcal{Z}}}(t))\right]~.
\end{split}
\end{equation} 
The above expression for the influence functional needs to be examined carefully. We start with the first term in the influence phase obtained in eq.(\ref{QGInfluence.37.137}). We observe that the first term in eq.(\ref{QGInfluence.37.137}) has a $t-t'$ symmetry or a time-translational symmetry indicating a static term. However, the second term in eq.(\ref{QGInfluence.37.137}) breaks this symmetry leading to a non-static term. We shall therefore be more focussed on the static part of the influence functional in eq.(\ref{QGInfluence.38.138}). It is therefore required to define a static and a non-static auxiliary function, the analytical forms of which are given by
\begin{align}
\mathcal{A}^{\text{S.}}(t-t')&=\frac{4\hbar G}{\pi}\cosh 2\mathfrak{r}\int_0^{\omega_{\text{max}}} d\omega\hspace{0.5mm} \omega \cos (\omega(t-t'))\label{QGInfluence.39.139}\\
\mathcal{A}^{\text{N.S.}}(t+t')&=\frac{4\hbar G}{\pi}\sinh 2\mathfrak{r}\int_0^{\omega_{\text{max}}} d\omega\hspace{0.5mm} \omega\cos (\omega (t+t')-\varphi_\omega)\label{QGInfluence.40.140}
\end{align}
where S. and N.S. in the superscripts of the autocorrelation function $\mathcal{A}(t,t')$ denote the static and the non-static contributions. One important thing to notice is that we have already implemented the cut-off frequency $\omega_{\text{max}}$ to avoid divergences of the autocorrelation functions.

\noindent Making use of the Feynman-Vernon trick, it is now possible to obtain two new noise terms, $\mathcal{N}^{\text{S}.}$ corresponding to the static autocorrelation function and $\mathcal{N}^{\text{N.S.}} $ corresponding to the non-static autocorrelation function. Compared to the coherent case, we can make simple identification between the static noise term and $\mathcal{N}_0(t)$ as $\mathcal{N}^{\text{S.}}(t)=\sqrt{\cosh 2\mathfrak{r}}\mathcal{N}_0(t)$. Similarly the $\mathcal{A}^{\text{S}.}(t-t')$ is related to $\mathcal{A}^{\text{R}}_0(t-t')$ by $\mathcal{A}^{\text{S.}}(t-t')=\cosh 2\mathfrak{r}\mathcal{A}^{\text{R}}_0(t-t')$.
With the identification $\mathcal{A}^{\text{S}.}(t-t')=\cosh 2\mathfrak{r}\mathcal{A}_0(t-t')$ and setting $\varphi_\omega=0$, we can now obtain the analytical form of the transition probability (up to some constant factor) as
\begin{equation}\label{QGInfluence.41.141}
\begin{split}
&P_{\mathcal{z}}^{[\phi^\text{s}_i\rightarrow\phi^\text{s}_f]}=\int d\xi^\text{s}_id\acute{\xi}^\text{s}_id\xi^\text{s}_fd\acute{\xi}^\text{s}_f\phi^\text{s}_i(\xi^\text{s}_i)\phi_i^{\text{s}^*}(\acute{\xi}^\text{s}_i)\phi^{\text{s}^*}_f(\xi^\text{s}_f)\phi^\text{s}_f(\acute{\xi}^\text{s}_f)\int\tilde{\mathcal{D}}\mathcal{N}_0~e^{-\frac{1}{2}\int_0^Tdt\int_0^Tdt'\mathcal{A}^{\text{R}^{-1}}_0(t-t')\mathcal{N}_0(t)\mathcal{N}_0(t')}\\&\times\int\tilde{\mathcal{D}}\mathcal{N}^{\text{N.S.}}~e^{\frac{1}{2}\int_0^Tdt\int_0^Tdt'\left(\mathcal{A}^{\text{N.S.}}(t+t')\right)^{-1}\mathcal{N}^{\text{N.S.}}(t)\mathcal{N}^\text{N.S.}(t')}\int\left[\tilde{\mathcal{D}}\xi^\text{s}\right]_{\xi^\text{s}_i,0}^{\xi^\text{s}_f,T}\left[\tilde{\mathcal{D}}\acute{\xi}^\text{s}\right]_{\acute{\xi}^\text{s}_i,0}^{\acute{\xi}^\text{s}_f,T}\\&\times e^{\frac{im_0}{2\hbar}\int_{0}^{T}dt\left[\left[\dot{\xi}^{\text{s}^2}-\dot{\acute{\xi}}^{\text{s}^2}\right]-2\beta m_0^2\left[\dot{\xi}^{\text{s}^4}-\dot{\acute{\xi}}^{\text{s}^4}\right]+\frac{1}{2}\left(\sqrt{\cosh2\mathfrak{r}}\mathcal{N}_0(t)+\mathcal{N}^{\text{N.S.}}(t)\right)\left[\mathcal{Z}(t)-\acute{\mathcal{Z}}(t)\right]-\frac{m_0G}{4}\left[\mathcal{Z}(t)-\acute{\mathcal{Z}}(t)\right]\left[\dot{\mathcal{Z}}(t)+\dot{\acute{\mathcal{Z}}}(t)\right]\right]}~.
\end{split} 
\end{equation}
From the analytical form of the above transition probability, we can again implement the saddle-point approximation, and setting $\xi^{\text{s}}=\acute{\xi}^\text{s}$, we arrive at the stochastic Langevin-like geodesic deviation equation as 
\begin{equation}\label{QGInfluence.42.142}
\begin{split}
&\ddot{\xi}^\text{s}-\frac{\xi^\text{s}}{2}\left[\left[\ddot{\mathcal{N}}^{\text{N.S.}}(t)+\sqrt{\cosh 2\mathfrak{r}}\ddot{\mathcal{N}}_0(t)-\frac{m_0G}{c^5}\frac{d^5\xi^{\text{s}^2}}{dt^5}\right]\left[1+3\beta m_0^2\dot{\xi}^{\text{s}^2}\right]+\frac{3\beta m_0^3G }{c^5}\frac{d^4}{dt^4}\left[\frac{d\xi^{\text{s}^2}}{dt}\dot{\xi}^{\text{s}^2}\right]\right]\\&+3\beta m_0^2\left(\dot{\xi}^{\text{s}^3}+3\xi^\text{s}\dot{\xi}^\text{s}\ddot{\xi}^\text{s}\right)\left[\ddot{\mathcal{N}}^{\text{N.S.}}(t)+\sqrt{\cosh 2\mathfrak{r}}\ddot{\mathcal{N}}_0(t)--\frac{m_0G}{c^5}\frac{d^4}{dt^4}\left(\xi^{\text{s}^2}\right)\right]=0~.
\end{split}
\end{equation}
This is one of the most important results in our analysis. Our next aim is to solve eq.(\ref{QGInfluence.32.132}) and eq.(\ref{QGInfluence.42.142}) analytically and from there obtain the uncertainties in the measurement of the geodesic separation for the two cases.
\section{Quantum gravitational ``Memory" effect?}
We have already obtained the Langevin-like stochastic differential equations involving the geodesic separation (or the detector arm length) for the gravitons initially being in a coherent state and squeezed vacuum states. The most important observation one can make from eq.(s)(\ref{QGInfluence.32.132}) and eq.(\ref{QGInfluence.42.142}) are that the geodesic deviation equations have contributions from terms involving GUP effects as well as the gravitational noise. If all other contributions are dropped then the geodesic deviation equation in eq.(\ref{QGInfluence.32.132}) reduces to $\ddot{\xi}^{\text{s}}(t)-\frac{1}{2}\ddot{\bar{\mathfrak{h}}}(t)\xi^{\text{s}}(t)=0$ which is the classical geodesic deviation equation due to small gravitational fluctuations where the $\ddot{\bar{\mathfrak{h}}}(t)\xi^{\text{s}}(t)$ denotes the tidal acceleration. The higher order terms involving the fifth and fourth order derivative indicate towards terms due to gravitational radiation reaction \cite{Thorne,Burke,ChandrasekharPaul,
GravitationalRadiationReaction}. The gravitational radiation reaction terms are significantly smaller compared to the other terms and can be neglected in this analysis considering that the geodesic separation is measured using coarse-graining techniques which renders the higher derivatives to be negligible.
From eq.(\ref{QGInfluence.32.132}), we can write down the simplified equation that we shall be solving as
\begin{equation}\label{QGInfluence.43.143}
\begin{split}
&\ddot{\xi}^\text{s}-\frac{1}{2}\left[\left(\ddot{\bar{\mathfrak{h}}}(t)+\ddot{\mathcal{N}}_0(t)\right)\left(1+3\beta m_0^2\dot{\xi}^{\text{s}^2}\right)\right]\xi^\text{s}+3\beta m_0^2\left(\dot{\xi}^{\text{s}^3}+3\xi^\text{s}\dot{\xi}^\text{s}\ddot{\xi}^\text{s}\right)\left[\dot{\bar{\mathfrak{h}}}(t)+\dot{\mathcal{N}}_0(t)\right]=0~.
\end{split}
\end{equation}
The above equation now can be solved using an iterative procedure. We start by considering the base part of the equation as $\ddot{\xi}^{\text{s}}(t)=0$. A straightforward solution of the above equation reads $\xi^{\text{s}(0)}=\xi^{\text{s}}_0+\lambda^\text{s} t$. In principle $\lambda^\text{s}$ has the dimension of the velocity and can have maximum value $\lambda^{\text{s}}_{\text{max}}=c$. In the process of a graviton being absorbed and released, one can have the maximum value of $t$ to be $t_{\text{max}}=\frac{\xi^{\text{s}}_0}{c}$. The reason behind this consideration lies in the fact that the gravitational wave passes at the speed of light. The leading order quantum gravity correction is achieved by considering up to $\mathcal{O}(\bar{\mathfrak{h}},\mathcal{N}_0)$. In order to incorporate the effect of the generalized uncertainty principle infused in the detector part, we also need to consider up to $\mathcal{O}(\beta,\beta\bar{\mathfrak{h}},\beta \mathcal{N}_0)$. Continuing the iterative procedure and keeping terms up to $\mathcal{O}(\beta,\beta\bar{\mathfrak{h}},\beta \mathcal{N}_0)$, we arrive at the approximate analytical expression for the geodesic separation as
\begin{equation}\label{QGInfluence.44.144}
\begin{split}
\xi^{\text{s}}(t)\cong&[\xi^{\text{s}}_0+\lambda^{\text{s}} t]\left[1+\frac{1}{2}\left[1+3\beta m_0^2\lambda^{\text{s}^2}\right][\bar{\mathfrak{h}}(t)+\mathcal{N}_0(t)]\right]-\lambda^\text{s}[1+6\beta m_0^2\lambda^{\text{s}^2}]\int_0^t d\tilde{t}(\bar{\mathfrak{h}}(\tilde{t})+\mathcal{N}_0(\tilde{t})].
\end{split}
\end{equation}
The higher limit of integration in the above expression has a cut-off at $t=\frac{\xi^{\text{s}}_0}{c}$. 
We are now in a position to calculate the standard deviation in the measurement of the geodesic deviation $\xi^\text{(s)}(t)$. The standard deviation is obtained by calculating $\sigma(t)=\sqrt{\left\langle\left(\xi^{\text{s}}(t)-\left\langle\xi^{\text{s}}\right\rangle\right)^2\right\rangle}$ where it is possible to separate the standard deviation term as $\sigma^{\text{s}}(t)=\sigma^{\text{s}}_0(t)+\sigma^{\text{s}}_{\beta}(t)$. The standard deviation without any GUP correction is equal to 
\begin{equation}\label{QGInfluence.45.145}
\sigma^{\text{s}}_0(t)\sim\sqrt{2\pi}l_{\text{Pl}}\sim 10^{-35}\text{ m}~.
\end{equation}
which is same as the result obtained in \cite{QGravLett,QGravD}. One important thing to note that the linear time dependent contribution in $\xi^{\text{s}(0)}(t)$ contributes towards the calculation of the $\sigma^{\text{s}}_{\beta}(t)$ part. One can, however, drop the linear contribution from the standard deviation expression which results in continuous stacking up of the standard deviation which is a standard result while considering calculation of gravitational memory effect and does not truly imply any quantum gravitational effects. In order to truly extract the quantum gravitational behaviour one needs to get rid of the overall linear contribution which results in the expression for the dimensionless standard deviation part $\frac{\sigma^{\text{s}}_\beta(t)}{\sqrt{2\pi}l_{\text{Pl}}}$ as
\begin{equation}\label{QGInfluence.46.146}
\begin{split}
&\frac{\sigma^{\text{s}}_\beta(t)}{\sqrt{2\pi}l_{\text{Pl}}}\cong3\beta m_0^2 c^2\left[1-\frac{2\xi^{\text{s}}_0}{\left[1+\frac{c t}{\xi^{\text{s}}_0}\right]}\frac{\sin^2\left[\frac{\pi c t}{\xi^{\text{s}}_0}\right]}{\pi^2 c t}+\frac{4}{\pi^2 \left[1+\frac{c t}{\xi^{\text{s}}_0}\right]^2}\left[\gamma_\varepsilon-\text{Ci}\left[\frac{2\pi c t}{\xi_0}\right]+\ln\left[\frac{2\pi c t}{\xi_0}\right]\right]\right]
\end{split}
\end{equation}
where $\text{Ci}$ is the cosine integral function \cite{Gradshteyn} and $\gamma_\varepsilon$ is the Euler constant. The total dimensionless standard deviation then reads
\begin{equation}\label{QGInfluence.47.147}
\begin{split}
\frac{\sigma^{\text{s}}(t)}{\sigma^{\text{s}}_0}=&1+3\beta m_0^2 c^2\left[1-\frac{2\xi^{\text{s}}_0}{\left[1+\frac{c t}{\xi^{\text{s}}_0}\right]}\frac{\sin^2\left[\frac{\pi c t}{\xi^{\text{s}}_0}\right]}{\pi^2 c t}+\frac{4}{\pi^2 \left[1+\frac{c t}{\xi^{\text{s}}_0}\right]^2}\left[\gamma_\varepsilon-\text{Ci}\left[\frac{2\pi c t}{\xi_0}\right]+\ln\left[\frac{2\pi c t}{\xi_0}\right]\right]\right]~.
\end{split}
\end{equation}
In case of squeezed graviton states one can obtain the analytical expression for the standard deviation as 
\begin{equation}\label{QGInfluence.48.148}
\sigma_{\text{Sq.}}^{\text{s}}(t)=\sqrt{\cosh 2r}\sigma^{\text{s}}(t)
\end{equation}
 which in shall indicate that the correction in the standard deviation due the GUP correction will also get enhanced by the squeezing parameter. 
\subsection{Bound on the GUP parameter and important phenomenological aspects}
The signatures of any quantum nature of gravity need to be detected using gravitational wave detectors. As a result for correct theoretical estimations, we shall use the parameters from the existing or proposed gravitational wave observatories. The currently operating observatories like LIGO or VIRGO are all $L$ shaped. The arm length of LIGO is $\xi^{\text{s}}_0=4$ km whereas VIRGO has a arm length of $\xi^{\text{s}}_0=3$ km. The arm is a large Fabry-Perot cavity where at the both ends there are suspended mirrors. The mirror coating consists of fused Silica, $\text{SiO}_2$. where the mass of a single molecule of  Silica is $m_{\text{SiO}_2}= 10^{-25}$ kg. This $\text{SiO}_2$ layer is also known as the low-index layer. The high-index layer is made of Tantalum pentoxide and the mass of a single molecule of $\text{Ta}_2\text{O}_5$ is $m_{\text{Ta}_2\text{O}_5}=10^{-24}$ kg \cite{Accadiaetal}. We can now substitute the mass of the heavier molecule which is $m_{\text{Ta}_2\text{O}_5}=10^{-24}$ kg to obtain a bound on the dimensionless GUP parameter $\beta_0$. From the expression for the dimensionless standard deviation $\frac{\sigma^{\text{s}}(t)}{\sqrt{2\pi}l_{\text{Pl}}}$ in eq.(\ref{QGInfluence.47.147}), it is evident that the leading order term $\frac{\sigma^\text{s}_0(t)}{\sqrt{2\pi}l_{\text{Pl}}}$ must be greater than the subleading corrections coming due to the consideration of the generalized uncertainty principle framework. Hence, we obtain the inequality $\rho \beta m_0^2c^2\leq 1$ where the maximum value of $\rho$ is of the order of 10. Using $m_0=m_{\text{Ta}_2\text{O}_5}$, we obtain the value of $\beta m_0^2 c^2$ to be of the order of $10^{-33}$. It is then possible to write down an inequality involving the dimensionless GUP parameter as $\beta_0\leq 10^{31}$. The bound obtained on the GUP parameter in this case is weaker than the bound obtained in \cite{GUPResonantDetector2} where the authors have used resonant bar detectors interacting with gravitational waves to estimate the bound on the dimensionless GUP parameter. However, the bound obtained 
in this current analysis is stronger than the case of \cite{BoundsGUP} where the bound was estimated using real observational data.
With the maximum value of the dimensionless GUP parameter, we can now proceed towards analyzing the time-dependence of the dimensionless standard deviation $\frac{\sigma^{\text{s}}(t)}{\sqrt{2\pi}l_{\text{Pl}}}$. As has been already observed in \cite{QGravD,QGravLett}, $\sigma_0(t)=\sqrt{2\pi}l_{\text{Pl}}\sim 10^{-35}$ m which is way beyond the current detectability range. It was however argued in \cite{QGravD,QGravLett} that if the gravitons are in a thermal or a squeezed vacuum state, the standard deviation can be enhanced. The most significant enhancement was observed in case of squeezed states where $\sigma_{\text{Sq.}}(t)=\sqrt{\cosh2\mathfrak{r}}\sigma_0(t)$. This indicates that if the squeezing from the initial graviton states are very high, in such a scenario the quantum gravity signatures may be detectable. Such highly squeezed states can occur during the inflation and post-inflationary periods of the cosmos \cite{InflationSqueezed1,
InflationSqueezed2,InflationSqueezed3} especially in primordial gravitational waves. As primordial gravitational waves have frequencies in the range $0.0001-10$ Hz, they lie way beyond the current detectability range of the existing gravitational wave detectors. In order to investigate relic gravitational waves in such low frequency ranges, two new space based gravity wave detectors LISA (Laser Interferometer Space Antenna) and DECIGO (Decihertz Interferometer Gravitational Wave Observatory) have been proposed and is targeted to be operational in a matter of five to ten years.
%\begin{figure}
%\begin{center}
%\includegraphics[scale=0.35]{OTM.jpg}
%\caption{$\frac{\sigma(t)}{\sqrt{2\pi}l_p}$ vs $t$ plot for gravitational wave with maximum frequency $\omega_{\text{max}}\sim 1 \text{ Hz}$\label{F1}.}
%\end{center}
%\end{figure}
\begin{figure}[t!]
\begin{center}
\includegraphics[scale=0.5]{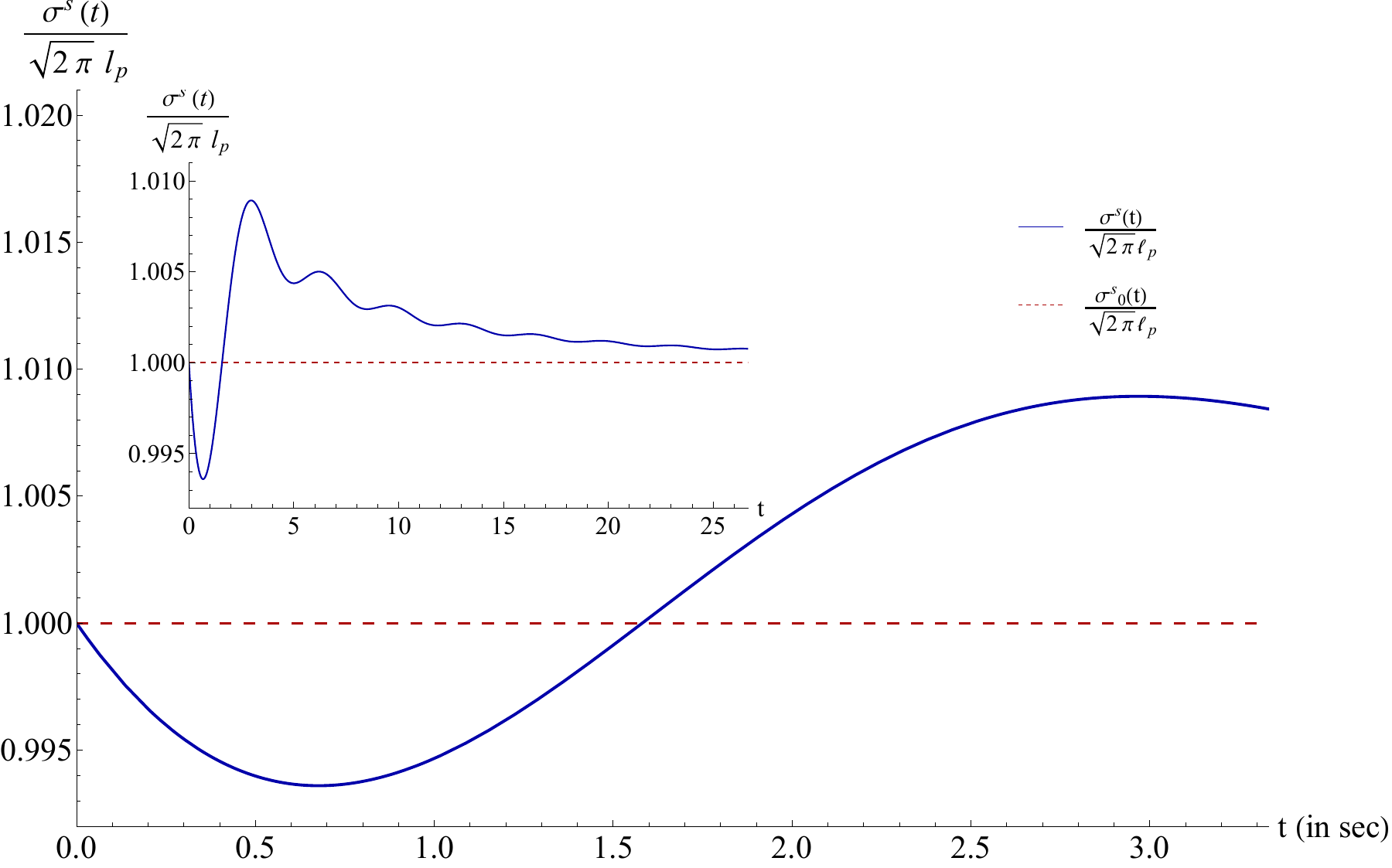}
\caption{The dimensionless standard deviation is plotted against time where the standard deviation without any $\beta$ correction is considered to be the reference line. The maximum frequency of the incoming gravitational wave is set to $\omega^{\text{max}}\sim 1$ Hz. The small inset image indicates the long-term behaviour of the GUP modified standard deviation where one observe a residual standard deviation which asymptotes towards $\frac{\sigma^\text{s}_0(t)}{\sqrt{2\pi}l_{\text{Pl}}}$ with increasing value of $t$.\label{Memory_OTM}}
\end{center}
\end{figure}
In case of primordial gravitational waves, the squeezing parameter $\mathfrak{r}$ can be as high as $\mathfrak{r}\sim 50-58$. With a squeezing $\mathfrak{r}\sim 58$, the enhancing factor $\sqrt{\cosh 2\mathfrak{r}}\sim\frac{1}{2}\exp[\mathfrak{r}]\sim 10^{24}$ \cite{KannoSodaTokuda}. In such a scenario $\sigma_{\text{Sq.}}\sim 10^{-11}$ m= 10 picometer which
will lie in the projected LISA sensitivity range ($10^{-12}$ m \cite{LISASensitivity}) \footnote{However, we find in \cite{QGravD} that the projected sensitivity of LISA is $10^{-18}$ m which requires a squeezing of only $\mathfrak{r}=42$.}. 
For the GUP correction to the standard deviation, we shall be using values from the projected parameters of LISA observatory where the geodesic separation is of the order of $\xi^{\text{s}}_0\sim10^6 \text{ km}$ and the maximum frequency is of the order of $\omega_{\text{max}}\sim 1\text{ Hz}$ \cite{QGravD}.  Using $\beta_0\sim 10^{31}$, one can obtain $\sigma^{\text{s}}_\beta(t)\sim 10^{-37}$ m at the time $t=t_{\text{max}}=10/3$ sec. For squeezed graviton state $\sigma^{\text{s}}_{\text{Sq.}_\beta}(t)\sim \sqrt{\cosh 2\mathfrak{r}}^{-37}$ and for $\mathfrak{r}\sim 56$, $\sigma^{\text{s}}_{\text{Sq.}_\beta}(t)\sim10^{-13}$ m which lies an order beyond the proposed LISA sensitivity range \cite{LISASensitivity}. With a squeezing of $\mathfrak{r}\sim58$, the correction in the standard deviation due to the generalized uncertainty principle lies in the LISA sensitivity range\footnote{Following \cite{QGravD} only $\mathfrak{r}\sim 44$ is enough for $\sigma^{\text{s}}_{\text{Sq.}_\beta}(t_{\text{max}})$ to be in the projected sensitivity range of LISA.}.  %We can observe that at $t=0$ and $t=t_{\text{max}}$ the GUP effect is negligible due to the existence of the square of the sinusoidal term and the maximum effect can be obtained if the standard deviation is measured in the middle of the range $0<t<t_{\text{max}}$. It can be shown that the optimal value of the standard deviation will be $\sigma_{\text{Squeezed}}(t\sim \frac{t_{\text{max}}}{2})\sim \sqrt{\cosh 2r} \left(10^{-34}+(10^{-36}-10^{-37})\right)\text{ m}$ where we have written the contribution due to the GUP effect as a range in the parentheses. 
One can find out the ratio $\frac{\sigma^{\text{s}}_{\text{Sq.}_\beta}(t)}{\sigma^{\text{s}}_{\text{Sq.}_0}}\sim10^{-2}$. Hence, for gravitons with high enough squeezing, the fluctuations due GUP effect can be observed. In order to truly investigate the behaviour of the GUP part of the standard deviation, we plot the dimensionless standard deviation $\frac{\sigma^{\text{s}}(t)}{\sqrt{2\pi}l_{\text{Pl}}}$ against the interaction time $t$ in Fig.(\ref{Memory_OTM}). As a base line, we have used the time-independent standard deviation $\sigma^{\text{s}}_0$ which refers to the $\frac{\sigma^{\text{s}}_0}{\sqrt{2\pi}l_{\text{Pl}}}$ line in Fig.(\ref{Memory_OTM}). To truly plot the time-dependence of the standard deviation, we have plotted $\frac{\sigma^{\text{s}}(t)}{\sqrt{2\pi}l_{\text{Pl}}}=\frac{\sigma^{\text{s}}_0+\acute{\sigma}^\text{s}_\beta(t)}{\sqrt{2\pi}l_{\text{Pl}}}$ where $\acute{\sigma}^\text{s}_\beta(t)=\sigma^\text{s}_\beta(t)-\sigma^{\text{s}}_\beta(0)$.
From Fig.(\ref{Memory_OTM}), it is evident that the standard deviation in presence of the generalized uncertainty principle framework becomes dependent with time. We observe that within $t\sim 10/3-4$ sec, the standard deviation becomes smaller and then higher than the base value of the standard deviation which is indicated by the red dashed line. What is important to observe that this enhanced deviation does not die out suddenly and asymptotes towards the red dashed line with higher values of the time $t$. This can be considered as a quantum gravitational ``memory" effect. The GUP effect makes sure that the fundamental minimal length effects are infused into the detector degrees of freedom and as a result after the interaction with gravitons the detector retains some of the gravitational fluctuations purely due to the consideration of generalized uncertainty principle framework. One thing to remember is if we have not got rid of the linear time dependent parts as a whole from the expression of the standard deviation the deviation would keep on increasing mimicking some results from standard gravitational wave memory effects. In our analysis, however, we have primarily considered the deviation which generates due to the consideration of low-energy quantum gravity theory which indeed is maximum during lower values of $t$. Now if such fluctuations are detected in future generation of gravitational wave observatories, it will not only indicate towards the existence of gravitons but also indicate towards the existence of a modified Heisenberg's uncertainty relation.
%\begin{figure}
%\begin{center}
%\includegraphics[scale=0.35]{OTM2.jpg}
%\caption{$\frac{\sigma(t)}{\sqrt{2\pi}l_p}$ vs $t$ plot for gravitational wave with maximum frequency $\omega_{\text{max}}\sim 1 \text{ Hz}$\label{F2}. The plot signifies that the time-dependent standard deviation asymptotes towards the value of the standard deviation obtained in \cite{QGravLett,QGravD}  (dotted line) with time.}
%\end{center}
%\end{figure}
\section{Discussion and conclusion}
In this chapter, we discuss a simply freely falling two-particle system with one of the particles being way heavier than the other one. To depict the trajectory of the smaller mass particle one needs to construct a Fermi-normal coordinate system where the origin of the coordinate system moves with the freely falling particle. This model depicts the arm of an `$L$' shaped gravitational wave observatory like LIGO or VIRGO. The analysis considers the interaction of the gravitational wave detector with an incoming gravitational wave while the small spacetime fluctuations are quantized where the detector degrees of freedom obey the generalized uncertainty principle \cite{KempfManganoMann}. As the detector arms are very long, it is prudent to use a one dimensional model while the gravitational wave is carrying a plus polarization only. Following the approach and methodology of \cite{QGravD,QGravLett}, we have considered a path integral approach to quantize the gravitational part of the model system. The model considers that the detector absorbs a graviton and after some time it emits a graviton and in this process both the final state of the detector and the graviton gets changed. However, in our analysis it is not important to consider the final state of the graviton and as a result all the final graviton states are summed over. One can now extend this analysis from a single graviton mode to a gravitational field via summing over all graviton mode frequencies which provides us with an action for the detector-graviton model system. Using a saddle-point approximation, one can then finally arrive at the geodesic deviation equation in this GUP infused quantum gravity framework. It is observed that the geodesic deviation equation gets infused by stochastic graviton noise fluctuations where the noise-fluctuations also get coupled to the GUP infused part of the equation. Calculating the standard deviation, we then arrive at an expression which is dependent on the total measurement time. This is a very interesting result in our analysis. For the gravitons being in squeezed states, we find out that the standard part as well as the GUP corrected part get enhanced by an exponential factor for very high value of the squeezing parameter. Then using the idea that the GUP corrected part can not surpass the standard deviation term, we also obtain a bound on the dimensionless GUP parameter which is found to be tighter than standard bounds which have been obtained using observational data from the existing gravitational wave observatories. Using the maximum value of the GUP parameter possible in our current analysis, we plot the time-dependent and dimensionless GUP parameter against time. We observe from the plot that due to the GUP infusion, the standard deviation first decreases and then increases and finally asymptotes towards the time-independent standard deviation with increasing values of time. This is an important observation and it indicates that the GUP infusion leads to a quantum gravitational ``memory"-like effect which if detected in future generation gravitational wave observatories, will be conclusive evidence for the existence of the signatures of quantum gravity along with the existence of an underlying generalized uncertainty principle.

\chapter{A quantum gravitational uncertainty relation}\label{C.4.OTM}
In the previous chapter, we investigated the effect of graviton interaction on a two-particle system where the geodesic separation between the two particles is infused by graviton-induced noise in a generalized uncertainty principle framework. We have already discussed in detail the effect of the consideration of a generalized uncertainty principle framework in a quantum gravitational scenario. Now, the fundamental question that can be asked is whether the quantization of linearized gravity itself leads to a quantum gravity modified uncertainty relation.
In a series of works \cite{QGravNoise,QGravLett,QGravD,
KannoSodaTokuda,KannoSodaTokuda2}, interaction of a gravitational wave detector with a gravitational wave have been considered where the gravitational fluctuations are quantized. These analyses lead towards the finding of small quantum gravity signatures in classical observable quantities, with one of them being the geodesic deviation. One important thing to note is that all of these models are based on low-energy quantum gravity theories. As has already been discussed in the earlier chapters, the geodesic deviation equations get infused with stochastic quantum gravitational fluctuation terms. In \cite{ChawlaParikh}, the authors have taken a much simpler model where they have considered that a point particle is freely falling under the effect of Earth's gravitational field. The freely falling particle is considered to be interacting with gravitational fluctuations (primarily from gravitational waves) while the quantum gravitational fluctuations are quantized. Instead of the standard analysis using a flat Minkowski background as has been done in \cite{QGravNoise,QGravLett,QGravD,KannoSodaTokuda,
KannoSodaTokuda2}, the authors in \cite{ChawlaParikh} have made use of a slightly curved background over which the gravitational fluctuations have been considered. For the slightly curved geometry, the post-Newtonian metric \cite{Weinberg} has been used with no off-diagonal components. Considering the gravitational fluctuations to be quantized, it was observed in \cite{ChawlaParikh} that aside from a few relativistic corrections (special and general), there is a stochastic noise fluctuation term in Newton's universal law for gravitational free-fall\footnote{Formulation of a quantum gravitational matter system has been done in a recent analysis \cite{ManicciaMontaniAntonini}}.

\noindent In this chapter\footnote{This chapter is based on the publication S. Sen and S. Gangopadhyay, ``\textit{Uncertainty principle from the noise of gravitons}", \href{https://link.springer.com/article/10.1140/epjc/s10052-024-12481-7}{Eur. Phys. J. C 84 (2024) 116}.}, we use the basic model proposed in \cite{ChawlaParikh} and inspect the product of the standard deviations in the position as well as the momentum operators when the entire free-fall is observed by an observer standing on the Earth. In order to truly extract quantum gravity effects, we have considered up to Newtonian correction term in the background geometry only. The uncertainty in the position has already been calculated and later argued for squeezed graviton states in \cite{ChawlaParikh}. Our analysis, therefore, boils down to calculating the uncertainty in the momentum for different graviton states and multiplying it by the corresponding standard deviation in position. Now, as has been argued and discussed extensively in chapters (\ref{C.2.OTM}) and (\ref{C.3.OTM}), any existing quantum field theories of gravity propose the existence of a minimal length in nature, which is of the order of the Planck length. The existence of the minimal length improves the Heisenberg uncertainty relation, which is called as the generalized uncertainty principle \cite{KempfManganoMann}. This form of the generalized uncertainty relation has been used quite extensively in \cite{MaggioreAlgebraic,ScardigliGedanken,AdlerSantiago,
AdlerSantiago2,GUPOscillator1,GUPOscillator2,
AliDasVagenas,MajumderGUP,GUPOpto1,GUPOpto2,GUPOpto3,
SGADAS,ScardigliCasadio,FengYangLiZu,
RMSBSG,GUPResonantDetector1,
CorpuscularGUP,GUPResonantDetector2,
GUPResonantDetector3,PetruzzeilloIlluminati,GUPOpto4,GUPOpto5,
GUPOpto6,OngJCAP,Vagnozzi,Culetu}. One of the primary problems with the current form of the generalized uncertainty principle is that the GUP parameter is an undetermined constant with no appropriate analytical expression. The next problem lies in the fact that the correction term to the Heisenberg uncertainty principle is independent of Planck's constant. A true quantum gravity correction must contain the Planck's constant as well as Newton's gravitational constant in a way that if one of the constants is taken to zero, the correction term vanishes. Our primary motivation in this work, therefore, lies in the investigation of an uncertainty relation when true quantum gravity effects are considered in the model and derive the uncertainty product and obtain the inequality from there. One important thing to remember is that the entire calculation considers Newtonian approximation, and as a result, further investigation is needed for a proper Lorentz-invariant uncertainty relation. The post-Newtonian corrections do not play an important role in the leading order of the uncertainty product and as a result have been neglected in our analysis. 

\noindent The chapter is organized as follows. We start with the basic discussion of the model proposed in \cite{ChawlaParikh} and discuss some of the important results. We then proceed to obtain the standard deviation in the position and momentum, and from there obtain the uncertainty product. We then propose to write down inequalities involving the uncertainty product for the graviton to be initially in a vacuum state, a squeezed vacuum state, and finally in a thermal state. We then investigate the uncertainty relation for the three cases and search for a universal behaviour, and check for any limiting form of the uncertainty product. Finally, we conclude our results. 
\section{Apple falling under Earth's gravitational field}
\noindent 
In this section, we start with the discussion and review of the model system proposed in \cite{ChawlaParikh}. The model considers a point particle freely falling under the effect of Earth's gravitational field while it is being observed by a terrestrial observer. The background of the model system then reads
\begin{equation}\label{Apple.1}
g_{\mu\nu}=g^{\oplus}_{\mu\nu}+h_{\mu\nu}
\end{equation}  
where $g^\oplus_{\mu\nu}$ denotes the metric due to the Earth and $h_{\mu\nu}$ denotes standard gravitational fluctuations. The line element solely due to the Earth's gravitational field in the post-Newtonian approximation takes the form \cite{Weinberg}
\begin{equation}\label{Apple.2}
\begin{split}
ds^2=&g^{\oplus}_{\mu\nu}dx^\mu dx^\nu\\
=&-\left(1+2\phi_\oplus+2\phi_\oplus^2+2\psi_\oplus+\cdots\right)dt^2+\left(\delta_{jk}-2\phi_\oplus\delta_{jk}+g_{jk}^{(4)}+\cdots\right)dx^jdx^k
\end{split}
\end{equation}
where it is evident that the contributions from all the off-diagonal components have been neglected as the slow rotation of Earth is being neglected in this model. In the above expression $\phi_\oplus^2,\psi_\oplus$, and $g^{(4)}_{jk}$ denote the leading order post Newtonian corrections and $\phi_\oplus$ denotes the Newtonian potential due to the Earth's gravitational field. It is now possible to write down the gravitational fluctuations in the transverse-traceless gauge. In the transverse-traceless gauge the particle part of the action takes the form
\begin{equation}\label{Apple.3}
\begin{split}
S_p=&m_0^{\text{p}}\int dt\left(\frac{1}{2}\dot{\xi}^{\text{p}^{2}}+\frac{1}{8}\dot{\xi}^{\text{p}^4}-\frac{3}{2}\phi_\oplus\dot{\xi}^{\text{p}^2}-\phi_\oplus-\frac{1}{2}\phi_\oplus^2-\psi_\oplus+\frac{1}{4}\ddot{\mathfrak{h}}_{jk}\xi^{\text{p}^j}\xi^{\text{p}^k}\right)
\end{split}
\end{equation}
where the coordinate for the freely falling particle with mass $m_0^{\text{p}}$ is given by $\mathfrak{Y}^{\text{p}^\mu}=\{t,\xi^{\text{p}^j}\}$. The action for the gravitational part is given by the Einstein-Hilbert action which up to $\mathcal{O}(\mathfrak{h}^2)$ reads (in the transverse-traceless gauge)
\begin{equation}\label{Apple.4}
S_{\text{EH}}=-\frac{1}{64\pi G}\int d^4 x~ \partial_{\mu}\mathfrak{h}_{ij}\partial^{\mu}\mathfrak{h}^{ij}.
\end{equation}
which is same as eq.(\ref{QGMinimal.6.38}). It is then possible to use the same mode decomposition of the fluctuation term $\mathfrak{h}_{jk}$ as in eq.(\ref{QGMinimal.14.46}), as (inside of a box with volume $V$ and length $L$)
\begin{equation}\label{Apple.5}
\mathfrak{h}_{jk}(t,\vec{x})=\frac{1}{l_\text{Pl}}\sum\limits_{\vec{k},s}q_s(\vec{k})e^{i\vec{k}\cdot\vec{x}}\epsilon^{s}_{jk}(\vec{k})
\end{equation}
where $\epsilon_{jk}^s(\vec{k})$ gives the polarization tensor corresponding to the gravitational fluctuation $\mathfrak{h}_{jk}$. Now, the model is of a freely falling particle, where the Earth's rotation is being neglected. Hence, it is quite natural to restrict the entire analysis to a single dimension which is the direction perpendicular to the Earth's surface or the $z$ direction. Further considering only plus polarization of the gravitational wave with a single mode, one can write down the Lagrangian for the model system as
\begin{equation}\label{Apple.6}
\begin{split}
L=&m_0^{\text{p}}\left(\frac{1}{2}\dot{z}^{\text{p}^2}+\frac{1}{8}\dot{z}^{\text{p}^4}-\frac{3}{2}\phi_\oplus\dot{z}^{\text{p}^2}-\phi_\oplus-\frac{1}{2}\phi_\oplus^2-\psi_\oplus\right)+\frac{1}{2}m\left(\dot{q}^2-\omega^2q^2\right)-\mathfrak{g}\dot{q}\dot{z}^\text{p}z^\text{p}
\end{split}
\end{equation}
where the effective mass is defined as $m\equiv\frac{V}{16\pi\hbar G^2}$ and the particle-field coupling constant reads $\mathfrak{g}=\frac{m_0^{\text{p}}}{2l_{\text{Pl}}}$, and the coordinate is aligned in a way such that $\xi^{\text{p}}_1=z^\text{p}$.  Obtaining the canonically conjugate momenta to $z^\text{p}$ and $q$, one can write down the Hamiltonian for the model system as \cite{ChawlaParikh}
\begin{equation}\label{Apple.7}
H\simeq\left(\frac{p^2}{2m}+\frac{\frac{\pi_z^{\text{p}^2}}{2}+\frac{\mathfrak{g}p\pi_z^\text{p} z^{\text{p}}}{m}}{m_0^{\text{p}}(1-3\phi_\oplus)}\right)\left(1-\frac{\mathfrak{g}^2z^{\text{p}^2}}{mm_0^{\text{p}}\left(1-3\phi_\oplus\right)}\right)^{-1}+\frac{1}{2}m\omega^2q^2+m_0^{\text{p}}\left(\phi_\oplus+\frac{\phi_\oplus^2}{2}+\psi_\oplus\right)
\end{equation}
where $\pi_z^\text{p}=\frac{\partial L}{\partial \dot{z}^\text{p}}$ and $p=\frac{\partial L}{\partial\dot{q}}$. In the above Hamiltonian, $\phi_\oplus=\phi_\oplus(z^\text{p})$ and $\psi_\oplus=\psi_\oplus(z^\text{p})$. With the basic model in hand, one can quantize the model by raising the phase space variables to operator status and implementing appropriate canonical commutation relation between them. As a path integral quantization is used in \cite{ChawlaParikh}, a direct canonical quantization technique is not required. Consider that the initial state of the graviton is given by $|\Psi^\text{g}\rangle$ where initially the state of the freely falling particle is $|\Phi_A^{\text{p}}\rangle$. Now, while touching the ground, if the final state of the particle is $|\Phi_B^\text{p}\rangle$ and the state of the graviton is $|\mathcal{F}^\text{g}\rangle$ at time $t=T$, then the transition probability for the system to go from an initial state $|\Psi^\text{g},\Phi_A^{\text{p}}\rangle$ to $|\mathcal{F}^\text{g},\Phi_B^{\text{p}}\rangle$ reads
\begin{equation}\label{Apple.8}
P^{\Psi^\text{g}}_{A\rightarrow B}=\sum_{|\mathcal{F}^\text{g}\rangle}\left|\langle \mathcal{F}_\text{g},\Phi^\text{p}_B|\hat{U}(T,0)|\Psi^\text{g},\Phi_A^{\text{p}}\rangle\right|^2
\end{equation}
where $T$ denotes the total time of interaction between the particle and the graviton. As before, all final states of the gravitons have been summed over as the dynamics of the particle is the point of investigation in this analysis. Now, making use of a path integral analysis, one can follow the methodology of \cite{QGravNoise,QGravLett,QGravD,ChawlaParikh} as well as the methodology of the previous chapter to write down the transition probability in the above equation in a path integral representation. Now summing over all graviton modes and making use of the Feynman-Vernon trick \cite{FeynmanVernon}, one can arrive at the analytical form of the transition probability as
\begin{equation}\label{Apple.9}
\begin{split}
&P^{\Psi^\text{g}}_{A\rightarrow B}=\int dz^\text{p}_id\acute{z}^\text{p}_idz^\text{p}_fd\acute{z}^\text{p}_f\Phi^\text{p}_i(z^\text{p}_i)\Phi_i^{\text{p}^*}(\acute{z}^\text{p}_i)\Phi^{\text{p}^*}_f(z^\text{p}_f)\Phi^\text{p}_f(\acute{z}^\text{p}_f)\int\tilde{\mathcal{D}}\mathcal{N}_{\text{q}}~e^{-\frac{1}{2}\int_0^Tdt\int_0^Tdt'\mathcal{A}^{-1}_{\Psi^\text{g}}(t,t')\mathcal{N}_{\text{q}}(t)\mathcal{N}_{\text{q}}(t')}\\&\times\int\left[\tilde{\mathcal{D}}z^\text{p}\right]_{z^\text{p}_i,0}^{z^\text{p}_f,T}\left[\tilde{\mathcal{D}}\acute{z}^\text{p}\right]_{\acute{z}^\text{p}_i,0}^{\acute{z}^\text{p}_f,T}~e^{\frac{im_0^{\text{p}}}{\hbar}\int_{0}^{T}dt\left[\left(\mathcal{I}^\text{p}(z)-\mathcal{I}^\text{p}(\acute{z})\right)+\frac{1}{4}\mathcal{N}_{\text{q}}(t)\left(\mathcal{X}(t)-\acute{\mathcal{X}}(t)\right)-\frac{m_0^{\text{p}}G}{8}\left(\mathcal{X}(t)-\acute{\mathcal{X}}(t)\right)\left(\dot{\mathcal{Z}}(t)+\dot{\acute{\mathcal{Z}}}(t)\right)\right]}
\end{split} 
\end{equation}
where the function $\mathcal{X}(t)$ is defined as  $\mathcal{X}(t)=\frac{d^2}{dt^2}(z^{\text{p}^2}(t))$ and $\mathcal{I}^{\text{p}}(z)$ is defined as
\begin{equation}\label{Apple.10}
\mathcal{I}^\text{p}(z)\equiv\frac{1}{2}\dot{z}^{\text{p}^2}-\frac{3}{2}\phi_\oplus(z^\text{p})\dot{z}^{\text{p}^2}-\phi_\oplus(z^\text{p})-\frac{1}{2}\phi^2_\oplus(z^\text{p})-\psi_\oplus(z^\text{p})~.
\end{equation}
In eq.(\ref{Apple.9}), $\mathcal{N}_{\text{q}}(t)$ is a random fluctuation term. The stochastic average of the random fluctuation term is defined as
\begin{equation}\label{Apple.11}
\llangle\mathcal{N}_{\text{q}}(t)\rrangle\equiv\int\mathcal{D}\mathcal{N}_{\text{q}}\exp\Bigr[-\frac{1}{2}\int_0^T\int_0^Tdt' dt''\mathcal{A}^{-1}_{\Psi^\text{g}}(t',t'')\mathcal{N}_{\text{q}}(t')\mathcal{N}_{\text{q}}(t'')\Bigr]\mathcal{N}_{\text{q}}(t)=0~.
\end{equation}
One can again obtain the two point correlator similar to eq.(\ref{QGInfluence.26.126}) as 
\begin{equation}\label{Apple.12}
\begin{split}
\llangle \mathcal{N}_{\text{q}}(t)\mathcal{N}_{\text{q}}(t')\rrangle&\equiv \int\mathcal{D}\mathcal{N}_{\text{q}}\exp\Bigr[-\frac{1}{2}\int_0^T\int_0^Tdt'' dt'''\mathcal{A}^{-1}_{\Psi^\text{g}}(t'',t''')\mathcal{N}_{\text{q}}(t'')\mathcal{N}_{\text{q}}(t''')\Bigr]\mathcal{N}_{\text{q}}(t)\mathcal{N}_{\text{q}}(t')\\&=\mathcal{A}_{\Psi^\text{g}}(t,t')~.
\end{split}
\end{equation}
Here, ``$\llangle\rrangle$" is used instead of ``$\langle\rangle$" to emphasize upon the fact that a stochastic average is being taken. Eq.(\ref{Apple.11}) combined with eq.(\ref{Apple.12}) portrays that $\mathcal{N}_{\text{q}}(t)$ is a stochastic noise fluctuation term. Making use of a saddle-poin approximation, it is now possible to find out the differential equation for $z^\text{p}$ from eq.(\ref{Apple.9}) as
\begin{equation}\label{Apple.13}
\ddot{z}^\text{p}=-g+\frac{1}{2}\ddot{\mathcal{N}}_{\text{q}}(t)z^\text{p}(t)
\end{equation}
where $g$ gives the acceleration due to the Earth's gravitational field and all other relativistic contributions have been dropped as well. From eq.(\ref{Apple.13}), it is possible to write down the quantum gravity modified Newton's equation as
\begin{equation}\label{Apple.14}
F=m_0^{\text{p}}\ddot{z}^\text{p}(t)\simeq m_0^{\text{p}}\left(-g+\frac{1}{2}\ddot{\mathcal{N}}_{\text{q}}(t)z^\text{p}(t)\right)
\end{equation}
which is the result obtained in \cite{ChawlaParikh}. However, we obtain an extra half factor in front of the noise term. In the next section we shall try to obtain the uncertainty product by individually obtaining the uncertainties in the position and momentum of the freely falling particle. 
\section{Quantum gravity modified uncertainty relation}
\noindent The first step towards obtaining the uncertainty product is to solve eq.(\ref{Apple.13}). We at first will reproduce the result for the standard deviation in position from \cite{ChawlaParikh}. In standard nonrelativistic classical mechanics, when a particle freely falls under the effect of gravity, then its trajectory is given by
\begin{equation}\label{QGUncertainty.15}
z^{\text{p}}_{\text{cl}}(t)=z^\text{p}_0-\frac{1}{2}gt^2
\end{equation}
where $z_0$ is the initial height from which the particle is being dropped and $g$ is the acceleration due to the Earth's gravitational field. If at time $\tau_g$, the particle touches the surface of the Earth then the classical value of $z$ becomes $z(\tau_g)=0$, and from eq.(\ref{QGUncertainty.15}), we can obtain the analytical expression for $\tau_g$ as $\tau_g=\sqrt{\frac{2z^\text{p}_0}{g}}$. The equation in eq.(\ref{QGUncertainty.15}), is a solution of the second order differential equation $\ddot{z}^{\text{p}}_{\text{cl}}(t)=0$. As we can observe from eq.(\ref{Apple.13}), the second-order ordinary differential equation has inputs from stochastic noise fluctuations due to the interaction between the particle and the graviton. Hence, it is possible to separate the solution of eq.(\ref{Apple.13}) into two parts
\begin{equation}\label{QGUncertainty.16}
z^{\text{p}}(t)\simeq z^{\text{p}}_{\text{cl}}(t)+z^{\text{p}}_{\text{QG}}(t)
\end{equation}
where $z^{\text{p}}_{\text{cl}}(t)$ is given by eq.(\ref{QGUncertainty.15}) and $z^{\text{p}}_{\text{QG}}(t)$ gives the quantum gravity corrected part. Using the above ansatz, and substituting it back in eq.(\ref{Apple.13}), we arrive at the differential equation involving $z^{\text{p}}_{\text{QG}}(t)$ as
\begin{equation}\label{QGUncertainty.17}
\ddot{z}^{\text{p}}_{\text{QG}}(t)\simeq \frac{1}{2}\ddot{\mathcal{N}}_{\text{q}}(t)z^{\text{p}}_{\text{cl}}(t)
\end{equation}
where the $\frac{1}{2}\mathcal{O}(z^{\text{p}}_{\text{QG}}(t) \mathcal{N}_{\text{q}}(t))$ term has been dropped as it contributes in the second-order of the graviton-noise fluctuation. Integrating the above equation with respect to time and doing integration by parts, one can arrive at the following relation\begin{equation}\label{QGUncertainty.18}
\begin{split}
\dot{z}^{\text{p}}_{\text{QG}}(t)&\simeq\frac{1}{2}\dot{\mathcal{N}}_{\text{q}}(t)z^{\text{p}}_{\text{cl}}(t)-\frac{1}{2}\mathcal{N}_{\text{q}}(t)\dot{z}^{\text{p}}_{\text{cl}}(t)+\frac{1}{2}\int_0^tdt'\mathcal{N}_{\text{q}}(t')\ddot{z}^{\text{p}}_{\text{cl}}(t)\\
&=\frac{1}{2}\dot{\mathcal{N}}_{\text{q}}(t)z^{\text{p}}_{\text{cl}}(t)+\frac{1}{2}gt\mathcal{N}_{\text{q}}(t)-\frac{g}{2}\int_0^tdt'\mathcal{N}_{\text{q}}(t')~.
\end{split}
\end{equation}
Integrating the above equation one final time and executing integration by parts, we arrive at the analytical part of the quantum gravity corrected part of $z^\text{p}(t)$ as
\begin{equation}\label{QGUncertainty.19}
\begin{split}
z^{\text{p}}_{\text{QG}}(t)&\simeq\frac{1}{2}\mathcal{N}_{\text{q}}(t)z^{\text{p}}_{\text{cl}}(t)+\int_0^tdt' gt'\mathcal{N}_{\text{q}}(t') -\frac{g}{2}\int_0^tdt'\int_0^{t'} dt''\mathcal{N}_{\text{q}}(t'')~.
\end{split}
\end{equation}
The above result is same as obtained in \cite{ChawlaParikh} up to an overall $\frac{1}{2}$ factor. Taking a stochastic average of $z^{\text{p}}_{\text{QG}}(t)$, we then obtain
\begin{equation}\label{QGUncertainty.20}
\begin{split}
\llangle z^{\text{p}}_{\text{QG}}(t)\rrangle&=\simeq\frac{1}{2}\llangle\mathcal{N}_{\text{q}}(t)\rrangle z^{\text{p}}_{\text{cl}}(t)+\int_0^tdt' gt'\llangle \mathcal{N}_{\text{q}}(t')\rrangle -\frac{g}{2}\int_0^tdt'\int_0^{t'} dt''\llangle\mathcal{N}_{\text{q}}(t'')\rrangle=0
\end{split}
\end{equation}
where we have made use of eq.(\ref{Apple.11}). The above result is a direct consequence of the fact that $\mathcal{N}_{\text{q}}(t)$ is a stochastic noise term. Taking stochastic average of eq.(\ref{QGUncertainty.16}), we obtain
\begin{equation}\label{QGUncertainty.21}
\llangle z^{\text{p}}(t)\rrangle=\llangle (z^{\text{p}}_{\text{cl}}(t)+z^{\text{p}}_{\text{QG}}(t))\rrangle=z^{\text{p}}_{\text{cl}}(t)~.
\end{equation}
One can now write down the variance in the position of the particle at time $t$ as
\begin{equation}\label{QGUncertainty.22}
\begin{split}
(\Delta z^{\text{p}}(t))^2=&\llangle z^\text{p}(t) z^\text{p}(t)\rrangle-\llangle z^\text{p}(t)\rrangle^2\\
=&\llangle z^{\text{p}}(t) z^\text{p}(t)\rrangle-z^\text{p}_{\text{cl}}(t)^2\\
=&\llangle z^{\text{p}}_{\text{QG}}(t) z^\text{p}_{\text{QG}}(t)\rrangle+\llangle z^{\text{p}}_{\text{cl}}(t) z^\text{p}_{\text{QG}}(t)\rrangle+\llangle z^{\text{p}}_{\text{QG}}(t) z^\text{p}_{\text{cl}}(t)\rrangle-z^\text{p}_{\text{cl}}(t)^2\\
=&\llangle z^{\text{p}}_{\text{QG}}(t) z^\text{p}_{\text{QG}}(t)\rrangle-z^\text{p}_{\text{cl}}(t)^2
\end{split}
\end{equation}
where in the second line and last line of the above equation, we have made use of eq.(\ref{QGUncertainty.20}). As we are mostly interested in obtaining the uncertainty relation while the particle touches the ground (as it maximizes the time of interaction between the particle and the gravitons), we need to write down the variance obtained in eq.(\ref{QGUncertainty.22}) at time $t=\tau_g=\sqrt{\nicefrac{2z_0^\text{p}}{g}}$. It is important to note that at time $t=\tau_g$, $z^{\text{p}}_{\text{cl}}(\tau_g)=0$. At time $t=\tau_g$, the variance in position reads \cite{ChawlaParikh}
\begin{equation}\label{QGUncertainty.23}
\begin{split}
\left(\Delta z^\text{p}(\tau_g)\right)^2=&\llangle z^\text{p}_\text{QG}(\tau_g) z^\text{p}_\text{QG}(\tau_g)\rrangle\\
=&g^2\int_0^{\tau_g} t' dt'\int_0^{\tau_g} \bar{t} d\bar{t} \llangle \mathcal{N}_{\text{q}}(t')\mathcal{N}_{\text{q}}(\bar{t})\rrangle-g^2\int_{0}^\tau dt'\int_0^{t'} dt''\int_0^{\tau_g}\bar{t}d\bar{t}\llangle \mathcal{N}_{\text{q}}(t'')\mathcal{N}_{\text{q}}(\bar{t})\rrangle\\
+&\frac{g^2}{4}\int_0^{\tau_g}dt'\int_0^{t'}dt''\int_{0}^{\tau_g} d\bar{t}\int_0^{\bar{t}}d\bar{\bar{t}}\llangle\mathcal{N}_{\text{q}}(t'')\mathcal{N}_{\text{q}}(\bar{\bar{t}})\rrangle~.
\end{split}
\end{equation}
In our analysis, we are more focused on the quantum gravity-induced noise fluctuations in the uncertainties of the position and momentum of the particle, and as a result, all  post-Newtonian contributions can be dropped from the Lagrangian in eq.(\ref{Apple.6}). This gives us the Lagrangian for the model system in the Newtonian approximation as 
\begin{equation}\label{QGUncertainty.24}
\begin{split}
L=&\frac{1}{2}m_0^{\text{p}}\dot{z}^{\text{p}^2}-m_0^{\text{p}}\phi_\oplus(z^\text{p})+\frac{1}{2}m\left(\dot{q}^2-\omega^2q^2\right)-\mathfrak{g}\dot{q}\dot{z}^\text{p}z^\text{p}.
\end{split}
\end{equation}
From the above Lagrangian, one can obtain the conjugate momentum to $z^{\text{p}}$ as
\begin{equation}\label{QGUncertainty.25}
\pi_z^\text{p}=\frac{\partial L}{\partial \dot{z}^\text{p}}=m_0^{\text{p}}\dot{z}^{\text{p}}-\mathfrak{g}\dot{q}z^\text{p}~.
\end{equation}
Our next aim is to calculate the variance in the momentum $\pi_z^{\text{p}}$ at time $t=\tau_g$. Substituting the form of $z^{\text{p}}$ from eq.(\ref{QGUncertainty.16}), we obtain the form of $\pi_z^{\text{p}}(\tau_g)$ as
\begin{equation}\label{QGUncertainty.26}
\begin{split}
\pi^\text{p}_z(\tau_g)&=m_0^{\text{p}}(\dot{z}^{\text{p}}_{\text{cl}}(\tau_g)+\dot{z}^{\text{p}}_{\text{QG}}(\tau_g))-\mathfrak{g}\dot{q}(\tau_g)\left(z^\text{p}_{\text{cl}}(\tau_g)+z^{\text{p}}_\text{QG}(\tau_g)\right)\\
&=m_0^{\text{p}}(\dot{z}^{\text{p}}_{\text{cl}}(\tau_g)+\dot{z}^{\text{p}}_{\text{QG}}(\tau_g))-\mathfrak{g}\dot{q}(\tau_g)z^{\text{p}}_\text{QG}(\tau_g)
\end{split}
\end{equation}
where in the last line of the above equation, we have used the result $z^\text{p}_{\text{cl}}(\tau_g)=0$\footnote{The expression $\dot{z}^{\text{p}}_{\text{cl}}(\tau_g)$ denotes $\dot{z}^{\text{p}}_{\text{cl}}(\tau_g)=\left.\frac{d z^{\text{p}}_{\text{cl}}(t)}{dt}\right.\rvert_{t\rightarrow\tau_g}$ and the similar thing holds true for $\dot{z}^{\text{p}}_\text{QG}(\tau_g)$.}. Some important observations are in order now. We already now that $\mathfrak{g}$ is the graviton-particle coupling constant. In the last term of eq.(\ref{QGUncertainty.26}), the coupling constant is multiplied with the quantum gravitation part $z^{\text{p}}_\text{QG}(\tau_g)$ of the position of the particle which is very small as well. Hence, as a result one can neglect the contribution from the last term in eq.(\ref{QGUncertainty.26}). Up to a good approximation, one can then recast the analytical expression for the momentum $\pi^{\text{p}}_z$ as 
\begin{equation}\label{QGUncertainty.27}
\pi^{\text{p}}_z(\tau_g)\simeq m_0\left(\dot{z}^{\text{p}}_{\text{cl}}(\tau_g)+\dot{z}^{\text{p}}_\text{QG}(\tau_g)\right)~.
\end{equation}
Hence, one can compute the square of the variance in the momentum variable to be
\begin{equation}\label{QGUncertainty.28}
\begin{split}
(\Delta\pi_z^{\text{p}}(\tau_g))^2=&m_0^{\text{p}^2}\left(\llangle\dot{z}^{\text{p}}(\tau_g)\dot{z}^{\text{p}}(\tau_g)\rrangle-\llangle\dot{z}^{\text{p}}(\tau_g)\rrangle^2\right)\\
=&m_0^{\text{p}^2} \left(\dot{z}_{\text{cl}}^{\text{p}^2}(\tau_g)+2\dot{z}^{\text{p}}_{\text{cl}}(\tau_g)\llangle \dot{z}^\text{p}_\text{QG}(\tau_g)\rrangle+\llangle \dot{z}^{\text{p}}_{\text{QG}}(\tau_g)\dot{z}^{\text{p}}_{\text{QG}}(\tau_g)\rrangle-\dot{z}_{\text{cl}}^{\text{p}^2}(\tau_g)\right)\\
=&m_0^{\text{p}^2}\left(2\dot{z}^{\text{p}}_{\text{cl}}(\tau_g)\llangle \dot{z}^{\text{p}}_\text{QG}(\tau_g)\rrangle+\llangle \dot{z}^{\text{p}}_\text{QG}(\tau_g)\dot{z}^{\text{p}}_\text{QG}(\tau_g)\rrangle\right)~.
\end{split}
\end{equation}
In order to obtain the second line of the above equation, we have made use of the result that $\llangle \dot{z}^\text{p}_{\text{QG}}(\tau_g)\rrangle=0$ which can easily be inferred from the fact that $\llangle \dot{\mathcal{N}}_q(t)\rrangle=\frac{d}{dt}\llangle\mathcal{N}_\text{q}(t)\rrangle=0$. It is then possible to finally recast eq.(\ref{QGUncertainty.28}) as
\begin{equation}\label{QGUncertainty.29}
(\Delta\pi^{\text{p}}_z(\tau_g))^2=m_0^{\text{p}^2}\llangle\dot{z}^{\text{p}}_\text{QG}(\tau_g)\dot{z}^{\text{p}}_\text{QG}(\tau_g)\rrangle~.
\end{equation}
With the analytical expressions for the variances in position and momentum of the particle from eq.(s)(\ref{QGUncertainty.23},\ref{QGUncertainty.29}), we are now in a position to calculate the uncertainty product while the gravitons are initially in a vacuum, a squeezed vacuum, and in a thermal state. 
\subsection{Graviton initially being in a vacuum state}
\noindent We start by simply considering the initial graviton state to be in a vacuum state. The primary quantity that we need to calculate are the two point noise-noise correlators and for the graviton being in a vacuum state, we obtain the analytical expression for the same as
 \begin{equation}\label{QGUncertainty.30}
\begin{split}
\llangle \mathcal{N}_\text{q}^0(t')\mathcal{N}_\text{q}^0(t'')\rrangle&=\mathcal{A}^{0}(t',t'')=\frac{4\hbar G}{\pi c^5}\int_0^\infty d\omega \omega \cos(\omega(t'-t''))~.
\end{split}
\end{equation}
As has also been discussed in Chapter(\ref{C.3.OTM}), the above frequency integral is indeed divergent. Hence, one needs to provide an upper bound to the integration limit to properly regularized the frequency integral. For a freely falling particle from a heigh $z_0^p$, it is possible to give an upper-bound to the frequency integral in eq.(\ref{QGUncertainty.30}), which can be set to $\omega_{\text{max}}=\frac{2\pi  c}{z_0^p}$ without any loss of generality. Now, the measurement is done for a finite amount of time $t=\tau_g$ which enables one to provide a lower limit of integration as $\omega_{\text{min}}=\frac{2\pi}{\tau_g}$. Substituting the lower and upper limits of integration in eq.(\ref{QGUncertainty.30}), it is possible to write down the autocorrelation function as
\begin{equation}\label{QGUncertainty.31}
\begin{split}
\mathcal{A}^0_{\text{q}}(t',t'')&=\frac{4\hbar G}{\pi c^5}\int_{\omega_{\text{min}}}^{\omega_{\text{max}}} d\omega \omega \cos\left(\omega (t'-t'')\right)\\
&=\frac{4\hbar G}{\pi c^5}\left(\frac{\cos(\omega(t'-t''))}{(t-t')^2}+\frac{\omega\sin(\omega(t'-t''))}{t'-t''}\right)\biggr\rvert_{\omega_{\text{min}}}^{\omega_{\text{max}}}~.
\end{split}
\end{equation}
If the graviton is in the vacuum state initially, then we can write down the variance in position form eq.(\ref{QGUncertainty.23}) as
\begin{equation}\label{QGUncertainty.32}
\begin{split}
\left(\Delta z^\text{p}(\tau_g)\right)^2=&\llangle z^\text{p}_\text{QG}(\tau) z^\text{p}_\text{QG}(\tau)\rrangle\\
=&g^2\int_0^{\tau_g} t' dt'\int_0^{\tau_g} \bar{t} d\bar{t} ~\mathcal{A}^{\text{q}}_0(t',\bar{t})-g^2\int_{0}^{\tau_g} dt'\int_0^{t'} dt''\int_0^{\tau_g}\bar{t}d\bar{t}~\mathcal{A}_\text{q}^0(t'',\bar{t})\\
+&\frac{g^2}{4}\int_0^{\tau_g}dt'\int_0^{t'}dt''\int_{0}^{\tau_g} d\bar{t}\int_0^{\bar{t}}d\bar{\bar{t}}~\mathcal{A}_\text{q}^0(t'',\bar{\bar{t}})~.
\end{split}
\end{equation} 
Substituting the analytical form of the autocorrelation function from eq.(\ref{QGUncertainty.32}) in the above expression, we arrive at the expression for the variance in the position of the particle
\begin{equation}\label{QGUncertainty.33}
\begin{split}
\left(\Delta z^\text{p}(\tau_g)\right)^2\simeq& \frac{4\hbar g^2 G}{\pi c^5}\left[\frac{9z_0^{\text{p}^2}}{16\pi^2c^2}\left[\cos\left(\frac{2\pi c\tau_g}{z_0^\text{p}}\right)-1\right]+\frac{1}{2}\tau_g^2\left[\text{Ci}(2\pi)-\text{Ci}\left(\frac{2\pi c\tau_g}{z_0^\text{p}}\right)\right]\right.\\&\left.+\frac{1}{8}\tau_g^2\ln\left(\frac{4\pi^2 z_0^{\text{p}^2}}{c^2\tau_g^2}\right)+\tau_g^2\ln\left(\frac{c\tau_g}{z_0^\text{p}}\right)\right]
\end{split}
\end{equation}
where `$\text{Ci}$' denotes the cosine integral function \cite{Gradshteyn}. One now needs to investigate the above analytical form of the variance in the position of the particle. Plotting the individual functions with respect to $\tau_g$, it is easily observable that the $\tau_g^2\ln\left(\frac{c\tau_g}{z_0^\text{p}}\right)$ function dominates all other contributions and as a result one can approximately express eq.(\ref{QGUncertainty.33}) as
\begin{equation}\label{QGUncertainty.34}
\left(\Delta z^\text{p}(\tau_g)\right)^2\simeq \frac{4\hbar G g^2 \tau_g^2}{\pi c^5}\ln\left(\frac{c\tau_g}{z_0^\text{p}}\right)=\frac{4g^2\tau_g^2l_{\text{Pl}}^2}{\pi c^2}\ln\left(\frac{c\tau_g}{z_0^\text{p}}\right)~.
\end{equation}
Up to a constant factor the above result matches with the outcome of \cite{ChawlaParikh}. We now proceed to obtain the analytical expression for the variance in the momentum of the particle when the graviton initially is in a vacuum state. The variance in the momentum variable is given by the expression $(\Delta\pi^{\text{p}}_z(\tau_g))^2=m_0^2\llangle\dot{z}^{\text{p}}_\text{QG}(\tau_g)\dot{z}^{\text{p}}_\text{QG}(\tau_g)\rrangle$. Now, we can substitute the analytical form of $\dot{z}^{\text{p}}_\text{QG}(\tau_g)$ from eq.(\ref{QGUncertainty.18}), and substitute it back in eq.(\ref{QGUncertainty.29}) to obtain the following analytical expression
\begin{equation}\label{QGUncertainty.35}
\begin{split}
(\Delta\pi^{\text{p}}_z(\tau_g))^2=&m_0^{\text{p}^2}\llangle\dot{z}^{\text{p}}_\text{QG}(\tau_g)\dot{z}^{\text{p}}_\text{QG}(\tau_g)\rrangle\\
=&\frac{m_0^{\text{p}^2}g^2\tau_g^2}{4}\llangle \mathcal{N}^0_\text{q}(\tau_g)\mathcal{N}^0_\text{q}(\tau_g)\rrangle-\frac{m_0^{\text{p}^2}g^2\tau_g}{2}\int_0^{\tau_g} dt' \llangle \mathcal{N}^0_{\text{q}}(t')\mathcal{N}^0_\text{q}(\tau_g)\rrangle\\+&\frac{m_0^{\text{p}^2}g^2}{4}\int_0^{\tau_g}\int_0^{\tau_g}dt'dt''\llangle \mathcal{N}^0_\text{q}(t')\mathcal{N}^0_{\text{q}}(t'')\rrangle
\end{split}
\end{equation}
where we have considered the graviton to be initially in a vacuum state. Combining eq.(\ref{QGUncertainty.30}) with eq.(\ref{QGUncertainty.31}), we know that $\llangle\mathcal{N}_{\text{q}}^0(t')\mathcal{N}_{\text{q}}^0(t'')\rrangle=\mathcal{A}^0_q(t',t'')$. We can therefore obtain the analytical form of the variance in the momentum as
\begin{equation}\label{QGUncertainty.36}
\begin{split}
(\Delta \pi_z^{\text{p}}(\tau_g))^2=&\frac{2m_0^{\text{p}^2}g^2l_{\text{Pl}}^2}{\pi c^2}\left[\frac{\pi^2c^2\tau_g^2}{z_0^{\text{p}^2}}+\cos\left[\frac{2\pi c\tau_g}{z_0^\text{p}}\right]-1-\pi^2+\text{Ci}[2\pi]-\text{Ci}\left[\frac{2\pi c\tau_g}{z_0^\text{p}}\right]+\ln\left[\frac{c\tau_g}{z_0^\text{p}}\right]\right]
\end{split}.
\end{equation}
As before, keeping the dominant contribution in the above equation, we can write down the final expression for the variance in the momentum as
\begin{equation}\label{QGUncertainty.37}
(\Delta \pi^{\text{p}}_z(\tau_g))^2\simeq\frac{2\pi m_0^{\text{p}^2}g^2l_{\text{Pl}}^2\tau_g^2}{z_0^{\text{p}^2}}~.
\end{equation}
With the variances of the position as well as the momentum in hand, we are now in a position to write down the square of the uncertainty product as
\begin{equation}\label{QGUncertainty.38}
\begin{split}
(\Delta z^{\text{p}}(\tau_g))^2(\Delta \pi_z^\text{p}(\tau_g))^2=\frac{8m_0^{\text{p}^2}g^4\tau_g^4l_{\text{Pl}}^4}{z_0^{\text{p}^2}c^2}\ln\left(\frac{c\tau_g}{z_0^\text{p}}\right)~.
\end{split}
\end{equation}
Defining $\Delta z^{\text{p}}\equiv\Delta z^{\text{p}}(\tau_g)$ and $\Delta \pi_z^{\text{p}}\equiv\Delta \pi_z^{\text{p}}(\tau_g)$, we can obtain the product of uncertainties in the position and the momentum of the particle from eq.(\ref{QGUncertainty.38}) as
\begin{equation}\label{QGUncertainty.39}
\Delta z^{\text{p}}\Delta\pi^\text{p}_z=\frac{2\sqrt{2}m_0^\text{p}g^2\tau_g^2l_{\text{Pl}}^2}{z^\text{p}_0 c}\sqrt{\ln \left(\frac{c\tau_g}{z^\text{p}_0}\right)}~.
\end{equation}
We shall now rearrange the right hand side of the above expression as 
\begin{equation}\label{QGUncertainty.40}
\begin{split}
\Delta z^{\text{p}}\Delta\pi^\text{p}_z=&\frac{2\sqrt{2}m_0^\text{p}g^2\tau_g^2l_{\text{Pl}}^2}{z^\text{p}_0 c}\sqrt{\ln \left(\frac{c\tau_g}{z^\text{p}_0}\right)}\\
=&\sqrt{\frac{2}{\pi^2}\ln \left(\frac{c\tau_g}{z^\text{p}_0}\right)}\left(\frac{2\pi m_0^2g^2l_{\text{Pl}}^2\tau_g^2}{z_0^{\text{p}^2}}\right)\frac{z_0^{\text{p}}}{m_0^{\text{p}}c}~.
\end{split}
\end{equation}
In the right hand side of the above equation, the term inside of the parenthesis is equal to the square of the variance in the momentum as can be seen from eq.(\ref{QGUncertainty.37}), and this helps us to recast eq.(\ref{QGUncertainty.40}) as
\begin{equation}\label{QGUncertainty.41}
\Delta z^\text{p}\Delta\pi_z^{\text{p}}=\sqrt{\frac{2}{\pi^2}\ln \left(\frac{c\tau_g}{z^\text{p}_0}\right)}\frac{z^\text{p}_0}{m_0^\text{p} c}(\Delta \pi^\text{p}_z)^2~.
\end{equation}
As both $z_0^\text{p}$ and $\tau_g$ are constants, we can define a new constant of the form $\beta\equiv\sqrt{\frac{2}{\pi^2}\ln \left(\frac{c\tau_g}{z^\text{p}_0}\right)}$, using which, we can express the uncertainty product in the above equation in a more compact form as
\begin{equation}\label{QGUncertainty.42}
\Delta z^\text{p}\Delta\pi_z^{\text{p}}=\sqrt{\frac{1}{\pi^2}\ln \left(\frac{2c^2}{gz^\text{p}_0}\right)}\frac{z^\text{p}_0}{m_0^\text{p} c}(\Delta \pi^\text{p}_z)^2=\frac{\beta z^\text{p}_0}{m_0^\text{p} c}(\Delta \pi^\text{p}_z)^2~.
\end{equation}
From the above equation, we can write down the following inequality with a particle with mass $m_0^\text{p}$ as 
\begin{equation}\label{QGUncertainty.43}
\Delta z^\text{p}\Delta\pi_z^{\text{p}}=\frac{\beta z^\text{p}_0}{m_0^\text{p} c}(\Delta \pi^\text{p}_z)^2\geq\frac{\beta l_{\text{Pl}}}{m_0^\text{p} c}(\Delta \pi^\text{p}_z)^2~.
\end{equation}
The reason behind using the inequality $z_0^{\text{p}}\geq l_{\text{Pl}}$ lies in the fact that in any quantum gravity theory, the minimum length that one can probe is equal to the Planck length. As a result, it is possible to consider that the minimum value of $z_0^{\text{p}}$ is equal to the Planck length $l_{\text{Pl}}=\sqrt{\frac{\hbar G}{c^3}}$ as well. Eq.(\ref{QGUncertainty.43}) provides a lower bound to the uncertainty product $\Delta z^{\text{p}}\Delta\pi^\text{p}$. If the particle or detector sized also obeys standard quantum mechanics, in that case $\Delta z^{\text{p}}\Delta\pi^\text{p}$ will obey the standard Heisenberg uncertainty principle in the absence of any quantum gravitational effects as
\begin{equation}\label{QGUncertainty.44}
\Delta z^{\text{p}}\Delta\pi_z^\text{p}\geq \frac{\hbar}{2}~.
\end{equation}
Now, the right hand side of eq.(\ref{QGUncertainty.43}) is a purely quantum gravitational contribution which is very small compared to $\hbar$. As a result, for a quantum detector in a quantum gravity set-up, we can combine \cite{ScardigliGedanken} the right hand sides of eq.(s)(\ref{QGUncertainty.43}) and (\ref{QGUncertainty.44}) to write down a quantum gravity modified uncertainty relation as 
\begin{equation}\label{QGUncertainty.45}
\begin{split}
\Delta z^{\text{p}}\Delta\pi_z^\text{p}\geq& \frac{\hbar}{2}+\frac{\beta l_{\text{Pl}}}{m_0^\text{p} c}(\Delta \pi^\text{p}_z)^2\\
\implies\Delta z^{\text{p}}\Delta\pi_z^\text{p}\geq&\frac{\hbar}{2}\left(1+\frac{\tilde{\beta} l_{\text{Pl}}}{\hbar m_0^\text{p} c}(\Delta \pi^\text{p}_z)^2\right)
\end{split}
\end{equation}
where in the last line of the above equation, we have defined $\tilde{\beta}\equiv 2\beta$. It is important to note from the above uncertainty relation that the quantum gravity induced correction term is dependent on the Planck's constant $\hbar$, Newton's gravitational constant $G$, and the speed of light $c$. This indicates that if we either take the limit $G\rightarrow 0$ or $\hbar\rightarrow 0$, the quantum gravity induced correction term from the right hand side of eq.(\ref{QGUncertainty.45}) vanishes. Now, in the $\hbar\rightarrow 0$ limit the system loses its quantumness and it is indeed observable from eq.(\ref{QGUncertainty.45}) that the right hand side of the uncertainty relation vanishes which is an expected result. In the $G\rightarrow 0$ limit, the quantum gravity correction term in the right hand side of the inequality in eq.(\ref{QGUncertainty.45}) vanishes only restoring the usual Heisenberg uncertainty relation in eq.(\ref{QGUncertainty.44}). These limits confirm that the uncertainty relation, observed in eq.(\ref{QGUncertainty.45}), is indicative of a true quantum gravity modified uncertainty relation. We have already discussed in Chapter(\ref{C.2.OTM}) and observed in eq.(\ref{Uncertainty.23}), for the well-known generalized uncertainty principle \cite{KempfManganoMann}, the coefficient corresponding to the variance in the momentum square depends on $G$ and $c$ and is independent of the Planck's constant. Now, for a quantum gravity modified uncertainty relation, the involvement of the Planck's constant as well as the Newton's gravitational constant are required. If the $\hbar$ is taken to zero which is indicative of the classical limit, we observe that the right hand side of eq.(\ref{Uncertainty.23}) does not vanish indicating the existence of an uncertainty product even in the classical limit. Hence, it is best suited to call this a Newtonian gravity correction to the Heisenberg uncertainty principle. However, we shall try to investigate whether it is possible to obtain the generalized uncertainty principle from eq.(\ref{QGUncertainty.45}). In our analysis the mass of the point-particle is larger than the Planck mass $m_{\text{Pl}}=\sqrt{\frac{\hbar c}{G}}$. The primary reason behind this consideration lies in the fact that for particles with masses higher than the Planck mass, the de Broglie wavelength becomes smaller than the Planck length, leading to a dominant classical particle behaviour for the freely-falling particle. Now, if the quantization is executed inside of a box with volume equal to the Planck volume, then for a particle with mass higher than the Planck mass it results in a microscopic or a quantum black hole. As a result throughout our analysis, $m_0^\text{p}\leq m_{\text{Pl}}$.  Hence, it is possible to further simplify the inequality from eq.(\ref{QGUncertainty.45}) in the Planck mass limit as
\begin{equation}\label{QGUncertainty.46}
\begin{split}
\Delta z^\text{p}\Delta \pi_z^\text{p}&\geq \frac{\hbar}{2}+\frac{\beta l_{\text{Pl}}}{m_0^\text{p}c}(\Delta \pi_z^\text{p})^2\\&\geq\frac{\hbar}{2}+\frac{\beta l_{\text{Pl}}}{m_{\text{Pl}}c}(\Delta \pi_z^\text{p})^2\\
&=\frac{\hbar}{2}+\frac{\beta G}{c^3}(\Delta\pi_z^\text{p})^2\\
\implies \Delta z^\text{p}\Delta\pi_z^\text{p} &\geq \frac{\hbar}{2}\left(1+\frac{\tilde{\beta}}{m_{\text{Pl}}^2c^2}(\Delta\pi_z^\text{p})^2\right)~.
\end{split}
\end{equation}
The final uncertainty relation is identical to the generalized uncertainty relation given in eq.(\ref{Uncertainty.23}). The constant $\tilde{\beta}$ is not undetermined anymore and it can also be considered as a true quantum gravitational derivation of the generalized uncertainty principle. It is important to note that the above uncertainty relation is obtained in an one dimensional model and one can indeed generalize it to three spatial dimensions.
%\begin{equation}\label{1.34}
%\begin{split}
%\Delta z\Delta \pi_z&\geq \frac{\hbar}{2}+\frac{\beta l_p}{m_0c}(\Delta \pi_z)^2\geq\frac{\hbar}{2}+\frac{\beta l_p}{m_pc}(\Delta \pi_z)^2\\
%&=\frac{\hbar}{2}+\frac{\beta G}{c^3}(\Delta\pi_z)^2\\
%\implies \Delta z\Delta\pi_z &\geq \frac{\hbar}{2}\left(1+\frac{\beta'}{m_p^2c^2}(\Delta\pi_z)^2\right)
%\end{split}
%\end{equation}
We shall now consider the equality condition from eq.(\ref{QGUncertainty.45}), which helps us to write down a quadratic equation in $\Delta\pi_z^\text{p}$ as
\begin{equation}\label{QGUncertainty.47}
\frac{\beta l_{\text{Pl}}}{m_0^\text{p} c}(\Delta \pi^\text{p}_z)^2-\Delta z^{\text{p}}\Delta\pi_z^\text{p}+ \frac{\hbar}{2}=0~.
\end{equation}
Solving the above equation, we obtain an analytical expression for $\Delta\pi_z^\text{p}$ as
\begin{equation}\label{QGUncertainty.48}
\Delta\pi_z^{\text{p}}=\frac{m_0^\text{p} c}{\tilde{\beta}l_{\text{Pl}}}\Delta z^\text{p}\pm \frac{m_0^\text{p} c}{\tilde{\beta}l_{\text{Pl}}}\sqrt{(\Delta z^\text{p})^2-\frac{\tilde{\beta}\hbar l_{\text{Pl}}}{m_0^\text{p}c}}~.
\end{equation}
As the uncertainty in the momentum $\Delta\pi_z^{\text{p}}$ cannot be a complex quantity, the term inside of the square root must be equal to or greater than zero. Hence, from eq.(\ref{QGUncertainty.48}), it is straightforward to interpret that 
\begin{equation}\label{QGUncertainty.49}
\begin{split}
\Delta z^\text{p}&\geq \sqrt{\frac{\tilde{\beta}\hbar l_{\text{Pl}}}{m_0^\text{p}c}}\implies \Delta z^\text{p}_{\text{min}}= \sqrt{\frac{\tilde{\beta}\hbar l_{\text{Pl}}}{m_0^{\text{p}}c}}~.
\end{split}
\end{equation}
Hence, we find out that the quantum gravity induced uncertainty relation indicates towards the existence of a fundamental minimal length scale. The minimum value of the uncertainty in $z^\text{p}$ can be obtained by setting $m_0$ to $m_{\text{Pl}}$. In this limit ($m_0\rightarrow m_p$), we obtain the minimum value of the uncertainty in position as
\begin{equation}\label{QGUncertainty.50}
\left.\Delta z^\text{p}_{\text{min}}\right\rvert_{m_0^{\text{p}}\rightarrow m_{\text{Pl}}}=\sqrt{\frac{\hbar\tilde{\beta}l_{\text{Pl}}}{m_\text{Pl} c}}=\sqrt{\tilde{\beta}} l_\text{Pl}~.
\end{equation}
The above minimum value of the uncertainty in the position is of the order of the Planck length, and is identical to the result obtained in \cite{KempfManganoMann}. It is evident from eq.(s)(\ref{QGUncertainty.49},\ref{QGUncertainty.50}) that $\left.\Delta z^\text{p}_{\text{min}}\right\rvert_{m_0^{\text{p}}\rightarrow m_{\text{Pl}}}\leq \Delta z^\text{p}_{\text{min}}$ $\forall$ $m_0^\text{p}\leq m_{\text{Pl}}$. Up to now, we are only concerned about the lower bound of the uncertainty relation. Now the uncertainty product in eq.(\ref{QGUncertainty.39}) can be arranged in a way such that the right hand side can be arranged in terms of the square of the variance in the position part
\begin{equation}\label{QGUncertainty.51}
\begin{split}
\Delta z^{\text{p}}\Delta\pi^\text{p}_z=&\frac{2\sqrt{2}m_0^\text{p}g^2\tau_g^2l_{\text{Pl}}^2}{z^\text{p}_0 c}\sqrt{\ln \left(\frac{c\tau_g}{z^\text{p}_0}\right)}\\=&\sqrt{\frac{\pi^2}{2\ln\left(\frac{c\tau_g}{z_0^\text{p}}\right)}}\frac{m_0^\text{p}c}{z_0^\text{p}}\frac{4g^2\tau_g^2l_{\text{Pl}}^2}{\pi c^2}\ln\left(\frac{c\tau_g}{z_0^\text{p}}\right)\\
\implies \Delta z^\text{p}\Delta\pi_z^\text{p}=&\frac{m_0^\text{p} c}{\beta z^{\text{p}}_0}(\Delta z^\text{p})^2~.
\end{split}
\end{equation}
It is evident as before that $z_0^{\text{p}}\geq l_{\text{Pl}}$ which enables us to write down an upper bound to the uncertainty product $\Delta z^\text{p}\Delta\pi_z^\text{p}$ as
\begin{equation}\label{QGUncertainty.52}
\Delta z^{\text{p}}\Delta\pi_z^\text{p}=\frac{m_0^\text{p} c}{\beta z_0^\text{p}}(\Delta z^\text{p})^2\leq \frac{m_0^\text{p} c}{\beta l_{\text{Pl}}}(\Delta z^\text{p})^2~.
\end{equation}
In our analysis we are considering that the mass of the point particle is smaller than or equal to the Planck mass which helps us to write down the maximum upper bound to the uncertainty product as
\begin{equation}\label{QGUncertainty.53}
\Delta z^{\text{p}}\Delta\pi_z^\text{p}\leq\frac{m_0^\text{p} c}{\beta z_0^\text{p}}(\Delta l_\text{Pl})^2\leq \frac{m_\text{Pl} c}{\beta l_{\text{Pl}}}(\Delta z^\text{p})^2~.
\end{equation}
As we are working with particles that have masses smaller than the Planck mass, we can get an even higher value of the upper bound simply by replacing $m_0$ with $m_p$ in the above inequality. We now have obtained a lower bound as well as an upper bound of the uncertainty product $\Delta z^\text{p}\Delta\pi_z^{\text{p}}$ from eq.(s)(\ref{QGUncertainty.45},\ref{QGUncertainty.53}). We can combine both of the inequalities to write down a bounded inequality of the form
\begin{equation}\label{QGUncertainty.54}
\frac{m_\text{Pl} c}{\beta l_{\text{Pl}}}(\Delta z^\text{p})^2\geq\Delta z^\text{p}\Delta \pi^\text{p}_z\geq \frac{\hbar}{2}\left(1+\frac{\tilde{\beta} l_{\text{Pl}}}{\hbar m_0^\text{p} c}(\Delta \pi_z^\text{p})^2\right)~.
\end{equation}
If the particle is considered to be of the mass such that $m_0^\text{p}<m_{\text{Pl}}$, then the above uncertainty relation takes the form
\begin{equation}\label{QGUncertainty.55}
\frac{m_\text{Pl} c}{\beta l_{\text{Pl}}}(\Delta z^\text{p})^2>\Delta z^\text{p}\Delta \pi_z^\text{p}\geq \frac{\hbar}{2}\left(1+\frac{\tilde{\beta} l_{\text{Pl}}}{\hbar m_0^\text{p} c}(\Delta \pi_z^\text{p})^2\right)~.
\end{equation}
The above result is the most important analytical outcome in this work and gives the analytical expression for a true quantum gravity modified uncertainty relation. We have already investigated the significance of the lower bound of the uncertainty product which indicates towards the existence of a minimum value of $\Delta z^\text{p}$ beyond which one cannot go further. This indicates that the minimum value of the uncertainty in the position can never go to zero. From the upper bound of the uncertainty principle, we can conclude a very important outcome. It is important to note that $\Delta z^\text{p}$ cannot go below a finite value which is of the order of the Planck length, however, it can become infinite. Even if $\Delta z^{\text{p}}\rightarrow\infty$ then also the upper bound relation will get satisfied as can be observed from eq.(\ref{QGUncertainty.53}). Hence, for a precise measurement of $\pi_z^\text{p}$ ($\Delta\pi_z^\text{p}=0$), $\Delta z^\text{p}$ will tend towards infinity where the upper bound criteria will be satisfied as has been discussed above. We can write down the expression of $\Delta\pi_z^{\text{p}}$ from eq.(\ref{QGUncertainty.48}) using eq.(\ref{QGUncertainty.49}) as
\begin{equation}\label{QGUncertainty.56}
\Delta \pi_z=\frac{m_0^\text{p} c}{\tilde{\beta} l_\text{Pl}}\Delta z^{\text{p}}\pm\frac{m_0^\text{p} c}{\tilde{\beta} l_{\text{Pl}}}\sqrt{(\Delta z^\text{p})^2-(\Delta z^\text{p}_{\text{min}})^2}~.
\end{equation}
 Multiplying both sides of the above equation with $\Delta z^{\text{p}}$, we can write down the from of eq.(\ref{QGUncertainty.56}) as
\begin{equation}\label{QGUncertainty.57}
\Delta z^\text{p}\Delta\pi_z^\text{p}=\frac{m_0^\text{p} c}{\tilde{\beta}l_{\text{Pl}}}(\Delta z^\text{p})^2\left(1\pm\sqrt{1-\frac{(\Delta z^\text{p}_{\text{min}})^2}{(\Delta z^\text{p})^2}}\right)~.
\end{equation}
In terms of $\tilde{\beta}$, the upper bound to the uncertainty product can be expressed as $\Delta z^\text{p}\Delta \pi_z^\text{p}\leq\frac{2m_\text{Pl} c}{\tilde{\beta} l_{\text{Pl}}}(\Delta z^\text{p})^2$. Now, we already know that $m_p \geq m_0$ and it is easy to understand that $2>\left(1\pm\sqrt{1-\frac{(\Delta z^\text{p}_{\text{min}})^2}{(\Delta z^\text{p})^2}}\right)$ which helps us to write down an inequality of the form
\begin{equation}\label{QGUncertainty.58}
\frac{2m_{\text{Pl}} c}{\tilde{\beta}l_\text{Pl}}(\Delta z^\text{p})^2>\frac{m^\text{p}_0 c}{\tilde{\beta}l_{\text{Pl}}}(\Delta z^\text{p})^2\left(1\pm\sqrt{1-\frac{(\Delta z^\text{p}_{\text{min}})^2}{(\Delta z^\text{p})^2}}\right)~.
\end{equation}
We can see from the above relation that the upper bound criteria always gets satisfied. If we consider minimum value for $\Delta z^\text{p}$, that is $\Delta z^\text{p}=\Delta z^\text{p}_\text{min}$, we obtain the value of the uncertainty product to be $\Delta z^\text{p}\Delta\pi_z^\text{p}=\hbar$ whereas the value of the upper bound reads $\frac{2m_{\text{Pl}} c}{\tilde{\beta}l_\text{Pl}}(\Delta z^\text{p}_\text{min})^2=\frac{2\hbar m_{\text{Pl}}}{m^\text{p}_0}$. As throughout our analysis $m_{\text{Pl}}\geq m_0^\text{p}$, it implies that $\frac{2\hbar m_{\text{Pl}}}{m^\text{p}_0}\geq \hbar$ indicating that the uncertainty upper bound holds true. The existence of this upper bound does indicate that if the position of the freely falling particle is measured to the best precision allowed by the theoretical model ($\Delta z^{\text{p}}=\Delta z^{\text{p}}_{\text{min}}$) then there exists an upper bound to the uncertainty in the measurement of the momentum. This maximum uncertainty in momentum for $\Delta z^{\text{p}}=\Delta z^{\text{p}}_{\text{min}}$ is obtained as
\begin{equation}\label{QGUncertainty.59}
\Delta{\pi}_{z_{\text{max}}}^\text{p}=2m_{\text{Pl}}\sqrt{\frac{\hbar c}{\tilde{\beta} m_0^\text{p} l_\text{Pl}}}~.
\end{equation}
Next, we shall investigate the quantum gravity modified uncertainty relations when the graviton is in a squeezed as well as in a thermal state.
\subsection{Gravitons initially being in a Squeezed vacuum state}
\noindent In this subsection, we shall investigate the scenario when the initial gravitation state is interacting with the freely falling particle while the graviton is in a squeezed state. The initial state of the graviton can be expressed as $|\psi^{\text{Sq.}}_\omega\rangle=\hat{S}(\zeta_{\omega})|0^\text{g}_\omega\rangle$ where $\hat{S}(\zeta_{\omega})$ is squeezing operator and is defined by $\hat{S}({\zeta_\omega})\equiv \exp\left[\frac{1}{2}\zeta_\omega^*\hat{a}^2-\zeta_\omega\hat{a}^{\dagger2}\right]$. The complex squeezing parameter $\zeta_\omega$ can be expressed as $\zeta_\omega=\mathfrak{r}_\omega \exp[i\varphi_\omega]$ with $\mathfrak{r}_\omega,\varphi\in\mathbb{R}$. The transition probability for the particle to go from an initial state $|\Phi^{\text{p}}_A\rangle$ to a final state $|\Phi^{\text{p}}_B\rangle$ with the graviton initially being in the state $|\psi^{\text{Sq.}}_\omega\rangle$ is given as
\begin{equation}\label{QGUncertainty.60}
\begin{split}
P^{\Psi_\omega^\text{Sq.}}_{A\rightarrow B}&=\int dz^\text{p}_id\acute{z}^\text{p}_idz^\text{p}_fd\acute{z}^\text{p}_f\Phi^\text{p}_A(z^\text{p}_i)\Phi_A^{\text{p}^*}(\acute{z}^\text{p}_i)\Phi^{\text{p}^*}_B(z^\text{p}_f)\Phi^\text{p}_B(\acute{z}^\text{p}_f)\\&\times\int\left[\tilde{\mathcal{D}}z^\text{p}\right]_{z^\text{p}_i,0}^{z^\text{p}_f,T}\left[\tilde{\mathcal{D}}\acute{z}^\text{p}\right]_{\acute{z}^\text{p}_i,0}^{\acute{z}^\text{p}_f,T}~e^{\frac{im_0^{\text{p}}}{\hbar}\int_{0}^{T}dt\left[\frac{1}{2}\left(\dot{z}^{\text{p}^2}-\dot{\acute{z}}^{\text{p}^2}\right)-\left(\phi_\oplus(z^\text{p})-\phi_\oplus(\acute{z}^\text{p})\right)\right]}F_{\zeta_\omega}\left[z^\text{p},\acute{z}^\text{p}\right]~.
\end{split} 
\end{equation}
%[\tilde{\mathcal{D}}z]_{z_i,0}^{z_f,T} [\tilde{\mathcal{D}}z']_{z_i',0}^{z_f',T}
The total influence functional for the model system can be expressed as 
\begin{equation}\label{QGUncertainty.61}
F_{\zeta_\omega}[z^\text{p},\acute{z}^\text{p}]=F_{0^g_\omega}[z^\text{p},\acute{z}^\text{p}]\exp\left[i\Phi_{\zeta_\omega}[z^\text{p},\acute{z}^\text{p}\right]
\end{equation}
 where the vacuum influence functional $F_{0^g_\omega}[z^\text{p},\acute{z}^\text{p}]$ reads
\begin{equation}\label{QGUncertainty.62}
\begin{split}
F_{0^g_\omega}[z^\text{p},\acute{z}^\text{p}]=&\exp\left[-\frac{\mathfrak{g}^2}{8\hbar m\omega}\int_0^Tdt\int_0^tdt' \left(\mathcal{X}(t)-\acute{\mathcal{X}}(t)\right)\left(\mathcal{X}(t')e^{-i\omega(t-t')}-\acute{\mathcal{X}}(t')e^{i\omega(t-t')}\right)\right]~.
\end{split}
\end{equation}
In the initial part of the analysis, we have kept the speed of light to be unity and it will be restored later while obtaining the final analytical form of the uncertainty product. The influence phase $i\Phi_{\zeta_\omega}[z^\text{p},\acute{z}^\text{p}]$ has the analytical form 
\begin{equation}\label{QGUncertainty.63}
\begin{split}
i\Phi_{\zeta_\omega}[z^{\text{p}},\acute{z}^\text{p}]&=\frac{\mathfrak{g}^2}{16\hbar m\omega}\int_0^T dt\int_0^T dt'\cos[\omega(t+t')-\varphi_\omega](\mathcal{X}(t)-\acute{\mathcal{X}}(t))(\mathcal{X}(t')-\acute{\mathcal{X}}(t'))\sinh 2\mathfrak{r}_\omega \\
&-\frac{\mathfrak{g}^2}{16\hbar m\omega}\int_0^Tdt\int_0^Tdt'\cos[\omega(t-t')]
(\mathcal{X}(t)-\acute{\mathcal{X}}(t))(\mathcal{X}(t')-\acute{\mathcal{X}}(t'))(\cosh 2\mathfrak{r}_\omega -1)
\end{split}
\end{equation} 
where the total interaction time is given by $T=\tau_g$. If $|\Phi^\text{p}_B\rangle$ denotes the final state of the freely falling particle when it touches the ground, in that case, we know $z_f^\text{p}=\acute{z}_f^\text{p}=0$. Similar to the case observed in Chapter(\ref{C.3.OTM}), we can again consider the squeezing parameter to be independent of the graviton mode frequency as $\mathfrak{r}_\omega=\mathfrak{r}$. One can also consider the squeezing phase to be independent of the mode frequency as well which is denoted by $\varphi_\omega=\varphi$. Summing over all graviton mode frequencies, one can write down the total influence functional for the graviton-particle model system as
\begin{equation}\label{QGUncertainty.64}
\begin{split}
&F_\zeta[z^\text{p},\acute{z}^\text{p}]\\=&F_{0^g}[z^\text{p},\acute{z}^\text{p}]e^{i\Phi_\zeta[z^\text{p},\acute{z}^\text{p}]}\\
=&\exp\left[-\frac{m_0^2G}{8\pi\hbar}\cosh 2\mathfrak{r}\int _0^\infty \omega d\omega\int_0^Tdt\int_0^Tdt'\cos(\omega(t-t'))\left(\mathcal{X}(t)-\acute{\mathcal{X}}(t)\right)\left(\mathcal{X}(t')-\acute{\mathcal{X}}(t')\right)\right.\\&\left.+
\frac{m_0^2G}{8\pi\hbar}\sinh 2r\int_0^\infty\omega d\omega\int_0^Tdt\int_0^Tdt'\cos(\omega(t+t')-\varphi)\left(\mathcal{X}(t)-\acute{\mathcal{X}}(t)\right)\left(\mathcal{X}(t')-\acute{\mathcal{X}}(t')\right)\right]\\\times&\exp\left[-\frac{im_0^2G}{8\pi\hbar}\int_0^Tdt \left(\mathcal{X}(t)-\acute{\mathcal{X}}(t)\right)\left(\dot{\mathcal{X}}(t)+\dot{\acute{\mathcal{X}}}(t)\right)\right]~.
\end{split}
\end{equation} 
%\begin{split}
%F_\zeta[z,z']&=F_0[z,z']e^{i\Phi_z[z,z']}\\
%&=\exp\biggr[-\frac{m_0^2G}{8\pi\hbar}\cosh 2r\int _0^\infty \omega d\omega\int_0^Tdt\int_0^Tdt'\\&\times\cos(\omega(t-t'))(X(t)-X'(t))(X(t')-X'(t'))\\&+
%\frac{m_0^2G}{8\pi\hbar}\sinh 2r\int_0^\infty\omega d\omega\int_0^Tdt\int_0^Tdt'\\&\times\cos(\omega(t+t')-\phi)(X(t)-X'(t))(X'(t)\\&-X'(t'))-\frac{im_0^2G}{8\pi\hbar}\int_0^Tdt (X(t)-X'(t))(\dot{X}(t)\\&+\dot{X}'(t))\biggr]~.
%\end{split}
As has been discussed earlier in subsection(\ref{C3.3.3.Squeezing}) of the previous chapter, the $cos(\omega(t-t'))$ preserves the time translational symmetry whereas the $\cos(\omega(t+t')-\varphi)$ is non-static in nature (as it does not preserve time translational symmetry). Based on the static and non-static natures of the terms, it is possible to define two auxiliary functions as
\begin{align}
\mathcal{A}^{\text{S.}}(t-t')&=\frac{4\hbar G}{\pi}\cosh 2\mathfrak{r}\int_0^{\infty} d\omega\hspace{0.5mm} \omega \cos (\omega(t-t'))\label{QGUncertainty.65}\\
\mathcal{A}^{\text{N.S.}}(t+t')&=\frac{4\hbar G}{\pi}\sinh 2\mathfrak{r}\int_0^{\infty} d\omega\hspace{0.5mm} \omega\cos (\omega (t+t')-\varphi)\label{QGUncertainty.66}
\end{align}
where in the above equations `$\text{S}.$' and `$\text{N.S}.$' in the superscript of the auxiliary functions indicate the static and non-static behaviour, and one can simply take the $\varphi\rightarrow 0$ limit here. As has been done in eq.(\ref{QGUncertainty.31}), we need to implement a similar upper bound as well as lower bound to the frequency integrals which helps to make the divergent functions to be finite and express eq.(s)(\ref{QGUncertainty.65},\ref{QGUncertainty.66}) as
\begin{align}
\mathcal{A}_\text{q}^{\text{S.}}(t-t')&=\frac{4\hbar G}{\pi}\cosh 2\mathfrak{r}\int_{\omega_{\text{min}}}^{\omega_{\text{max}}} d\omega\hspace{0.5mm} \omega \cos (\omega(t-t'))\label{QGUncertainty.67}\\
\mathcal{A}_\text{q}^{\text{N.S.}}(t+t')&=\frac{4\hbar G}{\pi}\sinh 2\mathfrak{r}\int_{\omega_{\text{min}}}^{\omega_{\text{max}}} d\omega\hspace{0.5mm} \omega\cos (\omega (t+t')-\varphi)\label{QGUncertainty.68}~.
\end{align}
The above auxiliary functions are related directly to the two-point graviton noise-noise correlators. Now if $\mathcal{N}_q^{\text{S.}}$ denote the graviton noise fluctuation term corresponding to the static part then $\mathcal{A}_\text{q}^\text{S.}(t-t')=\llangle \mathcal{N}_\text{q}^\text{S.}(t) \mathcal{N}_\text{q}^\text{S.}(t')\rrangle$ and the autocorrelation function $\mathcal{A}_\text{q}^\text{S.}(t-t')$ is directly related to $\mathcal{A}_\text{q}^0(t-t')$ (defined in eq.(\ref{QGUncertainty.31})) by the relation $\mathcal{A}_\text{q}^\text{S.}(t-t')=\cosh 2 \mathfrak{r}\mathcal{A}_\text{q}^\text{0}(t-t')$. It is also straightforward to make the identification that $\mathcal{N}^\text{S.}_{\text{q}}(t)=\sqrt{\cosh 2\mathfrak{r}}\mathcal{N}^0_{\text{q}}(t)$.
The transition probability for the model particle-graviton system then reads
\begin{equation}\label{QGUncertainty.69}
\begin{split}
&P^{\Psi^\text{Sq.}}_{A\rightarrow B}=\int dz^\text{p}_id\acute{z}^\text{p}_idz^\text{p}_fd\acute{z}^\text{p}_f\Phi^\text{p}_A(z^\text{p}_i)\Phi_A^{\text{p}^*}(\acute{z}^\text{p}_i)\Phi^{\text{p}^*}_B(z^\text{p}_f)\Phi^\text{p}_B(\acute{z}^\text{p}_f)\int\tilde{\mathcal{D}}\mathcal{N}_\text{q}^0e^{-\frac{1}{2}\int_0^Tdt\int_0^Tdt'\mathcal{A}^{0^{-1}}_\text{q}(t,t')\mathcal{N}^0_\text{q}(t)\mathcal{N}^0_\text{q}(t')}\\&\times\int\tilde{\mathcal{D}}\mathcal{N}^{\text{N.S.}}~e^{\frac{1}{2}\int_0^Tdt\int_0^Tdt'\left(\mathcal{A}^{\text{N.S.}}(t+t')\right)^{-1}\mathcal{N}^{\text{N.S.}}(t)\mathcal{N}^\text{N.S.}(t')}\int\left[\tilde{\mathcal{D}}\xi^\text{s}\right]_{\xi^\text{s}_i,0}^{\xi^\text{s}_f,T}\left[\tilde{\mathcal{D}}\acute{\xi}^\text{s}\right]_{\acute{\xi}^\text{s}_i,0}^{\acute{\xi}^\text{s}_f,T}\\&\times e^{\frac{im_0}{2\hbar}\int_{0}^{T}dt\left[\left[\dot{z}^{\text{p}^2}-\dot{\acute{z}}^{\text{p}^2}\right]-2\left(\phi_\oplus(z^\text{p})-\phi_\oplus(\acute{z}^\text{p})\right)+\frac{1}{2}\left(\sqrt{\cosh2\mathfrak{r}}\mathcal{N}_\text{q}^0(t)+\mathcal{N}_\text{q}^{\text{N.S.}}(t)\right)\left[\mathcal{X}(t)-\acute{\mathcal{X}}(t)\right]-\frac{m_0G}{4}\left[\mathcal{X}(t)-\acute{\mathcal{X}}(t)\right]\left[\dot{\mathcal{X}}(t)+\dot{\acute{\mathcal{X}}}(t)\right]\right]}
\end{split} 
\end{equation}
where $\mathcal{A}^0_\text{q}(t,t')=\mathcal{A}^0_\text{q}(t-t')$. One can now make use of the saddle-point approximation and obtain the Langevin-like differential equation involving $z^\text{p}$ as
\begin{equation}\label{QGUncertainty.70}
\begin{split}
\ddot{z}^\text{p}(t)+\frac{\partial\phi_{\oplus}}{\partial z^\text{p}}-\frac{1}{2}\left(\ddot{\mathcal{N}}_\text{q}^{\text{N.S}.}(t)+\sqrt{\cosh 2\mathfrak{r}}\ddot{\mathcal{N}}_\text{q}^0(t)\right)z^\text{p}(t)=0
\end{split}
\end{equation}
where $\phi_\oplus\equiv\phi_\oplus(z^\text{p})$, and we have neglected all higher order time derivatives in $z^\text{p}$ as they have very small contributions towards the overall dynamics of $z^\text{p}$.
As was argued in the previous chapter as well \cite{QGravD}, we can always focus on the static part of the noise fluctuation and neglect the non-static contribution. Proceeding with the same methodology presented for the vacuum case, we can arrive at the variance in the position term as (restoring the speed of light properly in the expression and setting $t=\tau_g$)
\begin{equation}\label{QGUncertainty.71}
\begin{split}
\left(\Delta z_{\text{Sq.}}^\text{p}(\tau_g)\right)^2\simeq& \frac{4\hbar g^2 G}{\pi c^5}\cosh 2\mathfrak{r}\left[\frac{9z_0^{\text{p}^2}}{16\pi^2c^2}\left[\cos\left(\frac{2\pi c\tau_g}{z_0^\text{p}}\right)-1\right]+\frac{1}{2}\tau_g^2\left[\text{Ci}(2\pi)-\text{Ci}\left(\frac{2\pi c\tau_g}{z_0^\text{p}}\right)\right]\right.\\&\left.+\frac{1}{8}\tau_g^2\ln\left(\frac{4\pi^2 z_0^{\text{p}^2}}{c^2\tau_g^2}\right)+\tau_g^2\ln\left(\frac{c\tau_g}{z_0^\text{p}}\right)\right]\\
\simeq & \frac{4 g^2\tau_g^2 l_{\text{Pl}}^2}{\pi c^2}\cosh 2\mathfrak{r}\ln\left(\frac{c\tau_g}{z_0^\text{p}}\right)
\end{split}
\end{equation}
where only the term which dominates over time are kept in the expression. Similarly the variance in the momentum parameter can be expressed as (where only the dominant contributions are kept)
\begin{equation}\label{QGUncertainty.72}
(\Delta \pi_{z_{\text{Sq.}}}^\text{p}(\tau_g))^2\simeq\frac{2\pi m_0^{\text{p}^2}g^2l_{\text{Pl}}^2\tau_g^2}{z_0^{\text{p}^2}}\cosh 2\mathfrak{r}~.
\end{equation} 
Defining $\Delta z^{\text{Sq.}}\equiv\Delta z_{\text{Sq.}}^\text{p}(\tau_g)$ and $\Delta \pi_z^{\text{Sq.}}\equiv\Delta \pi_{z_{\text{Sq.}}}^\text{p}(\tau_g)$, and using eq.(s)(\ref{QGUncertainty.71},\ref{QGUncertainty.72}), we can write down the uncertainty product $\Delta z^{\text{Sq.}}\Delta \pi_z^{\text{Sq.}}$ as
\begin{equation}\label{QGuncertainty.73}
\begin{split}
\Delta z^{\text{Sq.}}\Delta \pi_z^{\text{Sq.}}&=\frac{2\sqrt{2}m_0^\text{p}g^2\tau_g^2l_{\text{Pl}}^2}{z_0^\text{p} c}\cosh2\mathfrak{r}\sqrt{\ln \left(\frac{c\tau_g}{z_0^\text{p}}\right)}\\
&=\frac{2\pi m_0^{\text{p}^2}g^2\tau_g^2l_{\text{Pl}}^2}{z_0^\text{p} c}\cosh2\mathfrak{r}\frac{z_0^\text{p}}{m_0^\text{p} c}\sqrt{\frac{2}{\pi^2}\ln\left(\frac{c\tau_g}{z_0^\text{p}}\right)}\\
\implies \Delta z^{\text{Sq.}}\Delta \pi_z^{\text{Sq.}}&=\frac{\beta z_0^\text{p}}{m_0^\text{p} c}(\Delta \pi^\text{Sq.}_z)^2
\end{split}
\end{equation}
where in the last line of the above equation, we have made use of eq.(\ref{QGUncertainty.72}) which has a structure identical to eq.(\ref{QGUncertainty.42}). Following the arguments identical to the vacuum state case, we can write down the uncertainty relation as
\begin{equation}\label{QGUncertainty.74}
\Delta z^{\text{Sq.}}\Delta\pi_z^{\text{Sq.}}\geq\frac{\hbar}{2}\left(1+\frac{\tilde{\beta} l_\text{Pl}}{\hbar m_0^\text{p} c}(\Delta \pi_z^\text{Sq.})^2\right)~.
\end{equation}
We can again obtain the generalized uncertainty relation in the Planck-mass limit as before. Similar to the vacuum state case, we can also implement an uncertainty upper bound, which helps us to write down the complete uncertainty relation when the graviton is in a squeezed state as (for $m_0^\text{p}<m_{\text{Pl}}$)
\begin{equation}\label{QGUncertainty.75}
\frac{m_{\text{Pl}} c}{\beta l_{\text{Pl}}}(\Delta z^\text{Sq.})^2>\Delta z^\text{Sq.}\Delta \pi_z^\text{Sq.}\geq \frac{\hbar}{2}\left(1+\frac{\tilde{\beta} l_\text{Pl}}{\hbar m_0^\text{p} c} (\Delta\pi_z^\text{Sq.})^2\right)~.
\end{equation}
The above uncertainty relation is identical to the quantum gravity modified uncertainty relation obtained in eq.(\ref{QGUncertainty.55}). This is a very important outcome and we shall be investigating the uncertainty relation while the particle interacts with thermal gravitons. If the the uncertainty relations, obtained for the three cases are identical, it would be sufficient to claim that the quantum gravity modified uncertainty relation obtained in eq.(\ref{QGUncertainty.55}), is indeed universal in nature.
\subsection{Particle interacting with thermal gravitons}
\noindent We shall, in this subsection, consider the interaction of thermal gravitons with the freely falling point-particle. At first, we need to write down the Feynman-Vernon influence functional for the graviton initially being in a thermal state as
\cite{QGravD}
\begin{equation}\label{QGUncertainty.76}
F^{\text{Th.}}[z^\text{p},\acute{z}^\text{p}]=F_{0^g}[z^\text{p},\acute{z}^\text{p}]\exp\left[i\Phi^{\text{Th.}}[z^\text{p},\acute{z}^\text{p}]\right]
\end{equation}
where the vacuum influence functional $F_{0^g}[z^\text{p},\acute{z}^\text{p}]$ is obtained by summing over all possible mode frequencies of the single-mode vacuum influence functional $F_{0^g_\omega}[z^\text{p},\acute{z}^\text{p}]$ given in eq.(\ref{QGUncertainty.62}).
The influence phase for thermal gravitons is denoted by $i\Phi^{\text{Th.}}[z^\text{p},\acute{z}^\text{p}]$ (for $c=1$), which takes the analytical form
\begin{equation}\label{QGUncertainty.77}
\begin{split}
i\Phi^{\text{Th.}}[z^\text{p},\acute{z}^\text{p}]&=-\frac{m_0^2G}{4\pi\hbar}\int_0^\infty\frac{\omega d\omega}{e^{\frac{\hbar\omega}{k_BT_{G}}}-1}\int_0^T\int_0^T dtdt'[\mathcal{X}(t)-\acute{\mathcal{X}}(t)][\mathcal{X}(t')-\acute{\mathcal{X}}(t')]\cos(\omega(t-t'))
\end{split}
\end{equation}
with $T_G$ denoting the temperature of the thermal gravitons. It is easy to define an auxiliary function for the gravitons being in a thermal state after dimensional reconstruction as
\begin{equation}\label{QGUncertainty.78}
\begin{split}
\mathcal{A}^{\text{Th.}}(t,t')=\frac{8\hbar G}{\pi c^5}\int_0^\infty\frac{\omega d\omega}{e^{\frac{\hbar\omega}{k_BT}}-1}\cos(\omega(t-t'))~.
\end{split}
\end{equation}
The above frequency integral is convergent and as a result there is no need for a regularization by setting an finite upper and lower limit of integration as have been done for the previous two scenarios. This divergence free auxiliary function can be related to the noise-fluctuation terms corresponding to the thermal gravitons as $\mathcal{A}^{\text{Th.}}(t,t')=\llangle \mathcal{N}^{\text{Th.}}(t)\mathcal{N}^{\text{Th.}}(t')\rrangle$. One can finally write down the Langevin-like differential equation in $z^\text{p}$ as
\begin{equation}\label{QGUncertainty.79}
\ddot{z}^\text{p}(t)+\frac{\partial\phi_\oplus}{\partial z^\text{p}}-\frac{1}{2}\left(\ddot{\mathcal{N}}^0_\text{q}(t)+\ddot{\mathcal{N}}^{\text{Th.}}(t)\right)z^\text{p}(t)=0
\end{equation}
were all higher order time-derivatives in $z^\text{p}$ has been neglected. While calculating the standard deviation in position as well as momentum, we shall be focussing on the thermal noise fluctuation term. During the calculation of $\llangle z_{\text{QG}}^\text{p}(\tau_g)z_{\text{QG}}^\text{p}(\tau_g)\rrangle$, multiple unnecessary term came out which can be neglected by considering them as spurious contributions, and one can consider only the terms that dominate over time. It is then possible to write down the analytical expression for the variance of the position as
\begin{equation}\label{QGUncertainty.80}
\begin{split}
(\Delta z^{\text{Th.}})^2&\simeq\frac{2g^2\tau_g^2\hbar G}{\pi c^5}\ln\left[\sinh\left(\frac{\pi\tau k_B T_G}{\hbar}\right)\right]+\frac{5g^2G\tau_g^3k_BT_G}{ c^5}+\cdots\\
&\simeq \frac{2g^2\tau_g^2 l_{\text{Pl}}^2}{\pi c^2}\ln\left[\sinh\left(\frac{\pi\tau_g k_B T_G}{\hbar}\right)\right]
\end{split}
\end{equation}
where the contribution of the second term has been neglected. Now the second term in the first line of the above expression is independent of the Planck's constant which is very strange considering that the right hand side of eq.(\ref{QGUncertainty.80}) is coming purely due to quantum gravitational noise fluctuations. As a result we shall be considering the first term in the first line of the right hand side of eq.(\ref{QGUncertainty.80}). One can next obtain the variance in the momentum term as
\begin{equation}\label{QGUncertainty.81}
\begin{split}
(\Delta\pi_z^\text{Th.})^2=&\frac{\pi m_0^{\text{p}^2}g^2\tau_g^2 G k_B^2T_G^2}{3\hbar c^5}+\frac{2m_0^{\text{p}^2}\hbar Gg^2}{\pi c^5}-\frac{2m_0^{\text{p}^2}k_BT_GGg^2\tau_g\coth\left(\frac{\pi
\tau_g k_BT_G}{\hbar}\right)}{c^5}\\&-\frac{2m_0^{\text{p}^2}g^2\hbar G}{\pi c^5}\ln\left[\frac{\pi
\tau k_BT_G}{\hbar}\text{cosech}\left(\frac{\pi
\tau_g k_BT_G}{\hbar}\right)\right]~.
\end{split}
\end{equation}
Keeping the dominant contribution only and using the analytical expression for the Planck temperature $T_{\text{Pl}}=\sqrt{\frac{\hbar c^5}{Gk_B^2}}$, we can write down the analytical expression for the variance in momentum as
\begin{equation}\label{QGUncertainty.82}
\begin{split}
(\Delta\pi_z^{\text{Th.}})^2&\simeq\frac{\pi m_0^{\text{p}^2}g^2\tau_g^2 G k_B^2T_G^2}{3\hbar c^5}=\frac{\pi m_0^{\text{p}^2}g^2\tau_g^2 T_G^2}{3T_{\text{Pl}}^2}~.
\end{split}
\end{equation}
Multiplying the variance in the position from eq.(\ref{QGUncertainty.80}) with the variance in momentum from eq.(\ref{QGUncertainty.82}), we obtain the analytical expression for $(\Delta z^{\text{Th.}})^2(\Delta\pi_z^{\text{Th.}})^2$ as
\begin{equation}\label{QGUncertainty.83}
(\Delta z^{\text{Th.}})^2(\Delta\pi_z^{\text{Th.}})^2=
\frac{2m_0^{\text{p}^2}G^2k_B^2T_G^2g^4\tau_g^4}{3c^{10}}\ln\left[\sinh\left(\frac{\pi\tau_g k_B T_G}{\hbar}\right)\right]~.
\end{equation}
Taking a square root of the above expression, we arrive at the uncertainty product as
\begin{equation}\label{QGUncertainty.84}
\begin{split}
 \Delta z^{\text{Th.}}\Delta\pi_z^{\text{Th.}}&=\frac{\sqrt{2}m_0^\text{p}Gk_B T_G g^2 \tau_g^2}{\sqrt{3}c^5}\sqrt{\ln\left[\sinh\left(\frac{\pi\tau_g k_B T_G}{\hbar}\right)\right]}\\
&=\sqrt{\frac{6}{\pi^2}\ln\left[\sinh\left(\frac{\pi\tau k_B T_G}{\hbar}\right)\right]}\frac{\hbar}{m_0^{\text{p}}k_BT_G}(\Delta \pi_z^\text{Th.})^2
\end{split}
\end{equation}
where in the final line of the above equation, we have made use of eq.(\ref{QGUncertainty.82}). Defining a new constant, $\beta_{\text{Th.}}\equiv\sqrt{\frac{6}{\pi^2}\ln\left[\sinh\left(\frac{\pi\tau_g k_B T_G}{\hbar}\right)\right]}$, we can recast the above uncertainty product in a much simpler form as
\begin{equation}\label{QGUncertainty.85}
\begin{split}
 \Delta z^\text{Th.}\Delta\pi_z^\text{Th.}&=\frac{\beta_{\text{Th.}}\hbar}{m_0^\text{p}k_BT_G}(\Delta\pi_z^\text{Th.})^2~.
\end{split}
\end{equation}
The maximum value that the temperature can reach is equal to the Planck temperature $T_{\text{Pl}}=\sqrt{\frac{\hbar c^5}{Gk_B^2}}$ which implies that the temperature corresponding to the thermal state of the gravitons obey the inequality $T_G\leq T_{\text{Pl}}$. This inequality gives a natural lower bound to the uncertainty product in eq.(\ref{QGUncertainty.85}) as
\begin{equation}\label{QGUncertainty.86}
\begin{split}
\Delta z^\text{Th.}\Delta\pi_z^\text{Th.}=&\frac{\beta_{\text{Th.}}\hbar}{m_0^\text{p}k_BT_G}(\Delta\pi_z^\text{Th.})^2\geq\frac{\beta_{\text{Th.}}\hbar}{m_0^\text{p}k_BT_\text{Pl}}(\Delta\pi_z^\text{Th.})^2\\
\implies \Delta z^\text{Th.}\Delta\pi_z^\text{Th.}\geq&\frac{\beta_{\text{Th.}} l_\text{Pl}}{m_0^\text{p}c}(\Delta\pi_z^\text{Th.})^2~.
\end{split}
\end{equation} 
Again, if the detector (or particle) degrees of freedom obey standard Heisenberg uncertainty principle, then one can use similar arguments as have been done in the vacuum state case to give a lower bound to the uncertainty product as
\begin{equation}\label{QGUncertainty.87}
\Delta z^\text{Th.}\Delta\pi_z^\text{Th.}\geq\frac{\hbar}{2}\left(1+\frac{\tilde{\beta}_{\text{Th.}} l_\text{Pl}}{\hbar m_0^\text{p}c}(\Delta\pi_z^\text{Th.})^2\right)
\end{equation}
where $\tilde{\beta}_{\text{Th.}}\equiv 2{\beta}_{\text{Th.}}$.
Now, the free-fall time $\tau_g=\sqrt{\frac{2z_0^\text{p}}{g}}$ is constant leaving $\beta$ for the vacuum as well as the squeezed graviton state to be constant. However, $\beta_{\text{Th.}}$ also depends on the initial temperature of the thermal gravitons and for a fixed vale of $T_G$, $\beta_{\text{Th.}}$ is constant as well. It is then always possible to find a certain value of the temperature $T_G$ such that $\beta=\beta_{\text{Th.}}$. One can also recast the uncertainty product in eq.(\ref{QGUncertainty.85}) in terms of the variance in position from eq.(\ref{QGUncertainty.80}) as
\begin{equation}\label{QGUncertainty.88}
\begin{split}
\Delta z^\text{Th.}\Delta \pi_z^\text{Th.}&=\frac{m_0^\text{p}k_BT_G}{\beta_{\text{Th.}}\hbar}(\Delta z^\text{Th.})^2\leq \frac{m_0^\text{p}k_BT_\text{Pl}}{\beta_{\text{Th.}}\hbar}(\Delta z^\text{Th.})^2\\
&=\frac{m_0^\text{p} c}{\beta_{\text{Th.}} l_\text{Pl}}(\Delta z^\text{Th.})^2\leq\frac{m_\text{Pl} c}{\beta_{\text{Th.}} l_\text{Pl}}(\Delta z^{\text{Th.}})^2~. 
\end{split}
\end{equation}
Combing eq.(\ref{QGUncertainty.87}) with eq.(\ref{QGUncertainty.88}), we can write down the total uncertainty relation in a closed form as
\begin{equation}\label{QGUncertainty.89}
\frac{m_\text{Pl} c}{\beta_{\text{Th.}} l_\text{Pl}}(\Delta z^{\text{Th.}})^2> \Delta z^\text{Th.}\Delta\pi_z^\text{Th.}\geq\frac{\hbar}{2}\left(1+\frac{\tilde{\beta}_{\text{Th.}} l_\text{Pl}}{\hbar m_0^\text{p}c}(\Delta\pi_z^\text{Th.})^2\right)~.
\end{equation}
In the above equation, we have considered that the mass of the freely falling particle is smaller than the Planck mass. If the mass of the particle is equal to the Planck mass then one can modify the above closed form of the uncertainty relation where the the first ``$>$" sign will be replaced by ``$\geq$" symbol. In the Planck mass limit, the lower bound of the above uncertainty relation reduces to the standard generalized uncertainty relation. For $\beta=\beta_{\text{Th.}}$, eq.(\ref{QGUncertainty.89}) is exactly same to the uncertainty relations obtained for the vacuum and squeezed vacuum cases from eq.(s)(\ref{QGUncertainty.55},\ref{QGUncertainty.75}). Even if the constants are not equal to each other, the analytical structures remain exactly the same. Hence, we truly obtain an universal quantum gravity modified uncertainty relation with a lower bound as well as an upper bound.

\noindent For, all of the uncertainty relations obtained in eq.(s)(\ref{QGUncertainty.55},\ref{QGUncertainty.75}) and eq.(\ref{QGUncertainty.89}), the freely falling particle has a non-zero mass $m_0^\text{p}$ with the velocity of the particle being $\dot{z}^\text{p}(t)\ll c$. One can thus always consider $m_0^\text{p}$ to be the rest mass of the particle. The uncertainty relation for a massless particle can be achieved by just replacing the rest mass of the particle $m_0^\text{p}$ by $\frac{\hbar \omega_0}{c^2}$ where $\omega_0$ denotes the frequency of the massless particle.
It is then possible to write down the uncertainty relation for a massless particle as
\begin{equation}\label{QGUncertainty.90}
\frac{c^3}{\beta G}(\Delta z^\text{p})^2>\Delta z^\text{p}\Delta\pi_z^\text{p}\geq \frac{\hbar}{2}\left(1+\frac{\tilde{\beta} l_\text{Pl} c}{\hbar^2\omega_0}(\Delta\pi_z^\text{p})^2\right)~.
\end{equation}
\section{Discussion and conclusion}
\noindent In this work, we have used the model in \cite{ChawlaParikh}, where a freely falling particle under the effect of the Earth's gravitational field interacts with the background gravitational fluctuations, where the gravitational fluctuations are quantized. In our analysis, we have dropped any post-Newtonian contributions used in \cite{ChawlaParikh} and proceed with Newtonian correction terms, as the post-Newtonian corrections do not play any significant role in calculating the standard deviations in the position as well as the momentum of the particle. In \cite{ChawlaParikh}, the standard deviation in the position of the freely falling particle is calculated and evaluated when it touches the ground. It was observed that because of the noise induced by gravitons, the free-fall trajectory gets infused by graviton-noise fluctuations. In this work, we have made further consideration by calculating the uncertainty in the momentum of the particle at the time of touching the ground. This entire phenomena is being observed by a terrestrial observer. For a particle freely falling with mass $m_0^\text{p}$, we obtain the analytical expression for the uncertainty product when the gravitons are initially in a vacuum, squeezed vacuum, and thermal state. The lower bound to the uncertainty product is obtained by expressing the uncertainty product in terms of the variance in the momentum and then implementing the inequality that in a quantum gravity setting $z_0^\text{p}\geq l_{\text{Pl}}$ (where $z_0^\text{p}$ is the initial height from which the massive particle is being dropped) and combining it with the standard lower bound for the Heisenberg's uncertainty relation, we arrive at the quantum gravity modified uncertainty relation. The coefficient of the quantum gravity modified term depends on three of the fundamental constants, namely the Planck's constant $\hbar$, Newton's gravitational constant $G$, and the speed of light $c$. Hence, if either $\hbar$ or $G$ is taken to zero the term induced due to the noise of gravitons vanishes implying that the obtained correction indeed generates due to the consideration of quantum gravity in the model system. We also observe that in the Planck mass limit $m_0^\text{p}=m_{\text{Pl}}$, the quantum gravity modified uncertainty relation is reduced to the well-known analytical structure of the generalized uncertainty principle \cite{KempfManganoMann} where the GUP parameter is now pre-determined. This indeed serves as a true quantum gravitational derivation of the generalized uncertainty principle as well. It is also possible to express the uncertainty product in terms of the variance in the position which helps us to write down an uncertainty upper bound in the Planck limit ($z^{\text{p}}_0\rightarrow l_{\text{Pl}}$ and $m^{\text{p}}_0\rightarrow m_{\text{Pl}}$). We therefore obtained a closed-form expression for the uncertainty product when the graviton is initially in a vacuum state. The existence of a lower bound with a dependence on the variance of the momentum suggests that there is a lower bound to the measurement of the uncertainty in the position of the particle whereas the existence of an upper bound suggests the existence of an upper bound to the uncertainty in the momentum. In the next part of our calculation, we have considered the initial graviton state to be in a squeezed vacuum state and obtained the uncertainties in position and momentum. We find out that the uncertainties get enhanced by a $\sqrt{\cosh 2\mathfrak{r}}$ factor where $\mathfrak{r}$ denotes the real squeezing parameter. Analytically obtaining the lower and upper bounds to the uncertainty product, we obtain an identical uncertainty relation in eq.(\ref{QGUncertainty.75}) compared to the uncertainty relation eq.(\ref{QGUncertainty.55}) for the squeezed vacuum case. Finally, we have considered the interaction of the particles with thermal gravitons obtaining an uncertainty relation which is exactly similar to the structure of the uncertainty relations obtained in eq.(\ref{QGUncertainty.55},\ref{QGUncertainty.75}). In the case of thermal gravitons, the autocorrelation function needed no regularization as the frequency integrals were convergent. In order to obtain the uncertainty product, we needed to implement the temperature inequality $T_G\leq T_{\text{Pl}}$ and the dimensionless coefficient $\beta_{\text{Th.}}$ was found to be dependent on the initial temperature of the thermal gravitons. All three considerations do imply that the graviton-induced uncertainty relation obtained in this analysis is indeed a true quantum gravitational uncertainty relation which is universal in nature. In standard literature \cite{MaggioreAlgebraic,ScardigliGedanken,AdlerSantiago,
AdlerSantiago2,GUPOscillator1,GUPOscillator2,
AliDasVagenas,MajumderGUP,GUPOpto1,GUPOpto2,GUPOpto3,
SGADAS,ScardigliCasadio,FengYangLiZu,
RMSBSG,GUPResonantDetector1,
CorpuscularGUP,GUPResonantDetector2,
GUPResonantDetector3,PetruzzeilloIlluminati,GUPOpto4,GUPOpto5,
GUPOpto6,OngJCAP,Vagnozzi,Culetu} the correction term in the generalized uncertainty relation, is independent of the Planck's constant which is very unlikely for a quantum gravitational correction and the GUP parameter also is undefined. Our graviton-induced uncertainty relation gets rid of these problems. We have finally written down the uncertainty relation for a massless particle, which can be applied in cases of analyses where a massless scalar particle is considered.
%Finally, we have given a form of the uncertainty relation for a massless particle with a finite value of its associated frequency parameter.
%\begin{widetext}
%\begin{equation*}
%\begin{split}
%\sigma(t)=\sigma_0(t)+\sigma_\gamma(t)\cong\sqrt{2\pi}l_p\left(1+3\gamma m_0^2 c^2\biggr[1-\frac{2}{\pi(1+\frac{c t}{\xi_0})}\frac{\xi_0 \sin^2\left[\frac{\pi c t}{\xi_0}\right]}{\pi c t}+\frac{4}{\pi^2 \left(1+\frac{c t}{\xi_0}\right)^2}\left(\gamma_\varepsilon-\text{Ci}\left[\frac{2\pi c t}{\xi_0}\right]+\ln\left[\frac{2\pi c t}{\xi_0}\right]\right)\biggr]\right).
%\end{split}
%\end{equation*}
%\end{widetext}
\part{Quantum gravity phenomenology}
\chapter{Spontaneous emission of gravitons as a signature of quantum gravity}\label{C.5.OTM}
In the first part of this thesis, we have mainly focused on the path integral quantization and the theoretical aspects of a linearized quantum gravity theory. In this second part, we shall be more focused on a phenomenological perspective of the quanta of linearized gravity and its effect on the model system. We will also try to propose an experimental model based on which graviton detection may be possible in the near future. Now the graviton-detector interaction models have been quite thoroughly investigated in recent years \cite{MarlettoVedral,EntanglementQuantumGravity,
MarlettoVedral2,EntanglementQuantumGravity2,
EntanglementQuantumGravity3,QGravNoise,QGravLett,
QGravD,KannoSodaTokuda,KannoSodaTokuda2}. All these considerations primarily consider a flat background spacetime with small gravitational fluctuations, and then the gravitational fluctuations are quantized to incorporate quantum gravitational effects into the model system. As discussed in the earlier chapters as well, in \cite{QGravD,QGravNoise,QGravLett,KannoSodaTokuda,
KannoSodaTokuda2}, an interferometric detector setup is considered, which is modelled by a pair of freely falling point-particles where one of the particles is way more massive than the other particle\footnote{In a few recent works \cite{ParameterEstimation1,ParameterEstimation2}, investigations have been done for small parameter estimation where quantum metrological techniques have been implemented. Using these techniques as a baseline, a new generation of classical gravitational wave \cite{SabinBruschiAhmadiFuentes,HowlFuentes,MatthewMannAffshordi} as well as dark energy detectors \cite{HartleyKadingHowlFuentes} have been proposed, where a Bose-Einstein condensate serves the purpose of a detector. In \cite{TobarManikandanBeitelPikovski} quantum sensing techniques have been used to make a proposal for a single-graviton detector.}. This setup indeed mimics the arm of an `L' shaped interferometer detector. The geodesic deviation equation is then observed, which incorporates quantum gravitation noise fluctuation, giving the equation an overall stochastic behaviour. All these analyses revolve truly around a saddle-point approximation where the deviation in the classical equation of motion is observed as a result of the quantum gravity setting. The other important aspect of the analyses in \cite{QGravNoise,QGravD,QGravLett,KannoSodaTokuda,
KannoSodaTokuda2} is that only the gravitational fluctuation part is treated from a full quantum field theoretical viewpoint. In this chapter, we shall treat the linearized gravity part as well as the matter part quantum mechanically. Apart from the `L' shaped gravitational wave observatories, the most simplest gravitational wave detector, proposed initially was the Weber bar detector \cite{Weber,Weber2} proposed and made by J. Weber.   He was also the first person to create an experimental setup for gravitational wave observation. Now, in recent years, there have been several works investigating the interaction of quantum matter with classical gravitational wave fluctuations \cite{Fischer,Speliotopoulos,sg11,sg22,sg33,sg44,sg55} in various physical scenarios. Here, we work with a more interesting model where both the matter part and the gravitational wave part are treated quantum mechanically. In this model, we take the two-particle system from \cite{QGravLett,QGravD}, and place it inside a harmonic trap potential with a constant frequency. This model also mimics the resonant Weber-bar detectors, and we then execute the canonical quantization of both the matter as well as the gravitational fluctuation parts. We have then investigated the simple transition between two states, where the transition happens due to the existence of gravitons. This chapter is organized as follows.

\noindent In this chapter\footnote{This chapter is based on the publication S. Sen and S. Gangopadhyay, ``\textit{Quantum gravity signatures in gravitational wave detectors placed inside a harmonic trap potential}", \href{https://doi.org/10.1103/PhysRevD.110.026008}{Phys. Rev. D 110 (2024) 026008}.}, we start with the general model system and then move towards the canonical quantization of the two parts of the system. Next, we calculate the absorption and emission probabilities between two states of the joint model system and compare them with the classical counterpart, where the gravitational fluctuation is treated classically. Finally, we discuss some experimental aspects and conclude our results.
\section{The general model system and the canonical way of quantization}
The background model for the system in consideration is standard Minkowski background with gravitational fluctuations upon it, which is expressed as
\begin{equation}\label{QGSpontaneous.1}
g_{\mu\kappa}=\eta_{\mu\kappa}+h_{\mu\kappa}
\end{equation}
where $\eta_{\mu\kappa}$ is the Minkowski metric ($\mathop{\text{diag}}\{-1,1,1,1\}$) with $\mu,\kappa=\{0,1,2,3\}$. Here, we take the speed of light to be unity, which will be restored later using a dimensional analysis. In order to obtain the total Hamiltonian for the system, we first need to write down the Einstein-Hilbert part of the action which is given by the analytical form
\begin{equation}\label{QGSpontaneous.2}
\begin{split}
S_{\text{EH}}=\frac{1}{16\pi G}\int d^4x \sqrt{-g}R
\end{split}
\end{equation}
with $g$ denoting the determinant of the metric tensor $g_{\mu\kappa}$ and $R$ being the Ricci scalar. Using the background metric in eq.(\ref{QGSpontaneous.1}), and calculating the Ricci scalar up to second order in the background fluctuation, we can recast the Einstein-Hilbert action in eq.(\ref{QGSpontaneous.2}) as
\begin{equation}\label{QGSpontaneous.3}
\begin{split}
S_{\text{EH}}&\simeq \frac{1}{64\pi G}\int d^4 x\left(h_{\mu\kappa}\square h^{\mu\kappa}-h\square h+2h^{\mu\alpha}\partial_\mu\partial_\alpha h-2h_{\alpha\beta}\partial_\mu\partial^\beta h^{\alpha\mu}\right)
\end{split}
\end{equation} 
where in the above equation, we have made use of the fact that $\sqrt{-g}\simeq 1+\mathcal{O}(h^2)$ and this higher order contribution has been dropped. The perturbation term has the gauge symmetry given as
\begin{equation}\label{QGSpontaneous.4}
h_{\mu\kappa}=\bar{h}_{\mu\kappa}+\partial_\mu\zeta_\kappa+\partial_\kappa\zeta_\mu~.
\end{equation}
It is now possible to impose the transverse-traceless gauge conditions that get rid of all the redundant  degrees of freedom leaving only two intact. The transverse-traceless gauge condition can be expressed using a constant timelike vector $n_\alpha=\delta^0_{~\alpha}$ as
\begin{equation}\label{QGSpontaneous.5}
\begin{split}
\partial_\alpha{\bar{h}}^{\alpha\rho}=0~,~~\bar{h}^\alpha_{~\alpha}=0~,~~n_\alpha\bar{h}^{\alpha\beta}=0~.
\end{split}
\end{equation}
Making use of the transverse-traceless gauge, we can write down the Einstein-Hilbert action in eq.(\ref{QGSpontaneous.3}) as
\begin{equation}\label{QGSpontaneous.6}
\begin{split}
S_{\text{EH}}=-\frac{1}{64\pi G}\int d^4 x\hspace{0.5mm} \partial_\beta \bar{h}_{jk}\partial^\beta\bar{h}^{jk}
\end{split}
\end{equation}
where $\beta\in\{0,1,2,3\}$ and $j,k\in\{1,2,3\}$. The Einstein Hilbert part of the action remains same as the model discussed in Chapter(\ref{C.3.OTM}). Here, we are considering a two-particle system where  the system is freely falling. One of the mass of the particle is way higher than the other one and it is possible to consider the particle with the heavier mass to be on-shell. With respect to the world line of the particle with the heavier mass, it is then appropriate to describe the trajectory of the particle with the lighter mass using the Fermi-normal coordinates ($t,\xi^j$). The background spacetime metric can be expressed in terms of the Fermi normal coordinates as
\begin{equation}\label{QGSpontaneous.7}
\begin{split}
g_{00}(t,\xi)&=-1-R_{j0k0}(t,0)\xi^j\xi^k+\mathcal{O}(\xi^3)\\
g_{0j}(t,\xi)&=-\frac{2}{3}R_{0ijk}(t,0)\xi^i\xi^k+\mathcal{O}(\xi^3)\\
g_{ij}(t,\xi)&=\delta_{ij}-\frac{1}{3}R_{ikjl}(t,0)\xi^k\xi^l+\mathcal{O}(\xi^3)~.
\end{split}
\end{equation}
In the above equation $\xi=\sqrt{\xi_i\xi^i}$ denotes the spatial coordinate separation between two particles (or two end points of the resonant bar detector). In the expression of the metric elements, we observe that the Riemann curvature tensor depends only upon the coordinate time. The primary reason behind it lies in the fact that the the tensor is evaluated on a geodesic which is timelike. For small geodesic fluctuation over the Minkowski background, the Riemann tensor stays invariant under a gauge transformation. It is therefore important to note that under the change from the Fermi-normal coordinates to the transverse traceless gauge, the Riemann curvature tensor remains the same. As we have discussed earlier in Chapter(\ref{C.3.OTM}), if the coordinate of the particle corresponding to the smaller mass is denoted by $\mathfrak{Y}^{\alpha}=\{t,\xi^j\}$, then the relativistic action corresponding to the point particle reads
\begin{equation}\label{QGSpontaneous.8}
S_{\text{P}}=-m_{p}\int d\tau \sqrt{g_{\alpha\beta}\dot{\mathfrak{Y}}^\alpha\dot{\mathfrak{Y}}^\beta}
\end{equation}
where the the dot above $\mathfrak{Y}^\alpha$ denotes derivative with respect to the proper time $\tau$ of the particle. One important aspect of the above action is that it is reprametrization invariant and as a result it is always possible to change from the proper time of the particle to the coordinate time $t$. The above model, as already has been discussed in Chapter(\ref{C.3.OTM}), represents the arm of an interferometer detector where the dynamics of the heavier mass has been neglected. Now, if the entire detector system is placed inside of a harmonic trap potential, then it represents a resonating bar, the action of which reads
\begin{equation}\label{QGSpontaneous.9}
S_{\text{RD}}=-m_p\int dt\left(\sqrt{-g_{\alpha\beta}\frac{d\mathfrak{Y}^\alpha}{dt}\frac{d\mathfrak{Y}^\beta}{dt}}+\frac{1}{2}\omega_p^2g_{\alpha\beta}\mathfrak{Y}^\alpha\mathfrak{Y}^\beta\right)~.
\end{equation}
Using the analytical form of the metric from eq.(\ref{QGSpontaneous.7}) and substituting it back in the above equation, while keeping terms up to $\mathcal{O}(\xi^2)$, we can rewrite the action in eq.(\ref{QGSpontaneous.9}) as
\begin{equation}\label{QGSpontaneous.10}
\begin{split}
S_{\text{RD}}&=-m_p\int dt\left(\left(1+R_{i0j0}(t,0)\xi^i\xi^j-\delta_{jk}\dot{\xi}^j\dot{\xi}^k\right)^\frac{1}{2}-\frac{\omega_p^2}{2}\left(\left(1+R_{i0j0}(t,0)\xi^i\xi^j\right)t^2+\delta_{jk}\xi^j\xi^k\right)\right)\\
&\simeq -m_p\int dt \left(1+\frac{1}{2}R_{i0j0}(t,0)\xi^i\xi^j-\frac{1}{2}\delta_{ij}\dot{\xi}^i\dot{\xi}^j+\frac{\omega_p^2}{2}\delta_{ij}\xi^i\xi^j\right)
\end{split}
\end{equation}
where the terms that will not contribute towards the overall dynamics of the model system have been dropped\footnote{The gravitational wave detector has a very small interaction time with the gravity wave and as a result all contributions like $\mathcal{O}(t^3,\xi^2t^2)$ can be neglected.}. In the above equation the Riemann tensor $R_{i0j0}(t,0)$ in the transverse traceless gauge has the form
\begin{equation}\label{QGSpontaneous.11}
R_{i0j0}(t,0)=-\frac{1}{2}\ddot{h}_{ij}(t,0)~.
\end{equation} 
In eq.(\ref{QGSpontaneous.10}), one can get rid of the leading term as it has no contribution towards the overall dynamics of the detector and recast it as
\begin{equation}\label{QGSpontaneous.12}
S_{\text{RD}}\simeq  \frac{m_p}{2}\int dt \left(\delta_{ij}\dot{\xi}^i\dot{\xi}^j+\frac{1}{2}\ddot{h}_{ij}(t,0)\xi^i\xi^j-\omega_p^2\delta_{ij}\xi^i\xi^j\right)~.
\end{equation}
The next step is to quantizing the model system. For that, we shall decompose the gravitational fluctuation in the transverse traceless gauge in its discrete Fourier modes. Considering the entire model system to be kept inside of a box of length $L$, it is possible to express $\bar{h}_{ij}(t,\vec{x})$ as
\begin{equation}\label{QGSpontaneous.13}
\bar{h}_{ij}(t,\vec{x})=\frac{1}{l_{\text{Pl}}}\sum\limits_{q,\vec{k}}h_{q}(t,\vec{k})e^{i\vec{k}\cdot\vec{x}}\epsilon^q_{ij}(\vec{k})
\end{equation}
where $\epsilon^q_{ij}(\vec{k})$ denotes the polarization tensor annd $h_q(t,\vec{k})$ denotes the graviton mode function corresponding to the wave vector $\vec{k}=\frac{2\pi \vec{n}}{L}$ with $\vec{n}\in \mathbb{Z}^3$. Making use of the above mode decomposition in eq.(\ref{QGSpontaneous.13}), it is possible to write down the gauge fixed action corresponding to the detector-gravity wave system as
\begin{equation}\label{QGSpontaneous.14}
\begin{split}
S&=S_{\text{EH}}+S_{\text{RD}}\\
&=\frac{m}{2}\int dt\left[\dot{h}_q^2(t,\vec{k})-k^2 h^2_{q}(t,\vec{k})\right]+\frac{m_p}{2}\int dt \left[\delta_{ij}\dot{\xi}^i\dot{\xi}^j-\sum\limits_{q,\vec{k}}\frac{\dot{h}_q(t,\vec{k})}{\sqrt{\hbar G}}\epsilon^q_{ij}(\vec{k})\xi^i\xi^j-\omega_p^2\delta_{ij}\xi^i\xi^j\right]
\end{split}
\end{equation} 
where $c$ is set to unity. In the above equation $m=\frac{L^3}{16\pi\hbar G^2}$ denotes the effective mass corresponding to the gravitational wave part and $k^2=\vec{k}\cdot\vec{k}$. It is not necessary to consider three spatial directions for a resonant bar detector. For a resonant bar detector, the length is very large compared to the two perpendicular directions ($y$ and $z$ directions in our current scenario). As a result, we can restrict ourselves to one spatial direction only, along which the length of the resonant bar is aligned ($x$ direction in our case). If the gravitational wave is propagating along $z$ direction with the wave vector $\vec{k}$ and frequency $\omega=|\vec{k}|$, then the detector is placed in its polarization plane ($xy$-plane), and for further simplification, we can restrict the gravitational wave to carry plus polarization only. It is therefore possible to rewrite the action for the model system in a much simplified form as
\begin{equation}\label{QGSpontaneous.15}
\begin{split}
S&=\int dt\left[\frac{m}{2}\left(\dot{h}^2(t)-\omega^2h^2(t)\right)+\frac{m_p}{2}\left(\dot{\xi}^2(t)-\frac{2\mathcal{g}\dot{h}(t)\dot{\xi}(t)\xi(t)}{m_p}-\omega_p^2\xi^2(t)\right)\right]
\end{split}
\end{equation}
with $\xi(t)\equiv \xi^x(t)$, $\Re\left[h_+(t,k_z)\right]=h(t)$ (where fluctuations in the $y$ and $z$ directions have been neglected), and the detector field coupling constant being given as $\mathcal{g}=\frac{m_p}{2\sqrt{\hbar G}}$. 
From the action in eq.(\ref{QGSpontaneous.15}), it is possible to write down the total Lagrangian for the model system as
\begin{equation}\label{QGSpontaneous.16}
\begin{split}
L=\frac{m}{2}\left(\dot{h}^2(t)-\omega^2h^2(t)\right)+\frac{m_p}{2}\left(\dot{\xi}^2(t)-\omega_p^2\xi^2(t)\right)-\mathcal{g}\dot{h}(t)\dot{\xi}(t)\xi(t)~.
\end{split}
\end{equation}
Using the form of the Lagrangian in the above equation, it is now possible to write down the analytical forms of the conjugate momenta to $h(t)$ and $\xi(t)$ as
\begin{equation}\label{QGSpontaneous.17}
p_h=\frac{\partial L}{\partial \dot{h}}=m\dot{h}-\mathcal{g}\dot{\xi}\xi~,~~\pi_{\xi}=\frac{\partial L}{\partial \dot{\xi}}=m_p\dot{\xi}-\mathcal{g}\dot{h}\xi~.
\end{equation}
The Hamiltonian corresponding to the entire detector-harmonic oscillator system can be expressed using the conjugate momentum variables from eq.(\ref{QGSpontaneous.17}), and the Lagrangian from eq.(\ref{QGSpontaneous.16}) as
\begin{equation}\label{QGSpontaneous.18}
\begin{split}
H&=p_h\dot{h}+\pi_\xi\dot{\xi}-L\\
&=\frac{\frac{p_h^2}{2m}+\frac{\pi_\xi^2}{2m_p}+\frac{\mathcal{g}p_h\pi_\xi\xi}{mm_p}}{1-\frac{\mathcal{g}^2\xi^2}{mm_p}}+\frac{1}{2}m\omega^2h^2+\frac{1}{2}m_p\omega_p^2\xi^2~.
\end{split}
\end{equation}
Now, the graviton-detector coupling is very weak, and as a result it is possible to express the above analytical form of the Hamiltonian up to first order in the coupling constant as
\begin{equation}\label{QGSpontaneous.19}
H\simeq \frac{p_h^2}{2m}+\frac{\pi_\xi^2}{2m_p}+\frac{\mathcal{g}p_h\pi_\xi\xi}{mm_p}+\frac{1}{2}m\omega^2h^2+\frac{1}{2}m_p\omega_p^2\xi^2~.
\end{equation}
As has already been discussed in Chapter(\ref{C.3.OTM}), it is possible to write down the coupling constant as a time-dependent function in a way such that the coupling constant portrays an adiabatic turning on and off effect. One can write the coupling constant as $\mathcal{g}(t)\rightarrow \mathcal{g} \mathcal{f}(t)$, where $\mathcal{f}(t)=0$ for $t<t_i$ and $t>t_f$, and $\mathcal{f}(t_i\leq t\leq t_f)=1$. It is evident that the interaction of the gravity wave with the detector starts at time $t=t_i$ and continues up to the time $t=t_f$.
\subsection{Quantizing the model system}
The next step is quantizing the detector part as well as the gravitational wave part. To quantize the detector, both the detector phase space variables $\xi$ and $\pi_\xi$ needs to be raised to operator status, and one needs to implement a standard canonical commutation relation given by $\left[\hat{\xi},\hat{\pi}_\xi\right]=i\hbar$. Similarly for quantizing the gravitational wave part, one needs to raise the graviton mode function $h$ and its canonical conjugate momentum $\pi_h$ to operator status, and implement the canonical commutation relation $\left[\hat{h},\hat{\pi}_h\right]=i\hbar$. It is now possible to write down the Hermitian Hamiltonian operator using the Hamiltonian in eq.(\ref{QGSpontaneous.19}) as
\begin{equation}\label{QGSpontaneous.20}
\begin{split}
\hat{H}=\left(\frac{\hat{p}_h^2}{2m}+\frac{1}{2}m\omega^2\hat{h}^2\right)\otimes \hat{\mathbb{1}}_{\text{RD}}+\hat{\mathbb{1}}_{\text{GW}}\otimes\left(\frac{\hat{\pi}_\xi^2}{2m_p}+\frac{1}{2}m_p\omega_p^2\hat{\xi}^2\right)+\frac{\mathcal{g}}{2mm_p}\hat{p}_h\otimes\left(\hat{\xi}\hat{\pi}_\xi+\hat{\pi}_\xi\hat{\xi}\right)
\end{split}
\end{equation}
where $\hat{\mathbb{1}}_{\text{GW}}$ denotes the identity operator corresponding to the gravitational wave part and $\hat{\mathbb{1}}_{\text{RD}}$ denotes the identity operator corresponding to the resonant bar detecto part. From the form of the Hamiltonian operator in the above equation, it is evident that the Hamiltonian can be separated into a base part and an interaction part. Analytically, one can write down this separation as $\hat{H}=\hat{H}_{0}+\hat{H}_{\text{int}}$ where the base and the interaction Hamiltonians are defined as
\begin{align}
\hat{H}_0&=\left(\frac{\hat{p}_h^2}{2m}+\frac{1}{2}m\omega^2\hat{h}^2\right)\otimes \hat{\mathbb{1}}_{\text{RD}}+\hat{\mathbb{1}}_{\text{GW}}\otimes\left(\frac{\pi_\xi^2}{2m_p}+\frac{1}{2}m_p\omega_p^2\xi^2\right)\label{QGSpontaneous.21}\\
\hat{H}_{\text{int}}&=\frac{\mathcal{g}}{2mm_p}\hat{p}_h\otimes\left(\hat{\xi}\hat{\pi}_\xi+\hat{\pi}_\xi\hat{\xi}\right)\label{QGSpontaneous.22}~.
\end{align} 
If $\{\hat{b},\hat{b}^\dagger\}$ denote the ladder operators corresponding to the gravitational wave part, and $\{\hat{\zeta},\hat{\zeta}^\dagger\}$ denote the ladder operators corresponding to the detector part, then the position and their conjugate momentum operators can be expressed in terms of the ladder operators as
\begin{align}
\hat{h}=\sqrt{\frac{\hbar}{2 m\omega}}\left(\hat{b}+\hat{b}^\dagger\right)~&,~~\hat{p}_h=i\sqrt{\frac{m\hbar\omega}{2}}\left(\hat{b}^\dagger-\hat{b}\right);\label{QGSpontaneous.23}\\
\hat{\xi}=\sqrt{\frac{\hbar}{2m_p\omega_p}}\left(\hat{\zeta}+\hat{\zeta}^\dagger\right)~&,~~\hat{\pi}_\xi=i\sqrt{\frac{m_p\hbar\omega_p}{2}}\left(\hat{\zeta}^\dagger-\hat{\zeta}\right)\label{QGSpontaneous.24}
\end{align}
where from the commutation relation among the phase space operators, it is possible to obtain the commutation relation among the ladder operators as $\left[\hat{b},\hat{b}^\dagger\right]=\left[\hat{\zeta},\hat{\zeta}^\dagger\right]=1$\footnote{Now if $|0\rangle_{\text{GW}}$ denotes the zero particle state of the graviton part, and $|0\rangle_{\text{RD}}$ denotes the ground state of the detector, then $\hat{b}|0\rangle_{\text{GW}}=0$ and $\hat{\zeta}|0\rangle_{\text{RD}}=0$.}. For the number operators corresponding to the gravitons ($\hat{N}_{\text{GW}}=\hat{b}^\dagger\hat{b}$), and the number operator corresponding to the detector ($\hat{N}_{\text{RD}}=\hat{\zeta}^\dagger\hat{\zeta}$), it is possible to define their eigenstates as
\begin{equation}\label{QGSpontaneous.25}
\hat{N}_{\text{GW}}|n_G\rangle=n_G|n_G\rangle~,~~\hat{N}_{\text{RD}}|n_R\rangle=n_R|n_R\rangle
\end{equation}
where $n_G$ denotes the number of gravitons in the state $|n_G\rangle$, and $n_R$ gives the number of the excited state of the resonant bar detector. 
In terms of the raising and lowering operators, it is possible to express the Hamiltonian operator in eq.(\ref{QGSpontaneous.20}) as
\begin{equation}\label{QGSpontaneous.26}
\hat{H}=\hbar\omega\left(\hat{b}^\dagger\hat{b}+\frac{1}{2}\right)\otimes\hat{\mathbb{1}}_{\text{RD}}+\hat{\mathbb{1}}_{\text{GW}}\otimes \hbar\omega_p\left(\zeta^\dagger\zeta+\frac{1}{2}\right)-\frac{\hbar\mathcal{g}}{2mm_p}\sqrt{\frac{m\hbar \omega}{2}}(\hat{b}^\dagger-\hat{b})\otimes(\hat{\zeta}^{\dagger^2}-\hat{\zeta}^2)~.
\end{equation}
It is important to note that at the time of beginning of the interaction ($t=t_i$) of the gravitational wave with the resonant bar detector, the state can be separable into a tensor product of the graviton number state and the harmonic oscillator excited state as $|\psi_i\rangle=|n_{G_i}\rangle\otimes|n_{R_i}\rangle=|n_{G_i},n_{R_i}\rangle$. We also can safely assume that at the end of the interaction $t=t_f$, the final state of the system can be written as a tensor product state as well, $|\psi_f\rangle=|n_{G_f}\rangle\otimes|n_{R_f}\rangle=|n_{G_f},n_{R_f}\rangle$. It is important to note that $|\psi_i\rangle\neq |\psi_f\rangle$. In order to obtain the analytical form of the transition probability for system to go from a state $|\psi_i\rangle$ to state $|\psi_f\rangle$, we at first need to write down the interaction Hamiltonian in the interaction picture first, which reads
\begin{equation}\label{QGSpontaneous.27}
\begin{split}
\hat{H}_{\text{int}}^{I}(t)&=e^{\frac{i}{\hbar}\hat{H}_0t}\hat{H}_{\text{int}}e^{-\frac{i}{\hbar}\hat{H}_0t}\\
&=\frac{\mathcal{g}}{2mm_p}\hat{p}_h^I(t)\otimes \left(\hat{\xi}^I(t)\hat{\pi}^I_\xi(t)+\hat{\pi}^I_\xi(t)\hat{\xi}^I(t)\right)
\end{split}
\end{equation}
where all the phase space operators from eq.(s)(\ref{QGSpontaneous.23},\ref{QGSpontaneous.24}) are expressed in the interaction picture (in terms of their corresponding ladder operators) as
\begin{align}
\hat{\xi}^I(t)=\sqrt{\frac{\hbar}{2 m_p \omega_p}}\left(\hat{\zeta}e^{-i\omega_pt}+\hat{\zeta}^\dagger e^{i\omega_pt}\right)~&,~~\hat{\pi}^I_\xi(t)=i\sqrt{\frac{m_p\hbar\omega_p}{2}}\left(\hat{\zeta}^\dagger e^{i\omega_p t}-\hat{\zeta}e^{-i\omega_p t}\right)\label{QGSpontaneous.28}\\
\hat{h}^I(t)=\sqrt{\frac{\hbar}{2 m\omega}}\left(\hat{b}e^{-i\omega t}+\hat{b}^\dagger e^{i\omega t}\right)~&,~~\hat{p}^I_h(t)=i\sqrt{\frac{m\hbar\omega}{2}}\left(\hat{b}^\dagger e^{i\omega t}-\hat{b}e^{-i\omega t}\right)\label{QGSpontaneous.29}~.
\end{align}
With this basic model in hand, we shall now proceed towards calculating the transition probability of the system for going from the initial state $|\psi_i\rangle$ to some state $|\psi_f\rangle$. 
\section{The Fermi-Golden rule in quantum gravity}
In a linearized quantum gravity theory, we primarily work with perturbative models. In quantum mechanics, the Fermi golden rule gives the transition rate between two quantum states when there are small perturbations present in the system. It is possible to write down the unitary time-evolution operator corresponding to the interaction Hamiltonian in eq.(\ref{QGSpontaneous.27}) as
\begin{equation}\label{QGSpontaneous.30}
\begin{split}
\hat{U}^I(t,t_i)&=\mathcal{T}\left[\exp\left(-\frac{i}{\hbar}\int_{t_i}^tdt' \hat{H}^I_{\text{int}}(t')\right)\right]\\
&=1-\frac{i}{\hbar}\int_{t_i}^tdt' \hat{H}^I_{\text{int}}(t')+\left(-\frac{i}{\hbar}\right)^2\int_{t_i}^tdt'\int_{t_i}^{t'}dt'' \hat{H}^I_{\text{int}}(t') \hat{H}^I_{\text{int}}(t'')+\cdots
\end{split}
\end{equation}
where we have made use of the fact that the perturbation to the system is generated by the graviton interaction with the resonant bar detector system. As the coupling constant $\mathcal{g}$ in the interaction Hamiltonian is very small, we can truncate the series up to the first order in the interaction Hamiltonian in eq.(\ref{QGSpontaneous.30}). We can now calculate the transition amplitude of the model system for going from an initial state $|\psi_i\rangle$ to some final state $|\psi_f\rangle$ as $\langle\psi_f|\hat{U}^I(t,t_i)|\psi_i\rangle$. Now, it is important to note that $|\psi_f\rangle\neq |\psi_i\rangle$, and as a result $\langle \psi_f|\psi_i\rangle=0$. The transition probability then reads 
\begin{equation}\label{QGSpontaneous.31}
\begin{split}
\langle\psi_f|\hat{U}^I(t,t_i)|\psi_i\rangle&\simeq -\frac{i}{\hbar}\int_{t_i}^t dt'\langle \psi_f|\hat{H}^{I}_{\text{int}}(t')|\psi_i\rangle
\end{split}
\end{equation}
where we have dropped all higher order terms. Using the above equation, it is now possible to calculate the transition amplitude of the system for going from the state $|\psi_i\rangle=|n_{G_i},n_{R_i}\rangle$ to the final state $|\psi_f\rangle=|n_{G_f},n_{R_f}\rangle$ as
\begin{equation}\label{QGSpontaneous.32}
\begin{split}
P_{if}^{\text{QG}}(t)=&\left|\langle\psi_f|\hat{U}^I(t,t_i)|\psi_i\rangle\right|^2\\
\simeq&\frac{1}{\hbar^2}\left|\int_{t_i}^t dt'\langle \psi_f|\hat{H}^{I}_{\text{int}}(t')|\psi_i\rangle
\right|^2\\
=&\left|\frac{ig}{2mm_p}\sqrt{\frac{m\omega \hbar}{2}}\sqrt{(n_{G_i}+1)(n_{R_i}+2)(n_{R_i}+1)}\delta_{n_{G_f},n_{G_i}+1}\delta_{n_{R_f},n_{R_i}+2}\int_{t_i}^t dt'e^{i(\omega+2\omega_p)t'}\right.\\
&-\sqrt{(n_{G_i}+1)n_{R_i}(n_{R_i}-1)}\delta_{n_{G_f},n_{G_i}+1}\delta_{n_{R_f},n_{R_i}-2}\int_{t_i}^t dt'e^{i(\omega-2\omega_p)t'}\\&-\sqrt{n_{G_i}(n_{R_i}+2)(n_{R_i}+1)}\delta_{n_{G_f},n_{G_i}-1}\delta_{n_{R_f},n_{R_i}+2}\int_{t_i}^t dt'e^{-i(\omega-2\omega_p)t'}\\
&\left.+\sqrt{n_{G_i}n_{R_i}(n_{R_i}-1)}\delta_{n_{G_f},n_{G_i}-1}\delta_{n_{R_f},n_{R_i}-2}\int_{t_i}^t dt'e^{-i(\omega+2\omega_p)t'}\right|^2
\end{split}
\end{equation}
where the entire term inside of the modulus square gives the transition amplitude. The general approach now is to extend the limits of integration from $t_i\rightarrow-\infty$ to $t\rightarrow \infty$. Using these limits of integration, it is possible to recast the analytical form of the transition probability from eq.(\ref{QGSpontaneous.32}) as
\begin{equation}\label{QGSpontaneous.33}
\begin{split}
P_{if}^{\text{QG}}=&\frac{\hbar\omega\pi^2\mathcal{g}^2}{2mm_p^2}\left|\sqrt{(n_{G_i}+1)(n_{R_i}+2)(n_{R_i}+1)}\delta_{n_{G_f},n_{G_i}+1}\delta_{n_{R_f},n_{R_i}+2}~\delta(\omega+2\omega_p)\right.\\
&-\sqrt{(n_{G_i}+1)n_{R_i}(n_{R_i}-1)}\delta_{n_{G_f},n_{G_i}+1}\delta_{n_{R_f},n_{R_i}-2}~\delta(\omega-2\omega_p)\\
&-\sqrt{n_{G_i}(n_{R_i}+2)(n_{R_i}+1)}\delta_{n_{G_f},n_{G_i}-1}\delta_{n_{R_f},n_{R_i}+2}~\delta(-\omega+2\omega_p)\\
&\left.+\sqrt{n_{G_i}n_{R_i}(n_{R_i}-1)}\delta_{n_{G_f},n_{G_i}-1}\delta_{n_{R_f},n_{R_i}-2}~\delta(-\omega-2\omega_p)\right|^2
\end{split}
\end{equation}
where we have made use of the integral representation of the delta function given by $\delta(\omega_1-\omega_2)=\frac{1}{2\pi}\int_{-\infty}^\infty dx\hspace{0.5 mm} e^{i(\omega_1-\omega_2)x}$. The delta functions $\delta(\omega+2\omega_p)$ and $\delta(-\omega-2\omega_p)$, gives non vanishing contribution when $\omega=-2\omega_p$, which is an impossible scenario as both $\omega$ and $\omega_p$ are non-negative quantities. As a result the above two delta functions vanish, leading to the analytical form of the transition probability as
\begin{equation}\label{QGSpontaneous.34}
\begin{split}
P_{if}^{\text{QG}}=&\frac{\hbar\omega \pi^2\mathcal{g}^2}{2mm_p^2}\left(\sqrt{(n_{G_i}+1)n_{R_i}(n_{R_i}-1)}\delta_{n_{G_f},n_{G_i}+1}\delta_{n_{R_f},n_{R_i}-2}\right.\\&\left.+\sqrt{n_{G_i}(n_{R_i}+2)(n_{R_i}+1)}\delta_{n_{G_f},n_{G_i}-1}\delta_{n_{R_f},n_{R_i}+2}\right)^2\delta^2(\omega-2\omega_p)~.
\end{split}
\end{equation} 
One interesting thing to note from the above form of the transition probability is that the transition in the energy states of the gravitational wave detector always happens in steps of two whereas only a single graviton emission or absorption occurs. We shall now look at the absorption and emission probabilities respectively for the gravitational wave detector.
\subsection{The resonant absorption process} 
We consider that the detector is in its ground state and as a result $n_{R_i}=0$. For $n_{R_i}=0$, only the second term in the analytical form of the transition probability from eq.(\ref{QGSpontaneous.34}) connects implying that the second term only contributes in the overall transition probability provided $n_{R_f}=n_{R_i}+2=2$. For the transition probability to be truly non vanishing, the number of gravitons in the final state of the system has to be $n_{G_f}=n_{G_i}-1$. This implies that the detector jumps from its ground state to its second excited state by absorbing a single graviton. For the detector to be in the ground state, $n_{R_i}=0$, the only non vanishing contribution to the transition probability comes from the final state $|\psi_f\rangle=|n_{G_i}-1,2\rangle$ and the corresponding transition probability reads
\begin{equation}\label{QGSpontaneous.35}
P_{02}^{\text{QG}}=\frac{n_{G_i}\hbar\omega\pi^2\mathcal{g}^2}{mm_p^2}\delta^2(\omega-2\omega_p)=\frac{4n_{G_i}\hbar\omega G\pi^3}{L^3}\delta^2(\omega-2\omega_p)~.
\end{equation}
Here, we have made use of the analytical form of the effective mass $m$ and the coupling constant $\mathcal{g}$. We shall compare the absorption probability for the graviton case with the semiclassical case, where the gravitational wave behaves classically. In order to obtain the semiclassical limit, we can just set the generalized uncertainty principle parameter to zero in \cite{sg22,sg33} such that the detector phase space operators follow the standard Heisenberg uncertainty principle. If a periodic and linearly polarized gravitational wave is considered with the analytical form of the background fluctuation being given as (in the transverse-traceless gauge) $\bar{h}_{ij}(t)=2f_h\cos \omega t(\epsilon_\times\sigma_{ij}^1+\epsilon_+\sigma^3_{ij})$, the transition probability for the system to go from its ground state to the second excited state reads
\begin{equation}\label{QGSpontaneous.36}
P^{\text{SC}}_{02}=\frac{1}{2}\pi^2f_h^2\omega^2\epsilon_+^2\delta^2(\omega-2\omega_p)
\end{equation}
where the analysis is restricted to the $x$ direction only (the bar detector is placed along the $x$-direction) with $f_h$ being the amplitude of the gravitational fluctuation. As $\epsilon_+^2+\epsilon_{\times}^2=1$, for a plane polarized gravitational wave $\epsilon_+^2=1$, and this helps us to write down the transition probability for a classical gravitational wave from eq.(\ref{QGSpontaneous.36}) as
\begin{equation}\label{QGSpontaneous.37}
P^{\text{SC}}_{02}=\frac{1}{2}f_h^2\pi^2\omega^2\delta^2(\omega-2\omega_p)~.
\end{equation}
In order to truly compare the transition probability corresponding to the semiclassical case with that of the quantum gravitational scenario, we start by considering the total energy carried by a gravitational wave. If we consider that $d\mathcal{E}$ amount have energy has flown through the area $d\mathcal{A}$ in the time-interval $t=-\infty$ to $t=\infty$, then the analytical expression for the energy flux is given by \cite{Maggiore}
\begin{equation}\label{QGSpontaneous.38}
\frac{d\mathcal{E}}{d\mathcal{A}}=\frac{1}{32\pi G}\int_{-\infty}^\infty dt \left\langle \dot{h}_{jk}^{\text{TT}}\dot{h}_{jk}^{\text{TT}}\right\rangle
\end{equation}
where the $j,k$ indices are summed over and $\bar{h}_{jk}=h_{jk}^{\text{TT}}$\footnote{Here one needs to start from the expression $\dot{\bar{h}}_{jk}\dot{\bar{h}}^{jk} $ to approach the term inside of the temporal average in eq.(\ref{QGSpontaneous.38}) as $\dot{\bar{h}}_{jk}\dot{\bar{h}}^{jk}=g^{jl}g^{km}\dot{\bar{h}}_{jk}\dot{\bar{h}}_{lm}\simeq \delta^{jl}\delta^{km}\dot{\bar{h}}_{jk}\dot{\bar{h}}_{lm}+\mathcal{O}(h^3)\simeq \dot{\bar{h}}_{jk}\dot{\bar{h}}_{jk}$.}, which denotes gravitational fluctuation in the transverse traceless gauge. The temporal average is denoted using $\langle\cdots\rangle$ in the above equation. To arrive at the above analytical expression one needs to consider the entire system to reside inside of a sphere of radius $r$, where the energy is passing through a solid angle $d\Omega$ such that $d\mathcal{A}=r^2d\Omega$. As our analysis is restricted to the $x$ direction only, one can write down the expression for energy from eq.(\ref{QGSpontaneous.38}) as
\begin{equation}\label{QGSpontaneous.39}
\begin{split}
\mathcal{E}&=\int d\mathcal{E}=\frac{r^2}{32\pi G}\int d\Omega\int_{-\infty}^{\infty}dt\left\langle \dot{h}_{jk}^{\text{TT}}\dot{h}_{jk}^{\text{TT}}\right\rangle\\
&=\frac{r^2}{8G}\int_{-\infty}^\infty dt ~\langle \dot{h}_+^2(t)\rangle
\end{split}
\end{equation}
where, we have used the fact that $h^{\text{TT}}_{xx}(t)=h_+(t)$. There is a simple way to deal with the temporal average. One can simply execute the time integral first and then the average can be done which is nothing but an average over a constant parameter \cite{Maggiore}. If $h_+(t)$ has the analytical form $h_+(t)=2f_h\cos \omega t$, which is periodic in nature, then we can restrict ourselves from $t\in(-\infty,\infty)$ to a single time cycle. A single time cycle for the above gravitational wave ranges from $t=0$ to $t=\frac{2\pi}{\omega}$. The amount of energy passing through the sphere, in one time cycle, reads
\begin{equation}\label{QGSpontaneous.40}
\begin{split}
\mathcal{E}&=\frac{r^2f_h^2\omega^2}{2G}\int_{0}^{\frac{2\pi}{\omega}} dt \sin^2\omega t=\frac{\pi \omega r^2f_h^2}{2 G}~.
\end{split}
\end{equation}
\begin{figure}[t!]
\begin{center}
\includegraphics[scale=0.18]{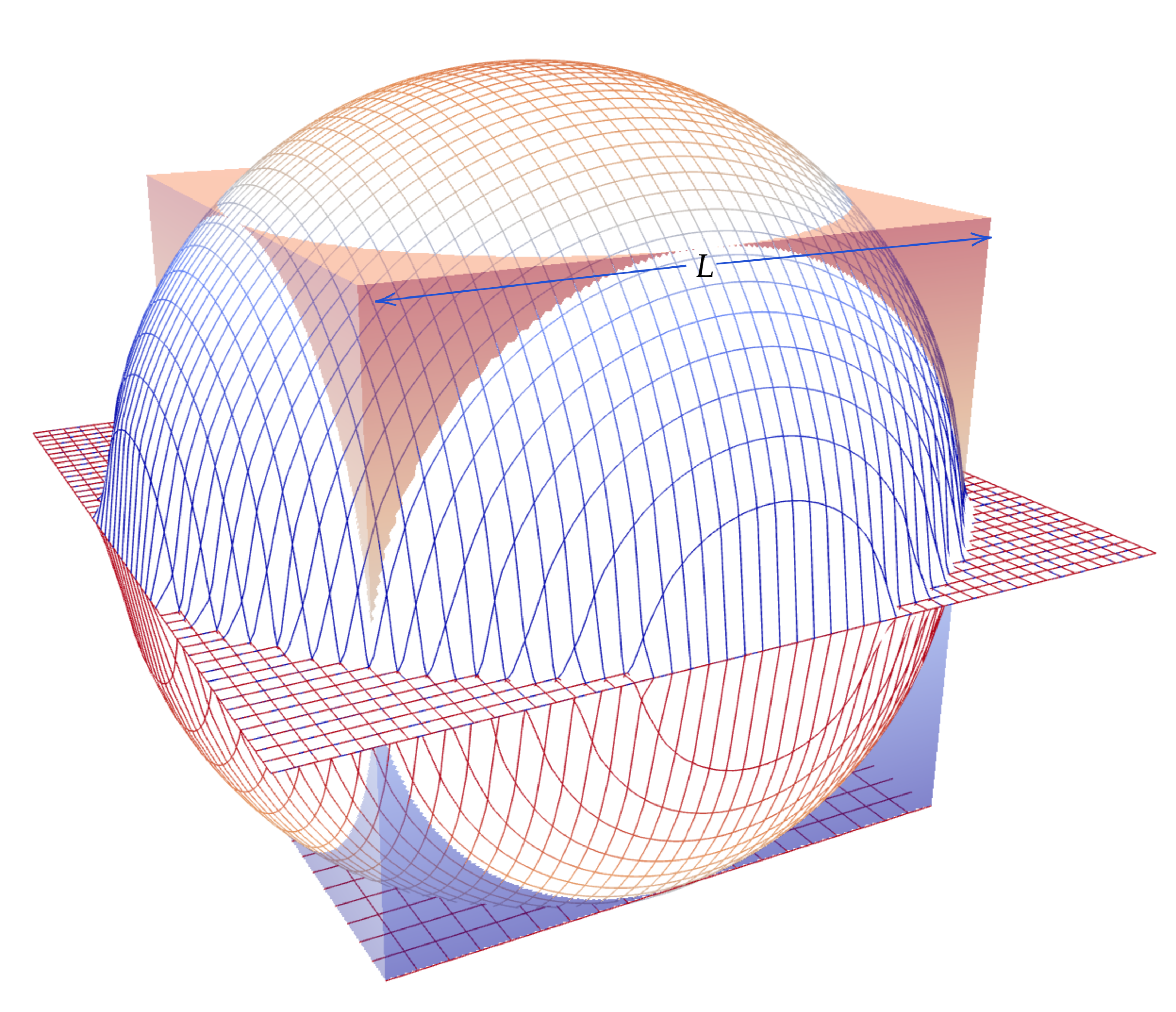}
\caption{A box of volume $L^3$ is compared with a sphere of radius $\frac{L}{\sqrt{2}}$ such that the surface area of both the sphere and the box are closest to each other signifying almost a similar amount of transfer of energy through the surface of both of the geometrical objects. \label{Embedding_OTM}}
\end{center}
\end{figure}
\noindent We have primarily considered here a one dimensional model but we can consider the energy to be passing through a three dimensional volume. The entire quantization has been done using a box of side $L$ but the energy transfer is being considered using a sphere. In order to establish a relation, we consider a sphere of radius $r=\frac{L}{\sqrt{2}}$ such that the midpoints of all the sides of the box lies on the surface of the sphere. The surface area of the box is $6L^2$ whereas for the sphere, it is $2\pi L^2$. It is very straightforward to notice that $2\pi L^2\simeq 6.28 L^2$, which is very close to the surface area of the box as has also been depicted in Fig.(\ref{Embedding_OTM}). Hence, the standard identification is $r=\frac{L}{\sqrt{2}}$, which helps us to recast eq.(\ref{QGSpontaneous.40}) as
\begin{equation}\label{QGSpontaneous.41}
\mathcal{E}=\frac{\pi \omega L^2 f_h^2}{4G}~.
\end{equation}
The above energy is carried by a classical gravitational wave with amplitude $f_h$ and frequency $\omega$ in a single time cycle through a sphere of radius $r=\frac{L}{\sqrt{2}}$. Now, if the gravitational wave is quantized, then the same amount of energy will be carried by $n_{G_i}$ number of gravitons with a mode frequency $\omega$, and the total energy can be represented by $\mathcal{E}=n_{G_i}\hbar \omega$. We can now equate this energy expression with the analytical form of total energy carried in eq.(\ref{QGSpontaneous.41}), which gives us the analytical expression for the amplitude $f_h$ as
\begin{equation}\label{QGSpontaneous.42}
f_h^2=\frac{4 n_{G_i}\hbar G}{\pi L^2}~.
\end{equation}
The next step is to find an analytical expression for $L$ when we are considering a single time cycle. The box of length $L$ in consideration simply implies the fact that the graviton-detector interaction and the corresponding consequences are confined inside of this finite volume $V=L^3$. Now, we have considered in eq.(\ref{QGSpontaneous.40}) that the amount of energy $\mathcal{E}$ is transferred within a time interval of $t\in[0,\frac{2\pi}{\omega}]$. In this time $\Delta t=t_f-t_i=\frac{2\pi}{\omega}$, a path of $\lambda_G=c\Delta t\rvert_{c\rightarrow1}=\frac{2\pi}{\omega}$ is travelled by the gravitational wave\footnote{The gravitational wave travels at the speed of light \cite{Maggiore}.}. It is therefore logical to consider a sphere of radius $\lambda_G$ and place the harmonic oscillator at the centre of the sphere, which results in the confinement of all the detector-gravitational wave interaction within a single time cycle. To relate the box volume $L$ with $\lambda_G$, we can stretch the box such that it fits on the sphere by joining four points in one side to a single point and also squeezing the points in the opposite plane and merging them in a single point which gives the value of $L$ to be $L\simeq \pi \lambda_G=\frac{2\pi^2}{\omega}$. The analytical form of $f_h^2$ in eq.(\ref{QGSpontaneous.42}) then becomes
\begin{equation}\label{QGSpontaneous.43}
f_h^2=\frac{n_{G_i}\omega^2\hbar G}{\pi ^5}\implies n_{G_i}\hbar\omega=\frac{f_h^2\pi^5}{\omega G}~.
\end{equation} 
Using the above equation along with the analytical form of $L$ in eq.(\ref{QGSpontaneous.35}), we arrive at the expression for the transition probability $P_{02}^{\text{QG}}$ as
\begin{equation}\label{QGSpontaneous.44}
P_{02}^{\text{QG}}=\frac{1}{2}\pi^2f_h^2\omega^2\delta^2(\omega-2\omega_p)
\end{equation}
which is identical to the transition probability for the semiclassical result $P^{\text{SC}}_{02}$ in eq.(\ref{QGSpontaneous.37}). This a remarkable result which shows that by using the energy-flux relation in eq.(\ref{QGSpontaneous.38}) and considering the gravitational wave to be a combination of $n_{G_i}$ number of gravitons, the semiclassical result (where the transition in a quantum detector is caused by a classical gravitational wave) can be reproduced exactly. This is an excellent analogy between the semiclassical treatment and the quantum gravity treatment. This resonant absorption of gravitons case is identical to the semiclassical results whereas the emission process packs a little bit more interesting outcomes.
\subsection{The spontaneous emission process}
In the previous subsection, we  have considered the transition of the resonant bar detector from the ground state to its second excited state, when resonant absorption of gravitons takes place. We now consider the opposite scenario, when the detector de-excites via the emission of gravitons. If the initial state of the system reads $|\psi_i\rangle=|n_{G_i},2\rangle$, then the transition probability will satisfy the non-vanishing condition provided that the final state is $|\psi_f\rangle=|n_{G_i}+1,0\rangle$. From eq.(\ref{QGSpontaneous.34}), we find the form of the transition probability to be 
\begin{equation}\label{QGSpontaneous.45}
P_{20}^{\text{QG}}=\left(n_{G_i}+1\right)\frac{4\hbar \omega G\pi^3}{L^3}\delta^2(\omega-2\omega_p)~.
\end{equation}
It is quite evident from the analytical form of the above de-excitation probability $P_{20}^{\text{QG}}$ that it is not equal to the transition probability $P_{02}^{\text{QG}}$ from eq.(\ref{QGSpontaneous.35}). However from the analyses \cite{sg11,sg22,sg33,sg44,sg55}, one can easily observe that in the Heisenberg's uncertainty principle scenario, $P_{02}^{\text{SC}}=P_{20}^{\text{SC}}$. If we now consider $\eta_{G_i}=0$, then from eq.(\ref{QGSpontaneous.35}), we find out that $P_{02}^{\text{QG}}=0$. In this same limit, we arrive at the analytical form of the transition probability $P_{20}^{\text{QG}}$ as
\begin{equation}\label{QGSpontaneous.46}
\begin{split}
P_{20}^{\text{QG}}\rvert_{n_{G_i}\rightarrow 0}&=\frac{4\hbar \omega G\pi^3}{L^3}\delta^2(\omega-2\omega_p)\\
&=\frac{\hbar G\omega^4 }{2\pi^3}\delta^2(\omega-2\omega_p)
\end{split}
\end{equation} 
where in the last line of the above equation, we have substituted the analytical expression for $L$. It is important to note that the initial state of the system in this $\eta_{G_i}\rightarrow 0$ limit goes to $|\psi_i\rangle=|0,2\rangle$. The initial state indicates that there are no gravitons initially, however, the detector descends to its ground state by spontaneously emitting a single graviton. This spontaneous emission of gravitons is a result of the quantum gravitational treatment, which results in a degenerate parametric down conversion process. In nature, if in a non-linear optical process a photon is spontaneously generated from two or more excitations, then it is called a parametric down conversion process. Here, we observe the exact gravitational analogue of the parametric down conversion process, which results in the asymmetry between the absorption and emission processes. In order to obtain the analytical form of the transition probability in eq.(\ref{QGSpontaneous.46}), we have assumed that the initial graviton  state is a vacuum state. Hence, it is quite logical to consider that the vacuum fluctuations surrounding the detector results in the spontaneous emission of a single graviton. We shall now look at some important phenomenological aspects of the model in the next subsection.
\subsection{Phenomenological aspects of the model}
The primary aim of resonant bar detectors (or large interferometric detectors inside harmonic trap potentials) is to capture gravitational wave signals. Till now the detected gravitational wave frequencies by the LIGO-VIRGO detectors lie in the range of $\omega\sim10^4$ Hz. For a resonant bar detector (detection of gravitational waves using a resonant bar detector has not yet been achieved), the detection frequency can be as high as $\omega\sim 10^4$ Hz as well. For $\omega=10^4$ Hz, the frequency of the bar detector must be $\omega_p=5\times 10^3$ Hz for the  resonance condition to get satisfied. It is then possible to obtain the analytical form of the spontaneous emission probability from eq.(\ref{QGSpontaneous.46}) to be $P_{20}^\text{QG}\rvert_{n_{G_i}\rightarrow 0}\simeq (10^{-73}~\text{sec}^{-2}) \delta^2(\omega-2\omega_p)$. The typical amplitude for a gravitational wave is $f_h\sim 10^{-21}$ and using the value of $f_h$, the semiclassical de-excitation probability takes the form $P_{20}^\text{SC}=(10^{-33}~\text{sec}^{-2}) \delta^2(\omega-2\omega_p)$. The transition probability for $n_{G_i}+1$ number of gravitons read $P_{20}^{\text{QG}}=(n_{G_i}+1)P_{20}^\text{QG}\rvert_{n_{G_i}\rightarrow 0}$. For a very large value of gravitons $n_{G_i}$ in the initial quantum state, $n_{G_i}+1\simeq n_{G_i}$. It is then possible to compare the semiclassical result with the quantum gravitational result which gives $n_{G_i}\sim 10^{40}$. This implies that $10^{40}$ gravitons carry the total energy $\mathcal{E}$ through the region, where the system is kept in the time interval $t\in[0,\frac{2\pi}{\omega}]$. One can now consider a single bar detector with a finite diameter to be a combination of a finite number of ``one dimensional" resonant bars. The ``one dimensional" bars can have diameter which can have the minimum value equal to the diameter of an atom. For an atom, its radius is approximately equal to $r_{\text{min}}\simeq3\times 10^{-10}$ m. Now for a resonant bar with a diameter of almost one meter, one can consider it to be a combination of $N$ number of  ``one dimensional" bar detectors with its width being of the order of the atomic diameter. It is then possible to estimate the maximum value of $N$ to be $N_{\text{max}}=\frac{\pi (0.5)^2}{\pi r_{\text{min}}^2}\sim 10^{18}$. Hence, a resonant bar detector with its diameter of the order of one meter can be considered as a combination of $N$ number of one dimensional cylinders, where the maximum value of $N$ is of the order of $10^{18}$ with the frequency of oscillation being $\omega_p$\footnote{It is important to note that all of the one dimensional resonant bar detectors have the same length and they are also composed of the same material as has been considered in our current analysis. This consideration results in the identical frequency for each of the one dimensional bar detectors.}. For spontaneous emission of single gravitons, the maximum total transition probability is just $N_{\text{max}}$ multiplied by $P_{20}^\text{QG}\rvert_{n_{G_i}\rightarrow 0}$, which gives the total transition probability to be $P_{{20}_{\text{tot}}}^{\text{QG}}=(10^{-55}~\text{sec}^{-2}) \delta^2(\omega-2\omega_p)$. Another important thing to note that when the resonant condition gets satisfied, the Dirac-delta function becomes infinite and as result significantly enhances the total transition probability in case of spontaneous emission of single gravitons by each of the one dimensional cylindrical detectors. In a realistic scenario, however, the integration limits does not extend from $-\infty$ to $\infty$, and therefore a significant amount of enhancement of the signal will be observed instead of a massive spike. After a spontaneous emission, the initial vacuum graviton state will now consist of a single graviton and a collection of such a large number of graviton from $N$ number of bar detectors, which shall create a combined perturbation. This combined perturbation can be considered to be similar to a gravitational fluorescence-like effect. In such a scenario, in all possible directions, graviton emissions shall occur. This effect will be so small that it will be almost impossible to detect using current experimental set-ups. The only plausible solution to this is to satisfy the resonance condition such that the transition probability becomes so high that even with absolutely minimum evidence of the classical gravitational wave detection, spontaneous emission in the detector shall occur. Another way is to increase the diameter to increase graviton emission during such spontaneous emission phenomena. It shall be also interesting to see if a large harmonic trap potential is created to embed interferometer detectors, which shall be more efficient in detecting spontaneous emission effects indicating a concrete existence of the quanta of linearized gravity.    
\section{Discussion and conclusion}
In this work, we consider a gravitational wave detector placed inside a harmonic trap potential with a fixed frequency, where this set-up mimics the case of a Weber bar detector. We start with the standard Einstein-Hilbert action corresponding to the gravitational fluctuations over the Minkowski spacetime. For the detector part, we consider the two-particle model, where the lighter mass trajectory is described by Fermi-normal coordinates with respect to the world-line of the particle with the heavier mass. Now the entire system is put in a harmonic oscillator potential, which mimics the scenario of a resonant bar detector or an interferometric detector placed inside of a harmonic trap potential. From the action of the model system, we have then written down the Hamiltonian for the model system, and quantize the model by raising the phase space variables corresponding to the gravitational wave as well the detector part to operator status and by implementing appropriate canonical commutation relations between the position operators and their conjugate momentum operator. We have then separated the base part and the interaction part of the Hamiltonian. With the interaction Hamiltonian in hand, we have then mainly focused on the scenario of the excitation and de-excitation of the gravitational wave detector via absorption or emission of gravitons. We have then obtained the transition amplitude and calculated the transition probability for the detector-graviton state to go from some initial tensor product state (where the tensor product is between the energy eigenstate of the detector and the number state of the graviton) to some final tensor product state. We consider the specific case of the detector to go from its ground state to the second excited state and vice-versa. We at first considered the case when the detector gets excited from its ground state and go to its second excited state via the absorption of a graviton. Making use of the energy flux relation for a classical gravitational wave and considering the gravitational wave to be combination of $n_{G_i}$ number of gravitons, we find out that the transition probability for the resonant absorption case in this quantum gravitational set-up is identical to the case of a classical gravitational wave interacting with the resonant bar detector. The distinct feature although comes from the case when we consider the de-excitation scenario, where the detector comes down from the second excited state to its ground state via the emission of a graviton. Now the important observation lies in the fact that if the initial graviton state is in a vacuum state and the initial state of the detector is its second energy eigenstate, even then the detector comes down to its ground state via the emission of a single graviton. This is purely a quantum gravitational phenomena and is absent in a semiclassical scenario. For the semiclassical case, both the transition and absorption probabilities are exactly the same. In this quantum gravity setting, if the initial graviton state is vacuum state, then the resonant absorption probability vanishes whereas the transition probability for the emission case gives a non-zero outcome. This phenomena is indicative of the spontaneous emission process of gravitons. This is the most important outcome of this analysis. If a spontaneous emission of gravitons is observed, which if happens at a very large scale, then it will generate small gravitational fluctuations in all possible directions. If such a phenomena is detected then it will be a stern indication of the existence of the quantum nature of gravity. In an analysis, where the gravitational perturbations are treated classically, it is not possible to obtain a spontaneous emission process. It is possible to increase the transition probability in the spontaneous emission of gravitons scenario via considering a resonant bar detector with a large enough diameter while the resonance condition is perfectly satisfied. Instead of a resonant bar detector, it is more prudent to use an interferometric detector placed inside of a harmonic trap potential as they are way more accurate in picking up gravitational wave signals. However, making such a large harmonic trap is a bit challenging. On the other hand, very small fluctuations in the Earth's gravitational field and other distinguished sources can generate small perturbations in the measurement, which needs to be properly taken care of while building a gravitational wave detector capable of picking up small background fluctuations generated by a collection of gravitons. In the next two chapters, we propose a Bose-Einstein condensate based graviton detector which may be able to pick up graviton signatures in very near future. 
\chapter{Bose-Einstein condensate as a probe to detect quantum nature of gravity}\label{C.6.OTM}
\begin{figure}[ht!]
\begin{center}
\includegraphics[scale=0.88]{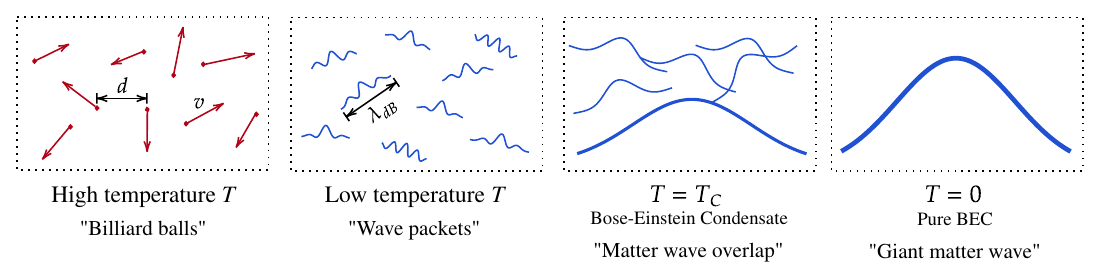}
\end{center}
\caption{A pictorial representation of the formation of a Bose-Einstein condensate.\label{BEC_OTM}}
\end{figure}

Prof. Satyendranath Bose in his seminal work \cite{SBosePlanck} derived the Planck's radiation law which led to the invention of the Bose statistics. Albert Einstein applied this new statistics into matter systems which led to the analytical models of an ideal Bose gas governed by the Bose statistics \cite{BoseEinsteinErste,BoseEinsteinZweite}.  Consider a system of bosonic particles inside a box of a fixed volume. Now at a very high temperature the gas molecules (atoms) behave like bullets or billiards ball executing an absolutely random motion. Now if the temperature of the system is lowered then the gas molecules start to lose energy and as a result the de Broglie wavelength of the particles starts to increase in length the matter-like behaviour starts dominating with a decreasing temperature. As has been shown in Fig.(\ref{BEC_OTM}), if the temperature is lowered to such a value ($T=T_C$) that the de Broglie wavelength is of the order of the inter-atomic separation, the matter-waves corresponding to each of the bosons start to superpose with each other. This phenomena is called a Bose-Einstein condensation and this new state of matter is termed as a Bose-Einstein condensate. At $T=0$, all the bosons occupy the ground state creating a pure Bose-Einstein condensate.
The Bose-Einstein condensate was first experimentally created in a gas of Rubidium atoms in 1995 \cite{1Nobel2001}, and later that year in a gas of Sodium atoms \cite{2Nobel2001}. In \cite{SabinBruschiAhmadiFuentes}, a quasi (1+1)-dimensional Bose-Einstein condensate at zero temperature has been considered, where the boundary conditions were considered to be fluctuating in nature. They have proposed a gravitational wave detector using there model of the Bose-Einstein condensate. In \cite{Schutzhold}, the interaction between a non-relativistic Bose-Einstein condensate with a classical gravitational wave has been considered. Later the idea proposed in \cite{SabinBruschiAhmadiFuentes} was extended to a (3+1)-dimensional model of the relativistic Bose-Einstein condensate where the gravitational wave has a template similar to a gravitational wave burst \cite{MatthewMannAffshordi,MatthewMannAffshordi2}. In \cite{MatthewMannAffshordi,MatthewMannAffshordi2}, the quantum Fisher information between two nearby single mode phonon states have been calculated where the quantum Fisher information is a quantifier of the information content of the gravitational wave in the Bose-Einstein condensate.  Very recently, in \cite{HartleyKadingHowlFuentes}, an experimental dark energy detection model using a Bose-Einstein condensate has also been proposed. The primary aim of any such phenomenological approach is to pinpoint the most fundamental aspects of a rigorous theoretical model and then unearth certain important experimental and application based accepts of that fundamental property. This chapter investigates the profoundness of the idea that an exotic state of matter like a Bose-Einstein condensate will be able to pick up signatures of quantum gravity better than other existing models. This chapter is organized as follows. 

\noindent In this chapter\footnote{This chapter is based on the publication S. Sen and S. Gangopadhyay, ``\textit{Probing the quantum nature of gravity using a Bose-Einstein condensate}", \href{https://link.aps.org/doi/10.1103/PhysRevD.110.026014}{Phys. Rev. D 110 (2024) 026014}.}, we shall start with the basics of quantum metrology where we shall discuss some important quantum metrological techniques required for understanding the phenomenological aspects of a relativistic Bose-Einstein condensate in a flat background with quantum gravitational fluctuations on it. . We then move on to analytical understanding of a Bose-Einstein condensation while treated in non-relativistic setting. Next, we shall discuss the relativistic BEC-graviton interaction model and will analytically calculate the quantum gravity signatures in the BEC and discuss about its phenomenological aspects. Finally, making use of the quantum metrological techniques we shall investigate the response of Bose-Einstein condensate to noise fluctuations generated due to the quantum nature of linearized gravity and compare it with the sensitivity curve of the upcoming space based gravitational wave observatory.

\section{Quantum metrology}\label{6.1}
In this section, we shall primarily investigate the quantum metrological techniques required for our current analysis. The required estimation parameter for a gravitational wave detection (in our case gravitons) is its amplitude. To correctly measure the amplitude one needs to estimate the error or the standard deviation in the measurement of this estimation parameter. In order to truly estimate the standard deviation, we start with the derivation of the Cram\'{e}r-Rao bound involving the classical Fisher information\footnote{For a concise discussion see S. L. Braunstein and C. M. Caves, ``\textit{Statistical Distance and the Geometry of Quantum States}", \href{https://link.aps.org/doi/10.1103/PhysRevLett.72.3439}{Phys. Rev. Lett. 72 (1994) 3439}.}.
\subsection{Cram\'{e}r-Rao bound and the Fisher information}\label{6.1.1}
The primary aim of this subsection is to distinguish two nearby quantum states based on the estimation of a parameter. One can start by considering the space of density operators where $\hat{\rho}(\vartheta)$ gives a curve on it with $\vartheta$ being the parameter. Here we follow the derivation given in \cite{BraunsteinCaves}. One can now consider a space made of density operator which depends on the parameter $\vartheta$. Hence, distinguishing between two neighbouring density operators is dependent solely on the precision of determining the parameter $\vartheta$. Here, one  needs to use generalized measurements \cite{GeneralizedMeasurement1,GeneralizedMeasurement2} which obeys the ground rules of quantum mechanics.

Consider a set of non-negative, Hermitian operators $\hat{\mathcal{M}}(\zeta)$ such that
\begin{equation}\label{CramerRao.1}
\int d\zeta \hat{\mathcal{M}}(\zeta)=\hat{\mathbb{1}}
\end{equation}
with $\zeta$ labelling the results of the measurement, then such a measurement is also termed as a `generalized measurement'. For a result (or outcome due to the measurement) $\zeta$, the probability density provided the parameter value is $\vartheta$ reads
\begin{equation}\label{CramerRao.2}
p(\zeta|\vartheta)=\text{tr}\left[\hat{\mathcal{M}}(\zeta)\hat{\rho}(\vartheta)\right]~.
\end{equation}
For $\mathfrak{N}$ number of independent measurements, consider that the outcomes are $\zeta_1,\zeta_2,\cdots, \zeta_{\mathfrak{N}}$. It is therefore possible to estimate the value of the parameter $\vartheta$ via an estimation parameter dependent on this $\mathfrak{N}$ number of independent parameters defined as $\vartheta_{\text{E}}\equiv\vartheta_{\text{E}}(\zeta_1,\zeta_2,\cdots,\zeta_{\mathfrak{N}})$. The deviation of the parameter $\vartheta$ from the estimated parameter $\vartheta_{\text{E}}$ is given as
\begin{equation}\label{CramerRao.3}
\delta\vartheta\equiv\frac{\vartheta_{\text{E}}}{\left|\frac{d\langle\vartheta_{\text{E}}\rangle_\vartheta }{d\vartheta}\right|}-\vartheta
\end{equation}
where the $\frac{d\langle\vartheta_{\text{E}}\rangle_\vartheta }{d\vartheta}$ term is introduced to equate the unit of the estimation parameter with the parameter $\vartheta$. Then the estimation parameter $\vartheta_\text{E}$ can be recast as
\begin{equation}\label{CramerRao.4}
\vartheta_{\text{E}}=(\vartheta+\delta\vartheta)\left|\frac{d\langle\vartheta_{\text{E}}\rangle_\vartheta }{d\vartheta}\right|~.
\end{equation}
In order to obtain the Cram\'{e}r-Rao bound, we start by writing down a standard identity as
\begin{equation}\label{CramerRao.5}
\begin{split}
\int d\zeta_1\cdots d\zeta_\mathfrak{N}~ p(\zeta_1|\vartheta)\cdots p(\zeta_\mathfrak{N}|\vartheta)\left(\vartheta_{\text{E}}-\langle \vartheta_{\text{E}}\rangle_\vartheta\right)=0~.
\end{split}
\end{equation}
Defining a new quantity $\Delta\vartheta_{\text{E}}\equiv \vartheta_{\text{E}}-\langle \vartheta_{\text{E}}\rangle_\vartheta$ and differentiating both sides of the above equation with respect to the parameter $\vartheta$, we arrive at the following relation
\begin{equation}\label{CramerRao.6}
\begin{split}
&\int d\zeta_1 \cdots d\zeta_{\mathfrak{N}}~ p(\zeta_1|\vartheta)\cdots p(\zeta_\mathfrak{N}|\vartheta)\left(\frac{1}{p(\zeta_1|\vartheta)}\frac{\partial p(\zeta_1|\vartheta)}{\partial\vartheta}+\cdots+\frac{1}{p(\zeta_\mathfrak{N}|\vartheta)}\frac{\partial p(\zeta_\mathfrak{N}|\vartheta)}{\partial\vartheta}\right)\Delta\vartheta_{\text{E}}\\&-\int d\zeta_1 \cdots d\zeta_{\mathfrak{N}}~ p(\zeta_1|\vartheta)\cdots p(\zeta_\mathfrak{N}|\vartheta)\frac{\partial\langle \vartheta_{\text{E}}\rangle_\vartheta }{\partial\vartheta}=0~.
\end{split}
\end{equation}
Defining $\mu_\vartheta(d\zeta)\equiv d\zeta_1 \cdots d\zeta_{\mathfrak{N}}~ p(\zeta_1|\vartheta)\cdots p(\zeta_\mathfrak{N}|\vartheta)$, we can write the above equation in a much more simplified form as
\begin{equation}\label{CramerRao.7}
\int \mu_\vartheta(d\zeta) \Delta\vartheta_{\text{E}}\sum_{k=1}^\mathfrak{N}\frac{\partial\ln\left[p(\zeta_k|\vartheta)\right]}{\partial \vartheta}=\frac{\partial\langle \vartheta_{\text{E}}\rangle_\vartheta }{\partial\vartheta}
\end{equation}
where we have made use of the relation $\int \mu_\vartheta(d\zeta)=1$. In order to obtain the Cram\'{e}r-Rao bound, we need to make use of the Cauchy-Schwarz inequality. Now, we need to make use of the H\"{o}lder's inequality \cite{Holder} (although initially found in 1888 by Leonard James Rogers) and from there obtain the Cram\'{e}r-Rao bound. 
Consider two vector-valued functions $\mathcal{F}=\{\mathcal{F}_1,\mathcal{F}_2,\cdots,\mathcal{F}_N\}$ and $\mathcal{G}=\{\mathcal{G}_1,\mathcal{G}_2,\cdots,\mathcal{G}_N\}$ on a measurable subset $K$ of $\mathbb{R}^N$, then the inequality has the form given by
\begin{equation}\label{CramerRao.8}
\begin{split}
\int_K \mu (d\xi)\sum_{j=1}^N|\mathcal{F}_j(\xi) \mathcal{G}_j(\xi)|\leq \left[\int_K \mu(d\xi)\sum_{i=1}^N |\mathcal{F}_i(\xi)|^p \right]^{\frac{1}{p}}\left[\int_K \mu(d\xi')\sum_{j=1}^N |\mathcal{G}_j(\xi')|^q \right]^{\frac{1}{q}}
\end{split}
\end{equation}
where $p,q\in [1,\infty)$. For the derivation of the Cram\'{e}r-Rao bound one needs to set $p=q=2$ in the above equation. Making use of eq.(\ref{CramerRao.8}), one can write down an inequality for the left hand side of eq.(\ref{CramerRao.7}) as
\begin{equation}\label{CramerRao.9}
\begin{split}
\int \mu_\vartheta (d\zeta)\Delta \vartheta_{\text{E}}\sum_{k=1}^{\mathfrak{N}}\frac{\partial\ln\left[p(\zeta_k|\vartheta)\right]}{\partial \vartheta}\leq \left[\int \mu_\vartheta(d\zeta)\sum_{k=1}^\mathfrak{N}\left(\frac{\partial\ln\left[p(\zeta_k|\vartheta)\right]}{\partial \vartheta}\right)^2\right]^{\frac{1}{2}}\left[\int \mu_\vartheta(d\zeta)(\Delta\vartheta_{\text{E}})^2\right]^{\frac{1}{2}}~.
\end{split}
\end{equation}
One can simplify the first term inside of the parenthesis in the right hand side of the above equation in the following way
\begin{equation}\label{CramerRao.10}
\begin{split}
\int \mu_\vartheta(d\zeta)\sum_{k=1}^\mathfrak{N}\left[\frac{\partial\ln\left[p(\zeta_k|\vartheta)\right]}{\partial \vartheta}\right]^2&=\int \mu_\vartheta(d\zeta)\left[\left[\frac{1}{p(\zeta_1|\vartheta)}\frac{\partial p(\zeta_1|\vartheta)}{\partial\vartheta}\right]^2\cdots+\left[\frac{1}{p(\zeta_\mathfrak{N}|\vartheta)}\frac{\partial p(\zeta_\mathfrak{N}|\vartheta)}{\partial\vartheta}\right]^2\right]\\
&=\sum_{k=1}^{\mathfrak{N}}\int d\zeta_k\frac{1}{p(\zeta_k|\vartheta)}\left(\frac{\partial p(\zeta_k|\vartheta)}{\partial\vartheta}\right)^2\\
&=\mathfrak{N}\mathcal{I}(\vartheta)
\end{split}
\end{equation}
where the classical Fisher information $\mathcal{I}(\vartheta)$ is defined as
\begin{equation}\label{CramerRao.11}
\mathcal{I}(\vartheta)\equiv\int d\zeta \frac{1}{p(\zeta|\vartheta)}\left(\frac{\partial p(\zeta|\vartheta)}{\partial \vartheta}\right)^2~.
\end{equation}
Making use of eq.(\ref{CramerRao.7}) and taking square of the both sides of eq.(\ref{CramerRao.9}), we can write down the form of H\"{o}lder's inequality as
\begin{equation}\label{CramerRao.12}
\begin{split}
\left(\frac{\partial\langle \vartheta_{\text{E}}\rangle_\vartheta }{\partial\vartheta}\right)^2\leq \mathfrak{N}\mathcal{I}(\vartheta)\langle \left(\Delta\vartheta_\text{E}\right)^2\rangle_\vartheta~.
\end{split}
\end{equation} 
Using the expression of $\Delta\vartheta_\text{E}$ and the form of $\vartheta_\text{E}$ from eq.(\ref{CramerRao.4}), we can write down the above inequality as
\begin{equation}\label{CramerRao.13}
\begin{split}
\left(\frac{\partial\langle \vartheta_{\text{E}}\rangle_\vartheta }{\partial\vartheta}\right)^2\leq& \mathfrak{N}\mathcal{I}(\vartheta)\left(\langle\vartheta_{\text{E}}^2\rangle_\vartheta-\langle\vartheta_{\text{E}}\rangle_\vartheta^2\right)\\
=&\mathfrak{N}\mathcal{I}(\vartheta)\left(\langle (\Delta \vartheta)^2\rangle_\vartheta+\langle (\Delta \delta\vartheta)^2\rangle_\vartheta
+2\langle\vartheta\delta\vartheta\rangle_\vartheta
-2\langle\vartheta\rangle_\vartheta\langle\delta\vartheta\rangle_\vartheta
\right)\left(\frac{\partial\langle \vartheta_{\text{E}}\rangle_\vartheta }{\partial\vartheta}\right)^2~.
\end{split}
\end{equation}
In order to obtain the final form of the Cram\'{e}r-Rao bound, we take the $\delta\vartheta\rightarrow 0$ limit which indicates that the value of the parameter approaches the value of the estimation parameter. Hence, in the $\delta\vartheta\rightarrow 0$ limit, we can recast the inequality in eq.(\ref{CramerRao.13}) as
\begin{equation}\label{CramerRao.14}
\left\langle \left(\Delta \vartheta\right)^2\right\rangle_\vartheta\geq \frac{1}{\mathfrak{N}\mathcal{I}(\vartheta)}
\end{equation}
which is the well known form of the Cram\'{e}r-Rao bound\footnote{Our final form of the bound slightly differs from the structure obtained in \cite{BraunsteinCaves} where the left hand side of the inequality reads $\langle (\delta\vartheta)^2\rangle_\vartheta$. Here instead we obtain the square of the standard deviation in $\vartheta$ as the left hand side of the inequality obtained in eq.(\ref{CramerRao.14}).}.
Now, if the parameter $\vartheta$ in consideration belongs to a quantum mechanical system then one needs to modify the Cram\'{e}r-Rao bound by replacing the classical Fisher information by the quantum Fisher information. The quantum Fisher information is defined as \cite{BraunsteinCaves}
\begin{equation}\label{CramerRao.15}
\mathcal{H}(\vartheta)=\mathop{\mathrm{max}}\limits_{\{\hat{\mathcal{M}}(\zeta)\}}\mathcal{I}(\vartheta)~.
\end{equation}
It is important to note that all of the measurements is taken into consideration while writing eq.(\ref{CramerRao.15}). One can then write down the new Cram\'{e}r-Rao bound for a quantum mechanical system as
\begin{equation}\label{CramerRao.16}
 \left\langle \left(\Delta \vartheta\right)^2\right\rangle_\vartheta\geq \frac{1}{\mathfrak{N}\mathcal{I}(\vartheta)}\geq\frac{1}{\mathfrak{N}\mathcal{H}(\vartheta)}~.
\end{equation}
\subsection{Fidelity and quantum Fisher information}\label{6.1.2}
From the discussion in the previous subsection, we have found out that the uncertainty in a parameter is related to the quantum Fisher information by the inequality in eq.(\ref{CramerRao.16}). As has already been discussed, that the space made of density parameters is dependent upon this parameter $\vartheta$. The first task therefore is to wrote down a distance between two quantum states or density operators.
\subsubsection{Fidelity} 
The measure of distance between two density operators $\hat{\rho}_1$ and $\hat{\rho}_2$ is given by an entity called Fidelity which is defined as
\begin{equation}\label{Fidelity.1}
\mathcal{F}_{\text{D}}(\rho_1,\rho_2)=\text{tr}\left[\sqrt{\sqrt{\hat{\rho_1}}\hat{\rho}_2\sqrt{\hat{\rho_1}}}\right]
\end{equation}
This Fidelity is indeed symmetric in its input is considered to be a good measure of distance between two quantum states \cite{NielsenChuang}. Fidelity is a measure of overlap between the density operators $\hat{\rho_1}$ and $\hat{\rho}_2$ and thus indicates how much one state can influence other state. One can also make use of the square of the quantity in the right hand side of eq.(\ref{Fidelity.1}) and a term it as the new Fidelity as
\begin{equation}\label{Fidelity.2}
\mathcal{F}(\rho_1,\rho_2)=\left(\text{tr}\sqrt{\sqrt{\hat{\rho_1}}\hat{\rho}_2\sqrt{\hat{\rho_1}}}\right)^2~.
\end{equation}
Here, both $\mathcal{F}_{\text{D}}(\rho_1,\rho_2)$ and $\mathcal{F}(\rho_1,\rho_2)$ are good measures of distance between the two density operators $\hat{\rho}_1$ and $\hat{\rho}_2$ but we shall make use of the definition in eq.(\ref{Fidelity.2}) for our current analysis. Our next aim is to express the quantum Fisher information in terms of the fidelity between two nearby quantum states.
\begin{figure}
\begin{center}
\includegraphics[scale=0.3]{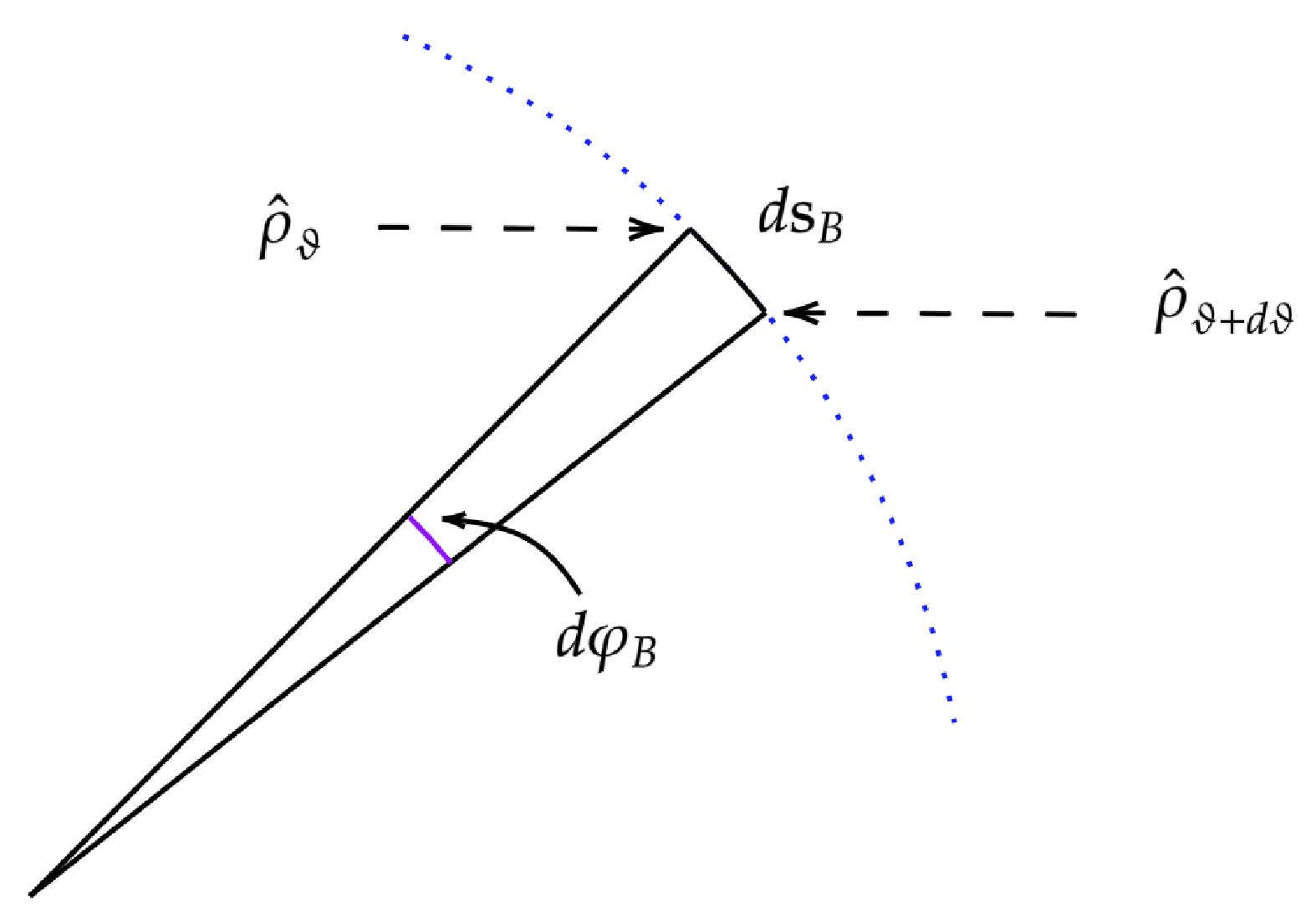}
\caption{Two nearby density matrices are placed on the perimeter of an unit circle, where $d\varphi_B$ denotes the Bures angle and $ds_B$ denotes the geometric distance also known as the Bures distance between the two nearby quantum states $\hat{\rho}_\vartheta$ and $\hat{\rho}_{\vartheta+d\vartheta}$.\label{Fig.Bures}}
\end{center}
\end{figure}
\subsubsection{Quantum Fisher information in terms of fidelity between two quantum states}
To express the quantum Fisher information in terms of the Fidelity we need to look at the geometric representation of the space of density operators as points on the perimeter of a unit circle. As can be seen from Fig.(\ref{Fig.Bures}), the two nearby density operators are separated by a distance $ds$ on the unit circle and the two rays corresponding to the two densioty operators make an angle $d\varphi_B$ at the center of the unit circle. This angle $d\varphi_B$ is termed as the Bures angle. The Bures angle is related to the fidelity between the operators $\hat{\rho}_\vartheta$ and $\hat{\rho}_{\vartheta+d\vartheta}$ by the relation \cite{BengtssonZyczkowski}
\begin{equation}\label{Bures.1}
\cos d\varphi_B=\sqrt{\mathcal{F}(\rho_\vartheta,\rho_{\vartheta+d\vartheta})}~.
\end{equation}
Now from Fig.(\ref{Fig.Bures}), it is easy to observe that $d\varphi_B$ is very small as the density operators are very close to each other and therefore up to $\mathcal{O}(d\varphi_B^2)$, one can expand the left hand side of the above equation as \footnote{For a detailed calculation obtained in the above equation one should see \cite{Hubner}.}
\begin{equation}\label{Bures.2}
\begin{split}
\cos d\varphi_B&\simeq 1-\frac{d\varphi_B^2}{2}+\mathcal{O}(d\varphi_B^4)=\sqrt{\mathcal{F}(\rho_\vartheta,\rho_{\vartheta+d\vartheta})}\\
\implies d\varphi_B&=\sqrt{2-2\sqrt{\mathcal{F}(\rho_\vartheta,\rho_{\vartheta+d\vartheta})}}
\end{split}
\end{equation}
where for writing down the last line in the above equation all of the higher order contributions have been dropped off. Now, the Bures metric $ds_B^2$ between the two nearby density operators is defined by the relation \cite{Bures_Metric1,Bures_Metric2,Bures_Metric3,Bures_Metric4}
\begin{equation}\label{Bures.3}
ds_B^2=\frac{1}{4}\mathcal{H}(\vartheta)d\vartheta^2
\end{equation}
where $\mathcal{H}(\vartheta)$ denotes the quantum Fisher information defined in eq.(\ref{CramerRao.16}). From Fig.(\ref{Fig.Bures}), we can write down $ds_B=R_C\sin d\varphi_B$ where the radius of the unit circle $R_C=1$. As $d\varphi_B$ is very small, up to first order in $d\varphi_B$ we can write down the following relation
\begin{equation}\label{Bures.4}
\begin{split}
ds_B&=\sin d\varphi_B\simeq d\varphi_B-\mathcal{O}(d\varphi_B^3)\\
\implies d\varphi_B&=\frac{1}{2}\sqrt{\mathcal{H}(\vartheta)}d\vartheta=ds_B
\end{split}
\end{equation}
where in the last line of the above equation we have truncated all of the higher order terms and made use of eq.(\ref{Bures.3}). Hence, comparing eq.(\ref{Bures.4}) with eq.(\ref{Bures.2}), we arrive at the following relation \cite{BraunsteinCaves,Bures_Metric3}
\begin{equation}\label{Bures.5}
\mathcal{H}(\vartheta)=\frac{8\left(1-\sqrt{\mathcal{F}(\rho_\vartheta,\rho_{\vartheta+d\vartheta})}\right)}{d\vartheta^2}~.
\end{equation}
The above equation gives the form of the quantum Fisher information in terms of Fidelity between two nearby quantum states dependent on the parameter $\vartheta$.
\subsection{Covariance matrix approach for Gaussian states of an \textit{n}-mode bosonic system}\label{6.1.3}
In this subsection, we shall briefly describe the covariance matrix formalism and finally calculate the covariance matrix corresponding to as single mode of an $n$-mode bosonic system. This formalism has been used to investigate entanglement generation and sharing in quantum field theoretic models \cite{CovarianceMatrix1,CovarianceMatrix2,CovarianceMatrix3}.

\noindent We start by defining the generalized phase space operators corresponding to the $k$-th mode of an $n$-mode bosonic system as
\begin{equation}\label{Covariance.1}
\hat{x}_{k}^B\equiv\sqrt{\frac{\hbar}{2 m_B\omega_B}}\left(\hat{a}^B_{k}+{\hat{a}^B_{k}}^\dagger\right)~,~~\hat{p}^B_k\equiv i\sqrt{\frac{m_B\hbar\omega_B}{2}}\left({\hat{a}^B_{k}}^\dagger-\hat{a}^B_{k}\right)
\end{equation}
with $m_B$ denoting the mass of the field and $\omega_B$ denoting the phonon frequency. In the above relation the creation ($\hat{a}^B_{j}$) and annihilation operators (${\hat{a}^B_{k}}^\dagger$) are defined as
\begin{equation}\label{Covariance.2}
\hat{a}^B_k|p_k^B\rangle=\sqrt{p_k^B}|p_k^B-1\rangle~,~~{\hat{a}^B_k}^\dagger |p_k^B\rangle=\sqrt{p^B_k+1}|p^B_k+1\rangle
\end{equation}
where $p^B_k$ denotes the number of bosonic particles in the $k$-th energy state and the vacuum state corresponding to the $k$-th energy mode is defined as $\hat{a}^B_k|0^B_k\rangle\equiv 0$. Now the creation and annihilation operators satisfy the following commutation relations
\begin{equation}\label{Covariance.3}
[\hat{a}^B_{j},{\hat{a}^B_{k}}^\dagger]=\delta_{jk}~,~~\left[\hat{a}^B_{j},{\hat{a}^B_{k}}\right]=[{\hat{a}^B_{j}}^\dagger,{\hat{a}^B_{k}}^\dagger]=0~.
\end{equation}
Using the above commutation relations between the ladder operators, one can write down the commutation relations between the generalized position and momentum operators in eq.(\ref{Covariance.1}) as
\begin{equation}\label{Covariance.4}
[\hat{x}^B_j,\hat{p}^B_k]=i\hbar\delta_{jk}~,~~[\hat{x}^B_j,\hat{x}^B_k]=[\hat{p}^B_j,\hat{p}^B_k]=0
\end{equation}
where $j,k=1,\ldots,n$.
For an $n$ mode bosonic system, it is then possible to write down a column matrix of the form
\begin{equation}\label{Covariance.5}
\mathbb{r}_{2n}=\begin{pmatrix}
\sqrt{\frac{m\omega_B}{\hbar}}\hat{x}_1^B\\
\frac{1}{\sqrt{m\hbar\omega_B}}\hat{p}_1^B\\
\vdots\\
\sqrt{\frac{m\omega_B}{\hbar}}\hat{x}_n^B\\
\frac{1}{\sqrt{m\hbar\omega_B}}\hat{p}_n^B
\end{pmatrix}
\end{equation}
where $2n$ in the suffix of $\mathbb{r}$ denotes the number of elements of  the column matrix.
One can define a commutation relation between the new operator valued matrix $\mathbb{r}$ and its transpose as 
\begin{equation}\label{Covariance.6}
\begin{split}
[\mathbb{r}_{2n},\mathbb{r}^T_{2n}]&\equiv \mathbb{r}_{2n}\mathbb{r}^T_{2n}-\left(\mathbb{r}_{2n}\mathbb{r}^T_{2n}\right)^T\\
&=\mathop{\oplus}\limits_{k=1}^ni\sigma_2
\end{split}
\end{equation}
where the commutation relations in eq.(\ref{Covariance.4}) have been used and $\sigma_2$ denotes the second Pauli spin matrix. For a single mode bosonic system one then obtains the result $[\mathbb{r}_{2},\mathbb{r}^T_{2}]=i\sigma_2$. Denoting the column matrix $\mathbb{r}_{2n}$ just by $\mathbb{r}$, it is possible to write down the covariance matrix corresponding to an $n$-mode bosonic system with density matrix $\hat{\rho}$ as
\begin{equation}\label{Covariance.7}
\Sigma_{jk}=\frac{1}{2}\langle\{\mathbb{r}_j,\mathbb{r}_k\}\rangle-\langle\mathbb{r}_j\rangle\langle \mathbb{r}_k\rangle
\end{equation}
where the expectation is taken with respect to the state of the system (here the density matrix is concerned). We shall now try to obtain the analytical form of the density matrix of a single mode ($j$-th mode) of an $n$-mode bosonic system in thermal equilibrium. The density matrix reads
\begin{equation}\label{Covariance.8}
\hat{\rho}_j=\frac{e^{-\beta\hat{H}_j}}{\mathop{\text{tr}}\left[e^{-\beta \hat H}\right]}
\end{equation}  
where $\beta=\frac{1}{k_B T}$ with $T$ being the temperature and $k_B$ being the Boltzmann constant and $\hat{H}_j=\hbar\omega_B\hat{a}_j^\dagger\hat{a}_j$ with $\omega_B$ being the frequency of each of the phonon modes. Substituting the form of the Hamiltonian in the above equation, the density matrix can simply be expressed as
\begin{equation}\label{Covariance.9}
\begin{split}
\hat{\rho}_j&=\frac{\sum\limits_{n_j^B=0}^\infty |n_j^B\rangle\langle n_j^B|e^{-\beta{\hat{a}_j^B}^\dagger\hat{a}_j^B}}{\sum\limits_{l_j^B=0}^\infty\langle l_j^B|e^{-\beta{\hat{a}_j^B}^\dagger\hat{a}_j^B }|l_j^B\rangle}\\
&=(1-e^{-\beta})\sum\limits_{n_j^B=0}^\infty |n_j^B\rangle\langle n_j^B|e^{-\beta n_k^B}~.
\end{split}
\end{equation}
Defining a new temperature-dependent constant parameter, $\mathcal{N}(T)\equiv\frac{1}{\exp[\beta(T)\hbar\omega_B]-1}$, one can recast the one mode form of the density matrix as
\begin{equation}\label{Covariance.10}
\hat{\rho}_j=\frac{1}{\mathcal{N}+1}\sum\limits_{n_j^B=0}^\infty \left(\frac{\mathcal{N}}{\mathcal{N}+1}\right)^{n_j^B}|n_j^B\rangle\langle n_j^B|~.
\end{equation}
The above expression for the single-mode density matrix of an $n$-mode bosonic system will be necessary when calculating the covariance matrix corresponding to the Bose-Einstein condensate. From the form of the density matrix in the above equation, it is easy to check that 
\begin{equation}\label{Covariance.11}
\langle \mathbb{r}_j\rangle=\text{tr}\left[\mathbb{r}_j\hat{\rho}_j\right]=0,~\langle \mathbb{r}_j^T\rangle=0~.
\end{equation}
Hence, the form of the covariance matrix with all of its elements for a single mode reads 
\begin{equation}\label{Covariance.12}
\Sigma=\frac{1}{2}\langle \{\mathbb{r},\mathbb{r}^T\}\rangle=\frac{1}{2}\text{tr}\left[\{\mathbb{r},\mathbb{r}^T\}\hat{\rho}\right]~.
\end{equation}
For the $j$-th sector of the covariance matrix, one needs to consider only the $\mathbb{r}_j$ column matrix with $\mathbb{r}_j=\begin{pmatrix}
\sqrt{\frac{m_B \omega_B}{\hbar}}\hat{x}_j^B\\
\frac{1}{\sqrt{m_B\hbar\omega_B}}\hat{p}_j^B
\end{pmatrix}$ and its transpose matrix $\mathbb{r}_j^T$. For the $j$-th sector the anti-commutator of $\mathbb{r}_j$ and $\mathbb{r}_j^T$ reads
\begin{equation}\label{Covariance.13}
\begin{split}
\{\mathbb{r}_j,\mathbb{r}_j^T\}&\equiv\mathbb{r}_j\mathbb{r}_j^T+\left(\mathbb{r}_j\mathbb{r}_j^T\right)^T=\begin{pmatrix}
\frac{2m_B\omega_B}{\hbar}{\hat{x}_j^B}^2&&\frac{1}{\hbar}\left(\hat{x}_j^B\hat{p}_j^B+\hat{p}_j^B\hat{x}_j^B\right)\\
\frac{1}{\hbar}\left(\hat{x}_j^B\hat{p}_j^B+\hat{p}_j^B\hat{x}_j^B\right)&&\frac{2}{m_B\hbar\omega_B}{\hat{p}_j^B}^2
\end{pmatrix}~.
\end{split}
\end{equation}
Using the analytical form of the anti-commutator from the above equation and the mode expansion of the position and momentum operators from eq.(\ref{Covariance.1}), it is now possible to obtain the elements covariance matrix corresponding to the $j$-th bosonic mode from eq.(\ref{Covariance.12}). We start by computing the $\{0,0\}$ component of the covariance matrix.
\begin{equation}\label{Covariance.14}
\begin{split}
\Sigma_{j}^{00}&=\frac{1}{2}\text{tr}\left[\frac{2m_B\omega_B}{\hbar}{\hat{x}_j^B}^2\hat{\rho}_j\right]\\
&=\frac{1}{2}\text{tr}\left[\left({\hat{a}_j^B}^2+{\hat{a}_j^{B\dagger}}+2\hat{a}_j^{B\dagger}\hat{a}_j^B+1\right)\hat{\rho}_j\right]\\
&=\frac{1}{2}\sum\limits_{n_j^B=0}^\infty\langle n_j^B|\left(2\hat{a}_j^{B\dagger}\hat{a}_j^B+1\right)\frac{1}{\mathcal{N}+1}\sum\limits_{l_j^B=0}^\infty \left(\frac{\mathcal{N}}{\mathcal{N}+1}\right)^{l_j^B}|l_j^B\rangle\langle l_j^B|n_j^B\rangle~.
\end{split}
\end{equation}
Now, as $\frac{\mathcal{N}}{\mathcal{N}+1}$ is always less than one, it is possible to execute the infinite sum in the last line of the above equation and then we can recast the $\{0,0\}$ element of the density matrix as
\begin{equation}\label{Covariance.15}
\Sigma_j^{00}=\frac{2\mathcal{N}+1}{2}~.
\end{equation}
Similarly one can obtain the other components of the covariance matrix as
\begin{equation}\label{Covariance.16}
\Sigma_{j}^{01}=\Sigma_{j}^{10}=0, \text{ and } \Sigma_j^{11}=\frac{2\mathcal{N}+1}{2}~.
\end{equation}
Making use of eq.(s)(\ref{Covariance.15},\ref{Covariance.16}), we can write down the covariance matrix corresponding to the $j$-th sector of an $n$-mode bosonic system as
\begin{equation}\label{Covariance.17}
\Sigma_j(T)=\frac{2\mathcal{N}(T)+1}{2}\begin{pmatrix}
1&&0\\
0&&1
\end{pmatrix}~.
\end{equation}
Now for absolute zero temperature, $T=0$, $\beta(T)\rightarrow\infty$ and as a result $\mathcal{N}(T)\rightarrow 0$. At absolute zero temperature an ideal Bose-gas will create a giant matter wave packet which is also known as pure Bose-Einstein condensate. As a result at absolute zero temperature, we obtain the covariance matrix corresponding to a single mode of a Bose-Einstein condensate as
\begin{equation}\label{Covariance.18}
\Sigma_{\text{BEC}}(0)=\frac{1}{2}\begin{pmatrix}
1&&0\\
0&&1
\end{pmatrix}
\end{equation}
where the $j$ subscript is omitted.

\section{Nonrelativistic Bose-Einstein condensate}
     In this section, we shall be more inclined towards understanding the low-temperature behaviour of non-relativistic gas of bosons in the presence and absence of inter-atomic interactions. At first, we shall start by considering an ideal gas system.
     \subsection{Ideal Bose gas}
We start by considering an ideal gas where the inter-molecular interaction is negligible. Considering the $j$-th boson to have an energy of $\mathcal{E}_j$, the number of particles in the ideal Bose gas can be expressed as \cite{PathriaBeale}
\begin{equation}\label{Bose.1}
N_B\equiv\sum\limits_{j}\frac{1}{\exp\left[\frac{\mathcal{E}_j-\mu}{k_BT}\right]-1}
\end{equation}
where the above equation is also known as the Bose-Einstein distribution function. In the above equation $\mu$ gives the chemical potential for the system of particles. For $n$ particles are kept inside a volume $V_\Sigma$ in three spatial dimensions, then the volume of the phase space available to the members of the system is given by the expression
\begin{equation}\label{Bose.2}
\mathcal{V}^n_\omega=\idotsint d^{3n}\mathcal{q}~d^{3n}\mathcal{p}
\end{equation}
and the expression for the number of microstates $N_{\Sigma}(p_M)$ which are available to a single free particle where $\mathcal{p}_M\leq p$ reads
\begin{equation}\label{Bose.3}
\begin{split}
N_{\Sigma}(p_M)&\simeq \frac{1}{h^3}\idotsint\limits_{\mathcal{p}\leq p}d^3\mathcal{q}d^3\mathcal{p}=\frac{1}{h^3}\frac{4\pi p_M^3 V_\Sigma }{3}~.
\end{split}
\end{equation}
with $m$ being the mass and $E$ being the energy of the particle.
One can then easily obtain from eq.(\ref{Bose.3}), the number of microstates lying in the momentum range $\{p,p+dp\}$ as
\begin{equation}\label{Bose.4}
f(p)dp=\frac{dN_{\Sigma}(p)}{dp}dp=\frac{V_{\Sigma}}{h^3}4\pi p^2 dp~.
\end{equation}
If $m_B$ is the mass of the particle with $\mathcal{E}$ being its energy then the momentum reads $p=\sqrt{2m_B \mathcal{E}}$ and it is possible to recast eq.(\ref{Bose.4}) as
\begin{equation}\label{Bose.5}
f(p) dp=g(\mathcal{E})d\mathcal{E}=\frac{2\pi V_{\Sigma}}{h^3}\left(2m_B\right)^{\frac{3}{2}}\mathcal{E}^{\frac{1}{2}}d\mathcal{E}~.
\end{equation}
It is important to note that the above weight factor becomes zero for the energy level with zero energy. Now in a quantum mechanical calculation it is not correct to give zero weight to a particular energy level. One can therefore just take out the zero energy level contribution as a whole. The equation of state for an ideal Bose gas is now given by
\begin{equation}\label{Bose.6}
\begin{split}
\frac{PV_{\Sigma}}{k_B T}&=-\sum\limits_{j}\ln\left[1-\mathcal{z}e^{-\beta \mathcal{E}_j}\right]\\
&=-\sum\limits_{\mathcal{j}\neq 0}\ln\left[1-\mathcal{z}e^{-\beta \mathcal{E}_j}\right]-\ln\left[1-\mathcal{z}\right]
\end{split}
\end{equation}
where $\mathcal{z}=e^{\beta\mu}$ is the fugacity of the ideal gas,  $\beta=\frac{1}{k_B T}$, and $\mathcal{E}_0=0$. For a continuous spectrum of the energy $\mathcal{E}$, it is possible to recast eq.(\ref{Bose.6}) as
\begin{equation}\label{Bose.7}
\begin{split}
\frac{P}{k_B T}&=-\frac{2\pi(2m)^{\frac{3}{2}}}{h^3}\int_0^{\infty}d\mathcal{E}\hspace{0.5mm}\mathcal{E}^{\frac{1}{2}}\ln\left[1-\mathcal{z}e^{-\beta\mathcal{E}}\right]-\frac{1}{V_{\Sigma}}\ln[1-\mathcal{z}]
\end{split}
\end{equation}
where as the weight factor is zero at $\mathcal{E}=0$, we can write the integration limits from zero to infinity. In the next subsection, we shall investigate the condensation procedure and the analytical form of the critical temperature for the formation of a Bose-Einstein condensate. Hence, if the system temperature is lowered below $T_{\mathcal{C}}$, condensation will occur, where the system will be a mixture of a normal phase in the excited state with number of particles equal to $N_{B_{\text{max}}}^\mathcal{e}$ and a Bose-Einstein condensate phase in the ground state, consisting of $N_B^0=N_B-N_{B_{\text{max}}}^\mathcal{e}$ number of particles.
\subsection{Bose-Einstein condensation and the critical temperature}
The number of particles in an ideal Bose gas can be recast as
\begin{equation}\label{Bose.8}
\begin{split}
N_B&=\sum\limits_j\frac{1}{\mathcal{z}^{-1}e^{\beta \mathcal{E}_j}-1}\\
&=\sum\limits_{j\neq 0} \frac{1}{z^{-1}e^{\beta\mathcal{E}_j}-1}+\frac{\mathcal{z}}{1-\mathcal{z}}~.
\end{split}
\end{equation}
Again going to the continuous energy spectrum consideration, $N_B$ can be rewritten as
\begin{equation}\label{Bose.9}
\begin{split}
N_B&=\frac{2\pi V_{\Sigma}\left(2m_B\right)^{\frac{3}{2}}}{h^3}\int_0^\infty d\mathcal{E}\hspace{0.5 mm}\mathcal{E}^{\frac{1}{2}}\frac{1}{\mathcal{z}^{-1}e^{\beta\mathcal{E}}-1}+\frac{\mathcal{z}}{1-\mathcal{z}}~.
\end{split}
\end{equation}
If $N_B^0$ denotes the number of particles in the ground state of the system then it is easy to identify from the above equation that
\begin{equation}\label{Bose.10}
N_B^0=\frac{\mathcal{z}}{1-\mathcal{z}}\implies \mathcal{z}=\frac{N_B^0}{N_B^0+1}~.
\end{equation}
The above expression implies that $0\leq \mathcal{z}\lesssim 1$ where $\mathcal{z}\simeq 1$ for very large values of $N_B^0$. It is then possible to recast eq.(\ref{Bose.9}) as
\begin{equation}\label{Bose.11}
\begin{split}
\frac{N_B^\mathcal{e}}{V_{\Sigma}}=\frac{N_B-N_B^0}{V_\Sigma}=\frac{2\pi\left(2m_B\right)^{\frac{3}{2}}}{h^3}\int_0^\infty d\mathcal{E}\hspace{0.5 mm}\mathcal{E}^{\frac{1}{2}}\frac{1}{\mathcal{z}^{-1}e^{\beta\mathcal{E}}-1}
\end{split}
\end{equation} 
with $N_B^{\mathcal{e}}$ denoting the number of particles in the excited state of the bosonic system. Making a change of variables $\kappa_\mathcal{E}=\beta \mathcal{E}$, we can recast the above expression as
\begin{equation}\label{Bose.12}
\begin{split}
\frac{N_B^\mathcal{e}}{V_{\Sigma}}&=\frac{2\pi\left(2m_Bk_BT\right)^{\frac{3}{2}}}{h^3}\int_0^\infty d\kappa_\mathcal{E}\hspace{0.5 mm}\kappa_\mathcal{E}^{\frac{1}{2}}\frac{1}{\mathcal{z}^{-1}e^{\kappa_\mathcal{E}}-1}\\
&=\frac{2\pi\left(2m_Bk_BT\right)^{\frac{3}{2}}}{h^3}\Gamma(3/2)\mathcal{g}_\frac{3}{2}(\mathcal{z})
\end{split}
\end{equation}
with $\mathcal{g}_\Xi(\mathcal{z})$ denote the Bose-Einstein distribution functions, the analytical forms of which are given as \cite{PathriaBeale}
\begin{equation}\label{Bose.13}
\begin{split}
\mathcal{g}_\Xi(\mathcal{z})&=\frac{1}{\Gamma(\zeta)}\int_0^\infty \frac{d\xi\hspace{0.5 mm}\xi^{\Xi-1}}{\mathcal{z}^{-1}e^{\xi}-1}\\&=\mathcal{z}+\frac{\mathcal{z}^2}{2^{\Xi}}+\frac{\mathcal{z}^3}{3^\Xi}+\cdots~.
\end{split}
\end{equation}
We can further simplify the expression in eq.(\ref{Bose.13}) and write it as
\begin{equation}\label{Bose.14}
\begin{split}
\frac{N_B^\mathcal{e}}{V_{\Sigma}}=\frac{1}{\lambda_B^3}\mathcal{g}_{\frac{3}{2}}(\mathcal{z})
\end{split}
\end{equation}
where the constant $\lambda_B$ is defined as $\lambda_B\equiv\frac{h}{\left(2\pi m_B k_B T\right)^\frac{1}{2}}$~.

\noindent It is easy to check from eq, that with increasing value of $\mathcal{z}$ the value of the Bose-Einstein function $\mathcal{g}_\Xi(\mathcal{z})$ increases. Now the maximum value of the fugacity is unity and as a result $\mathcal{g}_{\Xi}(\mathcal{z})$ obeys the inequality $\mathcal{g}_{\Xi}(\mathcal{z})\leq \mathcal{g}_\Xi(1)$. For $\Xi=\frac{3}{2}$, we have the inequality
\begin{equation}\label{Bose.15}
\mathcal{g}_\frac{3}{2}(\mathcal{z})\leq \mathcal{g}_\frac{3}{2}(1)=\zeta(3/2)
\end{equation}
where $\zeta(3/2)=\mathcal{g}_\frac{3}{2}(1)$ denotes the Riemann zeta function with argument $\frac{3}{2}$ which has the numerical value approximately equals to 2.61238. Hence, using the inequality in eq.(\ref{Bose.15}), it is possible to give a bound on the number of particles in the excited state of the system from eq.(\ref{Bose.14}) as
\begin{equation}\label{Bose.16}
\begin{split}
N_B^{\mathcal{e}}&\leq \frac{V_{\Sigma}}{\lambda_B^3}\zeta(3/2)\\\implies N_B^{\mathcal{e}}&\leq \frac{(2\pi m_Bk_B T)^\frac{3}{2}}{h^3}V_\Sigma\hspace{0.5 mm}\zeta(3/2)~.
\end{split}
\end{equation} 
If the total number of particles $N_B$ is greater than the right hand side of the above expression then the all the excited states combined will exactly have $N_{B_{\text{max}}}^\mathcal{e}=\frac{(2\pi m_Bk_B T)^\frac{3}{2}}{h^3}V_\Sigma\hspace{0.5 mm}\zeta(3/2)$ number of atoms while the rest of the particles have no other option but to occupy the ground state of the bosonic system. The number of particles occupying the ground state will be then given by the expression
\begin{equation}\label{Bose.17}
\begin{split}
N_B^0&=N_B-N_{B_{\text{max}}}^{\mathcal{e}}\\
&=N_B-\frac{(2\pi m_Bk_B T)^\frac{3}{2}}{h^3}V_\Sigma\hspace{0.5 mm}\zeta(3/2)~.
\end{split}
\end{equation}
This interesting event of the gathering of a large number of bosons in the ground state of an ideal gas is macroscopic and is termed as the Bose-Einstein condensation. For the Bose-Einstein condensate to occur, the number of particles required is given by the inequality
\begin{equation}\label{Bose.18}
N_B>\frac{(2\pi m_Bk_B T)^\frac{3}{2}}{h^3}V_\Sigma\hspace{0.5 mm}\zeta(3/2)~.
\end{equation}
If the number of particles is fixed in an ideal Bose gas with confinement volume $V_{\Sigma}$ being a constant then the above equation gives a condition involving the temperature of the system as
\begin{equation}\label{Bose.19}
T<\frac{h^2}{2\pi m_B k_B}\left(\frac{N_B}{V_\Sigma\hspace{0.5 mm}\zeta(3/2)}\right)^\frac{2}{3}~.
\end{equation}
The above inequality has a very important physical significance. If for an ideal Bose gas the confinement volume and the number of particles is kept fixed then for temperature less than $T_\mathcal{C}=\frac{h^2}{2\pi m_B k_B}\left(\frac{N_B}{V_\Sigma\hspace{0.5 mm}\zeta(3/2)}\right)^\frac{2}{3}$, Bose-Einstein condensation will happen. This temperature $T_{\mathcal{C}}$ is known as the critical temperature for the Bose-Einstein condensation and this macroscopic new state of matter is also known as a Bose-Einstein condensate. 
\section{Relativistic Bose-Einstein condensate}
In the previous section, we have discussed in detail the formation of the Bose-Einstein condensate in a nonrelativistic ideal Bose gas. In this section, we shall start with the basic formalism of a relativistic Bose-Einstein condensate in a quantum field theory approach and eventually proceed towards the consideration of the interaction of Bose-Einstein condensate with graviton fluctuations. 
\subsection{Background model and the relativistic Lagrangian}
In this current approach, we consider the background to be a flat Minkowski background with some small fluctuations on it. The background metric can then be expressed in the following way
\begin{equation}\label{QGRBEC.1}
g_{\alpha\beta}=\eta_{\alpha\beta}+h_{\alpha\beta}
\end{equation}
where $\alpha,\beta=\{0,1,2,3\}$ and $\eta_{\alpha\beta}$ is given by
\begin{equation}\label{QGRBEC.2}
\eta_{\alpha\beta}=\text{diag}\{-1,1,1,1\}.
\end{equation}
At first, we shall obtain the form of Einstein-Hilbert action which will dictate the dynamics of the time-dependent gravitational fluctuation part. The Einstein-Hilbert action takes the form given as
\begin{equation}\label{QGRBEC.3}
S_{\text{EH}}=\frac{c^3}{16\pi G}\int d^4x \sqrt{-g}R
\end{equation}
where $g=\text{det}\left[g_{\mu\nu}\right]$, and $R$ denotes the Ricci scalar. For convenience and to avoid cluttering in the expressions, we can set $c=1$ which can be restored later. We will keep only terms second order in the perturbation term and the Einstein-Hilbert action in eq.(\ref{QGRBEC.3}) can be recast in the following form
\begin{equation}\label{QGRBEC.4}
S_{\text{EH}}\simeq \frac{1}{64\pi G}\int d^4x \left(h_{\alpha\beta}\square h^{\alpha\beta}-h\square h+2 h^{\alpha\beta}\partial_{\alpha}\partial_{\beta}-2h_{\alpha\rho}\partial_\beta\partial^\rho h^{\alpha\beta}\right)~.
\end{equation}
Using the gauge freedoms of the fluctuation term, one can express $h_{\alpha\beta}$ as
\begin{equation}\label{QGRBEC.5}
h_{\alpha\beta}=\bar{h}_{\alpha\beta}+\partial_\alpha\xi_{\beta}+\partial_{\beta}\xi_{\alpha}.
\end{equation}
We shall now make use of the transverse-traceless gauge condition as
\begin{equation}\label{QGRBEC.6}
\partial_{\alpha}\bar{h}^{\alpha\beta}=0,~\bar{h}^\alpha_{~\alpha}=0,~k_{\alpha}\bar{h}^{\alpha\beta}=0
\end{equation}
where $k_\alpha$ denotes a constant timelike vector which can be expressed as $k_\alpha=\delta^0_{~\alpha}$. In this transverse traceless gauge, the form of the Einstein-Hilbert action presented in eq.(\ref{QGRBEC.4}) can be recast in form given as
\begin{equation}\label{QGRBEC.7}
S_{\text{EH}}=-\frac{1}{8\kappa_G^2}\int d^4x~\partial_\mu\bar{h}_{jk}\partial^\mu\bar{h}^{jk}  
\end{equation}
where we have defined $\kappa_G\equiv \sqrt{8\pi G}$ with $\mu=\{0,1,2,3\}$, and $j,k=\{1,2,3\}$. Our primary aim is to understand the response of a relativistic Bose-Einstein condensate to the quanta of linearized gravity or the gravitons. The first step is to conduct a box of volume $V_G$ where we can decompose the gravitational fluctuations in the transverse-traceless gauge into its discrete Fourier modes which is given as 
\begin{equation}\label{QGRBEC.8}
\bar{h}_{jl}(t,\vec{x})=\frac{2\kappa_G}{\sqrt{V_G}}\sum\limits_{\vec{k},q} h_q(t,\vec{k})e^{i\vec{k}.\vec{x}}\epsilon^q_{jl}(\vec{k})
\end{equation}
where $h_q(t,\vec{k})$ denotes the mode functions, $\epsilon^q_{jl}(\vec{k})$ denotes the polarization tensor. As $\bar{h}_{jl}(t,\vec{x})$ is a real quantity one can write down
$\bar{h}_{jl}(t,\vec{x})=\bar{h}^*_{jl}(t,\vec{x})$. One can now make use of the Fourier mode decomposition of $\bar{h}_{jl}(t,\vec{x})$ and substitute it back in the form of the Einstein-Hilbert action in eq.(\ref{QGRBEC.7}) and obtain the simplified form of the action as 
\begin{equation}\label{QGRBEC.9}
S_{\text{EH}}=\frac{1}{2}\sum\limits_{\vec{k},q}\int dt\left(\left|\dot{h}_q(t,\vec{k})\right|^2-k^2\left|h_q(t,\vec{k})\right|^2\right)
\end{equation} 
where we have made use of the fact that $\bar{h}_{jl}(t,\vec{x})$ is a real quantity. With the form of the Einstein-Hilbert action in  the transverse traceless gauge while the gravitational fluctuations are decomposed into individual Fourier modes, we are now in a position to investigate the action for a system of interacting bosons. We start by considering a bosonic complex scalar field theory with a self interaction term whose Lagrangian density is given by\footnote{Here $B$ in the suffix of the Lagrangian density and the coupling constant denotes bosons.}
\begin{equation}\label{QGRBEC.10}
\mathcal{L}_B=\nabla_\alpha\phi^\dagger(t,\vec{x})\nabla^\alpha \phi(t,\vec{x})+m_B^2\phi^\dagger(t,\vec{x})\phi(t,\vec{x})+\lambda_B\left(\phi^\dagger(t,\vec{x})\phi(t,\vec{x})\right)^2
\end{equation}
where $\phi(t,\vec{x})$ denotes the complex scalar field, $m_B$ denotes the mass of the field and $\lambda_B$ denotes the coupling constant corresponding to the self interaction term. As $\phi$ is a scalar field, the covariant derivative on $\phi$ can reduced to just a partial derivative which helps us to write eq.(\ref{QGRBEC.10}) in a simple form as
\begin{equation}\label{QGRBEC.11}
\mathcal{L}_B=g^{\alpha\beta}\partial_\alpha\phi^\dagger(t,\vec{x})\partial_\beta\phi(t,\vec{x})+m_B^2\phi^\dagger(t,\vec{x})\phi(t,\vec{x})+\lambda_B\left(\phi^\dagger(t,\vec{x})\phi(t,\vec{x})\right)^2~.
\end{equation}
Our primary aim is to obtain the effective quantum action for the Goldstone field. The Goldstone fields refer to the phase of the complex scalar field $\phi(t,\vec{x})$ in eq.(\ref{QGRBEC.11}). Hence, as a first step, we express $\phi(t,\vec{x})$ as
\begin{equation}\label{QGRBEC.12}
\phi(t,\vec{x})=\varphi(t,\vec{x})\exp\left[i\chi(t,\vec{x})\right]
\end{equation}
where $\varphi(t,\vec{x}),\chi(t,\vec{x})\in\mathbb{R}$. Substituting eq.(\ref{QGRBEC.12}) in eq.(\ref{QGRBEC.11}), we can obtain the Lagrangian density in terms of $\varphi(t,\vec{x})$ and $\chi(t,\vec{x})$ as\footnote{We have set $\varphi=\varphi(t,\vec{x})$ and $\chi=\chi(t,\vec{x})$ in the expressions to avoid cluttering in the expressions.}
\begin{equation}\label{QGRBEC.13}
\mathcal{L}_B=g^{\alpha\beta}\partial_{\alpha}\varphi\partial_\beta\varphi+\varphi^2g^{\alpha\beta}\partial_\alpha\chi\partial_\beta\chi+m_B^2\varphi^2+\lambda_B\varphi^4~.
\end{equation}
In order to obtain the effective quantum action for the Goldstone field, we follow the effective average action formalism described in \cite{DTSon} and later followed by \cite{MatthewMannAffshordi,MatthewMannAffshordi2}. In order to integrate out the ``\textit{high-frequency fields}" or the amplitude $\varphi$, one needs to simply minimize the action for the theory ($S_B=\int d^4x\mathcal{L}_B$) with respect to $\varphi$. The minimization done at the level of the Lagrangian density does not alter the minimization of the effective action for the system. We can write down the modified action as
\begin{equation}\label{QGRBEC.14}
S_B^M[\phi(t,\vec{x})]=\mathop{\text{min}}\limits_{\{\varphi(t,\vec{x})\}}S_B[\varphi(t,\vec{x}) e^{i\chi(t,\vec{x})}]~.
\end{equation}
If one defines the partition function for the theory then in the presence of an external source term $J(t,\vec{x})$ one can express $\mathcal{Z}[J]$ as
\begin{equation}\label{QGRBEC.15}
\mathcal{Z}[J]=\int \mathcal{D}\phi_i \exp\left[\frac{iS_B[\phi]}{\hbar}+\frac{i}{\hbar}\int d^4x J(t,\vec{x})\varphi(t,\vec{x})\right]
\end{equation} 
where $\phi_i$ is the elementary complex scalar field and $\varphi$ is considered to play no major role in the dynamics of the system. From the partition function of the theory one can write down a new functional as
\begin{equation}\label{QGRBEC.16}
\mathcal{W}[J]=-i\ln[\mathcal{Z}[J]]~.
\end{equation}
Taking the Legendre transformation of the functional $\mathcal{W}[J]$, we get the quantum effective action as
\begin{equation}\label{QGRBEC.17}
\mathcal{S}_{\text{Eff}}[\varphi]=\mathcal{W}[J]-J\varphi~.
\end{equation}
At the minima, it is possible to write down the effective action as
\begin{equation}\label{QGRBEC.18}
\mathcal{S}^M_{\text{Eff}}[\varphi]=\mathop{\text{min}}\limits_{\{\varphi\}}\mathcal{S}_{\text{Eff}}[\varphi]=\mathcal{W}[0]
\end{equation}
where the Schwinger functional for $J=0$ is equal to $\mathcal{W}[0]=-E_0\tau$ with $E_0$ denoting the ground state energy of the system and $\tau$ denoting the time interval. This shows that minimization with respect to the field amplitude $\varphi(t,\vec{x})$ results in the system to be in its ground energy state which is a condition required for Bose-Einstein condensation and as a result makes it an unavoidable step for writing down the correct action for a Bose-Einstein condensate. Taking a derivative of the Lagrangian density in eq.(\ref{QGRBEC.13}) with respect to $\varphi$ and then setting it equal to zero we obtain the minimization condition as
\begin{equation}\label{QGRBEC.19}
\begin{split}
\frac{\partial \mathcal{L}_B}{\partial \varphi}=2\varphi g^{\alpha\beta}\partial_\alpha\varphi\partial_\beta\varphi+2m_B^2\varphi+4\lambda_{B}\varphi^3&=0\\
\implies 2\varphi\left(g^{\alpha\beta}\partial_\alpha\varphi\partial_\beta\varphi+m_B^2
+2\lambda_B\varphi^2\right)&=0~.
\end{split}
\end{equation}
As $\varphi$ is an arbitrary real scalar field, it is possible to obtain condition that minimizes the Lagrangian density $\mathcal{L}_B$ as
\begin{equation}\label{QGRBEC.20}
\varphi^2=-\frac{1}{2\lambda_B}\left(g^{\alpha\beta}\partial_\alpha\chi\partial_\beta\chi+m_B^2\right)~.
\end{equation} 
We shall now make use of the above condition to minimize the Lagrangian density in eq.(\ref{QGRBEC.11}) and recast it as
\begin{equation}\label{QGRBEC.21}
\begin{split}
\mathcal{L}^M_{B}=\mathop{\text{min}}\limits_{\{\varphi\}}\mathcal{L}_B&=g^{\alpha\beta}\partial_\alpha\varphi\partial_\beta\varphi-\frac{1}{2\lambda_B}\left(g^{\mu\nu}\partial_\mu\chi\partial_\nu\chi+m_B^2\right)\left(g^{\alpha\beta}\partial_\alpha\chi\partial_\beta\chi+m_B^2\right)+\lambda_B\left(\varphi^2\right)^2\\
&=g^{\alpha\beta}\partial_\alpha\varphi\partial_\beta\varphi-\frac{1}{4\lambda_B}\left(g^{\alpha\beta}\partial_\alpha\chi\partial_\beta\chi+m_B^2\right)^2\\
&=g^{\alpha\beta}\partial_\alpha\varphi\partial_\beta\varphi+\mathcal{L}_{\text{Goldstone}}
\end{split}
\end{equation}
where we have defined the Lagrangian density corresponding to the Goldstone field as 
\begin{equation}\label{QGRBEC.22}
\mathcal{L}_{\text{Goldstone}}\equiv-\frac{1}{4\lambda_B}\left(g^{\alpha\beta}\partial_\alpha\chi\partial_\beta\chi+m_B^2\right)^2~.
\end{equation}
 As we are primarily focussed on to the Goldstone part of the Lagrangian density, we can also drop the dynamical term corresponding to the $\varphi$ term in the Lagrangian density in eq.(\ref{QGRBEC.21}) as it will have almost negligible contribution in the dynamics of the Goldstone field. Taking care of the overall minus sign in the Goldstone-Lagrangian density in eq.(\ref{QGRBEC.22}), we can write down the Lagrangian density for a relativistic Bose-Einstein condensate in curved spacetime as\footnote{One can also simply replace $\lambda_{B}$ by $-\lambda_B$ in the starting Lagrangian density in eq.(\ref{QGRBEC.10}) to directly get the Lagrangian density for the Bose-Einstein condensate after the minimization condition is applied.} 
 \begin{equation}\label{QGRBEC.23}
\mathcal{L}^{G}_{\text{BEC}}=\frac{1}{4\lambda_B}\left(g^{\alpha\beta}\partial_\alpha\chi\partial_\beta\chi+m_B^2\right)^2
\end{equation}   
where the superscript $G$ denotes the system is in a gravitational background. The action of the system corresponding to the above Lagrangian density takes the form
\begin{equation}\label{QGRBEC.24}
S^G_{\text{BEC}}=\int d^4x\sqrt{-g}\mathcal{L}^G_{\text{BEC}}
\end{equation}
with the determinant of the metric given by $g=\mathop{\text{det}}[g_{\alpha\beta}]$. One can now write down the  Goldstone field as a combination of two parts as
\begin{equation}\label{QGRBEC.25}
\chi(t,\vec{x})=-\sigma_Bt+\pi_B(t,\vec{x})
\end{equation}
where $\sigma_B$ is a constant and $\pi_B(t,\vec{x})\in \mathbb{R}$ denotes the pseudo-Goldstone bosons. Our primary aim is to investigate the dynamics of $\pi_B(t,\vec{x})$. We know that $x^0=t$ and as a result $x_0=g_{0\kappa}x^{\kappa}$. Now for a Minkowski background with gravitational fluctuations in the transverse traceless gauge, $g_{00}=\eta_{00}$ and $g_{0i}=0~\forall i\in\{1,2,3\}$. Hence, $x_0=\eta_{00}x^0=-t$. As a result, it is possible to recast eq.(\ref{QGRBEC.25}) as
\begin{equation}\label{QGRBEC.26}
\chi(t,\vec{x})=\sigma_B x_\kappa\delta^\kappa_{~0}+\pi_B(t,\vec{x})~.
\end{equation}
Now we shall write down the background on which we are investigating the Bose-Einstein condensate Lagrangian.
In the transverse-traceless gauge the, it is possible to write down the background metric as
\begin{equation}\label{QGRBEC.27}
\begin{split}
g_{\alpha\beta}=\eta_{\alpha\beta}+h_{\alpha\beta}
=&
\begin{pmatrix}
-1&&0&&0&&0\\
0&&1&&0&&0\\
0&&0&&1&&0\\
0&&0&&0&&1
\end{pmatrix}+\begin{pmatrix}
0&&0&&0&&0\\
0&&h_+(t)&&h_{\times}(t)&&0\\
0&&h_\times(t)&&-h_{+}(t)&&0\\
0&&0&&0&&0
\end{pmatrix}\\
=&\begin{pmatrix}
-1&&0&&0&&0\\
0&&1+h_+(t)&&h_{\times}(t)&&0\\
0&&h_\times(t)&&1-h_{+}(t)&&0\\
0&&0&&0&&1
\end{pmatrix}~.
\end{split}
\end{equation}
 The determinant of the metric in the above equation is given as
\begin{equation}\label{QGRBEC.28}
g=\mathop{\text{det}}[g_{\alpha\beta}]=-\left(1-\left(h_+^2(t)+h_\times^2(t)\right)\right)~.
\end{equation}
In our current analysis, we shall consider that the incoming gravitational wave has plus polarization only, which implies $h_\times(t)=0$. If we now drop any $\mathcal{O}(h^2)$ contributions in $\sqrt{-g}$ then we obtain $\sqrt{-g}\simeq 1$. Substituting the form of the Goldstone field in eq.(\ref{QGRBEC.26}) in the action given in eq.(\ref{QGRBEC.24}), and making use of the analytical form of $\sqrt{-g}$, we can recast the action as
\begin{equation}\label{QGRBEC.29}
\begin{split}
S^G_{\text{BEC}}&=\int d^4x\sqrt{-g}\mathcal{L}^G_{\text{BEC}}\simeq\int d^4x\mathcal{L}^G_{\text{BEC}}\\
&=\frac{1}{4\lambda_B}\int d^4x\left(g_{\alpha\beta}\left(\sigma_B\delta^\alpha_{~0}+\partial^\alpha \pi_B(t,\vec{x})\right)\left(\sigma_B\delta^\beta_{~0}+\partial^\beta \pi_B(t,\vec{x})\right)+m_B^2\right)^2\\
&=\frac{1}{4\lambda_B}\int d^4x\left(g_{00}\sigma_B^2+2g_{0\beta}\sigma_B\partial^{\beta}\pi_B(t,\vec{x})+g_{\alpha\beta}\partial^\alpha\pi_B(t,\vec{x})\partial^\beta\pi_B(t,\vec{x})+m_B^2\right)^2\\
&=\frac{1}{4\lambda_B}\int d^4x\left(-\sigma_B^2+2\sigma_Bg_{00}g^{0\alpha}\partial_\alpha\pi_B(t,\vec{x})+g_{\alpha\beta}\partial^\alpha\pi_B(t,\vec{x})\partial^\beta\pi_B(t,\vec{x})+m_B^2\right)^2~.
\end{split}
\end{equation}
In the above equation $\partial_0\pi_B(t,\vec{x})=\dot{\pi}(t,\vec{x})$, and using this result we can recast eq.(\ref{QGRBEC.29}) as
\begin{equation}\label{QGRBEC.30}
S^G_{\text{BEC}}=\frac{1}{4\lambda_B}\int d^4x\left(-\left(\dot{\pi}_B-\sigma_B\right)^2+g_{ij}\partial^i\pi_B\partial^j\pi_B+m_B^2\right)^2~.
\end{equation}
Following the assumption made in \cite{MatthewMannAffshordi}, we also drop any derivatives of $\pi(t,\vec{x})$ higher than second order because third or higher order derivatives result in terms that will have minimal contributions to the dynamics of the pseudo-Goldstone bosons. The resulting action reads
\begin{equation}\label{QGRBEC.31}
\begin{split}
S^G_{\text{BEC}}\simeq&\frac{1}{2\lambda_B}\int d^4x\left(\left(3\sigma_B^2-m_B^2\right)\dot{\pi}_B^2-\left(\sigma_B^2-m_B^2\right)g_{ij}\partial^i\pi_B\partial^j\pi_B\right)\\+&\frac{1}{4\lambda_B}\int d^4x \left(\sigma_B^2-m_B^2\right)^2-\frac{1}{\lambda_B}\int d^4x\left(\sigma_B^2-m_B^2\right)\sigma_B\dot{\pi}_B~.
\end{split}
\end{equation}
In the above equation, the second integral in the right hand side does not contribute to the dynamics of the pseudo-Goldstone bosons as a result one can easily drop this term. The third integral in the right hand side of the above equation can be expressed as a  total derivative term given by $\frac{1}{\lambda_B}\int d^4x\frac{d}{dt}\left[\left(\sigma_B^2-m_B^2\right)\sigma_B\pi_B\right]=\frac{1}{\lambda_B}\int d^3x\left(\sigma_B^2-m_B^2\right)\sigma_B\pi_B$ which is a boundary contribution and can be dropped\footnote{In our original paper, S. Sen and S. Gangopadhyay \href{https://link.aps.org/doi/10.1103/PhysRevD.110.026014}{Phys. Rev. D 110 (2024) 026014}, there is a plus sign in front of the third integral in eq.(\ref{QGRBEC.31}) which should be a minus sign but as this term vanishes it does not affects the main conclusion of our paper.}. Here the underlying assumption is that the Goldstone field  vanishes at the boundary. The resultant form of the action takes the form as
\begin{equation}\label{QGRBEC.32}
\begin{split}
S^G_{\text{BEC}}&\simeq\frac{1}{2\lambda_B}\int d^4x\left(\left(3\sigma_B^2-m_B^2\right)\dot{\pi}_B^2(t,\vec{x})-\left(\sigma_B^2-m_B^2\right)g_{ij}\partial^i\pi_B(t,\vec{x})\partial^j\pi_B(t,\vec{x})\right)\\
&=\frac{3\sigma_B^2-m_B^2}{2\lambda_B}\int d^4x\left(\dot{\pi}^2_B(t,\vec{x})-\frac{\sigma_B^2-m_B^2}{3\sigma_B^2-m_B^2}g_{ij}\partial^i\pi_B(t,\vec{x})\partial^j\pi_B(t,\vec{x})\right)~.
\end{split}
\end{equation} 
The Lagrangian density from the above equation reads
\begin{equation}\label{QGRBEC.33}
\mathcal{L}_{\text{GB}}=\dot{\pi}^2_B(t,\vec{x})-\frac{\sigma_B^2-m_B^2}{3\sigma_B^2-m_B^2}g_{ij}\partial^i\pi_B(t,\vec{x})\partial^j\pi_B(t,\vec{x})
\end{equation}
where $\text{GB}$ in the subscript of $\mathcal{L}$ stands for Goldstone bosons. For a flat background the above Lagrangian reduces to the form
\begin{equation}\label{QGRBEC.34}
\begin{split}
\mathcal{L}^\eta_{\text{GB}}=\dot{\pi}^2_B(t,\vec{x})-\frac{\sigma_B^2-m_B^2}{3\sigma_B^2-m_B^2}\partial_i\pi_B(t,\vec{x})\partial^i\pi_B(t,\vec{x})
\end{split}
\end{equation}
Extremizing the action in eq.(\ref{QGRBEC.32}) wile the whole set-up is considered to be placed in a flat background, we obtain the Euler-Lagrange equation for the Lagrangian density in eq.(\ref{QGRBEC.34}) as
\begin{equation}\label{QGRBEC.35}
\partial_t^2\pi_B(t,\vec{x})-\frac{\sigma_B^2-m_B^2}{3\sigma_B^2-m_B^2}\nabla^2\pi_B(t,\vec{x})=0.
\end{equation} 
Assuming a solution of the form $\pi_B(t,\vec{x})\propto \exp[-i\omega_Bt+i\vec{k}_B\cdot\vec{x}]$, we obtain the dispersion relation corresponding to the pseudo-Goldstone bosons as
\begin{equation}\label{QGRBEC.36}
\begin{split}
\omega_B^2&=\frac{\sigma_B^2-m_B^2}{3\sigma_B^2-m_B^2}k_B^2\\
\implies c_S&= \frac{\omega_B}{k_B}=\sqrt{\frac{\sigma_B^2-m_B^2}{3\sigma_B^2-m_B^2}}
\end{split}
\end{equation}
where $c_S$ denotes the speed of sound. Even in a background with small gravitational fluctuations, the above relation gives the leading order solution and we can approximate the dispersion relation as $\omega_B\simeq c_S k_B$. Making use of the dispersion relation and defining a new constant $\bar{\gamma}_B\equiv\frac{3\sigma_B^2-m_B^2}{2\lambda_B}$, we can recast the action in eq.(\ref{QGRBEC.32}) as
\begin{equation}\label{QGRBEC.37}
\begin{split}
S^G_{\text{BEC}}=\bar{\gamma}_B\int d^4x\left(\dot{\pi}_B^2(t,\vec{x})-c_S^2g_{ij}\partial^i\pi_B(t,\vec{x})\partial^j\pi_B(t,\vec{x})\right)~.
\end{split}
\end{equation}
We shall now assume that the pseudo-Goldstone bosons can be separated into a time-dependent and spatial part as
\begin{equation}\label{QGRBEC.38}
\pi_B(t,\vec{x})=\sum_{\vec{k}_B}\pi_{\vec{k}_B}(t,\vec{x})=\sum_{k_B}\psi_{\vec{k}_B}(t)\exp\left[i\vec{k}_B\cdot \vec{x}\right]
\end{equation}
where $\sum_{\vec{k}_B}\equiv\sum_{k_{B_x}}\sum_{k_{B_y}}\sum_{k_{B_z}}$ and we consider the total Goldstone boson to be a combination of single mode Goldstone bosons, where the summation includes all possible phonon mode frequencies. As discussed earlier, $\pi(t,\vec{x})$ is a real number and as a result it is equal to its complex conjugate. Hence, we can obtain the following relation
\begin{equation}\label{QGRBEC.39}
\begin{split}
\pi_B(t,\vec{x})&=\pi^*_B(t,\vec{x})\\
\implies \sum_{\vec{k}_B}\psi_{\vec{k}_B}(t)\exp\left[i\vec{k}_B\cdot \vec{x}\right]&=\sum_{\vec{k}'_B}\psi^*_{\vec{k}'_B}(t)\exp\left[-i\vec{k}'_B\cdot \vec{x}\right]
\\&=\sum_{\vec{k}''_B}\psi^*_{-\vec{k}''_B}(t)\exp\left[i\vec{k}''_B\cdot \vec{x}\right]
\end{split}
\end{equation}
where in the last line of the above equation we have made the substitution $k_B'\rightarrow-k_B''$. Again setting $k_B''=k_B$, we can recast the above equation as
\begin{equation}\label{QGRBEC.40}
\sum_{\vec{k}_B}\psi_{\vec{k}_B}(t)\exp\left[i\vec{k}_B\cdot \vec{x}\right]=\sum_{\vec{k}_B}\psi^*_{-\vec{k}_B}(t)\exp\left[i\vec{k}_B\cdot \vec{x}\right]
\end{equation}
which gives us the relation $\psi_{\vec{k}_B}(t)=\psi^*_{-\vec{k}_B}(t)$ $\forall~ {k}_{B_x},{k}_{B_y},{k}_{B_z}\in\mathbb{R}$. Substituting eq.(\ref{QGRBEC.38}) in eq.(\ref{QGRBEC.37}), we can write down the form of the action as
\begin{equation}\label{QGRBEC.41}
\begin{split}
S^G_{\text{BEC}}&=\bar{\gamma}_B\int d^4x\left(\dot{\pi}_B(t,\vec{x})\dot{\pi}_B^*(t,\vec{x})-c_S^2g_{ij}\partial^i\pi_B(t,\vec{x})\partial^j\pi_B^*(t,\vec{x})\right)\\
&=\bar{\gamma}_B\int d^4x\sum_{\vec{k}_B,\vec{k}'_B}\left(\dot{\psi}_{\vec{k}_{B}}(t)e^{i\vec{k}_B\cdot\vec{x}}\dot{\psi}^*_{\vec{k}'_{B}}(t)e^{-i\vec{k}'_B\cdot\vec{x}}-c_s^2g_{ij}k_B^i{k'}_B^j\psi_{\vec{k}_B}(t)e^{i\vec{k}_B\cdot\vec{x}}\psi^*_{\vec{k}'_B}(t)e^{-i\vec{k}'_B\cdot\vec{x}}\right)\\
&=\bar{\gamma}_B\sum_{\vec{k}_B,\vec{k}'_B}\int dt\int d^3x ~e^{i\left(\vec{k}_B-\vec{k}'_B\right)\cdot \vec{x}}\left(\dot{\psi}_{\vec{k}_B}(t)\dot{\psi}^*_{\vec{k}'_B}(t)-c_S^2(\eta_{ij}+\bar{h}_{ij})k_B^i{k'}_B^j\psi_{\vec{k}_B}(t)\psi^*_{\vec{k}'_B}(t)\right)~.
\end{split}
\end{equation}
From the form of the metric in eq.(\ref{QGRBEC.27}), it is easy to infer that the metric has only time dependence and no spatial dependence. Hence, it is quite straightforward to write that $\bar{h}_{ij}(t,\vec{x})\rightarrow \bar{h}_{ij}(t,0)$ ($\eta_{ij}$ $\forall~i,j\in\{1,2,3\}$ is constant) and $g_{ij}=\eta_{ij}+\bar{h}_{ij}(t,0)$. Again, we consider that the Bose-Einstein condensate is kept inside a cubical box with side $L_B$ which implies that it has a volume $V_B=L_B^3$ and as a result one can write down the following relation, $\int d^3x \exp\left[i\left(\vec{k}_B-\vec{k}'_B\right)\cdot\vec{x}\right]=V_B~\delta_{\vec{k}_B,\vec{k}'_B}$\footnote{The Kronecker-delta $\delta_{\vec{k}_B,\vec{k}'_B}$ is defined as $\delta_{\vec{k}_B,\vec{k}'_B}\equiv \delta_{k_B^x,{k'}_B^x}\delta_{k_B^y,{k'}^y_B}\delta_{k^z_B,{k'}^z_B}$.}. Using the Kronecker-delta relation in eq.(\ref{QGRBEC.41}) and defining a new constant $\gamma_B\equiv V_B\bar{\gamma}_B$, we can rewrite the action for the system as 
\begin{equation}\label{QGRBEC.42}
\begin{split}
S^G_{\text{BEC}}&=\gamma_B\sum_{\vec{k}_B}\int dt \left(\dot{\psi}_{\vec{k}_B}(t)\dot{\psi}_{\vec{k}_B}^*(t)-c_S^2\left[\eta_{ij}+\frac{2\kappa_G}{\sqrt{V_G}}\sum_{\vec{k},q}h_q(t,\vec{k})\epsilon^q_{ij}(\vec{k})\right]k_B^ik_B^j\psi_{\vec{k}_B}(t)\psi_{\vec{k}_B}^*(t)\right)\\
&=\gamma_B\int dt\left[\sum_{\vec{k}_B}\left|\dot{\psi}_{\vec{k}_B}(t)\right|^2-c_S^2\left[\eta_{ij}+\frac{2\kappa_G}{\sqrt{V_G}}\sum_{\vec{k},q}h_q(t,\vec{k})\epsilon^q_{ij}(\vec{k})\right]\sum_{\vec{k}_B}k_B^ik_B^j\left|\psi_{\vec{k}_B}(t)\right|^2\right]~.
\end{split}
\end{equation} 
In the above equation, we made use of the mode decomposition for $\bar{h}_{ij}(t,\vec{x})$ from eq.(\ref{QGRBEC.8}) when $\bar{h}_{ij}$ does not depend on the spatial coordinates. The action for the system can be obtained by combining eq.(\ref{QGRBEC.9}) with eq.(\ref{QGRBEC.42}), and it reads
\begin{equation}\label{QGRBEC.43}
\begin{split}
S&=S_{\text{EH}}+S^G_{\text{BEC}}\\&=\frac{1}{2}\sum\limits_{\vec{k},q}\int dt\left(\left|\dot{h}^q(t,\vec{k})\right|^2-k^2\left|h_q(t,\vec{k})\right|^2\right)\\
&+\gamma_B\int dt\left[\sum_{\vec{k}_B}\left|\dot{\psi}_{\vec{k}_B}(t)\right|^2-c_S^2\left[\eta_{ij}+\frac{2\kappa_G}{\sqrt{V_G}}\sum_{\vec{k},q}h_q(t,\vec{k})\epsilon^q_{ij}(\vec{k})\right]\sum_{\vec{k}_B}k_B^ik_B^j\left|\psi_{\vec{k}_B}(t)\right|^2\right]~.
\end{split}
\end{equation}
One now needs to extremize the action to obtain the equation of motions corresponding to $\psi_{\vec{k}_B}(t)$ and $h_q(t,\vec{k})$. To obtain the equation of motion corresponding to $\psi_{\vec{k}_B}(t)$, we calculate $\frac{\delta S}{\delta \psi^*_{\vec{k}'_B}(t')}$ which is given by
\begin{equation}\label{QGRBEC.44}
\begin{split}
\frac{\delta S}{\delta \psi^*_{\vec{k}'_B}(t')}=&-\gamma_B\int dt \Biggr[-\sum_{\vec{k}_B}\ddot{\psi}_{\vec{k}_B}(t) \delta_{\vec{k}_B,\vec{k}'_B}\delta(t-t')-c_S^2\Biggr[\eta_{ij}+\frac{2\kappa_G}{\sqrt{V_G}}\sum_{\vec{k},q}h_q(t,\vec{k})\epsilon^q_{ij}(\vec{k})\Biggr]\\&\times\sum_{\vec{k}_B}k_B^ik_B^j\psi_{\vec{k}_B}(t)\delta_{\vec{k}_B,\vec{k}'_B}\delta(t-t')\Biggr]
\end{split}
\end{equation}
where we have dropped the total derivative term with the assumption in mind that $\psi_{\vec{k}_B}(t)$ (as well as  its complex conjugate) vanishes at the boundary. Implementing the principle of least action (setting $\delta S/\delta \psi_{\vec{k}_B}(t)=0$) and replacing $t'$ with $t$, and $\vec{k}'_{B}$ with $\vec{k}_{B}$, we arrive at the equation of motion for the time dependent part corresponding to the pseudo-Goldstone boson with frequency $\omega_B=c_Sk_B$ as
\begin{equation}\label{QGRBEC.45}
\ddot{\psi}_{\vec{k}_B}(t)+\left[\eta_{ij}+\frac{2\kappa_G}{\sqrt{V_G}}\sum_{\vec{k},q}h_q(t,\vec{k})\epsilon^q_{ij}(\vec{k})\right]k_B^ik_B^j\psi_{\vec{k}_B}(t)=0~.
\end{equation} 
Again extremizing the  action, with respect to ${h^*}^q(t,\vec{k})$ by setting $\delta S/\delta {h^*}^q(t,\vec{k})=0$, we arrive at the equation of motion of $h_q(t,\vec{k})$ as
\begin{equation}\label{QGRBEC.46}
\ddot{h}_q(t,\vec{k})+k^2h_q(t,\vec{k})=-\frac{4\gamma_B\kappa_Gc_S^2}{\sqrt{V}_G}{\epsilon^q_{ij}}^*(\vec{k})\sum_{\vec{k}_B}k_B^ik_B^j\left|\psi_{\vec{k}_B}(t)\right|^2~.
\end{equation}
\section{Quantization of the gravitational fluctuation}
In this subsection, we shall quantize the gravitational fluctuation term in order to incorporate effects of linearized quantum gravity into the model system. The first step to quantization is to raise the Fourier mode function, $h_q(t,\vec{k})$ in eq.(\ref{QGRBEC.8}) to operator status and then implement a commutation relation between the gravitational fluctuation and its canonically conjugate variable (in the phase space). One can rewrite eq.(\ref{QGRBEC.8}) as
\begin{equation}\label{QGRBEC.47}
\hat{\bar{h}}_{ij}(t,\vec{x})=\frac{2\kappa_G}{\sqrt{V_G}}\sum\limits_{\vec{k},q} \hat{h}_q(t,\vec{k})e^{i\vec{k}\cdot\vec{x}}\epsilon^q_{ij}(\vec{k})
\end{equation}
where $\hat{h}_q(t,\vec{k})$ are now operators. In order to properly quantize the theory, we need  to investigate the model system in the interaction picture which helps us to write $\hat{h}_q^I(t,\vec{k})$ as
\begin{equation}\label{QGRBEC.48}
\hat{h}_q^I(t,\vec{k})=f_k(t)\hat{a}_q(\vec{k})+f_k^*(t)\hat{a}^\dagger_q(-\vec{k})
\end{equation}
where $f_k(t)$ denotes the mode function with $k=|\vec{k}|$. The mode functions satisfy the following normalization condition given by
\begin{equation}\label{QGRBEC.49}
\begin{split}
-if_k(t)\mathop{\partial_t}\limits^{\leftrightarrow}f_k^*(t)&=1\\
\implies -i\left(f_k(t)\dot{f}_k^*(t)-\dot{f}_k(t)f_k^*(t)\right)&=1~.
\end{split}
\end{equation}
The raising and lowering operators are defined by 
\begin{equation}\label{QGRBEC.50}
\begin{split}
\hat{a}_q(\vec{k})|n_{\vec{k}}\rangle=\sqrt{n_{\vec{k}}}|n_{\vec{k}}-1_{\vec{k}}\rangle,~\hat{a}^\dagger_q(\vec{k})|n_{\vec{k}}\rangle=\sqrt{n_{\vec{k}}+1}|n_{\vec{k}}+1_{\vec{k}}\rangle 
\end{split}
\end{equation}
where $|n_{\vec{k}}\rangle$ denotes the $n$ graviton state corresponding to the mode frequency $\omega_{\vec{k}}=c|\vec{k}|$. The vacuum graviton state is defined as
\begin{equation}\label{QGRBEC.51}
\hat{a}_q(\vec{k})|0\rangle=0~.
\end{equation}
The commutation relation satisfied by the graviton ladder operators are
\begin{equation}\label{QGRBEC.52}
\left[\hat{a}_q(\vec{k}),\hat{a}_{q'}(\vec{k}')\right]=0,~\left[\hat{a}^\dagger_q(\vec{k}),\hat{a}^\dagger_{q'}(\vec{k}')\right]=0,~\text{and }\left[\hat{a}_q(\vec{k}),\hat{a}^\dagger_{q'}(\vec{k}')\right]=\delta_{q,q'}\delta_{\vec{k},\vec{k}'}~.
\end{equation}
As we have represented the gravitational fluctuation by a discrete sum of its individual Fourier modes, the Kronecker-deltas appear in the commutation relation. If one proceeds with an integral representation where a continuum of all possible modes are considered then Kronecker-deltas will be replaced by Dirac-delta functions. The quantum gravitational contribution in the mode function (operator) $\hat{h}_q^I(t,\vec{k})$ of the fluctuation term $\hat{\bar{h}}_{ij}(t,\vec{x})$ is obtained by subtracting the expectation value $\langle\hat{h}_q^I(t,\vec{k})\rangle$ from $\hat{h}_q^I(t,\vec{k})$ itself. Hence, the quantum gravitational fluctuation reads \cite{KannoSodaTokuda}
\begin{equation}\label{QGRBEC.53}
\delta\hat{h}^I_q(t,\vec{k})=\hat{h}_q^I(t,\vec{k})-\langle \hat{h}_q^I(t,\vec{k})\rangle
\end{equation}
where the expectation value, $\langle\hat{h}_q^I(t,\vec{k})\rangle$, can be identified with the classical gravitational mode function defined by $h_q^{\text{cl}}(t,\vec{k})$. Hence, it is possible to write down $\hat{h}_q^I(t,\vec{k})$ as
\begin{equation}\label{QGRBEC.54}
\hat{h}_q^I(t,\vec{k})=h_q^{\text{cl}}(t,\vec{k})+\delta\hat{h}_q^I(t,\vec{k})~.
\end{equation}
One important thing to understand is that eq.(s)(\ref{QGRBEC.45},\ref{QGRBEC.46}) are both coupled differential equations and as a result if $h_q(t,\vec{k})$ is raised to operator status, $\psi_{\vec{k}_B}(t)$ will have contributions from $\hat{h}_q(t,\vec{k})$ term as well and as a result, we can write $\psi_{\vec{k}_B}(t)$ as $\hat{\psi}_{\vec{k}_B}(t)$. Eq.(\ref{QGRBEC.46}), in a quantum gravity set-up takes the from
\begin{equation}\label{QGRBEC.55}
\begin{split}
\ddot{\hat{h}}_q(t,\vec{k})+k^2\hat{h}_q(t,\vec{k})&=-\frac{4\gamma_B\kappa_Gc_S^2}{\sqrt{V}_G}{\epsilon^q_{ij}}^*(\vec{k})\sum_{\vec{k}_B}k_B^ik_B^j\left|\hat{\psi}_{\vec{k}_B}(t)\right|^2\\
\implies\left(\frac{\partial^2}{\partial t^2}+k^2\right)\hat{h}_q(t,\vec{k})&=-\frac{4\gamma_B\kappa_Gc_S^2}{\sqrt{V}_G}{\epsilon^q_{ij}}^*(\vec{k})\sum_{\vec{k}_B}k_B^ik_B^j\left|\hat{\psi}_{\vec{k}_B}(t)\right|^2
\end{split}
\end{equation}
where in the above equation $\left|\hat{\psi}_{\vec{k}_B}(t)\right|^2$ is given by $\left|\hat{\psi}_{\vec{k}_B}(t)\right|^2=\hat{\psi}^\dagger_{\vec{k}_B}(t)\hat{\psi}_{\vec{k}_B}(t)$. Our primary aim is to solve eq.(\ref{QGRBEC.55}) and obtain a solution for $\hat{h}_q(t,\vec{k})$. Here, $\hat{h}_q^I(t,\vec{k})$ satisfies the standard equation of motion given by
\begin{equation}\label{QGRBEC.56}
\ddot{\hat{h}}^I_q(t,\vec{k})+k^2\hat{h}^I_q(t,\vec{k})=0~.
\end{equation} 
The Green's function $G_k(t-t')$ corresponding to eq.(\ref{QGRBEC.55}) is defined by
\begin{equation}\label{QGRBEC.57}
(\partial_t^2+k^2)G_k(t-t')=\delta(t-t')~.
\end{equation}
The solution $\hat{h}_{q}(t,\vec{k})$ in terms of the Green's function then reads
\begin{equation}\label{QGRBEC.58}
\hat{h}_q(t,\vec{k})=h_q^I(t,\vec{k})-\frac{4\gamma_B\kappa_Gc_S^2}{\sqrt{V}_G}{\epsilon^q_{ij}}^*(\vec{k})\int_0^t dt' G_k(t-t')\sum_{\vec{k}_B}k_B^ik_B^j\left|\hat{\psi}_{\vec{k}_B}(t')\right|^2.
\end{equation}
The Fourier transform of $G_k(t-t')$ and the integral representation of the delta function read
\begin{equation}\label{QGRBEC.59}
G_k(t-t')=\frac{1}{2\pi}\int d\zeta ~e^{-i\zeta (t-t')}G_k(\zeta),~\delta(t-t')=\frac{1}{2\pi}\int d\zeta' ~e^{-i\zeta'(t-t')}.
\end{equation}
Substituting the expressions of $G_k(t-t')$ and $\delta(t-t')$ in the Fourier space from the above equation in eq.(\ref{QGRBEC.57}), we obtain the form of $G_k(\zeta)$ as
\begin{equation}\label{QGRBEC.60}
G_k(\zeta)=-\frac{1}{\zeta^2-k^2}~.
\end{equation}
Inserting the form of $G_k(\zeta)$ in the Fourier space expression of $G_k(t-t')$, we can re-express $G_k(t-t')$ as
\begin{equation}\label{QGRBEC.61}
G_k(t-t')=-\frac{1}{2\pi} \int d\zeta\frac{e^{-iq(t-t')}}{\zeta^2-k^2}
\end{equation}
which has  poles at $\zeta=\pm k$. Here, we are more interested in the retarded Green's function which is obtained by pushing the poles by an amount $-i\eta$ in the complex plane and after performing the contour integral, taking the $\eta\rightarrow 0$ limit. This retarded Green's function then reads
\begin{equation}\label{QGRBEC.62}
G^R_k(t-t')=\frac{\sin(k(t-t'))}{k}~.
\end{equation}
Using the form of the Green's function obtained in the above equation, we can finally write down the solution of $\hat{h}_q(t,\vec{k})$ as
\begin{equation}\label{QGRBEC.63}
\hat{h}_q(t,\vec{k})=h_q^{\text{cl}}(t,\vec{k})+\delta\hat{h}_q^I(t,\vec{k})-\frac{4\gamma_B\kappa_Gc_S^2}{\sqrt{V}_G}{\epsilon^q_{ij}}^*(\vec{k})\sum_{\vec{k}_B}k_B^ik_B^j\int_0^t dt' \frac{\sin(k(t-t'))}{k}\left|\hat{\psi}_{\vec{k}_B}(t')\right|^2.
\end{equation}
Now we write down the dynamical equation of motion corresponding to the time-dependent part of the Goldstone boson from eq.(\ref{QGRBEC.45}) in a quantum gravitational set-up as
\begin{equation}\label{QGRBEC.64}
\ddot{\hat{\psi}}_{\vec{k}_B}(t)+\left[\eta_{ij}+\frac{2\kappa_G}{\sqrt{V}_G}\sum\limits_{\vec{k},q}\hat{h}_q(t,\vec{k})\epsilon^q_{ij}(\vec{k})\right]k_B^i k_B^j\hat{\psi}_{\vec{k}_B}(t)=0~.
\end{equation}
Substituting the analytical form of $\hat{h}_q(t,\vec{k})$ in the above equation, we obtain the full dynamical equation involving $\hat{\psi}_{\vec{k}_B}(t)$ as
\begin{equation}\label{QGRBEC.65}
\begin{split}
&\ddot{\hat{\psi}}_{\vec{k}_B}(t)+c_S^2\Biggr[\eta_{ij}+\frac{2\kappa_G}{\sqrt{V}_G}\sum\limits_{\vec{k},q}\left(h_q^{\text{cl}}(t,\vec{k})+\delta\hat{h}_q^I(t,\vec{k})\right){\epsilon^q_{ij}}\Biggr]k_B^ik_B^j \hat{\psi}_{\vec{k}_B}(t)\\&-\sum\limits_{\vec{k},q}\Biggr(\frac{8\gamma_B\kappa_G^2c_S^4}{V_G}{\epsilon^q_{lm}}^*(\vec{k})\sum_{\vec{k}'_B}{k'}_B^l{k'}_B^m\int_0^t dt' \frac{\sin(k(t-t'))}{k}\left|\hat{\psi}_{{\vec{k}'}_B}(t')\right|^2\Biggr)\epsilon^q_{ij}(\vec{k})k_B^ik_B^j \hat{\psi}_{\vec{k}_B}(t)=0.
\end{split}
\end{equation}
Our aim is to solve the above equation of motion perturbatively and obtain a solution of $\hat{\psi}_{\vec{k}_B}(t)$.  The gravitational fluctuation in eq.(\ref{QGRBEC.47}) with the decomposition in eq.(\ref{QGRBEC.54}) can be recast as
\begin{equation}\label{QGRBEC.66}
\begin{split}
\hat{\bar{h}}_{ij}(t,\vec{x})&=\frac{2\kappa_G}{\sqrt{V}_G}\sum\limits_{\vec{k},q}\left(h^{\text{cl}}_q(t,\vec{k})+\delta\hat{h}^I_q(t,\vec{k})\right)e^{i\vec{k}\cdot\vec{x}}\epsilon^q_{ij}(\vec{k})\\
&=h^{\text{cl}}_{ij}(t,\vec{x})+\delta\hat{\mathcal{N}}_{ij}(t,\vec{x})
\end{split}
\end{equation}
where $h^{\text{cl}}_{ij}(t,\vec{x})$ and $\delta \hat{\mathcal{N}}_{ij}(t,\vec{x})$ are defined as
\begin{equation}\label{QGRBEC.67}
\begin{split}
h^{\text{cl}}_{ij}(t,\vec{x})&\equiv\frac{2\kappa_G}{\sqrt{V}_G}\sum\limits_{\vec{k},q}h^{\text{cl}}_q(t,\vec{k})e^{i\vec{k}\cdot\vec{x}}\epsilon^q_{ij}(\vec{k})\\\delta \hat{\mathcal{N}}_{ij}(t,\vec{x})&\equiv\frac{2\kappa_G}{\sqrt{V}_G}\sum\limits_{\vec{k},q}\delta\hat{h}^I_q(t,\vec{k})e^{i\vec{k}\cdot\vec{x}}\epsilon^q_{ij}(\vec{k})~.
\end{split}
\end{equation}
In the above equation $h^{\text{cl}}_{ij}(t,\vec{x})$ denotes the classical part of the gravitational fluctuation whereas $\delta\hat{\mathcal{N}}_{ij}(t,\vec{x})$ is purely a quantum gravitational term is similar to that of a noise or stochastic term. The motivation for calling $\delta\hat{\mathcal{N}}_{ij}(t,\vec{x})$ a stochastic term or a noise term will be clarified later. 
Using the above definitions, we can recast eq.(\ref{QGRBEC.65}) as
\begin{equation}\label{QGRBEC.68}
\begin{split}
&\ddot{\hat{\psi}}_{\vec{k}_B}(t)+c_S^2\left(\eta_{ij}+h^{\text{cl}}_{ij}(t,0)+\delta\hat{\mathcal{N}}_{ij}(t,0)\right)k_B^ik_B^j \hat{\psi}_{\vec{k}_B}(t)-\frac{8\gamma_B\kappa_G^2c_S^4}{V_G}\sum\limits_{\vec{k}}\left(\sum\limits_q{\epsilon^{q*}_{lm}}(\vec{k})\epsilon^q_{ij}(\vec{k})\right)\\\times&\sum_{\vec{k}'_B}{k'}_B^l{k'}_B^m\int_0^t dt' \frac{\sin(k(t-t'))}{k}\left|\hat{\psi}_{\vec{k}'_B}(t')\right|^2k_B^ik_B^j \hat{\psi}_{\vec{k}_B}(t)=0.
\end{split}
\end{equation}
In order to further simplify eq.(\ref{QGRBEC.68}), one needs to execute the summation over all possible polarizations $q$ and it is done by introducing a projection tensor defined by
\begin{equation}\label{QGRBEC.69}
\mathcal{P}^{\text{R}}_{ij}=\delta_{ij}-\frac{k_ik_j}{k^2}~.
\end{equation}
The summation inside of the parenthesis in eq.(\ref{QGRBEC.68}) can then be expressed in terms of the projection tensors defined in eq.(\ref{QGRBEC.69}) as
\begin{equation}\label{QGRBEC.70}
\begin{split}
\sum\limits_q{\epsilon^{q*}_{lm}}(\vec{k})\epsilon^q_{ij}(\vec{k})=\frac{1}{2}\left(\mathcal{P}^{\text{R}}_{li}\mathcal{P}^{\text{R}}_{mj}+\mathcal{P}^{\text{R}}_{lj}\mathcal{P}^{\text{R}}_{mi}-\mathcal{P}^{\text{R}}_{lm}\mathcal{P}^{\text{R}}_{ij}\right)~.
\end{split}
\end{equation}
The next logical step is to consider a continuum of $k$ for the graviton modes which is implemented via replacing the summation over $\vec{k}$ by a volume integral in the $k$ space. One important thing to remember is that $|\vec{k}|\leq\frac{\Omega_{\text{M}}}{c}$ with $\Omega_\text{M}$ denoting the cut-off frequency for gravitational waves. To solve the integrals, $\int^{\Omega_{\text{M}}} d^3k$, we at first convert the volume-integral in Cartesian coordinate system to spherical polar coordinate system as $\int_0^{\Omega_{\text{M}}}k^2dk\int_0^{\pi}\sin\theta d\theta\int_0^{2\pi}d\phi.$ In order to solve the angular integrals, we need to solve some standard integrations. 
\begin{equation}\label{QGRBEC.71}
\begin{split}
\int d\Omega&=\int_0^{\pi} d\theta\sin\theta\int_0^{2\pi}d\phi=4\pi
\end{split}
\end{equation}
which is a standard result. Next, we shall calculate the integral $\int d\Omega \frac{k^ik^j}{k^2}$ and to solve this, we execute a term by term analysis. In spherical polar coordinates, $k^x=k\cos\phi\sin\theta, k^y=k\sin\phi\sin\theta$, and $k^z=k\cos\theta$. A trivial calculation shows that $\int d\Omega \frac{k^ik^j}{k^2}=0$ when $i\neq j$. Now for the cases where $i=j$, we obtain
\begin{equation}\label{QGRBEC.72}
\begin{split}
\int d\Omega \frac{k^x k^x}{k^2}&=\int_0^{\pi} d\theta\sin\theta\int_0^{2\pi}d\phi \cos^2\phi\sin^2\theta=\frac{4\pi}{3}~,\\
\int d\Omega \frac{k^y k^y}{k^2}&=\int_0^{\pi} d\theta\sin\theta\int_0^{2\pi}d\phi \sin^2\phi\sin^2\theta=\frac{4\pi}{3}~,\\
\int d\Omega \frac{k^z k^z}{k^2}&=\int_0^{\pi} d\theta\sin\theta\int_0^{2\pi}d\phi \cos^2\theta=\frac{4\pi}{3}~.
\end{split} 
\end{equation}  
From eq.(\ref{QGRBEC.72}), we can express the second angular integral in a compact form as
\begin{equation}\label{QGRBEC.73}
\int d\Omega \frac{k^ik^j}{k^2}=\frac{4\pi}{3}\delta^{ij}
\end{equation}
with $\delta^{ij}$ denoting the Kronecker-delta. Our next aim is to calculate the integral $\int d\Omega \frac{k^lk^mk^ik^j}{k^4}$. For this case we find out that $\int d\Omega \frac{k^lk^mk^ik^j}{k^4}=\frac{4\pi}{15}$ when $l=m$ and $i=j$, or $l=i$ and $m=j$, or $l=j$ and $m=i$, and zero for all the other case. Hence, we can write down the final angular integral in a compact form as
\begin{equation}\label{QGRBEC.74}
\begin{split}
\int d\Omega \frac{k^lk^mk^ik^j}{k^4}=\frac{4\pi}{15}\left(\delta^{lm}\delta^{ij}+\delta^{li}\delta^{mj}+\delta^{lj}\delta^{mi}\right)~.
\end{split}
\end{equation}  
Taking the volume integral in the $k$ space for the expression in eq.(\ref{QGRBEC.70}) and making use of the angular integrations from eq.(s)(\ref{QGRBEC.71},\ref{QGRBEC.73},\ref{QGRBEC.74}), we obtain the following result 
\begin{equation}\label{QGRBEC.75}
\begin{split}
\sum\limits_{\substack{{|\vec{k}|\leq\Omega_{\text{M}}}}}\sum\limits_q{\epsilon^{q*}_{lm}}(\vec{k})\epsilon^q_{ij}(\vec{k})\rightarrow &\frac{V_G}{2(2\pi)^3}\int^{\Omega_{\text{M}}} d^3\vec{k}\left(\mathcal{P}^{\text{R}}_{li}\mathcal{P}^{\text{R}}_{mj}+\mathcal{P}^{\text{R}}_{lj}\mathcal{P}^{\text{R}}_{mi}-\mathcal{P}^{\text{R}}_{lm}\mathcal{P}^{\text{R}}_{ij}\right)\\
=&\frac{V_G}{2(2\pi)^3}\int_0^{\Omega_{\text{M}}}k^2 dk\int d\Omega\left(\mathcal{P}^{\text{R}}_{li}\mathcal{P}^{\text{R}}_{mj}+\mathcal{P}^{\text{R}}_{lj}\mathcal{P}^{\text{R}}_{mi}-\mathcal{P}^{\text{R}}_{lm}\mathcal{P}^{\text{R}}_{ij}\right)\\
=&\frac{8\pi V_G}{10(2\pi)^3}\left(\delta_{li}\delta_{mj}+\delta_{lj}\delta_{mi}-\frac{2}{3}\delta_{lm}\delta_{ij}\right)\int_0^{\Omega_{\text{M}}}k^2 dk~.
\end{split}
\end{equation}
Using the continuum limit and substituting the above integration in in eq.(\ref{QGRBEC.68}), we obtain the following dynamical equation for $\hat{\psi}_{\vec{k}_B}(t)$ as
\begin{equation}\label{QGRBEC.76}
\begin{split}
&\ddot{\hat{\psi}}_{\vec{k}_B}(t)+c_S^2\left(\eta_{ij}+h^{\text{cl}}_{ij}(t,0)+\delta\hat{\mathcal{N}}_{ij}(t,0)\right)k_B^ik_B^j \hat{\psi}_{\vec{k}_B}(t)-\frac{4\gamma_B\kappa_G^2c_S^4}{5\pi^2}\left[\delta_{li}\delta_{mj}+\delta_{lj}\delta_{mi}-\frac{2}{3}\delta_{lm}\delta_{ij}\right]\\\times&\sum_{\vec{k}'_B}{k'}_B^l{k'}_B^m\int_0^t dt' \int_0^{\Omega_{\text{M}}}dkk\sin(k(t-t'))\left|\hat{\psi}_{\vec{k}'_B}(t')\right|^2k_B^ik_B^j \hat{\psi}_{\vec{k}_B}(t)=0.
\end{split}
\end{equation}
Summing over all possible $i,j,l$, and $m$ values and executing the $k$ integral, we can recast the above equation as
\begin{equation}\label{QGRBEC.77}
\begin{split}
&\ddot{\hat{\psi}}_{\vec{k}_B}(t)+c_S^2\left(\eta_{ij}+h^{\text{cl}}_{ij}(t,0)+\delta\hat{\mathcal{N}}_{ij}(t,0)\right)k_B^ik_B^j \hat{\psi}_{\vec{k}_B}(t)
-\xi_B\sum\limits_{\vec{k}_B'}\left[(\vec{k}_B\cdot\vec{k}_B')^2-\frac{1}{3}k_B^2{k_B'}^2\right]\\\times&\int_0^t dt'\left[\frac{\sin[\Omega_\text{M}(t-t')]}{(t-t')^2}-\frac{\Omega_{\text{M}}\cos[\Omega_\text{M}(t-t')]}{t-t'}\right]\left|\hat{\psi}_{\vec{k}'_B}(t')\right|^2 \hat{\psi}_{\vec{k}_B}(t)=0.
\end{split}
\end{equation}
where we have defined $\xi_B\equiv\frac{4\gamma_B\kappa_G^2c_S^4}{5\pi^2}$. Observing the structure of the above equation, it is intuitive to understand that $\hat{\psi}_{\vec{k}_B}(t)$ will consist of three parts which can be expressed as
\begin{equation}\label{QGRBEC.78}
\hat{\psi}_{\vec{k}_B}(t)=\psi^{(0)}_{\vec{k}_B}(t)+\psi_{\vec{k}_B}^{h}(t)+\hat{\psi}^{\delta\mathcal{N}}_{\vec{k}_B}(t)
\end{equation}
where $\psi^{0}_{\vec{k}_B}(t)$ denotes the solution when no gravitational fluctuation is present, $\psi^{(0)}_{\vec{k}_B}(t)$ denotes the part of the solution which is induced due to the classical part of the gravitational fluctuation, and $\hat{\psi}^{\delta\mathcal{N}}_{\vec{k}_B}(t)$ denotes the quantum-gravitational contribution to the time-dependent part of the pseudo-Goldstone bosons. Substituting eq.(\ref{QGRBEC.78}), in eq.(\ref{QGRBEC.77}), we obtain the unperturbed equation of motion in the form
\begin{equation}\label{QGRBEC.79}
\ddot{\psi}_{\vec{k}_B}^{(0)}(t)+c_S^2\eta_{ij}k_B^ik_B^j\psi^{(0)}_{\vec{k}_B}(t)=0
\end{equation}
where $\eta_{ij}k_B^ik_B^j=k_B^2$ and $c_S^2k_B^2=\omega_B^2$. The above equation has a standard solution of the form given by
\begin{equation}\label{QGRBEC.80}
\psi_{\vec{k}_B}^{(0)}(t)=a^{(0)}_B\exp[-i\omega_B t]+b^{(0)}_{B} \exp[i\omega_B t]~.
\end{equation}
In the above solution, $\exp[i\omega_B t]$ describes the negative energy modes of the solution and it is quite intuitive to get rid of the negative energy modes by setting $b_B^{(0)}=0$. This choice for $b_B^{(0)}$ helps us to obtain the value of the coefficient corresponding to the positive energy mode solution to be $a_B^{(0)}=1$. Hence, one obtains $\psi_{\vec{k}_B}^{(0)}(t)=\exp[-i\omega_Bt]$. The first order equation of motion corresponding to the classical spacetime fluctuation reads
\begin{equation}\label{QGRBEC.81}
\ddot{\psi}^{h}_{\vec{k}_B}(t)+\omega_B^2\psi^h_{\vec{k}_B}=-c_S^2h_{jl}^{\text{cl}}(t,0)k_B^jk_B^l\psi^{(0)}_{\vec{k}_B}(t)+\mathcal{F}_B(t)
\end{equation}
where $\mathcal{F}_B(t)$ is defined as
\begin{equation}\label{QGRBEC.82}
\begin{split}
\mathcal{F}_B(t)\equiv&\xi_B\sum\limits_{k_B'}\left[(\vec{k}_B\cdot\vec{k}_B')^2-\frac{1}{3}k_B^2{k_B'}^2\right]\psi^{(0)}_{\vec{k}_B}(t)\\\times&\int_0^t dt'\left[\frac{\sin[\Omega_\text{M}(t-t')]}{(t-t')^2}-\frac{\Omega_{\text{M}}\cos[\Omega_\text{M}(t-t')]}{t-t'}\right]\left|\psi^{(0)}_{\vec{k}'_B}(t')\right|^2~.
\end{split}
\end{equation}
Using the solution of $\psi^{(0)}_{\vec{k}'_B}(t)$, it is easy to infer that $\left|\psi^{(0)}_{\vec{k}'_B}(t')\right|^2=1$. Now, converting the summation over all possible $\vec{k}_B'$, into a continuum of all possible phonon modes and executing the $t'$ integral, we can recast eq.(\ref{QGRBEC.82}) as
\begin{equation}\label{QGRBEC.83}
\mathcal{F}_B(t)=\frac{V_B\xi_B}{(2\pi)^3}\exp[-i\omega_Bt]\int d^3\vec{k}'_B\left[\left(\vec{k}_B\cdot\vec{k}'_B\right)^2-\frac{1}{3}k_B^2{k'_B}^2\right]\left(\Omega_\text{M}-\frac{\sin(\Omega_{\text{M}}t)}{t}\right)
\end{equation} 
with $V_B$ denting the volume of the box inside which the condensate is prepared. As we are considering a system where gravitons are interacting with the Bose-Einstein condensate, it is quite intuitive to set $V_B=V_G$. 

\begin{figure}
\begin{center}
\includegraphics[scale=0.4]{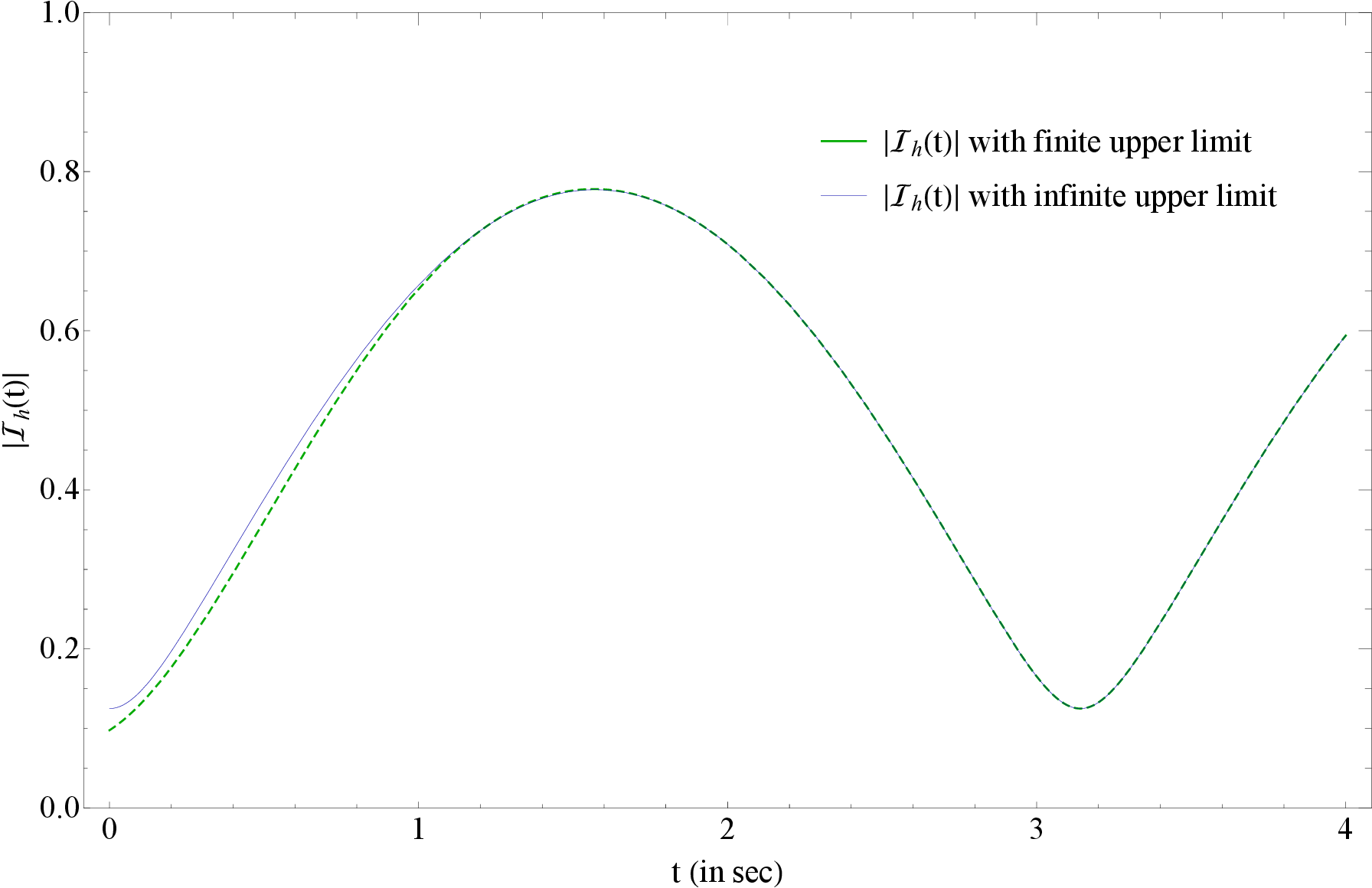}
\caption{Plot of $|\mathcal{I}_h(t)|$ against $t$ when the upper limit of integration in $\mathcal{I}_h(t)$ is finite versus the case when the upper limit of integration is infinite. \label{Infinite_Limit_OTM}}
\end{center}
\end{figure}

To solve, eq.(\ref{QGRBEC.81}), we need to simplify the equation by considering vanishing contributions from all of the cross polarization terms that is $\epsilon_{ij}^{\times}(\vec{k})=0~\forall ~i,j\in\{x,y,z\}$. The polarization tensor components corresponding to plus polarization only are $\epsilon_{xx}^{+}(\vec{k})=-\epsilon_{yy}^{+}(\vec{k})$\footnote{The primary reason behind this assumption lies in the fact that we are working in the transverse-traceless gauge.} and $\epsilon_{zz}^{+}(\vec{k})=0$. All other $\epsilon^{+}_{jl}(\vec{k})$ components for all   $j\neq l$ are zero. Using the expression of $h^{\text{cl}}_{ij}(t,\vec{x})$ in eq.(\ref{QGRBEC.67}), we can deduce the following relations
\begin{equation}\label{QGRBEC.84}
\begin{split}
h^{\text{cl}}_{xx}(t,0)=\frac{2\kappa_G}{\sqrt{V}_G}\sum\limits_{\vec{k}}h^{\text{cl}}_+(t,\vec{k})\epsilon^+_{xx}(\vec{k})&=-\frac{2\kappa_G}{\sqrt{V}_G}\sum\limits_{\vec{k}}h^{\text{cl}}_+(t,\vec{k})\epsilon^+_{yy}(\vec{k})=-h^{\text{cl}}_{yy}(t,0)~,\\
h^{\text{cl}}_{zz}(t,0)&=0~.
\end{split}
\end{equation}
which is the required condition for a classical gravitational wave fluctuation in the transverse-traceless gauge. Using the above relations, we can recast eq.(\ref{QGRBEC.81}) as
\begin{equation}\label{QGRBEC.85}
\begin{split}
\ddot{\psi}^h_{\vec{k}_B}(t)+\omega_B^2\psi^h_{\vec{k}_B}=-c_S^2h_{\text{cl}}(t,0)({k_{B}^x}^2-{k_B^y}^2)\exp[-i\omega_Bt]+\mathcal{F}_B(t)
\end{split}
\end{equation}
where we have used the definition $h_{\text{cl}}(t,0)\equiv h^{\text{cl}}_{xx}(t,0)=-h^{\text{cl}}_{yy}(t,0)$. Using a new definition $k_0^2\equiv \frac{{k_{B}^x}^2-{k_B^y}^2}{k_B^2}$, we can rewrite the above equation in a very simplified form given as
\begin{equation}\label{QGRBEC.86}
\ddot{\psi}^h_{\vec{k}_B}(t)+\omega_B^2\psi^h_{\vec{k}_B}=-\omega_B^2 k_0^2h_{\text{cl}}(t,0)\exp[-i\omega_Bt]+\mathcal{F}_B(t)~.
\end{equation}
Again the Green's function for the above differential equation is defined as
\begin{equation}\label{QGRBEC.87}
\left(\frac{d^2}{dt^2}+\omega_B^2\right)G^{(1)}(t-t')=\delta(t-t')~.
\end{equation}
Standard computational procedure shows that the analytical form of the Green's functions are (retarded Green's function is only considered)
\begin{equation}\label{QGRBEC.88}
G^{(1)}(t-t')=\frac{1}{\omega_B}\sin(\omega_B(t-t'))\Theta[t-t'].
\end{equation}
Using the above Green's function the analytical form of $\psi^h_{\vec{k}_B}(t)$ satisfying eq.(\ref{QGRBEC.86}) reads
\begin{equation}\label{QGRBEC.89}
\psi^h_{\vec{k}_B}(t)=a^h_B \exp[-i\omega_B t]+b^h_B \exp[i\omega_B t]+\int_{-\infty}^{\infty}dt'G^{(1)}(t-t')\left[-\omega_B^2 k_0^2h_{\text{cl}}(t',0)e^{-i\omega_Bt'}+\mathcal{F}_B(t')\right]
\end{equation}
where, without loss of any generality, we can set the undefined constants to be equal to zero, that is $a_B^h=b_B^h=0$. The above relation can be further simplified using the analytical form of $\mathcal{F}_B(t)$ in eq.(\ref{QGRBEC.83}). At first we convert the three integral $\int d^3\vec{k}'_B$ from Cartesian to spherical polar coordinates in eq.(\ref{QGRBEC.83}), and assume that the angle between $\vec{k}_B$ and $\vec{k}'_B$ is $\theta$. We can the recast eq.(\ref{QGRBEC.83}) as
\begin{equation}\label{QGRBEC.90}
\begin{split}
\mathcal{F}_B(t)&=\frac{V_B\xi_Be^{-i\omega_Bt}}{(2\pi)^2}\int_0^\infty dk_B'{k'}^2_B\int_0^\pi d\theta\sin\theta\left[k_B^2{k'}^2_B\cos^2\theta-\frac{1}{3}k_B^2{k'_B}^2\right]\left(\Omega_\text{M}-\frac{\sin(\Omega_{\text{M}}t)}{t}\right)\\
&=0~.
\end{split}
\end{equation}
Using the above equation, we can recast eq.(\ref{QGRBEC.89}) as
\begin{equation}\label{QGRBEC.91}
\psi^h_{\vec{k}_B}(t)=-\omega_B k_0^2\int_{-\infty}^{t}dt'
\exp\left[-i\omega_Bt'\right]\sin\left(\omega_B(t-t')\right)h_{\text{cl}}(t',0)~.
\end{equation}
With the forms of $\psi_{\vec{k}_B}^{(0)}(t)$ and $\psi^h_{\vec{k}_B}(t)$ obtained we now finally need to determine the analytical form of the operator part of $\hat{\psi}_{\vec{k}_B}(t)$ which carries the graviton signature. The equation of motion corresponding to the operator part of the time-dependent pseudo-Goldstone boson from eq.(\ref{QGRBEC.77}) reads
\begin{equation}\label{QGRBEC.92}
\ddot{\hat{\psi}}_{\vec{k}_B}^{\delta\mathcal{N}}(t)+\omega_B^2\hat{\psi}_{\vec{k}_B}^{\delta\mathcal{N}}(t)\simeq-c_S^2~\delta\hat{\mathcal{N}}_{ij}(t,0)k_B^ik_B^j\psi_{\vec{k}_B}^{(0)}(t)~.
\end{equation}
We have already made use of the relation that $\epsilon^+_{xx}(\vec{k})=-\epsilon^+_{yy}(\vec{k})$, and $\epsilon^+_{zz}(\vec{k})=0$. Applying these conditions in the expression of $\delta\hat{\mathcal{N}}_{ij}$ from eq.(\ref{QGRBEC.67}), we obtain the following results
\begin{equation}\label{QGRBEC.93}
\delta\hat{\mathcal{N}}_{xx}(t,0)=-\delta\hat{\mathcal{N}}_{yy}(t,0), \text{ and }\delta\hat{\mathcal{N}}_{zz}(t,0)=0~.
\end{equation} 
The above relation again helps us to write down eq.(\ref{QGRBEC.92}) in a much simpler form as
\begin{equation}\label{QGRBEC.94}
\ddot{\hat{\psi}}_{\vec{k}_B}^{\delta\mathcal{N}}(t)+\omega_B^2\hat{\psi}_{\vec{k}_B}^{\delta\mathcal{N}}(t)\simeq-\omega_B^2k_0^2\delta\hat{\mathcal{N}}(t,0)\exp[-i\omega_Bt]
\end{equation}
where $\delta\hat{\mathcal{N}}(t,0)\equiv \delta\hat{\mathcal{N}}_{xx}(t,0)=-\delta\hat{\mathcal{N}}_{yy}(t,0)$. The Green's function for the above equation of motion is exactly same to $G^{(1)}(t-t')$ given in eq.(\ref{QGRBEC.88}). The solution for eq.(\ref{QGRBEC.94}) reads
\begin{equation}\label{QGRBEC.95}
\hat{\psi}^{\delta\mathcal{N}}_{\vec{k}_B}(t)=-\omega_Bk_0^2\int_{-\infty}^t dt'\exp[-i\omega_Bt']\sin(\omega_B(t-t'))\delta\hat{N}(t',0)~.
\end{equation}
Substituting the values of $\psi_{\vec{k}_B}^{(0)}(t)$, $\psi_{\vec{k}_B}^h(t)$, and $\hat{\psi}^{\delta\mathcal{N}}_{\vec{k}_B}(t)$ from eq.(s)(\ref{QGRBEC.80},\ref{QGRBEC.91}), and eq.(\ref{QGRBEC.95}) (with proper coefficient choices) in eq.(\ref{QGRBEC.78}), we obtain the first-order perturbative solution to the time-dependent part of the pseudo Goldstone bosons as
\begin{equation}\label{QGRBEC.96}
\hat{\psi}_{\vec{k}_B}(t)=\exp[-i\omega_B t]-\omega_Bk_0^2\int_{-\infty}^t dt'\exp[-i\omega_Bt']\sin(\omega_B(t-t'))\left(h_{\text{cl}}(t,0)+\delta\hat{N}(t',0)\right)~.
\end{equation} 
In order to proceed further, we consider a gravitational wave template for the classical gravitational wave as $h_{\text{cl}}(t,0)=\varepsilon e^{-\frac{t^2}{\tau^2}}\sin(\Omega_0 t)$ where $\varepsilon$ denotes the amplitude of the gravitational wave, $\Omega_0$ denotes its frequency, and $\tau$ denotes the single measurement time for the gravitational wave. To simplify our analysis, we need to consider the analytical form of the operator part of $\hat{\psi}_{\vec{k}_B}(t)$ given in eq.(\ref{QGRBEC.95}). We already know that $\delta\hat{\mathcal{N}}(t',0)$ is a stochastic term, and as a result can not be integrated. The easiest way to deal with this problem is to apply the Markovian-like approximation, where we consider that the time-dependence of the Goldstone bosons do not depend on the the time evolution of the noise operator rather it simply depends on the stochastic operator at the final time (which is $t$ in eq.(\ref{QGRBEC.95}) denoted by $\delta\hat{\mathcal{N}}(t,0)$. It is also quite intuitive to assume that $\delta\hat{\mathcal{N}}(t,0)=0$ for $t<0$ as before $t=0$ there is no interaction of the BEC with the gravitational wave. We can then recast eq.(\ref{QGRBEC.96}) as
\begin{equation}\label{QGRBEC.97}
\begin{split}
\hat{\psi}_{\vec{k}_B}(t)=&\exp[-i\omega_B t]
-\omega_Bk_0^2~\int_{-\infty}^t dt'e^{-i\omega_Bt'}\sin(\omega_B(t-t'))h_{\text{cl}}(t',0)\\&-\omega_Bk_0^2\int_{0}^t dt'e^{-i\omega_Bt'}\sin(\omega_B(t-t'))\delta\hat{\mathcal{N}}(t',0)\\
\implies \hat{\psi}_{\vec{k}_B}(t)\simeq&\exp[-i\omega_B t]-\varepsilon\omega_Bk_0^2\int_{-\infty}^t dt'e^{-\frac{{t'}^2}{\tau^2}-i\omega_Bt'}\sin(\Omega_0 t)\sin(\omega_B(t-t'))\\
&-\omega_Bk_0^2~\delta\hat{\mathcal{N}}(t,0)\int_{0}^t dt'e^{-i\omega_Bt'}\sin(\omega_B(t-t')).
\end{split}
\end{equation}
For the second integral in the above equation the upper limit of integration can be extended to $t\rightarrow\infty$ because of the existence of the Gaussian decay factor in the integrand. Hence, the second integral can simply be reduced in the following way
\begin{equation}\label{QGRBEC.98}
\begin{split}
\mathcal{I}_{h}&\equiv\int_{-\infty}^\infty dt'e^{-\frac{{t'}^2}{\tau^2}-i\omega_Bt'}\sin(\Omega_0 t)\sin(\omega_B(t-t'))\\
&=-\frac{\sqrt{\pi}\tau}{4}\left(\exp\left[-\frac{\tau^2}{4}(\Omega_0-2\omega_B)^2\right]-\exp\left[-\frac{\tau^2}{4}(\Omega_0+2\omega_B)^2\right]\right)\exp[i\omega_B t]~.
\end{split}
\end{equation} 
One can use $\omega_B=1$ Hz, $\Omega_0=2$ Hz, and $\tau=1$ sec and plot the absolute value of $\mathcal{I}_h(t)$ against the time $t$ in Fig.(\ref{Infinite_Limit_OTM}) for the cases when the upper limit of integration is finite and the scenario when the upper limit of integration is infinite.
From Fig.(\ref{Infinite_Limit_OTM}), we observe that the finite limit case is very close to the infinite limit case when comparing the absolute values of the integral $\mathcal{I}_h(t)$ and therefore the approximation made in eq.(\ref{QGRBEC.98}) is quite valid.

\noindent The second integral in the last line of eq.(\ref{QGRBEC.97}) can be simplified to take the form as
\begin{equation}\label{QGRBEC.99}
\begin{split}
\mathcal{I}_{\delta\mathcal{N}}&\equiv\delta\hat{\mathcal{N}}(t,0)\int_{0}^t dt'e^{-i\omega_Bt'}\sin(\omega_B(t-t'))\\
&=\frac{\delta\hat{\mathcal{N}}(t,0)}{4\omega_B}\left((1+2i\omega_B t)\exp[-i\omega_B t]-\exp[i\omega_B t]\right)~.
\end{split}
\end{equation} 
Substituting the results of the integrations in eq.(s)(\ref{QGRBEC.98},\ref{QGRBEC.99}) back in eq.(\ref{QGRBEC.97}), we obtain the solution of $\hat{\psi}_{\vec{k}_B}(t)$ as
\begin{equation}\label{QGRBEC.100}
\begin{split}
\hat{\psi}_{\vec{k}_B}(t)=&\exp\left[-i\omega_B t\right]-\varepsilon\omega_Bk_0^2\mathcal{I}_h-\omega_Bk_0^2\mathcal{I}_{\delta\mathcal{N}}\\
=&\left(1-\frac{k_0^2}{4}\delta\hat{N}(t,0)(1+2i\omega_B t)\right)\exp\left[-i\omega_Bt\right]\\&+\left(\frac{\sqrt{\pi}\varepsilon\omega_B \tau k_0^2}{4}\left(e^{-\frac{\tau^2}{4}(\Omega_0-2\omega_B)^2}-e^{-\frac{\tau^2}{4}(\Omega_0+2\omega_B)^2}\right)+\frac{k_0^2}{4}\delta\hat{N}(t,0)\right)\exp\left[i\omega_Bt\right]\\
=&\hat{\alpha}^B(t)\exp[-i\omega_B t]+\hat{\beta}^B(t)\exp[i\omega_B t]
\end{split}
\end{equation}
where $\hat{\alpha}^B(t)$ and $\hat{\beta}^B(t)$ denotes the graviton-noise induced Bogoliubov operators. If the measurement time is corresponding to a single mode of the Bose-Einstein condensate is $t=\tau_{\text{M}}$\footnote{It is quite straightforward to infer that $\tau_{\text{M}}=\tau$ but for the moment we kept the nomenclatures for the two measurement  time to be different.} then without loss of any generality, one can set $t=\tau_{\text{M}}$ in the Bogoliubov coefficients which indicates that the entire effect of the noise-term can be encaptured at the final measurement time and does not depend upon the entire duration of measurement. Hence, one can write down the analytical form of $\hat{\psi}_{\vec{k}_B}(t)$ as
\begin{equation}\label{QGRBEC.101}
\begin{split}
\hat{\psi}_{\vec{k}_B}(t)\simeq&\hat{\alpha}^B(\tau_{\text{M}})\exp[-i\omega_B t]+\hat{\beta}^B(\tau_{\text{M}})\exp[i\omega_B t]
\end{split}
\end{equation}
where we can decompose the operator valued Bogoliubov coefficients as
\begin{align}
\hat{\alpha}^B(\tau_{\text{M}})&\equiv\alpha^B+\delta\hat{\alpha}^B(\tau_{\text{M}})= 1-\frac{\varepsilon_B}{4\varepsilon}(1+2i\omega_B\tau_{\text{M}})\delta\hat{\mathcal{N}}(\tau_{\text{M}},0)\label{QGRBEC.102}\\
\hat{\beta}^B(\tau_{\text{M}})&\equiv\beta^B+\delta\hat{\beta}^B(\tau_{\text{M}})=\frac{\sqrt{\pi}\varepsilon_B\omega_B\tau}{4}\left(e^{-\frac{\tau^2}{4}(\Omega_0-2\omega_B)^2}-e^{-\frac{\tau^2}{4}(\Omega_0+2\omega_B)^2}\right)+\frac{\varepsilon_B}{4\varepsilon}\delta\hat{\mathcal{N}}(\tau_{\text{M}},0)\label{QGRBEC.103}
\end{align}
with the smallness parameter $\varepsilon_B$ being defined as
\begin{equation}\label{QGRBEC.104}
\varepsilon_B\equiv\varepsilon\frac{{k_B^x}^2-{k_B^y}^2}{k_B^2}~.
\end{equation}
It is now possible to write down the quantum-gravity induced part of the operator-valued Bogoliubov coefficients as follows
\begin{equation}\label{QGRBEC.105}
\begin{split}
\delta \hat{\alpha}^B(\tau_{\text{M}})=\mathcal{c}_\alpha(\tau_{\text{M}})\delta\hat{\mathcal{N}}(\tau_{\text{M}},0)~,~~\delta \hat{\beta}^B(\tau_{\text{M}})=\mathcal{c}_\beta(\tau_{\text{M}})\delta\hat{\mathcal{N}}(\tau_{\text{M}},0)
\end{split}
\end{equation}
where the coefficients $\mathcal{c}_\alpha(\tau_{\text{M}})$ and $\mathcal{c}_\beta(\tau_{\text{M}})$ are defined as
\begin{equation}\label{QGRBEC.106}
\mathcal{c}_\alpha(\tau_{\text{M}})\equiv-\frac{\varepsilon_B}{4\varepsilon}(1+2i\omega_B\tau_{\text{M}})~,~~\mathcal{c}_\beta(\tau_{\text{M}})\equiv \frac{\varepsilon_B}{4\varepsilon}~.
\end{equation}
The important thing to note is that because of the quantized nature of the  gravitational wave the Bogoliubov coefficients are not just numbers but they have operator-valued contributions from it .
\section{Covariance matrix of  a BEC influenced by gravitons}
We have already obtained the operator-valued Bogoliubov coefficients and in order to obtain the covariance matrix of a Bose-Einstein condensate with inherent phonon mode squeezing induced by graviton noise, we at first need to write down the standard single-mode covariance matrix of the BEC with phonon squeezing.
\subsection{Covariance matrix of a single mode of the  BEC with inherent phonon squeezing}
In eq.(\ref{Covariance.18}), we have already obtained the covariance matrix corresponding to the single mode of a Bose-Einstein condensate which is given by $\Sigma_{\text{BEC}}(0)=\frac{1}{2}\mathbb{1}_{2}$. If the squeezing parameter is given by $r_{\text{Sq.}}=r_B e^{i\zeta_B}$ then the raising and lowering operators under effect of squeezing transform as
\begin{align}
\hat{S}(r_{\text{Sq.}})\hat{a}^B\hat{S}^\dagger(r_{\text{Sq.}})&=\hat{a}^B\cosh r_B+{\hat{a}^B}^\dagger \exp[i\zeta_B]\sinh r_B\label{QGCovariance.1}\\
\hat{S}(r_{\text{Sq.}}){\hat{a}^B}^\dagger\hat{S}^\dagger(r_{\text{Sq.}})&={\hat{a}^B}^\dagger\cosh r_B+\hat{a}^B \exp[-i\zeta_B]\sinh r_B\label{QGCovariance.2}
\end{align}
where the squeezing operators are defined as
\begin{equation}\label{QGCovariance.3}
\hat{S}(r_{\text{Sq.}})=\exp\left[r_{\text{Sq.}}^*{\hat{a}^B}^2-r_{\text{Sq.}}{{\hat{a}^B}^\dagger}^2\right]~,~~\hat{S}^\dagger(r_{\text{Sq.}})=\exp\left[r_{\text{Sq.}}{{\hat{a}^B}^\dagger}^2-r_{\text{Sq.}}^*{\hat{a}^B}^2\right]~.
\end{equation}
From the above equation, it is easy to observe that $\hat{S}(r_{\text{Sq.}})=\hat{S}^\dagger(-r_{\text{Sq.}})$ and vice versa. Action of the squeezing operators on the column vector $\mathbb{r}=\begin{pmatrix}
\sqrt{\frac{m_B \omega_B}{\hbar}}\hat{x}^B\\
\frac{1}{\sqrt{m_B\hbar\omega_B}}\hat{p}^B
\end{pmatrix}$, gives the following expression
\begin{equation}\label{QGCovariance.4}
\begin{split}
\hat{S}(r_{\text{Sq.}})\mathbb{r}\hat{S}^\dagger(r_{\text{Sq.}})&=\begin{pmatrix}
\sqrt{\frac{m_B \omega_B}{\hbar}}\hat{S}(r_{\text{Sq.}})\hat{x}^B\hat{S}^\dagger(r_{\text{Sq.}})\\
\frac{1}{\sqrt{m_B\hbar\omega_B}}\hat{S}(r_{\text{Sq.}})\hat{p}^B\hat{S}^\dagger(r_{\text{Sq.}})
\end{pmatrix}\\
&=\begin{pmatrix}
\cosh r_B+\cos\zeta_B\sinh r_B && \sin\zeta_B\sinh 2r_B\\
\sin\zeta_B\sinh 2r_B && \cosh r_B-\cos\zeta_B\sinh r_B 
\end{pmatrix}
\begin{pmatrix}
\sqrt{\frac{m_B \omega_B}{\hbar}}\hat{x}^B\\
\frac{1}{\sqrt{m_B\hbar\omega_B}}\hat{p}^B
\end{pmatrix}\\
&=\mathbb{S}_{\text{Sq.}}(r_\text{Sq.})\mathbb{r}
\end{split}
\end{equation}
where $\mathbb{S}_{\text{Sq.}}(r_\text{Sq.})$ denotes the squeezing matrix defined as
\begin{equation}\label{QGCovariance.5}
\mathbb{S}_{\text{Sq.}}(r_\text{Sq.})\equiv\begin{pmatrix}
\cosh r_B+\cos\zeta_B\sinh r_B && \sin\zeta_B\sinh r_B\\
\sin\zeta_B\sinh r_B && \cosh r_B-\cos\zeta_B\sinh r_B 
\end{pmatrix}~.
\end{equation}
Hence, if the phonon modes of a Bose-Einstein condensate are squeezed by a squeezing parameter $r_{\text{Sq.}}$ then the single mode squeezed BEC covariance matrix reads
\begin{equation}\label{QGCovariance.6}
\begin{split}
\Sigma_{\text{Sq.}}(0)&=\mathbb{S}_{\text{Sq.}}(r_\text{Sq.})\Sigma_{\text{BEC}}(0)\mathbb{S}^T_{\text{Sq.}}(r_\text{Sq.})\\&=\frac{1}{2}\begin{pmatrix}
\cosh 2r_B+\cos\zeta_B\sinh 2r_B && \sin\zeta_B\sinh 2r_B\\
\sin\zeta_B\sinh 2r_B && \cosh 2r_B-\cos\zeta_B\sinh 2r_B 
\end{pmatrix}~.
\end{split}
\end{equation}
\subsection{Covariance matrix of the BEC under graviton interaction}
We shall now write down the expression of the covariance matrix of a Bose-Einstein condensate with squeezed phonon modes when it starts interacting with a gravitational wave. The covariance matrices are the matrix or phase space representation of the Gaussian states. As discussed in \cite{AhmadiBruschiFuentes}, if an unitary transformation in the Hilbert space is generated by a quadratic Hamiltonian then in the phase space the same action can be described by a symplectic matrix. Using the Bogoliubov coefficients one can construct a symplectic matrix of the form
\begin{equation}\label{QSymplectic.1}
\mathcal{S}_{\mathcal{M}}=\begin{pmatrix}
\mathcal{M}_{11}&&\mathcal{M}_{12}&&\cdots\\
\mathcal{M}_{21}&&\mathcal{M}_{22}&&\cdots\\
\vdots&&\vdots&&\ddots
\end{pmatrix}
\end{equation}
where for an $n$-mode bosonic system the $(2n\times 2n)$ symplectic matrix $\mathcal{S}_{\mathcal{M}}$ satisfies
\begin{equation}\label{QSymplectic.2}
\mathcal{S}_{\mathcal{M}}\Omega \mathcal{S}_{\mathcal{M}}^T=\Omega
\end{equation}
with $\Omega=\mathop{\oplus}\limits_{j=1}^n\Omega_j$ and $\Omega_j=-i\sigma_2$. The matrices $\mathcal{M}_{ij}$ are constructed out of the Bogoliubov coefficients as
\begin{equation}\label{QSymplectic.3}
\mathcal{M}_{ij}=\begin{pmatrix}
\Re\left(\alpha^B_{ij}-\beta^B_{ij}\right)&&\Im\left(\alpha^B_{ij}+\beta^B_{ij}\right)\\
-\Im\left(\alpha^B_{ij}-\beta^B_{ij}\right)&&\Re\left(\alpha^B_{ij}+\beta^B_{ij}\right)
\end{pmatrix}
\end{equation}
with $\Re$ denoting the real part and $\Im$ denoting the imaginary part. Under this transformation by the  symplectic matrix, the initial covariance matrix $\Sigma_i$ corresponding to the Gaussian matrix transforms as
\begin{equation}\label{QSymplectic.4}
\Sigma_f=\mathcal{S}_\mathcal{M}\Sigma_i\mathcal{S}_{\mathcal{M}}^T
\end{equation}
The transformation of the $j$-th mode of the Bose-Einstein condensate by the gravitational wave can be represented as \cite{MatthewMannAffshordi}
\begin{equation}\label{QSymplectic.5}
\Sigma_j^\mathcal{M}(\varepsilon_B)=\mathcal{M}_{jj}(\varepsilon_B)\Sigma_{\text{Sq.}}[0]\mathcal{M}_{jj}^T(\varepsilon_B)+\sum\limits_{k\neq j}\mathcal{M}_{jk}(\varepsilon_B)\mathcal{M}_{jk}^T(\varepsilon_B)
\end{equation}
where $\varepsilon_B$ is defined in eq.(\ref{QGRBEC.104}) and $\Sigma_{\text{Sq.}}[0]$ is obtained in eq.(\ref{QGCovariance.6}). One important observation that needs to be made is that in eq.(\ref{QSymplectic.3}) the Bogoliubov coefficients $\alpha^B_{ij}$ and $\beta^B_{ij}$ are classical. In our current analysis, the Bogoliubov coefficients are infused by random noise-fluctuations due to graviton interactions which lead to operator valued Bogoliubov coefficients in eq.(s)(\ref{QGRBEC.102},\ref{QGRBEC.103}).  Hence, it is quite straightforward to understand that the symplectic matrix in eq.(\ref{QSymplectic.3}) will have contributions from the graviton noise term. The modified operator valued matrix reads
\begin{equation}\label{QSymplectic.6}
\begin{split}
\doublehat{\mathcal{M}}_{ij}(\varepsilon_B)\equiv& \mathcal{M}_{ij}(\varepsilon_B)+\delta\doublehat{\mathcal{M}}_{ij}(\varepsilon_B)\\
=&\mathcal{M}_{ij}(\varepsilon_B)
+\begin{pmatrix}
\Re\left(\mathcal{c}_{\alpha_{ij}}-\mathcal{c}_{\beta_{ij}}\right)&&\Im\left(\mathcal{c}_{\alpha_{ij}}+\mathcal{c}_{\beta_{ij}}\right)\\
-\Im\left(\mathcal{c}_{\alpha_{ij}}-\mathcal{c}_{\beta_{ij}}\right)&&\Re\left(\mathcal{c}_{\alpha_{ij}}+\mathcal{c}_{\beta_{ij}}\right)
\end{pmatrix}\delta\hat{\mathcal{N}}(\tau_\text{M},0)~.
\end{split}
\end{equation}
Here, the `~$\doublehat{}$~' symbol over $\mathcal{M}$ indicates a matrix with stochastic elements. Hence, after the BEC has interacted with gravitons, the transformed covariance matrix from eq.(\ref{QSymplectic.5}) now takes the form
\begin{equation}\label{QSymplectic.7}
\doublehat{\Sigma}_j^\mathcal{M}(\varepsilon_B)=\doublehat{\mathcal{M}}_{jj}(\varepsilon_B)\Sigma_{\text{Sq.}}[0]\doublehat{\mathcal{M}}_{jj}^T(\varepsilon_B)+\sum\limits_{k\neq j}\doublehat{\mathcal{M}}_{jk}(\varepsilon_B)\doublehat{\mathcal{M}}_{jk}^T(\varepsilon_B)~.
\end{equation}
For this $n$-node bosonic system no two modes influence one-another and as a result the Bogoliubov coefficients also don't involve two modes. It is therefore safe to write down for a classical gravitational wave \cite{MatthewMannAffshordi}, $\alpha_{ij}^B=\delta_{ij}\alpha^B$ and $\beta_{ij}^B=\delta_{ij}\beta^B$. Same thing holds true for the graviton-noise induced Bogoliubov coefficients and we can express them as
\begin{equation}\label{QSymplectic.8}
\hat{\alpha}_{ij}=\delta_{ij}\hat{\alpha}^B,~\hat{\beta}_{ij}=\delta_{ij}\hat{\beta}^B~.
\end{equation}
As from eq.(s)(\ref{QGRBEC.102},\ref{QGRBEC.103}), we already know the analytical form of the stochastic Bogoliubov coefficients, it is possible to analytically obtain the form of the matrix $\doublehat{\mathcal{M}}_{ij}(\varepsilon_B)$ from eq.(\ref{QSymplectic.6}) and for a single mode of the BEC one can set $i=j$ where $j\in\{1,\cdots,n\}$. For simplicity, we can set $i=j=1$. The analytical form of $\doublehat{\mathcal{M}}_{11}(\varepsilon_B)$ is given by
\begin{equation}\label{QSymplectic.9}
\doublehat{\mathcal{M}}_{11}(\varepsilon_B)=
\begin{pmatrix}
1-\frac{\varepsilon_B}{2\varepsilon}\left(\delta\hat{\mathcal{N}}(\tau_{\text{M}},0)+\frac{\sqrt{\pi}\varepsilon\omega_B\tau}{2}\mathcal{C}^{\omega_B}_{\Omega_0}(\tau)\right)&&-\frac{\varepsilon_B}{2\varepsilon}\omega_B\tau_\text{M}\delta\hat{\mathcal{N}}(\tau_{\text{M}},0)\\
\frac{\varepsilon_B}{2\varepsilon}\omega_B\tau_\text{M}\delta\hat{\mathcal{N}}(\tau_{\text{M}},0)&&1+\frac{\sqrt{\pi}\varepsilon_B\omega_B\tau}{4}\mathcal{C}^{\omega_B}_{\Omega_0}(\tau)
\end{pmatrix}
\end{equation}
where $\mathcal{C}_{\Omega_0}^{\omega_B}(\tau)$ is defined as
\begin{equation}\label{QSymplectic.10}
\mathcal{C}_{\Omega_0}^{\omega_B}(\tau)=\left(e^{-\frac{\tau^2}{4}(\Omega_0-2\omega_B)^2}-e^{-\frac{\tau^2}{4}(\Omega_0+2\omega_B)^2}\right)~.
\end{equation}
\section{Quantum gravitational Fisher information}\label{QGFI}
In this section, we shall calculate the quantum gravity modified Fisher information and then write down the quantum gravity modified Cram\'{e}r-Rao bound. At first, we shall discuss the methodology for writing the quantum Fisher information between two nearby covariance matrices.
\subsection{Quantum Fisher information for two covariance matrices}\label{6.7.1}
We have discussed in subsection (\ref{6.1.3}) that it is easy to deal with covariance matrices for Gaussian states. At first we shall follow the methodology of \cite{AhmadiBruschiFuentes} to express the Quantum Fisher information when the quantum states are represented by covariance matrices. In subsection (\ref{6.1.2}), we have discussed how to express the quantum Fisher information in terms of fidelity between two nearby density matrices. Hence, the first step is to write down the analytical expression for the fidelity between two quantum states when the quantum states are represented by covariance matrices. One can write down the fidelity or overlap between two covariance matrices $\Sigma_A$ and $\Sigma_B$ as \cite{UhlmannFidelity}
\begin{equation}\label{FidelityCovariance.1}
\mathcal{F}(\Sigma_A,\Sigma_B)=\frac{1}{\sqrt{\Xi_\Sigma+\Delta_\Sigma}-\sqrt{\Xi_\Sigma}}
\end{equation} 
for a single mode bosonic system where the analytical forms of $\Xi_\Sigma$ and $\Delta_\Sigma$ are given as
\begin{align}
\Xi_\Sigma(\Sigma_A,\Sigma_B)&=\frac{1}{4}\det\left[\Sigma_A+\frac{i}{2}\mathcal{O}\right]\det\left[\Sigma_B+\frac{i}{2}\mathcal{O}\right]\label{FidelityCovariance.2}\\
\Delta_\Sigma(\Sigma_A,\Sigma_B)&=\frac{1}{4}\det\left[\Sigma_A+\Sigma_B\right]
\end{align}
with $\mathcal{O}$ being defined as $\mathcal{O}\equiv\mathop{\oplus}\limits_{j=1}^ni\sigma_2$ and for a single mode bosonic system $\mathcal{O}=i\sigma_2$. Here, we shall briefly discuss the methodology presented in \cite{AhmadiBruschiFuentes} and obtain the analytical expression for the quantum Fisher information when the covariance matrix formalism is being considered. If $\vartheta$ is the small parameter under consideration then it is always possible to expand $\Sigma$ and as a result $\mathcal{F}$ and the quantum Fisher information $\mathcal{H}(\vartheta)$ in powers of $\vartheta$. We have observed from eq.(\ref{QSymplectic.4}) that the covariance matrix transforms via the action of the symplectic matrix $\mathcal{S}_\mathcal{M}$ in eq.(\ref{QSymplectic.1}). For a single mode consideration the matrix $\mathcal{S}_\mathcal{M}$ is equal to $\mathcal{M}_{jj}$ corresponding to the $j$-th mode of the $n$-mode bosonic system and the related transformation equation in eq.(\ref{QSymplectic.4}) takes the form $\Sigma_f=\mathcal{M}_{jj}\Sigma_i\mathcal{M}^T_{jj}$ where one needs to remember that no two modes of the bosonic system influence one another and $\mathcal{M}_{jj}$ is a $2\times 2$ matrix. If the uncertainty relation is still preserved in this process then the Bogoliubov coefficients can be expressed as \cite{AhmadiBruschiFuentes}
\begin{equation}\label{FidelityCovariance.3}
\begin{split}
\alpha_{jk}^B(\vartheta)&\simeq \alpha_{jk}^{B^{(0)}}+\vartheta \alpha_{jk}^{B^{(1)}}+\vartheta^2 \alpha_{jk}^{B^{(2)}}\\
\beta_{jk}^B(\vartheta)&\simeq \vartheta \beta_{jk}^{B^{(1)}}+\vartheta^2 \beta_{jk}^{B^{(2)}}
\end{split}
\end{equation}
where the expansions are kept up to $\mathcal{O}(\vartheta^2)$. It is possible to express all the first and second order moments of the column matrix $\mathbb{r}$ in powers of the small estimation parameter $\vartheta$ and as a result it is possible to express the covariance matrix in powers of this estimation parameter as
\begin{equation}\label{FidelityCovariance.4}
\begin{split}
\Sigma_f(\vartheta)\simeq \Sigma_f^{(0)}+\vartheta \Sigma_f^{(1)}+\vartheta^2\Sigma_f^{(2)}~.
\end{split}
\end{equation}
For notational simplicity, one can get rid of the $f$ in the subscript of $\Sigma$.
For a single quantum state, its overlap with itself is maximum and it is then safe to say that $\mathcal{F}\left(\Sigma(\vartheta),\Sigma(\vartheta)\right)=1$. Following the analysis in \cite{UhlmannFidelity}, one can also impose a second criteria that is given by $\frac{\partial \mathcal{F}(\Sigma(\vartheta),\Sigma(\vartheta+d\vartheta))}{\partial \vartheta}\rvert_{d\vartheta=0}\simeq 0$ which implies that for very small change in the estimation parameter the rate of change of the Fidelity is negligible. This can also be understood intuitively quite well. It is then possible to expand the Fidelity $\mathcal{F}(\Sigma(\vartheta),\Sigma(\vartheta+d\vartheta))$ using the above two conditions as 
\begin{equation}\label{FidelityCovariance.5}
\mathcal{F}(\Sigma(\vartheta),\Sigma(\vartheta+d\vartheta))\simeq 1-\mathcal{F}^{(2)}_\Sigma d\vartheta^2+\mathcal{O}(\vartheta^2d\vartheta+\vartheta d\vartheta^2)
\end{equation}
with the $\mathcal{F}_\Sigma^{(2)}$ term given as $\mathcal{F}_\Sigma^{(2)}=\mathcal{F}_\mathcal{E}^{(2)}+\mathcal{F}_{\mathcal{C}}^{(2)}$. The $(2)$ in the superscript denotes that it is a coefficient of $\mathcal{O}(\vartheta^2,\vartheta d\vartheta,d\vartheta^2)$ term. If squeezed states are in consideration then the $\mathcal{F}_\mathcal{E}^{(2)}$ term is dependent on the displacement of the squeezed state \cite{AhmadiBruschiFuentes}. In our current analysis, the squeezed phonon states have zero displacement and as a result $\mathcal{F}_\mathcal{E}^{(2)}=0$. The analytical form of $\mathcal{F}_\mathcal{C}^{(2)}$ reads
\begin{equation}\label{FidelityCovariance.6}
\begin{split}
\mathcal{F}_\Sigma^{(2)}=\mathcal{F}^{(2)}_{\mathcal{C}}=\frac{1}{2}\left(\Sigma_{11}^{(0)}\Sigma_{22}^{(2)}-2\Sigma_{12}^{(0)}\Sigma_{12}^{(2)}+\Sigma_{11}^{(2)}\Sigma_{22}^{(0)}\right)+\frac{1}{8}\left(\Sigma_{11}^{(1)}\Sigma_{22}^{(1)}-2\Sigma_{12}^{(1)}\Sigma_{12}^{(1)}\right)
\end{split}
\end{equation}
where $\Sigma_{ij}^{(k)}$ symbol denotes the ${i.j}$-th element of the matrix (for a $2\times 2$ matrix $i,j\in 1,2$) and $k$ denotes the order of the coefficient as can be easily understood from the expression in eq.(\ref{FidelityCovariance.4}). Now substituting the expansion of the Fidelity in eq.(\ref{FidelityCovariance.5}) and substituting it back in the expression for the quantum Fisher information in eq.(\ref{Bures.5}), one obtains
\begin{equation}\label{FidelityCovariance.7}
\mathcal{H}(\vartheta)\simeq 4\mathcal{F}_\Sigma^{(2)}=4\mathcal{F}_{\mathcal{C}}^{(2)}~.
\end{equation} 
The above equation gives the analytical expression for the quantum Fisher information for two nearby covariance matrices.
\subsection{The quantum gravitational Fisher information}\label{6.7.2}
Substituting the matrix $\doublehat{\mathcal{M}}(\varepsilon_B)$ from eq.(\ref{QSymplectic.9})\footnote{Here, we have removed the mode numbering from the subscript of $\doublehat{\mathcal{M}}$ as our entire analysis is focussed about the single mode of a Bose-Einstein condensate.} in eq.(\ref{QSymplectic.7}), one obtains the analytical form of the the covariance matrix for a single mode of the Bose-Einstein condensate after the gravitons have interacted with it. Now the covariance matrix after graviton interaction will also have influence from the quantum gravitational noise term of which the matrix elements can be written as $\doublehat{\Sigma}^\mathcal{M}_{ij}(\varepsilon_B)=\doublehat{\Sigma}^\mathcal{M^{(0)}}_{ij}+\varepsilon_B\doublehat{\Sigma}^\mathcal{M^{(1)}}_{ij}+\varepsilon_B^2\doublehat{\Sigma}^\mathcal{M^{(2)}}_{ij}$ $\forall ~i,j\in{1,2}$. The zeroth order coefficient is always independent of the graviton induced noise and as a result $\doublehat{\Sigma}^\mathcal{M^{(0)}}_{ij}=\Sigma^\mathcal{M^{(0)}}_{ij}$. We now assume that eq.(\ref{FidelityCovariance.6}) still holds true for a quantum gravity set-up. in such a scenario, it is possible to write down the quantum gravity modified Fisher information using the matrix elements of $\doublehat{\Sigma}^\mathcal{M}(\varepsilon_B)$ as
\begin{equation}\label{QGFI.1}
\hat{\mathcal{H}}(\varepsilon_B)=\frac{1}{2}\left(\doublehat{\Sigma}_{11}^{\mathcal{M}^{(0)}}\doublehat{\Sigma}_{22}^{\mathcal{M}^{(2)}}-2\doublehat{\Sigma}_{12}^{\mathcal{M}^{(0)}}\doublehat{\Sigma}_{12}^{\mathcal{M}^{(2)}}+\doublehat{\Sigma}_{11}^{\mathcal{M}^{(2)}}\doublehat{\Sigma}_{22}^{\mathcal{M}^{(0)}}\right)+\frac{1}{8}\left(\doublehat{\Sigma}_{11}^{\mathcal{M}^{(1)}}\doublehat{\Sigma}_{22}^{\mathcal{M}^{(1)}}-2\doublehat{\Sigma}_{12}^{\mathcal{M}^{(1)}}\doublehat{\Sigma}_{12}^{\mathcal{M}^{(1)}}\right)~.
\end{equation} 
In our current analysis and using the analytical forms of the components of the matrix, the quantum gravity modified Fisher information reads
\begin{equation}\label{QGFI.2}
\begin{split}
\hat{\mathcal{H}}(\varepsilon)&\simeq\mathcal{H}^{(0)}(\varepsilon)+\frac{\delta\hat{\mathcal{N}}(\tau_{\text{M}},0)}{32\varepsilon}\mathcal{H}^{(1)}(\varepsilon)+\frac{\left(\delta\hat{\mathcal{N}}(\tau_{\text{M}},0)\right)^2}{16\varepsilon^2}\mathcal{H}^{(2)}(\varepsilon)
\end{split}
\end{equation}
where the analytical forms of the coefficients are given by $\mathcal{H}^{(0)}(\varepsilon)$, $\mathcal{H}^{(1)}(\varepsilon)$, and $\mathcal{H}^{(2)}(\varepsilon)$ as
\begin{equation}\label{QGFI.3}
\begin{split}
\mathcal{H}^{(0)}(\varepsilon)&=\frac{1}{64}\pi\omega_B^2\tau^2
\left(e^{2\omega_B\Omega_0\tau^2}-1\right)^2e^{-\frac{\tau^2}{2}(\Omega_0+2\omega_B)^2}\left(1+\cosh 4r_B+(1-3\cos 2\zeta_B)\sinh^22r_B\right)\\
\mathcal{H}^{(1)}(\varepsilon)&=\sqrt{\pi}\omega_B\tau
\left(e^{2\omega_B\Omega_0\tau^2}-1\right)e^{-\frac{\tau^2}{4}(\Omega_0+2\omega_B)^2}\bigr(2\cosh^22r_B+(1-3\cos 2\zeta_B)\sinh^22r_B\\&+6\omega_B\tau_{\text{M}}\sin\zeta_B\sinh 4r_B\bigr)\\
\mathcal{H}^{(2)}(\varepsilon)&=1+\cosh 4r_B-\frac{\sinh^2 2r_B}{2}(3+\cos 2\zeta_B)+\omega_B\tau_\text{M}\Bigr(\sin \zeta_B\bigr(3\sinh 4r_B+2\sinh^2 2r_B\\& \times(\cos\zeta_B+\omega_B\tau_{\text{M}}\sin\zeta_B)\bigr)+4\omega_B\tau_{\text{M}}\cosh^22r_B\Bigr)~.
\end{split}
\end{equation}
In eq.(\ref{QGFI.2}) if $\delta\hat{\mathcal{N}}(\tau_{\text{M}},0)$ is set to zero then the quantum gravity modified Fisher information $\hat{\mathcal{H}}(\varepsilon)$ becomes equal to the quantum Fisher information $\mathcal{H}^{(0)}(\varepsilon)$. The analytical form of $\mathcal{H}^{(0)}(\varepsilon)$ from the above equation matches exactly with the result obtained in \cite{MatthewMannAffshordi} which serves as a nice consistency check. The important thing to observe is that the quantum gravity modified Fisher information has stochastic contributions due to graviton interaction. Therefore it is more prudent to call it as the ``quantum gravitational Fisher information" or the QGFI instead of calling it quantum gravity modified Fisher information. As it is $\hat{\mathcal{H}}(\varepsilon)$ can not give rise to any meaningful physical quantities hence, we take an average with respect to initial graviton states (also can be considered as a stochastic average). After taking the stochastic average with respect to graviton states, one can recast eq.(QGFI.2) as
\begin{equation}\label{QGFI.4}
\begin{split}
\llangle \hat{\mathcal{H}}(\varepsilon)\rrangle&\equiv\langle\Psi_\text{Graviton}|\hat{\mathcal{H}}(\varepsilon)|\Psi_\text{Graviton}\rangle\\
&=\mathcal{H}^{(0)}(\varepsilon)+\frac{\mathcal{H}^{(2)}(\varepsilon)}{32\varepsilon^2}\langle\Psi_\text{Graviton}| \{\delta\hat{\mathcal{N}}(\tau_{\text{M}},0),\delta\hat{\mathcal{N}}(\tau_\text{M},0)\}|\Psi_\text{Graviton}\rangle\\
\implies \llangle\hat{\mathcal{H}}(\varepsilon)\rrangle&=\mathcal{H}^{(0)}(\varepsilon)+\frac{\mathcal{H}^{(2)}(\varepsilon)}{32\varepsilon^2}\llangle \{\delta\hat{\mathcal{N}}(\tau_{\text{M}},0),\delta\hat{\mathcal{N}}(\tau_\text{M},0)\}\rrangle~.
\end{split}
\end{equation}
With this expectation value of the quantum gravitational Fisher information, we can write down the quantum gravity modified Cram\'{e}r-Rao bound as the stochastic average of the QGFI is a number for some fixed values of the parameter. As discussed earlier, $\delta\hat{\mathcal{N}}(\tau_{\text{M}},0)$ is a noise term and therefore the one point correlator $\llangle \delta\hat{\mathcal{N}}(\tau_{\text{M}},0)\rrangle=0$ resulting in the vanishing of the term in the expression of $\llangle\hat{\mathcal{H}}(\varepsilon)\rrangle$ in eq.(\ref{QGFI.4}). In order to obtain the analytical form of the expectation value of the QGFI, we need the analytical form of the two-point correlator when the initial gravitons are in the squeezed state.
\subsubsection{Squeezed graviton states and the two-point correlator}
\begin{figure}
\begin{center}
\includegraphics[scale=0.4]{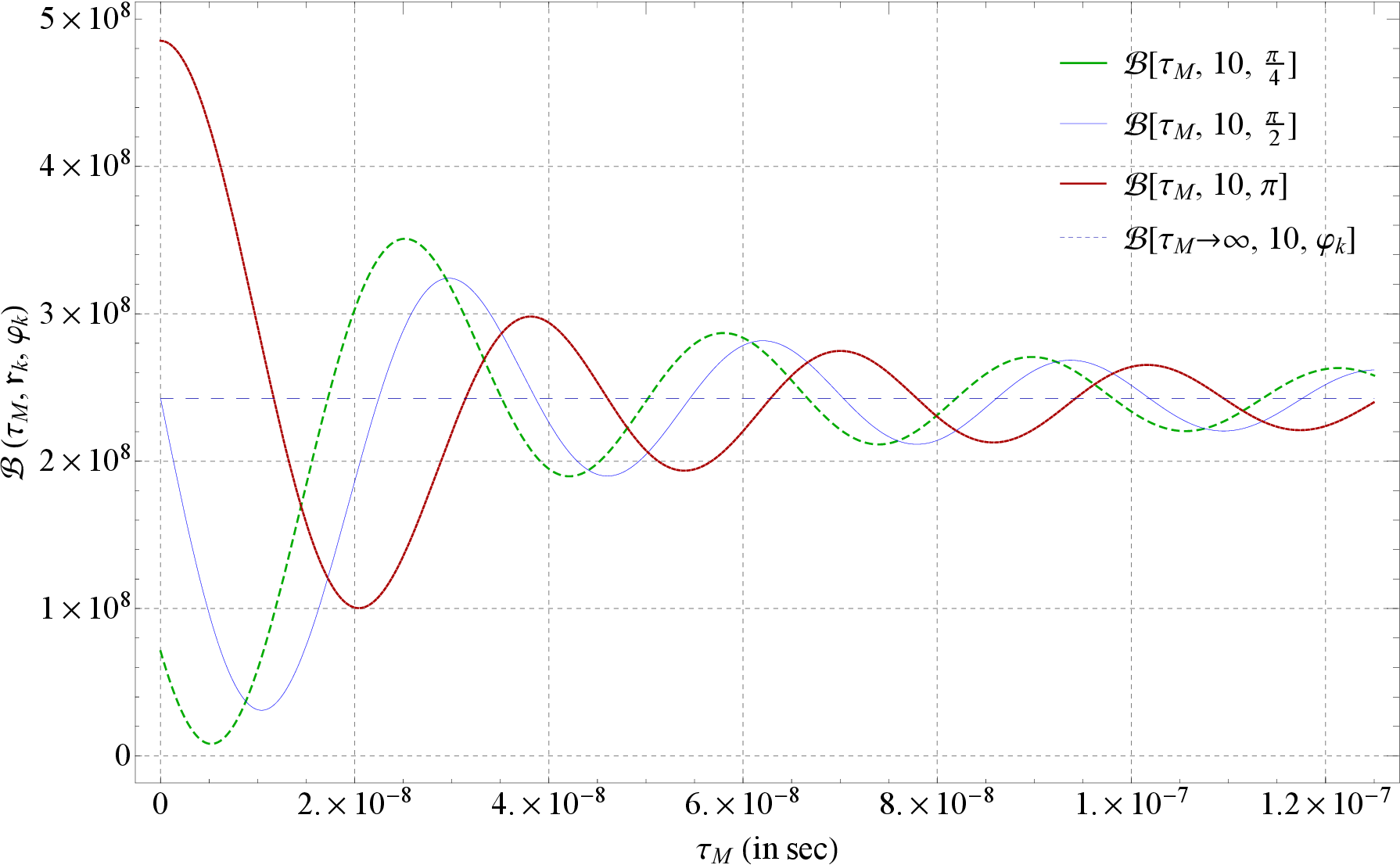}
\caption{We plot the function $\mathcal{B}(t,\mathfrak{r}_k,\varphi_k)$ against $t$ when $\mathfrak{r}_k=10$ and $\Omega_{\text{M}}=10^8$ Hz for the values of the squeezing angle $\varphi_k=\left\{\frac{\pi}{4},\frac{\pi}{2},\pi\right\}$. We have plotted $\mathcal{B}(t\rightarrow\infty,10,\varphi_k)$ as a reference line to have a proper understanding of the fluctuations.\label{Fluctuation_B_OTM} }
\end{center}
\end{figure}

Using the definition of the noise fluctuation from eq.(\ref{QGRBEC.67}), we obtain the analytical form of the two point correlator for arbitrary graviton states as
\begin{equation}\label{NoiseSqueezed.1}
\begin{split}
\llangle\{\delta \hat{\mathcal{N}}_{ij}(\tau,0)\delta\hat{\mathcal{N}}_{kl}(\tau',0)\}\rrangle&=\frac{4\kappa_G^2}{V_G}\sum\limits_{\vec{k},q}\sum\limits_{\vec{k}',q'}\epsilon_{ij}^q(\vec{k})\epsilon_{lk}^{q'}(\vec{k}')\llangle \{\delta\hat{h}^I_{q'}(\tau,\vec{k}),\delta\hat{h}^I_q(\tau',\vec{k}')\}\rrangle~.
\end{split}
\end{equation} 
To simplify the above expression, we need to find out the expression for the two-point correlator when the gravitons are in the squeezed state. The squeezing and displacement operators read
\begin{align}
\hat{S}_G(\mathfrak{r}^{\text{Sq.}})&=\exp\left[\frac{1}{V_G}\sum\limits_{\vec{k},q}\left({\mathfrak{r}^{\text{Sq.}}_k}^*\hat{a}_q(\vec{k})\hat{a}_q(-\vec{k})+\mathfrak{r}^{\text{Sq.}}_k\hat{a}^\dagger_q(\vec{k})\hat{a}^\dagger_q(-\vec{k})\right)\right]\label{NoiseSqueezed.2}\\
\hat{\mathcal{D}}_G(\mathfrak{b})&=\exp\left[\frac{1}{V_G}\sum\limits_{\vec{k},q}\left(\mathfrak{b}_k\hat{a}^\dagger_q(\vec{k})-\mathfrak{b}_k\hat{a}^\dagger_q(\vec{k})\right)\right]\label{NoiseSqueezed.3}
\end{align}
where the squeezing parameter corresponding to the mode with frequency $\omega=c k$ is given as $\mathfrak{r}_k^{\text{Sq.}}=\mathfrak{r}_ke^{i\varphi_k}$. It is important to note that both $\hat{S}_G(\mathfrak{r}^{\text{Sq.}})$ and $\hat{\mathcal{D}}_G(\mathfrak{b})$ are unitary in nature. Under the action of the above operators, the displaced squeezed state of the graviton reads
\begin{equation}\label{NoiseSqueezed.4}
\begin{split}
|\mathfrak{r}^{\text{Sq.}},\mathfrak{b}\rangle=\hat{S}_G(\mathfrak{r}^{\text{Sq.}})\hat{\mathcal{D}}_G(\mathfrak{b})|0\rangle~.
\end{split}
\end{equation}
The squeezed mode function $f_k^{\text{Sq.}}(t)$ in terms of the Minkowski mode functions $f_k(t)=\frac{1}{\sqrt{2k}}\exp[-i k t]$ reads
\begin{equation}\label{NoiseSqueezed.5}
f_k^{\text{Sq.}}(t)=f_k(t)\cosh\mathfrak{r}_k-e^{-i\varphi_k}f_k^*(t)\sinh \mathfrak{r}_k~.
\end{equation}
We already know the mode decomposition of $\hat{h}^I_q(t,\vec{k})=f_k(t)\hat{a}_q(\vec{k})+f_k^*(t)\hat{a}_q^\dagger(-\vec{k})$ and the quantum gravity fluctuation reads $\delta\hat{h}_q^I(t,\vec{k})=\hat{h}_q^I(t,\vec{k})-\llangle \hat{h}_q^I(t,\vec{k})\rrangle$. Let us now define a new operator
\begin{equation}\label{NoiseSqueezed.6}
\begin{split}
\hat{\mathcal{A}}_{\vec{k},q}(\mathfrak{r}^{\text{Sq.}},\mathfrak{b})\equiv \hat{\mathcal{D}}_G^\dagger(\mathfrak{b})\hat{S}_G^\dagger(\mathfrak{r}^{\text{Sq.}})\hat{a}_q(\vec{k})\hat{S}_G(\mathfrak{r}^{\text{Sq.}})\hat{\mathcal{D}}_G(\mathfrak{b})
\end{split}
\end{equation}
Making use of the analytical expressions for the squeezing as well as the displacement operators we can write down the expressions for $\hat{\mathcal{A}}_{\vec{k},q}(\mathfrak{r}^{\text{Sq.}},\mathfrak{b})$ and $\hat{\mathcal{A}}^\dagger_{-\vec{k},q}(\mathfrak{r}^{\text{Sq.}},\mathfrak{b})$ as
\begin{align}
\hat{\mathcal{A}}_{\vec{k},q}(\mathfrak{r}^{\text{Sq.}},\mathfrak{b})&=\left(\hat{a}_q(\vec{k})+\mathfrak{b}_k\right)\cosh \mathfrak{r}_k-\left(\hat{a}^\dagger_q(-\vec{k})+\mathfrak{b}^*_{k}\right)e^{i\varphi_k}\sinh \mathfrak{r}_k~,\label{NoiseSqueezed.7}\\
\hat{\mathcal{A}}^\dagger_{-\vec{k},q}(\mathfrak{r}^{\text{Sq.}},\mathfrak{b})&=\left(\hat{a}^\dagger_q(-\vec{k})+\mathfrak{b}^*_k\right)\cosh \mathfrak{r}_k-\left(\hat{a}_q(\vec{k})+\mathfrak{b}_{k}\right)e^{-i\varphi_k}\sinh \mathfrak{r}_k\label{NoiseSqueezed.8}
\end{align}
where $k$ remains independent of the sign in front of $\vec{k}$.
Using eq.(s)(\ref{NoiseSqueezed.7},\ref{NoiseSqueezed.8}) and making use of eq.(\ref{NoiseSqueezed.5}), we obtain the following analytical expression
\begin{equation}\label{NoiseSqueezed.9}
\begin{split}
&\hat{\mathcal{D}}_G^\dagger(\mathfrak{b})\hat{S}_G^\dagger(\mathfrak{r}^{\text{Sq.}})\delta\hat{h}_q^I(t,\vec{k})\hat{S}_G(\mathfrak{r}^{\text{Sq.}})\hat{\mathcal{D}}_G(\mathfrak{b})\\=&\hat{\mathcal{D}}_G^\dagger(\mathfrak{b})\hat{S}_G^\dagger(\mathfrak{r}^{\text{Sq.}})\hat{h}_q^I(t,\vec{k})\hat{S}_G(\mathfrak{r}^{\text{Sq.}})\hat{\mathcal{D}}_G(\mathfrak{b})-\langle 0|\hat{\mathcal{D}}_G^\dagger(\mathfrak{b})\hat{S}_G^\dagger(\mathfrak{r}^{\text{Sq.}})\hat{h}_q^I(t,\vec{k})\hat{S}_G(\mathfrak{r}^{\text{Sq.}})\hat{\mathcal{D}}_G(\mathfrak{b})|0\rangle\\
=&f_k(t)\left(\hat{\mathcal{A}}_{\vec{k},q}(\mathfrak{r}^{\text{Sq.}},\mathfrak{b})-\langle \hat{\mathcal{A}}_{\vec{k},q}(\mathfrak{r}^{\text{Sq.}},\mathfrak{b})\rangle\right)+f_k^*(t)\left(\hat{\mathcal{A}}^\dagger_{-\vec{k},q}(\mathfrak{r}^{\text{Sq.}},\mathfrak{b})-\langle \hat{\mathcal{A}}^\dagger_{-\vec{k},q}(\mathfrak{r}^{\text{Sq.}},\mathfrak{b})\rangle\right)\\
=&f_k^{\text{Sq.}}(t)\hat{a}_q(\vec{k})+{f_k^{\text{Sq.}}}^*(t)\hat{a}_q^\dagger(-\vec{k})~.
\end{split}
\end{equation}
The raising and lowering operators satisfy the commutation relation as
\begin{equation}\label{NoiseSqueezed.10}
\left[\hat{a}_q(\vec{k}),\hat{a}^\dagger_q(-\vec{k})\right]=\delta_{q,q'}\delta_{\vec{k},-\vec{k}'}~.
\end{equation}
We can then obtain the analytical form of the two-point correlator $\llangle \{\delta\hat{h}^I_{q'}(\tau,\vec{k}),\delta\hat{h}^I_q(\tau',\vec{k}')\}\rrangle$ as
\begin{equation}\label{NoiseSqueezed.11}
\begin{split}
\llangle \{\delta\hat{h}^I_{q'}(\tau,\vec{k}),\delta\hat{h}^I_q(\tau',\vec{k}')\}\rrangle&=\langle \mathfrak{r}^\text{Sq.},\mathfrak{b} |\{\delta\hat{h}^I_{q'}(\tau,\vec{k}),\delta\hat{h}^I_q(\tau',\vec{k}')\}|\mathfrak{r}^\text{Sq.},\mathfrak{b} \rangle\\
&=\left(f_k^{\text{Sq.}}(\tau){f_{k'}^{\text{Sq.}}}^*(\tau')+f_{k'}^{\text{Sq.}}(\tau'){f_k^{\text{Sq.}}}^*(\tau)\right)\delta_{q,q'}\delta_{\vec{k},-\vec{k}'}\\
&=\delta_{q,q'}\delta_{\vec{k}+\vec{k}',0}\mathcal{Q}_{\delta h}(\tau,\tau',\vec{k})
\end{split}
\end{equation}
where the function $\mathcal{Q}_{\delta h}(\tau,\tau',\vec{k})$ is defined as
\begin{equation}\label{NoiseSqueezed.12}
\mathcal{Q}_{\delta h}(\tau,\tau',\vec{k})=f_k^{\text{Sq.}}(\tau){f_{k'}^{\text{Sq.}}}^*(\tau')+f_{k'}^{\text{Sq.}}(\tau'){f_k^{\text{Sq.}}}^*(\tau)~.
\end{equation}
Using the expression for the squeezed mode function in eq.(\ref{NoiseSqueezed.5}), we can simplify the above expression as
\begin{equation}\label{NoiseSqueezed.13}
\begin{split}
\mathcal{Q}_{\delta h}(\tau,\tau',\vec{k})&=2\Re\left[f_k^{\text{Sq.}}(\tau){f_{k'}^{\text{Sq.}}}^*(\tau')\right]\\
&=\frac{1}{k}\left(\cosh 2\mathfrak{r}_k\cos(k(\tau-\tau'))-\sinh 2 \mathfrak{r}_k\cos(k(\tau+\tau')-\varphi_k)\right)~.
\end{split}
\end{equation}
Using the analytical form of $\llangle \{\delta\hat{h}^I_{q'}(\tau,\vec{k}),\delta\hat{h}^I_q(\tau',\vec{k}')\}\rrangle$ from eq.(\ref{NoiseSqueezed.11}) in eq.(\ref{NoiseSqueezed.1}), we can finally write down the expression for the two point noise-noise correlator as 
\begin{equation}\label{NoiseSqueezed.14}
\begin{split}
\llangle\{\delta \hat{\mathcal{N}}_{ij}(\tau,0)\delta\hat{\mathcal{N}}_{kl}(\tau',0)\}\rrangle&=\frac{4\kappa_G^2}{V_G}\sum\limits_{\vec{k},q}\epsilon^q_{ij}(\vec{k})\epsilon_{lk}^q(-\vec{k})\mathcal{Q}_{\delta h}(\tau,\tau',\vec{k})~.
\end{split}
\end{equation}
We now go to the continuous frequency limit and in such a scenario, we can directly make use of eq.(\ref{QGRBEC.75}), we can simplify the above equation as
\begin{equation}\label{NoiseSqueezed.15}
\begin{split}
\llangle\{\delta \hat{\mathcal{N}}_{ij}(\tau,0)\delta\hat{\mathcal{N}}_{kl}(\tau',0)\}\rrangle&=\frac{2\kappa_G^2}{5\pi^2}\left(\delta_{ik}\delta_{jl}+\delta_{il}\delta_{jk}-\frac{2}{3}\delta_{ij}\delta_{kl}\right)\int_0^{\Omega_{\text{M}}}dk k^2\mathcal{Q}_{\delta h}(\tau,\tau',\vec{k})~.
\end{split}
\end{equation}
For $\tau=\tau'$, we can recast the above equation after executing the $k$ integral as
\begin{equation}\label{NoiseSqueezed.16}
\begin{split}
\llangle\{\delta \hat{\mathcal{N}}_{ij}(\tau,0)\delta\hat{\mathcal{N}}_{kl}(\tau,0)\}\rrangle=&\frac{\kappa_G^2\Omega_{\text{M}}^2}{5\pi^2}\left(\delta_{ik}\delta_{jl}+\delta_{il}\delta_{jk}-\frac{2}{3}\delta_{ij}\delta_{kl}\right)\biggr(\cosh 2\mathfrak{r}_k+\frac{\sinh 2\mathfrak{r}_k}{2\Omega_{\text{M}}^2\tau^2}\bigr(\cos\varphi_k\\&-\cos(2\Omega_{\text{M}}\tau-\varphi_k)-2\Omega_{\text{M}}\tau\sin(2\Omega_{\text{M}}\tau-\varphi_k)\bigr)\biggr)~.
\end{split}
\end{equation}
For $i=j=k=l=1$, we obtain from the above equation
\begin{equation}\label{NoiseSqueezed.17}
\begin{split}
\llangle\{\delta \hat{\mathcal{N}}(\tau,0)\delta\hat{\mathcal{N}}(\tau,0)\}\rrangle&=\llangle\{\delta \hat{\mathcal{N}}_{11}(\tau,0)\delta\hat{\mathcal{N}}_{11}(\tau,0)\}\rrangle\\
&=\frac{4\kappa_G^2\Omega_{\text{M}}^2}{15\pi^2}\mathcal{B}(\tau,\mathfrak{r}_k,\varphi_k)
\end{split}
\end{equation}
where the function $\mathcal{B}(\tau,\mathfrak{r}_k,\varphi_k)$ is defined as
\begin{equation}\label{NoiseSqueezed.18}
\mathcal{B}(\tau,\mathfrak{r}_k,\varphi_k)\equiv\cosh 2\mathfrak{r}_k+\frac{\sinh 2\mathfrak{r}_k}{2\Omega_{\text{M}}^2\tau^2}\left(\cos\varphi_k-\cos(2\Omega_{\text{M}}\tau-\varphi_k)-2\Omega_{\text{M}}\tau\sin(2\Omega_{\text{M}}\tau-\varphi_k)\right)~.
\end{equation}
The important thing to observe is that the effect of squeezing in the graviton state is completely encaptured by this $\mathcal{B}(\tau,\mathfrak{r}_k,\varphi_k)$ function. One can now replace $\tau$ by $\tau_\text{M}$ in $\mathcal{B}(\tau,\mathfrak{r}_k,\varphi_k)$ and set $\tau_\text{M}\rightarrow 0$ limit to observe the value of the function initially. We find out that
\begin{equation}\label{NoiseSqueezed.19}
\lim\limits_{\tau_{\text{M}}\rightarrow0}\mathcal{B}(\tau_{\text{M}},\mathfrak{r}_k,\varphi_k)=\cosh 2\mathfrak{r}_k-\cos\varphi_k\sinh 2\mathfrak{r}_k~.
\end{equation}
The above equation shows that the function $\mathcal{B}$ will be minimum at $\varphi_k=0$ and maximum at $\varphi_k=\pi$. The cut-off frequency can be set to $\Omega_\text{M}\sim 10^8$ Hz while considering primordial gravitational waves generated during the inflationary period  \cite{KannoSodaTokuda}. For a graphical understanding, we plot $\mathcal{B}(\tau_{\text{M}},\mathfrak{r}_k,\varphi_k)$ against $\tau_{\text{M}}$ with a fixed value of the squeezing parameter $\mathfrak{r}_k=10$ and varying values of the squeezing angle $\varphi_k$ in Fig.(\ref{Fluctuation_B_OTM}). In the $\tau_{\text{M}}\rightarrow\infty$ limit $\mathcal{B}$ approaches the value $\cosh 2 \mathfrak{r}_k$ which can be approximated by $\frac{1}{2}\exp[2\mathfrak{r}_k]$ for very high value of  the graviton squeezing parameter. 
Fig.(\ref{Fluctuation_B_OTM}), implies that for sufficiently large value of the single measurement time by a single mode of the Bose-Einstein condensate $\mathcal{B}(\tau_{\text{M}},\mathfrak{r}_k,\varphi_k)$ becomes independent of the graviton squeezing angle. Although this function plays an important role while considering small time measurements.
\subsubsection{The quantum gravity modified Cram\'{e}r-Rao bound}
\begin{figure}
\begin{center}
\includegraphics[scale=0.35]{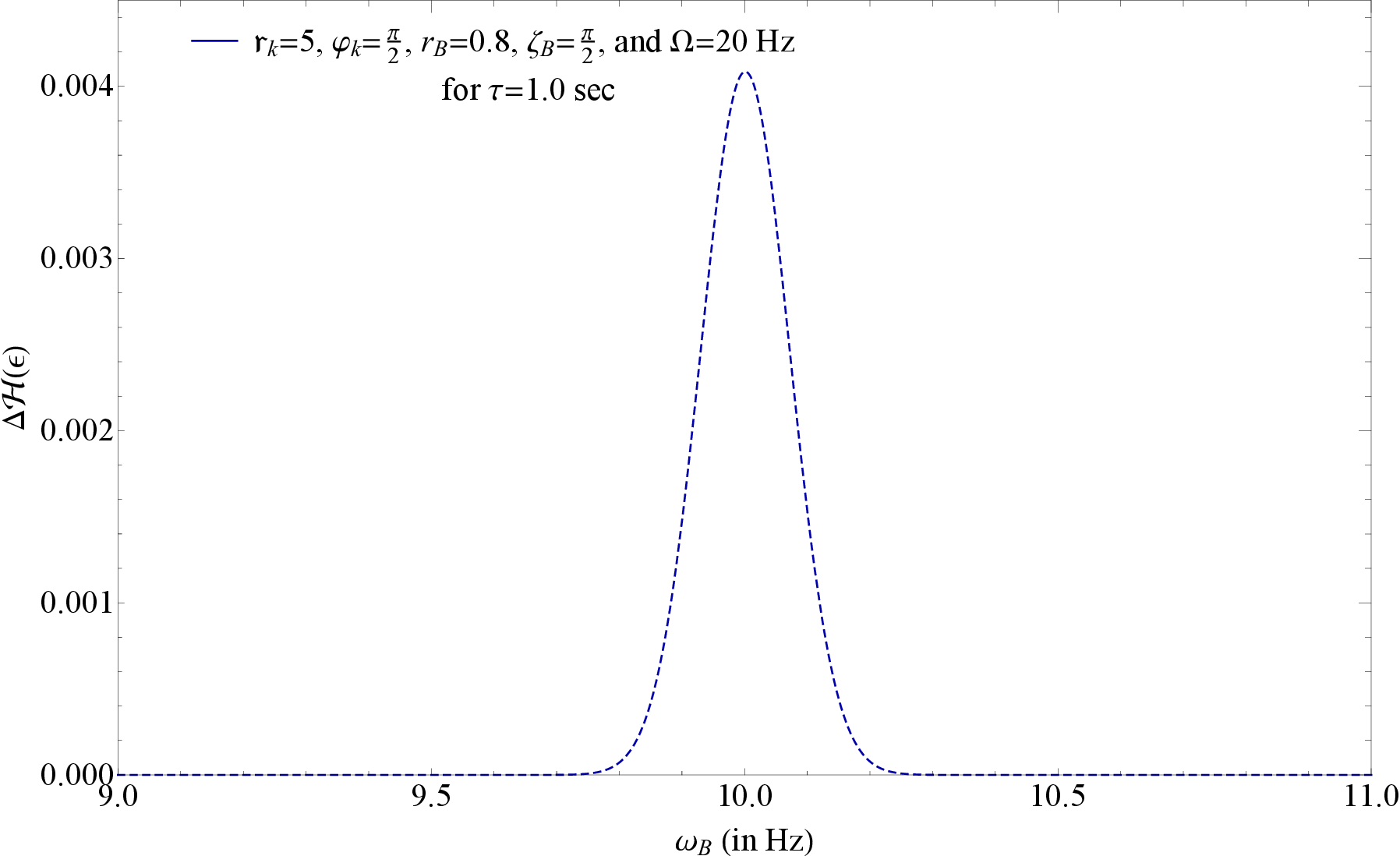}
\caption{Plot of $\Delta\mathcal{H}(\varepsilon)$ against the phonon mode frequency when a gravitational wave comes with a frequency of $\Omega_0=20$ Hz.\label{Resonance_QGFI_OTM}}
\end{center}
\end{figure}

\begin{figure}[t!]
\begin{center}
\includegraphics[scale=0.35]{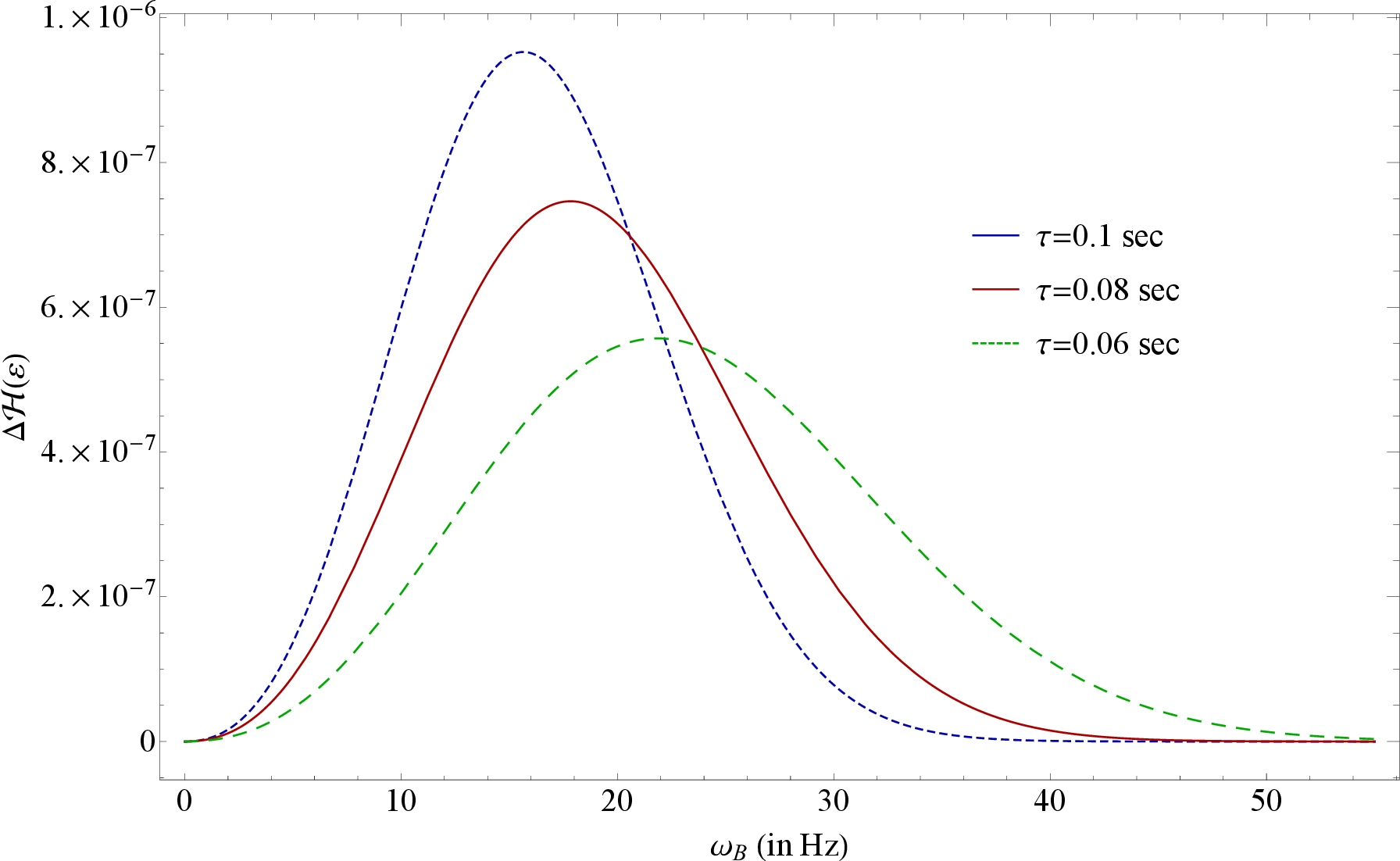}
\caption{Plot of $\Delta\mathcal{H}(\varepsilon)$ against the phonon mode frequency for a gravitational wave with a frequency of $\Omega_0=20$ Hz where the measurement time $\tau$ is fixed at different values.\label{Resonance_QGFI_Time_OTM}}
\end{center}
\end{figure}

With the form of the two-point noise correlator obtained in eq.(\ref{NoiseSqueezed.17}), we can recast the analytical expression for $\llangle\hat{\mathcal{H}}(\varepsilon)\rrangle$ from eq.(\ref{QGFI.4}) as
\begin{equation}\label{QGCramerRao.1}
\begin{split}
\llangle\hat{\mathcal{H}}(\varepsilon)\rrangle&=\mathcal{H}^{(0)}(\varepsilon)+\frac{\kappa_G^2\Omega_{\text{M}}^2}{120\varepsilon^2 \pi^2}\mathcal{B}(\tau_{\text{M}},\mathfrak{r}_k,\varphi_k)\mathcal{H}^{(2)}(\varepsilon)~.
\end{split}
\end{equation}
It is now quite logical to replace $\tau_{\text{M}}$ by $\tau$ in the above expression. Restoring the dimensions carefully and substituting the analytical form of $\kappa_G=\frac{8\pi\hbar G}{c^3}$ and using the $\tau_{\text{M}}\rightarrow\tau$ substitution, we can write down the expression for the stochastic average of QGFI as
\begin{equation}\label{QGCramerRao.2}
\llangle\hat{\mathcal{H}}(\varepsilon)\rrangle=\mathcal{H}^{(0)}(\varepsilon)+\frac{\hbar G\Omega_{\text{M}}^2}{15\pi\varepsilon^2 c^5}\mathcal{B}(\tau,\mathfrak{r}_k,\varphi_k)\mathcal{H}^{(2)}(\varepsilon)~.
\end{equation}
Using the analytical form of $\mathcal{H}^{(0)}(\varepsilon)$ and $\mathcal{H}^{(2)}(\varepsilon)$ from eq.(\ref{QGFI.3}), we can write down the above equation for the phonon squeezing angle $\zeta_B=\frac{\pi}{2}$\footnote{It is possible to squeeze the phonon modes at specific angles and is an experimentally achievable scenario \cite{SpecificPhononSqueezing,SpecificPhononSqueezing2}} as
\begin{equation}\label{QGCramerRao.3}
\begin{split}
\llangle \hat{\mathcal{H}}(\varepsilon)\rrangle&=\frac{1}{64}\pi\omega_B^2\tau^2
\left(e^{2\omega_B\Omega_0\tau^2}-1\right)^2e^{-\frac{\tau^2}{2}(\Omega_0+2\omega_B)^2}\left(1+\cosh 4r_B+4\sinh^22r_B\right)\\&+\frac{l_{\text{Pl}}^2\Omega_{\text{M}}^2}{30\pi\varepsilon^2c^2}\left(3+2\omega_B^2\tau^2+\cosh 4r_B+6\omega_B\tau\sinh 4r_B+6\omega_B^2\tau^2\cosh 4r_B\right)\\
&\times\left(\cosh 2\mathfrak{r}_k+\frac{\sinh 2\mathfrak{r}_k}{2\Omega_{\text{M}}^2\tau^2}\left(\cos\varphi_k-\cos(2\Omega_{\text{M}}\tau-\varphi_k)-2\Omega_{\text{M}}\tau\sin(2\Omega_{\text{M}}\tau-\varphi_k)\right)\right)~.
\end{split}
\end{equation}
We already know that $\mathcal{B}(\tau,\mathfrak{r}_k,\varphi_k)$ is a positive quantity and examining the above equation it is east to write down the inequality
\begin{equation}\label{QGCramerRao.4}
\llangle \hat{\mathcal{H}}(\varepsilon)\rrangle>\mathcal{H}^{(0)}(\varepsilon)~.
\end{equation}
Hence, from eq.(\ref{CramerRao.16}), it is straight forward to write down the quantum gravity modified Cram\'{e}r-Rao bound for $\mathfrak{N}$ number of independent measurements as
\begin{equation}\label{QGCramerRao.5}
\begin{split}
\langle \left(\Delta\varepsilon_B\right)^2\rangle\geq\frac{1}{\mathfrak{N}\llangle \hat{\mathcal{H}}(\varepsilon)\rrangle}
\end{split}
\end{equation}
which is the primary guiding equation in our analysis. One important thing to remember is that the quantum Fisher information is not directly an observable quantity as a result there is no restriction on the quantum gravitational correction term based on the analytical value of the standard quantum Fisher information term. At extremely high squeezing, it may so happen that the quantum gravity correction leads the standard classical gravitational wave term. For single mode BEC, we can recast $\langle(\Delta\varepsilon_B)^2\rangle$ using eq.(\ref{QGRBEC.104}) as
\begin{equation}\label{QGCramerRao.6}
\langle(\Delta\varepsilon_B)^2\rangle=\langle(
\Delta\varepsilon)^2\rangle\left(\frac{{k_B^x}^2-{k_B^y}^2}{k_B^2}\right)^2~.
\end{equation}
In spherical polar coordinates we can write down $k_B^x$ and $k_B^y$ as $k_B^x=k_B\sin\theta_B\cos\phi_B$ and $k_B^y=k_B\sin\theta_B\sin\phi_B$. In order to consider all the single mode states of the BEC, we need to integrate the square of the uncertainty in $\varepsilon_B$ over the first quadrant and from eq.(\ref{QGCramerRao.6}), we can get
\begin{equation}\label{QGCramerRao.7}
\begin{split}
\int d\Omega_B \langle(\Delta\varepsilon_B)^2\rangle&=\int_0^{\frac{\pi}{2}}d\theta_B\sin\theta_B\int_0^{\frac{\pi}{2}}d\phi_B\sin^4\theta_B\cos^22\phi_B  \langle(\Delta\varepsilon)^2\rangle\\
&=\frac{2\pi}{15}\langle(\Delta\varepsilon)^2\rangle~.
\end{split}
\end{equation}
Hence, it is now possible to rewrite the inequality in eq.(\ref{QGCramerRao.5}) as
\begin{equation}\label{QGCramerRao.8}
\begin{split}
\langle(\Delta\varepsilon)^2\rangle\geq \frac{15}{2\pi\mathfrak{N}\llangle\hat{\mathcal{H}}(\varepsilon)\rrangle}~.
\end{split}
\end{equation}
We consider that $\mathfrak{N}$ multiple measurements of the system state has taken the time $\mathfrak{t}_{\text{Total}}$ then $\mathfrak{t}_{\text{Total}}=\mathfrak{N}\tau$. If one considers a single measurement then $\mathfrak{t}_{\text{Total}}=\tau$.                 In such a scenario the minimum value of the uncertainty in the measurement of the gravitational wave amplitude reads
\begin{equation}\label{QGCramerRao.9}
\begin{split}
\sqrt{\langle(\Delta\varepsilon)^2\rangle}_{\text{Min}}=\sqrt{\frac{15}{2\pi \llangle \hat{\mathcal{H}}(\varepsilon)\rrangle}}~.
\end{split}
\end{equation}
To properly understand the quantum gravitational effect on the Bose-Einstein condensate, we shall plot the minimum value of the uncertainty in the measurement of the gravitational amplitude with respect to the measurement time $\tau$. We consider the value of the phonon squeezing to be $r_B=0.8$ and the phonon squeezing angle to be $\zeta_B=\frac{\pi}{2}$ both of which are experimentally achievable. Then the frequency of the BEC phonon mode is set to $\omega_B=10$ Hz. In that case resonance shall occur for an incoming gravitational wave with frequency $\Omega_0=20$ Hz.
\begin{figure}
\begin{center}
\includegraphics[scale=0.4]{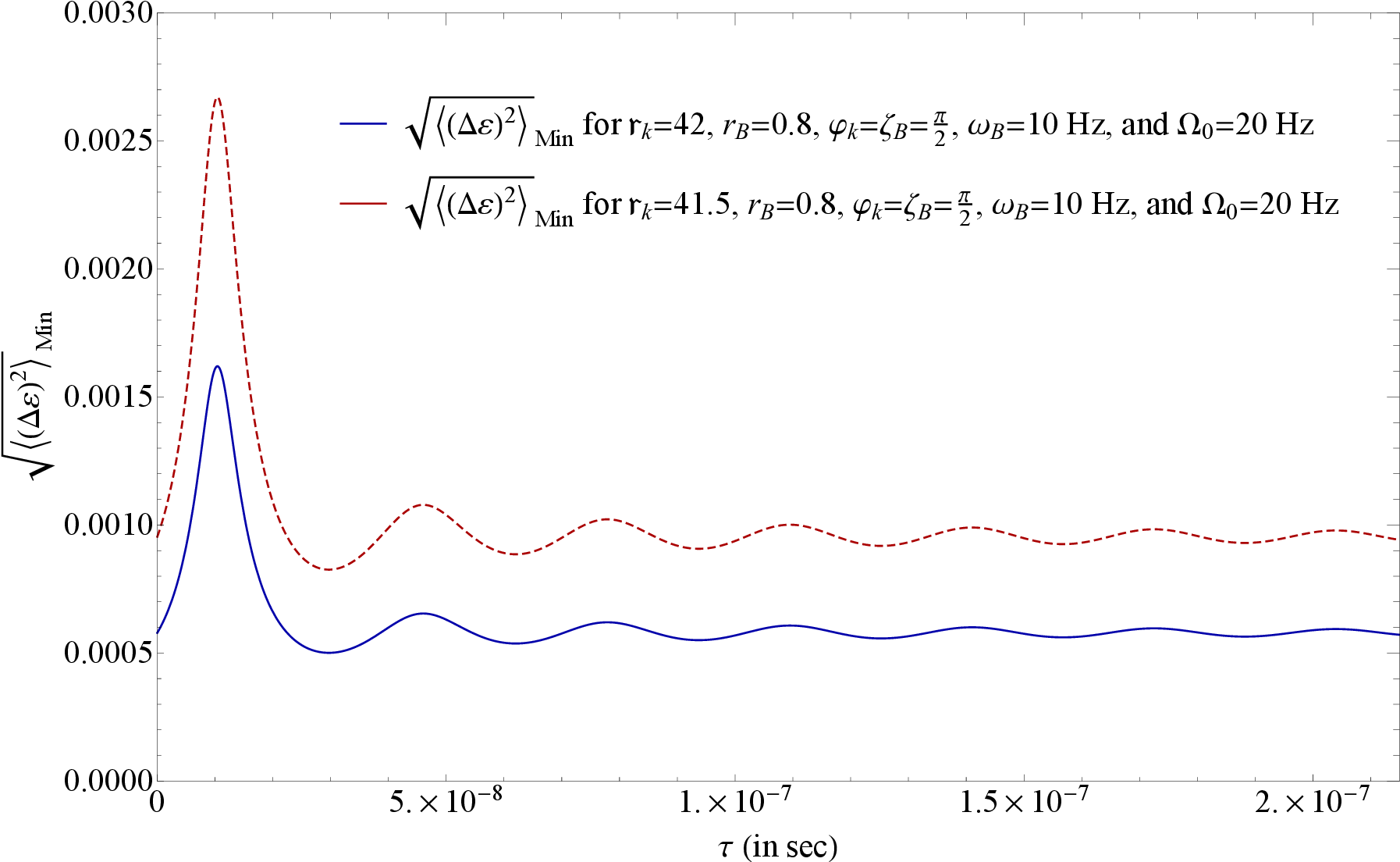}
\caption{$\sqrt{\langle(\Delta\varepsilon)^2\rangle}_{\text{Min}}$ vs $\tau$ plot for different values of the graviton squeezing parameter $\mathfrak{r}_k=\{41.5,42\}$. Other parameter values are set to $r_B=0.8$, $\varphi_k=\zeta_B=\frac{\pi}{2}$, $\omega_B=10$ Hz, and $\Omega_0=20$ Hz.\label{QGFI_Fluctuation_OTM}}
\end{center}
\end{figure}
From Fig.(\ref{QGFI_Fluctuation_OTM}), we observe that for very high graviton squeezing the uncertainty in $\varepsilon$ is quite low and for a lower value of the squeezing parameter the uncertainty in the measurement of $\varepsilon$ increases. In the standard classical gravitational wave case ($l_{\text{Pl}}=0$), we can find from eq.(\ref{QGCramerRao.2}) that at $\tau=0$, $\llangle \hat{\mathcal{H}}(\varepsilon)\rrangle=\mathcal{H}^{(0)}(\varepsilon)=0$ as a result $\sqrt{\langle(\Delta\varepsilon)^2\rangle}_{\text{Min}}$ becomes infinity indicating an impossible detection scenario for $\tau=0$. For the quantum gravity case, however, $\llangle \hat{\mathcal{H}}(\varepsilon)\rrangle\rvert_{\tau\rightarrow 0}\neq 0$ indicating a finite value of the uncertainty in the measurement of the gravitational wave amplitude. We observe in the $\tau\rightarrow 0$ limits that
\begin{equation}\label{QGCramerRao.10}
\begin{split}
\lim\limits_{\tau\rightarrow 0}\llangle \hat{\mathcal{H}}(\varepsilon)\rrangle&=\frac{l_{\text{Pl}}^2\Omega_{\text{M}}^2}{15\pi\varepsilon^2c^2}(1+\cosh^22r_B)\left(\cosh 2\mathfrak{r}_k-\cos\varphi_k\sinh 2\mathfrak{r}_k\right)~.
\end{split}
\end{equation}
Now the amplitude of a standard gravitational wave is $\varepsilon\sim10^{-21}$ and $\Omega_{\text{M}}\sim 10^{8}$ Hz for primordial gravitational waves. As a result when $\tau$ approaches zero, the value of the stochastic average of the QGFI when both the phonon and graviton squeezing are zero, reads $\lim\limits_{\tau\rightarrow 0}\llangle \hat{\mathcal{H}}(\varepsilon)\rrangle=\frac{2l_{\text{Pl}}^2}{15\pi\varepsilon^2c^2}\sim 10^{-30}$. The minimum value in the uncertainty in the measurement of $\varepsilon$ then takes the value $\sqrt{\langle(\Delta\varepsilon)^2\rangle}_{\text{Min}}\sim 10^{14}$. This value of the uncertainty is very high indicating a complete non-detection of any gravity wave signatures. For the uncertainty in $\varepsilon$ to be atleast or less than unity one needs to have atleast a graviton squeezing of $\mathfrak{r}_k\geq 34$. For a primordial gravitational wave which can generate due to inflationary models occurring due to a grand-unified theoretic model the squeezing can go as high as $\mathfrak{r}_k \simeq50$. For such high squeezing value $\sqrt{\langle(\Delta\varepsilon)^2\rangle}_{\text{Min}}\sim 10^{-7}$ indicating a better chance of detection of of a graviton signal. This phenomena indicates that even at the beginning of the experiment when the gravity wave-detector interaction has barely started there is a finite possibility of primordial gravitational wave from the inflationary time. This is a very bizarre result that goes against the standard intuition of gravitational wave detection scenarios. However, in a linearized quantum gravity model the quantum field exists throughout and there exists a quantum gravitational perturbation field around the Bose-Einstein condensate even at time $\tau=0$. Hence, if gravity-wave signatures are captured for very low measurement times then they for sure indicates towards the existence of the quanta of linearized gravity.
Another way to deal with this problem is to increase the phonon squeezing. In that scenario, the BEC will have more sensitivity towards the classical as well as quantum gravity signatures. 
\begin{figure}
\begin{center}
\includegraphics[scale=0.4]{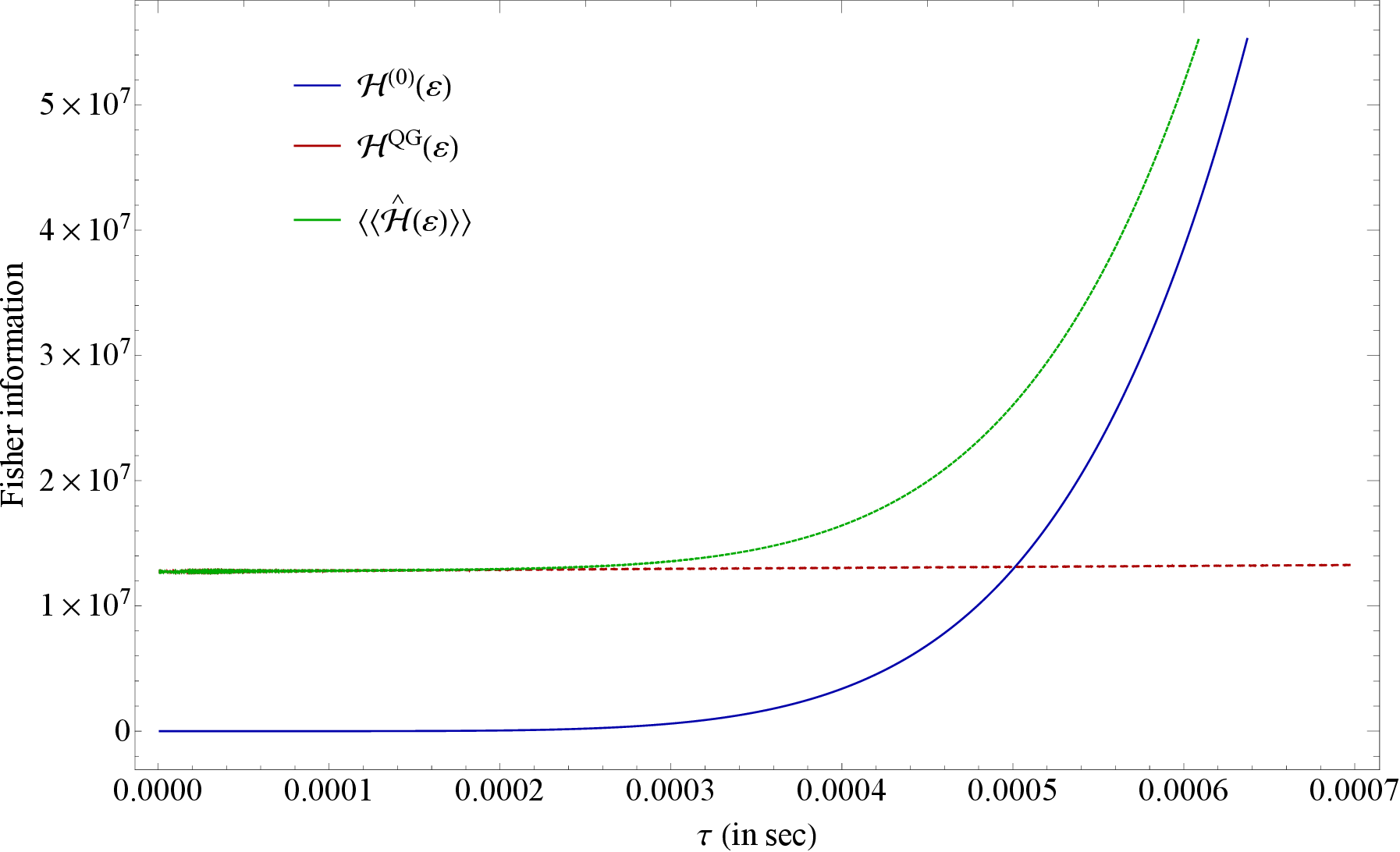}
\caption{We plot the quantum Fisher information as well as the quantum gravitational Fisher information and compare it against the quantum gravitational correction term of the quantum gravitational Fisher information where the parameter values are set to $r_B=12$, $\mathfrak{r}_k=20$, $\zeta_B=\varphi_k=\frac{\pi}{2}$, with the frequencies set at $\omega_B=10$ Hz and $\Omega_0=20$ Hz. \label{Comparison_FI_QGFI_OTM}}
\end{center}
\end{figure}
We plot $\mathcal{H}^{(0)}(\varepsilon)$, $\llangle\hat{\mathcal{H}}(\varepsilon)\rrangle$, $\mathcal{H}^{\text{QG}}(\varepsilon)$ against $\tau$ in Fig.(\ref{Comparison_FI_QGFI_OTM}) where $\mathcal{H}^{\text{QG}}(\varepsilon)$ is defined as
\begin{equation}\label{QGCramerRao.11}
\mathcal{H}^{\text{QG}}(\varepsilon)\equiv\frac{\hbar G\Omega_{\text{M}}^2}{15\pi\varepsilon^2 c^5}\mathcal{B}(\tau,\mathfrak{r}_k,\varphi_k)\mathcal{H}^{(2)}(\varepsilon)~.
\end{equation}
In Fig.(\ref{Comparison_FI_QGFI_OTM}), the parameter values are set to $r_B=12$, $\mathfrak{r}_k=20$, 
Some important observations are in order. We can observe from the figure that at around $\tau\simeq 5\times 10^{-4}$ sec $\mathcal{H}^{(0)}(\varepsilon)$ starts dominating over the quantum gravitational term $\mathcal{H}^{\text{QG}}(\varepsilon)$. The important point to note that the quantum gravity effects play a more significant role at earlier times. In general for a quantum mechanical observable there is always a restriction that a quantum gravity induced parameter should not lead over the standard term which is present in spite of quantum gravitational consideration. The quantum Fisher information is not a physically observable quantity and as a result the quantum gravitational correction term does not face the usual restrictions that is imposed while considering an observable. We also need to remember that $\sqrt{\langle(\Delta\varepsilon)^2\rangle}_{\text{Min}}\sim 10^{-4}$ initially ($\tau\rightarrow0$) for a graviton squeezing as low as $\mathfrak{r}_k=20$ when the phonon squeezing is high enough. This implies that even for relatively lower squeezing from the gravitons, at the starting of the experiment, the BEC will still respond provided the phonons are squeezed as well. 
If the squeezing from the gravitons is very low then for low $\tau$ values $\sqrt{\langle(\Delta\varepsilon)^2\rangle}_{\text{Min}}$ becomes very high which indicates that the gravitational wave signatures cannot be measured using a BEC. 
\begin{figure}
\begin{center}
\includegraphics[scale=0.4]{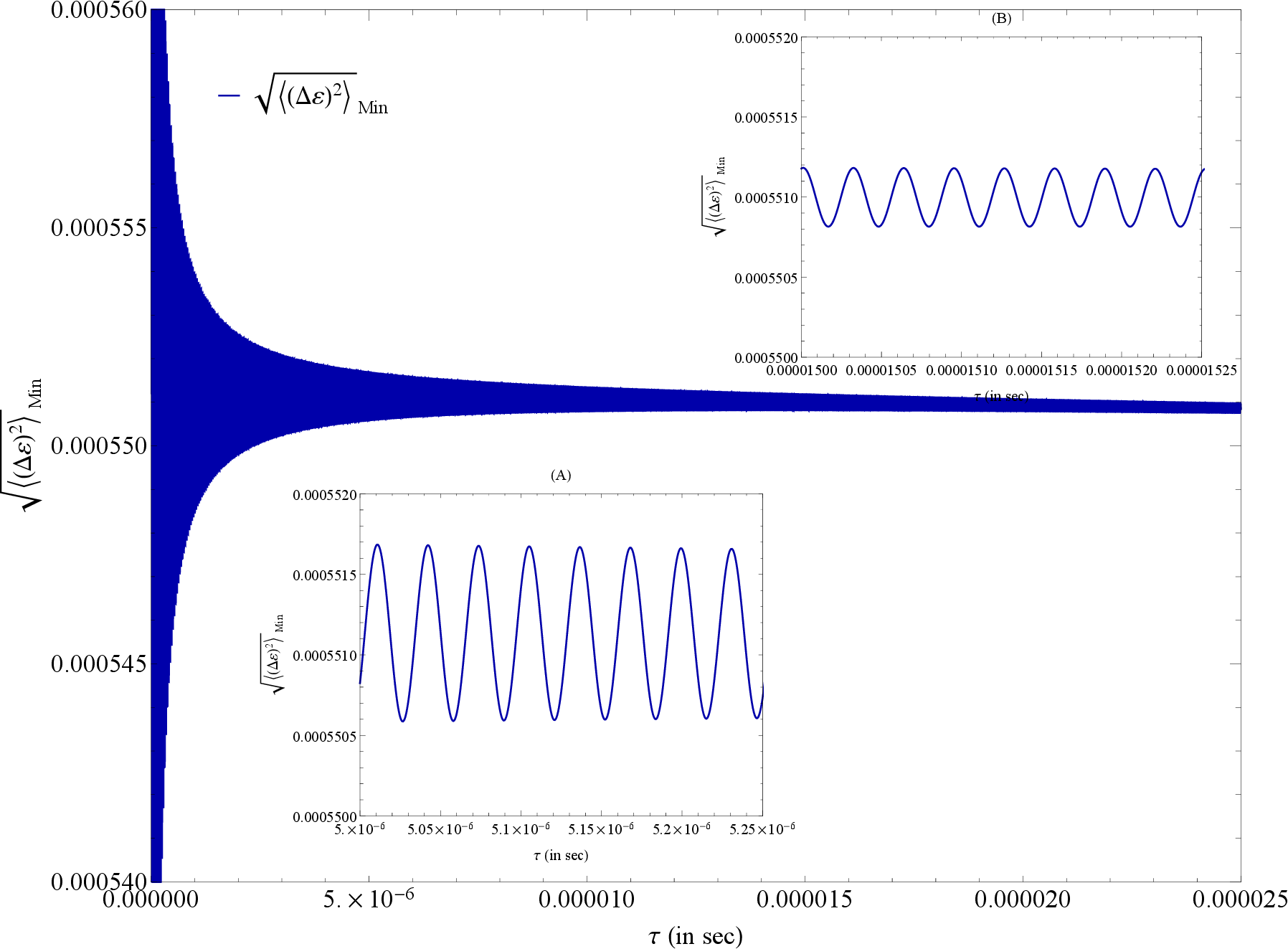}
\caption{$\sqrt{\langle(\Delta\varepsilon)^2\rangle}_{\text{Min}}$ vs $\tau$ plot for $\mathfrak{r}_k=42$, $r_B=0.83$, $\varphi_k=\zeta_B=\frac{\pi}{2}$, $\omega_B=10$ Hz, and $\Omega_0=20$ Hz.\label{Zoomed_Uncertainty_OTM}}
\end{center}
\end{figure}
We have next plotted $\sqrt{\langle(\Delta\varepsilon)^2\rangle}_{\text{Min}}$ against $\tau$ for $\mathfrak{r}_k=42$, $r_B=0.83$, $\varphi_k=\zeta_B=\frac{\pi}{2}$, $\omega_B=10$ Hz, and $\Omega_0=20$ Hz in Fig.(\ref{Zoomed_Uncertainty_OTM}). We find our that $\sqrt{\langle(\Delta\varepsilon)^2\rangle}_{\text{Min}}$ falls of with increase in the measurement time indicating a higher chance of detection of gravitational waves. From the inset plot (A), we observe that the amplitude of fluctuation is large in $1.5\times 10^{-5}$ sec$<\tau$ $<1.525\times 10^{-5}$ sec compared to the the range of $\tau$,  $5\times 10^{-6}$ sec $<\tau$ $<5.25\times 10^{-6}$ sec in the inset plot (B). 
\begin{figure}
\begin{center}
\includegraphics[scale=0.395]{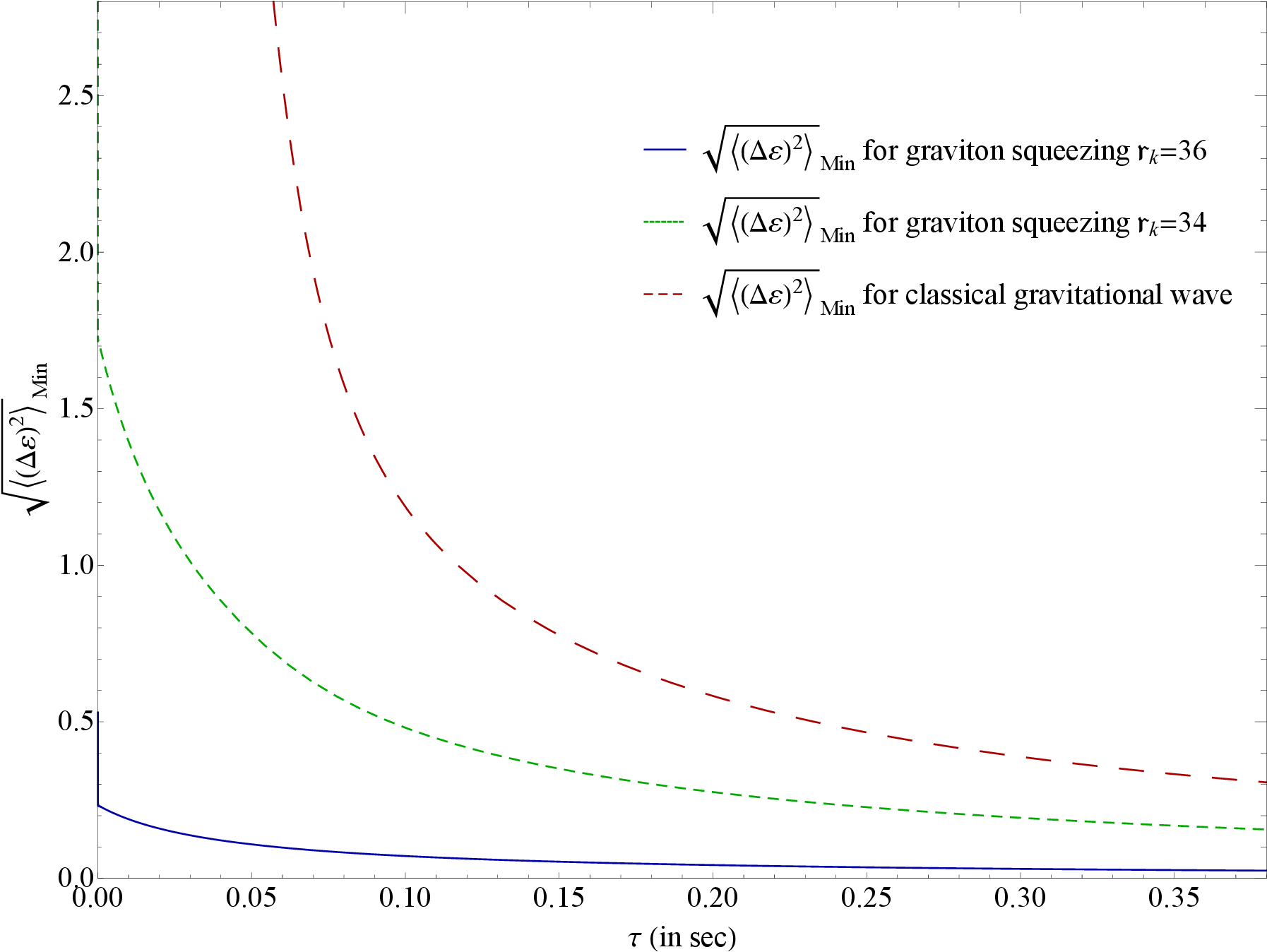}
\caption{$\sqrt{\langle(\Delta\varepsilon)^2\rangle}_{\text{Min}}$ vs $\tau$ plot for $r_B=0.83$, $\varphi_k=\zeta_B=\frac{\pi}{2}$, $\omega_B=10$ Hz, and $\Omega_0=20$ Hz. The graviton squeezing is set to different values of the parameter. \label{Graviton_Squeezing_OTM}}
\end{center}
\end{figure}
We now plot the minimum value of the uncertainty in the parameter $\varepsilon$ with respect to different values of the graviton squeezing parameter $\mathfrak{r}_k$ against the measurement time $\tau$ in Fig.(\ref{Graviton_Squeezing_OTM}). We observe that the uncertainty due to $\varepsilon$ goes to infinity for a classical gravitational wave and it decreases with increasing value of the graviton squeezing. Graviton signatures will arise as resonance pulses in the single mode of the Bose-Einstein condensate. Here, we shall primarily focus onto the standard deviation of the quantum gravitational Fisher information obtained in eq.(\ref{QGFI.2}). The standard deviation in the QGFI reads
\begin{equation}\label{Standard_Deviation.1}
\left(\Delta\mathcal{H}\right)^2=\llangle(\hat{\mathcal{H}}(\varepsilon)-\llangle\hat{\mathcal{H}}(\varepsilon)\rrangle)^2\rrangle~.
\end{equation}
We shall restrict ourselves up to the two-point noise-noise correlator and as a result, one can obtain the standard deviation in the quantum gravitational Fisher information as
\begin{equation}\label{Standard_Deviation.2}
\begin{split}
(\Delta\mathcal{H}(\varepsilon))^2&\simeq \frac{1}{2048\varepsilon^2}\llangle \{\delta\hat{\mathcal{N}}(\tau,0),\delta\hat{\mathcal{N}}(\tau,0)\}\rrangle (\mathcal{H}^{(1)}(\varepsilon))^2\\
&=\frac{l_{\text{Pl}}^2\Omega_{\text{M}}^2}{960\pi\varepsilon^2c^2}\mathcal{B}(\tau,\mathfrak{r}_k,\varphi_k)(\mathcal{H}^{(1)}(\varepsilon))^2\\
&=\frac{l_{\text{Pl}}^2\omega_B^2\Omega_{\text{M}}^2\tau^2}{960\varepsilon^2c^2}\left(e^{-\frac{\tau^2}{4}(\Omega_0-2\omega_B)^2}-e^{-\frac{\tau^2}{4}(\Omega_0+2\omega_B)^2}\right)^2\\&\times(2\cosh^22\mathfrak{r}_k+4\sinh^22\mathfrak{r}_k+6\omega_B\tau \sinh 4r)\mathcal{B}(\tau,\mathfrak{r}_k,\varphi_k)~.
\end{split}
\end{equation}
We shall now plot the standard deviation of the QGFI against the phonon mode frequency when the frequency of the incoming gravitational wave is $\Omega_0=20$ Hz with the squeezing parameter values fixed at $r_B=0.8$ and $\mathfrak{r}_k=5$, squeezing angles fixed at $\varphi_k=\zeta_B=\frac{\pi}{2}$, and the measurement time fixed at $\tau=1.0$ sec, in Fig.(\ref{Resonance_QGFI_OTM}).
From Fig.(\ref{Resonance_QGFI_OTM}), we observe that the uncertainty in the measurement of the quantum gravitational Fisher information hits a maximum value around the resonance point which is $\omega_B=\frac{\Omega_0}{2}=10$ Hz. The important thing to understand is that the uncertainty in the quantum gravitational Fisher information does not exist if the background geometry has no quantum field theoretical fluctuations. A more intricate behaviour of this standard deviation in the Fisher information can be observed when it is plotted for different values of the measurement time as can be seen from Fig.(\ref{Resonance_QGFI_Time_OTM}) where all other parameter values are kept fixed as has been used for the plot in Fig.(\ref{Resonance_QGFI_OTM}).
We observe from Fig.(\ref{Resonance_QGFI_Time_OTM}), that with an increase in the measurement time, the maxima of the standard deviation tends more and more towards the resonance frequency. The standard deviation in the QGFI can be increase to such an extent that its value becomes very high. In reality one can not measure quantum Fisher information directly and as a result it has in reality no direct consequences from an experimental point of view unlike the minimum value of the uncertainty in $\varepsilon$ which indeed is a physically measurable quantity. Another important thing to observe from Fig.(\ref{Resonance_QGFI_Time_OTM}) is that with a decrease in the measurement time the standard deviation in the QGFI shift away from the resonance point and its value gets gradually suppressed. The primary reason behind the diminishing amplitude lies in the fact that the standard deviation in the QGFI has a direct dependence on $\tau$ and for vanishing measurement time as can be seen from eq.(\ref{Standard_Deviation.2}), the standard deviation also vanishes. In the next section, we shall work with a more generalized structure of the noise fluctuation such one can take into consideration all possible phonon-mode frequencies.
\section{A general quantum noise fluctuation}
We shall now choose a more generalized profile for the noise fluctuation, which reads
\begin{equation}\label{Noise.1}
\delta\hat{\mathcal{N}}(t',0)\equiv e^{-\frac{t'^2}{\tau^2}}\cos (\Omega_0 t')\delta\mathcal{N}(t,0)~.
\end{equation}
Substituting the above noise template in the time-dependent solution of the Goldstone boson $\hat{\psi}_{\vec{k}_B}(t)$ and with standard simplifications, we obtain 
\begin{equation}\label{Noise.2}
\hat{\psi}_{\vec{k}_B}(t)=\hat{\alpha}^B_{\text{N}}\exp[-i\omega_B t]+\hat{\beta}^B_{\text{N}}\exp[i\omega_B t]
\end{equation}
where the Bogoliubov coefficients $\hat{\alpha}^B_{\text{N}}$ and $\hat{\beta}^B_{\text{N}}$ take the form
\begin{align}
\hat{\alpha}^B_{\text{N}}&=1-\frac{2\varepsilon_B}{\varepsilon}\sqrt{\pi}\omega_B\tau e^{-\frac{\Omega_0^2\tau^2}{4}}\delta\hat{\mathcal{N}}(\tau,0)\label{Noise.3}\\
\hat{\beta}^B_{\text{N}}&=\frac{\varepsilon_B}{\varepsilon}\sqrt{\pi}\omega_B\tau\left(e^{-\frac{\tau^2}{4}(\Omega_0-2\omega_B)^2}-e^{-\frac{\tau^2}{4}(\Omega_0+2\omega_B)^2}\right)\left(\frac{\varepsilon}{4}+\delta\mathcal{N}(\tau,0)\right)~.\label{Noise.4}
\end{align}
Following the procedure discussed in section (\ref{QGFI}), we can calculate the quantum gravitational Fisher information and eventually its stochastic average which is given as
\begin{equation}\label{Noise.5}
\begin{split}
\llangle \hat{\mathcal{H}}(\varepsilon)\rrangle=& \mathcal{H}_\varepsilon^{(0)}+\frac{\kappa_G^2\Omega_\text{M}^2\omega_B^2\tau^2}{15\varepsilon^2c^2}e^{-\frac{\tau^2}{2}(\Omega_0+2\omega_B)^2}\biggr((1+\cosh 4r_B+4\sinh^22r_B)\left(e^{2\omega_B\Omega_0\tau^2}+1\right)^2
\\&+4(7-\cosh4r_B)e^{2\omega_B^2\tau^2}e^{2\omega_B\Omega_0\tau^2}\biggr)\mathcal{B}(\mathfrak{r}_k,\varphi_k,\tau).
\end{split}
\end{equation}
The important thing to observe in this scenario is that in the limit $\tau$ going to zero the stochastic average of the quantum gravitational Fisher information vanishes which happens solely due to the consideration of a noise profile infused with a Gaussian decay term. This noise profile introduced in eq.(\ref{Noise.1}) now will help us to sum over all possible phonon mode frequencies.  At first, we can rewrite eq.(\ref{QGCramerRao.6}) as
\begin{equation}\label{Noise.6}
\frac{1}{\langle(
\Delta\varepsilon)^2\rangle}=\left(\frac{{k_B^x}^2-{k_B^y}^2}{k_B^2}\right)^2\frac{1}{\langle(\Delta\varepsilon_B)^2\rangle}~.
\end{equation}
Making use of the equality condition in eq.(\ref{QGCramerRao.10}) and summing over all possible phonon mode frequencies, we can rewrite eq.(\ref{Noise.6}) as
\begin{equation}\label{Noise.7}
\begin{split}
\frac{1}{\langle(
\Delta\varepsilon)^2\rangle_{\text{BEC}}^\text{Min}}=\sum\limits_{\vec{k}_B}\left(\frac{{k_B^x}^2-{k_B^y}^2}{k_B^2}\right)^2\mathfrak{N}\llangle\hat{\mathcal{H}}(\varepsilon)\rrangle~.
\end{split}
\end{equation}
Now the condensate is created inside of a box of volume $V_B=L_B^3$ where $L_B$ is the side length whereas there are $n_B$ number of atoms of the Bose gas that has contributed in the forming of the condensate phase and as a result, we can express $k_B$ analytically as $k_B=\frac{\pi n_B}{L_B}$. In the continuous picture,we can replace $\sum\limits_{\vec{k}_B}$ by
\begin{equation}\label{Noise.8}
\sum\limits_{\vec{k}_B}\rightarrow \frac{V_B}{(2\pi)^3}\int_0^{\infty}dk_B k_B^2\int d\Omega~.
\end{equation}
Using the continuous picture, we can recast eq.(\ref{Noise.7}) as
\begin{equation}\label{Noise.9}
\begin{split}
\frac{1}{\langle(\Delta\varepsilon)^2\rangle_{\text{BEC}}^\text{Min}}&=\frac{V_B}{(2\pi)^3}\int_0^{\infty}dk_B k_B^2\int_0^{\pi}d\theta_B \sin\theta_B\int_0^{2\pi}d\phi_B \sin^4\theta_B\cos^22\phi_B~\mathfrak{N}\llangle\hat{\mathcal{H}}(\varepsilon)\rrangle\\
&=\frac{2\pi\mathfrak{N}}{15}\int_0^{\infty}dn_B n^2_B\llangle\hat{\mathcal{H}}(\varepsilon)\rrangle\\&=\frac{\pi^4 \tau^2c_S^2 }{480L_B^2}\mathfrak{R}_1\mathcal{I}^B_1+\frac{2\Omega_\text{M}^2\tau^2\kappa_G^2\pi^2c_S^2}{15 \varepsilon^2L_B^2}\mathcal{B}(\tau,\mathfrak{r}_k,\varphi_k,)\mathcal{I}^B_2
\end{split}
\end{equation}
where the integrals $\mathcal{I}_1^B$ and $\mathcal{I}_1^B$ are defined as
\begin{align}
\mathcal{I}_1^B&=4\int_0^\infty dn_B \hspace{0.05cm}n_B^4 \exp\left[-\frac{\Omega_0^2\tau^2}{2}-\frac{2\pi^2c_S^2\tau^2n_B^2}{L_B^2}\right]\sinh^2\left[\frac{\pi  n_B \Omega_0c_S\tau^2}{L_B}\right]\label{Noise.10}\\
\mathcal{I}_2^B&=\mathfrak{R}_1\int_0^\infty dn_B\hspace{0.05cm}n_B^4 \exp\left[-\frac{\Omega_0^2\tau^2}{2}-\frac{2\pi^2c_S^2\tau^2n_B^2}{L_B^2}\right]\cosh^2\left[\frac{\pi  n_B \Omega_0c_S\tau^2}{L_B}\right]\nonumber\\&+\mathfrak{R}_2\exp\left[-\frac{\Omega_0^2\tau^2}{2}\right]\int_0^\infty dn_B\hspace{0.05cm}n_B^4\label{Noise.11}
\end{align}
with $\mathfrak{R}_1$ and $\mathfrak{R}_2$ being defined as
\begin{equation}
\mathfrak{R}_1=1+\cosh 4r_B+4\sinh^2 2r_B~,~~\mathfrak{R}_2=\frac{1}{2}\left(7-\cosh 4r_B\right)~.
\end{equation}
The second integral in eq.(\ref{Noise.11}) is divergent, which can be taken care of using two strategies. The first way is to provide a cut-off on the number of atoms that can exist inside of volume $V_B$ and the next one is to adjust the phonon squeezing in a way such that $\mathfrak{R}_2$ vanishes resulting in a vanishing of the divergent integral in $\mathcal{I}_2^B$. Now, $\mathfrak{R}_2$ vanishes for the squeezing value $r_B=\frac{1}{4}\cosh^{-1}7\simeq 0.658$. It is possible to control phonon  squeezing in an experimental scenario. Making use of a second order Raman scattering process, phonons can be squeezed and has been done in a crystal lattice \cite{PhononSqueezing1, PhononSqueezing2}. In the presence of an optical lattice potential where cold bosonic atoms are placed inside of the optical lattice \cite{OpticalLattice}, one can make use of pump probe detection scheme \cite{PumpProbeSqueezing} or second order Raman scattering \cite{PhononSqueezing1} for controlled squeezing of phonon modes. Hence, obtaining a squeezing of the phonons equal to $r_B=0.658$ is experimentally achievable\footnote{Squeezing is in general represented in decibels. The dimensionless squeezing parameter $r_B$ is related to the position squeezing parameter via the relation $s_B=-10\log_{10}(e^{-2r_B})$ \cite{Lvovsky}.}. Executing the integrals in eq.(s)(\ref{Noise.10},\ref{Noise.11}), we can simplify eq.(\ref{Noise.9}) as
\begin{equation}\label{Noise.12}
\begin{split}
\frac{1}{\mathfrak{N}\langle(\Delta\varepsilon)^2\rangle_{\text{BEC}}^\text{Min}}&=\frac{V_B\mathfrak{R}_1}{7680\sqrt{2\pi}c_S^3\tau^3}\left(3+6\Omega_0^2\tau^2+\Omega_0^4\tau^4-3 e^{-\frac{\Omega_0^2\tau^2}{2}}\right)\\&+\frac{\hbar G V_B\Omega_{\text{M}}^2\mathfrak{R}_1}{225\pi\varepsilon^2\sqrt{2\pi}c^5c_S^3\tau^3}\mathcal{B}(\tau,\mathfrak{r}_k,\varphi_k)\left(3+6\Omega_0^2\tau^2+\Omega_0^4\tau^4+3 e^{-\frac{\Omega_0^2\tau^2}{2}}\right)~.
\end{split}
\end{equation}
It is therefore possible to write down an inequality of the form
\begin{equation}\label{Noise.13}
\begin{split}
\frac{1}{\mathfrak{N}\langle(\Delta\varepsilon)^2\rangle_{\text{BEC}}}&\leq\frac{V_B\mathfrak{R}_1}{7680\sqrt{2\pi}c_S^3\tau^3}\left(3+6\Omega_0^2\tau^2+\Omega_0^4\tau^4-3 e^{-\frac{\Omega_0^2\tau^2}{2}}\right)\\&+\frac{l_{\text{Pl}}^2 V_B\Omega_{\text{M}}^2\mathfrak{R}_1}{225\pi\varepsilon^2c^2\sqrt{2\pi}c_S^3\tau^3}\mathcal{B}(\tau,\mathfrak{r}_k,\varphi_k)\left(3+6\Omega_0^2\tau^2+\Omega_0^4\tau^4+3 e^{-\frac{\Omega_0^2\tau^2}{2}}\right)~.
\end{split}
\end{equation}
This is one of the most important results in our analysis.
\subsection{Some important results}
We shall now focus towards a more restrictive analysis where we consider that the quantum gravitational part in the Fisher information can not surpass the standard quantum Fisher information. In standard experimental scenarios, the Bose-Einstein condensate is created in a single direction primarily where the perpendicular direction remain relatively small and it has been possible to construct a Bose-Einstein condensate with length $L_B\sim 10^{-3}$ m. Now $\tau$ is the time for a single measurement and as a result, it is possible to obtain $\tau$ as $\tau=\frac{L_B}{v_{\text{Max}}}=\frac{L_B}{c}\sim 10^{-11}$ sec. Now when the incoming gravitational wave frequency $\Omega_0$ is of the order of the cut-off frequency of primordial gravitational waves, that is $\Omega_0\sim \Omega_{\text{M}}\sim 10^8$ Hz then $\Omega_0\tau\sim 10^{-3}\ll 1$. If  the total observation time is $\tau_{\text{Obs.}}\sim\mathfrak{N}\tau$, in such a scenario it is possible to simplify eq.(\ref{Noise.13}) as
\begin{equation}\label{Noise.14}
\langle (\Delta\varepsilon)^2\rangle_{\text{BEC}}\gtrsim \frac{1024\sqrt{2\pi}c_S^3\tau^2}{\Omega_0^2V_B\tau_{\text{Obs.}}\mathfrak{R}_1}\left(1+\frac{1024 l_{\text{Pl}}^2\Omega_{\text{M}}^2}{50\pi\varepsilon^2c^2}\left(\frac{4}{3\Omega_0^2\tau^2}-1\right)\mathcal{B}(\tau,\mathfrak{r}_k,\varphi_k)\right)~.
\end{equation}
We can now give a lower bound to the minimum time of measurement as $\langle (\Delta\varepsilon)^2\rangle_{\text{BEC}}$ can not be negative and, we obtain
\begin{equation}\label{Noise.15}
\tau_{\text{Min}}\simeq \sqrt{\frac{2}{3\pi}}\frac{32 l_{\text{Pl}}\Omega_{\text{M}}}{\varepsilon c\Omega_0}\mathcal{B}(\tau,\mathfrak{r}_k,\varphi_k)~.
\end{equation}
Now, $\tau_{\text{Min}}$ hits its absolute lowest value when $\Omega_0\sim\Omega_{\text{M}}$ and the gravitons are in vacuum state with no inherent squeezing ($\mathcal{B}\rightarrow 1$). In such a scenario, we obtain the minimum measurement time to be
\begin{equation}\label{Noise.16}
\tilde{\tau}_{\text{Min}}\simeq\sqrt{\frac{2}{3\pi}}\frac{32 l_{\text{Pl}}\Omega_{\text{M}}}{\varepsilon c\Omega_0}\biggr\rvert_{\Omega_0\rightarrow\Omega_{\text{M}}}\simeq 1.59\times 10^{-22} \text{ sec}.
\end{equation}
%If the graviton squeezing is quite high then $\mathcal{B}\sim e^{2\mathfrak{r}_k}$ and from eq.(\ref{Noise.15}), we obtain 
%\begin{equation}\label{Noise.17}
%\mathfrak{r}_k=\frac{1}{2}\ln\left[\sqrt{\frac{3\pi}{2}}\frac{\varepsilon c\Omega_0\tau}{32 l_{\text{Pl}}\Omega_{\text{M}}}\right]~.
%\end{equation}
For a more stringent analysis, one can require that the quantum gravitational term in eq.(\ref{Noise.13}) must be less than or equal to the standard classical gravity wave induced term in the quantum Fisher information. From eq.(\ref{Noise.13}), we can write down the following expression
\begin{equation}\label{Noise.17}
\begin{split}
\mathcal{B}(\tau,\mathfrak{r}_k,\varphi_k)\leq\frac{15\pi\varepsilon^2c^2}{512 l_{\text{Pl}}^2\Omega_{\text{M}}^2}\left(\frac{3+6\Omega_0^2\tau^2+\Omega_0^4\tau^4-3 e^{-\frac{\Omega_0^2\tau^2}{2}}}{3+6\Omega_0^2\tau^2+\Omega_0^4\tau^4+3 e^{-\frac{\Omega_0^2\tau^2}{2}}}\right)~.
\end{split}
\end{equation}
Now for $\varphi_k=\frac{\pi}{2}$, we obtain the analytical form of 
\begin{equation}\label{Noise.18}
\mathcal{B}(\tau,\mathfrak{r}_k,\varphi_k)=\cosh 2\mathfrak{r}_k-\frac{\sinh 2\mathfrak{r}_k}{\Omega_\text{M}^2\tau^2}\left(2\Omega_\text{M}\tau\cos (2\Omega_\text{M}\tau)-\sin(2\Omega_\text{M}\tau)\right)~.
\end{equation}
At times $\tau\neq 0$, it provides a bound on the graviton-squeezing parameter $\mathfrak{r}_k$ but the discrepancy arises at time $\tau=0$. In the limit $\tau\rightarrow 0$, the right hand side of the inequality becomes zero whereas the left hand side reduces to $\cosh 2\mathfrak{r}_k$. From the equality one can find out the maximum value of $\mathfrak{r}_k$ possible to be a complex number which is an unphysical condition indicating that it is not very wise to implement such stringent bounds when gravitons are involved in the analysis. We therefore shall not abide by this upper bound  and we take this liberty keeping in mind the fact that the quantum Fisher information is not a true physical observable. Now the physical observable is indeed the amplitude and the standard deviation in the measurement of the amplitude and the inverse square root of the quantum gravitational Fisher information is related to it. In the next subsection, we shall investigate whether the BEC acts as a true graviton detector.
\subsection{Sensitivity curve of the BEC as a graviton detector}
\begin{figure}[t!]
\begin{center}
\includegraphics[scale=0.45]{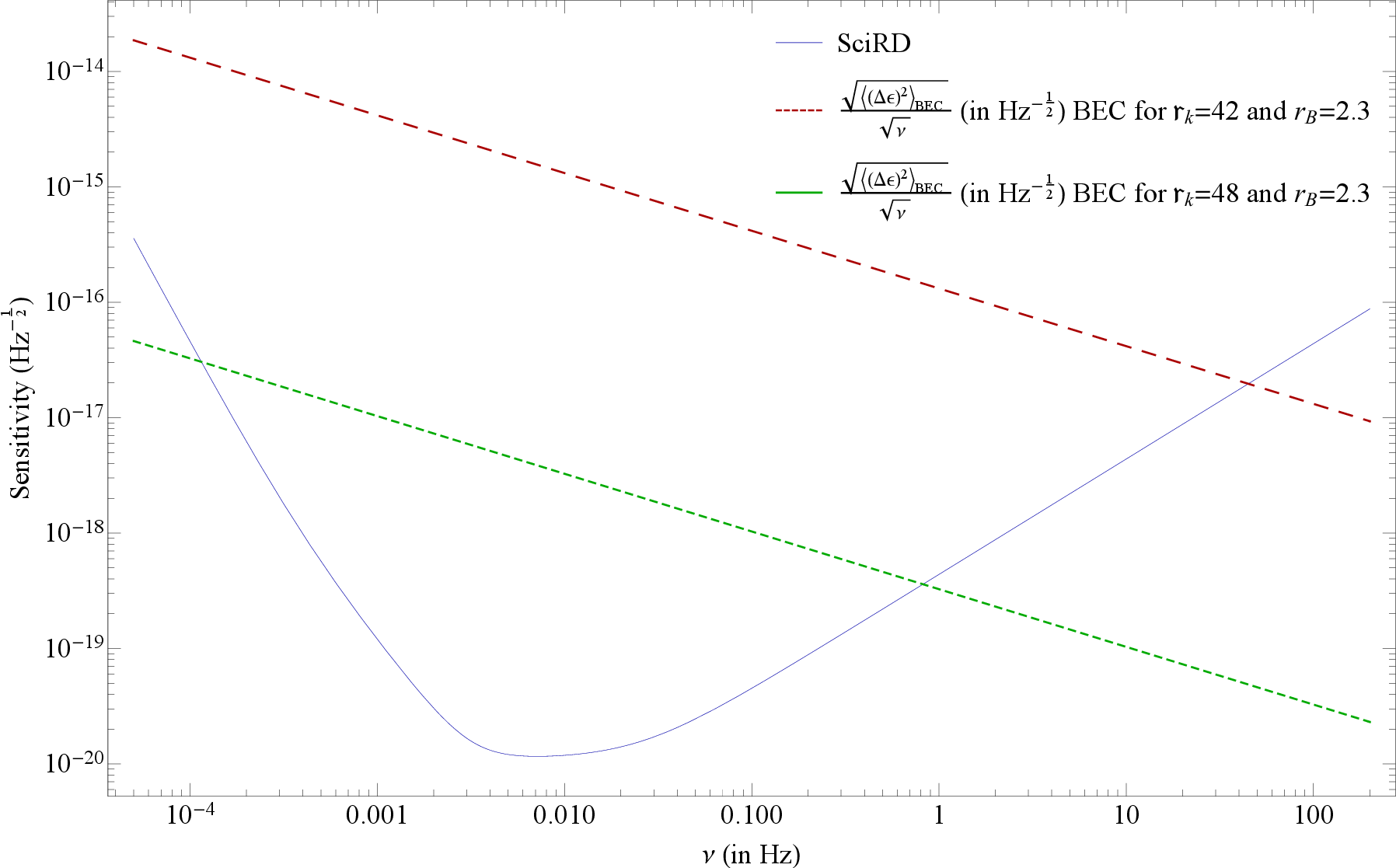}
\caption{The BEC sensitivity formula is plotted against the LISA sensitivity curve against different values of the incoming gravitational wave frequency.\label{Graviton_Detector_OTM}}
\end{center}
\end{figure}
In this subsection, we shall investigate whether it will be possible to detect graviton signatures while making use of a Bose-Einstein condensate. For reference, we shall make use of the proposed sensitivity curve for the space based LISA (Laser Interferometer Space Antenna) observatory. We make use of the sensitivity formula from the Scientific Research Document (SciRD) \cite{SciRD}, a detailed discussion of which can be found in \cite{SciRD2}. The sensitivity formula, presented in SciRD reads \cite{SciRD2}
\begin{align}
\mathcal{S}_{\text{h,SciRD}}(\nu)&=\frac{10}{3}\left(\frac{\mathcal{S}_I(\nu)}{(2\pi \nu)^4}+\mathcal{S}_{II}(\nu)\right)\mathcal{R}(\nu)~ \text{Hz}^{-1}\label{Sensitivity.1}\\
\mathcal{S}_I(\nu)&=5.76\time 10^{-48}\left(1+\frac{\nu_1^2}{\nu^2}\right) ~\text{sec}^{-4}.\text{ Hz}^{-1}\label{Sensitivity.2}\\
\mathcal{S}_{II}(\nu)&=3.6\times 10^{-41} \text{Hz}^{-1}\label{Sensitivity.3}\\
\mathcal{R}(\nu)&=1+\frac{\nu^2}{\nu_2^2}\label{Sensitivity.4}
\end{align}
with the value of $\nu_1$ and $\nu_2$ being $\nu_1=0.4$ mHz and $\nu_2=25$ mHz.
\begin{figure}[t!]
\begin{center}
\includegraphics[scale=0.45]{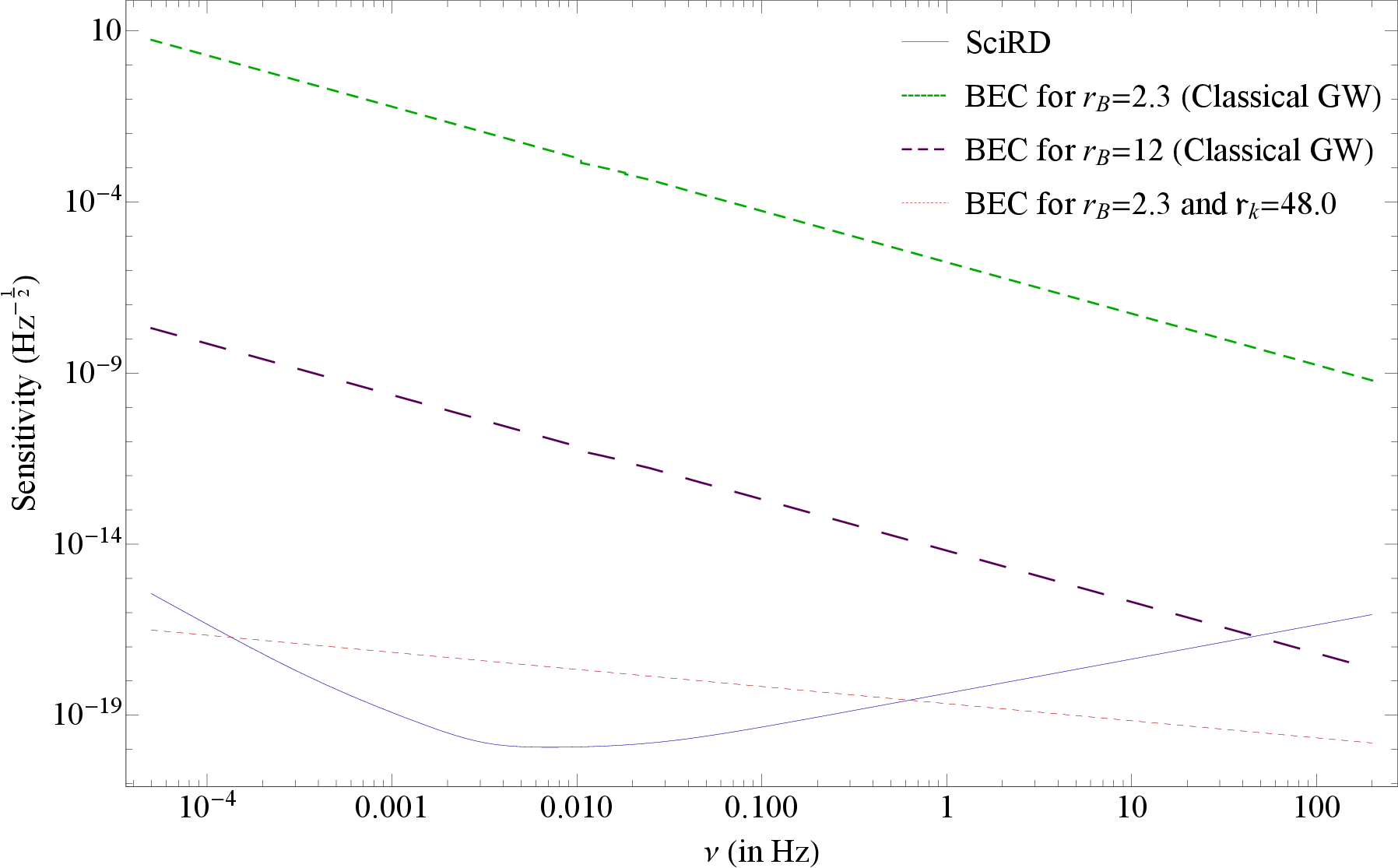}
\caption{Sensitivity versus frequency plot for the standard BEC-classical gravitational wave interaction case along with the BEC-graviton case with high graviton squeezing and then it is compared against the LISA-SciRD projected sensitivity curve.\label{Graviton_Detector_2_OTM}}
\end{center}
\end{figure}
For the expression for the sensitivity of the Bose-Einstein condensate towards incoming gravitational fluctuations, we make use of eq.(\ref{Noise.12}) and the sensitivity formula is given by $\frac{\sqrt{\langle(\Delta\varepsilon)^2\rangle_{\text{BEC}}^\text{Min}}\rvert_{\Omega_0\rightarrow\nu}}{\sqrt{\nu}}$ $\text{Hz}^{-\frac{1}{2}}$. Here, $\nu$ denotes the frequency of the incoming gravitational wave. We shall now plot the sensitivity formula corresponding to the Bose-Einstein condensate against incoming gravitational wave frequency and compare it with the sensitivity curve ($\sqrt{\mathcal{S}_{\text{h,SciRD}}(\nu)}$ $\text{Hz}^{-\frac{1}{2}}$) of the upcoming LISA observatory. We set $\tau=10^{-6}$ sec, $\tau_{\text{Obs.}}=10^2$ sec, speed of sound $c_S=0.012$ $\text{m}\cdot\text{sec}^{-1}$, and  $L_B=10^{-3}$ m. We also vary the squeezing of the gravitons to compare the sensitivity of the BEC to SciRD sensitivity formula \cite{SciRD,SciRD2} in Fig.(\ref{Graviton_Detector_OTM}). We observe from Fig.(\ref{Graviton_Detector_OTM}), we observe that the LISA observatory is most sensitive about in the $0.001-10$ Hz range which is the standard frequency range for primordial gravitational waves. As we can see from Fig.(\ref{Graviton_Detector_OTM}), for a significantly higher squeezing of the gravitons, the BEC becomes very sensitive towards incoming gravitational wave signals. Consider a future experimental scenario where a primordial gravitational wave has been detected by the space based LISA observatory. Now if a BEC based gravity wave detector (with minimal phonon squeezing) also senses gravity wave pulses, in such a scenario one can claim for sure that this fluctuation is generated solely due to the interaction of the gravitons with the Bose-Einstein condensate. In such a case the gravitons have to be highly squeezed as can be seen from Fig.(\ref{Graviton_Detector_OTM}). If it is possible to construct a meter long BEC ($L_B\sim 1-1.5$ m) which is extremely difficult to obtain experimentally, in such a scenario, it is possible to achieve similar sensitivity curve for graviton squeezing as low as $\mathfrak{r}_k=37-38$. In Fig.(\ref{Graviton_Detector_2_OTM}), we plot the projected LISA sensitivity formula ($\sqrt{\mathcal{S}_{\text{h,SciRD}}}(\nu)$ $\text{Hz}^{-\frac{1}{2}}$) alongside the BEC-gravity wave sensitivity model for classical as well as the quantum gravitational case. We observe from the figure that for a classical gravitational wave interaction with low phonon squeezing the BEC is not sensitive to incoming gravitational wave fluctuations and in order to do so one needs high phonon squeezing ($r_B=12$) as can be seen from Fig.(\ref{Graviton_Detector_2_OTM}). On the other hand for high graviton squeezing and low phonon squeezing the BEC is sensitive in the primordial gravitational wave frequency range. 
\begin{figure}[t!]
\begin{center}
\includegraphics[scale=0.42]{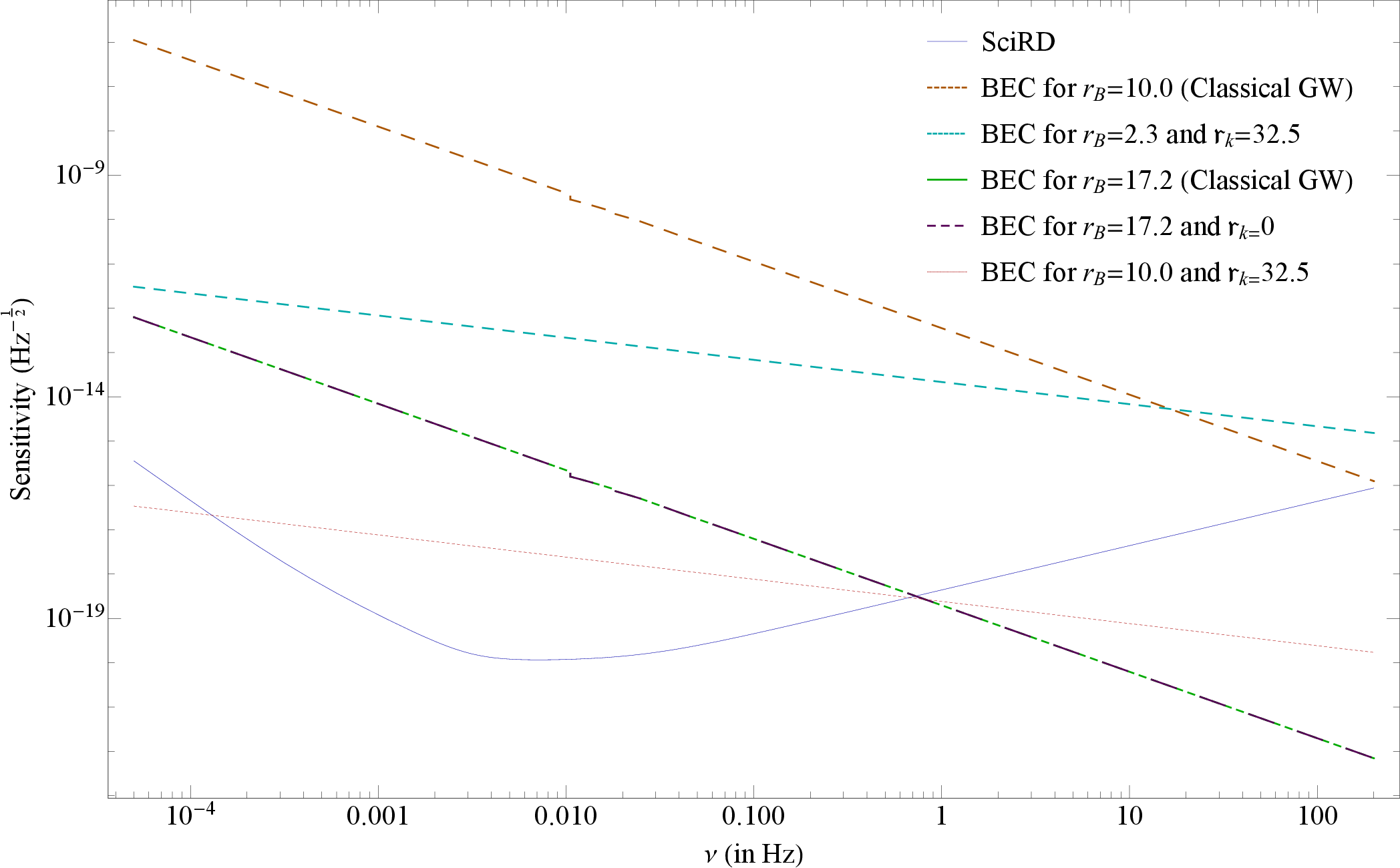}
\caption{Sensitivity is plotted against frequency and observed for the case when there is a balance between graviton and phonon mode squeezing.\label{Graviton_Detector_3_OTM}}
\end{center}
\end{figure}
Finally, in Fig.(\ref{Graviton_Detector_3_OTM}), we observe that for a high phonon squeezing ($r_B=17.2$) and no-graviton squeezing, a BEC has same sensitivity to classical gravitational fluctuation as well as a quantum gravitational scenario. We also observe that for high squeezing of the phonons the sensitivity of the detector significantly increases and it becomes sensitive in the 0.1-100.0 Hz range. The important observation lies in the fact that for a relatively lower squeezing of the phonon modes ($r_B=10.0$) for a graviton squeezing as low as $\mathfrak{r}_k=32.5$ the BEC based graviton detector becomes sensitive in the $0.001-10$ Hz range. Hence, the requirement of the graviton squeezing can be controlled by properly tuning the phonon mode squeezing. This helps in a balanced detection criteria for graviton-detection. One thing to remember from all the the Fig.(s)(\ref{Graviton_Detector_OTM},\ref{Graviton_Detector_2_OTM},\ref{Graviton_Detector_3_OTM}) is that the BEC is not sensitive to classical gravity waves unless the phonon modes are highly squeezed. Now, if such a scenario occurs in future where even with relatively lower phonon mode squeezing a gravitational wave signature is captured by the BEC based graviton detector when a simultaneous detection of primordial gravitational waves has been captured by LISA then it is a confirmation that the BEC based detector has been able to pick up graviton signatures. This implies that the Bose-Einstein condensate can serve as one of the best candidates for graviton detection. 
\section{Decoherence from the interacting phonon modes on the quantum gravitational Fisher information}
In this section, we start by considering a more realistic picture where interaction among the phonon modes has been considered. A simple way is to add a thermal bath to the BEC system. In case of a Gaussian state with single mode only, the BEC covariance matrix goes through a time evolution as
\begin{equation}\label{Dissipation.1}
\Sigma_D^{\text{G}}(t)=\Gamma_D(t)\hspace{0.1cm}\Sigma^\text{G}_0\hspace{0.1cm}\Gamma_D^T(t)+\Sigma_{\infty}^\text{G}(t)~.
\end{equation}
In the above equation $\Sigma_0^{\text{G}}$ denotes the covariance matrix for a single mode Gaussian state and $\Sigma_\infty^{\text{G}}$ denotes the covariance matrix for the Gaussian reservoir. In case of a single mode of the Bose-Einstein condensate we can replace $\Sigma_0^{\text{G}}$ by $\Sigma_{\text{Sq.}}(0)$, where $\Sigma_{\text{Sq.}}(0)$ is given in eq.(\ref{QGCovariance.6}). The dissipation factor $\Gamma_{D}(t)$ is given as $\Gamma_D(t)=\exp\left[-\frac{\gamma t}{2}\right]\mathbb{1}_{2\times 2}$ where $\gamma$ is the dissipation constant. The time-dependent covariance matrix corresponding to the Gaussian reservoir in eq.(\ref{Dissipation.1}) evolves as
\begin{equation}\label{Dissipation.2}
\Sigma^{\text{G}}_\infty(t)=\left(1-e^{-\gamma t}\right)\Sigma_\infty^{\text{G}}~.
\end{equation}
Using the analytical form of the dissipation factor $\Gamma_D(t)$ and the covariance matrix of the Gaussian reservoir from eq.(\ref{Dissipation.2}), we can write down the elements of the time dependent Gaussian covariance matrix from eq.(\ref{Dissipation.1}) as \cite{MPGThesis,PhononDecoherence}
\begin{equation}\label{Dissipation.3}
\Sigma^{\text{G}}_{ij}(t)=e^{-\gamma t}\Sigma^{\text{G}}_{0_{ij}}+\left(1-e^{-\gamma t}\right)\Sigma^{\text{G}}_{\infty_{ij}},~~\forall\hspace{0.15 cm} i,j\in\{1,2\}~.
\end{equation}
If $\mu_B(t)=\frac{1}{2\sqrt{\det\left[\Sigma(t)\right]}}$ denotes the purity of a quantum state (expressed by a covariance matrix) \cite{MatthewMannAffshordi,MPGThesis,PhononDecoherence} then for a single mode of a Bose-Einstein condensate, the elements corresponding to the covariance matrix $\Sigma^{\text{BEC}}(t)$ reads
\begin{equation}\label{Dissipation.4}
\begin{split}
\Sigma^{\text{BEC}}_{11}(t)&=\frac{1}{2\mu_B(t)}\left(\cosh 2r_B(t)+\cos\zeta_B(t)\sinh2r_B(t)\right)~,\\
\Sigma^{\text{BEC}}_{12}(t)&=\Sigma^{\text{BEC}}_{21}(t)=\frac{1}{2\mu_B(t)}\sin\zeta_B(t)\sinh 2r_B(t)~,\\
\Sigma^{\text{BEC}}_{22}(t)&=\frac{1}{2\mu_B(t)}\left(\cosh 2r_B(t)-\cos\zeta_B(t)\sinh2r_B(t)\right)~.
\end{split}
\end{equation}
Due to dissipation in the system, here $r_B$ and $\zeta_B$ both becomes time dependent. Initially, when the reservoir has been just connected to the bath then the purity, squeezing parameter and phase can be expressed as $\mu_B^0\equiv\mu_B(0),$ $r_B^0\equiv r_B(0)$, and $\zeta_B^0\equiv \zeta_B(0)$. The matrix element from eq.(\ref{Dissipation.4}) can then be expressed as
\begin{equation}\label{Dissipation.5}
\begin{split}
\Sigma^{\text{BEC}}_{0_{11}}&=\frac{1}{2\mu_B^0}\left(\cosh 2r_B^0+\cos\zeta_B^0\sinh2r_B^0\right)~,\\
\Sigma^{\text{BEC}}_{0_{12}}&=\Sigma^{\text{BEC}}_{0_{21}}=\frac{1}{2\mu_B^0}\sin\zeta_B^0\sinh 2r_B^0~,\\
\Sigma^{\text{BEC}}_{0_{22}}&=\frac{1}{2\mu_B^0}\left(\cosh 2r_B^0-\cos\zeta_B^0\sinh2r_B^0\right)~.
\end{split}
\end{equation}
For the Gaussian reservoir, the covariance matrix elements read
\begin{equation}\label{Dissipation.6}
\begin{split}
\Sigma^{\text{G}}_{\infty_{11}}&=\frac{1}{2\mu_\infty}\left(\cosh 2r_\infty+\cos\zeta_\infty\sinh2 r_\infty\right)~,\\
\Sigma^{\text{G}}_{\infty_{12}}&=\Sigma^{\text{G}}_{\infty_{21}}=\frac{1}{2\mu_\infty}\sin\zeta_\infty\sinh 2r_\infty~,\\
\Sigma^{\text{G}}_{\infty_{22}}&=\frac{1}{2\mu_\infty}\left(\cosh 2r_\infty-\cos\zeta_\infty\sinh2r_\infty\right)~.
\end{split}
\end{equation}
 where the purity, squeezing parameter and angle corresponding to the Gaussian reservoir are denoted by $\mu_\infty$, $r_\infty$, and $\zeta_\infty$. As has been discussed in \cite{MatthewMannAffshordi,MPGThesis}, it is quite natural to set the reservoir squeezing to zero which reduces the covariance matrix corresponding to the Gaussian matrix to $\Sigma_{\infty}^\text{G}=\frac{1}{\mu_\infty}\mathbb{1}_{2\times 2}$. The first step is to start with a finite reservoir squeezing and later set it to zero. Using eq.(s)(\ref{Dissipation.4}-\ref{Dissipation.6}), in eq.(\ref{Dissipation.2}), we obtain three equations as given below
 \begin{align}
 \frac{\cosh 2r_B(t)+\cos\zeta_B(t)\sinh2r_B(t)}{\mu_B(t)}&= \frac{e^{-\gamma t}}{\mu_B^0}\left(\cosh 2r_B^0+\cos\zeta_B^0\sinh2r_B^0\right)\nonumber\\
 &+\frac{\left(1-e^{-\gamma t}\right)}{\mu_\infty}\left(\cosh 2r_\infty+\cos\zeta_\infty\sinh2 r_\infty\right)\label{Dissipation.7}~,\\
 \frac{1}{2\mu_B(t)}\sin\zeta_B(t)\sinh 2r_B(t)&=\frac{e^{-\gamma t}}{2\mu_B^0}\sin\zeta_B^0\sinh 2r_B^0+\frac{\left(1-e^{-\gamma t}\right)}{2\mu_\infty}\sin\zeta_\infty\sinh 2r_\infty\label{Dissipation.8}~,
 \end{align}
 and
 \begin{equation}\label{Dissipation.9}
 \begin{split}
  \frac{\cosh 2r_B(t)-\cos\zeta_B(t)\sinh2r_B(t)}{\mu_B(t)}&= \frac{e^{-\gamma t}}{\mu_B^0}\left(\cosh 2r_B^0-\cos\zeta_B^0\sinh2r_B^0\right)\nonumber\\
 &+\frac{\left(1-e^{-\gamma t}\right)}{\mu_\infty}\left(\cosh 2r_\infty-\cos\zeta_\infty\sinh2 r_\infty\right)~.
 \end{split}
 \end{equation}
 Using eq.(s)(\ref{Dissipation.7},\ref{Dissipation.8},\ref{Dissipation.9}), and solving them we arrive at the time dependent evolution equation for the purity, squeezing parameter, and angle as\footnote{Our results are slightly different from the one used in \cite{PhononDecoherence}. The primary reason behind this fact is that the the covariance matrix elements derived in our current analysis are slightly different then used in \cite{PhononDecoherence}.} \cite{PhononDecoherence}
 \begin{align}
 \mu_B(t)&=\mu_B^0\left[e^{-2\gamma t}+\frac{2\mu_B^0}{\mu_\infty}e^{-\gamma t}\left(1-e^{-\gamma t}\right)\left[\cosh 2r_B^0\cosh 2r_\infty-\cos[\zeta_B^0-\zeta_\infty]\sinh 2r_B^0\sinh 2r_\infty\right]\right.\nonumber\\&\left.+\frac{{\mu_B^0}^2}{\mu_\infty^2}\left(1-e^{-\gamma t}\right)^2\right]^{-\frac{1}{2}}~,\label{Dissipation.10}\\
r_B(t)&=\frac{1}{2}\cosh^{-1}\left[\mu_B(t)\left[\frac{e^{-\gamma t}}{\mu_B^0}\cosh 2r_B^0+\frac{1-e^{-\gamma t}}{\mu_\infty}\cosh 2r_\infty\right]\right]~,\label{Dissipation.11}\\
\zeta_B(t)&=\tan^{-1}\left[\frac{\sin\zeta_B^0\sinh 2r_B^0+\frac{\mu_B^0}{\mu_\infty}\left(e^{\gamma t}-1\right)\sin \zeta_\infty \sinh 2r_\infty}{\cos\zeta_B^0\sinh 2r_B^0+\frac{\mu_B^0}{\mu_\infty}\left(e^{\gamma t}-1\right)\cos \zeta_\infty \sinh 2r_\infty}\right]\label{Dissipation.12}~.
 \end{align}
 We shall now set the reservoir squeezing parameter value to zero which reduces eq.(\ref{Dissipation.12}) to
 \begin{equation}\label{Dissipation.13}
 \zeta_B(t)=\tan^{-1}\left[\tan \zeta_B^0\right]=\zeta_B^0~.
 \end{equation}
 Hence, we find out that for zero squeezing of the reservoir, the squeezing angle remains fixed with time. Again using the condition $r_\infty=0$ in eq.(s)(\ref{Dissipation.10},\ref{Dissipation.11}), we obtain a much simpler expression for the time evolution of $\mu_B(t)$ and $r_B(t)$ as
 \begin{align}
 \mu_B(t)&=\mu_B^0\left[e^{-2\gamma t}+\frac{{\mu_B^0}^2}{\mu_\infty^2}\left(1-e^{-\gamma t}\right)^2+\frac{2\mu_B^0}{\mu_\infty}e^{-\gamma t}\left(1-e^{-\gamma t}\right)\cosh 2r_B^0\right]^{-\frac{1}{2}}\label{Dissipation.14}~,\\
r_B(t)&=\frac{1}{2}\cosh^{-1}\left[\mu_B(t)\left[\frac{e^{-\gamma t}}{\mu_B^0}\cosh 2r_B^0+\frac{1-e^{-\gamma t}}{\mu_\infty}\right]\right]~.\label{Dissipation.15}
 \end{align}
 Now, one can take the assumption that $r_B^0>\max\left[\frac{\mu_B^0}{\mu_\infty},\frac{\mu_\infty}{\mu_B^0}\right]$, one can obtain the analytical form of $t$ for which $\mu(t)$ reaches its minimum value to be \cite{MPGThesis,PhononDecoherence}
 \begin{equation}\label{Dissipation.16}
 t_{\text{min}}=\frac{1}{\gamma}\ln\left[\frac{{\mu_B^0}^2+\mu_\infty^2-2\mu_B^0\mu_\infty\cosh 2r_B^0}{{\mu_B^0}^2-\mu_B^0\mu_\infty\cosh 2r_B^0}\right]~.
 \end{equation}
The time $t_{\text{min}}$ can also be considered as the characteristic decoherence time corresponding to the squeezed single-mode state of the bosonic system. The primary change in this physical scenario is that the phonon squeezing parameter as well as the phase gets a direct time dependence. The first step is to again write down the covariance matrix an proceed from there. The stochastic average of the quantum gravitational Fisher information from eq.(\ref{QGFI.4}) in this set-up takes the form
\begin{equation}\label{Dissipation.17}
\begin{split}
\llangle\hat{\mathcal{H}}(\varepsilon)\rrangle&=\frac{1}{64\mu_B^2(t)}\pi\omega_B^2\tau^2\left(e^{2\omega_B\Omega_0\tau^2}
-1\right)^2e^{-\frac{\tau^2}{2}\left(\Omega_0+2\omega_B\right)^2}\left(1+\cosh 4r_B(\tau)+(1-3\cos 2\zeta_B)\right.\\&\times\left.\sinh^2 2r_B(\tau)\right)+\frac{l_{\text{Pl}}^2\Omega_{\text{M}}^2}{15\pi\varepsilon^2c^2\mu_B^2(\tau)}\Bigr(1-2\omega_B^2\tau^2+\cosh^22r_B(\tau)(1+6\omega_B^2\tau^2)\\&+ 6\omega_B\tau \cosh 2r_B(\tau) \sqrt{\cosh^2 2r_B(\tau)-1} \Bigr)\mathcal{B}(\tau,\mathfrak{r}_B,\varphi_k)~.
\end{split}
\end{equation} 
In the above equation, we have replaced the time $t$ by the single measurement time $\tau$ in $r_B(\tau)$ and $\mu_B(\tau)$. One now needs to identify the form of the decoherence factor $\gamma$. When the temperature is very low in the bosonic model system then Beliaev damping takes place which will take the analytical form at absolute zero as \cite{Beliaev,Beliaev2,Beliaev3}
\begin{equation}\label{Dissipation.18}
\gamma\simeq \frac{3}{640\pi}\frac{\hbar \omega_B^5}{m_B n_B c_S^5}
\end{equation}
with $m_B$ denoting the individual mass of the atoms and $n_B$ denoting the atomic number density. 
\begin{figure}
\begin{center}
\includegraphics[scale=0.4]{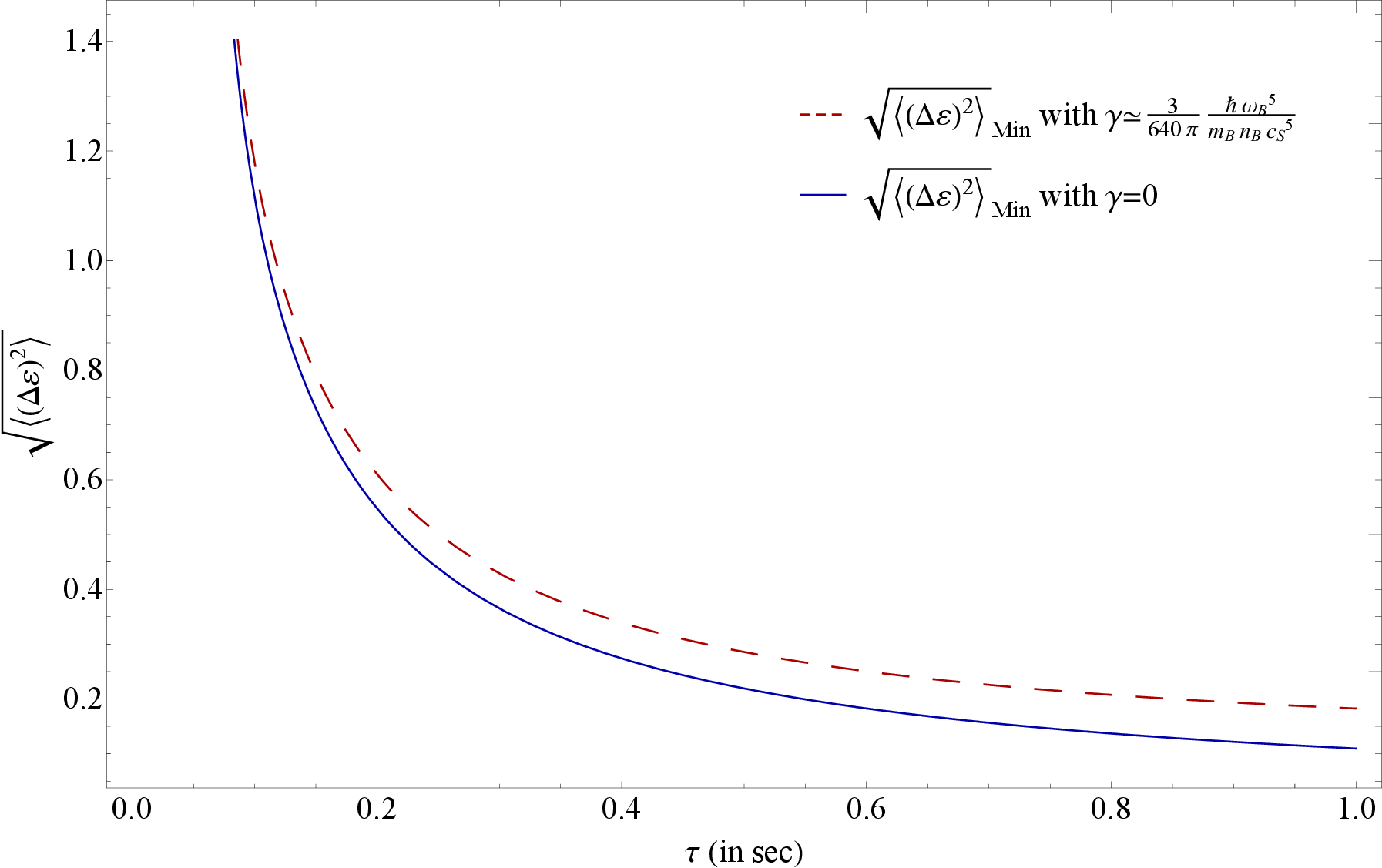}
\caption{Plot of $\sqrt{\langle(\Delta\varepsilon)^2\rangle}_\text{Min}$ versus $\tau$ (in sec) when the phonon modes are interacting against the standard non-interacting case.\label{Beliaev_OTM}}
\end{center}
\end{figure}
Now, it is straightforward to consider the atomic mass in the $10^{-25}$ kg and the speed of sound to be $c_S=1.2\times10^{-2}$ $\text{m}\cdot\text{sec}^{-1}$. For the number density being $n_B=7\times 10^{20} \text{ m}^{-3}$, the Beliaev damping factor has the value $\gamma=9.034 \times 10^{-19}$ $\text{sec}^{-1}$ which is significantly small. For such a small value of the damping coefficient, there is barely any change while measuring the uncertainty in $\varepsilon$ compared to the standard non-interacting model of the BEC. Now, if we consider a relatively low number density of atoms $n_B=10^3$ $\text{m}^{-3}$ the Beliaev damping takes the value $\gamma\simeq0.632$ which is way higher than the one presented in eq.(\ref{Dissipation.18}). We now take the phonon mode frequency to be $\omega_B=10$ Hz and the BEC to be initially pure ($\mu_B^0=1$). We also consider the reservoir to be pure, that is $\mu_\infty=1$. We set the graviton squeezing parameter to be $\mathfrak{r}_k=15.0$ and angle to be $\varphi_k=\frac{\pi}{2}$, whereas the phonon squeezing parameter is set to $r_B^0=0.83$ and angle to $\zeta_B^0=\frac{\pi}{2}$, and plot $\sqrt{\langle(\Delta\varepsilon)^2\rangle}_\text{Min}$ against the measurement time $\tau$ in Fig.(\ref{Beliaev_OTM}). From the plot in Fig.(\ref{Beliaev_OTM}), we observe that due to damping from the interacting phonon modes the fall of the minimum value in the standard deviation of the gravitational wave amplitude becomes slower compared to the non-interacting case with increasing measurement time. Indicating that phonon mode interaction results in a lower accuracy for detecting graviton signatures using a BEC. Although for very small measure time $\tau=10^{-6}-10^{-3}$ sec, there is almost no significant change due to this Beliaev damping and the difference is purely noticeable for higher values of the measurement time.
\section{Discussion and summary}
In this chapter, we have considered a relativistic Bose-Einstein condensate interacting with the quanta of linearized gravity. Here one starts with the complex scalar field Lagrangian with $\phi^4$ interaction in a curved background and extremizing the action with respect to the amplitude part of the complex scalar field. This gives one the relativistic Lagrangian corresponding to the Bose-Einstein condensate in a curved background \cite{DTSon,MatthewMannAffshordi}. From there one can write down the action corresponding to the time-dependent part of the pseudo-Goldstone bosons. The action corresponding to the gravitational fluctuation is obtained from the Einstein-Hilbert action. In order to incorporate quantum gravity effects into the system, we decompose the gravitational fluctuation term its individual Fourier modes and then raise the Fourier mode functions to operator status where the Fourier operator and its complex conjugate follow an appropriate canonical commutation relation. Because of the quantum gravity effects being incorporated in the model system the equation of motion corresponding to the time-dependent part of the Goldstone bosons become stochastic in nature where the stochastic nature is incorporated via the fluctuation part of the graviton mode operator. Hence, the equation of motion corresponding to the bosons become Langevin-like in nature. Solving the Langevin-like stochastic differential equation, it is possible to identify the Bogoliubov coefficients from the analytical form of the time-dependent part of the pseudo-Goldstone bosons. Now, due to the noise-fluctuations induced by the gravitons, the Bogoliubov coefficients also gets infused by the quantum gravitational noise terms and as a result they become operator valued in nature. Using the analytical forms of the operator valued Bogoliubov coefficients, we have then calculated the quantum Fisher information by making use of quantum metrological techniques. As a result of the noise induced by gravitons, the quantum Fisher information has contributions from the graviton noise terms. This information theoretic quantity is completely different from the quantum Fisher information and we call it as the quantum gravitational Fisher information or QGFI. In order to get a physical insight the stochastic average of the QGFI is taken with respect to a chosen graviton state and using this, we write down the quantum gravity modified Cram\'{e}r-Rao bound. From the quantum gravity modified Cram\'{e}r-Rao bound, we have obtained the minimum value in the uncertainty of the gravitational wave amplitude. The square of this uncertainty is proportional to the inverse of the stochastic average of the QGFI. From this expression, we get some very important insights. We find out that even for a very small measurement time it is possible for the BEC to pick up graviton signatures provided the squeezing of the gravitons are very high. It is important to note that a true non-perturbative analysis is required for such high squeezing graviton-BEC interaction analysis but the overall behaviour will remain the same. This scenario may seem unphysical  in nature as a zero time detection of gravitons is very bizarre. In this scenario, the field is present and the quantum behaviour of the gravitational fluctuations make it possible for a very small time detection scheme. Next, we have considered an more natural template of the noise term where the noise fluctuation is infused with a Gaussian decay term. This Gaussian decay gives the opportunity to consider all possible phonon mode frequencies and consider the effect of the entire BEC system as a whole. We observe that the stochastic average of the QGFI depends on the standard term which generated solely due to the interaction of the classical gravitational wave with the BEC, where as the term dependent of the square of the Planck length and the two point noise-noise correlator generates solely due to quantum gravity effects. Using the LISA projected sensitivity formula and comparing it with the sensitivity formula  obtained in our analysis, we observe that the quantum gravitational consideration always increases the sensitivity of the BEC and for a very high graviton squeezing with minimal squeezing from the phonons, the BEC becomes as sensitive as the projected LISA sensitivity curve in the $0.01-10$ Hz frequency range. The important thing to note is that for the case of a detection of a primordial gravitational wave along with a simultaneous detection by a BEC based future generation gravitational wave detector ensures the fact that graviton signatures have been captured. The primary reason behind this phenomena is that the BEC is sensitive to gravitational fluctuations when there is an amplifier like a graviton squeezing. Increasing the graviton squeezing does dramatically increase the  value of the graviton induced part in the expectation value of the QGFI making the BEC sensitive towards primordial gravitation fluctuations in the low frequency ranges. Hence, if a primordial gravitational wave is captured by the LISA observatory in future the simultaneous detection by a BEC based detector can happen only if gravitons have interacted with it. This makes the BEC as one of the best candidates for graviton detection. Finally, we have considered a more physical scenario where we have considered interaction between the phonon mode frequencies leading to an overall decoherence effect. We observe that while the phonon modes are interacting with each other the BEC becomes less sensitive towards the gravitational fluctuations than the standard case due to Beliaev damping at lower temperature Bose-Einstein condensates. In the next chapter, we extend the analysis of this chapter by incorporating a purely quantum gravitational model corresponding to the BEC part as well. In the next chapter we also propose a BEC based graviton detector which is based on the principle of atom interferometry. 
\chapter{Gravitational Bremsstrahlung from a Bose-Einstein condensate}\label{C.7.OTM}
In the previous chapter, we have discussed in details the model for a Bose-Einstein condensate and its interaction with gravitons, and we shall continue our analysis here and aim to propose an experimental model for future detection of gravitons. As has been discussed earlier, quantum gravity becomes dominant at very small length scales and to probe such lengths, one needs to access excessively high energy scales. The problem with building such high energy colliders lies in the fact that the radius of the colliders need to be of an unprecedented length. The highest energy reached by the Large Hadron collider (LHC) is 13 TeV whereas for probing the Planck length, one needs an energy of the level of $10^{16}$ TeV. Hence, making a particle accelerator just by using high energy colliders, is still an impossible task. Therefore, it is more efficient to look for ways to enhance the weak signatures of quantum gravity (if they exist) in an executable experimental set-up, which is possible to build with the current and most advanced engineering techniques known to the human kind. Recently, in \cite{MarlettoVedral,EntanglementQuantumGravity,
MarlettoVedral2,EntanglementQuantumGravity2,
EntanglementQuantumGravity3}, graviton detection via investigating the entanglement generation between two massive bodies have been claimed analytically. In these works, the entanglement is generated between two massive bodies due to graviton exchange, where two gravitons can superpose with each other (as the consideration is purely in a linearized quantum gravity model). This gravity mediated entanglement generation considers gravitational perturbations due to two massive coherent bodies, where it is still a challenge to prepare a coherent mass dual experimentally. Recently, in \cite{FragolinoSchutTorosBoseMazumdar}, the use of matter-wave interferometry to detect the decoherence due to dipole-dipole interaction in the quantum gravity induced entanglement of the masses has been proposed. In \cite{Haine}, a proposal for detecting quantum gravity signatures in quantum gases has been made as well. We have made use of the quantum metrological techniques in the previous chapter to find out the sensitivity of a Bose-Einstein condensate towards graviton fluctuations. We have found out that for high graviton squeezing and a moderate amount of phonon squeezing, the Bose-Einstein condensate (BEC) works as the best candidate for graviton detection. Here, \footnote{This chapter is based on the publication S. Sen and S. Gangopadhyay, ``\textit{Quantum nature of gravity in a Bose-Einstein condensate}", \href{https://doi.org/10.1103/PhysRevD.111.066002}{Phys. Rev. D 111 (2025) 066002}.}, we follow a density matrix approach for the BEC-graviton model system where the analytical form of the time-dependent part of the Goldstone bosons follow from the earlier chapter. By tracing out the gravitational field degrees of freedom, we then investigate the signs of graviton induced decoherence in the reduced density matrix which corresponds to the graviton infused state of the Bose-Einstein condensate. This chapter is organized as follows.

\noindent We start with the basic discussion of the model system that has been implemented in the current analysis. Next, the BEC part as well as the graviton part is quantized using canonical quantization techniques. Then we use the von-Neumann equation of motion corresponding to the density matrix of the graviton-BEC model system. Tracing out the field degrees of freedom, we then arrive at the reduced density matrix, where we search for graviton induced ``bremsstrahlung" in the Bose-Einstein condensate. We then make use of a maximally entangled momentum state of two Bose-Einstein condensates and based on the analytical form of the reduced density matrix, we propose an experimental model for atom laser based graviton detection, where we have made use of few subtle techniques (proposal) used in atom interferometry experiments.
\section{Background model and some important aspects}
The background metric is given by the standard Minkowski metric with gravitational fluctuations upon it as has been discussed in eq.(\ref{QGRBEC.1}), where we continue to follow the mostly positive signature for the metric tensor. The dynamics of the gravitational fluctuation can be obtained from the Einstein-Hilbert part of the action, the analytical form of which in the transverse-traceless gauge is given in eq.(\ref{QGRBEC.7}). The action for the complete model system is obtained by combining the Einstein-Hilbert action with the Bose-Einstein condensate action and the combined action follows from eq.(\ref{QGRBEC.43}) as
\begin{equation}\label{QGBremsstrahlung.1}
\begin{split}
S&=S_{\text{EH}}+S^G_{\text{BEC}}\\&=\frac{1}{2}\sum\limits_{\vec{k},q}\int dt\left(\left|\dot{h}^q(t,\vec{k})\right|^2-k^2\left|h_q(t,\vec{k})\right|^2\right)\\
&+\gamma_B\int dt\left[\sum_{\vec{k}_B}\left|\dot{\psi}_{\vec{k}_B}(t)\right|^2-c_S^2\left[\eta_{ij}+\frac{2\kappa_G}{\sqrt{V_G}}\sum_{\vec{k},q}h_q(t,\vec{k})\epsilon^q_{ij}(\vec{k})\right]\sum_{\vec{k}_B}k_B^ik_B^j\left|\psi_{\vec{k}_B}(t)\right|^2\right]
\end{split}
\end{equation}
where the discrete mode decomposition of the gravitational fluctuation term in the transverse traceless gauge as well as the mode decomposition of the Goldstone bosons reads
\begin{align}
\bar{h}_{jl}(t,\vec{x})&=\frac{2\kappa_G}{\sqrt{V_G}}\sum\limits_{\vec{k},q} h_q(t,\vec{k})e^{i\vec{k}.\vec{x}}\epsilon^q_{jl}(\vec{k})~,~~\pi_B(t,\vec{x})=\sum_{k_B}\psi_{\vec{k}_B}(t)\exp\left[i\vec{k}_B\cdot \vec{x}\right]~.\label{QGBremsstrahlung.2}
\end{align}
In eq.(\ref{QGBremsstrahlung.1}), the constant $\gamma_B$ and the speed of sound in the condensate $c_S$ have the analytical forms
\begin{equation}\label{QGBremsstrahlung.3}
\gamma_B=\frac{V_B}{2\lambda_Bc}\left(\frac{3\sigma_B^2}{c^2}-\frac{m_B^2c^2}{\hbar^2}\right)~,~~c_S=c\sqrt{\frac{\nicefrac{\sigma_B^2}{c^2}-\nicefrac{m_B^2c^2}{\hbar^2}}{\nicefrac{3\sigma_B^2}{c^2}-\nicefrac{m_B^2c^2}{\hbar^2}}}~.
\end{equation}
In the current analysis, we have taken the natural units that is $\hbar=c=1$. From the action of the system in eq.(\ref{QGBremsstrahlung.1}), it is easy to check that the action is in the $S=\int dt L$ form, and as a result one can easily read off the Lagrangian for the model system as
\begin{equation}\label{QGBremsstrahlung.4}
\begin{split}
L&=\frac{1}{2}\sum\limits_{\vec{k},q}\left(\left|\dot{h}^q(t,\vec{k})\right|^2-k^2\left|h_q(t,\vec{k})\right|^2\right)\\
&+\gamma_B\left[\sum_{\vec{k}_B}\left|\dot{\psi}_{\vec{k}_B}(t)\right|^2-c_S^2\left[\eta_{ij}+\frac{2\kappa_G}{\sqrt{V_G}}\sum_{\vec{k},q}h_q(t,\vec{k})\epsilon^q_{ij}(\vec{k})\right]\sum_{\vec{k}_B}k_B^ik_B^j\left|\psi_{\vec{k}_B}(t)\right|^2\right]~.
\end{split}
\end{equation}
In order to quantize the theory canonically, we need the conjugate variables to the variables $h_{q}(t,\vec{k})$, $h^*_q(t,\vec{k})$, $\psi_{\vec{k}_B}(t)$, and $\psi^*_{\vec{k}_B}(t)$. The conjugate variables corresponding to the Fourier mode function $h_{q}(t,\vec{k})$ and its complex conjugate $h^*_{q}(t,\vec{k})$ can be obtained from the Lagrangian in eq.(\ref{QGBremsstrahlung.4}) as
\begin{equation}\label{QGBremsstrahlung.5}
\mathfrak{P}_q(t,\vec{k})=\frac{\partial L}{\partial \dot{h}_{q}(t,\vec{k})}=\frac{1}{2}\dot{h}^*_{q}(t,\vec{k})~,~~\mathfrak{P}^*_q(t,\vec{k})=\frac{\partial L}{\partial \dot{h}_{q}(t,\vec{k})}=\frac{1}{2}\dot{h}_{q}(t,\vec{k})~.
\end{equation}
One can also find out the conjugate variables to $\psi_{\vec{k}_B}(t)$ and $\psi^*_{\vec{k}_B}(t)$ as
\begin{equation}\label{QGBremsstrahlung.6}
P_{\vec{k}_B}(t)=\frac{\partial L}{\partial \dot{\psi}_{\vec{k}_B}(t)}=\gamma_B\dot{\psi}^*_{\vec{k}_B}(t)~,~~P^*_{\vec{k}_B}(t)=\frac{\partial L}{\partial \dot{\psi}^*_{\vec{k}_B}(t)}=\gamma_B\dot{\psi}_{\vec{k}_B}(t)~.
\end{equation}
The total Hamiltonian of the system can then be obtained by using the above conjugate variables and the Lagrangian from eq.(\ref{QGBremsstrahlung.4}) in the following way
\begin{equation}\label{QGBremsstrahlung.7}
\begin{split}
H(t)&=\sum\limits_{\vec{k},q}\left(\mathfrak{P}_q(t,\vec{k})\dot{h}_q(t,\vec{k})+\mathfrak{P}^*_q(t,\vec{k})\dot{h}^*_q(t,\vec{k})\right)+\sum\limits_{\vec{k}_B}\left(P_{\vec{k}_B}(t)\dot{\psi}_{\vec{k}_B}(t)+P^*_{\vec{k}_B}(t)\dot{\psi}^*_{\vec{k}_B}(t)\right)-L\\
&=H_0(t)+H_{\text{int}}(t)
\end{split}
\end{equation} 
where the base Hamiltonian $H_0(t)$ and the interaction Hamiltonian $H_{\text{int}}(t)$ are defined as
\begin{align}
H_0(t)&=\sum\limits_{\vec{k},q}2\mathfrak{P}_q(t,\vec{k})\mathfrak{P}^*_q(t,\vec{k})+\frac{1}{2}\sum\limits_{\vec{k},q}k^2h_q(t,\vec{k})h^*_q(t,\vec{k})\nonumber \\&+ \frac{1}{\gamma_B}\sum\limits_{\vec{k}_B}P_{\vec{k}_B}(t)P^*_{\vec{k}_B}(t)+\gamma_Bc_S^2\sum\limits_{\vec{k}_B}k_B^2\psi_{\vec{k}_B}(t)\psi^*_{\vec{k}_B}(t)\label{QGBremsstrahlung.8}\\
H_{\text{int}}(t)&=\frac{2\kappa_G\gamma_Bc_S^2}{\sqrt{V_G}}\sum\limits_{\vec{k},q}\sum\limits_{\vec{k}_B}h_q(t,\vec{k})\epsilon^q_{ij}(\vec{k})k_B^ik_B^j\psi_{\vec{k}_B}(t)\psi^*_{\vec{k}_B}(t)~.\label{QGBremsstrahlung.9}
\end{align}
In order to quantize the model system, we raise the Fourier mode function corresponding to the gravitational wave modes as well as the wave vectors to operator status, and implement suitable canonical commutation relation between $\hat{h}_q(t,\vec{k})$, $\hat{\psi}_{\hat{\vec{k}}_B}(t)$\footnote{We denote $\hat{\psi}_{\hat{\vec{k}}_B}(t)$ by simply $\hat{\psi}_{\vec{k}_B}(t)$. The important difference between the early analysis with this one is that in the previous chapter the time dependent part of the Goldstone bosons became operator valued due to the fact that it got infused by the graviton noise term. However, here the phonon mode wave-vectors being operator valued gives the BEC part an inherent operatorial behaviour.} (and their Hermitian conjugate) and there conjugate operators. The conjugate operators are obtained by raising the conjugate variables in eq.(s)(\ref{QGBremsstrahlung.5},\ref{QGBremsstrahlung.6}) to operator status. The interaction Hamiltonian operator now reads
\begin{equation}\label{QGBremsstrahlung.10}
\hat{H}_{\text{int}}(t)=\frac{2\kappa_G\gamma_Bc_S^2}{\sqrt{V_G}}\sum\limits_{\vec{k},q}\hat{h}_q(t,\vec{k})\epsilon^q_{ij}(\vec{k})\sum\limits_{\vec{k}_B}\hat{k}_B^i\hat{k}_B^j\hat{\psi}_{\vec{k}_B}(t)\hat{\psi}^*_{\vec{k}_B}(t)~.
\end{equation}
In the above expression, the analytical form of the graviton mode operator $\hat{h}_q(t,\vec{k})$ can be taken directly from eq.(\ref{QGRBEC.63}) of Chapter(\ref{C.6.OTM}), and is given as follows\footnote{The only difference here is that the wave vectors corresponding to the BEC part are raised to operator status as well.}
\begin{equation}\label{QGBremsstrahlung.11}
\hat{h}_q(t,\vec{k})=h_q^{\text{cl}}(t,\vec{k})+\delta\hat{h}_q^I(t,\vec{k})-\frac{4\gamma_B\kappa_Gc_S^2}{\sqrt{V}_G}{\epsilon^q_{ij}}^*(\vec{k})\sum_{\vec{k}_B}\hat{k}_B^i\hat{k}_B^j\int_0^t dt' \frac{\sin(k(t-t'))}{k}\left|\hat{\psi}_{\vec{k}_B}(t')\right|^2.
\end{equation}
Here, $h^{\text{cl}}_q(t,\vec{k})$ is give by $h^{\text{cl}}_q(t,\vec{k})=\langle \hat{h}^I_q(t,\vec{k})\rangle$. Here the expectation is taken with respect to initial graviton states.
In the next section, we shall start by writing down the von-Neumann equation of motion corresponding to the density matrix of the joint Bose-Einstein condensate and graviton model system.
\section{The von-Neumann equation of motion for the density matrix of the system}
\noindent The primary focus of our analysis is to search for graviton induced decoherence in the Bose-Einstein condensate\footnote{It is important to note that, we are concerned here primarily with the momentum states of the bosons as the condensation is a phenomena occurring in the Fourier space, where the matter waves corresponding to the matter waves overlap with each other creating the Bose-Einstein condensate.}. To calculate the decoherence, we follow the methodology depicted in \cite{Bremsstrahlung}, and later adapted in \cite{KannoSodaTokuda,KannoSodaTokuda2}. We consider the density matrix corresponding to the BEC-graviton system to be denoted by $\hat{\rho}_{\text{S}}$, and for the model Hamiltonian in eq.(\ref{QGBremsstrahlung.7}), the von-Neumann equation of motion reads
\begin{equation}\label{QGBremsstrahlung.12}
\frac{\partial \hat{\rho}_{\text{S}}(t)}{\partial t}=-\frac{i}{\hbar}[\hat{H}(t),\hat{\rho}_{\text{S}}(t)]~.
\end{equation}
As we are working in the operator description where the operators are quantum mechanical, the Liouville equation of motion is termed as the  von-Neuman equation of motion. 
Now, the Liouville superoperator corresponding to the total Hamiltonian of the system in eq.(\ref{QGBremsstrahlung.7}) is defined as
\begin{equation}\label{QGBremsstrahlung.13}
\begin{split}
\hat{\mathfrak{L}}^{\text{Tot.}}_{\text{Su}}(t)\hat{\rho}_{\text{S}}(t)&\equiv-\frac{i}{\hbar}[\hat{H}(t),\hat{\rho}_\text{S}(t)]\\&=-\frac{i}{\hbar}[\hat{H}_0(t),\hat{\rho}_{\text{S}}(t)]-\frac{i}{\hbar}[\hat{H}_{\text{int}}(t),\hat{\rho}_{\text{S}}(t)]~.
\end{split}
\end{equation}
where in the last line of the above equation, we have made use of the fact that $\hat{H}(t)$ is a combination of the base Hamiltonian $\hat{H}_0(t)$ and the interaction Hamiltonian $\hat{H}_{\text{int}}(t)$. Using the definition of the Liouville superoperator from eq.(\ref{QGBremsstrahlung.13}), we can write down the Liouville superoperators corresponding to the base Hamiltonian and the interaction Hamiltonian as
\begin{equation}\label{QGBremsstrahlung.14}
\hat{\mathfrak{L}}_{\text{Su}}^0(t)\hat{\rho}_{\text{S}}(t)=-\frac{i}{\hbar}[\hat{H}_0(t),\hat{\rho}_{\text{S}}(t)]~,~~\hat{\mathfrak{L}}_{\text{Su}}^{\text{int}}(t)\hat{\rho}_{\text{S}}(t)=-\frac{i}{\hbar}[\hat{H}_{\text{int}}(t),\hat{\rho}_{\text{S}}(t)]~.
\end{equation}
Using the above expressions, we can recast eq.(\ref{QGBremsstrahlung.13}) as
\begin{equation}\label{QGBremsstrahlung.15}
\begin{split}
\hat{\mathfrak{L}}^{\text{Tot.}}_{\text{Su}}(t)\hat{\rho}_{\text{S}}(t)=\hat{\mathfrak{L}}^{0}_{\text{Su}}(t)\hat{\rho}_{\text{S}}(t)+\hat{\mathfrak{L}}^{\text{int}}_{\text{Su}}(t)\hat{\rho}_{\text{S}}(t)~.
\end{split}
\end{equation}
Here, we shall restrict ourselves to a single mode analysis only. It is now possible to express eq.(\ref{QGBremsstrahlung.12}) in terms of the Liouville superoperator corresponding to the total Hamiltonian of the system as
\begin{equation}\label{QGBremsstrahlung.16}
\frac{\partial \hat{\rho}_{\text{S}}(t)}{\partial t}=\hat{\mathfrak{L}}^{\text{Tot.}}_{\text{Su}}(t)\hat{\rho}_{\text{S}}(t)~.
\end{equation}
Our next aim is to solve the above von-Neumann equation of motion.
\subsection{Solving the von-Neumann equation of motion}
We start with the analytical form of the von-Neumann equation in eq.(\ref{QGBremsstrahlung.12})
\begin{equation}\label{QGBremsstrahlung.17}
\begin{split}
\frac{\partial \hat{\rho}_{\text{S}}(t)}{\partial t}&=-\frac{i}{\hbar}[\hat{H}(t),\hat{\rho}_{\text{S}}(t)]\\
\implies \hat{\rho}_{\text{S}}(t)&=-\frac{i}{\hbar}\int_{t_i}^t dt'[\hat{H}(t'),\hat{\rho}_{\text{S}}(t')]+\text{Const.}~.
\end{split}
\end{equation}
At $t=t_i$, the integral in the above equation vanishes and the constant is set to $\hat{\rho}_{\text{S}}(t_i)$. We can then rewrite the above expression as
\begin{equation}\label{QGBremsstrahlung.18}
\hat{\rho}_{\text{S}}(t)=\hat{\rho}_{\text{S}}(t_i)-\frac{i}{\hbar}\int_{t_i}^t dt'[\hat{H}(t'),\hat{\rho}_{\text{S}}(t')]~.
\end{equation}
Substituting the above expression for $\hat{\rho}_{\text{S}}(t)$ inside of the commutator bracket in the right hand side of the above equation, we obtain
\begin{equation}\label{QGBremsstrahlung.19}
\hat{\rho}_{\text{S}}(t)=\hat{\rho}_{\text{S}}(t_i)-\frac{i}{\hbar}\int_{t_i}^t dt'[\hat{H}(t'),\hat{\rho}_{\text{S}}(t_i)]+\left(-\frac{i}{\hbar}\right)^2\int_{t_i}^tdt'\left[\hat{H}(t'),\int_{t_i}^{t'}dt''[\hat{H}(t''),\hat{\rho}_\text{S}(t'')]\right]~.
\end{equation}
Repeating this process continuously, we can rewrite the above expression as
\begin{equation}\label{QGBremsstrahlung.20}
\begin{split}
\hat{\rho}_{\text{S}}(t)&=\hat{\rho}_{\text{S}}(t_i)-\frac{i}{\hbar}\int_{t_i}^t dt'[\hat{H}(t'),\hat{\rho}_{\text{S}}(t_i)]+\left(-\frac{i}{\hbar}\right)^2\int_{t_i}^tdt'\int_{t_i}^{t'}dt''[\hat{H}(t'),[\hat{H}(t''),\hat{\rho}_\text{S}(t_i)]]+\cdots\\
&=\hat{\rho}_{\text{S}}(t_i)-\frac{i}{\hbar}\int_{t_i}^t dt'[\hat{H}(t'),\hat{\rho}_{\text{S}}(t_i)]+\frac{1}{2}\left(-\frac{i}{\hbar}\right)^2\biggr[\int_{t_i}^tdt'\int_{t_i}^{t'}dt''[\hat{H}(t'),[\hat{H}(t''),\hat{\rho}_\text{S}(t_i)]]\\&+\int_{t_i}^t dt''\int_{t_i}^{t''}dt'[\hat{H}(t''),[\hat{H}(t'),\hat{\rho}_\text{S}(t_i)]]\biggr]+\cdots
\end{split}
\end{equation}
where in the last line of the above equation, we have made use of the fact that $t'$ and $t''$ are dummy indices and we have made a swapping between $t'$ and $t''$ while writing the second integral in the final term in the right hand side of the above equation. We now invoke the Heaviside theta functions to simplify the above equation
\begin{equation}\label{QGBremsstrahlung.21}
\begin{split}
\hat{\rho}_{\text{S}}(t)&=\hat{\rho}_{\text{S}}(t_i)-\frac{i}{\hbar}\int_{t_i}^t dt'[\hat{H}(t'),\hat{\rho}_{\text{S}}(t_i)]+\frac{i^2}{2\hbar^2}\biggr[\int_{t_i}^tdt'\int_{t_i}^{t'}\Theta(t'-t'')dt''[\hat{H}(t'),[\hat{H}(t''),\hat{\rho}_\text{S}(t_i)]]\\&+\int_{t_i}^t dt''\Theta(t''-t')\int_{t_i}^{t''}dt'[\hat{H}(t''),[\hat{H}(t'),\hat{\rho}_\text{S}(t_i)]]\biggr]+\cdots\\
&=\hat{\rho}_{\text{S}}(t_i)-\frac{i}{\hbar}\int_{t_i}^t dt'[\hat{H}(t'),\hat{\rho}_{\text{S}}(t_i)]+\frac{i^2}{2\hbar^2}\biggr[\int_{t_i}^tdt'\Theta(t'-t'')\int_{t_i}^{t}dt''[\hat{H}(t'),[\hat{H}(t''),\hat{\rho}_\text{S}(t_i)]]\\&+\int_{t_i}^t dt''\Theta(t''-t')\int_{t_i}^{t}dt'[\hat{H}(t''),[\hat{H}(t'),\hat{\rho}_\text{S}(t_i)]]\biggr]+\cdots\\
&=\hat{\rho}_{\text{S}}(t_i)-\frac{i}{\hbar}\int_{t_i}^t dt'[\hat{H}(t'),\hat{\rho}_{\text{S}}(t_i)]+\frac{i^2}{2\hbar^2}\int_{t_i}^tdt'\int_{t_i}^{t}dt''\left[\Theta(t'-t'')[\hat{H}(t'),[\hat{H}(t''),\hat{\rho}_\text{S}(t_i)]]\right.\\&+\left.\Theta(t''-t')[\hat{H}(t''),[\hat{H}(t'),\hat{\rho}_\text{S}(t_i)]]\right]+\cdots.
\end{split}
\end{equation}
Making use of the definition of the Liouville superoperator from eq.(\ref{QGBremsstrahlung.13}), it is possible to simply write down the above expression in a much simpler way as
\begin{equation}\label{QGBremsstrahlung.22}
\begin{split}
\hat{\rho}_{\text{S}}(t)&=\hat{\rho}_{\text{S}}(t_i)+\int_{t_i}^t dt'\hat{\mathfrak{L}}^{\text{Tot.}}_{\text{Su}}(t')\hat{\rho}_{\text{S}}(t_i)+\frac{1}{2}\int_{t_i}^tdt'\int_{t_i}^{t}dt''\mathcal{T}\left[\hat{\mathfrak{L}}^{\text{Tot.}}_{\text{Su}}(t')\hat{\mathfrak{L}}^{\text{Tot.}}_{\text{Su}}(t'')\right]\hat{\rho}_{\text{S}}(t_i)+\cdots\\
&=\mathcal{T}\left[e^{\int_{t_i}^tdt'\hat{\mathfrak{L}}^{\text{Tot.}}_{\text{Su}}(t')}\right]\hat{\rho}_{\text{S}}(t_i)
\end{split}
\end{equation}
where $\mathcal{T}[\cdots]$ denote time-ordering. Eq.(\ref{QGBremsstrahlung.22}) gives the solution to the von-Neumann equation of motion in eq.(\ref{QGBremsstrahlung.16}). Our next aim is to obtain the analytical form of the reduced density matrix of the system after the graviton part has been traced out.
\subsection{Reduced density matrix of the Bose-Einstein condensate}
The system density matrix $\hat{\rho}_{\text{S}}(t)$ denotes the combined state of the BEC-graviton system. Now, to truly investigate the graviton effects on the BEC part, we now need to trace over all the field degrees of freedom where the field refers to the graviton part. After tracing out the field degrees of freedom, one obtains the graviton influenced density matrix of the condensate at $t_f$ (some final time) as
\begin{equation}\label{QGBremsstrahlung.23}
\begin{split}
\hat{\rho}^{\mathcal{R}}_{\text{BEC}}(t_f)&=\text{tr}_\mathcal{F}\left[\hat{\rho}_{\text{S}}(t_f)\right]\\
&=\text{tr}_\mathcal{F}\left[\mathcal{T}\Bigr[e^{\int_{t_i}^{t_f}dt'\hat{\mathfrak{L}}^{\text{Tot.}}_{\text{Su}}(t')}\Bigr]\hat{\rho}_{\text{S}}(t_i)\right]
\end{split}
\end{equation}
where the $\mathcal{R}$ in the superscript of $\hat{\rho}^{\mathcal{R}}_{\text{BEC}}$ denotes that it is a reduced density matrix. It is possible to separate the time ordering into two parts, one corresponding to the BEC part and the other corresponding to the Field part as $\mathcal{T}=\mathcal{T}^{\text{BEC}}\mathcal{T}^{\mathcal{F}}$. Making use of the decomposition of the Liouville superoperator from eq.(\ref{QGBremsstrahlung.15}) in eq.(\ref{QGBremsstrahlung.23}), we can write down the reduced density matrix $\hat{\rho}^{\mathcal{R}}_{\text{BEC}}$ as
\begin{equation}\label{QGBremsstrahlung.24}
\begin{split}
\hat{\rho}^\mathcal{R}_{\text{BEC}}(t_f)&=\mathcal{T}^{\text{BEC}}\left[e^{\int_{t_i}^{t_f}dt'\hat{\mathfrak{L}}_{\text{Su}}^0(t')}\text{tr}_{\mathcal{F}}\left[\mathcal{T}^{\mathcal{F}}\left[e^{\int_{t_i}^{t_f}dt''\hat{\mathfrak{L}}_{\text{Su}}^{\text{int}}(t'')}\right]\hat{\rho}_{\text{S}}(t_i)\right]\right]~.
\end{split}
\end{equation}
Dealing with a time ordering of infinite number of Liouville superoperators is very difficult and it is possible to do a simplification of the $\mathcal{T}^{\mathcal{F}}\Bigr[e^{\int_{t_i}^{t_f}dt''\hat{\mathfrak{L}}_{\text{Su}}^{\text{int}}(t'')}\Bigr]\hat{\rho}_{\text{S}}(t_i)$ part from the above equation. Let us consider that the time interval $t_f-t_i$ is a combination of $N$ numbers of small and discrete time steps of value $\Delta t$. Then it is possible to express the time interval $t_f-t_i$ as $t_f-t_i=N\Delta t$. To come back to the continuum limit, one needs to take the $N\rightarrow\infty$ and $\Delta t\rightarrow 0$ limit simultaneously. Hence, $\mathcal{T}^{\mathcal{F}}\Bigr[e^{\int_{t_i}^{t_f}dt''\hat{\mathfrak{L}}_{\text{Su}}^{\text{int}}(t'')}\Bigr]\hat{\rho}_{\text{S}}(t_i)$ can be recast as
\begin{equation}\label{QGBremsstrahlung.25}
\begin{split}
&\mathcal{T}^{\mathcal{F}}\Bigr[e^{\int_{t_i}^{t_f}dt'\hat{\mathfrak{L}}_{\text{Su}}^{\text{int}}(t')}\Bigr]\hat{\rho}_{\text{S}}(t_i)\\&=\lim\limits_{\substack{{\Delta t\rightarrow 0}\\{N\rightarrow\infty}}}\exp\biggr[\Delta t\sum\limits_{i=0}^N\hat{\mathfrak{L}}_{\text{Su}}^{\text{int}}(t_i)+\frac{1}{2}(\Delta t)^2\sum\limits_{\substack{{j,k=0}\\{j>k}}}^N[\hat{\mathfrak{L}}_{\text{Su}}^{\text{int}}(t_j),\hat{\mathfrak{L}}_{\text{Su}}^{\text{int}}(t_k)]\biggr]\hat{\rho}_{\text{S}}(t_i)\\
&=\exp\left[\int_{t_i}^{t_f}dt\hspace{0.5mm} \hat{\mathfrak{L}}_{\text{Su}}^{\text{int}}(t)+\frac{1}{2}\int_{t_i}^{t_f}dt\int_{t_i}^{t_f}dt'\hspace{0.5mm}\Theta(t-t')\left[\hat{\mathfrak{L}}_{\text{Su}}^{\text{int}}(t),\hat{\mathfrak{L}}_{\text{Su}}^{\text{int}}(t')\right]\right]\hat{\rho}_{\text{S}}(t_i)
\end{split}
\end{equation} 
where to obtain the second line of the above equation, Baker-Campbell Hausdorff formula has been used.
We now need to calculate the commutator bracket $[\hat{\mathfrak{L}}_{\text{int}}(t),\hat{\mathfrak{L}}_{\text{int}}(t')]\hat{\rho}(t_i)$ to further simplify the expression in eq.(\ref{QGBremsstrahlung.25}). Using the analytical expression for the Liouville superoperator in eq.(\ref{QGBremsstrahlung.14}), it is possible to write down the analytical expression for the commutator bracket as
\begin{equation}\label{QGBremsstrahlung.26}
\begin{split}
[\hat{\mathfrak{L}}_{\text{Su}}^{\text{int}}(t),\hat{\mathfrak{L}}_{\text{Su}}^{\text{int}}(t')]\hat{\rho}_{\text{S}}(t_i)&=\hat{\mathfrak{L}}_{\text{Su}}^{\text{int}}(t)\hat{\mathfrak{L}}_{\text{Su}}^{\text{int}}(t')\hat{\rho}_{\text{S}}(t_i)-\hat{\mathfrak{L}}_{\text{Su}}^{\text{int}}(t')\hat{\mathfrak{L}}_{\text{Su}}^{\text{int}}(t)\hat{\rho}_{\text{S}}(t_i)\\
&=-\frac{1}{\hbar^2}\left[\hat{H}_{\text{int}}(t),[\hat{H}_{\text{int}}(t'),\hat{\rho}_{\text{S}}(t_i)]\right]+\frac{1}{\hbar^2}\left[\hat{H}_{\text{int}}(t'),[\hat{H}_{\text{int}}(t),\hat{\rho}_{\text{S}}(t_i)]\right]\\
&=-\frac{1}{\hbar^2}\left[\hat{H}_{\text{int}}(t),\hat{H}_{\text{int}}(t')\right]\hat{\rho}_{\text{S}}(t_i)+\frac{1}{\hbar^2}\hat{\rho}_{\text{S}}(t_i)\left[\hat{H}_{\text{int}}(t),\hat{H}_{\text{int}}(t')\right]\\
\implies [\hat{\mathfrak{L}}_{\text{Su}}^{\text{int}}(t),\hat{\mathfrak{L}}_{\text{Su}}^{\text{int}}(t')]\hat{\rho}_{\text{S}}(t_i)&=-\frac{1}{\hbar^2}\left[\left[\hat{H}_{\text{int}}(t),\hat{H}_{\text{int}}(t')\right],\hat{\rho}_{\text{S}}(t_i)\right]~.
\end{split}
\end{equation}
As we are working in natural units, the $\hbar^2$ from the last line in the right hand side of the above equation can directly be set to unity. All the dimensions will be reconstructed properly when we shall investigate the phenomenological aspects of the model. To completely obtain the full analytical expression corresponding to the left hand side of eq.(\ref{QGBremsstrahlung.26}), the first step is to obtain the commutator bracket between the two interaction Hamiltonians in the right hand side of eq.(\ref{QGBremsstrahlung.26}). Making use of the analytical form of the interaction Hamiltonian in eq.(\ref{QGBremsstrahlung.10}) (only single mode of the BEC has been considered), we can write down the commutation relation between the interaction Hamiltonians up to second order in the coupling constant $\kappa_G$ as
\begin{equation}\label{QGBremsstrahlung.27}
\begin{split}
&[\hat{H}_{\text{int}}(t),\hat{H}_{\text{int}}(t')]\\&\simeq\frac{4\kappa_G^2\gamma_B^2c_S^4}{V_G}\sum\limits_{\vec{k},q}\sum\limits_{\vec{k}',q'}\left[\hat{\delta h}^I_{q}(t,\vec{k})\epsilon^q_{ij}(\vec{k}),\hat{\delta h}^I_{q'}(t',\vec{k}')\epsilon^{q'}_{lm}(\vec{k}')\right]\hat{k}_B^i\hat{k}_B^j\hat{k}_B^l\hat{k}_B^m|\hat{\psi}_{\vec{k}_B}(t)|^2|\hat{\psi}_{\vec{k}_B}(t')|^2\\
&=\gamma_B^2c_S^4\left[\frac{2\kappa_G}{\sqrt{V_G}}\sum\limits_{\vec{k},q}\hat{\delta h}^I_{q}(t,\vec{k})\epsilon^q_{ij}(\vec{k}),\frac{2\kappa_G}{\sqrt{V_G}}\sum\limits_{\vec{k}',q'}\hat{\delta h}^I_{q'}(t',\vec{k}')\epsilon^{q'}_{lm}(\vec{k}')\right]\hat{k}_B^i\hat{k}_B^j\hat{k}_B^l\hat{k}_B^m|\hat{\psi}_{\vec{k}_B}(t)|^2|\hat{\psi}_{\vec{k}_B}(t')|^2.
\end{split}
\end{equation}
From eq.(\ref{QGRBEC.67}), we already know the analytical form of the graviton noise operator term as
\begin{equation}\label{QGBremsstrahlung.28}
\delta \hat{\mathcal{N}}_{ij}(t,\vec{x})=\frac{2\kappa_G}{\sqrt{V}_G}\sum\limits_{\vec{k},q}\delta\hat{h}^I_q(t,\vec{k})e^{i\vec{k}\cdot\vec{x}}\epsilon^q_{ij}(\vec{k})~.
\end{equation}
For $\vec{x}=0$, we can obtain from the above expression
\begin{equation}\label{QGBremsstrahlung.29}
\delta \hat{\mathcal{N}}_{ij}(t,0)\equiv\frac{2\kappa_G}{\sqrt{V}_G}\sum\limits_{q}\sum\limits_{\substack{{\vec{k}}\\{|\vec{k}|\leq\Omega_{\text{M}}}}}\delta\hat{h}^I_q(t,\vec{k})\epsilon^q_{ij}(\vec{k})
\end{equation}
where $\Omega_{\text{M}}\sim 10^8$ Hz gives the ultra-violet cut-off frequency as also has been used in the earlier chapter. Using eq.(\ref{QGBremsstrahlung.29}), it is possible to recast the commutator bracket in eq.(\ref{QGBremsstrahlung.27}) in terms of graviton noise fluctuations as
\begin{equation}\label{QGBremsstrahlung.30}
\begin{split}
[\hat{H}_{\text{int}}(t),\hat{H}_{\text{int}}(t')]&=\gamma_B^2c_S^4\left[\delta \hat{\mathcal{N}}_{ij}(t,0),\delta\hat{\mathcal{N}}_{lm}(t',0)\right]\hat{k}_B^i\hat{k}_B^j\hat{k}_B^l\hat{k}_B^m|\hat{\psi}_{\vec{k}_B}(t)|^2|\hat{\psi}_{\vec{k}_B}(t')|^2.
\end{split}
\end{equation}
The commutator between the two interaction Hamiltonians has contributions from the graviton noise fluctuation terms in the quadratic order, and as a result any graviton noise contribution from the modulus square of the time dependent solution of the Goldstone bosons, $|\hat{\psi}_{\vec{k}_B}(t)|^2$, can be neglected as they are of $\mathcal{O}(\delta\mathcal{N}^3)$ or higher. The commutator between the graviton-noise terms is just a number, and therefore it is possible to recast eq.(\ref{QGBremsstrahlung.26}) using eq.(\ref{QGBremsstrahlung.30}) as
\begin{equation}\label{QGBremsstrahlung.31}
\begin{split}
&[\hat{\mathfrak{L}}_{\text{Su}}^{\text{int}}(t),\hat{\mathfrak{L}}_{\text{Su}}^{\text{int}}(t')]\hat{\rho}_{\text{S}}(t_i)=-\gamma_B^2c_S^4[\delta\hat{\mathcal{N}}_{ij}(t,0),\delta\hat{\mathcal{N}}_{lm}(t',0)]\left[\hat{k}_B^i\hat{k}_B^j\hat{k}_B^l\hat{k}_B^m|\hat{\psi}_{\vec{k}_B}(t)|^2|\hat{\psi}_{\vec{k}_B}(t')|^2,\hat{\rho}_{\text{S}}(t_i)\right]~.
\end{split}
\end{equation}
In order to write the expressions in a much simpler form, it is now possible to define two new operators given as
\begin{align}
\hat{\mathfrak{K}}^{ij}_+(t)\hat{\rho}_{\text{S}}(t_i)&\equiv\hat{k}_B^i\hat{k}_B^j|\hat{\psi}_{\vec{k}_B}(t)|^2\hat{\rho}_{\text{S}}(t_i)~,\label{QGBremsstrahlung.32}\\
\hat{\mathfrak{K}}^{ij}_-(t)\hat{\rho}_{\text{S}}(t_i)&\equiv\hat{\rho}_{\text{S}}(t_i)\hat{k}_B^i\hat{k}_B^j|\hat{\psi}_{\vec{k}_B}(t)|^2~.\label{QGBremsstrahlung.33}
\end{align}
Making use of eq.(s)(\ref{QGBremsstrahlung.32},\ref{QGBremsstrahlung.33}), we can rewrite eq.(\ref{QGBremsstrahlung.31}) in terms of the newly defined operators $\hat{\mathfrak{K}}^{ij}_+(t)$ and $\hat{\mathfrak{K}}^{ij}_-(t)$ as
\begin{equation}\label{QGBremsstrahlung.34}
\begin{split}
[\hat{\mathfrak{L}}_{\text{Su}}^{\text{int}}(t),\hat{\mathfrak{L}}_{\text{Su}}^{\text{int}}(t')]\hat{\rho}_{\text{S}}(t_i)=-\gamma_B^2c_S^4[\delta\hat{\mathcal{N}}_{ij}(t,0),\delta\hat{\mathcal{N}}_{lm}(t',0)]\left(\hat{\mathfrak{K}}^{ij}_+(t)\hat{\mathfrak{K}}^{lm}_+(t')-\hat{\mathfrak{K}}^{lm}_-(t')\hat{\mathfrak{K}}^{ij}_-(t)\right)\hat{\rho}_{\text{S}}(t_i)\hspace{0.5mm}.
\end{split}
\end{equation}
Few important observations are in order. First, the noise-noise commutator is a number and can be taken out of the field trace in eq.(\ref{QGBremsstrahlung.24}), and at the same time both the operators $\hat{\mathfrak{K}}^{ij}_+(t)$ and $\hat{\mathfrak{K}}^{ij}_-(t)$ are also completely independent of any contributions from the gravitational field and achieves the same fate as the noise-noise commutator term. Hence, using eq.(s)(\ref{QGBremsstrahlung.25},\ref{QGBremsstrahlung.34}), it is possible to rewrite eq.(\ref{QGBremsstrahlung.24}) as
\begin{equation}\label{QGBremsstrahlung.35}
\begin{split}
\hat{\rho}^{\mathcal{R}}_{\text{BEC}}(t_f)=&\mathcal{T}^{\text{BEC}}\biggr[e^{\int_{t_i}^{t_f}dt\hat{\mathfrak{L}}_{\text{Su}}^0(t)-\frac{\gamma_B^2c_S^4}{2}\int_{t_i}^{t_f}dt\int_{t_i}^{t_f}dt'\Theta(t-t')\left[\delta\hat{\mathcal{N}}_{ij}(t,0),\delta\hat{\mathcal{N}}_{lm}(t',0)\right] \left(\hat{\mathfrak{K}}^{ij}_+(t)\hat{\mathfrak{K}}^{lm}_+(t')-\hat{\mathfrak{K}}^{lm}_-(t')\hat{\mathfrak{K}}^{ij}_-(t)\right)}\\&\times\text{tr}_{\mathcal{F}}\Bigr[e^{\int_{t_i}^{t_f}dt\hat{\mathfrak{L}}_{\text{Su}}^{\text{int}}(t)}\hat{\rho}_{\text{S}}(t_i)\Bigr]\biggr]~.
\end{split}
\end{equation}
At time $t=t_i$, the interaction between the gravitational wave as well as the Bose-Einstein condensate has started and as result, one can quite safely take the initial density matrix of the system to be separable where we can express it as a tensor product between the density matrix corresponding to the BEC part as well as the graviton part. We can then write down the total density matrix for the system at $t=t_i$\footnote{The initial time $t_i$ can also be set to zero without any significant loss of generality.} as
\begin{equation}\label{QGBremsstrahlung.36}
\hat{\rho}_{\text{S}}(t_i)=\hat{\rho}_{\text{BEC}}(t_i)\otimes\hat{\rho}_{\mathcal{F}}(t_i)~.
\end{equation} 
We now define a completely new quantity $\hat{\mathcal{W}}[\mathfrak{K}_+,\mathfrak{K}_-]$ containing the trace over the field part in eq.(\ref{QGBremsstrahlung.35}) as
\begin{equation}\label{QGBremsstrahlung.37}
\hat{\mathcal{W}}[\mathfrak{K}_+,\mathfrak{K}_-]\equiv\text{tr}_{\mathcal{F}}\Bigr[e^{\int_{t_i}^{t_f}dt\hat{\mathfrak{L}}_{\text{Su}}^{\text{int}}(t)}\hat{\rho}_{\text{S}}(t_i)\Bigr]~.
\end{equation}
Using the decomposition of the density matrix from eq.(\ref{QGBremsstrahlung.36}), we can simplify eq.(\ref{QGBremsstrahlung.37}) as
\begin{equation}\label{QGBremsstrahlung.38}
\begin{split}
\hat{\mathcal{W}}[\mathfrak{K}_+,\mathfrak{K}_-]&=\text{tr}_{\mathcal{F}}\Bigr[e^{\int_{t_i}^{t_f}dt\hat{\mathfrak{L}}_{\text{Su}}^{\text{int}}(t)}\hat{\rho}_{\mathcal{F}}(t_i)\Bigr]\hat{\rho}_{\text{BEC}}(t_i)\\
&=\left\langle e^{\int_{t_i}^{t_f}dt\hat{\mathfrak{L}}_{\text{Su}}^{\text{int}}(t)}\right\rangle_{\mathcal{F}}\hat{\rho}_{\text{BEC}}(t_i)\\
&\simeq\left(1+\int_{t_i}^{t_f}dt \left\langle \hat{\mathfrak{L}}_{\text{Su}}^{\text{int}}(t)\right\rangle_{\mathcal{F}}+\frac{1}{2}\int_{t_i}^{t_f}dt \int_{t_i}^{t_f}dt' \left\langle \hat{\mathfrak{L}}_{\text{Su}}^{\text{int}}(t)\hat{\mathfrak{L}}_{\text{Su}}^{\text{int}}(t')\right\rangle_{\mathcal{F}}+\cdots\right) \hat{\rho}_{\text{BEC}}(t_i)\\
&\simeq \hat{\rho}_{\text{BEC}}(t_i)+\frac{1}{2}\int_{t_i}^{t_f} dt\int_{t_i}^{t_f} dt'\left \langle\hat{\mathfrak{L}}_{\text{Su}}^{\text{int}}(t)\hat{\mathfrak{L}}_{\text{Su}}^{\text{int}}(t')\right\rangle_{\mathcal{F}}\hat{\rho}_{\text{BEC}}(t_i)
\end{split}
\end{equation}
where in the penultimate line of the above equation, we have made use of the fact that the Liouville superoperator is directly related to the quantum gravitational noise fluctuation, and as a result any one-point correlator will vanish. Here, we have also neglected any higher order contributions in the Liouville superoperator corresponding to the interaction part of the Hamiltonian as they will contribute towards higher order graviton noise correlators. 
Using the definition of the Liouville superoperators in eq.(\ref{QGBremsstrahlung.14}) and the analytical form of the interaction Hamiltonian in eq.(\ref{QGBremsstrahlung.10}) along with the graviton noise fluctuation term in eq.(\ref{QGBremsstrahlung.29}), we can express $\left\langle\hat{\mathfrak{L}}_{\text{Su}}^{\text{int}}(t)\hat{\mathfrak{L}}_{\text{Su}}^{\text{int}}(t')\right\rangle_{\mathcal{F}}\hat{\rho}_{\text{BEC}}(t_i)$ as 
\begin{equation}\label{QGBremsstrahlung.39}
\begin{split}
&\left \langle\hat{\mathfrak{L}}_{\text{Su}}^{\text{int}}(t)\hat{\mathfrak{L}}_{\text{Su}}^{\text{int}}(t')\right\rangle_{\mathcal{F}}\hat{\rho}_{\text{BEC}}(t_i)\\=&-\gamma_B^2c_S^4\left[\left\langle \delta\hat{\mathcal{N}}_{ij}(t,0)\delta\hat{\mathcal{N}}_{lm}(t',0)\right\rangle_{\mathcal{F}}\hat{\mathfrak{K}}^{ij}_+(t)\hat{\mathfrak{K}}^{lm}_+(t')-\left\langle \delta\hat{\mathcal{N}}_{lm}(t',0)\delta\hat{\mathcal{N}}_{ij}(t,0)\right\rangle_{\mathcal{F}}\hat{\mathfrak{K}}^{lm}_-(t')\hat{\mathfrak{K}}^{ij}_+(t)\right.\\
-&\left.\left\langle \delta\hat{\mathcal{N}}_{ij}(t,0)\delta\hat{\mathcal{N}}_{lm}(t',0)\right\rangle_{\mathcal{F}}\hat{\mathfrak{K}}^{ij}_-(t)\hat{\mathfrak{K}}^{lm}_+(t')+\left\langle \delta\hat{\mathcal{N}}_{lm}(t',0)\delta\hat{\mathcal{N}}_{ij}(t,0)\right\rangle_{\mathcal{F}}\hat{\mathfrak{K}}^{ij}_-(t)\hat{\mathfrak{K}}^{lm}_-(t')\right]\hat{\rho}_{\text{BEC}}(t_i)~.
\end{split}
\end{equation}
Making use of the expression in eq.(\ref{QGBremsstrahlung.39}), and substituting it back in eq.(\ref{QGBremsstrahlung.38}), we arrive at the expression
\begin{equation}\label{QGBremsstrahlung.40}
\begin{split}
&\hat{\mathcal{W}}[\mathcal{K}_+,\mathcal{K}_-]=\exp\left[-\frac{\gamma_B^2c_S^4}{2}\int_{t_i}^{t_f}dt\int_{t_i}^{t}dt'\left\langle\left\{\delta\hat{\mathcal{N}}_{ij}(t,0),\delta\hat{\mathcal{N}}_{lm}(t',0)\right\}\right\rangle_{\mathcal{F}}\right.\\&\times\left[\hat{\mathfrak{K}}^{ij}_+(t)\hat{\mathfrak{K}}^{lm}_+(t')+\hat{\mathfrak{K}}^{ij}_-(t)\hat{\mathfrak{K}}^{lm}_-(t')-\hat{\mathfrak{K}}^{ij}_-(t)\hat{\mathfrak{K}}^{lm}_+(t')-\hat{\mathfrak{K}}^{lm}_-(t')\hat{\mathfrak{K}}^{ij}_+(t)\right]\\&\left.-\frac{\gamma_B^2c_S^4}{2}\int_{t_i}^{t_f}dt\int_{t_i}^{t}dt'\left[\delta\hat{\mathcal{N}}_{ij}(t,0),\delta\hat{\mathcal{N}}_{lm}(t',0)\right]\Bigr(\hat{\mathfrak{K}}^{lm}_-(t')\hat{\mathfrak{K}}^{ij}_+(t)-\hat{\mathfrak{K}}^{ij}_-(t)\hat{\mathfrak{K}}^{lm}_+(t')\Bigr)\right]\hat{\rho}_{\text{BEC}}(t_i)~.
\end{split}
\end{equation}
Before writing down the expression for the reduced density matrix in eq.(\ref{QGBremsstrahlung.35}), we need to define two new operators that shall help us to write down the expression for the reduced density matrix in a compact form. The two new operators are defined as
\begin{align}
\hat{\mathfrak{K}}^{ij}_{c}(t)\hat{\rho}&\equiv\left[\hat{k}^i_B\hat{k}^j_B|\hat{\psi}_{\vec{k}_B}(t)|^2,\hat{\rho}\right]=\left(\hat{\mathfrak{K}}^{ij}_{+}(t)-\hat{\mathfrak{K}}^{ij}_{-}(t)\right)\hat{\rho}\label{QGBremsstrahlung.41}\\
\hat{\mathfrak{K}}^{ij}_{a}(t)\hat{\rho}&\equiv\left\{\hat{k}^i_B\hat{k}^j_B|\hat{\psi}_{\vec{k}_B}(t)|^2,\hat{\rho}\right\}=\left(\hat{\mathfrak{K}}^{ij}_{+}(t)+\hat{\mathfrak{K}}^{ij}_{-}(t)\right)\hat{\rho}~.\label{QGBremsstrahlung.42}
\end{align}
Making use of the operator redefinitions in eq.(s)(\ref{QGBremsstrahlung.41},\ref{QGBremsstrahlung.42}), and eq.(\ref{QGBremsstrahlung.40}), we can rewrite the analytical expression for the reduced density matrix corresponding to the BEC part in eq.(\ref{QGBremsstrahlung.35}) as
\begin{equation}\label{QGBremsstrahlung.43}
\begin{split}
&\hat{\rho}^{\mathcal{R}}_{\text{BEC}}(t_f)\\=&~\mathcal{T}^{\text{BEC}}\left[\exp\left(\int_{t_i}^{t_f}dt\hat{\mathfrak{L}}_{\text{Su}}^0(t)-\frac{\gamma_B^2c_S^4}{2}\int_{t_i}^{t_f}dt\int_{t_i}^{t}dt'\left[\delta\hat{\mathcal{N}}_{ij}(t,0),\delta\hat{\mathcal{N}}_{lm}(t',0)\right] \hat{\mathfrak{K}}^{ij}_c(t)\hat{\mathfrak{K}}^{lm}_a(t')\right)\right.\\&\times\left.\exp\left(-\frac{\gamma_B^2c_S^4}{2}\int_{t_i}^{t_f}dt\int_{t_i}^{t}dt'\left\langle\left\{\delta\hat{\mathcal{N}}_{ij}(t,0),\delta\hat{\mathcal{N}}_{lm}(t',0)\right\}\right\rangle_{\mathcal{F}} \hat{\mathfrak{K}}^{ij}_c(t)\hat{\mathfrak{K}}^{lm}_c(t')\right)\hat{\rho}_{\text{BEC}}(t_i)\right]\\
=&\mathcal{T}^{\text{BEC}}\biggr[e^{\int_{t_i}^{t_f}dt\hat{\mathfrak{L}}_{\text{Su}}^0(t)-\frac{\gamma_B^2c_S^4}{2}\int_{t_i}^{t_f}dt\int_{t_i}^{t}dt'[\delta\hat{\mathcal{N}}_{ij}(t,0),\delta\hat{\mathcal{N}}_{lm}(t',0)] \hat{\mathfrak{K}}^{ij}_c(t)\hat{\mathfrak{K}}^{lm}_a(t')}\hat{\tilde{\mathcal{W}}}(t_f,t_i)\hat{\rho}_{\text{BEC}}(t_i)\biggr]
\end{split}
\end{equation}
where a new operator $\hat{\tilde{\mathcal{W}}}(t_f,t_i)$ is defined as
\begin{equation}\label{QGBremsstrahlung.44}
\begin{split}
\hat{\tilde{\mathcal{W}}}(t_f,t_i)\hat{\rho}\equiv\exp\left(-\frac{\gamma_B^2c_S^4}{2}\int_{t_i}^{t_f}dt\int_{t_i}^{t}dt'\left\langle\left\{\delta\hat{\mathcal{N}}_{ij}(t,0),\delta\hat{\mathcal{N}}_{lm}(t',0)\right\}\right\rangle_{\mathcal{F}} \hat{\mathfrak{K}}^{ij}_c(t)\hat{\mathfrak{K}}^{lm}_c(t')\right)\hat{\rho}~.
\end{split}
\end{equation}
Eq.(s)(\ref{QGBremsstrahlung.43},\ref{QGBremsstrahlung.44}) are one of the most important results in this analysis. We start with the standard calculation of the $\left[\delta\hat{\mathcal{N}}_{ij}(t,0),\delta\hat{\mathcal{N}}_{lm}(t',0)\right]$  commutator when the mode decomposition of the graviton mode operator are done with the mode functions referring to graviton vacuum state. Using eq.(\ref{QGBremsstrahlung.29}), we can write down the above commutation relation as
\begin{equation}\label{QGBremsstrahlung.45}
\begin{split}
&\left[\delta\hat{\mathcal{N}}_{ij}(t,0),\delta\hat{\mathcal{N}}_{lm}(t',0)\right]=\frac{4\kappa_G^2}{V_G}\sum\limits_{\vec{k},q}\sum\limits_{\vec{k}',q'}\left[\delta\hat{h}^{I}_q(t,\vec{k})\epsilon^q_{ij}(\vec{k}),\delta\hat{h}^{I}_{q'}(t',\vec{k}')\epsilon^{q'}_{lm}(\vec{k}')\right]\\\
&=\frac{4\kappa_G^2}{V_G}\sum\limits_{\vec{k},q}\sum\limits_{\vec{k}',q'}\left[\left(\hat{h}^{I}_q(t,\vec{k})-h^{\text{cl}}_q(t,\vec{k})\right)\epsilon^q_{ij}(\vec{k}),\left(\hat{h}^{I}_{q'}(t',\vec{k}')-h^{\text{cl}}_{q'}(t',\vec{k}')\right)\epsilon^{q'}_{lm}(\vec{k}')\right]
\end{split}
\end{equation}
where we have made use of the expression of $\delta\hat{h}^I_q(t,\vec{k})$ from eq.(\ref{QGRBEC.54}). We can further simplify the above expression as
\begin{equation}\label{QGBremsstrahlung.46}
\begin{split}
\left[\delta\hat{\mathcal{N}}_{ij}(t,0),\delta\hat{\mathcal{N}}_{lm}(t',0)\right]&=\frac{4\kappa_G^2}{V_G}\sum\limits_{\vec{k},q}\sum\limits_{\vec{k}',q'}\epsilon^q_{ij}(\vec{k})\epsilon^{q'}_{lm}(\vec{k}')\left[\hat{h}^{I}_q(t,\vec{k}),\hat{h}^{I}_{q'}(t',\vec{k}')\right]~.
\end{split}
\end{equation}
 In order to to obtain the analytical form of the above commutator bracket, we use the mode decomposition of $\hat{h}^I_q(t,\vec{k})$ from eq.(\ref{QGRBEC.48}), which helps us to write down the bracket as
 \begin{equation}\label{QGBremsstrahlung.47} 
 \begin{split}
 \left[\hat{h}^{I}_q(t,\vec{k}),\hat{h}^{I}_{q'}(t',\vec{k}')\right]&=\delta_{q,q'}\delta_{\vec{k},-\vec{k}'}\left(f_k(t)f^*_{k'}(t')-f_k^*(t)f_{k'}(t')\right)\\
 &=-i\frac{\sin(k(t-t'))}{k}\delta_{q,q'}\delta_{\vec{k},-\vec{k}'}~.
 \end{split}
\end{equation}   
Substituting the above result back in eq.(\ref{QGBremsstrahlung.46}), and after a little bit of simplification, we arrive at the following expression for the commutator bracket in eq.(\ref{QGBremsstrahlung.46}) as
\begin{equation}\label{QGBremsstrahlung.48}
\begin{split}
\left[\delta\hat{\mathcal{N}}_{ij}(t,0),\delta\hat{\mathcal{N}}_{lm}(t',0)\right]&=-i\zeta_{ijlm}(t,t')
\end{split}
\end{equation}
where the analytical form of $\zeta_{ijlm}(t,t')$ reads
\begin{equation}\label{QGBremsstrahlung.49}
\begin{split}
\zeta_{ijlm}(t,t')\equiv& \frac{2\kappa_G^2}{5\pi^2c^2}\left(\delta_{ij}\delta_{lm}+\delta_{im}\delta_{jl}-\frac{2}{3}\delta_{ij}\delta_{lm}\right) \left(\frac{\sin(\Omega_\text{M}(t-t'))}{(t-t')^2}-\frac{\Omega_\text{M}\cos(\Omega_\text{M}(t-t'))}{t-t'}\right).
\end{split}
\end{equation}
The important thing to note is that, we have restored $c$ in the expression for $\zeta_{ijlm}(t,t')$ where $\zeta_{ijlm}(t,t')$ is indeed a number.
  \section{Maximally entangled BEC state and graviton induced bremsstrahlung}
With the basic analytical form of the reduced density matrix for the Bose-Einstein condensate part in eq.(\ref{QGBremsstrahlung.43}), we start by considering two Bose-Einstein condensates, where the initial state of the system is maximally entangled. The maximally entangled state of the joint BEC system takes the form 
\begin{equation}\label{Supercondensate.1}
|\psi_{\text{BEC}}\rangle=\frac{1}{\sqrt{2}}\left(|\vec{k}_{B_1}\rangle\otimes|0_{B_2}\rangle+|0_{B_1}\rangle\otimes|\vec{k}_{B_2}\rangle\right)
\end{equation}
where the state is considered at time $t=t_i=0$. 
The action of the $\hat{k}_B^j$ operator on the state $|\vec{k}_{B_f}\rangle$\footnote{Here, the state $|\vec{k}_{B_f}\rangle$ denotes the state of the Bose-Einstein condensate corresponding to the $f^\text{th}$ BEC system.} gives $\hat{k}_B^j|\vec{k}_{B_f}\rangle=k_{B_f}^j|\vec{k}_{B_f}\rangle$, where $f=\{1,2\}$ and $j$ being the spatial index, takes the values $j\in\{1,2,3\}$. A Bose-Einstein condensate is formed by superposing matter waves corresponding to each bosons in the Fourier space, and as a result it is quite natural to denote the BEC state by the transverse wave-number, which is related to the number of particles in the condensate. Here, the states are eigenstates of the transverse wave-number operator. We here therefore consider eigenstates corresponding to individual phonon mode frequencies instead of the position eigenstates, which will not be able to properly identify the decoherence effects generated due to interacting gravitons with BEC. We now need to write down the initial state corresponding to the entire model system. Now, if the state of the quantum gravitational field (which can be treated as the environment here) is given by $|h_G\rangle$ (at time $t=0$), in that case, for the interaction between the quantum gravitational field and the condensate system, it is possible to write down the initial state of the system as
\begin{equation}\label{Supercondensate.2}
\begin{split}
|\psi_{{\text{S}}_i}\rangle=&\frac{1}{\sqrt{2}}\left(|\vec{k}_{B_1}\rangle\otimes|0_{B_2}\rangle+|0_{B_1}\rangle\otimes|\vec{k}_{B_2}\rangle\right)\otimes|h_G\rangle\\
=&|\psi_{\text{BEC}}\rangle\otimes |h_G\rangle~.
\end{split}
\end{equation}
Initially, the BEC-graviton state is considered to be separable in nature, and we shall investigate whether after a finite time interval, the states remain separable. The state $|\psi_{\text{S}_i}\rangle$ must be normalized, which gives the following standard normalization conditions
\begin{equation}\label{Supercondensate.3}
\langle \vec{k}_{B_f}| \vec{k}_{B_f}\rangle=\langle 0_{B_f}| 0_{B_f}\rangle=1 \text{ and }\langle h_G| h_G\rangle=1~.
\end{equation}
We can obtain the density matrix  of the system at time $t=t_i$ (where $t_i$ can be set to zero) as
\begin{equation}\label{Supercondensate.4}
\hat{\rho}_{\text{S}_i}=|\psi_{\text{S}_i}\rangle\langle \psi_{\text{S}_i}|~.
\end{equation}
The reduced density matrix corresponding to the BEC part can be obtained by tracing over all quantum gravitational field degrees of freedom as
\begin{equation}\label{Supercondensate.5}
\begin{split}
\hat{\rho}^\mathcal{R}_{\text{BEC}}(0)&=\text{tr}_{\mathcal{F}}\left[\hat{\rho}_{\text{S}_i}\right]=\hat{\rho}^{\mathcal{R}}_{\text{B}_{11}}+\hat{\rho}^{\mathcal{R}}_{\text{B}_{12}}+\hat{\rho}^{\mathcal{R}}_{\text{B}_{21}}+\hat{\rho}^{\mathcal{R}}_{\text{B}_{22}}
\end{split}
\end{equation} 
and the components of the reduced density matrix are given by
\begin{align}
\hat{\rho}^{\mathcal{R}}_{\text{B}_{11}}&=\frac{1}{2}\left(|\vec{k}_{B_1}\rangle\otimes|0_{B_2}\rangle\right)\left(\langle \vec{k}_{B_1}|\otimes\langle 0_{B_2}|\right)\label{Supercondensate.6}\\
\hat{\rho}^{\mathcal{R}}_{\text{B}_{12}}&=\frac{1}{2}\left(|\vec{k}_{B_1}\rangle\otimes|0_{B_2}\rangle\right)\left(\langle 0_{B_1}|\otimes\langle \vec{k}_{B_2}|\right)\label{Supercondensate.7}\\
\hat{\rho}^{\mathcal{R}}_{\text{B}_{21}}&=\frac{1}{2}\left(|0_{B_1}\rangle\otimes|\vec{k}_{B_2}\rangle\right)\left(\langle \vec{k}_{B_1}|\otimes\langle 0_{B_2}|\right)\label{2.41c}\\
\hat{\rho}^{\mathcal{R}}_{\text{B}_{22}}&=\frac{1}{2}\left(|0_{B_1}\rangle\otimes|\vec{k}_{B_2}\rangle\right)\left(\langle 0_{B_1}|\otimes\langle \vec{k}_{B_2}|\right)\label{Supercondensate.9}~.
\end{align}
It is still possible to express the reduced density matrix of the system $\hat{\rho}^{\mathcal{R}}_{\text{BEC}}(0)$ as $\hat{\rho}_{\text{BEC}}(0)=|\psi_{\text{BEC}}\rangle\langle \psi_{\text{BEC}}|$, where the state $|\psi_{\text{BEC}}\rangle$ is given in eq.(\ref{Supercondensate.1}).
The system can be thought of as a Bose-Einstein condensate with a weak-coupling constant kept inside of a harmonic trap potential, where the two different paths are generated by two atom laser beams which remains coherent throughout the experiment. The entanglement between the atom laser beams are executed at source after their production from the trapped condensate. If the coupling between the BEC and graviton states is created, then the graviton state starts to change, and after a finite time $t_f$, the state of the BEC-graviton system reads
\begin{equation}\label{Supercondensate.10}
\begin{split}
|\psi_{\text{S}_f}\rangle&=\frac{1}{\sqrt{2}}|\vec{k}_{B_1},0_{B_2}\rangle\otimes |h_G;\vec{k}_{B_1}\rangle+\frac{1}{\sqrt{2}}|0_{B_1},\vec{k}_{B_2}\rangle\otimes|h_G;\vec{k}_{B_2}\rangle
\end{split}
\end{equation}
where $|A_1,A_2\rangle\equiv |A_1\rangle \otimes|A_2\rangle$, and $|h_G;\vec{k}_{B_f}\rangle$ denotes that the graviton state is modified by the $f^\text{th}$ atomic beam. The coupling between the graviton and the BEC is evident from the structure of the interaction Hamiltonian in eq.(\ref{QGBremsstrahlung.10}). We can now look at the reduced density matrix of the system by writing down the final density matrix of the system as $\hat{\rho}_{\text{S}_f}=|\psi_{\text{S}_f}\rangle\langle\psi_{\text{S}_f}|$, and taking trace over all gravitational field degrees of freedom. The resulting reduced density matrix takes the form
\begin{equation}\label{Supercondensate.11}
\begin{split}
\hat{\rho}^{\mathcal{R}}_{\text{BEC}}(t_f)&=\hat{\rho}^{\mathcal{R}}_{\text{B}_{11}}+\hat{\rho}^{\mathcal{R}}_{\text{B}_{22}}+e^{i\Delta_{\text{IF}}(t_f)}\hat{\rho}^{\mathcal{R}}_{\text{B}_{12}}+e^{-i\Delta_{\text{IF}}^*(t_f)}\hat{\rho}^{\mathcal{R}}_{\text{B}_{21}}~.
\end{split}
\end{equation}
In the above equation, $\Delta_{\text{IF}}(t_f)$ denotes the influence functional as it captures the effect of the gravitons on the BEC system, and its analytical form is given by 
\begin{equation}\label{Supercondensate.12}
e^{i\Delta_{\text{IF}}(t_f)}\equiv  \langle h_G;\vec{k}_{B_2}|h_G;\vec{k}_{B_1}\rangle~.
\end{equation}
The influence functional consists of a real  part and an imaginary part and the imaginary part leads to the decoherence effect. This imaginary part is also called the decoherence function. Between the two coherent atomic laser beams, the decoherence function introduces entanglement loss. To obtain the decoherence function, we start with the definition of the operator $\hat{\tilde{\mathcal{W}}}(t_f,0)$ from eq.(\ref{QGBremsstrahlung.44}), and investigate its action upon the reduced density matrix at time $t=0$ corresponding to the BEC part as
\begin{equation}\label{Supercondensate.13}
\begin{split}
&\hat{\tilde{\mathcal{W}}}(t_f,0)\hat{\rho}^{\mathcal{R}}_{\text{BEC}}(0)\\=&\exp\left[-\frac{\gamma_B^2c_S^4}{2}\int_{t_i}^{t_f}dt\int_{t_i}^{t}dt'\left\langle\left\{\delta\hat{\mathcal{N}}_{ij}(t,0),\delta\hat{\mathcal{N}}_{lm}(t',0)\right\}\right\rangle_{\mathcal{F}} \hat{\mathfrak{K}}^{ij}_c(t)\hat{\mathfrak{K}}^{lm}_c(t')\right]\hat{\rho}^{\mathcal{R}}_{\text{BEC}}(0)\\
\simeq&\left[1-\frac{\gamma_B^2c_S^4}{2}\int_{t_i}^{t_f}dt\int_{t_i}^{t}dt'\left\langle\left\{\delta\hat{\mathcal{N}}_{ij}(t,0),\delta\hat{\mathcal{N}}_{lm}(t',0)\right\}\right\rangle_{\mathcal{F}} \hat{\mathfrak{K}}^{ij}_c(t)\hat{\mathfrak{K}}^{lm}_c(t')+\cdots\right]\hat{\rho}^{\mathcal{R}}_{\text{BEC}}(0)
\end{split}
\end{equation}
where we have made use of the fact that the two-point noise-noise correlator is a contribution coming due to the graviton interaction, and as a result it is small.
The action of the $\hat{\mathfrak{K}}^{ij}_c(t)\hat{\mathfrak{K}}^{lm}_c(t')$ operator on the reduced density matrix $\hat{\rho}^{\mathcal{R}}_{\text{BEC}}(0)$
reads
\begin{equation}\label{Supercondensate.14}
\begin{split}
&\hat{\mathfrak{K}}^{ij}_c(t)\hat{\mathfrak{K}}^{lm}_c(t')\hat{\rho}^{\mathcal{R}}_{\text{BEC}}(0)\\=&\left(k_{B_1}^ik_{B_1}^j|\psi_{\vec{k}_{B_1}}(t)|^2-k_{B_2}^ik_{B_2}^j|\psi_{\vec{k}_{B_2}}(t)|^2\right)\left(k_{B_1}^lk_{B_1}^m|\psi_{\vec{k}_{B_1}}(t')|^2-k_{B_2}^lk_{B_2}^m|\psi_{\vec{k}_{B_2}}(t')|^2\right)\left[\hat{\rho}^{\mathcal{R}}_{\text{B}_{12}}+\hat{\rho}^{\mathcal{R}}_{\text{B}_{21}}\right]
\end{split}
\end{equation}
where it is evident that the action of the $\hat{\mathfrak{K}}^{ij}_c(t)\hat{\mathfrak{K}}^{lm}_c(t')$ operator makes the diagonal components of the reduced density matrix to disappear. Using the above relation it is then possible to write down the expression in eq.(\ref{Supercondensate.13}) (after recombining back all the higher order contributions) as
\begin{equation}\label{Supercondensate.15}
\begin{split}
\hat{\tilde{\mathcal{W}}}(t_f,0)\hat{\rho}^{\mathcal{R}}_{\text{BEC}}(0)&=\hat{\rho}^{\mathcal{R}}_{\text{B}_{11}}+\hat{\rho}^{\mathcal{R}}_{\text{B}_{22}}+e^{-\Gamma(t_f)}\hat{\rho}^{\mathcal{R}}_{\text{B}_{12}}+e^{-\Gamma(t_f)}\hat{\rho}^{\mathcal{R}}_{\text{B}_{21}}
\end{split}
\end{equation}
where $\Gamma(t_f)$ has the analytical form
\begin{equation}\label{Supercondensate.16}
\begin{split}
\Gamma(t_f)&=\frac{\gamma_B^2c_S^4}{2}\int_0^{t_f}dt\int_{0}^{t}dt'\langle\{\delta\hat{\mathcal{N}}_{ij}(t,0),\delta\hat{\mathcal{N}}_{lm}(t',0)\}\rangle_{\mathcal{F}}\Delta\left[k_B^ik_B^j|\psi_{\vec{k}_B}(t)|^2\right]\Delta\left[k_B^lk_B^m|\psi_{\vec{k}_B}(t')|^2\right]
\end{split}
\end{equation}
with the definition of $\Delta\left[k_B^ik_B^j|\psi_{\vec{k}_B}(t)|^2\right]$ being given as
\begin{equation}\label{Supercondensate.17}
\begin{split}
\Delta\left[k_B^ik_B^j|\psi_{\vec{k}_B}(t)|^2\right]\equiv k_{B_1}^ik_{B_1}^j|\psi_{\vec{k}_{B_1}}(t)|^2-k_{B_2}^ik_{B_2}^j|\psi_{\vec{k}_{B_2}}(t)|^2~.
\end{split}
\end{equation}
If the symmetry property of the anti-commutator is implemented, then it is possible to rewrite eq.(\ref{Supercondensate.16}) as
\begin{equation}\label{Supercondensate.18}
\Gamma(t_f)=\frac{\gamma_B^2c_S^4}{4}\int_0^{t_f}dt\int_{0}^{t_f}dt'\langle\{\delta\hat{\mathcal{N}}_{ij}(t,0),\delta\hat{\mathcal{N}}_{lm}(t',0)\}\rangle_{\mathcal{F}}\Delta\left[k_B^ik_B^j|\psi_{\vec{k}_B}(t)|^2\right]\Delta\left[k_B^lk_B^m|\psi_{\vec{k}_B}(t')|^2\right]
\end{equation}
and the analytical form of $\psi_{\vec{k}_{B_j}}(t)$ reads $\psi_{\vec{k}_{B_j}}(t)\simeq\alpha^{B}_{j}e^{-i\omega_{B_j}t}+\beta^{B}_{j}e^{i\omega_{B_j}t}$ for $j\in\{1,2\}$. The analytical forms of the Bogoliubov coefficients are given in eq.(s)(\ref{QGRBEC.102},\ref{QGRBEC.103}), where the quantum gravity corrections have been dropped. The reason behind dropping the graviton noise fluctuations lies in the fact that the decoherence factor is already second order in the noise term, and as a result the graviton noise contributions from the Bogoliubov coefficients will result in higher order noise contributions, which is very small and can be neglected in principle. Now, we already have obtained the form of the noise-noise commutator in eq.(\ref{QGBremsstrahlung.48}) and substituting this form of the commutator in the exponential term with the noise-noise commutator, which is acting on the $\hat{\tilde{\mathcal{W}}}(t_f,t_i)\hat{\rho}_{\text{BEC}}(t_i)$ factor in eq.(\ref{QGBremsstrahlung.43}), we obtain for $t_i=0$
\begin{equation}\label{Supercondensate.19}
\begin{split}
&\exp\left(\frac{i\gamma_B^2c_S^4}{2}\int_{0}^{t_f}dt\int_{0}^{t}dt'\zeta_{ijlm}(t,t') \hat{\mathfrak{K}}^{ij}_c(t)\hat{\mathfrak{K}}^{lm}_a(t')\right)\hat{\tilde{\mathcal{W}}}(t_f,0)\hat{\rho}^{\mathcal{R}}_{\text{BEC}}(0)\\
=&\hat{\rho}^{\mathcal{R}}_{\text{B}_{11}}+\hat{\rho}^{\mathcal{R}}_{\text{B}_{22}}+e^{-\Gamma(t_f)+i\Lambda(t_f)}\hat{\rho}^{\mathcal{R}}_{\text{B}_{12}}+e^{-\Gamma(t_f)-i\Lambda(t_f)}\hat{\rho}^{\mathcal{R}}_{\text{B}_{21}}~.
\end{split}
\end{equation}
The analytical form of $\hat{\tilde{\mathcal{W}}}(t_f,0)\hat{\rho}^{\mathcal{R}}_{\text{BEC}}(0)$ is obtained earlier in eq.(\ref{Supercondensate.15}), and the form of the $\Lambda(t_f)$ factor reads
\begin{equation}\label{Supercondensate.20}
\begin{split}
\Lambda(t_f)&=\frac{\gamma_B^2c_S^4}{2}\int_0^{t_f}dt\int_{0}^{t}dt'\zeta_{ijlm}(t,t')\Delta\left[k_B^ik_B^j|\psi_{\vec{k}_B}(t)|^2\right]\Delta^{+}\left[k_B^lk_B^m|\psi_{\vec{k}_B}|(t')|^2\right]
\end{split}
\end{equation} 
with $\Delta^{+}\left[k_B^lk_B^m|\psi_{\vec{k}_B}|(t')|^2\right]$ being defined as
\begin{equation}\label{Supercondensate.21}
\Delta^{+}\left[k_B^lk_B^m|\psi_{\vec{k}_B}|(t')|^2\right]\equiv k_{B_1}^lk_{B_1}^m|\psi_{\vec{k}_{B_1}}(t')|^2+k_{B_2}^lk_{B_2}^m|\psi_{\vec{k}_{B_2}}(t')|^2~.
\end{equation}
Defining $\Delta(t_f)\equiv \Lambda(t_f)+i\Gamma(t_f)$, it is possible to recast the final expression from eq.(\ref{Supercondensate.19}), as
\begin{equation}\label{Supercondensate.22}
\begin{split}
\hat{\tilde{\rho}}^{\mathcal{R}}_{\text{BEC}}(t_f)=\hat{\rho}^{\mathcal{R}}_{\text{B}_{11}}+\hat{\rho}^{\mathcal{R}}_{\text{B}_{22}}+e^{i\Delta(t_f)}\hat{\rho}^{\mathcal{R}}_{\text{B}_{12}}+e^{-i\Delta^*(t_f)}\hat{\rho}^{\mathcal{R}}_{\text{B}_{21}}
\end{split}
\end{equation}
which has a structure identical to the reduced density matrix in eq.(\ref{Supercondensate.11}).
\begin{figure}
\begin{center}
\includegraphics[scale=0.35]{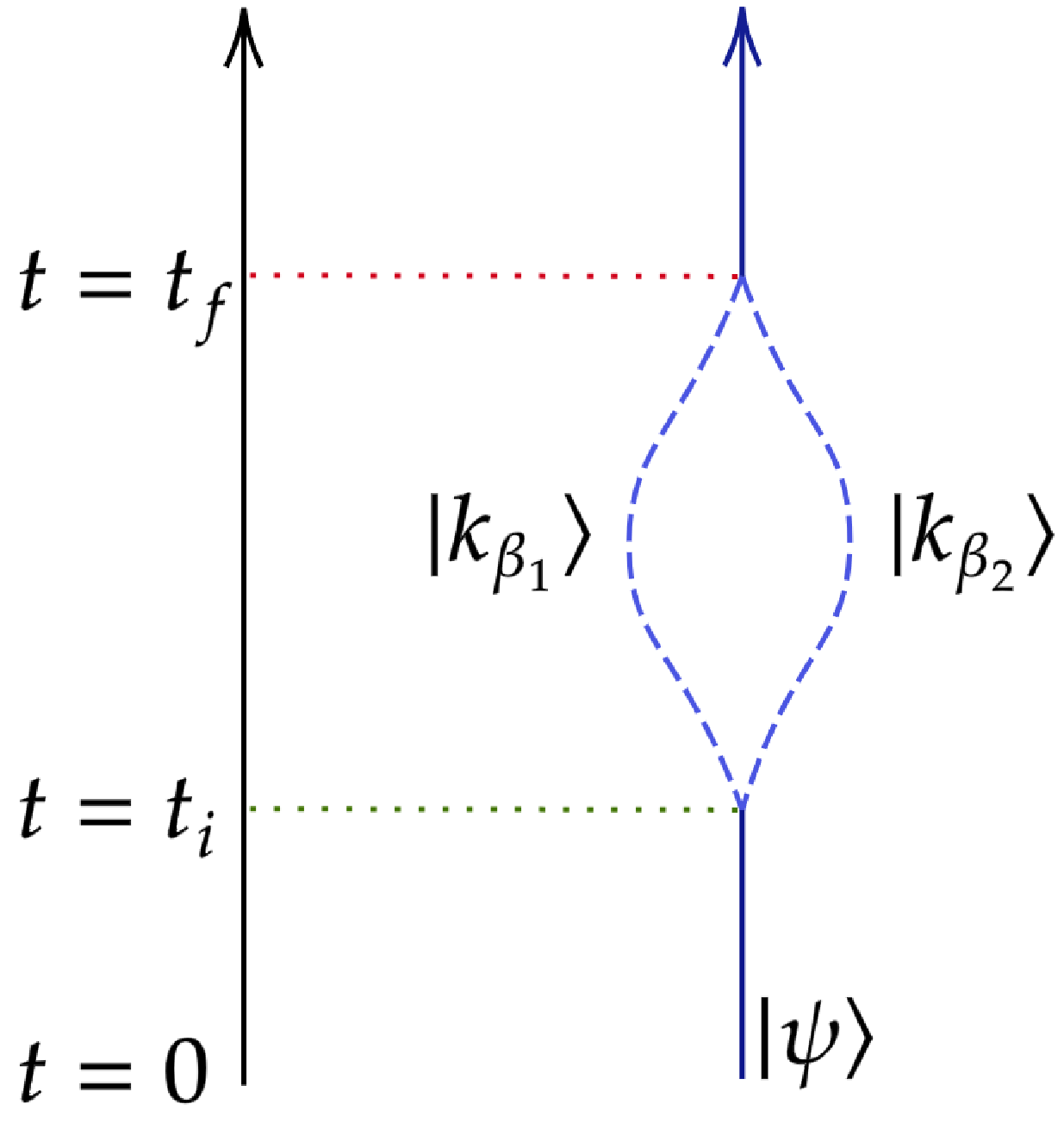}
 \caption{The state and the evolution of the two coherent BEC sources has been depicted, where just after the  formation of the BEC at $t=0$, the state goes into a superposition state in the time-interval $t\in[t_i,t_f]$. \label{Superposition_State_OTM}}
\end{center}
\end{figure}
Eq.(\ref{Supercondensate.22}) is very important in our analysis. We see that after a finite time interval $t_f-t_i=t_f$ ($t_i=0$), the final density matrix of the BEC $\hat{\tilde{\rho}}^{\mathcal{R}}_{\text{BEC}}(t_f)$ has contributions from the graviton noise term, where the off-diagonal terms of the density matrix gets modified by the exponential factor. The phase part of the exponential term is completely dependent on the graviton noise-noise commutator whereas the exponential term is dependent on the two-point graviton noise-noise correlator (expectation of the anti-commutator). As a result, even after a very small time, it is impossible to express the density matrix as an outer product of two quantum states. This phenomena is a result of the interaction of the gravitons with the Bose-Einstein condensate. This BEC-graviton entangled system is termed as a``\textit{Bose-Einstein supercondensate}" in our work. We shall now explore whether it is possible to detect quantum gravity signatures after a Bose-Einstein supercondensate is formed. We can safely ignore the action of the Liouville superoperator corresponding to the base Hamiltonian on the density matrix in eq.(\ref{Supercondensate.22}) and proceed with the reduced density matrix $\hat{\tilde{\rho}}^{\mathcal{R}}_{\text{BEC}}(t_f)$. The primary reason behind such a consideration lies in the fact that the entire effect of the graviton interaction is captured in eq.(\ref{Supercondensate.22}). From Fig.(\ref{Superposition_State_OTM}),  we see a schematic diagram of two single-mode phonon states which are separated by some frequency difference in the Fourier space in a way such that $|\vec{k}_{B_1}|\neq |\vec{k}_{B_2}|$ (in the diagram $k_\beta=k_B$), they remain in a superposition state in the time-interval $t_i<t<t_f$ . Here, we have primarily focused onto entangled state. 
\subsection{Decoherence rate corresponding to squeezed graviton states}
This decoherence analysis is valid for both superposition and entangled states. Here, we are considering entangled states as they result in more distinctive features. From eq.(\ref{Supercondensate.18}), one can easily find out that $\Gamma(t_f)$ gives the decoherence rate, and to obtain the decoherence rate, we at first need to calculate the two point noise-noise correlator $\langle \{\delta\hat{\mathcal{N}}_{ij}(t,0),\delta\hat{\mathcal{N}}_{lm}(t',0)\}\rangle_{\mathcal{F}}$ corresponding to squeezed graviton states. The analytical form of the two-point correlator in the presence of graviton squeezed states has already been obtained in eq.(\ref{NoiseSqueezed.15}), where to analytically obtain the same, we have made use of the summation condition given by $\sum_q\epsilon^{q*}_{ij}(\vec{k})\epsilon^{q}_{lm}(\vec{k})=\frac{1}{2}\left(\mathcal{P}^{\text{R}}_{il}\mathcal{P}^{\text{R}}_{jm}+\mathcal{P}^{\text{R}}_{im}\mathcal{P}^\text{R}_{jl}-\mathcal{P}^\text{R}_{ij}\mathcal{P}^{\text{R}}_{lm}\right)$ from eq.(\ref{QGRBEC.70}) with the projection tensor $\mathcal{P}^{\text{R}}_{ij}$ being given by $\mathcal{P}^\text{R}_{ij}=\delta_{ij}-\frac{k_ik_j}{k^2}$, and finally the angular integrals have been executed, given by $\int d\Omega=4\pi$, $\int d\Omega \frac{k^ik^j}{k^2}=\frac{4\pi}{3}\delta^{ij}$, and $\int d\Omega \frac{k^ik^jk^lk^m}{k^4}=\frac{4\pi}{15}\left(\delta^{ij}\delta^{lm}+\delta^{il}\delta^{jm}+\delta^{im}\delta^{jl}\right)$. The analytical form of the two-point correlator reads
 \begin{equation}\label{Supercondensate.23}
\begin{split}
\langle \{\delta\hat{\mathcal{N}}_{ij}(t,0),\delta\hat{\mathcal{N}}_{lm}(t',0)\}\rangle_{\mathcal{F}}&=\frac{2\kappa_G^2}{5\pi^2}\left(\delta_{ik}\delta_{jl}+\delta_{il}\delta_{jk}-\frac{2}{3}\delta_{ij}\delta_{kl}\right)\int_0^{\Omega_{\text{M}}}dk k^2\mathcal{Q}_{\delta h}(t,t',\vec{k})\\&=\frac{2\kappa_G^2}{5\pi^2c^2}\left(\delta_{il}\delta_{jm}+\delta_{im}\delta_{jl}-\frac{2}{3}\delta_{ij}\delta_{lm}\right)\left\langle\{\delta\hat{\mathcal{N}}(t,0),\delta\hat{\mathcal{N}}(t',0)\}\right\rangle_{\mathcal{F}}
\end{split} 
\end{equation}  
where the analytical form of $\left\langle\{\delta \hat{\mathcal{N}}(t,0),\delta \hat{\mathcal{N}}(t',0)\}\right\rangle_{\mathcal{F}}$ is obtained as
\begin{equation}\label{Supercondensate.24}
\begin{split}
&\left\langle\{\delta \hat{\mathcal{N}}(t,0),\delta \hat{\mathcal{N}}(t',0)\}\right\rangle_{\mathcal{F}}\\&=\cosh 2\mathfrak{r}_k\left[\frac{-1+\cos(\Omega_\text{M} (t-t'))+\Omega_\text{M}(t-t')\sin(\Omega_\text{M} (t-t'))}{(t-t')^2}\right]\\&+\sinh 2\mathfrak{r}_k\left[\frac{\cos\varphi_k+\cos(\Omega_\text{M} (t+t')-\phi_k)+\Omega_\text{M}(t+t')\sin(\Omega_\text{M} (t+t')-\varphi_k)}{(t+t')^2}\right] ~.
\end{split}
\end{equation}
From Fig.(\ref{Superposition_State_OTM}), it is evident that in the time interval $t_f-t_i$, the two states in the Fourier space gets slightly displaced, which is equivalent to a frequency shift. Similar case will occur when we are considering entangled states. For the BEC-graviton interaction case, it can be assumed that the overall frequency shift in the frequencies of the two coherent states (initially maximally entangled with each other) is proportional to the frequency of the incoming gravitational wave up to some constant number. For this scenario, as has already been discussed in Chapter(\ref{C.6.OTM}), the resonance condition gets satisfied when the frequency of the BEC is half of the frequency of the incoming gravitational wave ($\omega_B=\frac{\Omega_0}{2}$). The difference in frequency is kept such that it is less than resonance frequency of the BEC, that is $\Delta\omega_B=|\omega_{B_2}-\omega_{B_1}|\leq\frac{\Omega_0}{2}$. As discussed in Chapter(\ref{C.6.OTM}), $k_B=\frac{\omega_B}{c_S}$ with $c_S$ being the speed of sound. We assume simple analytical forms of the modulus of the wave vectors $k_{B_1}$ and $k_{B_2}$ as
$k_{B_1}=\frac{\omega_B}{c_S}-\frac{\Omega_0}{4c_S}$ and $k_{B_2}=\frac{\omega_B}{c_S}+\frac{\Omega_0}{4c_S}$. It is important to consider that we have considered maximum separation for the two paths in the Fourier space by considering maximum separation in the frequencies. With the form of the noise-noise correlator, we also need the analytical form of $\Delta\left[k_B^ik_B^j|\psi_{{\vec{k}_B}}(t)|^2\right]$ to obtain the analytical form of $\Gamma(t_f)$ from eq.(\ref{Supercondensate.18}). The time-dependent part corresponding to the Goldstone bosons can now be expressed as (for reference see Chapter(\ref{C.6.OTM})), $\hat{\psi}_{{\vec{k}_B}}(t)=\hat{\alpha}^Be^{-i\omega_B t}+\hat{\beta}^Be^{i\omega_B t}$ where the operator valued Bogoliubov coefficients ($\{\hat{\alpha}^B,\hat{\beta}^B\}$) can be expressed as a combination of the standard Bogoliubov coefficient in the semiclassical analogy, and a noise fluctuation term induced by the gravitons. The important aspect to notice from the analytical form of the decoherence factor in eq.(\ref{Supercondensate.18}) is that the decoherence factor is already second order in the noise fluctuation term, and as a result it is possible to neglect any noise contributions in the operator-valued Bogoliubov coefficients. Here, we consider the classical gravitational wave fluctuation to have an analytical form $h_{\text{cl}}(t,0)=\varepsilon e^{-\frac{t^2}{\tau^2}}\sin(\Omega_0 t)$ for which the operator valued Bogoliubov coefficients have the forms $\hat{\alpha}^B=\alpha^B+\mathcal{O}(\delta\hat{\mathcal{N}})$ and $\hat{\beta}^B=\beta^B+\mathcal{O}(\delta\hat{\mathcal{N}})$ with $\alpha^B=1$ and $\beta^B=\mathcal{O}(\varepsilon)$ \footnote{The analytical form of $\beta^B$ from eq.(\ref{QGRBEC.103}) reads $\beta^B=\frac{\sqrt{\pi}\varepsilon_B\omega_B\tau}{4}\left(e^{-\frac{\tau^2}{4}(\Omega_0-2\omega_B)^2}-e^{-\frac{\tau^2}{4}(\Omega_0+2\omega_B)^2}\right)$.} \cite{MatthewMannAffshordi}, where $\varepsilon$ denotes the amplitude of the gravitational wave. It is then simply possible to replace the operator valued Bogoliubov coefficients by the standard Bogoliubov coefficients corresponding to the classical gravitational wave, and express the $\Delta\left[k_B^ik_B^j|\psi_{{\vec{k}_B}}(t)|^2\right]$ term as
\begin{equation}\label{Supercondensate.25}
\begin{split}
&\Delta\left[k_B^ik_B^j|\psi_{{\vec{k}_B}}(t)|^2\right]\\&=\frac{\omega_{B_1}^i\omega_{B_1}^j}{c_S^2}\left({\alpha^{B}_1}^2+{\beta^{B}_1}^2+2\alpha^{B}_1\beta^{B}_1\cos(2\omega_{B_1}t)\right)-\frac{\omega_{B_2}^i\omega_{B_2}^j}{c_S^2}\left({\alpha^{B}_2}^2+{\beta^{B}_2}^2+2\alpha^{B}_2\beta^{B}_2\cos(2\omega_{B_2}t)\right)~.
\end{split}
\end{equation}
The easiest way to proceed with the above analytical form is by setting one of the Bogoliubov coefficients to zero. Now, as $\Gamma(t_f)\sim \mathcal{O}(\delta\hat{\mathcal{N}}^2)$ and $\beta^B\sim\mathcal{O}(\varepsilon)$ (with $\varepsilon$ being a very small quantity), we can as a whole ignore the $\mathcal{O}(\varepsilon \delta\hat{\mathcal{N}}^2)$ kind of contributions and set it to zero neglecting the overall contribution from the negative energy modes to be zero. This indeed is a quite logical step, which leaves us with $\alpha^B=1$ and $\beta^B\simeq 0$. The important observation is that with such a consideration, the difference term in eq.(\ref{Supercondensate.25}) becomes time independent. For a more generalized and intricate analysis, one can keep both of the coefficients and to simplify it enough consider $\alpha^B_i\simeq\beta^B_i$ where $i\in\{1,2\}$. One can also consider $\alpha^B_1\simeq\alpha^B_2=\alpha_B$. It is easier to start with case where the Bogoliubov coefficients $\beta^B_i$ for $i\in\{1,2\}$ vanishes for which the decoherence rate from eq.(\ref{Supercondensate.18}) takes the form
\begin{equation}\label{Supercondensate.26}
\begin{split}
\Gamma(t_f)=\frac{8\gamma_B^2\omega_B^2\Omega^2_0}{3}\int_0^{t_f}dt\int_{0}^tdt'\left\langle \{\delta\hat{\mathcal{N}}(t,0),\delta\hat{\mathcal{N}}(t',0)\}\right\rangle_{\mathcal{F}}.
\end{split}
\end{equation}
\begin{figure}[t!]
\begin{center}
\includegraphics[scale=0.42]{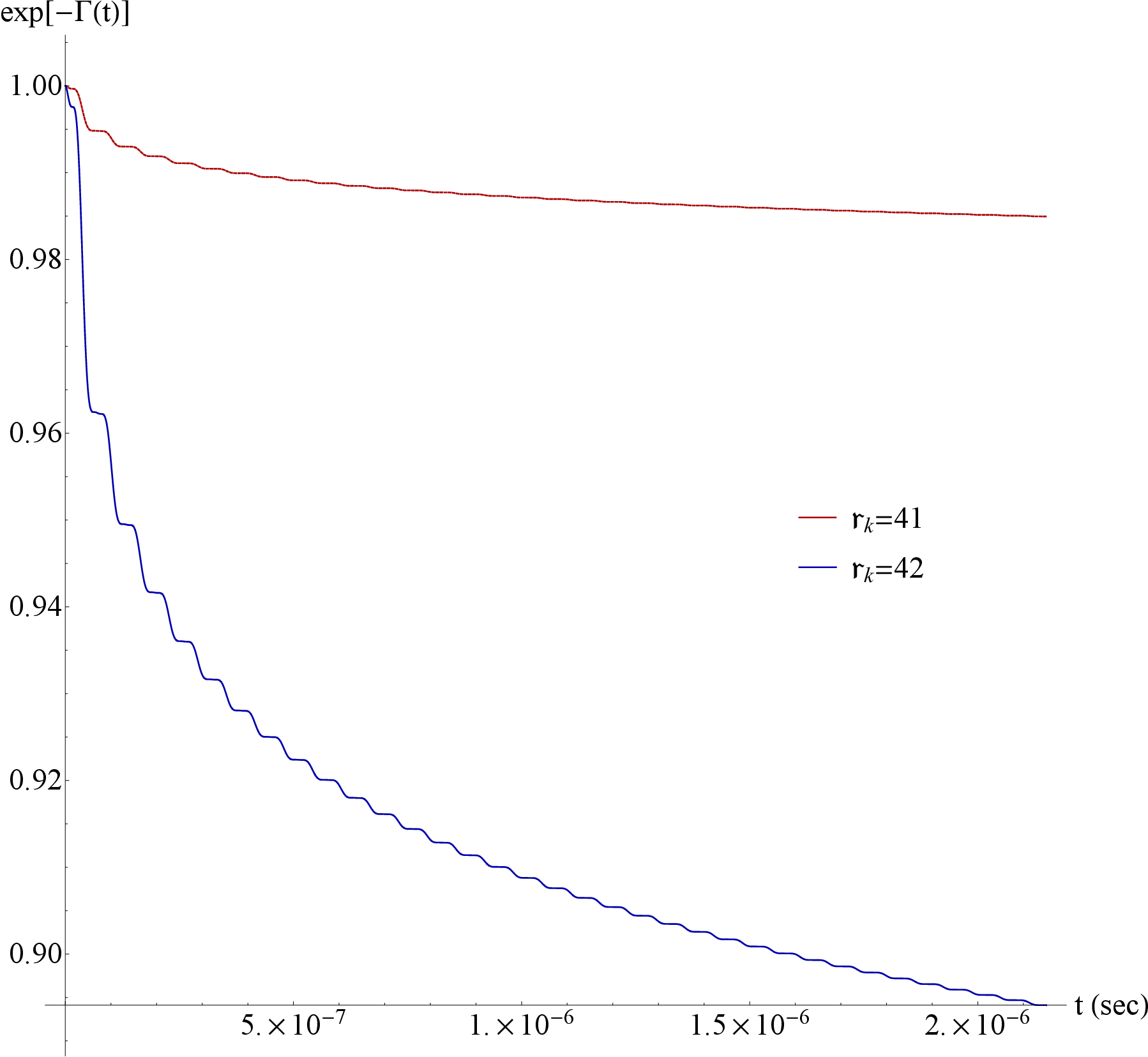}
\caption{The decoherence term $\exp(-\Gamma(t))$ is plotted against time $t$ for different values of the graviton squeezing. We observe that for higher values of $\mathfrak{r}_k$, the loss of coherence is faster compared to the case where the squeezing parameter has a lower value.\label{Decoherence_Factor_OTM}}
\end{center}
\end{figure}
In order to arrive at the above result, we have taken the angle between the two wave vectors $\vec{k}_{B_1}$ and $\vec{k}_{B_2}$ to be zero. The two analytical expression for the two-point correlator is given in eq.(\ref{Supercondensate.24}) for squeezed graviton states. It is claimed that squeezed graviton states are present in primordial gravitational waves which were generated during the time of inflation \cite{KannoSodaTokuda,KannoSodaTokuda2}. In general, the frequency of a primordial gravitational wave lies in the $10^{-4}-10$ Hz range. For convenience, we choose the incoming gravitational wave frequency to be $\Omega_0=1$ Hz. As a result the resonance frequency of the BEC shall lie at $\omega_B=\frac{\Omega_0}{2}=0.5$ Hz. It is also quite important to properly choose the value of the coupling constant $\lambda_B$ which can vary depending on whether the BEC is strongly or weakly coupled.  For a Bose-Einstein condensate with weakly interacting bosons \cite{LinWolfe}, the value of the coupling constant can be taken to be $\lambda_B\sim 10^{-7}$ which is very small. The next step is to estimate the constant $\gamma_B$ in eq.(\ref{Supercondensate.26}), which has the analytical form $\gamma_B\simeq \frac{L_B^3}{\lambda_B c}\left(\frac{m^2 c^2}{\hbar^2}\right)$ where $\sigma_B\sim \frac{m_B c^2}{\hbar}$. We consider that the condensate is quantized inside of a box of length $10^{-3}$ m, where the mass of a single bosonic atom is $m_B\sim 10^{-25}$ kg, it is then possible to estimate the value of the constant $\gamma_B$ to be $\gamma_B\sim 10^{24}-10^{25}$ sec. The numerical value of the pre-factor in eq.(\ref{Supercondensate.26}) for the above choices of the constant values reads $\frac{8\gamma_B^2\omega_B^2\Omega^2_0}{3}\sim 10^{48}-10^{50}$ $\text{sec}^{-2}$ where the frequency of the incoming gravitational wave is taken to be $\Omega_0=1$ Hz, and the frequency of the BEC system to be $\omega_B=0.5$ Hz. The next thing to investigate the order of magnitude of the two-point correlator in eq.(\ref{Supercondensate.26}) can simply be estimated by $\left\langle\{\delta\hat{\mathcal{N}}(t,0),\delta\hat{\mathcal{N}}(t',0)\}\right\rangle_{\mathcal{F}}\sim \frac{\hbar G}{c^5 }e^{2\mathfrak{r}_k}\mathcal{f}(t,t')\sim 10^{-65} \mathcal{f}(t,t')$ (for $\mathfrak{r}_k=24$) provided that the graviton squeezing parameter is high enough. In the above expression, the time dependence is bore by the $\mathcal{f}(t,t')$\footnote{The analytical form of $\mathcal{f}(t,t')$ can be obtained from eq.(\ref{Supercondensate.24}) by considering high squeezing of the gravitons as $\mathcal{f}(t,t')\simeq\frac{-1+\cos(\Omega_\text{M} (t-t'))+\Omega_\text{M}(t-t')\sin(\Omega_\text{M} (t-t'))}{2(t-t')^2}+\frac{\cos\varphi_k+\cos(\Omega_\text{M} (t+t')-\phi_k)+\Omega_\text{M}(t+t')\sin(\Omega_\text{M} (t+t')-\varphi_k)}{2(t+t')^2} $.} term which has the dimension $[\mathcal{f}(t,t')]=T^{-2}$. We are now in a position to obtain an analytically obtained numerical value of the decoherence rate $\Gamma(t_f)$. Considering the cut-off frequency of primordial gravitational waves to be $\Omega_\text{M}\sim 10^8$ Hz and the squeezing parameter value to be $\mathfrak{r}_k=24$, we can estimate the value of the decoherence factor $\Gamma(t_f)$ to be of the order of $\Gamma(t_f)\sim 10^{-16}$ while the value of $t_f$ is set to $t_f=2~\mu\text{s}$. 
The decoherence term in the off-diagonal elements of the density matrix can then be estimated to be $e^{-\Gamma(t_f)}\simeq 1-\Gamma(t_f)=1-10^{-16}$, where the decoherence due to quantum gravity effect can be found to be very small. However, when no graviton squeezing is present, the decoherence term can be estimated to be (for the above choices of the constants) $1-10^{-37}$, which is considerably smaller compared to the case where the gravitons are squeezed. The author in \cite{Schutzhold}, considered a nonrelativistic Bose-Einstein condensate to detect classical gravitational waves where the approximate estimation of the amplitude was found to be $1-N\varepsilon$, with $N$ denoting the number of atoms in the BEC and $\varepsilon$ having the value $\varepsilon\sim 10^{-21}$ (where the condensate was taken to be in a coherent state). In the discussion of \cite{Schutzhold}, it is quite clearly argued that by increasing the number of atoms in the condensate, it is not possible to overcome this gap of magnitude equivalent to twenty orders compared to the leading term. In \cite{MatthewMannAffshordi,MatthewMannAffshordi2} a relativistic BEC was considering and was considered as a detector of classical gravitational wave where phonon squeezing as well as parametric resonance was implemented to increase the sensitivity of the condensate towards incoming gravitational waves. The interesting aspect of our analysis lies in the fact that we have two parameters that can contribute towards the enhancement of the graviton induced decoherence, one being the graviton squeezing whereas the other being the phonon squeezing. For phonon squeezing of the order of $r_B\sim 8.5-9.0$, the decoherence term can be estimated by the numerical value $e^{-\Gamma(t_f)}\sim 1-10^{-1}$ to $1-10^{-2}$ which significantly compensates the gap with respect to the leading order contribution when $t_f$ is $1-2$ $\mu\text{s}$. We shall investigate the effect of the phonon squeezing on the decoherence in details in the next subsection. The substantial gap observed in case of \cite{Schutzhold} is hugely compensated by this phonon squeezing. One can also achieve the same result when the graviton squeezing is very high such that $\mathfrak{r}_k\sim 41-42$. For gravitons with very small squeezing along with minimal phonon squeezing, decoherence effect will be negligible. To observe the dependence of the decoherence term on the graviton squeezing parameter, we plot $\exp(-\Gamma(t_f))$ against the final measurement time in Fig.(\ref{Decoherence_Factor_OTM}). From Fig.(\ref{Decoherence_Factor_OTM}), we observe that with higher value of the graviton squeezing parameter, the exponential decay becomes more significant in this entangled BEC system. It is also important to notice that the decay pattern is wave-like where the decay slows down for a moment and then again it goes faster. The time interval of the slowing down of the decoherence remains same, however, the amount of decay after this slowing down becomes significantly smaller with the passing of time. This effect can be a direct consequence of graviton-noise fluctuations.  From Fig.(\ref{Decoherence_Factor_OTM}), we observe that for $\mathfrak{r}_k=42$, there is a $10\%$ loss of coherence in the time interval $t_f-t_i\simeq 2$ $\mu\text{s}$. For $\mathfrak{r}_k=41$, however, a $10\%$ loss of coherence occurs for a time interval as large as $4\times 10^7$ sec. This observation is indicative of the fact that graviton induced decoherence is very difficult to observe when the graviton states do not carry high squeezing. 
\begin{figure}[t!]
\begin{center}
\includegraphics[scale=0.48]{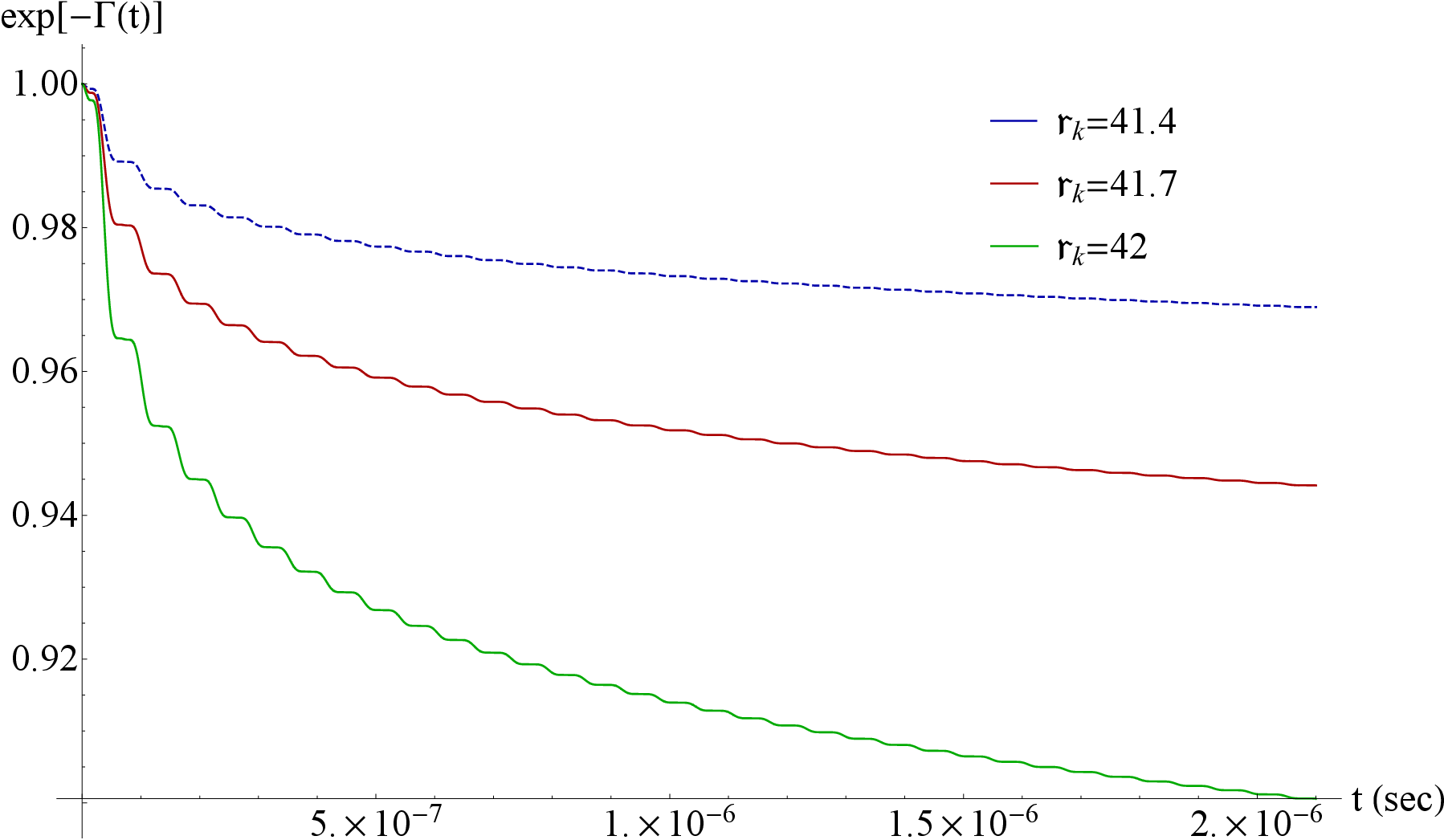}
\caption{We plot the decoherence term against the time $t$ when the difference term in eq.(\ref{Supercondensate.25}) has time dependence and both the Bogoliubov coefficients are non-vanishing.
\label{Decoherence_Factor_OTM_2}}
\end{center}
\end{figure}
We now consider the case when the difference term in eq.(\ref{Supercondensate.25}) is time dependent with $\alpha^B_i=\beta^B_i=\frac{1}{\sqrt{2}}$ for all $i=\{1,2\}$. We plot the decoherence term against the measurement time in Fig.(\ref{Decoherence_Factor_OTM_2}). For analytical simplicity, we use a Taylor series expansion of the square of the cosine term obtained after simplifying the expression in eq.(\ref{Supercondensate.25}). We observe that the fall of coherence is almost same and if noticed very carefully, then this scenario has a slightly slower decay rate than the time independent case considered in Fig.(\ref{Decoherence_Factor_OTM}). It is although important to notice that both analyses indicate towards identical conclusions. 

\noindent Up to now, we have been considering the case where the atoms of the Bose-Einstein condensate are weakly coupled. We now shift our focus towards the case where the condensate atoms are strongly coupled. It is important to remember that it is possible to construct a BEC in a strongly interacting Bose gas but for such a scenario segregating graviton-induced decoherence would be way more difficult as our investigation shall reveal below.  
\begin{figure}[t!]
\begin{center}
\includegraphics[scale=0.48]{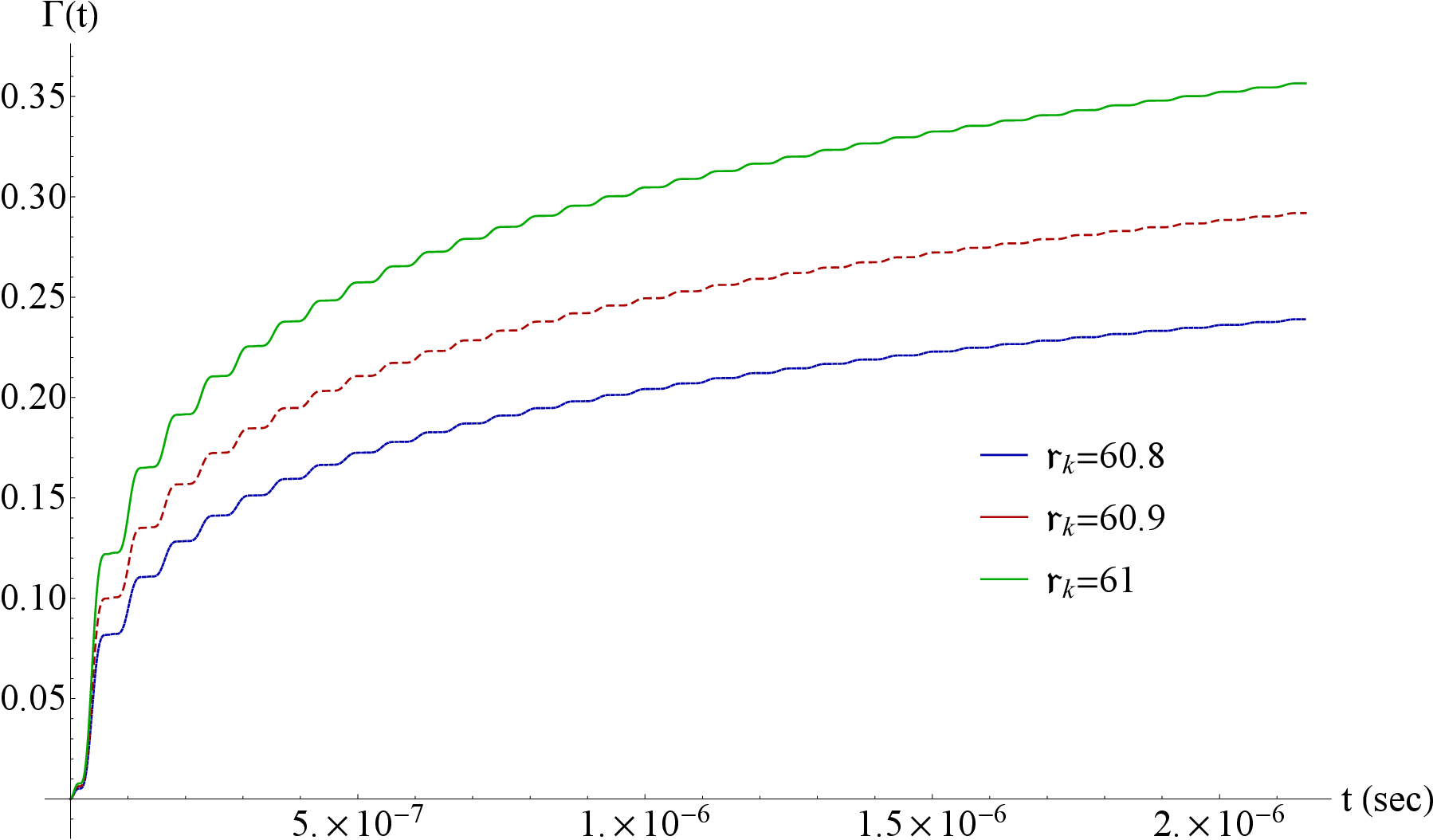}
\caption{We plot the decoherence rate $\Gamma(t)$ against the measurement time $t$ for a strongly coupled BEC with $\lambda_B=10$ and compare the case with varying graviton-squeezing parameter. \label{Decoherence_Rate_OTM}}
\end{center}
\end{figure}  
For a strongly coupled BEC system, the coupling constant can be estimated by the analytical form \cite{MatthewMannAffshordi}
\begin{equation}\label{Supercondensate.27}
\begin{split}
\lambda_B\sim \frac{m_B^4c_S^2c^3}{\rho_B\hbar^3}
\end{split}
\end{equation}
where $\rho_B=T_{00}$ is the energy density of the condensate with the energy momentum tensor being calculated using the expression $T_{\mu\nu}=\frac{1}{\sqrt{-g}}\frac{\delta S}{\delta g^{\mu\nu}}$. The action $S$ corresponds to the total action for the relativistic BEC and the classical gravitational wave. For a quantum gravitational set-up the change in the energy momentum tensor can be neglected which keeps the coupling constant intact. From the analysis in \cite{MatthewMannAffshordi}, $\rho_B$ can be estimated and the value for the coupling constant $\lambda_B$ comes out approximately to be $\lambda_B\sim10^{11}$. A standard calculation reveals that for such high value of the coupling constant, there is only a $2\times 10^{-5}\%$ loss of coherence in a time interval of $\Delta t=t_f-t_i\sim 10^4$ sec when the graviton squeezing is as high as $\mathfrak{r}_k=76$. This `back of the envelop' calculation reveals that detecting graviton signatures using a very strongly coupled BEC is next to impossible and one should consider using weakly coupled BEC systems. We instead consider a strong;y coupled BEC system where the coupling constant is $\lambda_B=10$ which is way higher than the coupling constant for the weak coupling case considered earlier. We plot the decoherence rate against measurement time in Fig.(\ref{Decoherence_Rate_OTM}) for different values of the graviton squeezing parameter. We find out that the decoherence rate increases significantly for very high graviton squeezing with the value of $\mathfrak{r}_k\sim 60-61$ or higher. We find out using analytical calculations that for $\mathfrak{r}_k=60.8$, the decoherence time is $t\simeq 29.275$ sec\footnote{The decoherence time indicates the time $t_d$ such that the amplitude becomes $\frac{1}{e}$ of its initial value or $\Gamma(t_d)$ becomes unity.}. For a slightly higher graviton squeezing, that is $\mathfrak{r}_k=60.9$, the decoherence time becomes  $t\simeq0.585$ sec, and  for $\mathfrak{r}_k=61$ the decoherence time becomes $t\simeq0.024$ sec. In Fig.(\ref{Decoherence_Rate_OTM}), the plot of the decoherence rate is depicted up to $t=2~\mu\text{s}$ as it properly showcases the wavy behaviour as was also visible in the nature of the decoherence term for the weak coupling case. It is evident that a BEC with a very low coupling constant is more sensitive towards incoming graviton signatures and loses coherence at a faster rate even with significantly lower graviton squeezing. The reason behind this behaviour lies in the fact that the strong coupling makes the atoms in the condensate comparatively less prone towards external perturbations whereas for a weak coupling case the atoms are more prone towards graviton induced bremsstrahlung in the  BEC. In the next subsection, we investigate the phonon squeezing effect on the graviton induced bremsstrahlung.

\subsection{Phonon squeezing as a tool to reduce the decoherence time}\label{Sub1}
\noindent Up to now, we have considered graviton induced bremsstrahlung from the Bose-Einstein supercondensate when there is no inherent squeezing of the phonon modes of the BEC. In this subsection, we shall investigate the case where the phonon modes of the condensate carry a inherent squeezing. The benefit of dealing with phonon squeezing is that it is a parameter that can be controlled experimentally whereas the graviton squeezing depends completely on the source from which it has been generated. In order to incorporate squeezing in the model, we start focussing on the transverse wave number operators $\hat{k}_B$. If the BEC system is quantized within a box of volume $V_B=L_B^3$, then the transverse wave number $k_B$ is related to $n_B$ (number of phonons) by the relation $k_B=\frac{\pi n_B}{L_{B}}$. This box of volume $V_B$ can be considered to be equal to the size of the harmonic trap inside of which the Bose-Einstein condensate is created. In the quantum description then one can write the relation $\hat{k}_B=\frac{\pi\hat{n}_B}{L_B}$ where one can easily identify $\hat{n}_{B}$ as the number operator. In terms of the phonon creation and annihilation operators ($\hat{a}_B$ and $\hat{a}_B^\dagger$) one can express the number operator as $\hat{n}_B=\hat{a}_B^\dagger\hat{a}_B$. Our primary aim is to obtain the analytical form of the eigenvalue $k_B$ corresponding to the state $|\vec{k}_B\rangle$ when the phonon modes are squeezed. The squeezing is generated analytically by the action of a squeezing operator of the form $\hat{S}(r_{\text{Sq.}})=\exp\left[\frac{1}{2}\left(r_{\text{Sq.}}^*\hat{a}_B^2-r_{\text{Sq.}}\hat{a}^{\dagger2}_B\right)\right]$. In the above expression $r_{\text{Sq.}}$ can be expressed as $r_{\text{Sq.}}=r_Be^{i\zeta_B}$ where $r_B$ gives the phonon squeezing parameter and $\zeta_B$ gives the phonon squeezing angle. The squeezing operator satisfies the condition $\hat{S}(r_{\text{Sq.}})\hat{S}^\dagger(r_{\text{Sq.}})=\hat{\mathbb{1}}$. The transformation of the phonon-creation and annihilation operators under the action of the squeezing operators read
\begin{equation}\label{QGSqueezed.1}
\begin{split}
\hat{S}^\dagger(r_{\text{Sq.}})\hat{a}_B\hat{S}(r_{\text{Sq.}})&=\hat{a}_B\cosh(r_B)-e^{i\zeta_B}\sinh(r_B)\hat{a}^\dagger_B\\
\hat{S}^\dagger(r_\text{Sq.})\hat{a}_B^\dagger\hat{S}(r_\text{Sq.})&=\hat{a}^\dagger_B\cosh(r_B)-e^{-i\zeta_B}\sinh(r_B)\hat{a}_B~.
\end{split}
\end{equation}
We shall now investigate the action of $\hat{k}_B$ on the squeezed state $|k_B^{\text{Sq.}}\rangle\equiv\hat{S}(r_{\text{Sq.}})|\vec{k}_B\rangle$\footnote{Where for convenience, we denote $|\vec{k}_B\rangle$ by $|k_B\rangle$.} which is given by
\begin{equation}\label{QGSqueezed.2}
\begin{split}
\hat{k}_B|k_B^{\text{Sq.}}\rangle=\frac{\pi\hat{a}^\dagger_B\hat{a}_B}{L_B}\hat{S}(r_{\text{Sq.}})|k_B\rangle&=\frac{\pi\hat{S}(r_{\text{Sq.}})\hat{S}^\dagger(r_{\text{Sq.}})\hat{a}^\dagger_B\hat{S}(r_{\text{Sq.}})\hat{S}^\dagger(r_{\text{Sq.}})\hat{a}_B}{L_B}\hat{S}(r_{\text{Sq.}})|k_B\rangle\\
&=\frac{\pi\hat{S}(r_{\text{Sq.}})}{L_B}\left(\hat{S}^\dagger(r_\text{Sq.})\hat{a}_B^\dagger\hat{S}(r_\text{Sq.})\right)\left(\hat{S}^\dagger(r_{\text{Sq.}})\hat{a}_B\hat{S}(r_{\text{Sq.}})\right)|k_B\rangle
\end{split}
\end{equation} 
where in the first line of the above equation, we have made use of the relation involving the squeezing operators, $\hat{S}(r_{\text{Sq.}})\hat{S}^\dagger(r_{\text{Sq.}})=\hat{\mathbb{1}}$. Using the transformation equation of the ladder operators in eq.(\ref{QGSqueezed.1}), we can recast eq.(\ref{QGSqueezed.2}) as
\begin{equation}\label{QGSqueezed.3}
\begin{split}
\hat{k}_B|k_B^{\text{Sq.}}\rangle&=\frac{\pi\hat{S}(r_\text{Sq.})}{L_B}\left[\hat{n}_B\cosh^2r_B+[\hat{n}_B+1]\sinh^2r_B-\frac{\sinh2r_B}{2}\left[e^{i\zeta_B}\hat{a}^{\dagger2}_B+e^{-i\zeta_B}\hat{a}_B^2\right]\right]|k_B\rangle~.
\end{split}
\end{equation}
\begin{figure}[t!]
\begin{center}
\includegraphics[scale=0.42]{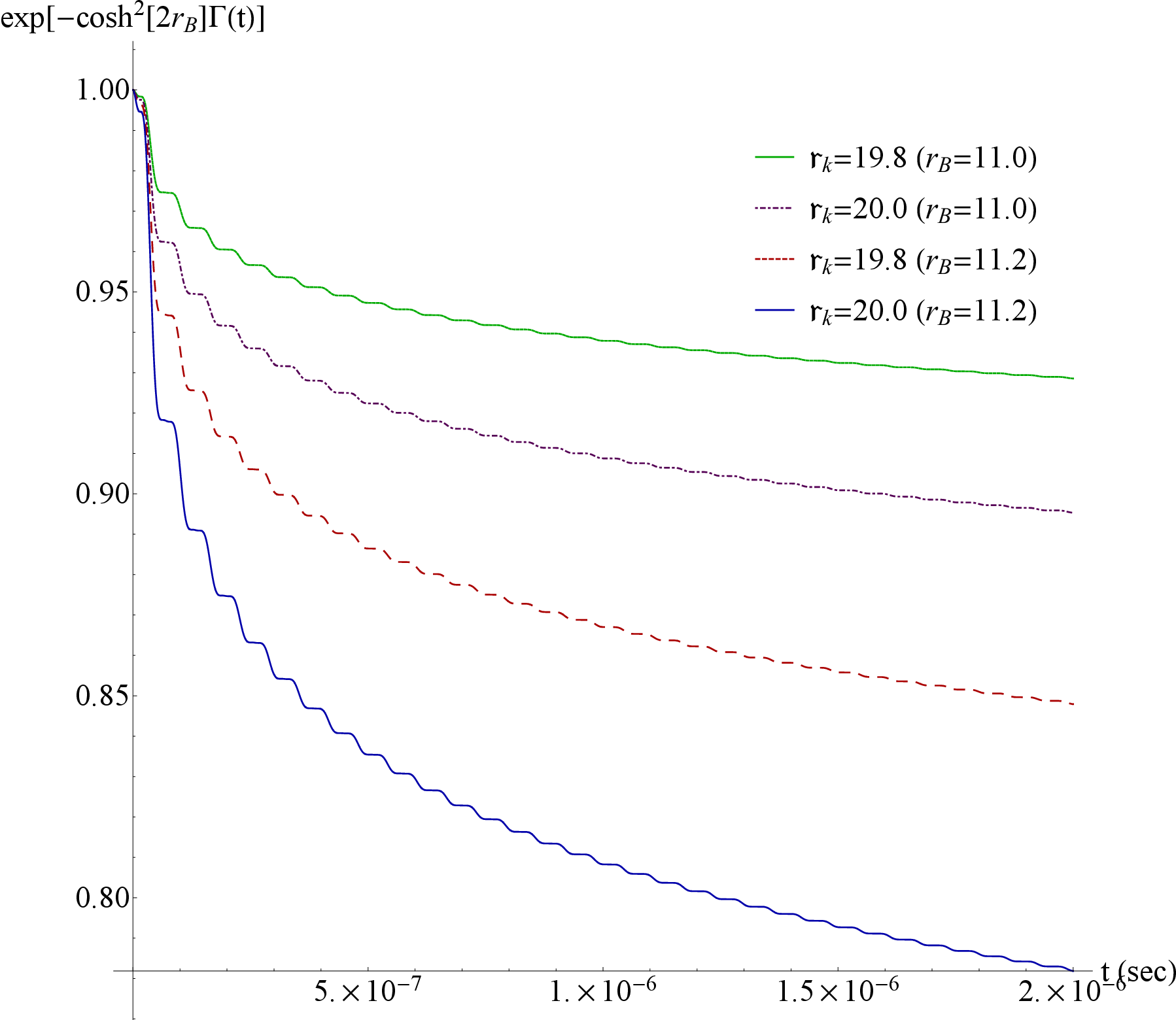}
\caption{The time dependence of the decoherence term $e^{-\Gamma^{\text{Sq.}}}=e^{-\cosh^2(2r_B)\Gamma(t_f)}$ is plotted for different values of the graviton and phonon squeezing parameters with $\mathfrak{r}_k\in\{19.8,20.0\}$ and $r_B\in \{11.0,11.2\}$.\label{Decoherence_Phonon_Squeezing_OTM}}
\end{center}
\end{figure}
In a Bose-Einstein condensate, the accumulation of bosons in the ground state of the system occurs in large numbers and as a result $n_B\gg 1$ even for a weakly coupled BEC system.
It is therefore possible to write down the action of the lowering operator on the BEC state $|k_B\rangle=\left|\frac{\pi n_B}{L_B}\right\rangle$ as $\hat{a}_B|k_B\rangle=\sqrt{n_B}\left|\frac{\pi(n_B-1)}{L_B}\right\rangle\simeq \sqrt{n_B}\left|\frac{\pi n_B}{L_B}\right\rangle=\sqrt{n_B}\left|k_B\right\rangle$. One can similarly obtain the action of the raising operator on the BEC state as $\hat{a}_B^\dagger|k_B\rangle= \sqrt{n_B+1}\left|\frac{\pi(n_B+1)}{L_B}\right\rangle\simeq \sqrt{n_B}|k_B\rangle$. It is then possible to recast eq.(\ref{QGSqueezed.3}) after a little bit of simplification as
\begin{equation}\label{QGSqueezed.4}
\begin{split}
\hat{k}_B|k^{\text{Sq.}}_B\rangle&\simeq \frac{\pi \hat{S}(r_\text{Sq.})}{L_B}n_B\left[ \cosh(2r_B)-\sinh(2r_B)\cos\zeta_B\right]|k_B\rangle.
\end{split}
\end{equation}
One can now set the phonon squeezing angle to $\zeta_B=\frac{\pi}{2}$. Phonon squeezing at specific angle is experimentally achievable and has been done in experimental set-ups\cite{SpecificPhononSqueezing,SpecificPhononSqueezing2}. 
For $\zeta_B=\frac{\pi}{2}$, one can recast eq.(\ref{QGSqueezed.4}) as
\begin{equation}\label{QGSqueezed.5}
\begin{split}
\hat{k}_B|k^{\text{Sq.}}_B\rangle&=\frac{\pi\cosh (2r_B)n_B}{L_B}\hat{S}(r_\text{Sq.})|k_B\rangle\\
&=\cosh(2r_B)k_B|k_B^{\text{Sq.}}\rangle
\end{split}
\end{equation}
where in the last line of the above equation, we have substituted $k_B=\frac{\pi n_B}{L_B}$. Again, the phonon mode frequency is related to the transverse wave number by the dispersion relation as $\omega_B=c_Sk_B$, and under the effect of squeezing, the frequency also transforms as $\omega_B^{\text{Sq.}}=\cosh (2r_B)\omega_B$. Hence, under the effect of phonon squeezing, the decoherence rate 
takes the form
\begin{equation}\label{QGSqueezed.6}
\begin{split}
\Gamma^{\text{Sq.}}(t_f)&=\frac{8\gamma_B^2\cosh^2(2r_B)\omega_B^2\Omega^2_0}{3}\int_0^{t_f}dt\int_{0}^tdt'\left\langle \{\delta\hat{\mathcal{N}}(t,0),\delta\hat{\mathcal{N}}(t',0)\}\right\rangle_{\mathcal{F}}\\
&=\cosh^2(2r_B)\Gamma(t_f)
\end{split}
\end{equation}
where $\Gamma(t_f)$ is given in eq.(\ref{Supercondensate.26}). In recent experimental scenarios, a phonon squeezing of $r_B=0.83$ has already been achieved which is equal to $7.2$ dB of phonon squeezing \cite{GuLiWuYang}. Using the semiclassical case with classical gravitational wave-BEC model, an upperbound of $r_B\approx 27$ was given in \cite{MatthewMannAffshordi}. The important aspect of our analysis lies in the fact that the graviton detection is directly related to the decoherence time. The faster the coherence loss, the easier it is to detect graviton signatures. As has been discussed earlier, for introducing a substantial amount of decay in the decoherence term, one needs graviton squeezing equal to $\mathfrak{r}_k=41$ or higher. Our primary aim here is to use the phonon squeezing parameter and tune the BEC part in a way that the need for such high graviton squeezing boils down to more plausible squeezing regime. From the expression of the decoherence rate in eq.(\ref{QGSqueezed.6}), we can easily argue that the decoherence term now goes as $e^{-\Gamma(t_f)}\rightarrow e^{-\cosh^2(2r_B)\Gamma(t_f)}$. We now plot this modified decoherence term against the total observation term in Fig.(\ref{Decoherence_Phonon_Squeezing_OTM}) for different values of the graviton and phonon squeezing parameters. For gravitons, we choose the squeezing parameter to take values $\mathfrak{r}_k\in\{19.8,20.0\}$, and for phonons, the squeezing parameter is considered to take values $r_B\in\{11.0,11.2\}$. We find out from Fig.(\ref{Decoherence_Phonon_Squeezing_OTM}) that if the phonon squeezing is $r_B=11.0$, then for graviton squeezing as low as $\mathfrak{r}_k=20.0$, there is approximately a $10\%$ loss of coherence over a time interval of $2~\mu\text{s}$, which is equivalent to the case when there was no phonon squeezing and the graviton squeezing was more than double ($\mathfrak{r}_k=42.0$). We further observe from Fig.(\ref{Decoherence_Phonon_Squeezing_OTM}) that for an increase in the phonon squeezing parameter $\Delta r_B=0.2$, the decoherence over the time interval $\Delta t\sim 2$ $\mu\text{s}$ goes as high as $15\%$ for graviton squeezing $\mathfrak{r}_k=19.8$ whereas for $\mathfrak{r}_k=20.0$ the decoherence becomes as high as $20\%$. This behaviour of the decoherence induced by gravitons indicate that via properly tampering the phonon squeezing, it is possible to detect graviton signatures with lower graviton squeezing. The phonon squeezing $r_B=11.0$ is equal to $s_B=-10\log_{10}(-2r_B)=95.54$ dB and a squeezing of the order of $100$ dB is extremely difficult to achieve experimentally but may be achievable in a matter decades time. In the next subsection, we shall investigate entanglement degradation due to graviton induced bremsstrahlung.
\subsection{Entanglement degradation as a                                                                                                                                                                                  quantum gravity signature}
\noindent We shall now investigate the entanglement negativity to investigate the entanglement in the Bose-Einstein ``supercondensate". Initially, the total density matrix of the system is represented by $\hat{\rho}_{\text{S}_i}$ at time $t=t_i$ in eq.(\ref{Supercondensate.10}), where the BEC part of the density matrix represents a maximally entangled state.
After a finite time, the density matrix is constructed via the state $|\psi_{\text{S}_f}\rangle$ in eq.(\ref{Supercondensate.10}) and taking a partial trace, we arrive at the reduced density matrix $\hat{\rho}^{\mathcal{R}}_{\text{BEC}}(t_f)$ in eq.(\ref{Supercondensate.11}).
Taking the transpose of the matrix $\hat{\rho}^{\mathcal{R}}_{\text{BEC}}(t_f)$ from eq.(\ref{Supercondensate.11}), we obtain 
\begin{equation}\label{QGNegativity.1}
\hat{\rho}^{\mathcal{R}^T}_{\text{BEC}}(t_f)=\frac{1}{2}\begin{pmatrix}
0&&0&&0&&e^{i\Delta_{\text{IF}}(t_f)}\\
0&&1&&0&&0\\
0&&0&&1&&0\\
e^{-i\Delta_{\text{IF}}^*(t_f)}&&0&&0&&0
\end{pmatrix}~.
\end{equation}
\begin{figure}[t!]
\begin{center}
\includegraphics[scale=0.42]{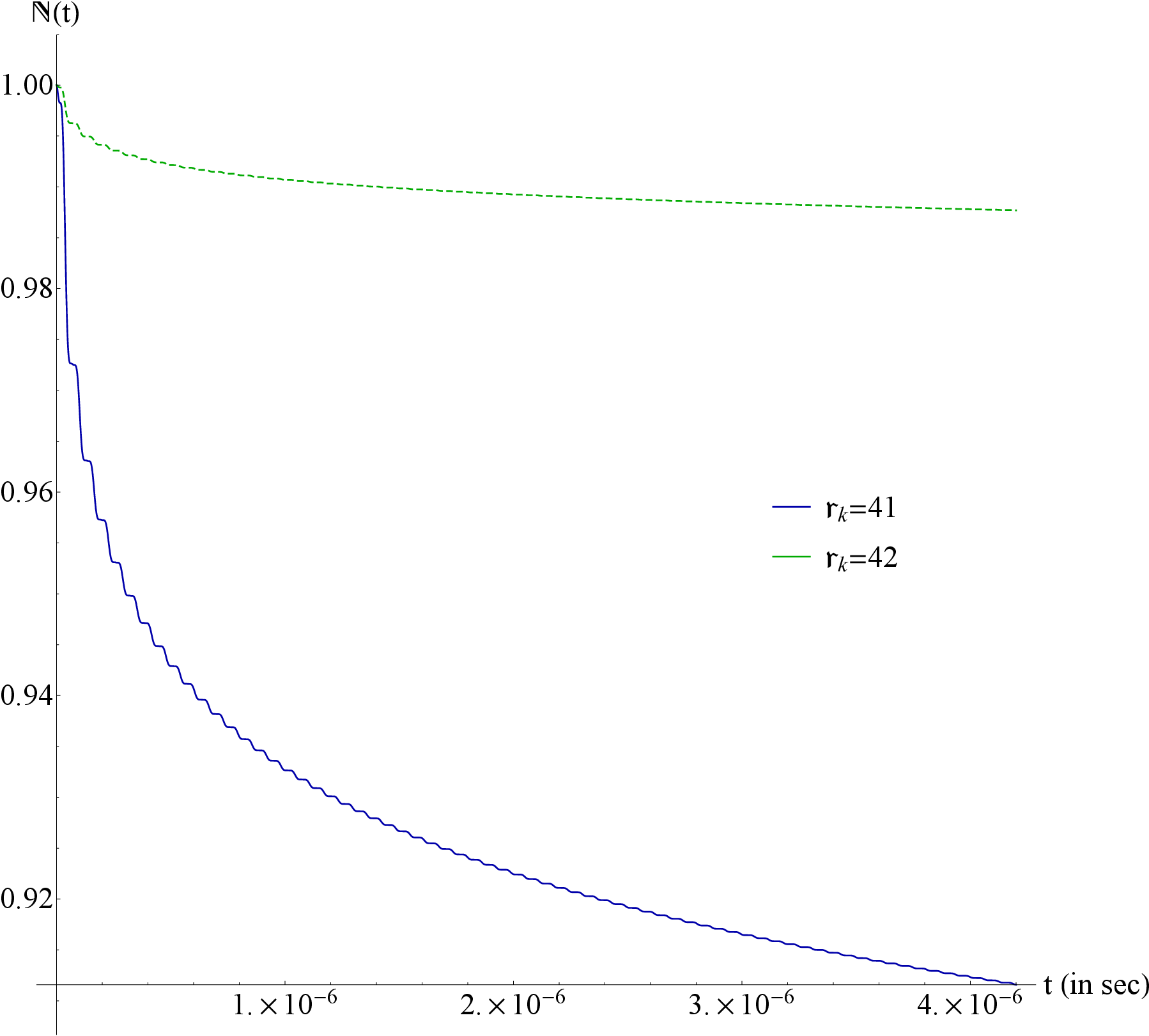}
\caption{The logarithmic negativity is plotted against total measurement time for different values of the graviton squeezing parameter. We observe a jiggly decay pattern of the entanglement negativity over time.\label{Negativity_OTM}}
\end{center}
\end{figure}
We can now obtain the eigenvalues corresponding to the matrix above as
\begin{equation}\label{QGNegativity.2}
\lambda^B_1=\lambda^B _2=\frac{1}{2},~\lambda^B_3=\frac{1}{2}e^{-\Gamma(t_f)}, \text{ and }\lambda^B_4=-\frac{1}{2}e^{-\Gamma(t_f)}~.
\end{equation}
Here, our aim is to investigate the time dependence of the logarithmic entanglement negativity which is defined by the following relation \cite{FuentesMann}
\begin{equation}\label{QGNegativity.3}
\begin{split}
\mathbb{N}\equiv\log_2||\hat{\rho}_S^T||=\log_2\left[1+\sum\limits_{\lambda^S_-}\left(\left|\lambda^S_-\right|-\lambda^S_-\right)\right]
\end{split}
\end{equation}
with $\lambda^S_-$ denoting the negative eigenvalue (or eigenvalues) for the partially transposed and reduced density matrix $\hat{\rho}^S$. As we have seen from eq.(\ref{QGNegativity.2}), the only negative eigenvalue is $\lambda_4^B$, and as a result this eigenvalue will contribute towards the calculation of the entanglement negativity. Hence, one can calculate the entanglement negativity corresponding to the reduced density matrix in eq.(\ref{Supercondensate.11}) as
\begin{equation}\label{QGNegativity.4}
\begin{split}
\mathbb{N}(t_f)&=\log_{2}\left[1+\left(\frac{e^{-\Gamma(t_f)}}{2}-\left(-\frac{e^{-\Gamma(t_f)}}{2}\right)\right)\right]=\log_2\left[1+e^{-\Gamma(t_f)}\right]~.
\end{split}
\end{equation}
In order to plot the negativity with respect to measurement time, we use the analytical form of the decoherence rate from eq.(\ref{Supercondensate.26}), and plot $\mathbb{N}(t)$ against $t$ in Fig.(\ref{Negativity_OTM}) for different values of the graviton squeezing parameter.
From Fig.(\ref{Negativity_OTM}), we observe that the entanglement negativity decays with increase in the measurement time and the rate of decay becomes higher with increase in the graviton squeezing parameter. This investigation confirms that the loss of entanglement from the Bose-Einstein supercondensate occurs via the emission of bremsstrahlung induced by the graviton noise. As for the case of the decay of the decoherence term, we also observe that the entanglement degradation slows down after a certain time interval and then again becomes steep. The entanglement loss is an indication of the fact that the supercondensate has formed. We shall now proceed to propose an experimental set-up that will help to detect graviton signatures in near future.
\section{A new graviton detector}\label{S4}
\begin{center}
\begin{figure}
\centering
\includegraphics[scale=0.297]{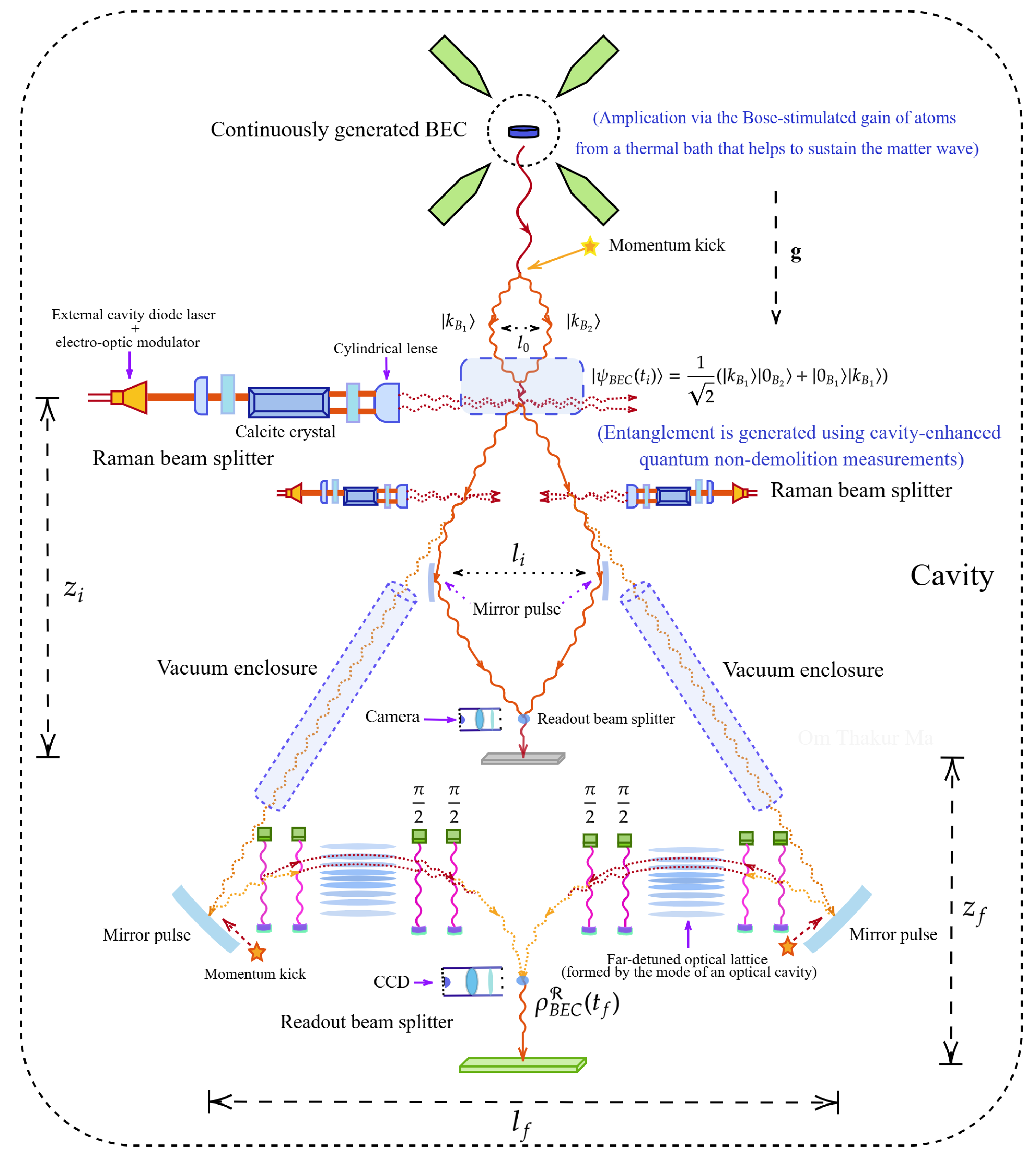}
\caption{A schematic diagram is presented for a Bose-Einstein condensate based graviton detector where a freely falling atom laser set-up  is placed inside a cavity under the effect of Earh's gravitational field. After the generation of maximal entanglement between two split coherent atom laser beams from a single source, they are recombined and further split into four coherent atom laser beams. They are then further recombined to create two atom interference patterns such that there is a finite time gap between the generation of the two patterns. The two patterns are then compared  to search for graviton induced decoherence in the Bose-Einstein ``supercondensate".\label{Graviton_Detector_Om_Thakur_Ma}}
\end{figure}
\end{center}

\noindent The first experimental creation of a Bose-Einstein condensate was first achieved in 1995 \cite{1Nobel2001,2Nobel2001}. The fundamental property of the state of a Bose-Einstein condensate is that it is coherent in nature. Now, if it is possible to outcouple few coherent condensate atoms from the respective harmonic trap potential (in which the condensate is formed), then the coherent atoms fall under the effect of gravity. Researchers at the Massachusetts Institute of Technology were the first to create such coherent freely falling atom beams, which are also termed as atom lasers, from trapped condensate of sodium atoms \cite{3Nobel2001}. One of the very distinctive features of atom lasers are that they interfere like electromagnetic waves, and create an interference pattern when two coherent atom laser beams interfere with each other but this entire phenomena occurs in the Fourier space.  This simply implies a very interesting thing about the atomic interference pattern which is indeed a distribution of the number of particles instead of the distribution of energy. One observes a no-particle zone, which is indicative of a destructive interference, and then a high particle density zone indicating a constructive interference. This gives the interference fringe pattern corresponding to the atom interference. The atom lasers create a set of atomic waves which propagates under the effect of gravity while remaining coherent with each other, which is one of the primary reasons for working with atom lasers in our analysis. It is now required to create atleast two coherent atom laser beams for the creation of atomic interference pattern, where advanced atom interferometry techniques needs to be used \cite{AtomInterferometry,AtomInterferometry2}. If any kind of external interaction creates decoherence in the joint atom laser beams, then it should be theoretically possible to measure this decoherence by investigating the atom interference pattern. For implementing our model, we at first need two sources of the BEC that are maximally entangled as also has been expressed in eq.(\ref{Supercondensate.1}). To proceed with the experimental modelling, one needs an experimental set-up that can create two or multiple coherent beams of atom laser from a single Bose-Einstein condensate. In an all-optical Bose-Einstein condensate, oppositely polarized components were outcoupled to produce two beams of coherent atom lasers \cite{DualbeamAtomLaser}. In \cite{MultibeamAtomLaser}, a novel techniques was experimentally implemented, where a single far-detuned laser was used to produce multiple coherent atomic beams. 
%For a primordial gravitational wave detection ($\sim 1$ Hz) one needs to attain a BEC in a length of $L_\beta=10^{-2}\text{m}$, with $c_s\sim 1 \mu\text{m}\cdot\text{sec}^{-1}$ with $n_\beta\sim1.6\times10^{3}$ number of atoms in it. Creating a 10cm long BEC will be challenging but it should be attainable in the course of a few years. 
Making use of an optical Raman transition, the Bose-Einstein condensate is outcoupled from a harmonic trap potential, where the outcoupling is done using two optical beams driving two-photon Raman-transition \cite{Raman1,Raman2}. This Raman-transition transfers the atoms to the untrapped state from the initial magnetically trapped state which results in the generation of continuous atom laser beam. Through the interaction of the outcoupled atoms with the individual atom laser beams, the splitting is initiated. This atom laser splitting can be directly related to the Bragg diffraction, where multiple coherent atom laser beams are created with slightly different momentum than each other from a single, continuously generated, atom laser beam which later helps in atom interference scenarios. In \cite{MultibeamAtomLaser}, five such coherent atom laser beams were generated, where each beam has very small difference in momentum which indeed is required for the implementation of our experimental proposal.
%Making use of a magnetic trap potential and Bragg diffraction from two optical standing wave gratings one can produce a continuous atom laser coherently split into multiple momentum states (in \cite{MultibeamAtomLaser} three such atom laser beams were observed). These momentum states have slightly different momentum values as required by our analysis as well.
In our work, while calculating the decoherence rate, the base frequency is taken to be the same. Later, alterations are made by $\frac{\Omega_0}{4}$ which can be adjusted properly in an experimental set-up. With this splitting, the next step is to create a maximally entangled coherent state to replicate the initial state of the Bose-Einstein condensate in eq.(\ref{Supercondensate.1}). Very recently, in \cite{NatureEntangled}, a matter wave interferometry has been done between two entangled matter waves inside of a high-fineness cavity. In two spatially separated traps, entanglement between two atoms can be achieved by the use of relative atom number squeezing \cite{NatureEntangled2,NatureEntangled3,ScienceEntangled1}. Similarly, such entanglement is experimentally generated by the use of Coulomb interactions \cite{NatureEntangled4,PRLEntangled} or direct collisional interactions \cite{NatureEntangled2,NatureEntangled3,
NatureEntangled5,NatureEntangled6,
ScienceEntangled1,ScienceEntangled2,ScienceEntangled3}. 
Entanglement between two momentum eigenstates has been achieved where internal entanglement was mapped onto the relative atom number \cite{PRLEntangled2}. In \cite{NatureEntangled}, external momentum eigenstates of different atoms were entangled where a cavity quantum-electrodynamical set-up was used where between the optical cavity and the atoms, strong collective coupling was implemented where experimentally $18.5$ dB of entanglement was achieved. A Mach-Zehnder type interferometer (in this process the matter waves (here atom lasers) are passed through the matter wave interferometer) was used for checking the output data and it was confirmed whether it is below the standard quantum limit. Using atom interferometers for classical gravitational wave detection has already been proposed In \cite{PRLGravWave,PRAGravWave}, the use of atomic sensors have been proposed for the detection of classical gravitational waves where the underlying principle is the standard matter wave interferometry. We are now in a position to propose a next generation ultra-cold atom laser based graviton detector which primarily uses a weakly coupled Bose-Einstein condensate as its source. A schematic diagram of the condensate based detector is given in Fig.(\ref{Graviton_Detector_Om_Thakur_Ma}). The set-up can be though of as a combination of two Mach-Zehnder type atom interferometers where the entire set-up is placed inside of a cavity under the effect of the Earth's gravitational field. We now carefully describe the experimental segments which shall perform specific tasks to complete the entire graviton detection scenario.
\begin{enumerate}
\item\textcolor{blue}{\textit{A continuous ultra-cold atom laser generation}}

\noindent The first step with any gravitational wave detector is that they need to be operated continuously. The primary reason behind this obligation lies in the fact that gravitational waves are generated as a result of astrophysical phenomena which can occur at any time, and we have no direct control on the time at which the gravitational fluctuation will arrive on Earth. In case of graviton detection, we have already discussed that graviton squeezing is required for the detection of significant amount of decoherence within a relatively reasonable time interval. Such squeezing can be found in primordial gravitational waves generated during the time of inflation which is a very rare phenomena. As a result, if one needs to use atom-lasers generated from BEC, one also need to continuously generate Bose-Einstein condensate. Using strontium atom, a continuous-wave Bose-Einstein condensate was generated which can be made to last for an extremely long time as has been demonstrated and reported in \cite{ContinuousBEC}. The experiment at its heart has a large reservoir and the reservoir is loaded with $\text{Sr}$ atoms. In the reservoir, there is a small but quite deep dimple where the Bose-Einstein condensate is formed. In this experiment, ${}^{84}\text{Sr}$ atoms from a steady-state and narrow-line magneto-optical harmonic trap are outcoupled continuously in a way such that they accumulate in a cross-beamed dipole trap forming this large reservoir. Inside of the dimple the atoms accumulate largely (increasing the phase space density), and because of the ongoing laser cooling process, the BEC is formed in a way such that it is in a steady state. The continuous-wave BEC is maintained by the use of constant gain. This constant gain is generated by atom-refilling while maintaining a high phase-space flux and scattering via Bose stimulation. This process is crucial for our continuous-matter wave interference device and we propose the use of this process for the generation of continuous wave Bose-Einstein condensate as can be seen from Fig.(\ref{Graviton_Detector_Om_Thakur_Ma}). However, our proposed model relies on the fact that continuous wave Bose-Einstein condensate is generated where the bosons are weakly coupled. Another important thing that needs to be considered is that we have considered the use of a pure Bose-Einstein condensate while doing the analytical calculations and a pure BEC is formed at the absolute zero temperature. This is not possible to implement in an experimental laboratory as no laser cooling technique can cool to absolute zero till now. Now, recent experimental developments have been able to generate a Bose-Einstein condensate where the cooling is done to piko-Kelvin order which is still extremely small and if used, will go quite well with our current theoretical model\footnote{For a more accurate description, one needs to consider Finite temperature field theoretical techniques.}. In order to create continuous atom laser beams from the continuous-wave Bose-Einstein condensate, an external energy source like a periodic radio frequency pulse can be used in order to stimulate the BEC. In this way, from the trap potential, a bunch of ultra-cold atoms get released creating an atom-laser which propagates by falling freely under the effect of gravity \cite{1Nobel2001,2Nobel2001,3Nobel2001}.  This technique, which helps in creating a continuously generated atom laser beam (denoted by the wave like red line in Fig.(\ref{Graviton_Detector_Om_Thakur_Ma})) from a BEC, is also known as ``output-coupling" \cite{ContinuousBEC} and needs to be implemented in our proposed experimental model. In the next step, the aim is to create the maximally entangled state of the BEC, $|\psi_{\text{BEC}}\rangle$ in eq.(\ref{Supercondensate.1}). 

\item \textcolor{blue}{\textit{Generation of the maximally entangled state $|\psi_\text{BEC}\rangle$}}

\noindent The first step towards the creation of a maximally entangled state is to create two atom laser beams which represents the states $|\vec{k}_{B_1}\rangle$ and $|{k}_{B_2}\rangle$ such that $\vec{k}_{B_1}\neq {k}_{B_2}$. Now momentum is related to the transverse wave number via the relation $\vec{p}_B=\hbar {k}_B$, which implies that the two coherent atom laser beams must carry different momentum. To entangle two atom laser beams, one can implement the methodology implemented in \cite{PRLEntangled2}. In our analysis, however, we primarily propose the use of the methodology cum technique implemented in \cite{NatureEntangled}. Making use of quantized momentum kicks, the freely falling atom laser beams are separated into two beams of coherent atom lasers carrying a slightly different momentum which is also required for the creation of the maximally entangled state in eq.(\ref{Supercondensate.1}). During the implementation of the separation, it should be kept in mind that the kicks should be such that the separation remains in the range $l_i\sim 10-20$ $\mu \text{m}$ such that the two beams remain coherent with each other before going through the recombination. A two-photon Raman transition can be implemented to initiate this splitting and recombination process. To entangle  the atoms, atomic probe lights are used insider of the cavity. In order to generate entanglement, one needs to implement quantum non-demolition (QND) based techniques of measurement \cite{nondemolition1,nondemolition2,nondemolition3}. The use of entangled atomic ensembles in atom interferometers allows the surpassing of restrictions due to the standard quantum limit (SQL). We propose the implementation of the entanglement generation techniques proposed in \cite{nondemolition1,nondemolition2,nondemolition3}, as has been depicted by the small red line after the recombination (before the first phase of splitting using Raman beam splitter) of the two wavy orange lines in Fig.(\ref{Graviton_Detector_Om_Thakur_Ma}).
\item \textcolor{blue}{\textit{First phase of interference}}

\noindent There are, in total, two phases of interference in our proposed experimental model. To create an interference, one needs two coherent atom laser beams which can be achieved by splitting the maximally entangled atom laser beam. For a coherent splitting, one can make use of Raman beam splitter. We can find a schematic diagram of the Raman beam splitter in Fig.(\ref{Graviton_Detector_Om_Thakur_Ma}). This beam splitter uses an electro optic modulator which is combined with an external cavity diode laser. Finally, the laser is passed through a mechanism with a calcite crystal and cylindrical lenses. This entire mechanism helps to split the atom laser beams, and in our current proposal, it will be able to separate the maximally entangled atom laser beam into two coherent atom laser beams. The schematic representation of the Raman beam splitter and some of the details can be found in \cite{Beamsplitter}. In order to implement a two phase interference, we need a total of four coherent atom laser beams. It is possible to achieve by splitting each of the two coherent beams into two more coherent atom laser beams via the use of two additional Raman beam splitters (as can be seen from Fig.(\ref{Graviton_Detector_Om_Thakur_Ma})). The nearest two atom laser beams are denoted by curvy dark orange lines whereas the other two distant beams are denoted by dotted and curvy orange lines. For the creation of first interference, the two nearest beams must be refocussed. We propose the implementation of a mirror pulse as has been used in \cite{NatureEntangled} for refocussing the two nearest atom laser beams with a maximum spatial separation of $l_i$ as can be seen from Fig.(\ref{Graviton_Detector_Om_Thakur_Ma}). The interference then occurs at a height $z_i$ from the splitting point using the first Raman beam splitter. In this experimental set-up $l_i$ can be similar or equal to $0.5$ mm and a good experimental set-up should restrict $l_i$ within $60$ $\mu\text{m}$. If $l_i$ is made higher then in the second phase of interference the maximum splitting between the furthest two beams $l_f$ will be very large. This $l_i$ is also equal to the separation between the initial mirror pulses for the first phase of interference. A general free fall time of and atom laser is equal to $t\sim20$ ms \cite{MultibeamAtomLaser}. For $20$ ms, if the atom laser beams remain coherent, then they have fallen almost close to $z_{\text{cl}}=\frac{1}{2}gt^2$ with $g$ denoting acceleration due to Earth's gravity. One can neglect the effects induced due to the Earth's rotation and other external influences. Again if quantum gravity effects are being considered, in that case, there should be graviton induced fluctuation in the total free fall length \cite{ChawlaParikh} and the free-fall distance will be modified by $z\sim z_{\text{cl}}\pm z_{\text{QG}}\sim  z_{\text{cl}}+\sqrt{\cosh(2\mathfrak{r}_k)}l_{\text{Pl}}\sim z_{\text{cl}}\pm e^{\mathfrak{r}_k} l_{\text{Pl}}$. If the graviton squeezing is as high as $\mathfrak{r}_k=42$, even then $z_{\text{QG}}\sim 10^{-17}$ which is negligible with respect to the classical free fall distance. One can neglect such a length scale in an experimental scenario as it is smaller than the radius of the nucleus of an atom. The classical value of the free fall distance for $t\approx 20$ ms, can be obtained as $z\approx 2$ mm. A more comfortable free fall distance would be about $z_i< 40-80$ $\mu$m so that the free fall time is equivalent to $t\sim 2-4 $ ms. This time gap, although is apparently quite small, is huge when graviton interactions are being considered. In this time interval the gravitons will introduce approximately a $23-24\%$ decoherence in the Bose-Einstein \textit{supercondensate} for graviton squeezing as high as $\mathfrak{r}_k=42$ and $r_B=0$. For an accurate experimental outcome, $z_i$ should be made so small that the free fall time is in the microsecond order. The only problem being that in such a case $z_i\sim 10^{-1} -10^{-2}$\AA~ and a  matter-wave interference set-up within such a small distance is quite difficult. Before discussing the second phase of interference, we need to consider a minute details in this first phase of interference.
\begin{enumerate}
\item \textcolor{blue}{\textit{Graviton interaction with the atom laser before $t_i$}}

\noindent In the experimental proposal made in Fig.(\ref{Graviton_Detector_Om_Thakur_Ma}), we have considered up to now that the graviton truly starts interacting with the matter waves after the maximally entangled BEC state is formed. However, as soon as the primordial gravitational wave reaches Earth, the interaction shall start immediately and the gravitons, being quanta of the linearized quantum field theory of gravity, is present all around and will therefore start interacting with the BEC as soon as it is generated in the harmonic trap in an real experimental scenario. Here our aim is to observe decoherence due to the emission of bremsstrahlung from the BEC supercondensate which is an entangled state generated out of a maximally entangled BEC state and a graviton state. It is therefore important to remember that any interaction of graviton with the BEC is inconsequential before the formation of the maximally entangled BEC states using QND techniques. This can be significant if the initial BEC graviton induces a decoherence compared to which the final decoherence becomes negligible. Consider that $\Delta t$ amount of time is required for the atom laser to come to the point of maximal entanglement generation and then at $t_0=t_i-\Delta$, the BEC is prepared. Denoting $t_0=t_i-\Delta t=0$ and considering that the gravitons start interacting with the condensate at time $t=0$, the initial state of the system can be written as $|\psi_{\text{S}}(t_0)\rangle=|\vec{k}_{B_i}\rangle\otimes|h_G,0\rangle$ where the zero in the graviton state indicates the time of interaction of the graviton with the condensate. As discussed above, the maximally entangled momentum state is formed after a time interval $\Delta t$, and the state becomes in this case $|\psi_{\text{S}_i}\rangle=\frac{1}{\sqrt{N}}\left(|\vec{k}_{B_1},0_{B_2}\rangle\otimes |h_G,\Delta t;\vec{k}_{B_1}\rangle+|0_{B_1},\vec{k}_{B_2}\rangle\otimes |h_G,\Delta t;\vec{k}_{B_2}\rangle\right)$ with $N$ being the normalization constant. After a time interval $t_f-t_i$ and after the formation of the maximally entangled state, the final state of the system is given by $|\psi_{\text{S}_i}\rangle=\frac{1}{\sqrt{N}}\left[|\vec{k}_{B_1},0_{B_2}\rangle\otimes |h_G,t_f-t_i+\Delta t;\vec{k}_{B_1}\rangle+|0_{B_1};\vec{k}_{B_2}\rangle\otimes |h_G,t_f-t_i+\Delta t,\vec{k}_{B_2}\rangle\right]$. Using the expression in eq.(\ref{Supercondensate.12}), one can express the influence functional $\Delta_{\text{IF}}(t_f+\Delta t)$ as $e^{i\Delta_{\text{IF}} (t_f+\Delta t)}=\langle h_G,t_f-t_i+\Delta t;\vec{k}_{B_2}|h_G,t_f-t_i+\Delta t,\vec{k}_{B_1}\rangle$. For the decoherence factor $\Gamma(t_f)$ in eq.(\ref{Supercondensate.16}), it will now become $\Gamma(t_f+\Delta t)$ and the lower limit of the integrations in eq.(\ref{Supercondensate.16}) will be substituted by $-\Delta t$. For a careful study, $\Delta t $ can be kept and the decoherence rate as well as the decoherence term can be carefully calculated. However, if $\Delta t$ is made so small that it approaches zero, in that case one can proceed with the theoretical model presented in our work without the need of the consideration of the small time $\Delta t$. Another way to deal with this issue is to shield the initial part of the experiment up to the generation of the maximally entangled BEC, such that any fluctuations due to linearized gravity do not alter the system. Although this can be close to an impossible task experimentally, there have been some proposals that deal with the implementation of a system which will be able to shield a system (in our case the initial part of the system up to the formation of the maximally entangled BEC) from incoming gravitational fluctuations \cite{GWShielding0,GWShielding1,GWShielding2}. In \cite{GravitonAbsorption}, a recent proposal is made regarding stimulated absorption of gravitons. This absorption is achieved via the continuous sensing of quantum mechanical jumps. This graviton or gravitational wave shielding is a very complicated procedure which is very difficult to implement experimentally, and as a result, we propose on the reduction of the initial time $\Delta t$ rather than going for a graviton shielding process. The next part is the second phase of interference between the remaining two atom laser beams (denoted by dotted and wavy orange lines in Fig.(\ref{Graviton_Detector_Om_Thakur_Ma})).
\end{enumerate}
\item \textcolor{blue}{\textit{Second phase of interference}}

\noindent In the second phase of interference the primary elements are the two remaining atom laser beams which were not refocussed to create the first atom interference pattern. 
The decoherence tend to increases with time as has been observed in Fig.(\ref{Decoherence_Factor_OTM}), and as a result, our primary aim is to increase the time interval $t_f-t_i$ which indicates the increasing of the time interval between the first splitting of the atom laser using a Raman beam splitter and the second phase of interference at time $t=t_f$. It is possible to use three unique ways for the creation of the second phase of interference however the first technique is the most efficient among the three of the unique methods.

\begin{enumerate}
\item \textcolor{blue}{\textit{Far detuned optical lattice:}} Using far detuned optical lattice laser, one can increase the time interval $t_f-t_i$ in a way such that the final two atom lasers remain coherent for a significantly large amount of time. Initially, after the second phase of splitting, the two distant atom laser beams are passed through a vacuum enclosure and they are shielded from any external electromagnetic or other noise fluctuations. Using two mirror pulses, separated by a distance $l_f$, the atom laser beams are given a thrust upwards using momentum kicks so that they follow a short projectile motion as can also be seen from Fig.(\ref{Graviton_Detector_Om_Thakur_Ma}). In order to prevent further coherence loss, the projectile path can also be shielded by vacuum enclosures. The atom are then carefully loaded into higher-intensity regions of the standing waves corresponding to an optical lattice laser. Now, following the technique in \cite{NatureStableAtomLaser}, optical lattice lasers are created from the fundamental modes of a vertically oriented Fabry-Perot cavity. The matter waves following the projectile path are then loaded in the highest intensity regions of the standing waves corresponding to the optical lattice laser. In order to redirect and recombine the launched matter wave packets, two $\pi/2$ pulses are used as can be seen from Fig.(\ref{Graviton_Detector_Om_Thakur_Ma}). It is easy to observe that in total eight $\pi/2$ pulses are used to create the final two atom laser beams, which creates the second interference pattern. It is possible to observe matter wave interference even at this level but for additional clarity, we used the final two recombined wave packets or atom lasers. After the final interference, at a distance $z_f$ below the first interference phenomena, a read out beam splitter is used to analyze the final interference pattern. The spatial separation between the centre of the two mirror pulse can be made as large as $l_f\sim 500$ $\mu\text{m}$ to $2$ mm \cite{Beamsplitter}. The distance $l_f$ is maintained in a way so that the coherence between the two atom lasers is not lost. It is important to note that $l_f$ should not be so large that the coherence between the two atom laser beams is lost. Inside of the far detuned optical lattice, the launched matter wave packets can sustain coherence up to 60 to 70 seconds and after this time they are further launched which are then recombined using the $\pi/2$ pulses  \cite{NatureStableAtomLaser}. In order to further enhance the decoherence effect the distance $z_f$ between the first and second phases of interference can be set in a way such that $z_f>z_i$, and without any loss of generality, we propose to keep $z_f$ to be as high as $2$ mm. If gravitons interact with the atom lasers creating the Bose-Einstein supercondensate, where the gravitons carry an inherent squeezing of $\mathfrak{r}_k=42$, we estimate a $39\%$ loss of coherence. This $39\%$ loss of coherence in the second phase is equivalent to a $20\%$ loss of coherence compared to the first phase of coherence. Here, we restrain ourselves from considering decoherence induced by atomic ensemble dephasing, however, in an experimental implementation ,all such factors need to be considered carefully. Here, our major focus is on the gravitational bremsstrahlung induced decoherence from the Bose-Einstein supercondensate. 
\item \textcolor{blue}{\textit{Standard free-fall under the effect of gravity:}} The second way is by standard free fall of the remaining distant pair of atom laser beams. Using mirror pulses, the two freely falling atom laser beams are then recombined to create the second interference pattern at a distance $z_f$ below the first interference generation. In order to increase the decoherence effect the final time of interference $t_f$ must be increased. The only way to increase $t_f$ here is by increasing $z_f$. This does introduce some problems. One of the problems being that if the free fall distance is made unnecessarily large then it is possible that the coherence loss starts to occur as the matter wave packets cannot sustain coherence for such a large path of travel. In that case, an upperbound on the $z_f$ value is imposed to truly investigate graviton induced decoherence. Even if $z_f$ is made reasonable, in that case, $t_f$ can be so small that in this smaller $t_f-t_i$ time interval substantial amount of graviton induced decoherence does not occur which may not be tractable just by comparing the two interference patterns side by side. One of the benefits of this method lies in the fact that there is less chances of alterations by external factors is present here as no far-detuned optical lattice is used to increase the interaction time.
\item \textcolor{blue}{\textit{Doppler compensation:}} As a more physically motivated technique one can also consider a Doppler compensation technique where the second phase of interference is kept movable. As a result, $z_f$ can be varied (vertical sliding mechanism) and to properly focus the two atom laser beams, one also need to keep the distance between the final two mirror pulses $l_f$ to be movable (horizontal sliding mechanism). 
Instead of introducing the optical lattice set-up, it is more logical the proceed with the freely falling model discussed for this Doppler compensation technique. If graviton induced decoherence is introduced and it is measured, that by providing a certain blue shift to the entire moving part of the mechanism (giving an upward thrust), and there is no loss of coherence then it will indeed confirm that graviton induced decoherence has occurred. Calculating the velocity required for the sufficient Doppler blue shift, it may be possible to estimate the graviton induced decoherence via the emission of  Bremsstrahlung from the \textit{supercondensate}. It will although be very difficult to provide such a small velocity to mitigate the loss of coherence when the graviton induced decoherence occurs for a very short period of time. It is therefore more appropriate for a future direction to this experimental proposal. 
\end{enumerate}
\item{\textcolor{blue}{\textit{Comparison of the interference patterns}}}

\noindent The final step is to compare the two interference patterns and observe for coherence loss in the second pattern compared to the first interference pattern. The atom interference pattern gives a distribution for the atoms in the atom lasers. For a very distinct interference pattern one observe bright and dark zones repeatedly where the bright zone indicates accumulation of atoms whereas the dark zone indicates a less or no accumulation of particles. Now, if graviton induced decoherence occurs in the system, then one will observe a more scattered interference pattern where there will be a comparatively less number of particles in the previously mentioned bright zones whereas more accumulation of particles will occur in the dark zones making it a bit brighter. If the two patterns are compared side by side and a more even distribution is observed in the second interference pattern compared to the first one, then it will confirm the detection of gravitons as well as the Bose-Einstein supercondensate in this proposed future generation gravitational wave detector in Fig.(\ref{Graviton_Detector_Om_Thakur_Ma}). Compared to the results in Chapter(\ref{C.6.OTM}), this model is much more robust and accurate as a classical gravitational fluctuation is unable to produce similar decoherence effects even if the phonons are highly squeezed.
\end{enumerate}
One can substantially increase the viability of such an atom interferometer by implementing momentum squeezing techniques \cite{NatureEntangled} which may be able to amplify further the decoherence effects induced by gravitons.

\noindent Unlike the previous BEC based graviton detector in Chapter(\ref{C.6.OTM}), the current model does not strictly require the space based LISA observatory to confirm whether a primordial gravitational wave has arrived or not. However, the confirmation by LISA and as well as our BEC based graviton detector ensures that the decoherence is not induced by other external sources and at the same time it confirms the existence of a Bose-Einstein supercondensate.

\section{Discussion and conclusion}\label{S5}

\noindent In this work, we consider the interaction of gravitons on a maximally entangled two mode state of a Bose-Einstein condensate, where we primarily consider the case where the bosons are weakly coupled. Taking the model Lagrangian for the relativistic BEC-gravitational wave system (derived in Chapter(\ref{C.6.OTM})), we write down the action for the model system. From the model Lagrangian, we then obtain the conjugate momentum variables corresponding to the `position' variables of the BEC as well as the gravity wave part and using them to write down the Hamiltonian for the system. The model system is then quantized by raising the phase space variables corresponding to the BEC part as well as the gravitational fluctuation part to operator status and implementing suitable canonical commutation relations between the canonically conjugate operators. We take a density matrix approach and make use of the quantum mechanical ``Liouville equation of motion" (or the von-Neuman equation) to investigate the time evolution of the system density matrix under the effect of the interaction part of the Hamiltonian operator in the interaction picture. In terms of the initial system density matrix, the solution of the density matrix at the final time is then obtained. In order to truly investigate the effect of gravitons on the Bose-Einstein condensate, we trace over all gravitational field degrees of freedom from the final solution of the system density matrix which lands us at the reduced density matrix corresponding to the BEC part of the system. We start with a maximally entangled BEC state and construct the system state at time $t=t_i$ by executing an outer product with the graviton state. The system density matrix is then constructed out of this state of the system. After a finite time evolution, we observe that the initial density matrix is not separable into a BEC part and a graviton part which is a result of the noise induced by gravitons. This mixed BEC-graviton states is termed as a Bose-Einstein `supercondensate' in our work. As a result of the graviton induced bremsstrahlung from the Bose-Einstein supercondensate, we observe an overall entanglement degradation as well as a time dependent loss of coherence. We observe that exponential decay signifying the loss of coherence becomes substantially large if the graviton squeezing parameter keeps on increasing. For example, a $2\%$ coherence loss is observed from Fig.(\ref{Decoherence_Factor_OTM}) in a time interval of $2~\mu\text{s}$ when the graviton squeezing parameter has the value $\mathfrak{r}_k=42$. We also observe a $10\%$ loss in the entanglement negativity in a $4~\mu\text{s}$ time interval for a graviton squeezing of the value $\mathfrak{r}_k=42$ which confirms an entanglement loss over time. We have then investigated the time dependence of the decoherence term in the presence of phonon squeezing. We observe that for a phonon squeezing of $r_B=11.0$, an identical $10\%$ coherence loss over a $2\mu$s  time interval is observed for the value of the graviton squeezing parameter $\mathfrak{r}_k=20.0$. For a phonon squeezing of $r_B=11.2$, an astonishing $20\%$ coherence loss is observed in the same time interval where the graviton squeezing is kept fixed to the previous value. This is a very significant result as it provides one with a true experimentally controllable parameter. Now, even if the graviton squeezing from the primordial gravitational waves are not very high, one can also aim to detect graviton signatures just by tuning the phonon squeezing parameter. Making use of the analytical results of our work, we finally implement a BEC based graviton detector where two Mach-Zehnder type interferometers are considered. For generating the interference, atom lasers, generated from a continuous-wave Bose-Einstein condensates, are used and then four coherent maximally entangled atom laser beams are generated. These beams are then recombined to create two interference patterns such that there is substantial time interval between the two interferences. A schematic diagram of the experimental proposal is depicted in Fig.(\ref{Graviton_Detector_Om_Thakur_Ma}). The experiment stands on the principle that if the time between the two interference patterns is substantially large then due to the graviton induced noise, a significant loss of coherence will be observed in the final interference pattern (located at the bottom of the experimental set-up), in comparison to the first one. The fundamental reason behind such an observation is that the bottom most interference has higher time to interact with the gravitons leading to more loss of coherence via the emission of bremsstrahlung compared to the first one. Making use of the methodology in \cite{NatureStableAtomLaser}, one can increase the time interval to $60-70$ seconds by the use of far-detuned optical lattice laser. The benefit of the experimental model proposed in our set-up is that all the individual experimental components exists in the most advanced experimental laboratories of the world and they just need to be carefully combined to create our experimental model. We hope that if the model is organized and graviton exists, then it will be a matter of a decade for detecting graviton signatures using our proposed experimental model.

\chapter{Summary and conclusion}\label{C.8.OTM}
In this thesis, we have considered simple model systems where the background metric consists of small gravitational fluctuations which are quantized. In the first part of the thesis, ``\textit{The fundamental minimal length in quantum gravity}", we have primarily made use of the path integral quantization of the small gravitational fluctuations and probed into several fundamental aspects of a linearized quantum gravity theory whereas in the next part ``Quantum gravity phenomenology", we have made use of the canonical quantization technique for quantizing the small gravitational fluctuations over the background spacetime and investigated several important and unique phenomenological aspects of linearized quantum gravity model.

\noindent In Chapter(\ref{C.3.OTM}) of part I of the thesis, titled ``\textit{The minimal length scale correction in linearized quantum gravity}", we start with the basic model of a two-particle systems where the background is considered to be flat with small gravitational fluctuations upon it. In any existing models of quantum gravity, there is an existence of a fundamental minimal length in nature which is incorporated into a system by modifying the Heisenberg uncertainty principle and this modified uncertainty relation is also termed as the generalized uncertainty principle. In order to incorporate the effects of a fundamental minimal length, we implement this generalized uncertainty principle in the detector degrees of freedom where the arm length of the detector is represented by the geodesic separation between the two massive particles. To truly model the arm of an interferometric detector, one needs to consider the massive particle following a time-like trajectory as the origin, and define a Fermi-Normal coordinate system where the geodesic separation is defined in this coordinate system. Making use of a path integral approach the transition probability for the system to go from an initial state to some final state is calculated where the final graviton state is summed over. Executing the path integral over the detector momentum, and with a substantial amount of analytical simplification, we arrive at the transition probability for the detector-graviton model system in the generalized uncertainty principle framework. The influence of the gravitons on the detector part is captured truly by the Feynman-Vernon influence functional and the aim of the analysis boils down to the minute evaluation of the Feynman-Vernon influence functional when all of the graviton mode frequencies are being considered for different initial states of the gravitons. Using the Feynman-Vernon trick, one can write down the Feynman-Vernon influence functional in a way such that it involves effects from stochastic terms which are defined as graviton induced noise terms \cite{QGravD}. The reason behind calling them the noise terms lies in the fact that the one-point correlator vanishes whereas the two-point correlator gives a real number. Extremizing the action with respect to the geodesic separation, we arrive at the quantum gravity modified geodesic deviation equation in a generalized uncertainty principle framework. These differential equations are Langevin-like in nature as they include effects of the quantum gravitational noise fluctuations, and we find out that they include terms corresponding to the generalized uncertainty principle where the GUP modified elements get coupled to the quantum gravitational noise-fluctuations. From the equation of motion, we obtain the perturbative analytical expression for the geodesic separation as well and from there obtain the standard deviation in the geodesic separation which now depends on time. We observe that the GUP modified standard deviation first decrease then increases than the base value of the standard deviation and eventually asymptotes towards the base value of the standard deviation. This analysis is very important as well as striking, as it indicates towards a residual increase in the uncertainty of the standard deviation which indicates towards a lingering effect of the graviton noise fluctuation and can be considered as a graviton induced ``memory effect". In case of the gravitons being in a squeezed state, we find out that such effects may be detectable in future generation of space base gravitational wave observatories indicating towards the existence of the generalized uncertainty relation as well as a ``memory"-like effect induced by the noise of gravitons.

\noindent In the next chapter of part I, titled ``\textit{A quantum gravitational uncertainty relation}", we ask a more fundamental question. We investigate whether the uncertainty principle get modified while considering quantization of linearized gravity. We use the model proposed in \cite{ChawlaParikh}, which considers a particle freely-falling under the effect of the gravitational field of the Earth where it is interacting with incoming gravitational fluctuations. This entire phenomena is being observed by an observer sitting on the surface of the Earth. To incorporate quantum gravitational effects in the model system, the gravitational fluctuations are quantized using the path integral quantization technique. Extremizing the final transition probability of the particle-graviton detector system with respect to the distance of the particle from the surface of Earth, one arrives at the quantum gravity modified Newton's equation of motion. From the above equation, one can obtain an analytical expression for the height of the particle from the surface of the Earth which gets infused by the graviton noise fluctuations. It is now possible to get the standard deviation of the position as well as the momentum for the particle at the time of touching the surface of the Earth. The uncertainty product is then investigated at the time of the particle touching the ground as it maximizes the interaction time between the graviton and the freely-falling particle. Multiplying the uncertainties in the position as well as the momentum, we arrive at the quantum gravity modified uncertainty product. In a quantum gravity set-up, the minimum distance is always grater than or equal to the Planck length. This inequality helps us to write down a quantum gravitational uncertainty relation by combining the lower bound from the Heisenberg uncertainty principle with the quantum gravitational inequality obtained for the uncertainty product. Expressing the uncertainty product in terms of the variance of the position and implementing the Planckian limits for the mass as well as the height of the particle, we also have obtained an uncertainty upper bound. We calculate this uncertainty relation for the graviton initially being in a vacuum, a squeezed vacuum, and finally in a thermal state, and obtained the exact same closed form of the uncertainty relation. This indicates that the uncertainty relation obtained in our analysis is indeed universal. The quantum gravity correction term in the lower bound is proportional to the variance in the momentum whereas for the upper bound, it is proportional to the variance in the position. For the coefficient of the variance of the momentum it depends on the Planck's constant, Newton's gravitational constant, and the speed of light. Hence, the correction obtained here are true quantum gravitational correction. In the Planck mass limit, we arrive at the proposed analytical form of the generalized uncertainty principle. This analysis not only gives a close quantum gravity modified uncertainty relation but also gives a derivation of the generalized uncertainty principle \cite{KempfManganoMann}. This is indeed a very fundamental result revealing that even for a linearized quantum theory of gravity, the Heisenberg uncertainty principle gets modified.

\noindent With the investigations of the intricate fundamental aspects of a linearized quantum theory of gravity, we put our focus on the phenomenological aspects of a linearized quantum gravity model in part II of the thesis. In Chapter(\ref{C.5.OTM}) titles, ``\textit{Spontaneous emission of gravitons as a signature of quantum gravity}", we consider the two-particle model system discussed in Chapter(\ref{C.3.OTM}) when the entire model system is placed inside of a harmonic trap potential. This model mimics the structure of a resonant bar detector or an interferometric detector placed inside a harmonic trap potential. The gravitational fluctuations as well as the detector phase space variables are now quantized where we make use of a canonical quantization technique by raising the phase space variables to operator status. The next step towards quantization is the implementation of suitable canonical commutation relation between the graviton as well as the detector position operator and their canonically conjugate momentum operators. We now investigate a simple scenario where the initial tensor product state of the detector-graviton model system goes to some final state as a result of graviton detector interaction. We find out that the detector always makes a jump of two states (excitation or de-excitation) while absorbing or emitting a single graviton which can be easily understood by inspecting the analytical form of the transition probability. If one considers the incoming gravitational wave as a combination of a finite number of gravitons, then making use of the energy flux relation for a gravitational wave, it is possible to observe that the transition probability in the resonant absorption case in this quantum gravitational set-up is identical to the resonant absorption case when the gravitational wave is treated classically. However, the difference occurs in the emission scenario. We observe that even if there are no gravitons in the initial state, the detector drops to its ground state by spontaneously emitting a single graviton. This spontaneous emission of gravitons makes the transition probability different from the semi-classical case where the gravitational fluctuations are treated classically. If it is possible to create a coherent resonant bar source such that all such spontaneous emissions occur simultaneously, then one can observe a gravitational ``fluorescence"-like effect where the gravitons emit in all possible directions creating a very faint background gravitational fluctuation. Building such a detector where the resonance condition is exactly satisfied is very difficult making the spontaneous emission scenario to be quite difficult in an advanced experimental set-up.  However, a detection of such an effect is an absolute evidence of the existence of gravitons in nature.

\noindent Based on the difficulties of detecting graviton signatures, we consider a completely new model where a Bose-Einstein condensate at absolute zero temperature is modelled as a graviton detector. In Chapter(\ref{C.6.OTM}) titled, ``\textit{Bose-Einstein condensate as a probe to detect quantum nature of gravity}", we investigate the sensitivity of a relativistic Bose-Einstein condensate towards incoming graviton fluctuations. One can here consider the Lagrangian density for a complex scalar field theory with a quadratic self-interaction term where the system is placed in a curved background. Extremizing the action with respect to the amplitude of the complex scalar field, we obtain the Lagrangian density for a Bose-Einstein condensate from which one can write down the action for a relativistic Bose-Einstein condensate in a curved spacetime and combining this action with the Einstein-Hilbert action, one arrives at the action for the model detector system. To incorporate quantum gravity effects in the model system, the gravitational fluctuations are decomposed into its individual Fourier modes. Now, the phase part of the complex scalar field can be expressed as a combination of a purely time-dependent part and the pseudo-Goldstone bosons. Extremizing the action with respect to the time-dependent part of the Goldstone bosons as well as the gravitational mode functions, we arrive at two dynamical equations corresponding to the time-dependent part of the pseudo-Goldstone bosons as well as the gravitational mode functions.
The gravitational fluctuations are then quantized by raising the individual Fourier mode functions to operator status and implementing canonical commutation relation between the mode operators and their conjugate operators. 
We now, analytically solve the equation of motion corresponding to the time-dependent part of the pseudo-Goldstone bosons and from the analytical solution, we obtain the graviton-noise infused analytical form of the time-dependent part of the Goldstone bosons. From the solution it is now possible to identify the  Bogoliubov coefficients where the coefficients get infused by graviton noise fluctuations. This is a very important result. From this graviton noise infused Bogoliubov coefficients, one can construct ``symplectic" matrices that transforms the covariant matrix of a single mode of the Bose-Einstein condensate and produces the gravitational fluctuation infused covariant matrix of the Bose-Einstein condensate. Making use of quantum metrological techniques, one can obtain the quantum Fisher information out of the elements of the transformed covariance matrix which now get infused by graviton noise fluctuations. This new quantum Fisher information is termed as the quantum gravitational Fisher information. From the analytical expression of the quantum gravitational Fisher information, it is easy to observe that the expectation value of this information theoretic quantity with respect to initial graviton states is always higher than the standard quantum Fisher information for a classical gravitational wave. This observation helps us to write down a quantum gravity modified Cram\'{e}r-Rao bound from which we can obtain the minimum value of in the measurement of the standard deviation in the gravitational wave amplitude. The standard deviation in the amplitude of the gravitational wave becomes finite for gravitons with high squeezing even for very low measurement time which indicates that if a BEC picks up gravitational fluctuations for a very small measurement time, gravitons indeed have been detected. The standard deviation in the amplitude divided by the square root of the gravitational wave frequency gives the sensitivity formula of the Bose-Einstein condensate towards gravitational fluctuations. Summing over all phonon mode frequencies for a noise term with a Gaussian decay term we observe that for high graviton squeezing, the condensate is sensitive towards graviton fluctuations where the frequency lies in the range of a primordial gravitational wave. In case of a classical gravitational wave however, for similar phonon squeezing, the condensate is not at all sensitive towards incoming graviton fluctuations. This implies that in future generation of space based gravitational wave detector, if a primordial gravitational wave fluctuation is captured then a simultaneous detection by a Bose-Einstein condensate based detector with minimal squeezing of the phonons ensures the detection of gravitons by the condensate. One can also reduce the requirement of very high graviton squeezing by tuning the phonon mode squeezing accordingly.

\noindent Based on the results in Chapter(\ref{C.6.OTM}), we propose an advanced Bose-Einstein condensate based graviton detector in Chapter(\ref{C.7.OTM}) titled, ``\textit{Gravitational Bremsstrahlung from a Bose-Einstein condensate}". We now quantize the condensate part of the model system by raising the phonon wave vectors to operator status and implementing canonical commutation relation between the phase space canonically conjugate operators. We use the von-Neumann equation of motion involving the density matrix of the Bose-Einstein condensate (BEC)-graviton model detector system. Solving the von-Neumann equation of motion and tracing out the graviton degrees of freedom, we arrive at the reduced density matrix corresponding to the BEC part of the model system which now carries the influence of the graviton noise fluctuations. We observe that the off-diagonal terms of the density matrix get modified by a time-dependent term which includes a pure phase part whereas the other part is a pure time dependent decay term dependent on the two-point graviton noise-noise correlator. This decay term introduces decoherence between two superposing BEC momentum states. Instead of superposition states, we consider a maximally entangled BEC momentum state and observe graviton induced decoherence, via the emission of graviton induced bremsstrahlung. Based on this analytical outcome, we propose a two-phase atom-interferometer for picking up the signature of gravitons. It is also important to note that the initial tensor product state of the graviton and the entangled BEC state gets mixed as well as entangled over time due to graviton induced decoherence which is termed as a Bose-Einstein ``supercondensate". The experimental model proposed in Chapter(\ref{C.7.OTM}) uses existing experimental techniques where a two-phase atom-interferometer is used which uses atom-lasers generated from a continuous wave Bose-Einstein condensate created inside of a harmonic trap potential. The atom-lasers plays the part of the matter waves which create a clear interference pattern if the two matter waves are coherent to each other. Between the two interference patterns, there is a finite time interval which ensures a lengthened graviton interaction time with the atom-lasers after the first phase of interference. Now, if graviton induces decoherence then the second interference pattern will be more distributed than the first interference patter leading to a detection of gravitons as well as a Bose-Einstein supercondensate. We hope that if the experimental model is successfully implemented, it will be able to detect graviton signatures within a matter of a decades time. This work concludes the results in our thesis.

\noindent Hence, in this thesis, we start with a model detector system where the matter part obeys the generalized uncertainty principle and find out that the standard deviation in position holds the signature of graviton noise fluctuation for a long period of time. Using similar model we then find out that a quantum gravity modified model indeed implements a new uncertainty relation where the lower bound indicates the existence of a minimal length in nature and upper bound indicates the existence of a maximum value to the upper-bound of the variance in momentum. The first part of the thesis, therefore, points at fundamental aspects of quantum gravity. In the second part, we find out significant features of quantum gravitational model which is implementable in an experimental scenario and based on our findings, we propose a very advance Bose-Einstein condensate based graviton detector, and we hope that these new methodologies for detection of graviton signatures, if implemented, will be sufficient to prove the existence of gravitons in nature. One can now investigate several other aspects based on the findings of this thesis. Some of the fundamental analyses would be investigate same models in the presence of a dynamical electromagnetic field where there is presence of photon-graviton interaction. Using a path integral approach to investigate the same scenario would be a connected work as well. We shall also investigate the implications of graviton detection while using a BEC based detector in an acceleration radiation scenario. These analyses shall ensure significant theoretical as well experimental prospects of our analytical model and extend the works presented in this thesis.
%\noindent Finally, we comment on some of the future directions of this work. One of the works being analytical observation of a BEC-graviton interaction model in the presence of an electromagnetic field. We also aim to use a path integral technique for modelling this BEC-graviton detector model. 

\pagebreak
\begin{center}
\includegraphics[scale=0.65]{OTM.pdf}\\
\vspace{6cm}
\includegraphics[scale=0.75]{Om_Thakur_Ma.pdf}
\end{center}
\end{document}